\documentclass[arguments]{aastex63}

\usepackage{amsmath}
\submitjournal{The Astrophysical Journal}

\shorttitle{MSPES. VI. Pulsar Polarization}
\shortauthors{Mitra, Melikidze \& Basu}

\begin{document}

\title{Meterwavelength Single-pulse Polarimetric Emission Survey. VI. Towards understanding the Phenomenon of Pulsar Polarization in Partially Screened Vacuum Gap model.}

\author[0000-0002-9142-9835]{Dipanjan Mitra}
\affiliation{National Centre for Radio Astrophysics, Tata Institute of Fundamental Research, Pune 411007, India.}
\affiliation{Janusz Gil Institute of Astronomy, University of Zielona G\'ora, ul. Szafrana 2, 65-516 Zielona G\'ora, Poland.}

\author[0000-0003-1879-1659]{George I. Melikidze}
\affiliation{Janusz Gil Institute of Astronomy, University of Zielona G\'ora, ul. Szafrana 2, 65-516 Zielona G\'ora, Poland.}
\affiliation{Evgeni Kharadze Georgian National Astrophysical Observatory, 0301 Abastumani, Georgia.}

\author[0000-0003-1824-4487]{Rahul Basu}
\affiliation{Janusz Gil Institute of Astronomy, University of Zielona G\'ora, ul. Szafrana 2, 65-516 Zielona G\'ora, Poland.}

\begin{abstract}
We have observed 123 pulsars with periods longer than 0.1 seconds in the 
Meterwavelength Single-pulse Polarimetric Emission Survey. In this work a 
detailed study of polarization behaviour of these pulsars have been carried 
out. We were able to fit the rotating vector model to the polarization position
angle sweeps in 68 pulsars, and in 34 pulsars the emission heights could be
measured. In all cases the radio emission was constrained to arise below 10\% 
of the light cylinder radius. In pulsars with low spindown energy loss, 
$\dot{E}<10^{34}$ ergs s$^{-1}$, we found the mean fractional linear 
polarization of the individual times samples in single pulses to be around 0.57
(57\%) which is significantly larger than the fractional linear polarization of
0.29 (29\%) obtained from the average profiles. On the other hand the mean 
fractional circular polarization of the individual time samples in single 
pulses is around 0.08 (8\%), similar to the measurements from the average 
profiles. To explain the observed polarization features, we invoke the 
partially screened vacuum gap model of pulsars, where dense spark associated 
plasma clouds exist with high pair plasma multiplicity, with significant 
decrease of density in the regions between the clouds, that are dominated by 
iron ions. The coherent radio emission is excited by curvature radiation from 
charge bunches in these dense plasma clouds and escape as linearly polarized 
waves near cloud boundaries. We suggest that the circular polarization arises 
due to propagation of waves in the low pair multiplicity, ion dominated 
inter-cloud regions.
\end{abstract}


\section{Introduction}
\noindent
The pulsar magnetosphere is composed of dense electron-positron pair plasma 
embedded in strong magnetic fields, and can be divided into regions of open and
closed magnetic field lines (\citealt{1969ApJ...157..869G}). A mix of relativistically charged particles, 
comprising of pair plasma as well as beams of positrons and iron ions, stream
outward along the open magnetic field lines (\citealt{1975ApJ...196...51R,CR77}). The coherent radio emission in 
pulsars arises due to growth of plasma instabilities in the relativistically 
streaming plasma \citep[see e.g.][]{1975ARA&A..13..511G}. In most cases the 
radio emission from pulsars is seen to exhibit high levels of linear 
polarization as well as significant circular polarization (e.g. \citealt{1975ApJ...196...83M,1998MNRAS.301..235G}). 
The nature of 
polarization is determined by the processes by which the radio emission is 
generated and the subsequent propagation effects in the pulsar magnetosphere. A 
major theoretical challenge is to understand these physical mechanisms that 
give rise to the observed polarization features \citep[see e.g.][for a 
review]{1995JApA...16..137M}.

Radio polarization behaviour of pulsars can be quantified using two different 
schemes of measurements : the average profiles (hereafter AP), where the Stokes
parameters are obtained by averaging a large number of pulses; and single 
pulses (hereafter SP) where the Stokes parameters are obtained collating from
individual bright pulses. The AP studies are sufficient for understanding 
several aspects of polarization behaviour in normal pulsars ($P >$ 50 msec). 
The most prominent behaviour is the polarization position angle (PPA) of the 
linear polarization showing a characteristic S-shaped traverse across the pulse
window and can be interpreted in terms of the rotating vector model (RVM) which
was first proposed by \citet{1969ApL.....3..225R}. According to the RVM the 
swing of the PPA arises when the observer's line of sight (LOS) cuts across the 
dipolar magnetic field line plane and is determined by the LOS geometry, 
characterised by the angle between the rotation axis and magnetic axis, 
$\alpha$, and the angle between the rotation axis and the LOS during closest 
approach, $\beta$. Thus finding suitable RVM fits to the observed the PPA 
traverse in principle should constrain the LOS geometry of the pulsar 
\citep{2001ApJ...553..341E,1997A&A...324..981V}. However, in all practical 
purposes only the shape of the PPA traverse and it's steepest gradient point 
(hereafter SGP) can be used to the extent of distinguishing between central and
tangential cuts across the pulsar emission beams and not the detailed geometry, 
since the angles $\alpha$ and $\beta$ obtained from the RVM fits are found to 
be highly correlated. 

AP measurement of linear polarization PPAs in conjunction with total intensity 
profiles can also be used to determine the radio emission heights, utilizing 
the relative shifts that arise between them due to aberration-retardation 
effects \citep[hereafter A/R,][]{1991ApJ...370..643B}. The A/R method puts 
tight constraints on the emission height to be below 10\% of the light cylinder 
distance. The A/R effect have been 
studied recently for a large sample of pulsars around 1 GHz (\citealt{2023MNRAS.520.4801J}), 
although for lower and higher than 1 GHz frequency bands 
relatively small sample of pulsars have been studied (see \cite{2017JApA...38...52M} 
and references therein). APs have also 
been used to estimate the nature of emission beam shape at radio frequencies. 
In most works the beam is considered to have a core-cone structure, with a 
central core and nested conal components, and follows a distinct radius to 
frequency mapping where lower frequencies are found to arise at progressively 
higher heights from the neutron star surface \citep[see 
e.g.][]{1983ApJ...274..333R,1993ApJ...405..285R,1999A&A...346..906M,
2002ApJ...577..322M}. Using relatively fewer pulsars 
\citet{1981A&A...100..107M} showed that depolarization of AP is higher for 
older, longer period pulsars compared to shorter period, younger pulsars. 
Subsequent works on larger samples have found that the degree of linear 
polarization ($\%L$) correlates with the spindown energy loss ($\dot{E}$), with
low $\dot{E}$ pulsars having $\%L$ of about 20\% and high $\dot{E}$ pulsars 
about 80\% \citep{2008MNRAS.391.1210W,2018MNRAS.474.4629J,2016ApJ...833...28M,
2019MNRAS.489.1543O,2022arXiv221111849P}. A change in the beam shape properties
between high and low $\dot{E}$ pulsars has also been suggested by 
\cite{2007MNRAS.380.1678K}.

The SP studies on the other hand reveal dynamic emission features like mode 
changing, subpulse drifting, nulling, microstructures, periodic amplitude 
modulation, etc., \citep{1986ApJ...301..901R,2006A&A...445..243W,
2015ApJ...806..236M,2016ApJ...833...29B,2020ApJ...889..133B,
2023arXiv230104067S}. Mode changing, nulling and subpulse drifting are 
usually associated with low energetic pulsars, while periodic amplitude 
modulations are seen in more energetic systems \citep{2020ApJ...889..133B}. 
For example, the SP phenomenon of subpulse drifting shows a prominent 
dependence on the spindown energy loss and is only seen in pulsars with 
$\dot{E} < 5\times10^{32}$ ergs~s$^{-1}$ \citep{2016ApJ...833...29B,
2019MNRAS.482.3757B}. The PPA distribution obtained from SP in low $\dot{E}$ 
pulsars often show the presence of orthogonal polarization modes (OPM) where 
two orthogonal RVM tracks, separated by 90$\degr$~in phase, can be identified. 
In some cases they are highly disordered and do not resemble a S-shaped curve 
corresponding to RVM. The high $\dot{E}$ pulsars on the hand generally have a 
single PPA track that can mostly be identified with the RVM 
\citep{2016ApJ...833...28M}.

The preceding discussion highlights that a distinct trend has emerged from 
recent works, involving both AP and SP measurements, suggesting a change in 
radio emission properties as a function of $\dot{E}$. At the higher $\dot{E}$ 
range certain pulsar radio emission properties appear to be quite different 
from the lower energetic population. In this work we further explore whether 
the SP fractional linear polarization changes with $\dot{E}$ in a manner 
similar to their behaviour in APs. Either the depolarization of APs in low 
$\dot{E}$ pulsars arises due to all SPs having intrinsically low fractional 
linear polarization; or it is possible that the individual SPs have high 
fractional linear polarization that reduces during the averaging process. 
Understanding such trends in the general pulsar population requires 
observations of a large number of pulsars with high sensitivity detection of 
polarized SPs.
 
A large number of studies exist in the literature that have been devoted 
towards understanding the polarization properties in pulsars. The radio 
emission is generated in the inner magnetosphere where a dense pair plasma is 
present in very high magnetic fields. In such systems it can be shown that 
there exist two distinct plasma modes that are linearly polarized, the 
extraordinary (X) mode, with the electric vector being perpendicular to the 
propagation vector and the local magnetic field, and ordinary (O) mode, with
the electric vector lying in the plane of the propagation vector and the local 
magnetic field, respectively \citep[see ][]{1985FizPl..11..531V,
1986FizPl..12.1233L,1986ApJ...302..120A}. Generally, the observed OPMs in 
pulsars are associated with the naturally occurring orthogonal wave modes that 
are excited by a suitable emission mechanism arising in the pair plasma. The 
two modes split in the plasma during propagation due to their different 
refractive indices, and travel with different phase velocities, and eventually 
emerge as OPMs \citep{1979AuJPh..32...61M,2014ApJ...794..105M}. The circular 
polarization arises either due to an intrinsic emission process or as a result 
of propagation effects \citep{1982PASA....4..365A,1991MNRAS.253..377K,
2010MNRAS.403..569W}. Observations show that individual OPMs are elliptically 
polarized, even when very high level of linear polarization is seen in SPs, and
strongly favour the propagation effects being the underlying cause. However, 
the theoretical premise where both the emission mechanism and the propagation 
effect is addressed in a self consistent manner is still missing.

The Meterwavelength single-pulse polarimetric emission survey (MSPES) was
conducted using the Giant Meterwave Radio Telescope (GMRT) 
primarily to increase the available sample for AP/SP studies 
\citep[hereafter PaperI,][]{2016ApJ...833...28M}. 
The GMRT consists of a Y-shaped array of 30 antennas of
45 m diameter, with 14 antennas located within a central one
square kilometer area and the remaining 16 antennas placed
along the three arms, which have a maximum distance of 27
km (\citealt{1991CuSc...60...95S}). The MSPES observations were carried out at two different
frequencies centered around 325 and 610 MHz with a
bandwidth of 16.67 MHz at each frequency (see PaperI for details). 
These observations 
contributed towards detailed characterisation of the subpulse drifting 
behaviour in pulsars \citep{2016ApJ...833...29B,2019MNRAS.482.3757B}, and were 
instrumental in demonstrating that subpulse drifting arises from a partially 
screened vacuum gap \citep[PSG,][]{2003A&A...407..315G,2020MNRAS.496..465B,
2022ApJ...936...35B,2023arXiv230312229B}. The highly polarized single pulses 
\citep{2009ApJ...696L.141M,2023MNRAS.tmpL..22M} and variation in pulsar 
spectral nature across the emission beam \citep{2021ApJ...917...48B,
2022ApJ...927..208B} have provided compelling evidence that the coherent radio 
emission is excited by curvature radiation from charge bunches. In this paper 
we use the MSPES observations to measure the radio emission heights in pulsars
using APs, and further compare the SP polarization properties with AP. In 
section \ref{sec2}, the RVM fits to the PPA in all relevant cases have been 
shown and the radio emission heights have been found using A/R shifts. In 
section \ref{sec3} we have shown the variation of the SP polarization 
properties with $\dot{E}$. In Section~\ref{sec4} we use the PSG model and 
coherent curvature radiation mechanism to understand the origin of linear and 
circular polarization properties of normal pulsars, followed by the conclusion 
presented in section~\ref{sec5}.

\section{RVM fits and Emission heights using A/R method} \label{sec2}

\begin{figure}
\gridline{\fig{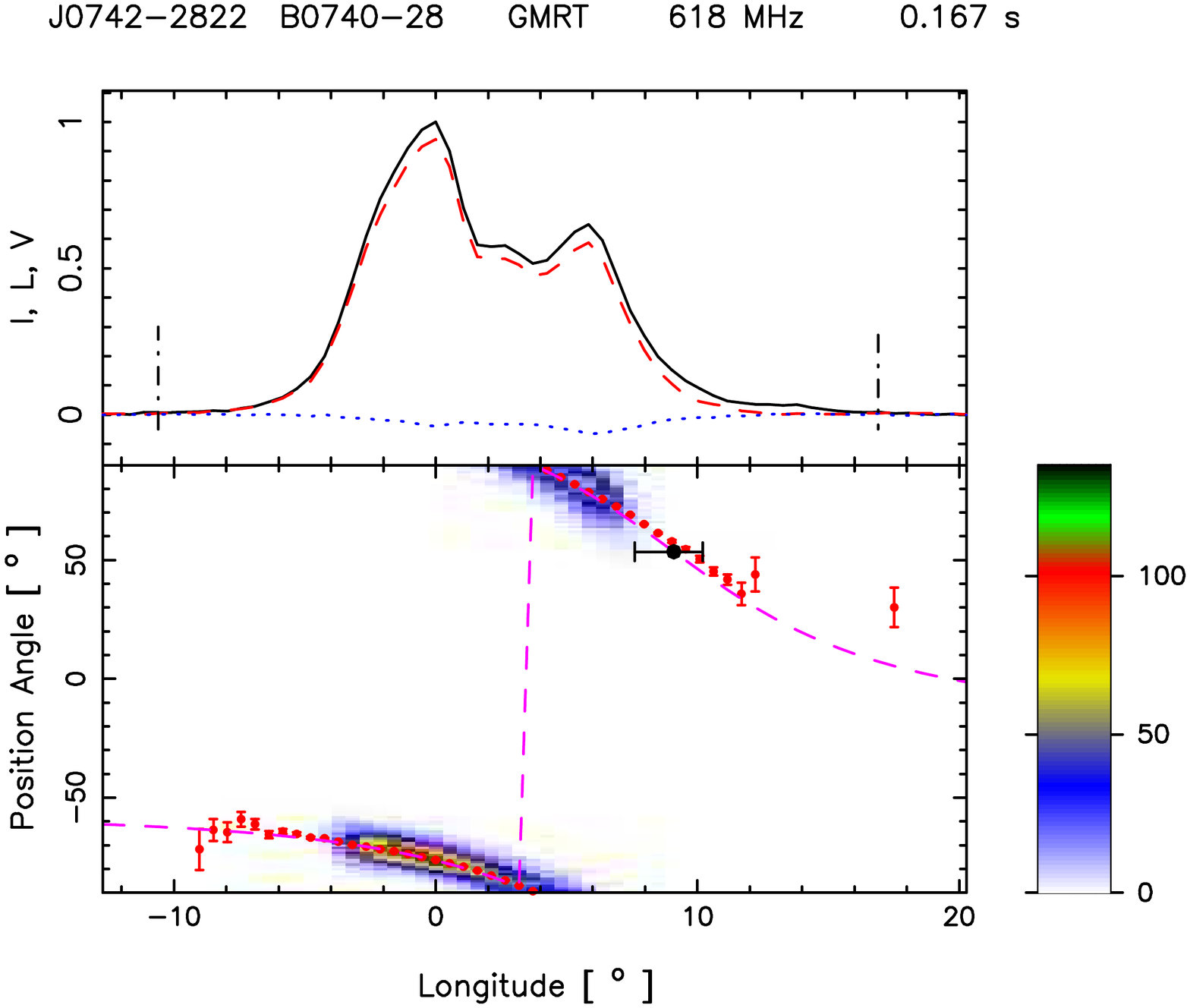}{0.32\textwidth}{}
	  \fig{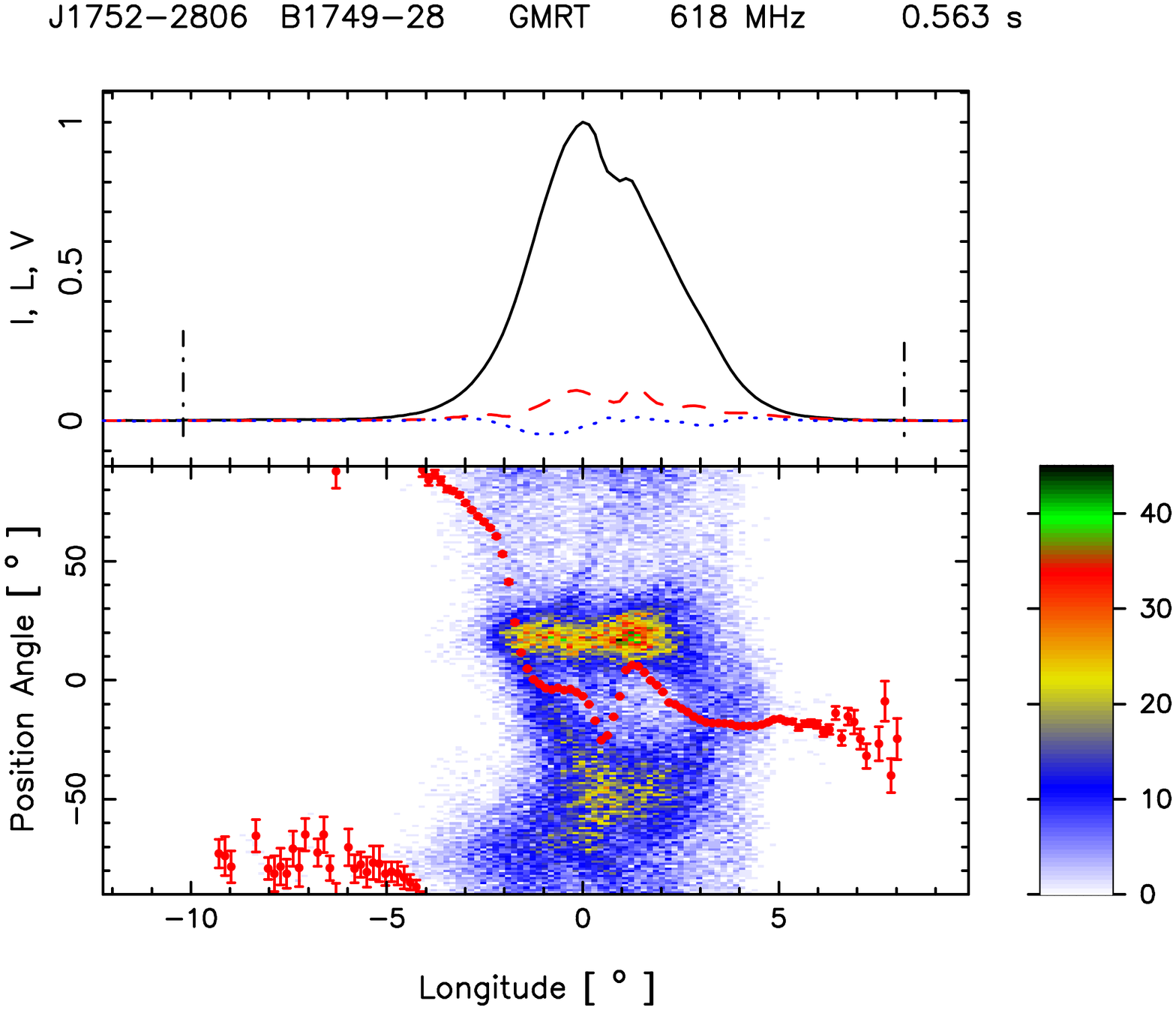}{0.32\textwidth}{}
          \fig{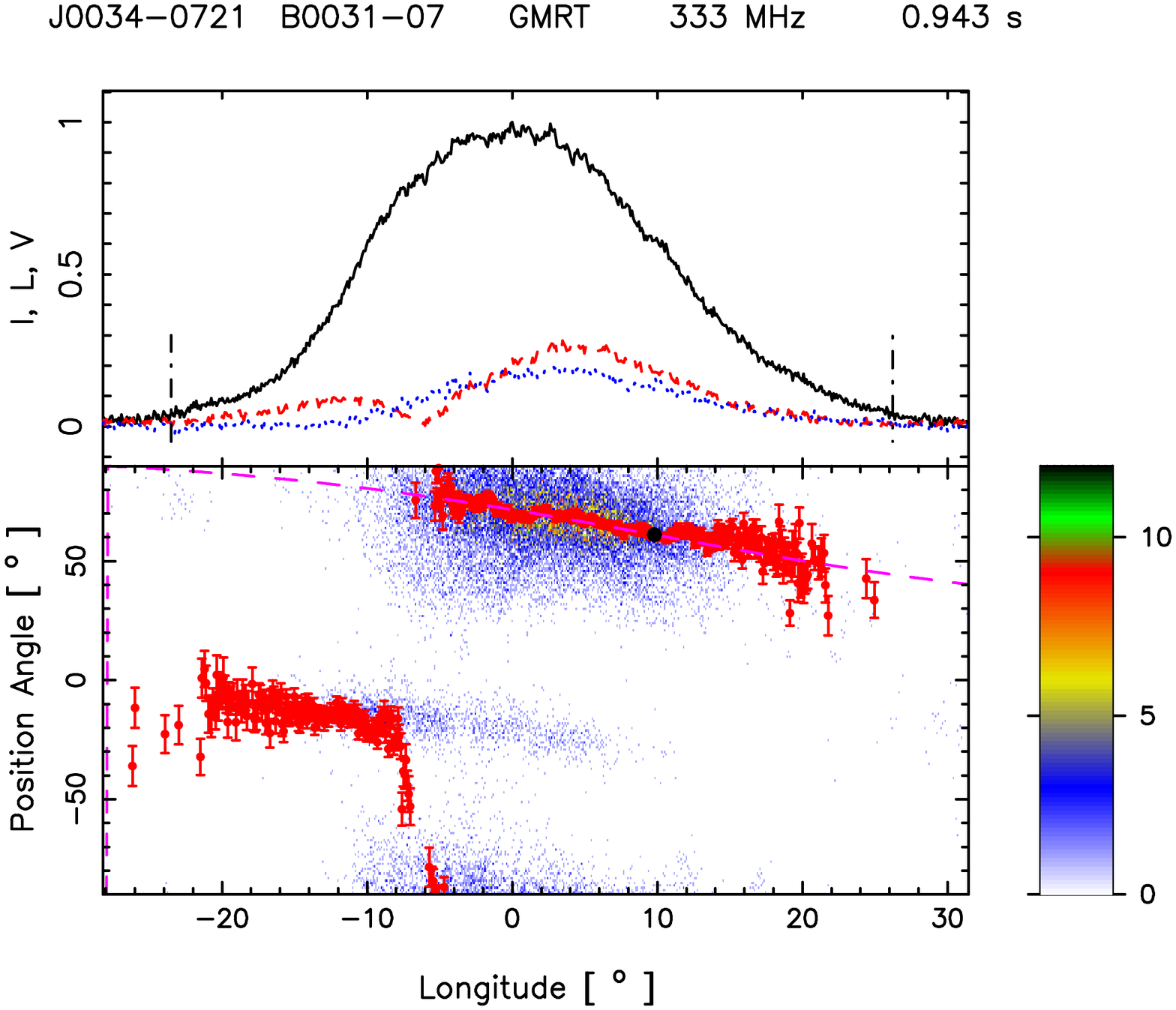}{0.32\textwidth}{}
         }
\gridline{\leftfig{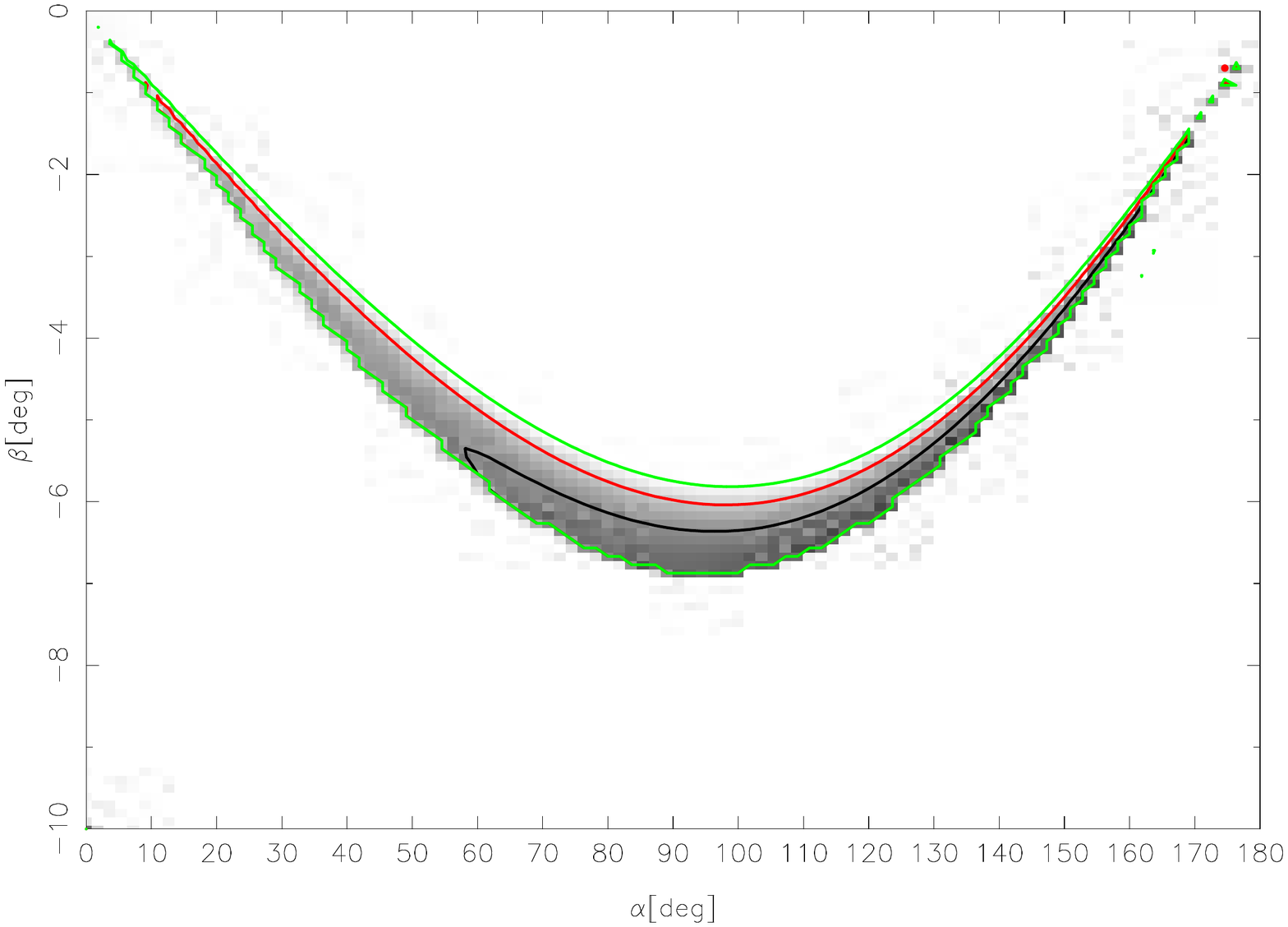}{0.32\textwidth}{}
          \rightfig{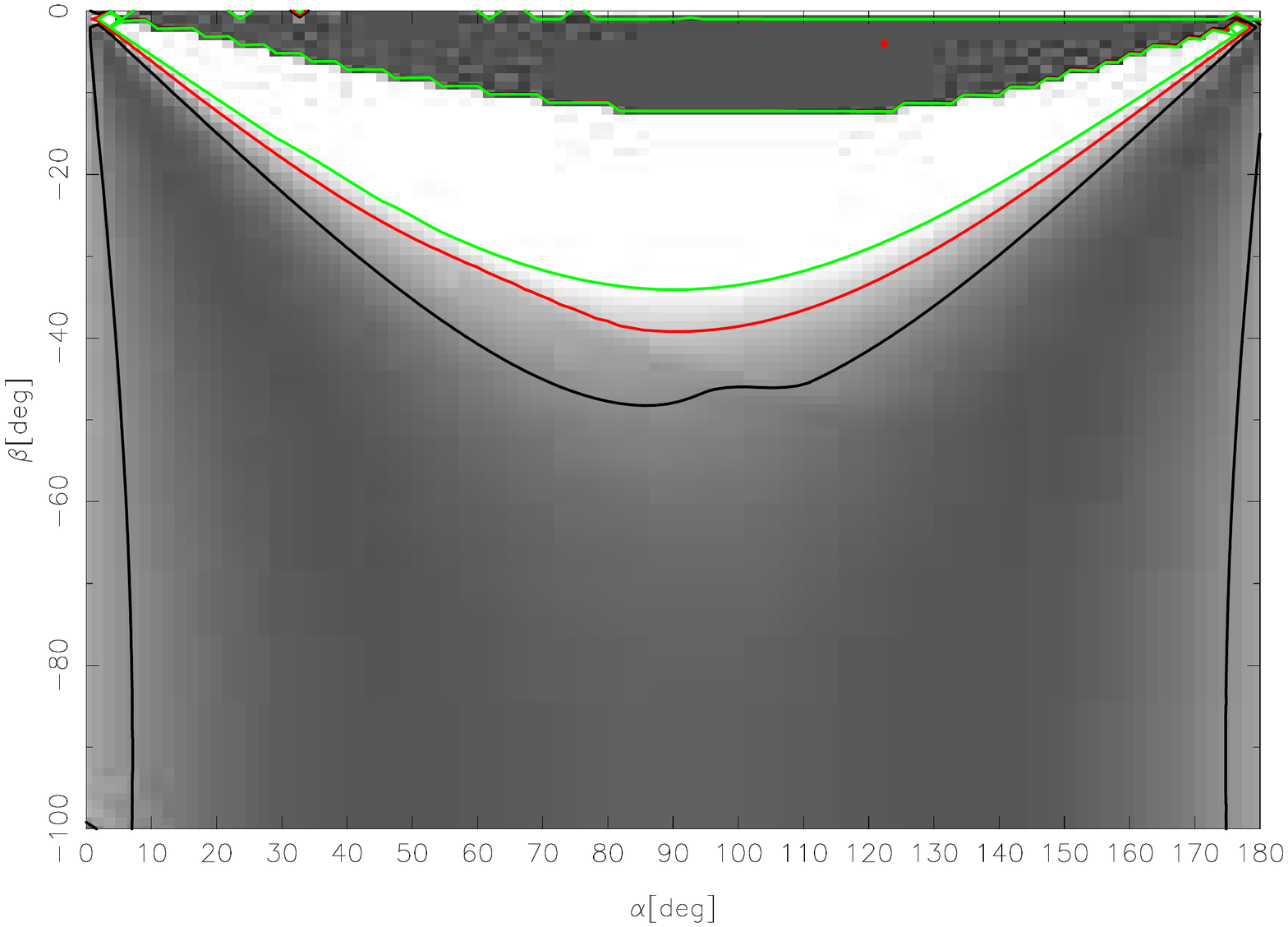}{0.32\textwidth}{}
	 }
\gridline{\fig{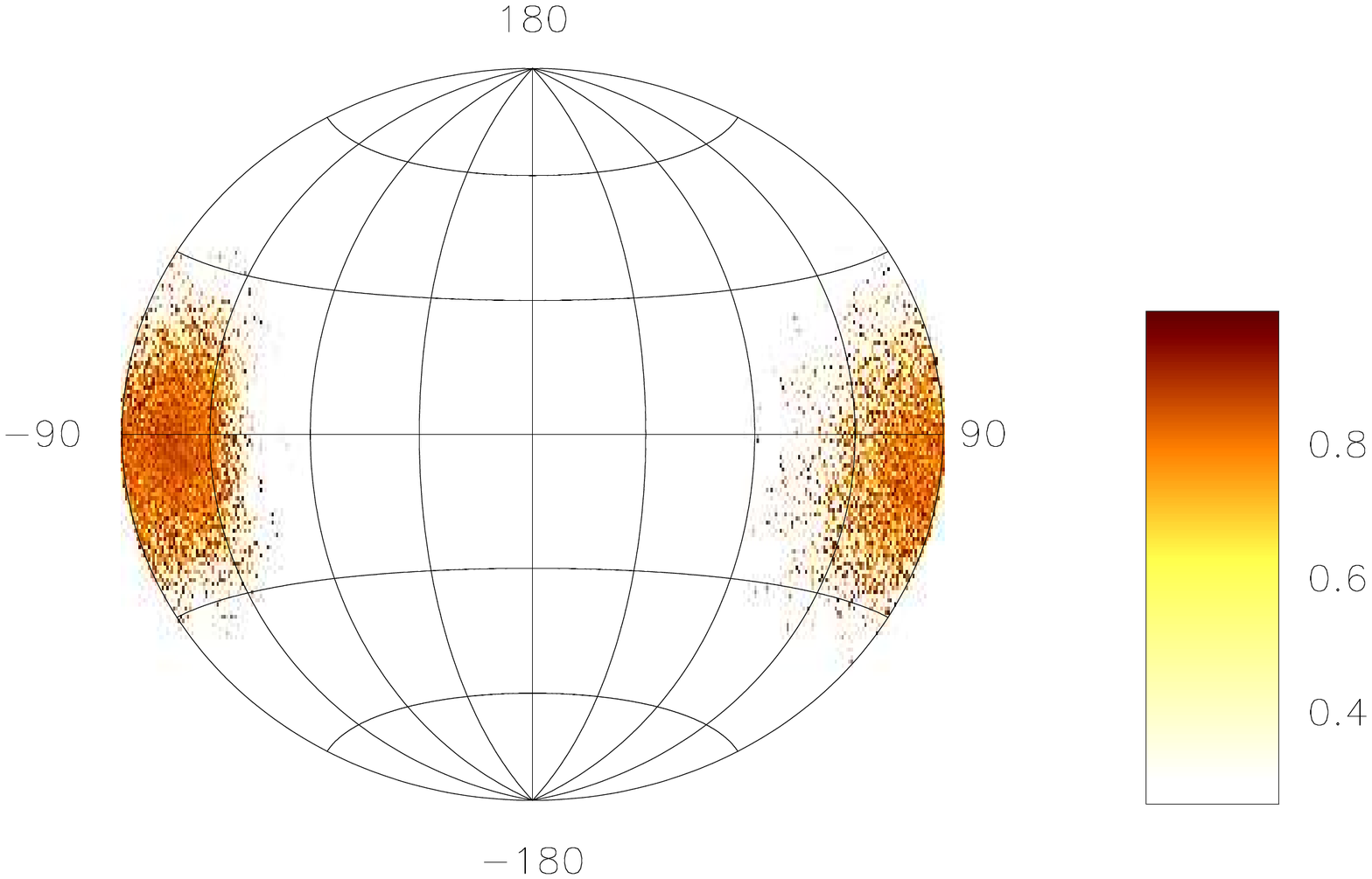}{0.32\textwidth}{}
	  \fig{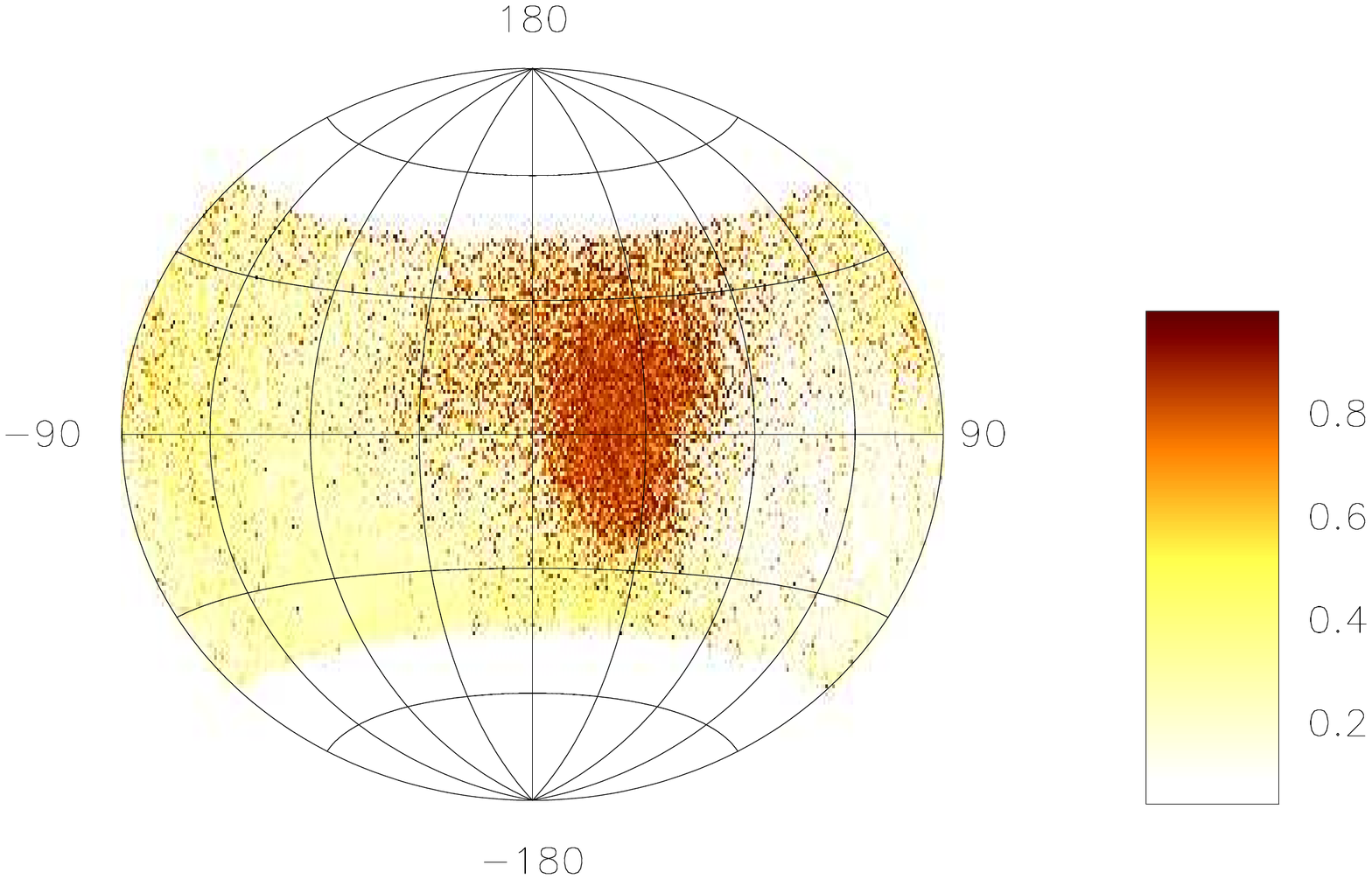}{0.32\textwidth}{}
          \fig{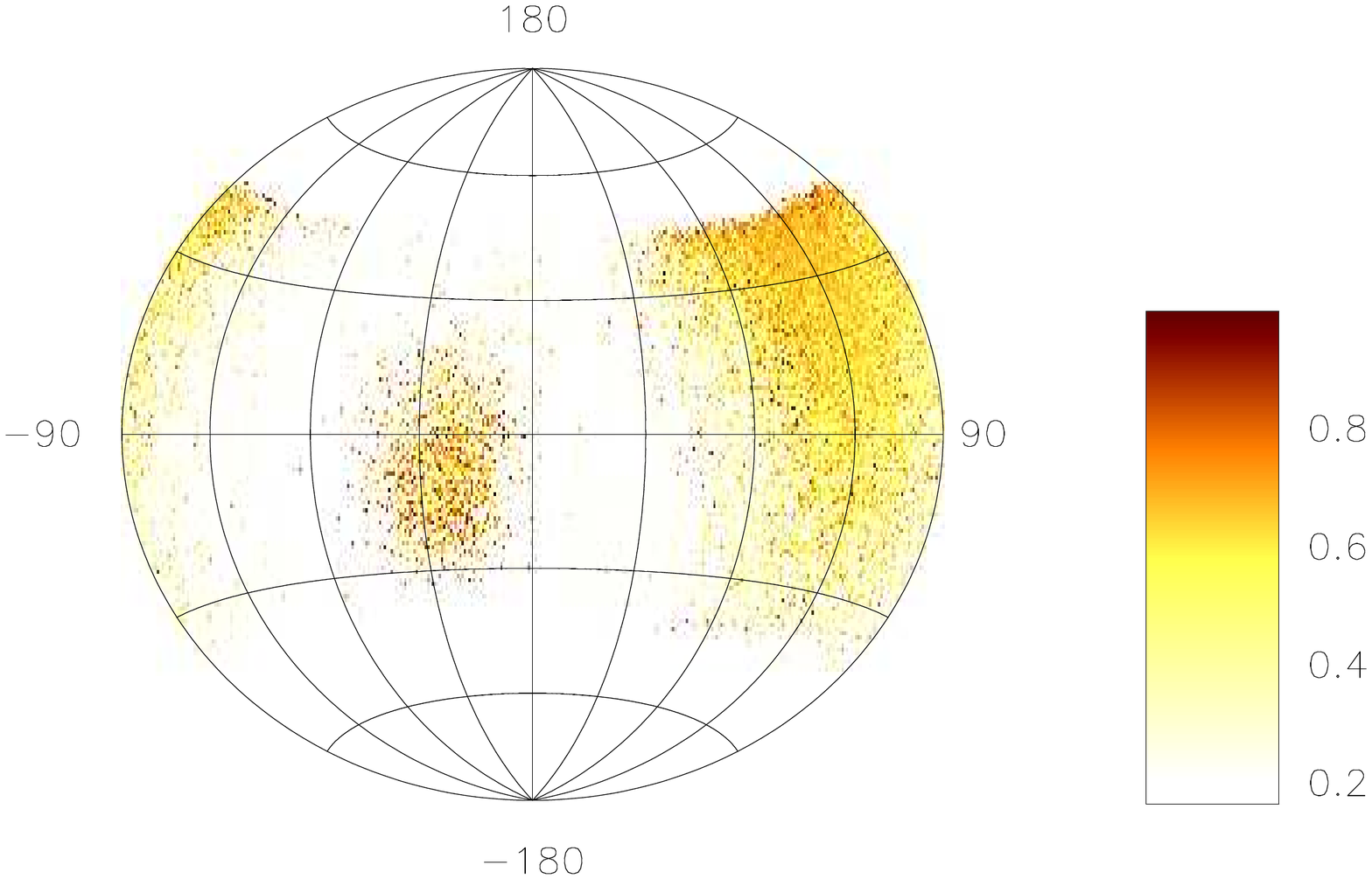}{0.32\textwidth}{}
	 }
\caption{The typical AP and SP polarization properties of three pulsars with 
different $\dot{E}$ values are shown. Left column represents the highly 
polarized PSR J0742-2822 with $\dot{E}=1.43\times10^{35}$ ergs~s$^{-1}$, middle 
column corresponds to PSR J1752-2806 with $\dot{E}=1.8\times10^{33}$ 
ergs~s$^{-1}$ and the right column shows the polarization behaviour of PSR 
J0034-0721 with $\dot{E}=1.92 \times 10^{31}$ ergs~s$^{-1}$. Top panels (upper 
window) show the average profile with total intensity (Stokes I; solid black
lines), total linear polarization (dashed red line) and circular polarization 
(Stokes V; dotted blue line). Top panels (lower window) also show the single 
pulse PPA distribution (colour scale) along with the average PPA (red error 
bars). The RVM fits to the average PPA (dashed pink line) in case of 
PSR J0742-2822 and PSR J0034-0721 is also shown in this plot. No suitable RVM 
fit was possible for the disordered PPA distribution of PSR J1752-2806. Middle 
panels show the $\chi^2$ contours for the parameters $\alpha$ and $\beta$ 
obtained from RVM fits. Bottom panels show the Hammer-Aitoff projection of the 
polarized time samples with the colour scheme representing the fractional 
polarization level. (The complete set of figures (88 figures) are 
	available in the appendix ~\ref{supp}).}
\label{fig1}
\end{figure}

The RVM gives an estimate of the PPA, $\Psi$, as a function of the pulse 
longitude, $\phi$, in terms of the geometrical angles $\alpha$ and $\beta$ as :
\begin{equation}
\Psi = \Psi_{\circ} + \\ 
\tan^{-1}{\left(\frac{\sin{\alpha} \sin{(\phi-\phi_{\circ})}}
{\sin{(\alpha + \beta)} \cos{\alpha} - \sin{\alpha} \cos{(\alpha + \beta)}\cos{(\phi-\phi_{\circ})}}\right)} \\ 
\label{eq1}
\end{equation}
Here, $\Psi_{\circ}$ and $\phi_{\circ}$ are the phase offsets of $\Psi$ and 
$\phi$ respectively. At the longitude $\phi_{\circ}$, where the PPA is 
$\Psi_{\circ}$, the rotation axis and the magnetic axis lie in the same plane 
and the PPA traverse goes through an inflexion point. The steepest gradient of 
the PPA traverse also corresponds to this inflexion point, and has a dependence
$\mid d\Psi/d\phi\mid_{max}=\sin{\alpha}/\sin{\beta}$. We have carried out 
suitable fits to the PPA, using Eq.~\ref{eq1}, in all pulsars of the MSPES 
sample that exhibits smoothly varying monotonic RVM like behaviour, by 
minimizing $\chi^2$. Examples of RVM fits to the PPA are shown in 
Fig.~\ref{fig1} (see pink dashed line in lower window of top panels). The 
estimated values of the angles $\alpha$ and $\beta$ from the RVM fits are 
highly correlated and cannot be used to constrain the pulsar LOS geometry 
\citep{1997A&AS..126..121V,2001ApJ...553..341E,2004A&A...421..215M}. This is 
also evident from the $\chi^2$ variations as a function of $\alpha$ and $\beta$
in Fig.~\ref{fig1} (see middle panels). The above analysis also allows for the 
determination of $\Psi_{\circ}$ and $\phi_{\circ}$, as indicated in 
Fig.~\ref{fig1} (see black error bar in lower window of top panels). 

Despite their inadequacy in constraining the LOS geometry, the RVM fits to the
PPA can yield robust estimates of radio emission heights by taking into account 
the shifts due to the A/R effect. If the radio emission region arises close to 
the stellar surface, less than 10\% of the light cylinder distance, then due to
pulsar rotation, the aberration and retardation causes a positive shift, 
$\Delta\phi$, of the phase of the steepest gradient point, $\psi_{\circ}$, from
the center of the profile. \citet{1991ApJ...370..643B} showed that if the 
emission across the pulse profile arises at a height $h$, it can be related to 
this shift as :
\begin{equation}
h = \cfrac{c P \Delta\phi}{1440}~~~ \mathrm{km}, 
\label{eq2}
\end{equation}
here $c$ is the velocity of light and $P$ the rotation period of the pulsar.
It was further shown by \citet{2020MNRAS.491...80P} that these estimates using 
A/R shifts remain valid up to heights less than 20\% of the light cylinder 
radius, even when deviations from a star centered static dipole structure, due 
to perturbations introduced by rotating plasma, are taken into account. A 
number of studies in the literature have used the A/R method to find estimates 
of the emission height in the normal pulsar population, and showed the radio 
emission to arise below 10\% of the light cylinder radius 
\citep{1991ApJ...370..643B,1997A&A...324..981V,2004A&A...421..215M,
2008MNRAS.391.1210W,2011ApJ...727...92M}. 
 
The estimation of $h$ using Eq.~\ref{eq2} has certain underlying assumptions
regarding the relevant quantities that require careful considerations. 
$\Delta\phi$ is obtained from the difference between $\phi_{\circ}$ and the 
center of the total intensity profile, $\phi_c$. RVM fits to the PPA yields 
$\phi_{\circ}$, while $\phi_c$ is obtained by identifying the phases 
corresponding to the leading and trailing edges of the profile, $\phi_l$ and 
$\phi_t$, respectively, such that $\phi_c = \phi_l + (\phi_t -\phi_l)/2$, and 
$\Delta\phi = \phi_{\circ} - \phi_c$. The first assumption for the above scheme
to be applicable is for the emission height to be constant across the entire 
pulse profile. This can be considered reasonable if the estimated RVM clearly 
fits the observed PPA traverse, providing a robust estimate for $\phi_{\circ}$.
On the other hand, if the emission height shows large variations across the 
profile, the PPA traverse will show differential height dependent A/R effect 
and deviate from the RVM nature \citep{2004hell.conf..205M}. The second 
assumption is connected with the estimates of $\phi_{l}$ and $\phi_t$, which 
are expected to be associated with the last open magnetic field lines on either
side of the LOS cut across the emission beam. In practical purposes $\phi_{l}$ 
and $\phi_t$ are identified from the profile edges lying above the detection 
threshold, as specified by the background noise levels. If the profiles are 
detected with high signal to noise levels, the above assumption is more 
reasonable, although the pulsar emission can become significantly weaker at the
profile edges and may still lead to inaccuracies. In more severe cases the 
estimated $\Delta\phi$ has negative values which is incompatible with the A/R 
predictions. In most reported works the estimated $h$ from positive 
$\Delta\phi$ were around a few hundred kilometers, with the measured 
$\Delta\phi$ values varying by an order of magnitude between faster and slower 
pulsars, as expected from Eq.~\ref{eq2}, thereby highlighting the efficacy of 
the A/R method. In this context a negative estimate for $\Delta{\phi}$ can 
certainly be attributed to the improper identification of the last open 
magnetic field line in the pulsar profile. 

We have used the A/R method to estimate the emission heights of the pulsars 
observed in MSPES. We identified $\phi_l$ and $\phi_t$ from the profile edges
at 5 times the noise level, while $\phi_{\circ}$ was obtained from the RVM fits
to the PPA, see Table~\ref{tab1}. The errors in estimating $\phi_{t}$ and 
$\phi_l$ were calculated using the prescription of \citet[][see equation 
4]{1997MNRAS.288..631K}. The error in $\phi_{\circ}$ was obtained by computing
the variation of $\phi_{\circ}$ from various combinations of $\alpha$ and 
$\beta$ lying within two times the minimum $\chi^2$ value obtained from the 
fit. This leads to asymmetric errors in $\phi_{\circ}$ (see Fig.~\ref{fig1}), 
however in Table~\ref{tab1} only the average value is reported, which is 
further used for the error estimate of $\Delta{\phi}$ and $h$.

\startlongtable
\begin{deluxetable}{cccccccccc}
\tablenum{1}
\tablecaption{The measurements of 68 pulsars from the MSPES sample where it was
possible to obtain suitable RVM fits to the PPA traverse. Column 2 to 4 reports
the observing frequency in MHz, Jname and period in seconds. Column 5 presents 
the minimum $\chi^2$ of the fits, Column 6 to 8 presents the phase of the 
trailing edge, $\phi_t$, the leading edge, $\phi_l$, of the pulsar profile and 
the phase of the steepest gradiant point, $\phi_{\circ}$. Column 9 reports the 
the shift between midpoint of the profile and the SG point, $\Delta \phi$, 
while Column 10 shows our estimate of the emission height, $h$, using 
Eq.~\ref{eq2}. {\bf Note that generally errors in $\phi_{\circ}$ are asymmetric, however
here only the average value is quoted, which is further used to find the error in 
$\Delta \phi$ and $h$.}
\label{tab1}}
\tablewidth{0pt}
\tablehead{  
       & \colhead{Freq} & \colhead{PSR} & \colhead{Period} &  \colhead{$\chi^2$} & \colhead{$\phi_l$} & \colhead{$\phi_t$} & \colhead{$\phi_{\circ}$} & \colhead{$\Delta \phi$} & \colhead{$h$} \\
  & \colhead{(MHz)} &     & \colhead{(sec)}   &  &\colhead{($^{\circ}$)} & \colhead{($^{\circ}$)} & \colhead{($^{\circ}$)} & \colhead{($^{\circ}$)} & \colhead{(km)}}
\startdata
  1 & 333 &  J0034-0721    &   0.943    &     2.5 &-23.5$\pm$0.1 &26.2$\pm$0.1 &9.8$\pm$50 & --- & --- \\
  2 & 618 &  J0151-0635    &   0.833    &     1.3 &-6.6$\pm$0.1 &31.2$\pm$0.1 &12$\pm$2 &-0.3$\pm$2 & --- \\
  3 & 333 &  J0304+1932    &   1.387    &     1.3 &-16.5$\pm$0.1&4.9$\pm$0.1 &-5.3$\pm$0.1 &0.5$\pm$0.1 & 144$\pm$29 \\
    & 618 &  J0304+1932    &   1.387    &     1.3 &-14.4$\pm$0.1&5.1$\pm$0.1 &-3.8$\pm$0.1 &0.8$\pm$0.1 & 230$\pm$29 \\
  4 & 333 &  J0452-1759    &   0.549    &     3.2 &-17.6$\pm$0.1&17.3$\pm$0.1&3.8$\pm$1.0&3.9$\pm$1     & 457$\pm$114 \\
    & 618 &  J0452-1759    &   0.549    &     1.2 &-14.6$\pm$0.2&14.3$\pm$0.2&3.2$\pm$2.0&4.0$\pm$2     & 457$\pm$228 \\
  5 & 333 &  J0525+1115    &   0.354    &     23.5&-3.2$\pm$0.2&19.7$\pm$0.2&10.2$\pm$2.0&3.0$\pm$2     & 221$\pm$147 \\
    & 618 &  J0525+1115    &   0.354    &     1.9 &-13.2$\pm$0.3&3.5$\pm$0.3&-1.2$\pm$3.0&3.6$\pm$3     & 265$\pm$221 \\
  6 & 333 &  J0528+2200    &   3.746    &     1.5 &-5.0$\pm$0.1&19.9$\pm$0.1&7.6$\pm$0.5&0.2$\pm$0.5     & --- \\
  7 & 333 &  J0543+2329    &   0.246    &     1.2 &-12.0$\pm$0.3&27.2$\pm$0.3&27.9$\pm$8.0&20$\pm$8     & 1025$\pm$410 \\
    & 618 &  J0543+2329    &   0.246    &     0.9 &-11.2$\pm$0.3&24.8$\pm$0.3&24.6$\pm$6&17.8$\pm$6     & 912$\pm$307 \\
  8 & 333 &  J0614+2229    &   0.335    &     1.02 &-8.1$\pm$0.3&11.5$\pm$0.3&3.2$\pm$15&1.5$\pm$15     & --- \\
    & 618 &  J0614+2229    &   0.335    &     0.8  &-5.2$\pm$0.3&5.9$\pm$0.3&9.5$\pm$10 & 9.1$\pm$10     & --- \\
  9 & 333 &  J0629+2415    &   0.477    &     10.5  &-7.8$\pm$0.2&16.2$\pm$0.2&-0.1$\pm$1 & -4.0$\pm$1     & --- \\
    & 618 &  J0629+2415    &   0.477    &     13.6  &-7.8$\pm$0.2&11.9$\pm$0.2&1.0$\pm$2 & -1.0$\pm$2     & --- \\
 10 & 333 &  J0630-2834    &   1.245    &     1.7  & -35.8$\pm$0.1 & 38.1$\pm$0.1 & 2.7$\pm$1 & 1.6$\pm$1  & 415$\pm$259 \\
    & 618 &  J0630-2834    &   1.245    &     1.5  & -23.6$\pm$0.1 & 29.9$\pm$0.1 & 2.7$\pm$1 & -1.0$\pm$1 & --- \\
 11 & 333 &  J0659+1414    &   0.384    &     0.9  & -16.1$\pm$0.2 & 14.9$\pm$0.2 & 19.4$\pm$15 & 20$\pm$15&1600$\pm$1200\\
    & 618 &  J0659+1414    &   0.384    &     1.6  & -13.7$\pm$0.2 & 13.6$\pm$0.2 & 15.6$\pm$30 &  ---    & ---\\
 12 & 333 &  J0729-1836    &   0.510    &     0.9  & -5.8$\pm$0.2  & 18.9$\pm$0.2 & 8.6$\pm$2 & 2$\pm$2    & ---   \\
    & 618 &  J0729-1836    &   0.510    &     1.0  & -3.9$\pm$0.2  & 15.7$\pm$0.2 & 8.0$\pm$3 & 2$\pm$3    &   ---   \\
 13 & 618 &  J0742-2822    &   0.167    &     9.1  & -10.6$\pm$0.5 & 16.9$\pm$0.5 & 9.1$\pm$1 & 3$\pm$1    &104$\pm$35\\
 14 & 333 &  J0846-3533    &   1.116    &     1.0  & -22.2$\pm$0.1  & 10.1$\pm$0.1 & -6.8$\pm$1 & -0.7$\pm$2    &---\\
    & 618 &  J0846-3533    &   1.116    &     2.4  & -19.5$\pm$0.1  &  6.8$\pm$0.1 & -5.6$\pm$1 & 0.7$\pm$2    &---\\
 15 & 618 &  J0905-4536    &   0.988    &     1.8  & -6.6$\pm$0.1  &  53.8$\pm$0.1 & 24.9$\pm$8 & 1.3$\pm$8    &---\\
 16 & 618 &  J0922+0638    &   0.431    &     8.8  & -14.2$\pm$0.2  &  6.6$\pm$0.2 & 3.4$\pm$1 & 7.2$\pm$1    &646$\pm$90\\
 17 & 333 &  J0944-1354    &   0.570    &     74.0  & -2.8$\pm$0.2  &  5.7$\pm$0.2 & 1.4$\pm$0.5 & 0.0$\pm$0.5    &---\\
    & 618 &  J0944-1354    &   0.570    &      6.3  & -3.1$\pm$0.2  &  4.0$\pm$0.2 & 0.8$\pm$0.1 & 0.2$\pm$0.2    &---\\
 18 & 333 &  J0953+0755    &   0.253    &      20.6 & -172$\pm$0.4  &  84.4$\pm$0.4 & -11$\pm$8 & 33$\pm$8    &1793$\pm$421\\
    & 618 &  J0953+0755    &   0.253    &      22.6 &  -171$\pm$0.4 &  66.3$\pm$0.4 & -4$\pm$4 &  48$\pm$4    & 2530 $\pm$210  \\
 19 & 618 &  J0959-4809    &   0.670    &      1.6 &  -54$\pm$0.2 &  9.1 $\pm$0.2   & -23$\pm$3&  -0.5$\pm$3  &  ---  \\
 20 & 618 &  J1034-3224    &   1.151    &      2.3 &  -39$\pm$0.1 &  50.1$\pm$0.1   & 11.4$\pm$9&5.9$\pm$9    &  --- \\
 21 & 618 &  J1041-1942    &   1.386    &      1.6 &  -2.3$\pm$0.1 &  15.7 $\pm$0.1   & 7.3$\pm$0.3&  0.6$\pm$0.3  & 173  $\pm$ 86 \\
 22 & 333 &  J1136+1551    &   1.188    &      84 &   -7.3$\pm$0.1 &  12.1 $\pm$0.1   &  4.2$\pm$0.5 & 1.8$\pm$0.5& 445   $\pm$ 124   \\
    & 618 &  J1136+1551    &   1.188    &      25 &   -3.2$\pm$0.1 &  9.6 $\pm$0.1   &  3.1$\pm$2.5 & 0$\pm$2.5&    ---    \\
 23 & 333 &  J1239+2453    &   1.382    &      9.4&   -3.2$\pm$0.1 &  15.1 $\pm$0.1   &  6.7$\pm$0.5 & 0.75$\pm$0.5& 215  $\pm$ 143   \\
    & 618 &  J1239+2453    &   1.382    &      12.4&   -2.3$\pm$0.1 &  13.3 $\pm$0.1   &  5.7$\pm$0.5 & 0.2$\pm$0.5&   ---    \\
 24 & 618 &  J1321+8323    &   0.670    &      3.2 &   -4.5$\pm$0.1 &  12.8 $\pm$0.1   &  7.7$\pm$9 & 2$\pm$9      &   ---    \\
 25 & 333 &  J1328-4357    &   0.533    &      1.4 &   -3.0$\pm$0.2 &  12.0 $\pm$0.2   &  4.1$\pm$1 & -0.4$\pm$1      &       ---      \\
    & 618 &  J1328-4357    &   0.533    &      13.1 &   -6.7$\pm$0.2 &  6.4 $\pm$0.2   &  -1.1$\pm$1 & -0.9$\pm$1      &   ---     \\
 26 & 333 &  J1328-4921    &   1.479    &      10.5 &   -15.1$\pm$0.1 &  2.7 $\pm$0.1   &  -6.5$\pm$1 & -0.3$\pm$1      &   ---     \\
    & 618 &  J1328-4921    &   1.479    &       1.5 &   -13.5$\pm$0.1 &  2.3 $\pm$0.1   &  -6.1$\pm$1 & -0.5$\pm$1      &   ---     \\
 27 & 333 &  J1507-4352    &   0.287    &       1.1 &   -7.1 $\pm$0.3 &  4.9 $\pm$0.3    &  1.9$\pm$1 & 3$\pm$1      &  179 $\pm$60     \\
    & 618 &  J1507-4352    &   0.287    &       0.8 &   -5.6 $\pm$0.3 &  4.9 $\pm$0.3    &  2.0$\pm$1 & 2.3$\pm$1    &  138 $\pm$60     \\
 28 & 333 &  J1527-3931    &   2.418    &       6.1 &   -2.1 $\pm$0.1 &  5.7 $\pm$0.1    &  1.7$\pm$0.5 & -0.1$\pm$0.5    &      ---       \\
    & 618 &  J1527-3931    &   2.418    &       1.6 &   -1.5 $\pm$0.1 &  4.7 $\pm$0.1    &  1.5$\pm$0.5 & -0.1$\pm$0.5    &      ---       \\
 29 & 333 &  J1555-3134    &   0.518    &       1.2 &   -20.4 $\pm$0.2 &  6.4 $\pm$0.2    &  -7.4$\pm$5 & -0.7$\pm$5    &      ---       \\
    & 618 &  J1555-3134    &   0.518    &       2.3 &   -4.4 $\pm$0.2 &  19.2 $\pm$0.2    &  13.1$\pm$10 & 5.7$\pm$10    &      ---       \\
 30 & 333 &  J1559-4438    &   0.257    &       3.1 &   -25.3 $\pm$0.3 &  16.4 $\pm$0.3    &  -6.5$\pm$1 & -2.1$\pm$1    &      ---       \\
    & 618 &  J1559-4438    &   0.257    &       11.1 &   -28.1 $\pm$0.3 &  15.7 $\pm$0.3    &  -5.3$\pm$1 & 1.0$\pm$1    &      ---       \\
 31 & 333 &  J1603-2531    &   0.283    &       0.9 &   -4.8 $\pm$0.3 &  5.8 $\pm$0.3    &   8.3$\pm$20 &    $\pm$     &      ---       \\
    & 618 &  J1603-2531    &   0.283    &       1.3 &   -7.3 $\pm$0.3 &  7.0 $\pm$0.3    &   13.7$\pm$55 &    $\pm$     &      ---       \\
 32 & 618 &  J1700-3312    &   1.358    &       1.0 &   -1.9 $\pm$0.1 &  11.5 $\pm$0.1    &   5.5$\pm$0.5 & 0.7$\pm$0.5& 198     $\pm$ 141      \\
 33 & 333 &  J1703-3241    &   1.212    &       5.2 &   -10.1 $\pm$0.1 &  10.6 $\pm$0.1    &   -0.5$\pm$0.5 & -0.75$\pm$0.5     &      ---  \\
    & 618 &  J1703-3241    &   1.212    &       7.8 &   -4.6 $\pm$0.1 &  12.3 $\pm$0.1    &   4.0$\pm$0.5 & 0.15$\pm$0.5     &      ---       \\
 34 & 333 &  J1709-1640    &   0.653    &       2.3 &   -11.7 $\pm$0.1 &  6.6 $\pm$0.1    &   -1.7$\pm$0.5 & 0.8$\pm$0.5     & 108   $\pm$68   \\
    & 618 &  J1709-1640    &   0.653    &       1.8 &   -10.2 $\pm$0.1 &  5.9 $\pm$0.1    &   -0.3$\pm$0.5 & 1.9$\pm$0.5     & 258   $\pm$68   \\
 35 & 618 &  J1709-4429    &   0.103    &       1.2 &   -28.0 $\pm$0.8 &  22.0 $\pm$0.8    &   5.2$\pm$15 & 8$\pm$15     &       ---         \\
 36 & 333 &  J1720-2933    &   0.620    &       1.3 &   -9.7 $\pm$0.1 &  15.8 $\pm$0.1    &   2.8$\pm$2 & -0.3$\pm$2     &       ---         \\
    & 618 &  J1720-2933    &   0.620    &       1.3 &   -15.1 $\pm$0.1 &  7.7 $\pm$0.1    &   -3.2$\pm$2 & 0.5$\pm$2     &       ---         \\
 37 & 333 &  J1722-3712    &   0.236    &       1.5 &   -9.4 $\pm$0.4 &  14.6 $\pm$0.4    &   -7.2$\pm$10 & -9.8$\pm$10     &   ---         \\
    & 618 &  J1722-3712    &   0.236    &       0.7 &   -10.1 $\pm$0.4 &  6.4 $\pm$0.4    &   4.7$\pm$6 & 6.5$\pm$6     &       ---         \\
 38 & 618 &  J1733-3716    &   0.338    &       1.4 &   -5.6 $\pm$0.3 &  48.9 $\pm$0.3    &   29.2$\pm$8 & 7.5$\pm$8     &       ---         \\
 39 & 333 &  J1740+1311    &   0.803    &       7.0 &   -15.9 $\pm$0.1 &  11.3 $\pm$0.1    &   -1.0$\pm$0.5 & 1.3$\pm$0.5   &   217$\pm$ 83   \\
    & 618 &  J1740+1311    &   0.803    &       20.5 &   -6.3 $\pm$0.1 &  19.8 $\pm$0.1    &   8.7$\pm$0.5 & 1.9$\pm$0.5     &   317 $\pm$ 83 \\
 40 & 333 &  J1741-0840    &   2.043    &       1.2 &    -17.1 $\pm$0.1 &  3.3 $\pm$0.1    &   -6.4$\pm$0.5 & 0.5$\pm$0.5     &          ---            \\
 &618 &  J1741-0840    &   2.043    &       1.1 &    -15.2 $\pm$0.1 &  2.6 $\pm$0.1    &   -5.9$\pm$0.5 & 0.4$\pm$0.5     &          ---            \\
 41 & 618 &  J1741-3927    &   0.512    &       2.7 &    -9.1 $\pm$0.1 &  12.2 $\pm$0.1    &   4.4$\pm$3.2 & 2.8$\pm$3.2     &            ---            \\
 42 & 333 &  J1751-4657    &   0.742    &       92.5 &    -5.1 $\pm$0.1 &  9.3 $\pm$0.1    &   2.4$\pm$0.5 & 0.3$\pm$0.5     &          ---            \\
    & 618 &  J1751-4657    &   0.742    &       23.5 &    -3.8 $\pm$0.1 &  8.5 $\pm$0.1    &   1.6$\pm$0.5 & -0.7$\pm$0.5     &            ---            \\
 43 & 333 &  J1801-0357    &   0.921    &       8.7 &    -7.0 $\pm$0.1 &  5.9 $\pm$0.1    &   -0.6$\pm$0.5 &  0.0$\pm$0.5     &            ---            \\
    & 618 &  J1801-0357    &   0.921    &       4.1 &    -7.0 $\pm$0.1 &  3.6 $\pm$0.1    &   -1.7$\pm$1.0 &  0.0$\pm$0.5     &            ---            \\
 44 & 333 &  J1801-2920    &   1.082    &       3.6 &    -11.9 $\pm$0.1 &  11.6 $\pm$0.1    &   0.2$\pm$0.5 &  0.4$\pm$0.5     &            ---            \\
    & 618 &  J1801-2920    &   1.082    &       5.4 &    -2.1 $\pm$0.1 &  18.1 $\pm$0.1    &   8.5$\pm$0.5 &  0.5$\pm$0.5     &            ---            \\
 45 & 333 &  J1807-0847    &   0.164    &       1.6 &    -18.3 $\pm$0.5 &  31.4 $\pm$0.5    &   -11.9$\pm$12 &  -18$\pm$12     &            ---            \\
    & 618 &  J1807-0847    &   0.164    &       13.7 &    -16.7 $\pm$0.5 &  15.1 $\pm$0.5    &  -6.9$\pm$8   &  -6$\pm$8     &            ---            \\
 46 & 333 &  J1808-0813    &   0.876    &       1.8  &    -10.4 $\pm$0.1 &  10.9 $\pm$0.1    &  -9.9$\pm$5   &  -6.2$\pm$5     &            ---            \\
    & 618 &  J1808-0813    &   0.876    &       0.9  &    -11.1 $\pm$0.1 &  4.1 $\pm$0.1    &  0.3$\pm$1   &   3.8$\pm$1     &       693     $\pm$ 182       \\
 47 & 333 &  J1816-2650    &   0.593    &       1.3  &    -30.6 $\pm$0.1 &  9.3 $\pm$0.1    &  -8.5$\pm$6   &   2.1$\pm$6     &               ---           \\
    & 618 &  J1816-2650    &   0.593    &       1.3  &    -3.2 $\pm$0.1 &  26.4 $\pm$0.1    &  15.8$\pm$7   &   4.2$\pm$7     &               ---           \\
 48 & 618 &  J1820-0427    &   0.598    &       79.3  &    -7.6 $\pm$0.2 &  8.8 $\pm$0.2    &  -0.1$\pm$1   &   -0.7$\pm$1     &               ---           \\
 49 & 333 &  J1822-2256    &   1.874    &        1.4  &    -15.9 $\pm$0.1 &  10.1 $\pm$0.1    &  -6.9$\pm$5   &   -4$\pm$5     &               ---           \\
    & 618 &  J1822-2256    &   1.874    &        1.1  &    -11.1 $\pm$0.1 &  4.9 $\pm$0.1    &  -4.2$\pm$8   &   -1$\pm$8     &               ---           \\
 50 & 333 &  J1823-3106    &   0.284    &        5.2  &    -12.3 $\pm$0.3 &  10.1 $\pm$0.3    &   3.0$\pm$2   &   4.1$\pm$2     &    242       $\pm$ 118          \\
    & 618 &  J1823-3106    &   0.284    &        0.9  &    -8.6 $\pm$0.3 &  6.4 $\pm$0.3    &   1.7$\pm$2   &   2.8$\pm$2     &    165       $\pm$ 118          \\
 51 & 333 &  J1834-0426    &   0.290    &        4.7  &    -56.1 $\pm$0.3 &  69.3 $\pm$0.3    &   16.4$\pm$10   &   9.8$\pm$10     &               ---              \\
    & 618 &  J1834-0426    &   0.290    &        4.0  &    -21.9 $\pm$0.3 &  103.7 $\pm$0.3    &   58.5$\pm$6   &   17.6$\pm$6     &    1063       $\pm$ 362          \\
 52 & 618 &  J1835-1106    &   0.166    &        0.6  &    -9.3 $\pm$0.5 &  10.4 $\pm$0.5    &   -1.7$\pm$12   &   -2.25$\pm$12     &               ---              \\
 53 & 333 &  J1841+0912    &   0.381    &        1.5  &    -7.3 $\pm$0.2 &  4.5 $\pm$0.2    &   0.7$\pm$0.5   &   2.1$\pm$0.5     &     167        $\pm$ 40             \\
    & 618 &  J1841+0912    &   0.381    &        1.2  &    -7.8 $\pm$0.2 &  5.9 $\pm$0.2    &   -0.6$\pm$0.3   &   0.7$\pm$0.3     &     47        $\pm$ 23             \\
 54 & 618 &  J1842-0359    &   1.840    &        1.2  &    -62.2 $\pm$0.1 &  10.6 $\pm$0.1    &   -26.1$\pm$1   &   -0.3$\pm$0.5     &               ---                \\
 55 & 333 &  J1900-2600    &   0.612    &        95  &    -26.3 $\pm$0.1 &  21.9 $\pm$0.1    &   -3.9$\pm$2  &   -1.7$\pm$2     &               ---                \\
    & 618 &  J1900-2600    &   0.612    &        22  &    -38.0 $\pm$0.1 &  5.3 $\pm$0.1    &   -19.2$\pm$2  &   -2.8$\pm$2     &               ---                \\
 56 & 333 &  J1901-0906    &   1.782    &        1.8  &    -12.4 $\pm$0.1 &  2.5 $\pm$0.1    &   -5.1$\pm$4  &   -0.2$\pm$4     &               ---                \\
    & 618 &  J1901-0906    &   1.782    &        14.1  &   -10.2 $\pm$0.1 &  1.7 $\pm$0.1    &   -2.9$\pm$6  &   1.3$\pm$6     &               ---                \\
 57 & 333 &  J1910+0358    &   2.330    &         1.9  &   -65.6 $\pm$0.1 &  6.1 $\pm$0.1    &   -31.3$\pm$1  &   -1.5$\pm$1     &               ---                \\
    & 618 &  J1910+0358    &   2.330    &         1.8  &   -59.4 $\pm$0.1 &  6.2 $\pm$0.1    &   -27.8$\pm$1  &   -1.1$\pm$1     &               ---                \\
 58 & 333 &  J1917+1353    &   0.195    &         1.9  &   -11.6 $\pm$0.5 &  13.4 $\pm$0.5    &   7.6$\pm$2  &   6.7$\pm$2     &    272          $\pm$   81           \\
    & 618 &  J1917+1353    &   0.195    &         1.5  &   -8.4 $\pm$0.5 &  10.7 $\pm$0.5    &   8.2$\pm$4  &   7.0$\pm$4     &    333          $\pm$   162           \\
 59 & 618 &  J1919+0134    &   1.604    &         1.5  &   -14.6 $\pm$0.1 &  2.8 $\pm$0.1    &   -5.3$\pm$9  &   0.6$\pm$9     &                 ---                 \\
 60 & 333 &  J1946+1805    &   0.441    &         1.5  &   -22.1 $\pm$0.2 &  25.5 $\pm$0.2    &   -5.8$\pm$70  &      $\pm$      &                 ---                 \\
    & 618 &  J1946+1805    &   0.441    &         1.2  &   -18.3 $\pm$0.2 &  20.9 $\pm$0.2    &   -3.6$\pm$40  &      $\pm$      &                 ---                 \\
 61 & 333 &  J2006-0807    &   0.581    &         2.3  &   -37.1 $\pm$0.2 &  31.6 $\pm$0.2    &   -3.6$\pm$3  &    -0.85  $\pm$ 3      &                 ---                 \\
    & 618 &  J2006-0807    &   0.581    &         5.9  &   -31.5 $\pm$0.2 &  26.6 $\pm$0.2    &   -5.5$\pm$8  &    -3     $\pm$ 8      &                 ---                 \\
 62 & 618 & J2046+1540     &   1.136    &         1.0  &   -12.0 $\pm$0.1 &   3.2 $\pm$0.1    &   -4.6$\pm$3  &    -0.2     $\pm$ 3      &                 ---                 \\
 63 & 333 & J2048-1616     &   1.961    &         36   &   -20.0 $\pm$0.1 &   6.8 $\pm$0.1    &   -6.7$\pm$0.1  &     0.1     $\pm$ 0.1      &                 ---                 \\
    & 618 & J2048-1616     &   1.961        &     13.4  &   -15.7 $\pm$0.1 &   3.1 $\pm$0.1    &   -5.9$\pm$0.5  &     0.4     $\pm$ 0.5      &                 ---                 \\
 64 & 333 & J2144-3933     &   8.510        &     273.7 &   -1.1 $\pm$0.1 &   1.2 $\pm$0.1    &   -0.1$\pm$0.5  &     0.0     $\pm$ 0.5      &                 ---                 \\
 65 & 333 & J2305+3100     &   1.576        &     1.5   &   -5.7 $\pm$0.1 &   5.3 $\pm$0.1    &    1.3$\pm$0.5  &     1.5     $\pm$ 0.5      &    492             $\pm$   164              \\
 66 & 333 & J2317+2149     &   1.445        &     18.3   &   -4.0 $\pm$0.1 &   4.9 $\pm$0.1    &    0.0$\pm$0.5  &     -0.5     $\pm$ 0.5      &                 ---                 \\
    & 618 & J2317+2149     &   1.445        &     3.4   &   -4.2 $\pm$0.1 &   3.8 $\pm$0.1    &    -0.1$\pm$0.5  &      0.1     $\pm$ 0.5      &                 ---                 \\
 67 & 333 & J2330-2005     &   1.643        &     147.9   &   -2.6 $\pm$0.1 &   7.2 $\pm$0.1    &    2.3$\pm$0.5  &      0.0     $\pm$ 0.5      &                 ---                 \\
    & 618 & J2330-2005     &   1.643        &      74.4   &   -1.8 $\pm$0.1 &   6.6 $\pm$0.1    &    2.0$\pm$0.5  &      -0.4     $\pm$ 0.5      &                 ---                 \\
 68 & 333 & J2346-0609     &   1.181        &       1.2   &   -3.1 $\pm$0.1 &   20.2 $\pm$0.1    &    8.6$\pm$0.5  &       0.0     $\pm$ 0.5      &                 ---                 \\
    & 618 & J2346-0609     &   1.181        &       1.1   &   -2.2 $\pm$0.1 &   16.7 $\pm$0.1    &    8.2$\pm$0.5  &       0.9     $\pm$ 0.5      &    221             $\pm$    123             \\
\enddata
\end{deluxetable}

\begin{figure}
\epsscale{0.7}
\plotone{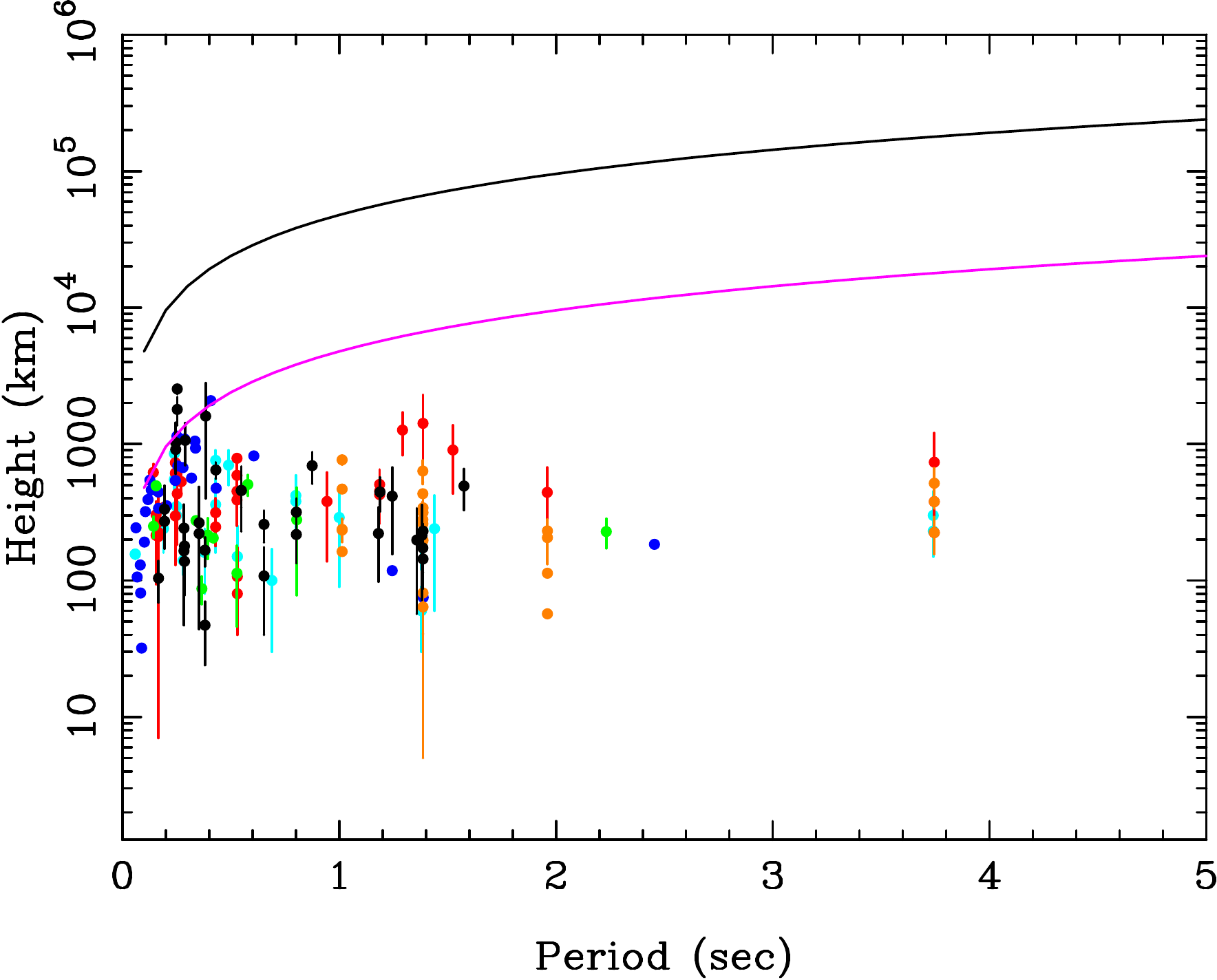}
\caption{The distribution of radio emission heights with the pulsar period, 
estimated using the A/R shifts of PPA traverse within the pulsar profile. 
A total of 71 pulsars are shown in the plot, including the latest measurements
from MSPES sample (shown in black) as well as other assorted sources in the literature 
(red  from \citealt{1997A&A...324..981V}, green from \citealt{2011ApJ...727...92M},
blue from \citealt{2008MNRAS.391.1210W}, cyan from BCW, orange from \citealt{2004A&A...421..215M}). Only 
data points for which estimated heights are positive and larger than the error 
bar are shown in the plot. The black points correspond to the MSPES sample of 
Table~\ref{tab1}. The variation of the light cylinder radius (black line) as 
well as 10\% of the light cylinder radius (pink line) with period, are also
shown for comparison.
\label{fig2}}
\end{figure}

Table~\ref{tab1} reports the results from 68 pulsars in the MSPES sample where 
RVM fits to the PPA traverse were possible, including 47 pulsars where such 
studies could be carried out at both observing frequencies, 333 MHz and 618 
MHz. The estimates of $h$ using Eq.~\ref{eq2} were obtained in 34 pulsars, 
where $\Delta\phi$ was positive and larger than the measurement errors. In 
Fig.~\ref{fig2} we present the distribution of $h$ with $P$, combining our 
latest measurements with other previously reported estimates from the 
literature. A total of 71 pulsars are included in Fig.~\ref{fig2}, with several
cases having multiple measurements of $h$. The figure also shows the variation 
of the light cylinder radius (black line) as well as 10\% of the light cylinder
distance (pink line) with period. All estimates of $h$ are below 10\% of the 
light cylinder radius and across the entire period range the radio emission 
height appears to be constant with a mean value of roughly 500 km. The emission
height found using the A/R method corresponds to the location where the 
radiation detaches from the pulsar magnetosphere with the linear polarization 
features being frozen in at this height (except for changes due to interstellar
Faraday rotation).

\section{Single pulse polarization property}\label{sec3}

\begin{figure}
\epsscale{0.7}
\plotone{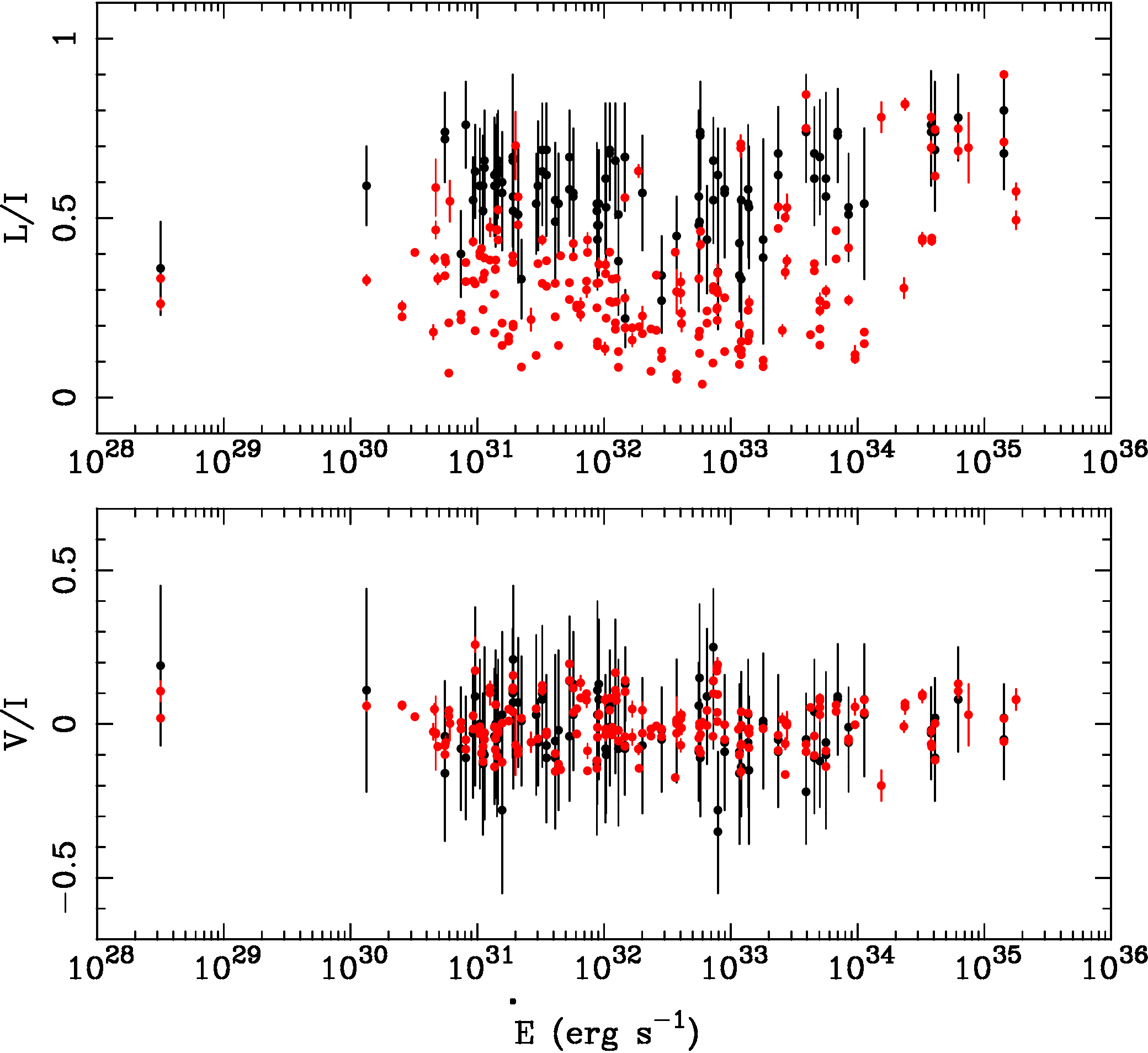}
\caption{The upper panel shows the variation of fractional linear polarization 
in AP (red error bars) as well as the mean of the SP linear fractional 
polarization distribution (black error bars) as a function of $\dot{E}$, for 
the pulsars observed in MSPES. The corresponding measurements of the fractional
circular polarization is shown in the lower panel. The measurements at both
observing frequencies, 333 MHz and 618 MHz, are used in these plots.}
\label{fig3}
\end{figure}

\begin{figure}
\gridline{\fig{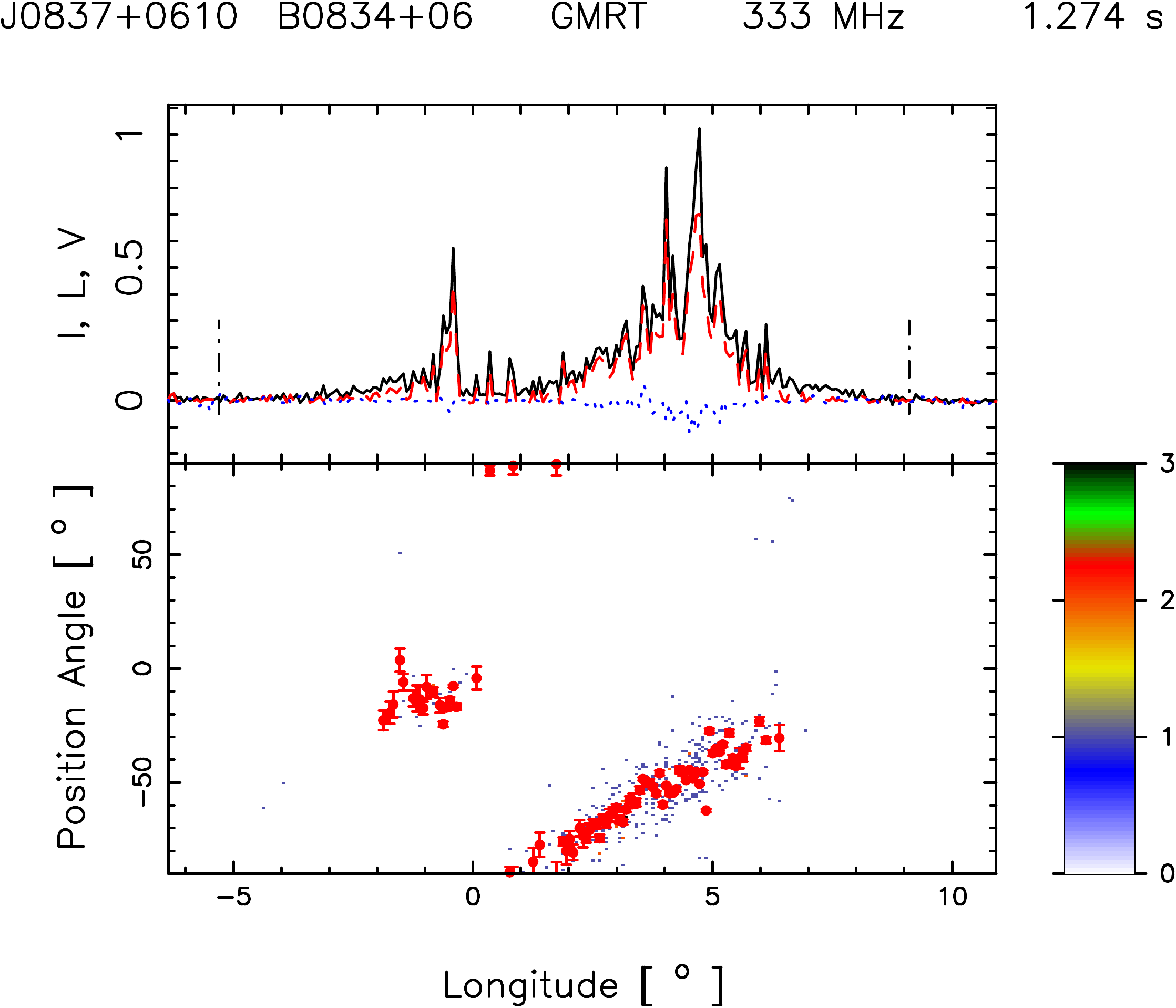}{0.43\textwidth}{}
          \fig{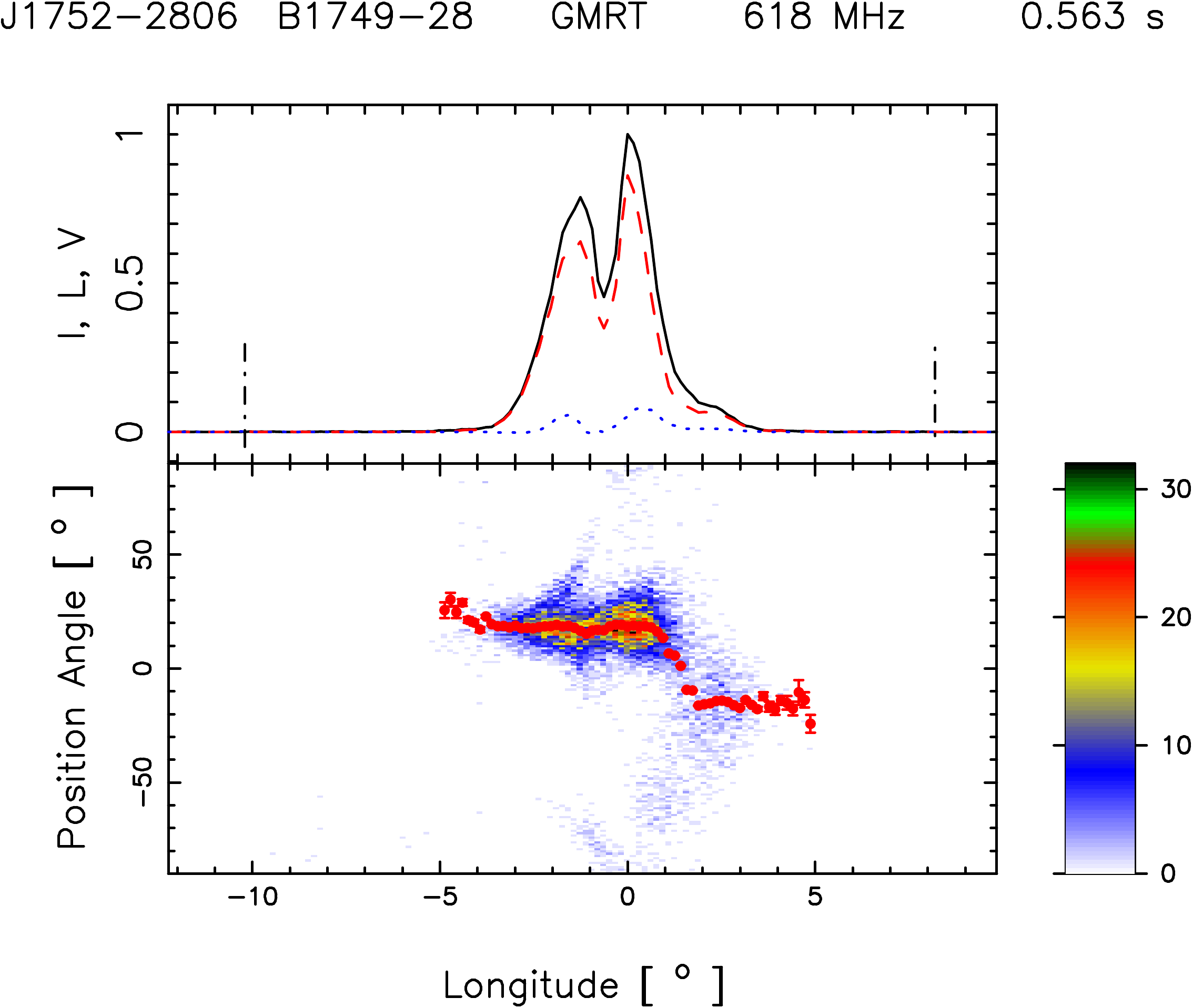}{0.43\textwidth}{}
         }
\gridline{\fig{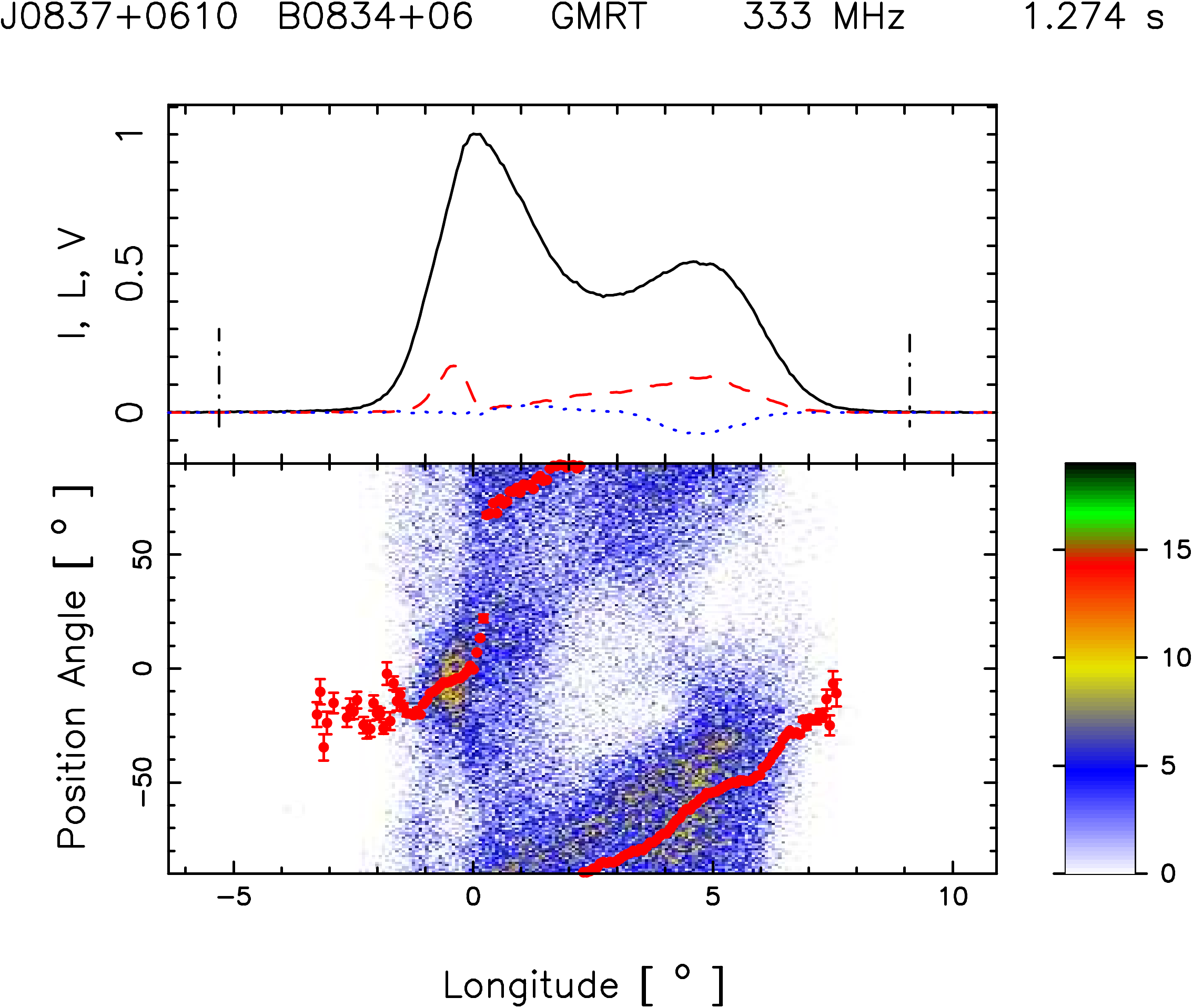}{0.43\textwidth}{}
          \fig{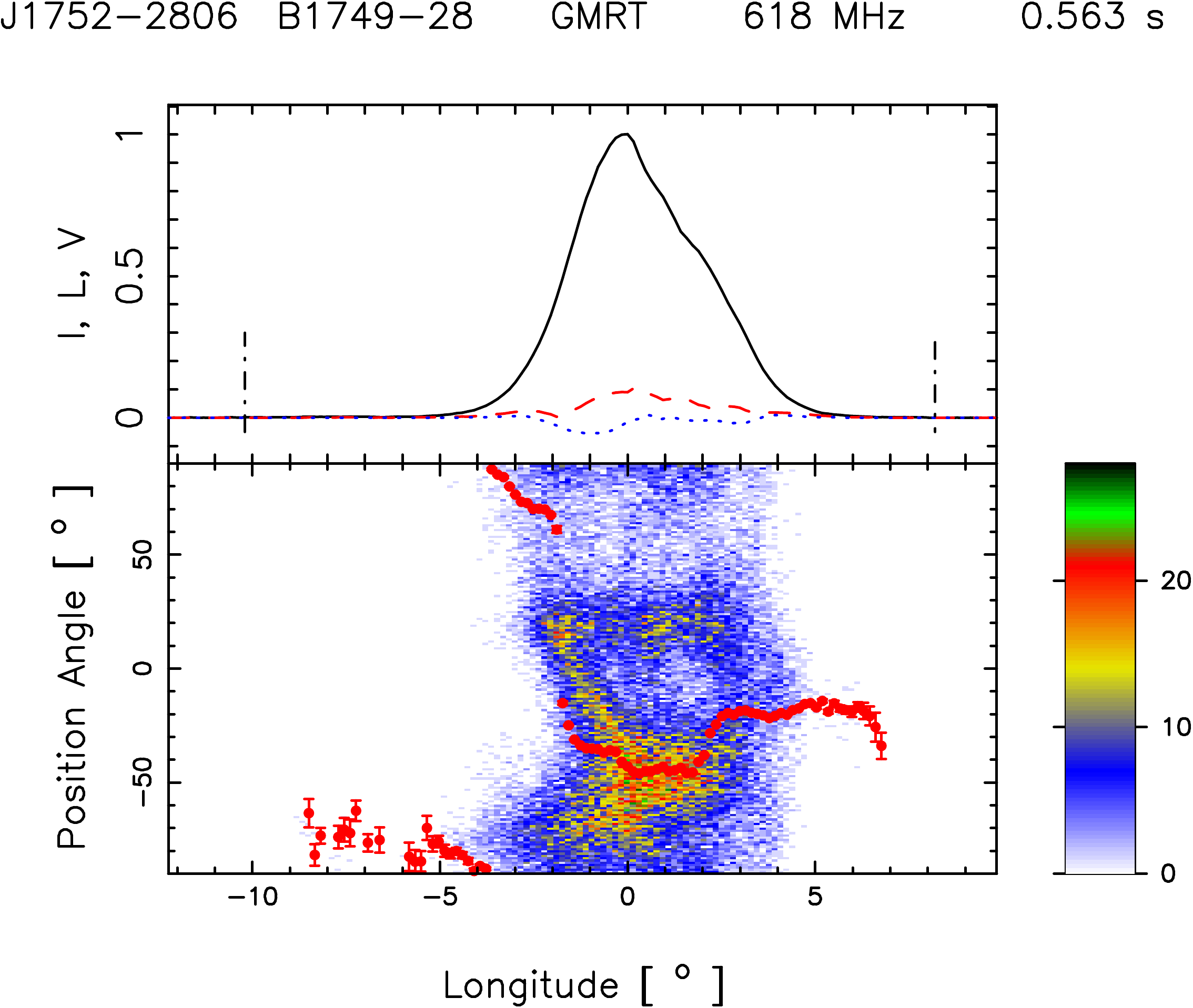}{0.43\textwidth}{}
         }
\caption{See caption of Fig~\ref{fig1} for a description of the quantities 
shown in the figure. The top panels present the polarization behaviour of 
highly polarized time samples with linear polarization levels greater than 0.8,
for PSR J0837+0610 (left panel) and PSR J1752-2806 (right panel). The bottom 
panels show the corresponding polarization behaviour for weakly polarized time 
samples with fractional linear polarization levels less than 0.7. The PPA 
distributions in the top panels are clustered around a narrow track, while they
show significantly larger scatter in the bottom panels.} 
\label{fig3a}
\end{figure}

We have used the Hammer-Aitoff projection of the Poincar\'{e} sphere to study 
the distribution of the Stokes parameters for the time samples in the SPs (see 
bottom panel of Fig~\ref{fig1}). This was implemented by selecting time samples
of each SP showing significant detection of linear polarization, i.e. 
polarization intensity being greater than three times the baseline noise rms. 
The level of polarization is indicated by the colour scheme in each figure and 
these estimates were carried out only in pulsars with sufficient number of 
significant samples ($>$90) for further statistical analysis. The fractional 
linear and circular polarization in SP measurements of every pulsar show a wide
spread in their distributions, and their mean and rms is reported in 
Table~\ref{tab2}. The mean levels represent the typical values of the 
polarization fraction in the SP time sample and are compared with the AP 
properties. In PaperI we found a general tendency for the evolution of the AP 
fractional linear polarization in the pulsar population, with generally younger
and more energetic pulsars showing higher polarization levels compared to 
slower, older pulsars. We have extended these comparisons to the SP 
polarization measurements as shown in Fig.~\ref{fig3}. The upper panel of the 
figure shows the mean fractional linear polarization levels from SP (black 
error bars) along with the fractional linear polarization from AP (red error 
bars) as a function of $\dot{E}$, while the bottom panel shows the 
corresponding results for the circular polarization. In pulsars with $\dot{E} <
10^{34}$ ergs~s$^{-1}$ the fractional linear polarization from SP measurements 
has mean value of 0.57$\pm$0.15 which is significantly higher than the 
estimates of 0.29$\pm$0.14 from APs. However, in more energetic pulsars, 
$\dot{E} > 10^{34}$  ergs~s$^{-1}$, the fractional linear polarization levels 
from both SP and AP appear to be comparable at around 0.75, although we were 
unable to derive any statistically significant inference from the MSPES 
observations due to insufficient sample size in this regime. The mean 
fractional circular polarization are at low levels and roughly similar across 
the entire $\dot{E}$ range, with mean values of 0.08$\pm$0.2 for the SP 
measurements and 0.06$\pm$0.23 for the AP measurements, respectively. The 
estimated errors are greater than the mean fractional circular polarization 
since the measurements show large spread within the pulsar population. 

The difference in the linear polarization between the SP and AP levels can be 
associated with the process of estimating the AP polarization from the SPs. 
This involves first averaging the Stokes parameters $U$ and $Q$ over all SPs 
and then estimating the linear polarization for every phase bin in the pulse 
window. Finally, the linear polarization is average across the emission window 
of the profile. In the process there is significant depolarization in the APs
arising from incoherent summation of the highly polarized time samples spread 
out in the PPAs, and adding time samples with weaker polarization level. A 
comparison between the SP and AP fractional polarization levels provides 
several important clues about the radio emission process. Fig~\ref{fig3} (top
panel) clearly shows that the differences between the mean SP fractional linear
polarization between the low and high energetic pulsars are less prominent 
compared to the AP measurements. This indicates that the intrinsic emission 
process primarily produces highly polarized radio emission at shorter 
timescales. There are more time samples in SPs with low fractional polarization
for lower $\dot{E}$ pulsars, suggesting depolarization before the emission 
detaches from the pulsar magnetosphere. On the other hand the circular 
polarization level seems to be independent of either the averaging process or 
pulsar energetics.  

A spread in the PPA distribution of SPs is seen in almost all pulsars. In 
certain cases that have high $\dot{E}$, the time samples in SPs have high 
levels of linear polarization and the corresponding PPAs are tightly 
distributed around a mean level. An example of such behaviour from PSR 
J0742$-$2822 is shown in the left panel of Fig.~\ref{fig1}. In case of pulsars 
with low $\dot{E}$ the linearly polarized time samples have a wider spread and 
their corresponding PPA traverse can show two different types of behaviour. In 
the first case the PPAs exhibit highly disordered patterns with wide spread in 
the distribution of their linear polarization level, such as the behaviour of 
PSR J1752$-$2806 shown in the middle panel of Fig.~\ref{fig1}. In the second 
case the PPAs are clustered around two parallel tracks, that are separated by 
90$\degr$~in phase, as seen in PSR J0034$-$0731~in the right panel of 
Fig.~\ref{fig1}. In several pulsars it was shown that the PPAs corresponding to 
the high linearly polarized time samples for all three categories follow the 
RVM nature \citep{2009ApJ...696L.141M,2023MNRAS.tmpL..22M}. A remarkable result
was obtained using the MSPES observations of PSR J1645$-$0317 with disordered 
PPA distribution by \citet{2023MNRAS.tmpL..22M}. It was seen that in a subset 
of the PPA distribution, comprising of only the high linearly polarized time 
samples, a S-shaped PPA traverse was recovered that followed the RVM. We have 
expanded these studies to other pulsars in the MSPES sample having disordered 
PPA distributions, with two examples shown in Fig.~\ref{fig3a} corresponding to
PSR J0837+0610 (left panel) and PSR J1752$-$2806 (right panel). The top panels
show PPA distributions comprising of time samples with fractional linear 
polarization levels greater than 0.8, and are tightly confined within a narrow 
region with little spread. The PPA distribution of PSR J0837+0610 clearly 
follows the RVM, but the RVM nature is still unclear in case of PSR 
J1752$-$2806. The bottom panels show the PPA distribution made up of time 
samples with weaker polarization signals at levels less than 0.7, and have a 
large spread. 

\startlongtable
\begin{deluxetable}{ccccccc}
\tablenum{2}
\tablecaption{The measurements of the average fractional SP polarization are
presented. Column 1, 2 and 3 correspond to the Jname, observing frequency and 
spindown energy loss as $\dot{E}_{31} = \dot{E}/10^{31}$ ergs~s$^{-1}$, 
respectively. Column 4 presents the number of time samples in the SP sequence 
that were above the detection threshold and used for the polarization 
estimates. Columns 5, 6 and 7 reports the mean fractional linear polarization, 
the mean circular polarization and mean of absolute circular polarization,
respectively. 
\label{tab2}}
\tablewidth{0pt}
\tablehead{ \colhead{PSR} & \colhead{Freq} & \colhead{$\dot{E_{31}}$} & \colhead{Nsamples} & \colhead{L/I} & \colhead{V/I} & \colhead{$\mid V \mid$/I}}

\startdata
J0034$-$0721 & 333 & 1.92 & 26292 & 0.52 $\pm$ 0.14 & 0.21 $\pm$ 0.24 & 0.24 $\pm$0.24 \\
             & 618 &      & 3702  & 0.56 $\pm$ 0.15 & 0.07 $\pm$ 0.28 & 0.20 $\pm$0.20 \\
J0151$-$0635 & 333 & 0.556&  162  & 0.74 $\pm$ 0.11 &-0.04 $\pm$ 0.18 & 0.10 $\pm$0.10 \\
             & 618 &      &  587  & 0.72 $\pm$0.12  &-0.16 $\pm$ 0.22 & 0.19 $\pm$0.19 \\
J0152$-$1637 & 333 & 8.88 & 2219  & 0.44 $\pm$0.14  & 0.11 $\pm$ 0.29 & 0.22 $\pm$0.22 \\
             & 618 &      & 1241  & 0.48 $\pm$0.15  & 0.03 $\pm$ 0.29 & 0.20 $\pm$0.20 \\
J0304+1932   & 333 & 1.91 & 16569 & 0.67 $\pm$0.14  & 0.11 $\pm$ 0.20 & 0.16 $\pm$0.16 \\
             & 618 &      & 7900  & 0.66 $\pm$0.24  & 0.10 $\pm$ 0.21 & 0.16 $\pm$0.16 \\
J0452$-$1759 & 333 & 137  & 20056 & 0.58 $\pm$0.18  & 0.03 $\pm$ 0.18 & 0.11 $\pm$0.11 \\
             & 618 &      & 2702  & 0.54 $\pm$0.15  &-0.06 $\pm$ 0.21 & 0.14 $\pm$0.14 \\
J0528+2200   & 333 & 3.01 & 8878  & 0.59 $\pm$0.18  &-0.06 $\pm$ 0.14 & 0.07 $\pm$0.07 \\
J0543+2329   & 333 & 4090 & 4038  & 0.74 $\pm$0.14  & 0.02 $\pm$ 0.13 & 0.07 $\pm$0.07 \\
             & 618 &      & 3543  & 0.69 $\pm$0.17  &-0.11 $\pm$ 0.14 & 0.10 $\pm$0.10 \\
J0614+2229   & 618 & 6240 &  281  & 0.78 $\pm$0.12  & 0.08 $\pm$ 0.17 & 0.12 $\pm$0.12 \\
J0629+2415   & 333 &  72.8& 6617  & 0.55 $\pm$0.12  & 0.25 $\pm$ 0.19 & 0.25 $\pm$0.24 \\
             & 618 &      &  169  & 0.66 $\pm$0.12  & 0.14 $\pm$ 0.22 & 0.16 $\pm$0.16 \\
J0630$-$2834 & 333 & 14.6 & 9427  & 0.67 $\pm$0.13  &-0.07 $\pm$ 0.16 & 0.11 $\pm$0.11 \\
             & 618 &      &35268  & 0.67 $\pm$0.15  &-0.08 $\pm$ 0.14 & 0.10 $\pm$0.10 \\ 
J0659+1414   & 333 &3810  & 1433  & 0.76 $\pm$0.15  &-0.02 $\pm$ 0.14 & 0.08 $\pm$0.08 \\
             & 618 &      & 451   & 0.74 $\pm$0.15  &-0.03 $\pm$ 0.15 & 0.08 $\pm$0.08 \\ 
J0729$-$1836 & 333 & 564  & 1392  & 0.56 $\pm$0.17  &-0.06 $\pm$ 0.23 & 0.17 $\pm$0.17 \\
             & 618 &      &  188  & 0.61 $\pm$0.24  &-0.10 $\pm$ 0.24 & 0.19 $\pm$0.19 \\
J0738$-$4042 & 333 & 121  & 3714  & 0.55 $\pm$0.13  &-0.14 $\pm$ 0.16 & 0.15 $\pm$0.19 \\
J0742$-$2822 & 333 &14300 &33200  & 0.68 $\pm$0.10  & 0.02 $\pm$ 0.11 & 0.06 $\pm$0.06 \\ 
             & 618 &      &16328  & 0.80 $\pm$0.11  &-0.05 $\pm$ 0.13 & 0.08 $\pm$0.08 \\
J0758$-$1528 & 618 & 20.1 &   97  & 0.57 $\pm$0.16  &-0.07 $\pm$ 0.22 & 0.16 $\pm$0.16 \\
J0820$-$1350 & 333 & 4.38 & 1320  & 0.54 $\pm$0.14  &-0.02 $\pm$ 0.26 & 0.19 $\pm$0.19 \\
J0837+0610   & 333 & 13   &50931  & 0.38 $\pm$0.15  &-0.02 $\pm$ 0.23 & 0.17 $\pm$0.17 \\
             & 618 &      & 6084  & 0.51 $\pm$0.14  &-0.08 $\pm$ 0.25 & 0.19 $\pm$0.19 \\
J0922+0638   & 333 & 697  & 2840  & 0.73 $\pm$0.13  & 0.09 $\pm$ 0.17 & 0.11 $\pm$0.11 \\
             & 618 &      & 2628  & 0.74 $\pm$0.12  & 0.08 $\pm$ 0.18 & 0.12 $\pm$0.12 \\
J0944$-$1354 & 333 & 0.963&  516  & 0.63 $\pm$0.13  & 0.09 $\pm$ 0.29 & 0.21 $\pm$0.21 \\
J0953+0755   & 333 & 56   &64345  & 0.56 $\pm$0.24  & 0.06 $\pm$ 0.14 & 0.10 $\pm$0.10 \\
             & 618 &      &84594  & 0.48 $\pm$0.24  &-0.09 $\pm$ 0.16 & 0.13 $\pm$0.13 \\
J1034$-$3224 & 333 & 79.1 & 1531  & 0.35 $\pm$0.16  &-0.35 $\pm$ 0.20 & 0.35 $\pm$0.35 \\
             & 618 &      &  614  & 0.62 $\pm$0.13  &-0.28 $\pm$ 0.19 & 0.26 $\pm$0.26 \\
J1041$-$1942 & 618 &  1.4 &  927  & 0.59 $\pm$0.17  & 0.00 $\pm$ 0.24 & 0.17 $\pm$0.17 \\
J1116$-$4122 & 333 & 37.4 &  176  & 0.45 $\pm$0.11  & 0.01 $\pm$ 0.20 & 0.12 $\pm$0.12 \\
J1136+1551   & 333 & 8.79 & 5572  & 0.54 $\pm$0.17  &-0.13 $\pm$ 0.23 & 0.20 $\pm$0.20 \\
             & 618 &      & 4946  & 0.52 $\pm$0.17  &-0.12 $\pm$ 0.21 & 0.17 $\pm$0.17 \\
J1239+2453   & 333 & 1.43 &33204  & 0.63 $\pm$0.16  &-0.09 $\pm$ 0.21 & 0.17 $\pm$0.17 \\
             & 618 &      & 6241  & 0.62 $\pm$0.16  &-0.05 $\pm$ 0.21 & 0.14 $\pm$0.14 \\
J1328$-$4921 & 333 & 0.745& 848   & 0.40 $\pm$0.12  &-0.08 $\pm$ 0.20 & 0.16 $\pm$0.16 \\
J1527$-$3931 & 333 & 5.33 & 1120  & 0.58 $\pm$0.13  &0.14  $\pm$ 0.21 & 0.18 $\pm$0.18 \\
             & 618 &      &  160  & 0.67 $\pm$0.13  &-0.04 $\pm$ 0.21 & 0.12 $\pm$0.12 \\
J1559$-$4438 & 317 & 237  & 788   & 0.68 $\pm$0.13  &-0.05 $\pm$ 0.21 & 0.14 $\pm$0.14 \\
             & 618 &      &20858  & 0.62 $\pm$0.12  &-0.09 $\pm$ 0.18 & 0.14 $\pm$0.14 \\
J1625$-$4048 & 618 & 0.134&   95  & 0.59 $\pm$0.11  &0.11  $\pm$ 0.33 & 0.23 $\pm$0.23 \\
J1645$-$0317 & 618 & 121  &36693  & 0.33 $\pm$0.19  &-0.01 $\pm$ 0.18 & 0.13 $\pm$0.13 \\
J1703$-$3241 & 333 & 1.46 & 3165  & 0.64 $\pm$0.14  & 0.02 $\pm$ 0.19 & 0.14 $\pm$0.14 \\
             & 618 &      & 3644  & 0.66 $\pm$0.14  &-0.04 $\pm$ 0.22 & 0.15 $\pm$0.15 \\
J1709$-$1640 & 333 & 89.4 &17632  & 0.57 $\pm$0.18  &-0.09 $\pm$ 0.17 & 0.14 $\pm$0.14 \\
             & 618 &      & 1881  & 0.58 $\pm$0.16  &-0.06 $\pm$ 0.22 & 0.15 $\pm$0.15 \\
J1720$-$2933 & 618 & 12.3 &  124  & 0.66 $\pm$0.16  & 0.09 $\pm$ 0.25 & 0.19 $\pm$0.19 \\
J1731$-$4744 & 333 &1130  &21120  & 0.54 $\pm$0.21  & 0.08 $\pm$ 0.18 & 0.14 $\pm$0.14 \\
             & 618 &      &25042  & 0.54 $\pm$0.21  & 0.03 $\pm$ 0.20 & 0.14 $\pm$0.14 \\
J1735$-$0724 & 333 &  65  & 4287  & 0.44 $\pm$0.15  & 0.09 $\pm$ 0.22 & 0.19 $\pm$0.19 \\
J1740+1311   & 333 &  11.1& 1805  & 0.68 $\pm$0.13  & 0.06 $\pm$ 0.18 & 0.13 $\pm$0.13 \\
             & 618 &      & 3384  & 0.69 $\pm$0.13  &-0.01 $\pm$ 0.19 & 0.12 $\pm$0.12 \\
J1741$-$0840 & 333 &  1.05& 1735  & 0.59 $\pm$0.15  & 0.00 $\pm$ 0.20 & 0.13 $\pm$0.13 \\
             & 618 &      &  838  & 0.59 $\pm$0.16  & 0.00 $\pm$ 0.21 & 0.12 $\pm$0.12 \\
J1741$-$3927 & 618 & 56.7 &  360  & 0.49 $\pm$0.14  & 0.15 $\pm$ 0.24 & 0.20 $\pm$0.20 \\
J1745$-$3040 & 333 & 849  & 144   & 0.51 $\pm$0.13  &-0.01 $\pm$ 0.13 & 0.12 $\pm$0.12 \\
             & 618 &      &4237   & 0.53 $\pm$0.15  &-0.06 $\pm$ 0.16 & 0.13 $\pm$0.13 \\
J1751$-$4657 & 333 &  9.11&19827  & 0.48 $\pm$0.14  & 0.13 $\pm$ 0.19 & 0.17 $\pm$0.17 \\
             & 618 &      & 5509  & 0.54 $\pm$0.15  & 0.08 $\pm$ 0.26 & 0.19 $\pm$0.19 \\
J1752$-$2806 & 333 & 180  &35973  & 0.39 $\pm$0.24  & 0.00 $\pm$ 0.18 & 0.13 $\pm$0.13 \\
             & 618 &      &53938  & 0.44 $\pm$0.28  & 0.01 $\pm$ 0.22 & 0.16 $\pm$0.16 \\
J1801$-$2920 & 333 & 10.3 &  313  & 0.61 $\pm$0.21  &-0.08 $\pm$ 0.24 & 0.18 $\pm$0.18 \\
J1817$-$3618 & 333 & 139  &  683  & 0.53 $\pm$0.17  &-0.15 $\pm$ 0.24 & 0.23 $\pm$0.23 \\ 
J1820$-$0427 & 333 & 117  &9973   & 0.34 $\pm$0.10  &-0.16 $\pm$ 0.23 & 0.24 $\pm$0.24 \\
             & 618 &      &3208   & 0.43 $\pm$0.13  &-0.09 $\pm$ 0.22 & 0.17 $\pm$0.17 \\
J1822$-$2256 & 618 &0.812 & 181   & 0.76 $\pm$0.12  &-0.11 $\pm$ 0.20 & 0.15 $\pm$0.15 \\
J1823+0550   & 333 &  2.1 & 555   & 0.51 $\pm$0.19  & 0.07 $\pm$ 0.21 & 0.16 $\pm$0.16 \\
J1823$-$3106 & 333 & 504  &2964   & 0.67 $\pm$0.16  &-0.12 $\pm$ 0.15 & 0.15 $\pm$0.15 \\
J1900$-$2600 & 333 & 3.52 &12408  & 0.62 $\pm$0.17  &-0.11 $\pm$ 0.21 & 0.17 $\pm$0.17 \\
             & 618 &      &2039   & 0.69 $\pm$0.13  &-0.07 $\pm$ 0.23 & 0.16 $\pm$0.16 \\
J1901$-$0906 & 333 & 1.14 &1671   & 0.64 $\pm$0.13  &-0.10 $\pm$ 0.21 & 0.17 $\pm$0.17 \\
             & 618 &      & 190   & 0.66 $\pm$0.14  &-0.03 $\pm$ 0.28 & 0.19 $\pm$0.19 \\
J1909+1102   & 333 & 457  & 638   & 0.61 $\pm$0.13  & 0.04 $\pm$ 0.17 & 0.14 $\pm$0.14 \\
             & 618 &      &  91   & 0.68 $\pm$0.13  &-0.11 $\pm$ 0.18 & 0.13 $\pm$0.13 \\
J1913$-$0440 & 333 & 28.5 &4960   & 0.27 $\pm$0.09  &-0.02 $\pm$ 0.13 & 0.09 $\pm$0.09 \\
             & 618 &      &7026   & 0.34 $\pm$0.11  &-0.05 $\pm$ 0.17 & 0.13 $\pm$0.13 \\ 
J1919+0021   & 333 & 14.7 & 783   & 0.22 $\pm$0.08  & 0.13 $\pm$ 0.12 & 0.14 $\pm$0.14 \\
J1921+2153   & 618 & 2.23 &36710  & 0.33 $\pm$0.11  & 0.01 $\pm$ 0.21 & 0.15 $\pm$0.15 \\
J1932+1059   & 333 & 393  &36794  & 0.74 $\pm$0.12  &-0.22 $\pm$ 0.17 & 0.22 $\pm$0.22 \\
             & 618 &      &20016  & 0.75 $\pm$0.15  &-0.05 $\pm$ 0.15 & 0.09 $\pm$0.09 \\
J1941$-$2602 & 333 & 57.5 & 118   & 0.74 $\pm$0.14  &-0.11 $\pm$ 0.18 & 0.13 $\pm$0.13 \\
             & 618 &      & 635   & 0.73 $\pm$0.15  &-0.11 $\pm$ 0.19 & 0.15 $\pm$0.15 \\
J1946+1805   & 333 & 1.11 &9953   & 0.52 $\pm$0.14  &-0.09 $\pm$ 0.18 & 0.14 $\pm$0.14 \\
             & 618 &      &4822   & 0.59 $\pm$0.14  &-0.13 $\pm$ 0.23 & 0.19 $\pm$0.19 \\
J2006$-$0807 & 333 & 0.927& 303   & 0.55 $\pm$0.12  &-0.03 $\pm$ 0.21 & 0.12 $\pm$0.12 \\
J2046$-$0421 & 333 & 1.57 &17322  & 0.57 $\pm$0.16  & 0.03 $\pm$ 0.27 & 0.21 $\pm$0.21 \\
             & 618 &      & 2805  & 0.60 $\pm$0.14  &-0.28 $\pm$ 0.27 & 0.30 $\pm$0.30 \\
J2048$-$1616 & 333 & 5.73 &114821 & 0.57 $\pm$0.18  & 0.13 $\pm$ 0.17 & 0.16 $\pm$0.16 \\
             & 618 &      & 68016 & 0.56 $\pm$0.17  & 0.03 $\pm$ 0.18 & 0.11 $\pm$0.11 \\
J2144$-$3933 & 333 &0.00318&14763 & 0.36 $\pm$0.13  & 0.19 $\pm$ 0.26 & 0.22 $\pm$0.22 \\
J2305+3100   & 333 & 2.92 & 5543  & 0.54 $\pm$0.14  & 0.03 $\pm$ 0.26 & 0.19 $\pm$0.19 \\
J2313+4253   & 333 & 10.4 & 8555  & 0.53 $\pm$0.15  &-0.10 $\pm$ 0.19 & 0.15 $\pm$0.15 \\
J2317+2149   & 333 &  1.37& 1564  & 0.62 $\pm$0.16  &-0.04 $\pm$ 0.22 & 0.14 $\pm$0.14 \\
             & 618 &      &  207  & 0.59 $\pm$0.14  &-0.01 $\pm$ 0.18 & 0.11 $\pm$0.11 \\
J2330$-$2005 & 333 &  4.12&40307  & 0.49 $\pm$0.16  &-0.11 $\pm$ 0.23 & 0.19 $\pm$0.19 \\
             & 618 &      & 1921  & 0.55 $\pm$0.16  &-0.55 $\pm$ 0.28 & 0.20 $\pm$0.20 \\
J2346$-$0609 & 333 &  3.26& 3035  & 0.63 $\pm$0.15  & 0.10 $\pm$ 0.22 & 0.16 $\pm$0.16 \\
             & 618 &      & 432   & 0.69 $\pm$0.13  & 0.08 $\pm$ 0.22 & 0.16 $\pm$0.16 \\
\enddata
\end{deluxetable}

\section{ A model for the formation of polarization properties of pulsars.}\label{sec4}

We have currently accrued significant observational constraints on the radio 
emission mechanism. The location of the source has been mapped out
to be well inside the magnetosphere, at heights below 10\% of the light cylinder radius 
(see section \ref{sec2}). 
In addition to the limitations on the height of emission 
the observed polarization behavior makes further demands on the nature of the mechanism.
The observed radio waves show high level of linear polarization with the polarization position
vector being either perpendicular or parallel to the plane in which the curved dipolar
magnetic field lines lie. The emission is also around 10-20~\% circularly 
polarized (see section ~\ref{sec2} and ~\ref{sec3}). 

In general, the polarization feature of a pulsar is
initially defined by the emission mechanism, and then evolves as the excited
wave mode propagate through the magnetosphere.
The radio emission should be generated due 
to some plasma processes and should correspond to the eigenmodes of the magnetospheric 
plasma, which consists of dense electron-positron pair plasma streaming 
relativistically outwards in the open magnetic field line region. Generally, 
pulsar radio emission mechanisms can be divided into two classes, maser 
mechanisms and antenna mechanisms \citep{1975ARA&A..13..511G,
1975ApJ...196...51R,2000ApJ...544.1081M,1991MNRAS.253..377K}. In maser 
mechanisms the radio emission is generated by certain plasma instabilities that
can excite plasma waves capable of escaping the pulsar magnetosphere. The 
antenna mechanisms on the other hand relies on the coherent curvature radiation
(CCR) being the main source of the radio waves. It is therefore essential to 
understand all possible instabilities that can develop in a relativistic pair 
plasma in pulsar magnetospheric conditions. 

In the radio emission region the magnetic field geometry is dipolar and is still very 
strong, i.e. the ratio between plasma and cyclotron frequencies $\sigma \ll 1$ here 
$\sigma = {\omega_{p}}/{\omega _{B}}=2\times 10^{-4}\times \kappa^{0.5}\mathcal{R}^{1.5} 
({P^{3}}/{\dot{P}_{-15}})^{0.25}$, $\mathcal{R}=r/R_{lc}$ with $r$ being the distance and $R_{lc}$ 
the light cylinder radius, and $\dot{P}_{-15}$ is the period derivative in the units of $10^{-15}$ s~s$^{-1}$
 (see Appendix~\ref{apex1}).
At the emission altitude, $r\simeq 50r_{\ast }$, where 
$r_{\ast} = 10^{6}$ cm is the neutron star radius, $\mathcal{R}\approx 0.01 
P^{-1}$. At this distance most of the plasma instabilities are suppressed and only the two-stream instability 
(i.e. the wave-particle interaction at the Cherenkov resonance), can develop. 
However, $\sigma$ increases with distance as $\mathcal{R}^{1.5}$ and some
other instabilities, like cyclotron and/or Cherenkov drift instabilities, can grow near the light cylinder. 

If the distribution functions of both plasma components, i.e.
electrons and positrons, are identical, the system of dispersion equations is 
divided into two parts, describing the so called ordinary (O-mode) and 
extraordinary (X-mode), orthogonally polarized modes of the magnetized pair 
plasma. We follow the the nomenclature of \citet{2003PhRvE..67b6407S} to 
describe the eigen modes in the plasma. The polarization vector of O-mode lies 
in the plane of $\vec{k}$ and $\vec{B}$ and it has a component along the 
ambient magnetic field as well as along the propagation vector, $\vec{k}$. The 
X-mode has a purely non-potential nature and its polarization vector is 
directed perpendicular to plane containing $\vec{k}$ and $\vec{B}$. There are 
two additional branches of the O-mode, $lt_{1}$-mode and $lt_{2}$-mode, both 
having a mixed longitudinal-transverse nature. In case of strictly parallel 
propagation with respect to the external magnetic field, i.e. 
$\vec{k}\parallel\vec{B}$, the $lt_{2}$-mode coincide with the Langmuir mode 
and have purely potential character. The second branch of O-mode, i.e. the
$lt_{1}$-mode, as well as X-mode have characteristics of purely transverse 
waves of arbitrary polarization. The phase velocity of $lt_{1}$-mode is always 
sub-luminal while $lt_{2}$-mode is super-luminal for relatively small values 
of the wave vectors and can be sub-luminal at higher frequencies. In case of 
the frequencies that fall into radio-band the X-mode has one sub-luminal branch
and it always has a non-potential nature. The direction of the polarization 
vectors of the X-mode and O-mode are specified by the plane containing 
$\vec{k}$ and the local magnetic field, while the polarization vector of 
observed radio waves are perpendicular to the plane of curved dipolar field 
lines. Therefore, the emission mechanism should distinguish this plane. In our 
opinion the only mechanism which can fulfill these requirements is the coherent
curvature radiation (CCR). There are indeed several observational evidences for
CCR to be the likely emission mechanism for radio emission in pulsars, like the
presence of highly polarized subpulses that follow the mean RVM 
\citep{2009ApJ...696L.141M}, spectral shape variation across the pulse profile 
\citep{2022ApJ...927..208B}, etc. Multiple studies have demonstrated that 
the stable charge bunches capable of exciting CCR can be associated with 
charge solitons, that can form due to the nonlinear growth of plasma 
instability in relativistically flowing non-stationary pulsar plasma 
\citep{1980Afz....16..161M,1998MNRAS.301...59A,2000ApJ...544.1081M,
2018MNRAS.480.4526L,2020MNRAS.497.3953R,2021A&A...649A.145M,
2022MNRAS.516.3715R}. 

\subsection{The PSG model and plasma parameters}
The non-stationary plasma flow in pulsars is generated in an inner accelerating
region (IAR) above the polar cap. \citet[][hereafter RS75]{1975ApJ...196...51R}
was the first to propose the existence of an IAR for pulsars where 
$\vec{\Omega}\cdot \vec{B} < 0$ above the polar cap, here $\vec{\Omega}$ is the
angular velocity of the neutron star and $\vec{B}$ is the local magnetic field.
As the outflowing plasma leaves along the open magnetic field lines, the 
surface ions, that are expected to have large binding energies, cannot be 
pulled out to screen the large co-rotational electric fields, leading to the 
formation of an inner vacuum gap (IVG) with large electric potential drop along
it. Further, due to the presence of strong and curved magnetic fields above the
polar cap electron-positron pairs can be produced in the IAR that are 
subsequently accelerated to relativistic speeds, such that pair cascades ensue 
to form a relativistic plasma flow \citep[][RS75]{1971ApJ...164..529S}. This 
process is called spark formation and in general several isolated sparks can 
exist above the polar cap depending on its size. The sparks grow laterally till
a size $h$ and screens the electric potential drop within the sparking region, 
with the next spark forming at a distance $h$ away. Thus a system of isolated 
sparks are produced, where the center of the spark and the boundary between two
sparks are associated with high and low potential drop, respectively. When 
there is enough charges produced during cascading, the plasma screens the 
entire electric field, thereby halting the sparking process. The plasma 
generated during sparking gradually flows out of the IAR, such that the 
electric potential drop reappears and the the sparking process repeats. This 
results in a non-stationary spark associated plasma flow and at emission 
heights of about 50$r_{\ast}$ the coherent radio emission is generated due to 
instabilities in this outflowing plasma. The emerging radiation associated with
each sparking column is seen as subpulses within the pulse emission window. The
sparks in the IAR undergo a gradual $\vec{E} \times \vec{B}$ drift motion at 
longer timescales that is observed as the phenomenon of subpulse drifting. 
During the sparking process the electric field in the IAR separates the 
electron-positron pairs, with the electrons accelerated towards the surface and
thus heating the polar cap. The positrons which form the primary beam 
accelerate away from the star with Lorentz factors $\gamma_p$, and produces 
additional cascades of secondary pairs with Lorentz factors $\gamma_s$.

The polar cap temperatures estimated from the IVG model of RS75 is higher than
the measured values, while the estimated subpulse drift speeds from the RS75 
model are faster than those derived from the observed drifting periodicities. 
This led \citet[][hereafter GMG03]{2003A&A...407..315G} to propose a modified 
IVG model, known as the partially screened gap (PSG), where they postulated the
presence of thermally ejected ions in 
the IAR that can partially screen the electric field, the possibility of which 
was also discussed earlier by \citet{CR77,CR80}. In the PSG model the number 
density of ions, $n_{ion}$, with respect to the Goldreich-Julien number 
density, $n_{GJ}$, is estimated as $n_{ion}/n_{GJ} \approx \exp\left[30(1 - 
T_i/T_s) \right]$, where $T_i$ is the temperature corresponding to the binding 
energy of $^{56}$Fe$_{26}$ ions and $T_s$ is the surface temperature. Since 
$n_{ion}$ has an exponential dependence on temperature, if $T_s$ goes slightly 
below $T_i$ almost vacuum like conditions will appear in the gap, while $T_s = 
T_i$ implies a fully screened gap. GMG03 suggested that  under thermostatic 
conditions the surface temperature of the polar cap is slightly lower than 
$T_i$, such that $T_s = T_i (1-\delta)$, where $\delta$ is a small temperature 
offset parameter, $\delta \approx 10^{-3}$. Thus unlike the IVG, the PSG never 
attains a fully vacuum state, and the potential difference in the gap is 
screened by a factor $\eta = 1 - n_{ion}/n_{GJ} \approx 0.1$, with ions usually
contributing about 90\% of $n_{GJ}$. The sparking process in the PSG commences 
in a manner similar to the IVG model where a set of tightly packed equidistant 
sparks are produced. The potential drop is maximum at the center of the spark,
where the pair multiplicity is largest, and reduces gradually towards the 
boundary between sparks, which is mostly filled up by diffusion of particles
from surrounding sparks, and can be approximated to have a multiplicity factor 
of unity. The maximum potential drop is obtained as $\eta \Delta V_{vac} = \eta 
(2\pi B_s/cP) h^2$, where $\Delta V_{vac}$ is the potential difference of the 
IVG, $h$ is the gap height and $B_s$ the value of surface magnetic field. The 
sparks grows in size radially with the temperature below it increasing, till it
reaches a maximum size when $T_s = T_i$ and the sparking process ceases. The 
sparking restarts once the the plasma column, also referred to as plasma cloud,
leaves the gap such that $T_s < T_i$ and once again the potential drop 
develops, thus giving rise to the non-stationary plasma flow. The lateral size 
of the sparks, $h_{\perp}$, and the total number of sparks, $N_{sp}$, in the 
gap at any given time can be estimated as \citep{2020MNRAS.492.2468M,
2020MNRAS.496..465B,2022ApJ...936...35B} :
\begin{eqnarray}
h_{\perp} \simeq 15 ~~\frac{T_6^2}{\eta b (\cos{\alpha_l})^{1/2}} {\dot{E}_{32}}^{-1/2} P^{-1} ~~~ {\rm m,} \nonumber\\
N_{sp} \simeq 20~\frac{\eta b^{1/2} \cos{\alpha_l}}{T_6^2} (\dot{E}_{32} P)^{1/2}.
\label{hnsp}
\end{eqnarray}
Here, $T_6 = T_s/10^6 K$, $b=B_s/B_d$, where $B_s$ is the non-dipolar field in
the surface and $B_d$ the equivalent dipolar field, $\alpha_l$ is the angle
between the local magnetic field and the rotation axis, and $\dot{E}_{32} = 
\dot{E}\times 10^{-32}$ erg~s$^{-1}$.

\begin{figure}
\epsscale{0.6}
\plotone{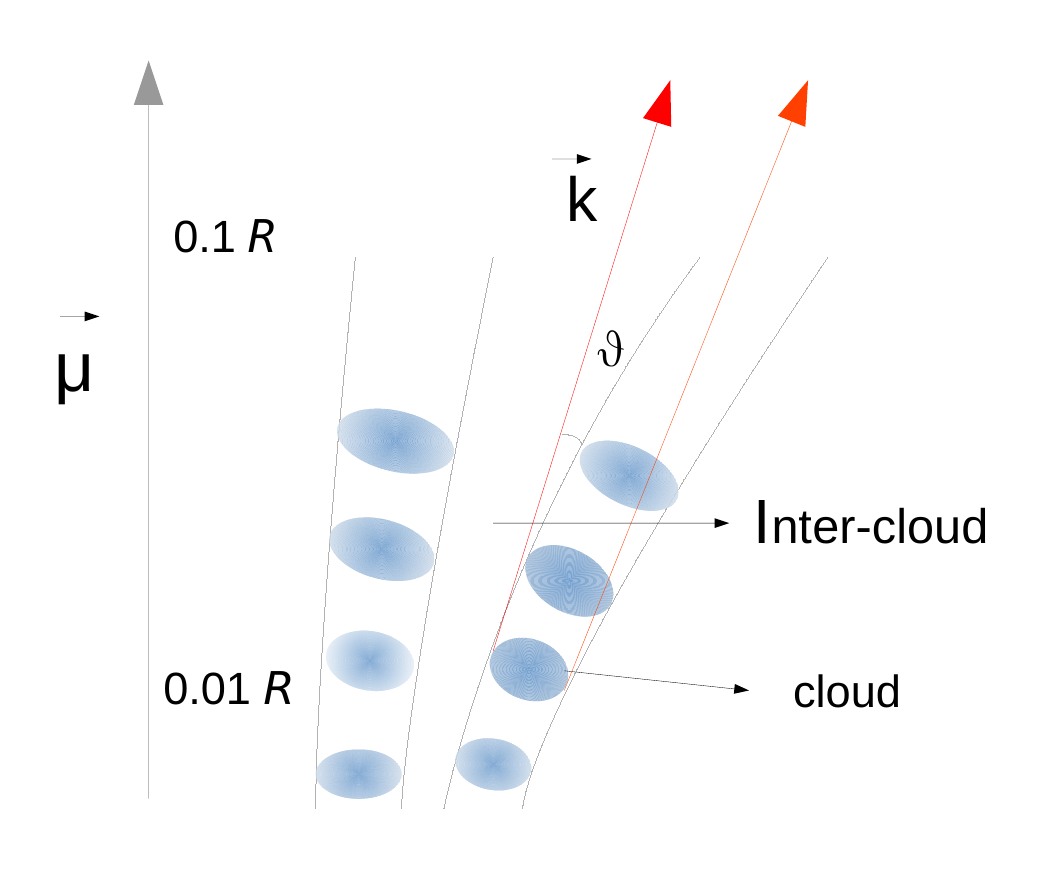}
\caption{The above cartoon shows the open magnetic field line of a pulsar 
between heights of 0.01$\mathcal{R}$ and 0.1$\mathcal{R}$, that correspond to 
radio emission region. The grey lines depict the dipolar magnetic field lines 
and the shaded blue correspond to plasma clouds. A series of plasma clouds 
move along the open magnetic field lines due to the non-stationary flow setup 
by the sparking discharges in the IAR. The white background correspond to the 
inter-cloud region where the plasma density is lower. CCR is excited in plasma 
clouds and the propagation vector $\vec{k}$ is depicted as solid red lines. The
waves escape to the inter-cloud region above 0.1$\mathcal{R}$ to eventually 
reach the observer. $\vec{k}$ makes an angle $\vartheta$ with respect to the local
	dipole magnetic field, $\vec{B}$, given by Eq.~\ref{theta}. $\vec{\mu}$ 
	correspond to the dipole magnetic axis.}
\label{car}
\end{figure}
     
GMG03 used observational constraints to estimate $\eta \approx 0.1$ and $T_s 
\approx 2 \times 10^6$ K. Associating $T_s$ with the bombardment of electrons 
at the polar cap surface, an estimate for the Lorentz factor of the positron 
beam can be obtained as $\gamma_{p}^{sp} \sim 10^6$, where the superscript 
`$sp$' specifies the sparking region. The positron beam further cascades and 
produces secondary pairs with multiplicity $\kappa^{sp}$. Several pair cascade 
simulations suggest $\kappa^{sp} \sim 10^4$, and this gives an estimate for the
secondary pair plasma Lorentz factor as $\gamma_{s}^{sp} \sim 
\gamma_p^{sp}/2\kappa^{sp}\sim 10^2$. Since the ions are much heavier, the 
corresponding Lorentz factor for the ions are $\gamma_{ion}^{sp} \approx 10^3$.
In the inter-spark region the potential has a minimum value $\Delta V_{min}$ 
and there is no pair discharge. GMG03 estimated $\Delta V_{min}$ which
accelerates the ion in the inter-spark region to characteristic Lorentz factor 
of $\gamma_{ion}^{isp}\sim10$ (see Appendix \ref{apex2a}), here superscript 
`$isp$' specifies the inter-spark region. In the inter-spark region the pair 
plasma is produced primarily from stray high energy photons and has 
$\kappa^{isp} \sim 1$ and $\gamma_{s}^{isp} \sim 10^2$. Each spark associated 
plasma column consists of intermittently outflowing dense plasma clouds along 
the open, curved, dipolar magnetic field lines. Thus the radio emission region 
alternates between dense plasma clouds and low density inter-cloud region (see 
Fig.~\ref{car}).

\begin{figure}
\gridline{\fig{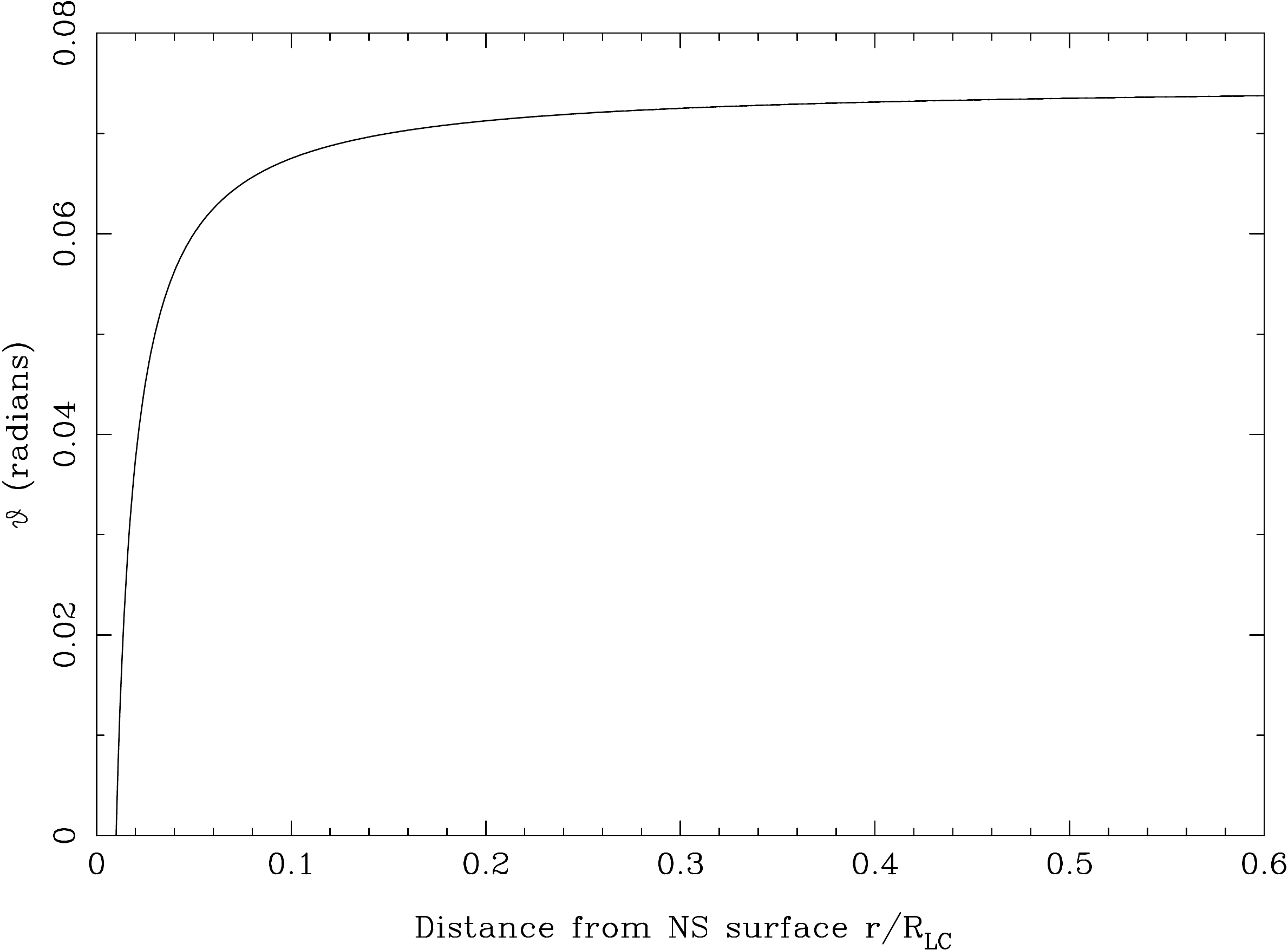}{0.45\textwidth}{}
          \fig{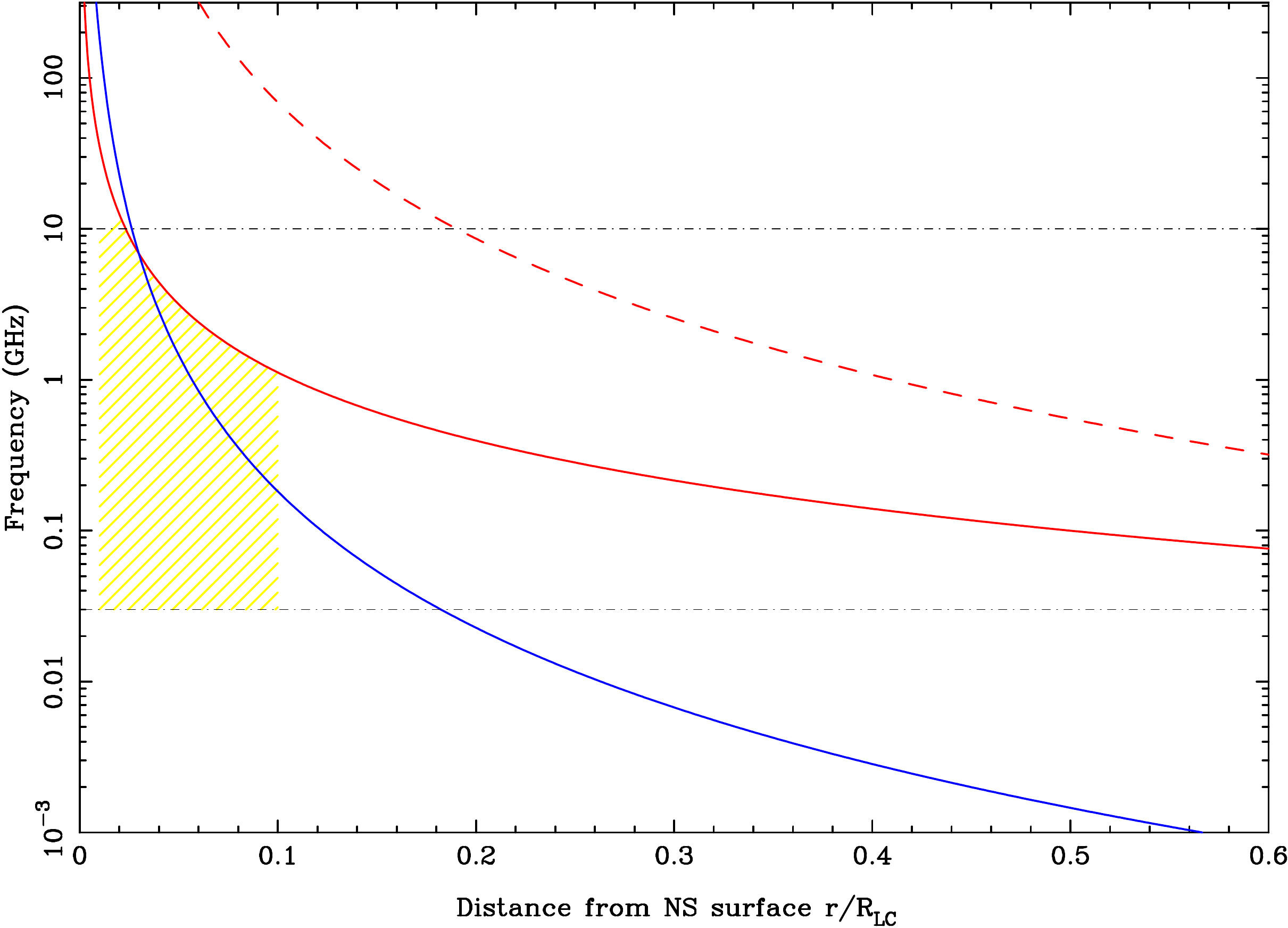}{0.45\textwidth}{}
         }
\caption{The angle between the propagation vector, $\vec{k}$, and the local 
magnetic field $\vec{B}$, is represented by $\vartheta$, and its variation with
altitude, $\mathcal{R}$, is shown in the left panel (see Eq.~\ref{theta}). The 
altitude is represented in terms of the light cylinder radius and we assume the
emission to originate at $\mathcal{R}_1=0.01$ where $\vartheta=0$. The right 
panel shows the altitude evolution of resonance frequency of the pair plasma, 
$\nu_{res,p}$ (dashed red line, Eq.~\ref{p3}), the resonance frequency of ions, 
$\nu_{res,ion}$ (blue line, Eq.~\ref{p3}) and the emission frequency excited by
CCR, $\nu_r$ (red line, Eq.~\ref{p4}). The shaded area (in yellow) represents 
the region from which the radio emission can be generated, with the radio 
frequency range typically between 20 MHz and 10 GHz (horizontal dashed lines).}
\label{fig4a}
\end{figure}

\subsection{The origin of the nature of pulsar polarization}
The CCR from charge bunches (or solitons) are excited inside the dense
pair-plasma clouds at distances of about 1\% of the light cylinder radius. The
Lorentz factors of solitons have the same order of magnitude as
$\gamma_s^{sp}$, with CCR exciting the $lt_1$-mode and $t$-mode at the radio
frequency range, $\nu_{cr} < 2 \sqrt{\gamma_s^{sp}} \nu_{p}^{sp}$. The plasma
modes are linearly polarized and oriented parallel and perpendicular to the
magnetic field planes, respectively. The power in the $t$-mode is seven times
smaller than the $lt_1$-mode. The refractive indices of the two modes are
different and they split and propagate as independent modes through the plasma
clouds. The $t$-mode has vacuum like properties in the dense plasma clouds and
can escape as electromagnetic X-mode to reach the observer. On the other hand
the magnetic field lines act like ducts for the $lt_1$-mode to propagate along
them, and it eventually undergoes Landau damping. As a result the $lt_1$-mode
cannot escape a homogeneous plasma. However, if large density gradients exist
such that in regions of lower pair multiplicity the condition $\nu_{cr} > 
2\sqrt{\gamma_s^{sp}} \nu_{p}^{sp}$ is satisfied, then the $lt_1$-mode is able
escape and reach the observer. This is precisely the nature of the 
non-stationary plasma flow in pulsars that originates from a PSG, and near the 
boundary of the clouds both the $t$-mode and $lt_1$-mode escape into the 
inter-cloud region as X-mode and O-mode, respectively.

In the wave generation region, located well inside the pulsar magnetosphere, 
the wave vector $\vec{k}$ is directed almost tangent to the magnetic field 
lines, i.e. the angle between them $\vartheta\simeq 0$, while the opening angle
of a single charge bunch is about $1/\gamma _{s}^{sp}$. As the magnetic field 
lines are curved, $\vartheta$ increases with distance as the waves propagate 
and when $\vartheta > 1/\gamma_{s}^{sp}$ the emission enters the inter-cloud 
region where the plasma density is lower. The resultant emission has 
contribution from a large number of charge bunches inside the dense plasma 
column, such that the angular size of the profile component, having typical 
widths of around 2\degr, can be associated with the width of the plasma column.
We use polar co-ordinates ($\mathcal{R}$, $\theta$) to specify the wave 
generation region, where $\mathcal{R}$ is defined in terms of light cylinder
radius. If the location of the wave generation point, where $\vec{k} \parallel 
\vec{B}_{1}$, is specified as ($\mathcal{R}_{1}$, $\theta_{1}$), such that 
$\vartheta_1=0$, then at an altitude $\mathcal{R}$ the angle $\vartheta$ is 
given as (see Appendix \ref{apex2b})
\begin{equation}
\vartheta =\frac{3}{4}\left[ \theta _{1}-2\sin^{-1} \left(
\frac{\mathcal{R}_{1}}{\mathcal{R}} \sin{\cfrac{\theta_1}{2}}\right) \right] .
\label{theta}
\end{equation}
The variation of $\vartheta$ with $\mathcal{R}$ obtained from Eq.~\ref{theta} 
is shown in Fig.~\ref{fig4a} (left panel). At an altitude of $\mathcal{R} = 
0.1R_{lc}$ (10\% of the light cylinder radius), $\vartheta \approx 0.07$ radian
or about $4^{\circ}$, which is greater than the profile component widths. This
clearly suggests that around 20\% of light cylinder radius $\vartheta$ is large 
enough for the emission to enter the inter-cloud region (see cartoon in 
Fig.~\ref{car}). 

Once the waves enter the inter-cloud region, both the X-mode and the O-mode 
propagate following the dispersion relation of the $t$-mode, $\omega = kc(1 - 
\omega_p^2/4\gamma^3\omega_B^2)$. The waves can undergo \v{C}erenkov as well
as cyclotron damping, with the resonance condition of the form,
\begin{equation}
\omega_{res} - \vec{k}.{\vec{v}} - \frac{1}{\gamma} \omega_B = 0
\label{p1}
\end{equation}
Here $\omega_{res}$ is the resonance frequency and $\vec{v}$ is the velocity of 
the resonant particles that are confined to move along the local magnetic 
field. Using the dispersion relation of the $t$-mode the resonance condition 
can be used to find the resonance frequency,  
\begin{equation}
\nu_{res} = \frac{\omega_B}{2\pi \gamma} \left(\frac{\vartheta^2}{2} 
+\frac{1}{2\gamma^2} - \frac{\omega_p^2}{4\gamma^3\omega_B^2}\right)^{-1}
\label{p2}
\end{equation}
where we have used $\vec{k}\cdot\vec{v} = k v \cos{\vartheta} = kc (1 - 
\vartheta^2/2 - 1/2\gamma^2)$. The resonance frequency of the pair plasma, 
$\nu_{res,p}$, and ions, $\nu_{res,ion}$, in the inter-cloud region as a 
function of $\mathcal{R}$ is given as
\begin{eqnarray}
\nu_{res,p} & = & \frac{1.69 \times 10^{5}}
{\mathcal{R}^3(2.45\times10^{-3} - 3.76 \times 10^{-16}\mathcal{R}^3)} ~~\rm{GHz}, \nonumber \\
\nu_{res,ion} & = & \frac{1.273 \times 10^{-6}}
{\mathcal{R}^3(7\times10^{-3} - 4.06 \times 10^{-8}\mathcal{R}^3)} ~~\rm{GHz}.
\label{p3}
\end{eqnarray}
Here, typical values of the parameters $P=\dot{P}_{-15}=1$ was used, and in the
inter-cloud region $\gamma_{s}^{isp}=300$, $\kappa^{isp} =1$ for the pair 
plasma, while $\gamma_{ion}^{isp}=10$ for ions. 

If the frequency of CCR, $\nu_r$, coincide with the resonance frequency 
then they will undergo damping. Inside the plasma cloud the condition for the 
emission of radio frequencies can be expressed in terms of $\mathcal{R}$ as,
\begin{equation}
\nu_r < 2\sqrt{\gamma_{s}^{sp}}\nu_p = 0.035/\mathcal{R}^{1.5},
\label{p4}
\end{equation}
where typical values of the pair plasma, $\kappa^{sp} = 10^4$ and 
$\gamma_{s}^{sp} = 300$, have been used. The change in resonance frequencies 
(Eq.~\ref{p3}) as well as the characteristic plasma frequency (Eq.~\ref{p4}), 
with $\mathcal{R}$, have been shown in Fig.~\ref{fig4a} (right panel), where 
the shaded area (in yellow) specifies the radio emission region within the 
magnetosphere. At distances above 0.1$\mathcal{R}$ the broadband pulsar 
radiation can detach from the dense plasma cloud and propagate in the 
inter-cloud region. The resonance conditions shown in Fig.~\ref{fig4a} indicate
that the plasma waves can travel a large distance in the inter-cloud region 
before crossing the resonance point. \citet{1982SvAL....8..369M} showed that in
a homogeneous plasma the waves generated inside the magnetosphere will undergo 
cyclotron damping and cannot escape the pulsar. However, when inhomogeneity is 
introduced in the plasma, as explained above, the resonance condition only 
becomes applicable higher up in the pulsar magnetosphere for the waves 
propagating in the inter-cloud region. In the upper magnetosphere the pair 
plasma density decreases significantly and hence the effect of damping will be 
much smaller. In this work we have used a a static dipole model, which is not 
adequate for describing the magnetic field structure in the upper 
magnetosphere, where magnetic sweep-back effects dominate. Therefore, a 
comprehensive analysis of the damping effects will require a more detailed 
model of the upper magnetosphere and will be addressed in future studies.

\begin{figure}
\epsscale{0.7}
\plotone{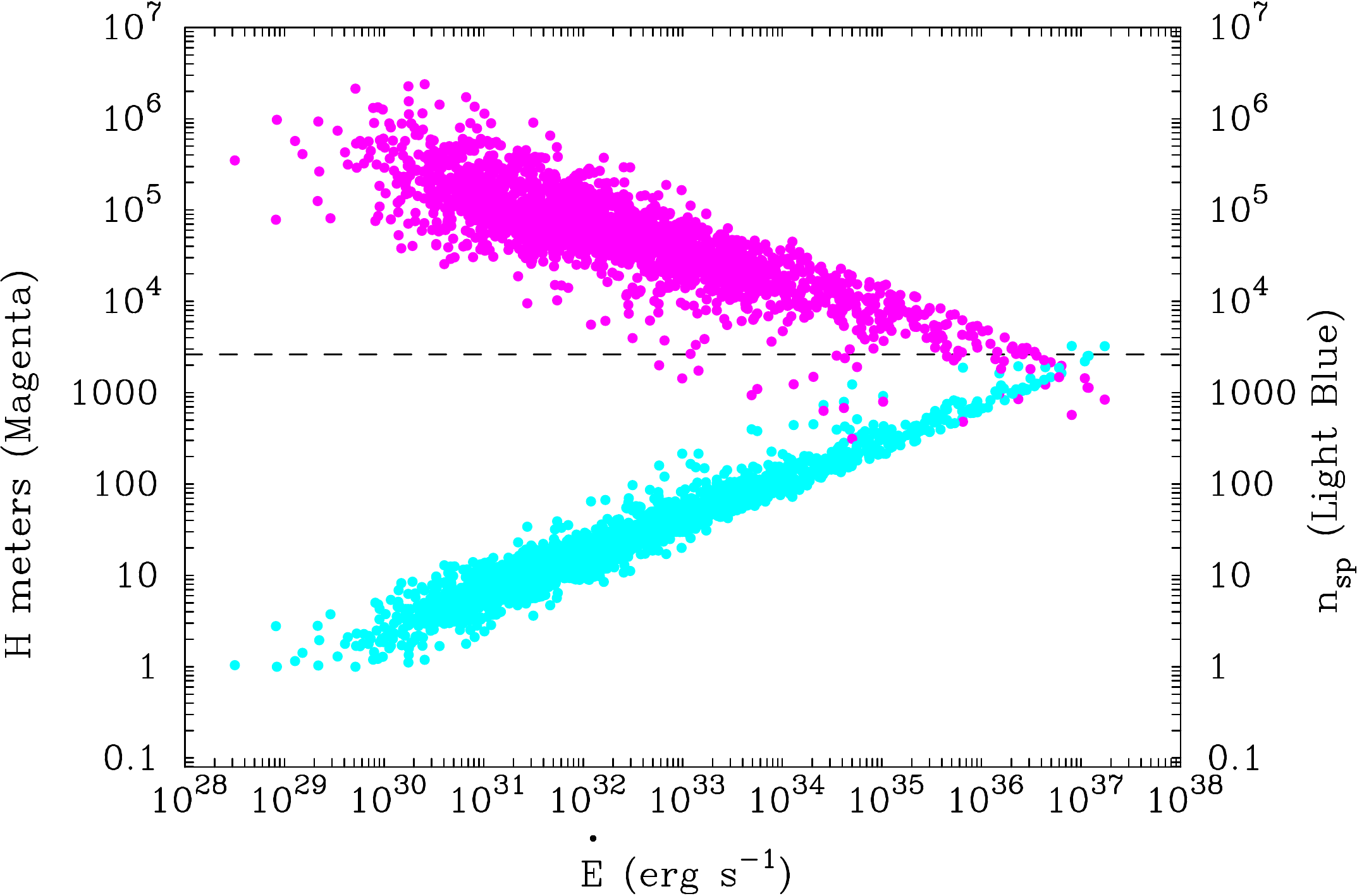}
\caption{The change in the inter-cloud separation with $\dot{E}$ is shown for 
pulsars with $P > 0.1$ s (Magenta). These estimates are carried out at the
typical emission height $r=50r_{\ast}$ (see Eq.~\ref{icdist}). The figure also 
shows the variation of number of sparks, $N_{sp}$, and as an extension the 
number of plasma columns in the open field line region, as a function of 
$\dot{E}$ (light blue). In the low $\dot{E}$ range the estimated $N_{sp}$ is 
less than unity in certain pulsars, and has been approximated as a single 
spark in the plot. The charge bunches in the plasma clouds emit within a narrow
cone with angular size $(3/2)(1/\gamma_{s}^{sp})$. The horizontal line (dashed)
shows the linear extension of the emission cone at $50r_{\ast}$.}
\label{fig4b}
\end{figure}

\subsection{Origin of Linear Polarization}

We propose that the observed linear polarization features in pulsars arise due
to incoherent addition of the X-mode and O-mode of CCR, excited by the large 
number of charge bunches in the plasma clouds. The $t$-mode can easily escape 
from the inter-cloud region and appear as the X-mode with high levels of linear
polarization (see \citealt{2022MNRAS.512.3589R}). The $lt_1$-mode, 
which is seven times stronger than $t$-mode, is 
mostly damped inside the plasma cloud, but can also escape as a dominant mode 
from the inter-cloud boundary region as the linearly polarized O-mode (\citealt{1986ApJ...302..120A,
2004ApJ...600..872G,2014ApJ...794..105M}). At any 
instant of time there are multiple plasma clouds that emit X-mode and O-mode 
with different intensities, that undergo incoherent averaging in the 
inter-cloud region. This process gives us the resultant linear polarization and 
the PPA distribution observed from pulsars and depends on the level of
inhomogeneity in the magnetospheric plasma.

A measure of the inhomogeneity in the plasma comes from the spark size, 
$h_{\perp}$, and number of sparks, $N_{sp}$, in the PSG. Eq.~\ref{hnsp} gives 
an estimate of $h_{\perp}$ and $N_{sp}$ as well as their variation in the 
pulsar population, specified by the dependence on $\dot{E}$. The inter-cloud 
separation, $H$, in the emission region $r$ can be calculated from $h_{\perp}$ 
at the surface $r_\ast$ as : 
\begin{equation}
H = b h_{\perp} \left(\frac{r}{r_\ast}\right)^{1.5}
\label{icdist}
\end{equation}
Fig.~\ref{fig4b} shows the variation of $H$ (magenta), estimated at $50r_\ast$,
and $N_{sp}$ (light blue) with $\dot{E}$ for normal pulsars with $P\ge0.1$ s. 
We have used typical parameters of a PSG, $\eta=0.1$, $b=10$, $T_6=2$, 
$\alpha_m = 45\degr$, for these estimates, and the pulsar parameters were 
obtained from the ATNF pulsar database \citep{2005AJ....129.1993M}. $N_{sp}$ 
increases significantly with $\dot{E}$, ranging from one expected spark at 
10$^{29}$ erg~s$^{-1}$ to around 10$^3$ sparks at 10$^{36}$ erg~s$^{-1}$, while 
$H$ has a decreasing trend ranging from 10$^6$ m to 10$^3$ m over the same 
range of $\dot{E}$. The charge bunches emit CCR tangential to the curved 
magnetic field lines over a narrow cone with angular extent 
$(3/2)(1/\gamma_{s}^{sp})$ rad. The size of the emission cone at $50r_\ast$ is 
around 2000 m (horizontal line in Fig.~\ref{fig4b}) and lies below $H$ for 
almost the entire population. 

In low $\dot{E}$ pulsars, with lower number of plasma clouds and larger 
separation between them, once the emission enters the inter-cloud region there 
is sufficient gap for the $lt_1$-mode to escape without getting damped by 
encountering another plasma cloud. For example PSR J0034$-$0721 shown in 
Fig.~\ref{fig1} has low $\dot{E} \sim 1.9 \times 10^{31}$ ergs/s and the estimated
$H \sim 1.3 \times 10^6$ m and $N_{sp} \sim 8$, and clearly
the gap size is orders of magnitude larger than the emission cone.
As a result both $t$-mode and $lt_1$-mode
can escape as X-mode and O-mode and give rise to the orthogonal PPA traverse 
that follow the RVM, similar to what is observed in PSR J0034$-$0721. 
In contrast, high $\dot{E}$ pulsars have large number of
closely packed clouds and have higher likelihood of absorbing the $lt_1$-mode. 
The situation here can be applied to PSR J0742$-$2822 of Fig.~\ref{fig1} with $\dot{E} = 1.4 \times 10^{35}$ ergs/s.
The estimated $H \sim 8400$ m is of the same order of the emission cone thus 
reducing the gap for escape of $lt_1$-mode and further $N_{sp} \sim 308$ is large.
Thus in this case only highly polarized time samples with a single PPA traverse 
following the RVM are observed. In certain intermediate cases 
it is possible for both $t$-mode and $lt_1$-mode to escape as X-mode and O-mode
from plasma clouds which add up incoherently and show large variation in the 
degree of polarization with significant spread in PPA distribution. 
PSR J1752$-$2806 shown in Fig.~\ref{fig1} with $\dot{E}=1.8 \times 10^{33}$ ergs/s correspond to such a case.
The estimated $H \sim 2 \times 10^{5}$ m falls in the middle range compared to above
two cases, and $N_{sp} \sim 63$. In this situation it appears that mode mixing
can occur as observed in Fig.~\ref{fig1}.
The above 
model also predicts that the PPA from highly polarized time samples should 
follow the RVM, that have been identified in certain pulsars 
\citep{2009ApJ...696L.141M,2023MNRAS.tmpL..22M}. Incoherent averaging of CCR 
due to a large number of coherently emitting single charge bunches has also 
been explored in earlier studies to explain the observed depolarization and 
corresponding PPA behaviour \citep{1985Ap&SS.109..381G,1990A&A...234..269G}. 
Our present scheme involves a more consistent model of plasma clouds emerging 
from sparks in a PSG, with several other implications, like high surface 
temperatures in the polar cap region, presence of non-dipolar fields near the 
stellar surface, subpulse drifting of sparks, etc., and have various degrees of
observational verifications. However, understanding the full spectrum of the 
observed polarization features using this model requires detailed simulations, 
which will be explored in a separate study.

\subsection{Circular Polarization as a propagation effect}
The circularly polarized waves require additional conditions that can break the
symmetry between the positive and negative components of the electron-positron 
pair plasma. This can likely be realised from the small differences in the 
distribution functions of electrons and positrons as well as the presence of 
small amounts of positively charged Iron ions in the plasma. \citet[][hereafter 
KMM91]{1991MNRAS.253..377K} used the difference in distribution functions of pair plasma to
explain the presence of circularly polarized modes. We use the standard 
textbook definitions to investigate the percentage circular polarization in 
plasma waves \citep[see e.g.][]{1967RvPP....3....1S}. We have used a 
co-ordinate system where one axis is directed along the propagation vector, 
$\vec{k}$, and electric vector is of the form $\vec{E}=E_{y}(ia_{x},1,ia_{z})$,
here $a_{x}$ governs the polarization of the waves. The dispersion relation in 
this system is given as 
\begin{eqnarray}
a_{x}^{2}-\Theta a_{x}-1 = 0,  \nonumber \\
\Theta =\cfrac{\left( \eta -1\right) +g^{2}}{g\eta \cos \vartheta }\sin
^{2}\vartheta ,  \label{3}
\end{eqnarray}
here, $\vartheta $ is the angle between $\vec{k}$ and $\vec{B}$, while $\eta$ 
and $g$ depend on the parameters of the plasma (see Appendix~\ref{apex3}). The 
solution of this equation has the form  
\begin{equation}
a_{x}=\frac{1}{2}\Theta \pm \sqrt{\left( \frac{1}{2}\Theta \right) ^{2}+1}.
\end{equation}
When $\Theta \ll 1$ then $a_{x}\simeq \pm 1$ and there is circular 
polarization, while $\Theta \gg 1$ implies either $a_{x}=0$ or $a_{x}\gg 1$ and
there are two linearly polarized waves. 

In the model proposed by KMM91 the radio emission is generated near the light 
cylinder, around the last open field lines, and the waves leave the 
magnetosphere immediately after their emission, such that their polarization 
properties are preserved. It is clear from Eq.~\ref{3} that $\Theta$ is 
directly dependent on $\vartheta$. As a result in this model the core 
component, located at the center of the profile, i.e. $\vartheta \ll 1$, should
show circular polarization, while the conal components that are located 
farther away are expected to be linearly polarized. However, this is not 
adequate to explain the observed circular polarization since the radio emission
is excited in the inner inner magnetosphere around 0.01$\mathcal{R}$. 

The PSG model gives rise to a system of spark associated plasma clouds where 
the waves are generated at the emission heights and subsequently enter the 
inter-cloud region above 0.1$\mathcal{R}$. The multiplicity factor of the 
pair-plasma density, $\kappa$, shows considerable variation across the cloud 
and inter-cloud regions, with maximum value at the center of the cloud that 
decreases monotonically to much lower values in the middle of the inter-cloud 
region. The Lorentz factor of the ions follow a similar profile, having highest
value at cloud center and the lowest in the middle of the inter-cloud region. 
The wave propagates in the inter-cloud region at an angle $\vartheta \sim 0.07$
rad, which remains constant over large distances in the inner magnetosphere 
(see Fig.~\ref{fig4a}). The parameter $\Theta$ has the following form in the 
inter-cloud region (see Appendix ~\ref{apex3})
\begin{equation}
\Theta =-\cfrac{I}{g}\sin^{2}\vartheta =1.8\times10^{10}~\mathcal{R}^3 \cfrac{\kappa^{isp}}{{(\gamma _{s}^{isp})^3}} \left(\cfrac{P^5}{\dot{P}_{-15}}\right)^{0.5} \omega_{G}.
\end{equation}
Here, $(\gamma_{s}^{isp})^{-3}=\langle 1/(\gamma_{s}^{isp})^3\rangle$ is the 
average Lorentz factor of the electron-positron component of the plasma, and 
$\omega_{G}$ is the radio frequency in GHz. Using typical parameters of the 
inter-cloud region at $\mathcal{R}=0.1$, i.e. $\gamma_{s}^{isp}\approx300$, 
$\kappa^{isp}=P=\dot{P}_{-15}=\omega_G=1$, and assuming $1/\gamma_{ion}^{isp} >
\sin\vartheta$, we obtain the absolute value $\mid\Theta\mid \approx 0.67$. 
We can estimate $\mid \Theta
\mid $ for the three pulsars shown in Fig.~\ref{fig1}, using similar values
for $\mathcal{R}\sim 0.1$, $\gamma _{s}^{isp}\approx 300$ and $\kappa
^{isp}=1$ and the observing frequency $\omega _{G}=0.6.$ Then for younger
pulsars PSR J0742$-$2822 (with $P=0.167$ s and $\dot{P}_{-15}=16.8$) and PSR
J1752$-$2806  (with $P=0.563$ s and $\dot{P}_{-15}=8.13$ )  value of $\mid
\Theta \mid $ is much less then one ($0.001$ and $0.03$
respectively). For relatively older pulsar PSR J0034$-$0721 (with $P=0.943$
s and $\dot{P}_{-15}=0.408$) $\mid \Theta \mid \sim 0.5$ . Thus, it can be
seen that the necessary condition for circular polarization  $\mid 
\Theta \mid \ll 1$ can be easily satisfied for each case.

The above discussion shows that the electromagnetic waves passing through the 
inter-cloud region will be elliptically polarized, with the degree of 
polarization depending on the values of $a_x$. If $a_x=1$ then the waves
should have 100\% circular polarization, while for either smaller or bigger 
values of $a_x$ the level of linear polarization increases. The observed levels
of circular polarization will ultimately depend on the multiplicity profile 
between the inter-cloud region and incoherent addition of the elliptically 
polarized emission from several plasma clouds, with the sign of the circular 
polarization depending on the viewing geometry.

\section{Conclusion} \label{sec5}
We have carried out a systematic analysis of the AP and SP polarization 
properties of the normal pulsar population observed in MSPES. We have estimated
the radio emission heights in several cases using the A/R method and found the 
radio emission to arise below 10\% of the light cylinder radius. The emission
heights appear to be constant across a wide period range, which is consistent 
with earlier studies \citep{2017JApA...38...52M,2022arXiv221111849P}. The SP
polarization behaviour show that in general the time samples have high levels 
of linear polarization. The depolarization in AP arises due to incoherent 
addition of polarized time samples from the SP, resulting in significant spread 
of the PPA distributions. The average percentage linear polarization from SP 
do not show significant variation across the $\dot{E}$ range, which is in 
sharp contrast with the AP polarization levels, where high $\dot{E}$ pulsars 
have much higher levels of linear polarization compared to lower $\dot{E}$ 
cases. The percentage of circular polarization on the other hand do not show
any significant variations between the SP and AP measurements.  

There is increasing observational evidence that the radio emission arises due
to CCR from charge bunches in the outflowing plasma. The charge bunches develop
in clouds of plasma formed due to sparking discharges in a PSG, and is 
primarily composed of dense electron-positron pair plasma, mixed with small
amounts of relativistic iron ions, with varying densities and Lorentz factor 
in the plasma clouds and inter-cloud regions. We have explored the origin of 
the observed polarization behaviour due to CCR from charge bunches in this 
plasma configuration. The relativistic charge bunches can excite the $t$-mode 
and $lt_1$-mode that have linear polarizations directed perpendicular and 
parallel to the curved dipolar magnetic field line planes, respectively. The 
resultant emission from the plasma cloud is due to incoherent averaging of 
emission from a large number of charge bunches. The $t$-mode and $lt_1$-mode 
escape into the inter-cloud region at around 10\% of the light cylinder radius,
where the pair multiplicity is much lower as the ions have smaller Lorentz
factors. The modes can propagate as electromagnetic waves corresponding to the 
X-mode and O-mode, and further incoherent averaging takes place. We show that 
the waves remains in the inter-cloud region with the angle between the 
propagation vector almost reaching a constant value, and due to propagation 
effects the modes attain elliptical polarization below 20\% of the light 
cylinder radius. As the plasma density decreases with height, the waves are
eventually able to escape the magnetosphere from the inter-cloud region and 
reach the observer. However, the magnetic field sweep-back effects become 
important at heights above 50\% of the light cylinder radius and introduce 
effects like cyclotron damping, which requires more detailed considerations.

\section*{Acknowledgments}
We thank the anonymous referee for helpful comments that improved the 
quality of the paper. We thank the staff of the GMRT that made these observations possible.
D.M. acknowledges the support of the Department of Atomic Energy, Government of
India, under project No. 12-R\&D-TFR-5.02-0700. D.M. acknowledges funding from 
the grant ``Indo-French Centre for the Promotion of Advanced Research - 
CEFIPRA'' grant IFC/F5904-B/2018. This work was supported by grant 
2020/37/B/ST9/02215 from the National Science Center, Poland.




\bibliography{References}{}

\begin{thebibliography}{}
\expandafter\ifx\csname natexlab\endcsname\relax\def\natexlab#1{#1}\fi
\providecommand{\url}[1]{\href{#1}{#1}}
\providecommand{\dodoi}[1]{doi:~\href{http://doi.org/#1}{\nolinkurl{#1}}}
\providecommand{\doeprint}[1]{\href{http://ascl.net/#1}{\nolinkurl{http://ascl.net/#1}}}
\providecommand{\doarXiv}[1]{\href{https://arxiv.org/abs/#1}{\nolinkurl{https://arxiv.org/abs/#1}}}

\bibitem[{{Allen} \& {Melrose}(1982)}]{1982PASA....4..365A}
{Allen}, M.~C., \& {Melrose}, D.~B. 1982, \pasa, 4, 365,
  \dodoi{10.1017/S1323358000021147}

\bibitem[{{Arons} \& {Barnard}(1986)}]{1986ApJ...302..120A}
{Arons}, J., \& {Barnard}, J.~J. 1986, \apj, 302, 120, \dodoi{10.1086/163978}

\bibitem[{{Asseo} \& {Melikidze}(1998)}]{1998MNRAS.301...59A}
{Asseo}, E., \& {Melikidze}, G.~I. 1998, \mnras, 301, 59,
  \dodoi{10.1046/j.1365-8711.1998.01990.x}

\bibitem[{{Basu} {et~al.}(2022{\natexlab{a}}){Basu}, {Melikidze}, \&
  {Mitra}}]{2022ApJ...936...35B}
{Basu}, R., {Melikidze}, G.~I., \& {Mitra}, D. 2022{\natexlab{a}}, \apj, 936,
  35, \dodoi{10.3847/1538-4357/ac8479}

\bibitem[{{Basu} {et~al.}(2020{\natexlab{a}}){Basu}, {Mitra}, \&
  {Melikidze}}]{2020ApJ...889..133B}
{Basu}, R., {Mitra}, D., \& {Melikidze}, G.~I. 2020{\natexlab{a}}, \apj, 889,
  133, \dodoi{10.3847/1538-4357/ab63c9}

\bibitem[{{Basu} {et~al.}(2020{\natexlab{b}}){Basu}, {Mitra}, \&
  {Melikidze}}]{2020MNRAS.496..465B}
---. 2020{\natexlab{b}}, \mnras, 496, 465, \dodoi{10.1093/mnras/staa1574}

\bibitem[{{Basu} {et~al.}(2021){Basu}, {Mitra}, \&
  {Melikidze}}]{2021ApJ...917...48B}
---. 2021, \apj, 917, 48, \dodoi{10.3847/1538-4357/ac0828}

\bibitem[{{Basu} {et~al.}(2022{\natexlab{b}}){Basu}, {Mitra}, \&
  {Melikidze}}]{2022ApJ...927..208B}
---. 2022{\natexlab{b}}, \apj, 927, 208, \dodoi{10.3847/1538-4357/ac5039}

\bibitem[{{Basu} {et~al.}(2023){Basu}, {Mitra}, \&
  {Melikidze}}]{2023arXiv230312229B}
---. 2023, arXiv e-prints, arXiv:2303.12229, \dodoi{10.48550/arXiv.2303.12229}

\bibitem[{{Basu} {et~al.}(2016){Basu}, {Mitra}, {Melikidze}, {Maciesiak},
  {Skrzypczak}, \& {Szary}}]{2016ApJ...833...29B}
{Basu}, R., {Mitra}, D., {Melikidze}, G.~I., {et~al.} 2016, \apj, 833, 29,
  \dodoi{10.3847/1538-4357/833/1/29}

\bibitem[{{Basu} {et~al.}(2019){Basu}, {Mitra}, {Melikidze}, \&
  {Skrzypczak}}]{2019MNRAS.482.3757B}
{Basu}, R., {Mitra}, D., {Melikidze}, G.~I., \& {Skrzypczak}, A. 2019, \mnras,
  482, 3757, \dodoi{10.1093/mnras/sty2846}

\bibitem[{{Blaskiewicz} {et~al.}(1991){Blaskiewicz}, {Cordes}, \&
  {Wasserman}}]{1991ApJ...370..643B}
{Blaskiewicz}, M., {Cordes}, J.~M., \& {Wasserman}, I. 1991, \apj, 370, 643,
  \dodoi{10.1086/169850}

\bibitem[{{Cheng} \& {Ruderman}(1977)}]{CR77}
{Cheng}, A.~F., \& {Ruderman}, M.~A. 1977, \apj, 214, 598,
  \dodoi{10.1086/155285}

\bibitem[{{Cheng} \& {Ruderman}(1980)}]{CR80}
---. 1980, \apj, 235, 576, \dodoi{10.1086/157661}

\bibitem[{{Everett} \& {Weisberg}(2001)}]{2001ApJ...553..341E}
{Everett}, J.~E., \& {Weisberg}, J.~M. 2001, \apj, 553, 341,
  \dodoi{10.1086/320652}

\bibitem[{{Gil} {et~al.}(2004){Gil}, {Lyubarsky}, \&
  {Melikidze}}]{2004ApJ...600..872G}
{Gil}, J., {Lyubarsky}, Y., \& {Melikidze}, G.~I. 2004, \apj, 600, 872,
  \dodoi{10.1086/379972}

\bibitem[{{Gil} {et~al.}(2003){Gil}, {Melikidze}, \&
  {Geppert}}]{2003A&A...407..315G}
{Gil}, J., {Melikidze}, G.~I., \& {Geppert}, U. 2003, \aap, 407, 315,
  \dodoi{10.1051/0004-6361:20030854}

\bibitem[{{Gil} \& {Rudnicki}(1985)}]{1985Ap&SS.109..381G}
{Gil}, J., \& {Rudnicki}, W. 1985, \apss, 109, 381, \dodoi{10.1007/BF00651284}

\bibitem[{{Gil} \& {Snakowski}(1990)}]{1990A&A...234..269G}
{Gil}, J.~A., \& {Snakowski}, J.~K. 1990, \aap, 234, 269

\bibitem[{{Ginzburg} \& {Zhelezniakov}(1975)}]{1975ARA&A..13..511G}
{Ginzburg}, V.~L., \& {Zhelezniakov}, V.~V. 1975, \araa, 13, 511,
  \dodoi{10.1146/annurev.aa.13.090175.002455}

\bibitem[{{Goldreich} \& {Julian}(1969)}]{1969ApJ...157..869G}
{Goldreich}, P., \& {Julian}, W.~H. 1969, \apj, 157, 869,
  \dodoi{10.1086/150119}

\bibitem[{{Gould} \& {Lyne}(1998)}]{1998MNRAS.301..235G}
{Gould}, D.~M., \& {Lyne}, A.~G. 1998, \mnras, 301, 235,
  \dodoi{10.1046/j.1365-8711.1998.02018.x}

\bibitem[{{Johnston} \& {Kerr}(2018)}]{2018MNRAS.474.4629J}
{Johnston}, S., \& {Kerr}, M. 2018, \mnras, 474, 4629,
  \dodoi{10.1093/mnras/stx3095}

\bibitem[{{Johnston} {et~al.}(2023){Johnston}, {Kramer}, {Karastergiou},
  {Keith}, {Oswald}, {Parthasarathy}, \& {Weltevrede}}]{2023MNRAS.520.4801J}
{Johnston}, S., {Kramer}, M., {Karastergiou}, A., {et~al.} 2023, \mnras, 520,
  4801, \dodoi{10.1093/mnras/stac3636}

\bibitem[{{Karastergiou} \& {Johnston}(2007)}]{2007MNRAS.380.1678K}
{Karastergiou}, A., \& {Johnston}, S. 2007, \mnras, 380, 1678,
  \dodoi{10.1111/j.1365-2966.2007.12237.x}

\bibitem[{{Kazbegi} {et~al.}(1991){Kazbegi}, {Machabeli}, \&
  {Melikidze}}]{1991MNRAS.253..377K}
{Kazbegi}, A.~Z., {Machabeli}, G.~Z., \& {Melikidze}, G.~I. 1991, \mnras, 253,
  377, \dodoi{10.1093/mnras/253.3.377}

\bibitem[{{Kijak} \& {Gil}(1997)}]{1997MNRAS.288..631K}
{Kijak}, J., \& {Gil}, J. 1997, \mnras, 288, 631,
  \dodoi{10.1093/mnras/288.3.631}

\bibitem[{{Lakoba} {et~al.}(2018){Lakoba}, {Mitra}, \&
  {Melikidze}}]{2018MNRAS.480.4526L}
{Lakoba}, T., {Mitra}, D., \& {Melikidze}, G. 2018, \mnras, 480, 4526,
  \dodoi{10.1093/mnras/sty2152}

\bibitem[{{Lominadze} {et~al.}(1986){Lominadze}, {Machabeli}, {Melikidze}, \&
  {Pataraia}}]{1986FizPl..12.1233L}
{Lominadze}, D.~G., {Machabeli}, G.~Z., {Melikidze}, G.~I., \& {Pataraia},
  A.~D. 1986, Fizika Plazmy, 12, 1233

\bibitem[{{Manchester} {et~al.}(2005){Manchester}, {Hobbs}, {Teoh}, \&
  {Hobbs}}]{2005AJ....129.1993M}
{Manchester}, R.~N., {Hobbs}, G.~B., {Teoh}, A., \& {Hobbs}, M. 2005, \aj, 129,
  1993, \dodoi{10.1086/428488}

\bibitem[{{Manchester} {et~al.}(1975){Manchester}, {Taylor}, \&
  {Huguenin}}]{1975ApJ...196...83M}
{Manchester}, R.~N., {Taylor}, J.~H., \& {Huguenin}, G.~R. 1975, \apj, 196, 83,
  \dodoi{10.1086/153395}

\bibitem[{{Manthei} {et~al.}(2021){Manthei}, {Ben{\'a}{\v{c}}ek}, {Mu{\~n}oz},
  \& {B{\"u}chner}}]{2021A&A...649A.145M}
{Manthei}, A.~C., {Ben{\'a}{\v{c}}ek}, J., {Mu{\~n}oz}, P.~A., \&
  {B{\"u}chner}, J. 2021, \aap, 649, A145, \dodoi{10.1051/0004-6361/202039907}

\bibitem[{{Melikidze} {et~al.}(2000){Melikidze}, {Gil}, \&
  {Pataraya}}]{2000ApJ...544.1081M}
{Melikidze}, G.~I., {Gil}, J.~A., \& {Pataraya}, A.~D. 2000, \apj, 544, 1081,
  \dodoi{10.1086/317220}

\bibitem[{{Melikidze} {et~al.}(2014){Melikidze}, {Mitra}, \&
  {Gil}}]{2014ApJ...794..105M}
{Melikidze}, G.~I., {Mitra}, D., \& {Gil}, J. 2014, \apj, 794, 105,
  \dodoi{10.1088/0004-637X/794/2/105}

\bibitem[{{Melikidze} \& {Pataraia}(1980)}]{1980Afz....16..161M}
{Melikidze}, G.~I., \& {Pataraia}, A.~D. 1980, Astrofizika, 16, 161

\bibitem[{{Melrose}(1979)}]{1979AuJPh..32...61M}
{Melrose}, D.~B. 1979, Australian Journal of Physics, 32, 61,
  \dodoi{10.1071/PH790061}

\bibitem[{{Melrose}(1995)}]{1995JApA...16..137M}
---. 1995, Journal of Astrophysics and Astronomy, 16, 137,
  \dodoi{10.1007/BF02714830}

\bibitem[{{Mikhailovskii} {et~al.}(1982){Mikhailovskii}, {Onishchenko},
  {Suramlishvili}, \& {Sharapov}}]{1982SvAL....8..369M}
{Mikhailovskii}, A.~B., {Onishchenko}, O.~G., {Suramlishvili}, G.~I., \&
  {Sharapov}, S.~E. 1982, Soviet Astronomy Letters, 8, 369

\bibitem[{{Mitra}(2017)}]{2017JApA...38...52M}
{Mitra}, D. 2017, Journal of Astrophysics and Astronomy, 38, 52,
  \dodoi{10.1007/s12036-017-9457-6}

\bibitem[{{Mitra} {et~al.}(2015){Mitra}, {Arjunwadkar}, \&
  {Rankin}}]{2015ApJ...806..236M}
{Mitra}, D., {Arjunwadkar}, M., \& {Rankin}, J.~M. 2015, \apj, 806, 236,
  \dodoi{10.1088/0004-637X/806/2/236}

\bibitem[{{Mitra} {et~al.}(2016){Mitra}, {Basu}, {Maciesiak}, {Skrzypczak},
  {Melikidze}, {Szary}, \& {Krzeszowski}}]{2016ApJ...833...28M}
{Mitra}, D., {Basu}, R., {Maciesiak}, K., {et~al.} 2016, \apj, 833, 28,
  \dodoi{10.3847/1538-4357/833/1/28}

\bibitem[{{Mitra} {et~al.}(2020){Mitra}, {Basu}, {Melikidze}, \&
  {Arjunwadkar}}]{2020MNRAS.492.2468M}
{Mitra}, D., {Basu}, R., {Melikidze}, G.~I., \& {Arjunwadkar}, M. 2020, \mnras,
  492, 2468, \dodoi{10.1093/mnras/stz3620}

\bibitem[{{Mitra} \& {Deshpande}(1999)}]{1999A&A...346..906M}
{Mitra}, D., \& {Deshpande}, A.~A. 1999, \aap, 346, 906.
\newblock \doarXiv{astro-ph/9904336}

\bibitem[{{Mitra} {et~al.}(2009){Mitra}, {Gil}, \&
  {Melikidze}}]{2009ApJ...696L.141M}
{Mitra}, D., {Gil}, J., \& {Melikidze}, G.~I. 2009, \apjl, 696, L141,
  \dodoi{10.1088/0004-637X/696/2/L141}

\bibitem[{{Mitra} \& {Li}(2004)}]{2004A&A...421..215M}
{Mitra}, D., \& {Li}, X.~H. 2004, \aap, 421, 215,
  \dodoi{10.1051/0004-6361:20034094}

\bibitem[{{Mitra} {et~al.}(2023){Mitra}, {Melikidze}, \&
  {Basu}}]{2023MNRAS.tmpL..22M}
{Mitra}, D., {Melikidze}, G.~I., \& {Basu}, R. 2023, \mnras,
  \dodoi{10.1093/mnrasl/slad022}

\bibitem[{{Mitra} \& {Rankin}(2002)}]{2002ApJ...577..322M}
{Mitra}, D., \& {Rankin}, J.~M. 2002, \apj, 577, 322, \dodoi{10.1086/342136}

\bibitem[{{Mitra} \& {Rankin}(2011)}]{2011ApJ...727...92M}
---. 2011, \apj, 727, 92, \dodoi{10.1088/0004-637X/727/2/92}

\bibitem[{{Mitra} \& {Seiradakis}(2004)}]{2004hell.conf..205M}
{Mitra}, D., \& {Seiradakis}, J.~H. 2004, in Hellenic Astronomical Society
  Sixth Astronomical Conference, ed. P.~{Laskarides}, 205.
\newblock \doarXiv{astro-ph/0401335}

\bibitem[{{Morris} {et~al.}(1981){Morris}, {Graham}, \&
  {Sieber}}]{1981A&A...100..107M}
{Morris}, D., {Graham}, D.~A., \& {Sieber}, W. 1981, \aap, 100, 107

\bibitem[{{Olszanski} {et~al.}(2019){Olszanski}, {Mitra}, \&
  {Rankin}}]{2019MNRAS.489.1543O}
{Olszanski}, T. E.~E., {Mitra}, D., \& {Rankin}, J.~M. 2019, \mnras, 489, 1543,
  \dodoi{10.1093/mnras/stz2172}

\bibitem[{{P{\'e}tri} \& {Mitra}(2020)}]{2020MNRAS.491...80P}
{P{\'e}tri}, J., \& {Mitra}, D. 2020, \mnras, 491, 80,
  \dodoi{10.1093/mnras/stz2974}

\bibitem[{{Posselt} {et~al.}(2022){Posselt}, {Karastergiou}, {Johnston},
  {Parthasarathy}, {Oswald}, {Main}, {Basu}, {Keith}, {Song}, {Weltevrede},
  {Tiburzi}, {Bailes}, {Buchner}, {Geyer}, {Kramer}, {Spiewak}, \& {Venkatraman
  Krishnan}}]{2022arXiv221111849P}
{Posselt}, B., {Karastergiou}, A., {Johnston}, S., {et~al.} 2022, arXiv
  e-prints, arXiv:2211.11849, \dodoi{10.48550/arXiv.2211.11849}

\bibitem[{{Radhakrishnan} \& {Cooke}(1969)}]{1969ApL.....3..225R}
{Radhakrishnan}, V., \& {Cooke}, D.~J. 1969, \aplett, 3, 225

\bibitem[{{Rahaman} {et~al.}(2020){Rahaman}, {Mitra}, \&
  {Melikidze}}]{2020MNRAS.497.3953R}
{Rahaman}, S.~M., {Mitra}, D., \& {Melikidze}, G.~I. 2020, \mnras, 497, 3953,
  \dodoi{10.1093/mnras/staa2280}

\bibitem[{{Rahaman} {et~al.}(2022{\natexlab{a}}){Rahaman}, {Mitra}, \&
  {Melikidze}}]{2022MNRAS.512.3589R}
---. 2022{\natexlab{a}}, \mnras, 512, 3589, \dodoi{10.1093/mnras/stac696}

\bibitem[{{Rahaman} {et~al.}(2022{\natexlab{b}}){Rahaman}, {Mitra},
  {Melikidze}, \& {Lakoba}}]{2022MNRAS.516.3715R}
{Rahaman}, S.~M., {Mitra}, D., {Melikidze}, G.~I., \& {Lakoba}, T.
  2022{\natexlab{b}}, \mnras, 516, 3715, \dodoi{10.1093/mnras/stac2264}

\bibitem[{{Rankin}(1983)}]{1983ApJ...274..333R}
{Rankin}, J.~M. 1983, \apj, 274, 333, \dodoi{10.1086/161450}

\bibitem[{{Rankin}(1986)}]{1986ApJ...301..901R}
---. 1986, \apj, 301, 901, \dodoi{10.1086/163955}

\bibitem[{{Rankin}(1993)}]{1993ApJ...405..285R}
---. 1993, \apj, 405, 285, \dodoi{10.1086/172361}

\bibitem[{{Ruderman} \& {Sutherland}(1975)}]{1975ApJ...196...51R}
{Ruderman}, M.~A., \& {Sutherland}, P.~G. 1975, \apj, 196, 51,
  \dodoi{10.1086/153393}

\bibitem[{{Shafranov}(1967)}]{1967RvPP....3....1S}
{Shafranov}, V.~D. 1967, Reviews of Plasma Physics, 3, 1

\bibitem[{{Shapakidze} {et~al.}(2003){Shapakidze}, {Machabeli}, {Melikidze}, \&
  {Khechinashvili}}]{2003PhRvE..67b6407S}
{Shapakidze}, D., {Machabeli}, G., {Melikidze}, G., \& {Khechinashvili}, D.
  2003, \pre, 67, 026407, \dodoi{10.1103/PhysRevE.67.026407}

\bibitem[{{Song} {et~al.}(2023){Song}, {Weltevrede}, {Szary}, {Wright},
  {Keith}, {Basu}, {Johnston}, {Karastergiou}, {Main}, {Oswald},
  {Parthasarathy}, {Posselt}, {Bailes}, {Buchner}, {Hugo}, \&
  {Serylak}}]{2023arXiv230104067S}
{Song}, X., {Weltevrede}, P., {Szary}, A., {et~al.} 2023, arXiv e-prints,
  arXiv:2301.04067, \dodoi{10.48550/arXiv.2301.04067}

\bibitem[{{Sturrock}(1971)}]{1971ApJ...164..529S}
{Sturrock}, P.~A. 1971, \apj, 164, 529, \dodoi{10.1086/150865}

\bibitem[{{Swarup} {et~al.}(1991){Swarup}, {Ananthakrishnan}, {Kapahi}, {Rao},
  {Subrahmanya}, \& {Kulkarni}}]{1991CuSc...60...95S}
{Swarup}, G., {Ananthakrishnan}, S., {Kapahi}, V.~K., {et~al.} 1991, Current
  Science, 60, 95

\bibitem[{{Volokitin} {et~al.}(1985){Volokitin}, {Krasnoselskikh}, \&
  {Machabeli}}]{1985FizPl..11..531V}
{Volokitin}, A.~S., {Krasnoselskikh}, V.~V., \& {Machabeli}, G.~Z. 1985, Fizika
  Plazmy, 11, 531

\bibitem[{{von Hoensbroech} \&
  {Xilouris}(1997{\natexlab{a}})}]{1997A&A...324..981V}
{von Hoensbroech}, A., \& {Xilouris}, K.~M. 1997{\natexlab{a}}, \aap, 324, 981

\bibitem[{{von Hoensbroech} \&
  {Xilouris}(1997{\natexlab{b}})}]{1997A&AS..126..121V}
---. 1997{\natexlab{b}}, \aaps, 126, 121

\bibitem[{{Wang} {et~al.}(2010){Wang}, {Lai}, \& {Han}}]{2010MNRAS.403..569W}
{Wang}, C., {Lai}, D., \& {Han}, J. 2010, \mnras, 403, 569,
  \dodoi{10.1111/j.1365-2966.2009.16074.x}

\bibitem[{{Weltevrede} {et~al.}(2006){Weltevrede}, {Edwards}, \&
  {Stappers}}]{2006A&A...445..243W}
{Weltevrede}, P., {Edwards}, R.~T., \& {Stappers}, B.~W. 2006, \aap, 445, 243,
  \dodoi{10.1051/0004-6361:20053088}

\bibitem[{{Weltevrede} \& {Johnston}(2008)}]{2008MNRAS.391.1210W}
{Weltevrede}, P., \& {Johnston}, S. 2008, \mnras, 391, 1210,
  \dodoi{10.1111/j.1365-2966.2008.13950.x}

\end{thebibliography}
\bibliographystyle{aasjournal}

\appendix
\section{Some useful definitions} 
\label{apex1}
The plasma frequency $\omega_p$ and cyclotron frequency $\omega_{B}$ in 
electron positron plasma is given as 
\begin{eqnarray}
\omega_p  & =  & \sqrt{ \frac{4\pi {q_e}^2 n }{m_e}}\\
          & =  & 6.4151\times10^{4}~\kappa^{0.5} \mathcal{R}^{-1.5}P^{-1.75} \dot{P}_{15}^{0.25}
\end{eqnarray}
Here the number density $n = \kappa n_{GJ}$, where $n_{GJ}$ is the 
Goldreich-Julien density.
\begin{eqnarray}
\omega_{B} & = & \frac{e B}{m_e c}\\
           & = & 3.1834\times 10^{8} ~\mathcal{R}^{-3} P^{-2.5} {\dot{P}_{15}}^{0.5} 
\end{eqnarray}
Estimates of the plasma frequency corresponding to ions, $\omega_{p,ion}$, and 
the cyclotron frequency, $\omega_{B,ion}$, using $^{56} Fe_{26}$,
\begin{eqnarray}
\omega _{p,ion} &=&1.02\times 10^{3} ~\mathcal{R}^{-1.5} P^{-1.75} {\dot{P}_{-15}}^{0.25} \\
\omega _{B,ion} &=&8\times 10^{4} ~\mathcal{R}^{-3} P^{-2.5} {\dot{P}_{-15}}^{0.5} 
\end{eqnarray}

\section{Lorentz factor of $^{56}$Fe$_{26}$ ions in inter-spark region}\label{apex2a}
In the inter-spark region the gap potential decreases and as a result the 
Lorentz factors of the outflowing plasma is lower. A minimum potential $\Delta 
V_{min}$ can be defined where the sparking discharge terminates and the Lorentz
factor of the ions $^{56}Fe_{26}$ can be written as, 
\begin{equation}
\gamma _{Fe}=\Delta V_{\min }\left( \cfrac{26q_{e}}{56m_{p}c^{2}}\right)
\end{equation}
In the PSG model an estimate of $\Delta V_{min} \approx 10^8$ V was obtained 
by \citet{2003A&A...407..315G}, and the minimum Lorentz factor required for 
pair production dominated by inverse Compton scattering at near threshold 
condition was (see appendix A.1 therein) \\
$\gamma _{Fe}= 10^{8} \left( \cfrac{26q_{e}}{56m_{p}c^{2}}\right) \sim 14$.

\section{Evolution of propagation angle} \label{apex2b}
At the emission region $\vec{k}$ is almost parallel to local $\vec{B}$. It
means that in the wave generation region, which is defined as a region which
corresponds to $r=r_{1}$ and $\theta =\theta _{1},$ angle between $k$ and 
$\vec{B}$ is about $0$, here $r$ is radial distance in the units of the
stellar radius. The angle $\zeta $ between $\vec{k}$ and magnetic axes 
$\vec{\mu}$ is $\zeta =\frac{3}{2}\theta_{1}=\frac{3}{2}\theta_{p}r_{1}^{0.5}$,
here $\theta_{p}=1.4477\times 10^{-2}P^{-0.5}$ is a coordinate of the last open
field line at $r_{1}=1$. The radio waves propagate along $\vec{k}$ (let us note
that this direction does not coincide with radius vector direction) and thus we
need to calculate the angle $\vartheta $ between $\vec{k}$ and local magnetic 
field direction at some altitude $(r,\theta)$. In spherical coordinates the 
equation of the line in the plane of the curved magnetic field plane can be 
expressed as
\begin{equation}
r\cos\theta = ar\sin\theta + b
\label{a21}
\end{equation}
Here at $r_1$, 
\begin{equation}
a = \cot \cfrac{3}{2}\theta_{1}
\end{equation}
\begin{equation}
b=r_{1}\left[\cos\theta_{1} -\cot\left(\frac{3\theta _{1}}{2}\right) \sin\theta_{1} \right] =r_{1}~\frac{\sin \left( \cfrac{\theta_{1}}{2}\right) }{\sin\left(\cfrac{3\theta_{1}}{2}\right) }
\end{equation}
Then,
\begin{equation}
\vartheta =\cfrac{3}{2}\left( \theta -\theta _{1}\right) = \frac{3}{2}\left[\frac{1}{2}\theta _{1}-\sin^{-1} \left(\frac{r_{1}}{r}\sin \cfrac{\theta_{1}}{2} \right) \right] 
\end{equation}
\begin{equation}
\vartheta =\frac{3}{4}\left[ \theta_{1} - 2\sin^{-1}\left( \frac{\mathcal{R}_{1}}{\mathcal{R}}\sin\cfrac{\theta_{1}}{2} \right) \right].
\end{equation}

\section{Estimation of the parameter $\Theta$} \label{apex3}

We consider a wave with electric field $\vec{E}$ where the 
propagation vector $\vec{k}$ is directed along $z-$axes and $\vartheta$ 
is the angle between $\vec{B}$ and $\vec{k}$.
To describe the wave polarization we introduce the following notation:

\begin{equation}
\vec{E}=E_{y}\left( ia_{x},1,ia_{z}\right) 
\end{equation}

where $ia_{x}$ and $ia_{z}$ are the ratio of the $x$ and $z$ components
of $\vec{E}$ to its $y$ component $E_y$.
Following \cite{1967RvPP....3....1S} and \cite{1991MNRAS.253..377K}
we need to define the coefficients of the equation that describes the wave
polarisation as

\begin{eqnarray}
\eta _{xx} &=&\cfrac{\varepsilon _{1}\eta }{\varepsilon _{1}\sin ^{2}\vartheta
+\eta \cos ^{2}\vartheta } \\
\eta _{xy} &=&-\eta _{yx}=i\cfrac{g\eta \cos \vartheta }{\varepsilon _{1}\sin
^{2}\vartheta +\eta \cos ^{2}\vartheta } \\
\eta _{yy} &=&\cfrac{\varepsilon _{2}\left( \varepsilon _{1}\sin ^{2}\vartheta
+\eta \cos ^{2}\vartheta \right) -g^{2}\sin ^{2}\vartheta }{\varepsilon _{1}\sin
^{2}\vartheta +\eta \cos ^{2}\vartheta }
\end{eqnarray}
For the case of plasma in the strong magnetic field $\varepsilon
_{1}=\varepsilon_{2}=1$. The coefficients of the permitivity tensor are%
\begin{eqnarray}
\eta &=&1-\sum\limits_{\alpha }\omega _{p_{\alpha }}^{2}\int \cfrac{%
f_{\alpha }\left( p\right) }{\gamma ^{3}}dp\cfrac{1}{\left( \omega -\vec{k}%
\vec{v}\right) ^{2}}\equiv 1-I \\
g &=&\sum\limits_{\alpha }\cfrac{\omega _{p_{\alpha }}^{2}}{\omega ^{2}}%
\int \cfrac{f_{\alpha }\left( p\right) dp}{\gamma }\left( \omega -kv\right)
A_{\alpha }^{-} \\
A_{\alpha }^{-} &=&\cfrac{1}{\omega -kv+\gamma ^{-1}\omega _{B\alpha }}-%
\cfrac{1}{\omega -kv-\gamma ^{-1}\omega _{B\alpha }}
\end{eqnarray}

Here $\alpha$ and $\gamma$ correspond to the species and Lorentz factor
of the species and $v$ is the velocity of the plasma. 

\begin{eqnarray}
 \eta _{xx} & = & \cfrac{1-I}{1-I\cos ^{2}\vartheta}\\
 \eta _{yy} & = & \cfrac{\sin ^{2}\vartheta +\eta \cos ^{2}\vartheta -g^{2}\sin ^{2}\vartheta }{1-I\cos^{2}\vartheta }\\
 \cfrac{\eta _{xx}-\eta _{yy}}{\eta _{yx}}& = & \cfrac{-I+g^{2}}{g\eta \cos \vartheta }\sin ^{2}\vartheta 
\end{eqnarray}

Thus the equation for $a_{x}$ is
\begin{equation}
a_{x}^{2}-\Theta a_{x}-1=0
\end{equation}
Here $\Theta$ is given by,
\begin{eqnarray}
\Theta = & \cfrac{\eta _{xx}-\eta _{yy}}{i\eta _{yx}} = & \cfrac{\left( \eta -1\right) +g^{2}}{g\eta \cos \vartheta }\sin^{2}\vartheta 
\end{eqnarray}

The solution is $a_{x}=\frac{1}{2}\Theta \pm \sqrt{\left( \frac{1}{2}\Theta \right)^{2}+1}$. Now if $\Theta \ll 1$ then $a_{x}\simeq \pm 1,$ so the 
polarization is circular. If $\Theta \gg 1$ then $a_{x}=0$ or $a_{x}\gg 1$. 
Thus there are two linearly polarised waves.

In the present case we apply the above theory to determine the effect
of propagation in the inter-cloud region where   
we assume that $\sin \vartheta \gg \left\langle \cfrac{1}{\gamma_{s}^{isp}}
\right\rangle$ and $\sin \vartheta <\left\langle \cfrac{1}{\gamma_{ion}^{isp}} 
\right\rangle.$
Further, in the low density inter-cloud plasma we can assume that the 
dispersion relation for all waves are electromagnetic, such that $\omega \approx kc$.
Thus one can write $\left( \omega ^{2}-k^{2}v^{2}\right)
=\left( \omega ^{2}-k^{2}v^{2}\cos ^{2}\vartheta \right) =\omega ^{2}\left(
\sin ^{2}\vartheta +\cfrac{1}{\gamma ^{2}}\right)$

Under this assumption $A_{\alpha}^{-}$ is dominated by the ion component, 
and we can estimate 
\begin{equation}
(A_{ion}^{isp})^{-}=8(\gamma _{ion}^{isp})^{3}
\cfrac{\omega _{B,ion}}{\omega ^{2}}.
\end{equation}
The angle $\vartheta $ in the inter-cloud region is almost constant and 
$\vartheta\approx 0.07$, and $g$ is dominated by the ion component while $I$ is
dominated by the pair plasma. Finally we can estimate $g$ and $I$ (see Appendix
\ref{apex1} for parameters) as
\begin{eqnarray}
g & = &2\left(\cfrac{\omega_{B,ion}~\omega_{p,{ion}}^{2}}{\omega^{3}}\right)\\ 
I & = & \cfrac{1}{(\gamma _{s}^{isp})^{3}}\cfrac{\omega _{p}^{2}}{\omega ^{2}}\cfrac{4}{\sin ^{4}\vartheta }
\end{eqnarray}
Therefore, $\Theta = -\cfrac{I}{g}\sin^{2}\vartheta  = 17.77~\kappa\mathcal{R}^{3} \cfrac{\omega}{(\gamma_{s}^{isp})^{3}} \left(\cfrac{ {P}^{5}}{\dot{P}_{-15}}\right)^{0.5}$, and introducing $\omega_G = \cfrac{\omega}{10^{9}}$ 
\begin{equation}
\Theta =-1.8\times10^{10} ~\kappa \mathcal{R}^{3} \cfrac{1}{(\gamma _{s}^{isp})^{3}}\left(\cfrac{P^{5}}{\dot{P}_{-15}}\right)^{0.5} \omega_{G}
\end{equation}

\section{Supplementary Figures}
\label{supp}

\begin{figure*}
\begin{center}
\begin{tabular}{cc}
{\mbox{\includegraphics[width=9cm,height=6cm,angle=0.]{J0034-0721_317MHz_16Feb14.dat.epn2a.pdf}}}&
{\mbox{\includegraphics[width=9cm,height=6cm,angle=0.]{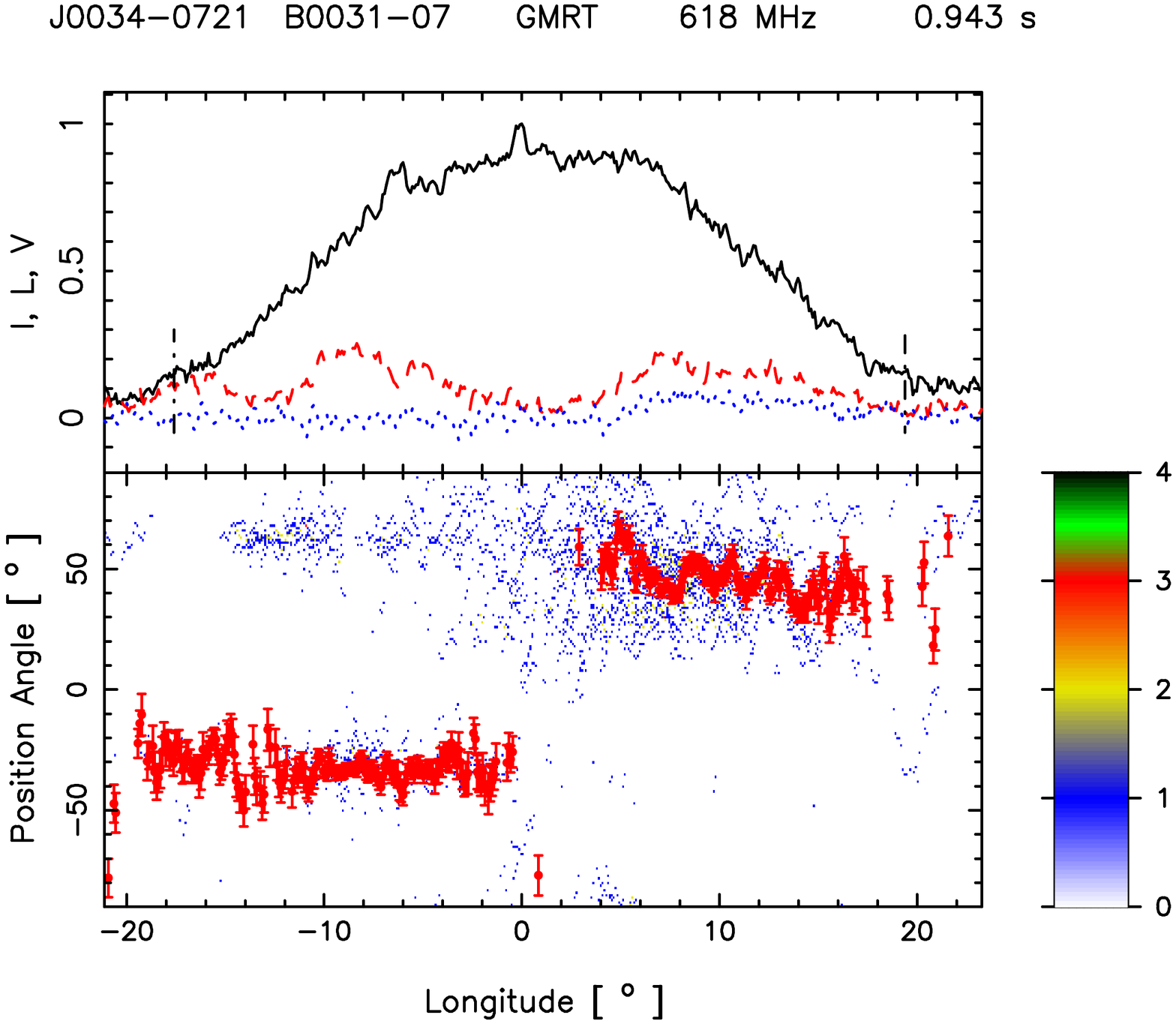}}}\\
{\mbox{\includegraphics[width=9cm,height=6cm,angle=0.]{J0034-0721_317MHz_16Feb14.dat.orderf.dat_chisq.pdf}}}&
\\
{\mbox{\includegraphics[width=9cm,height=6cm,angle=0.]{J0034-0721_317MHz_16Feb14.dat.epn2a.71.pdf}}}&
{\mbox{\includegraphics[width=9cm,height=6cm,angle=0.]{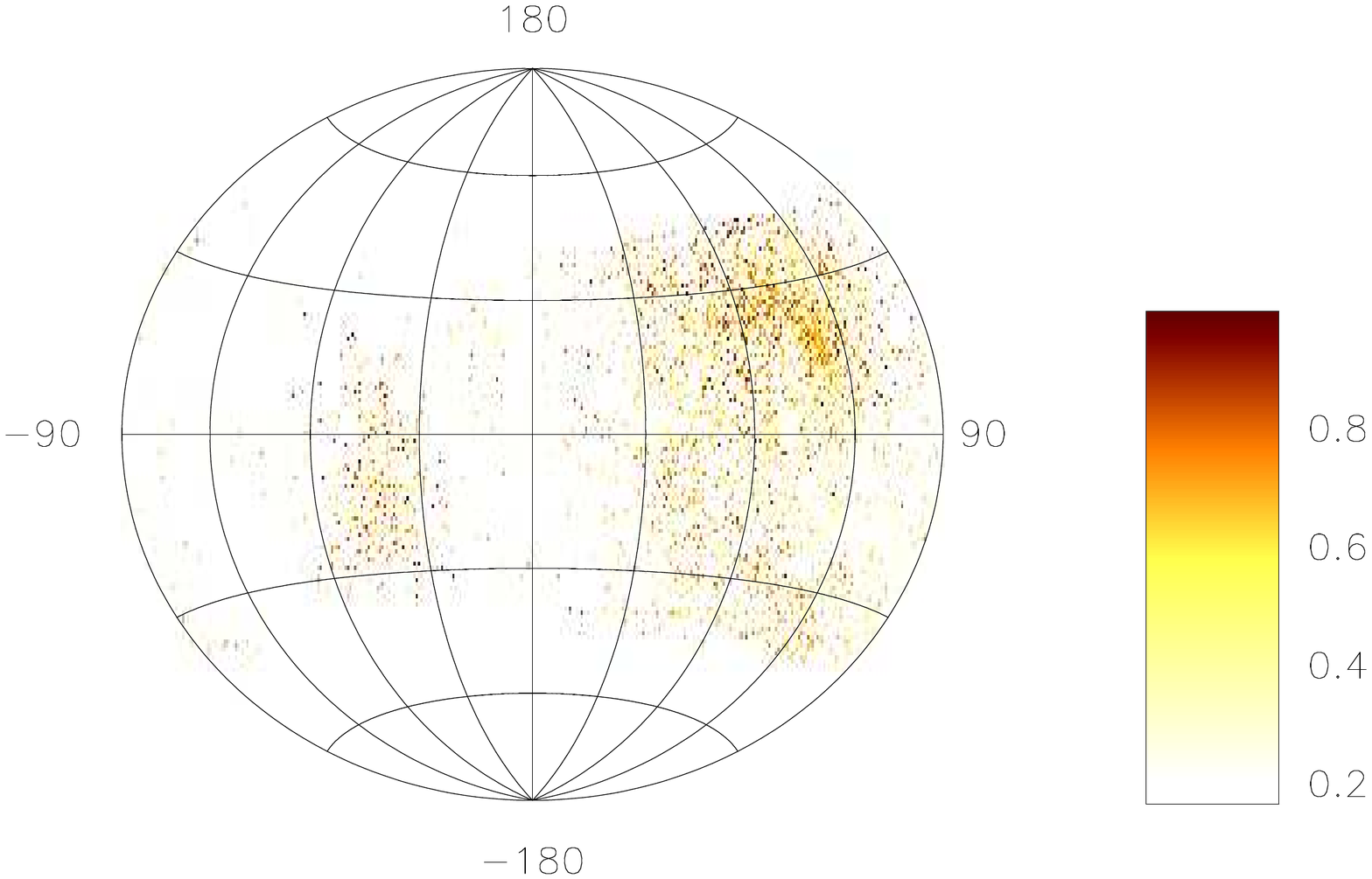}}}\\
\end{tabular}
\caption{ Top panel (upper window) shows the average profile with total
intensity (Stokes I; solid black lines), total linear polarization (dashed red
line) and circular polarization (Stokes V; dotted blue line). Top panel (lower
window) also shows the single pulse PPA distribution (colour scale) along with
the average PPA (red error bars). The RVM fits to the average PPA (dashed pink
line) is also shown in this plot. Middle panel show
the $\chi^2$ contours only for 333 MHz and the correlated parameters $\alpha$ and $\beta$ obtained from RVM
fits. Bottom panel shows the Hammer-Aitoff projection of the polarized time
samples with the colour scheme representing the fractional polarization level.}
\label{a1}
\end{center}
\end{figure*}


\begin{figure*}
\begin{center}
\begin{tabular}{cc}
{\mbox{\includegraphics[width=9cm,height=6cm,angle=0.]{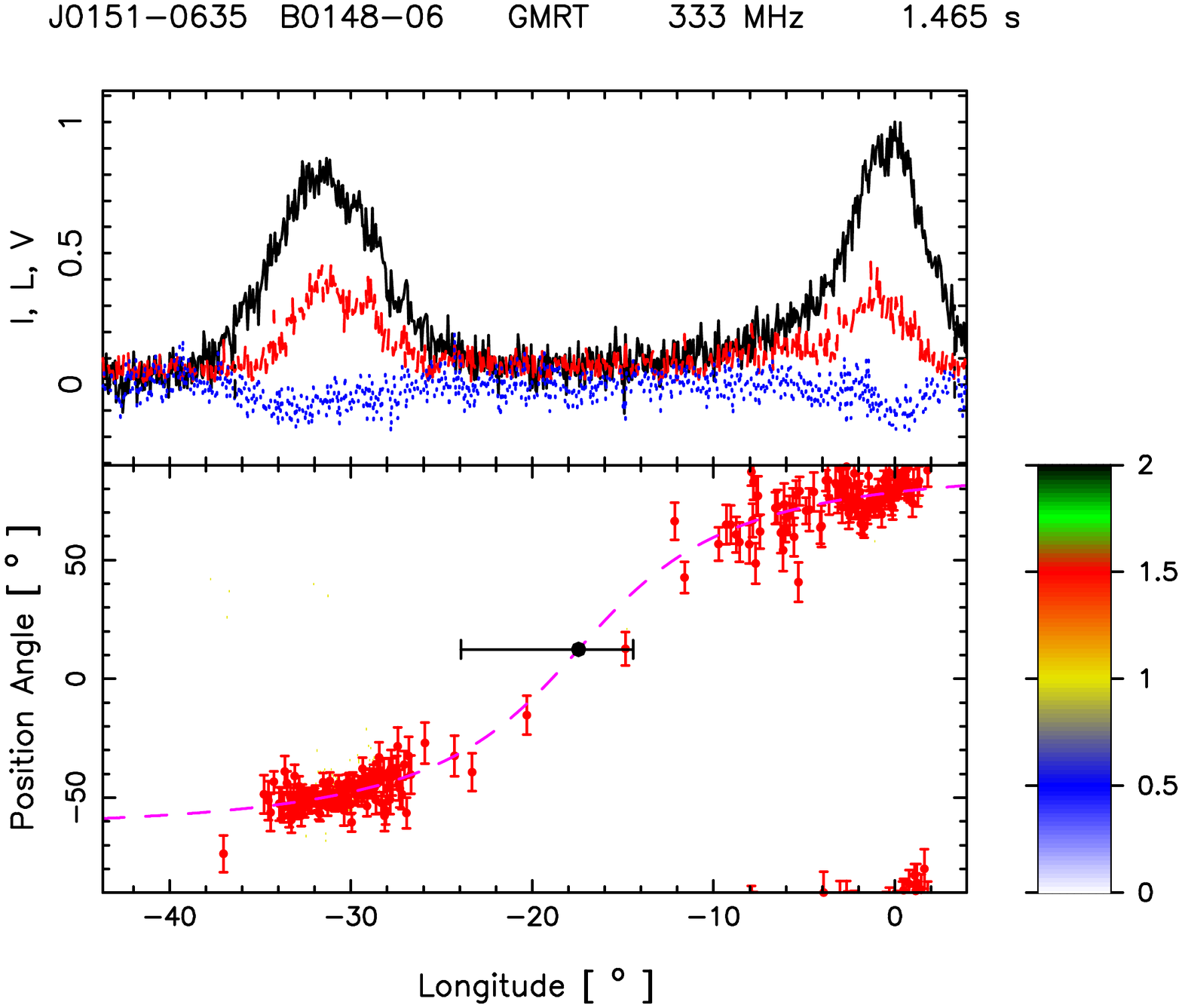}}}&
{\mbox{\includegraphics[width=9cm,height=6cm,angle=0.]{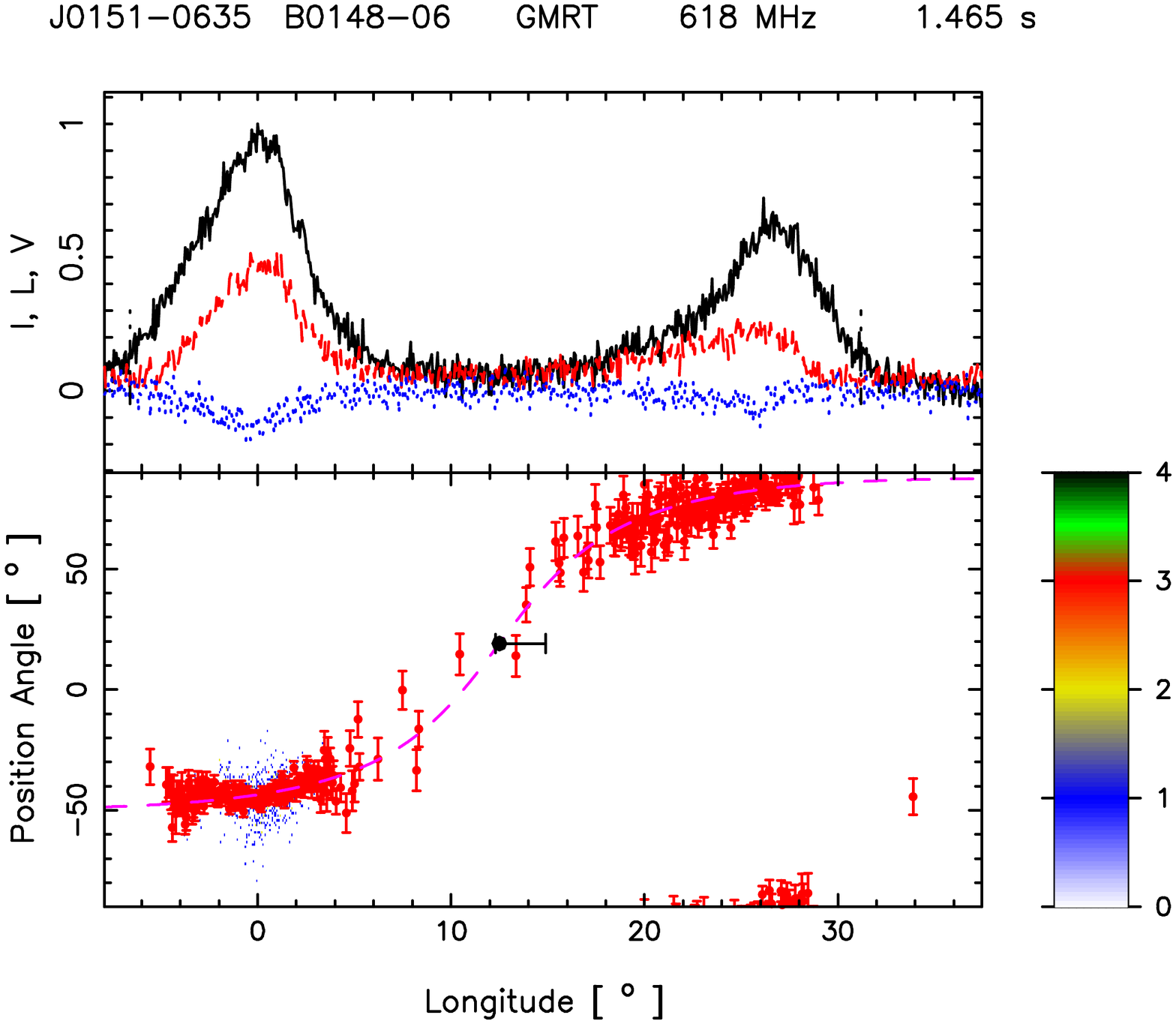}}}\\
{\mbox{\includegraphics[width=9cm,height=6cm,angle=0.]{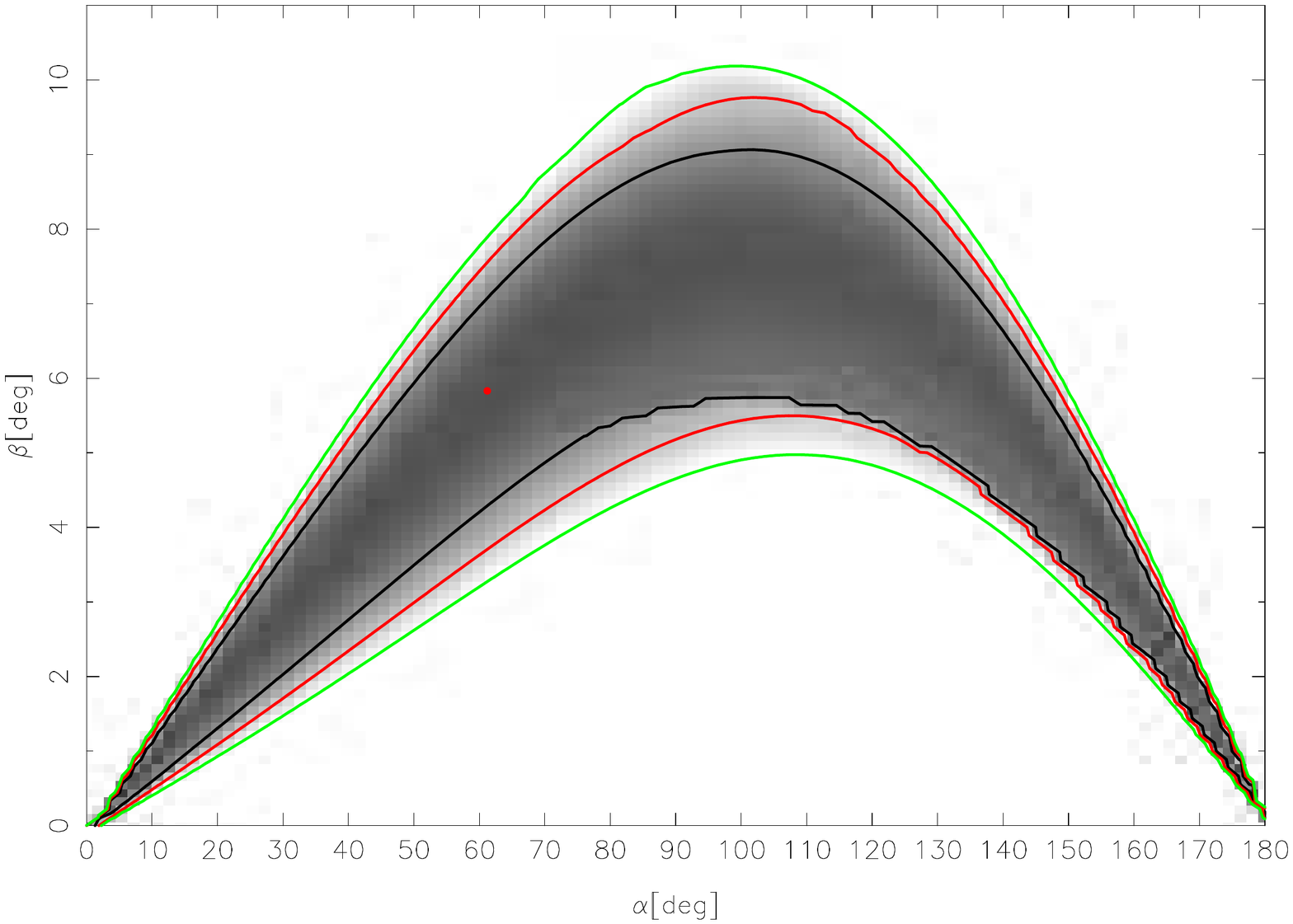}}}&
{\mbox{\includegraphics[width=9cm,height=6cm,angle=0.]{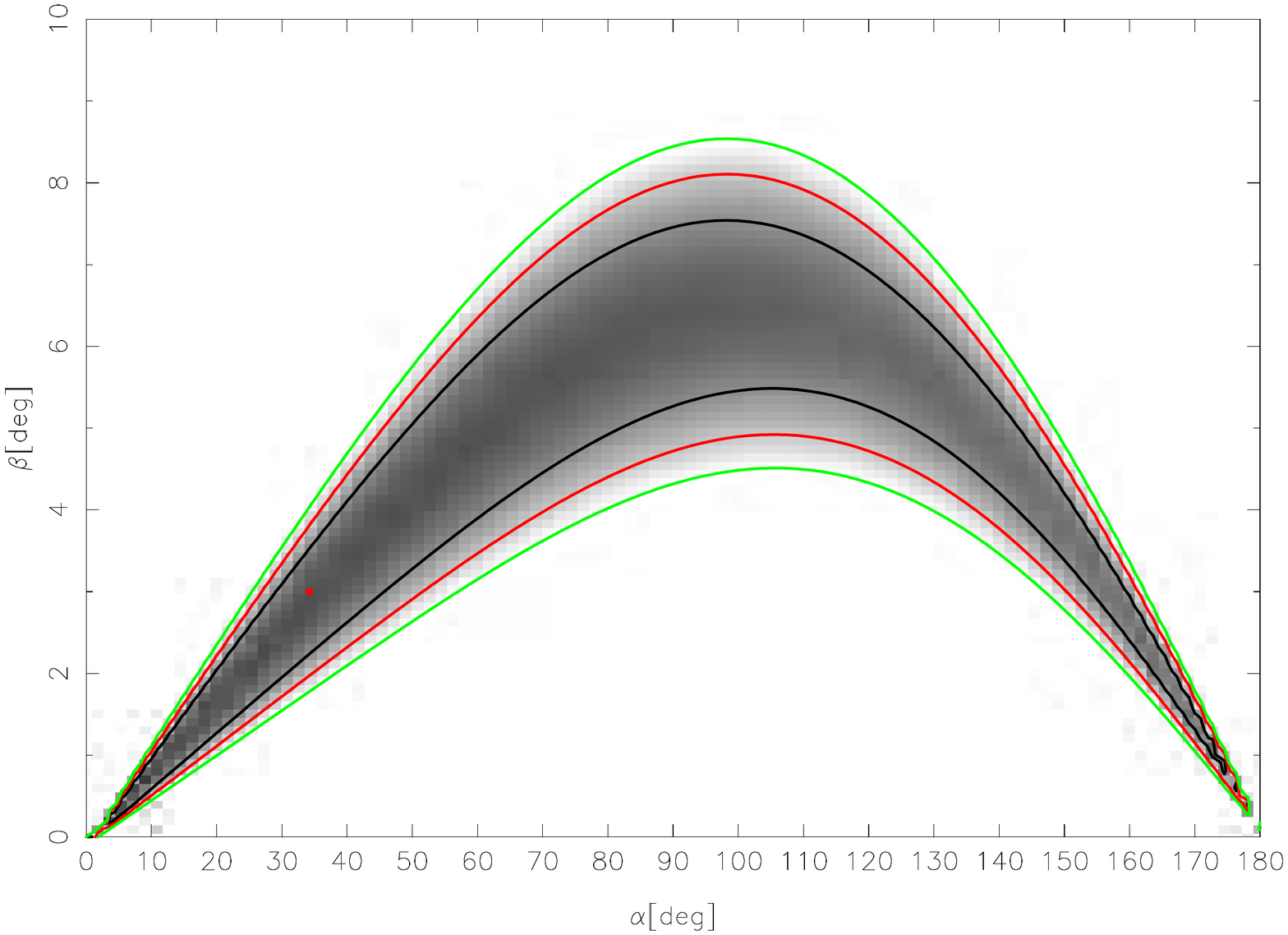}}}\\
{\mbox{\includegraphics[width=9cm,height=6cm,angle=0.]{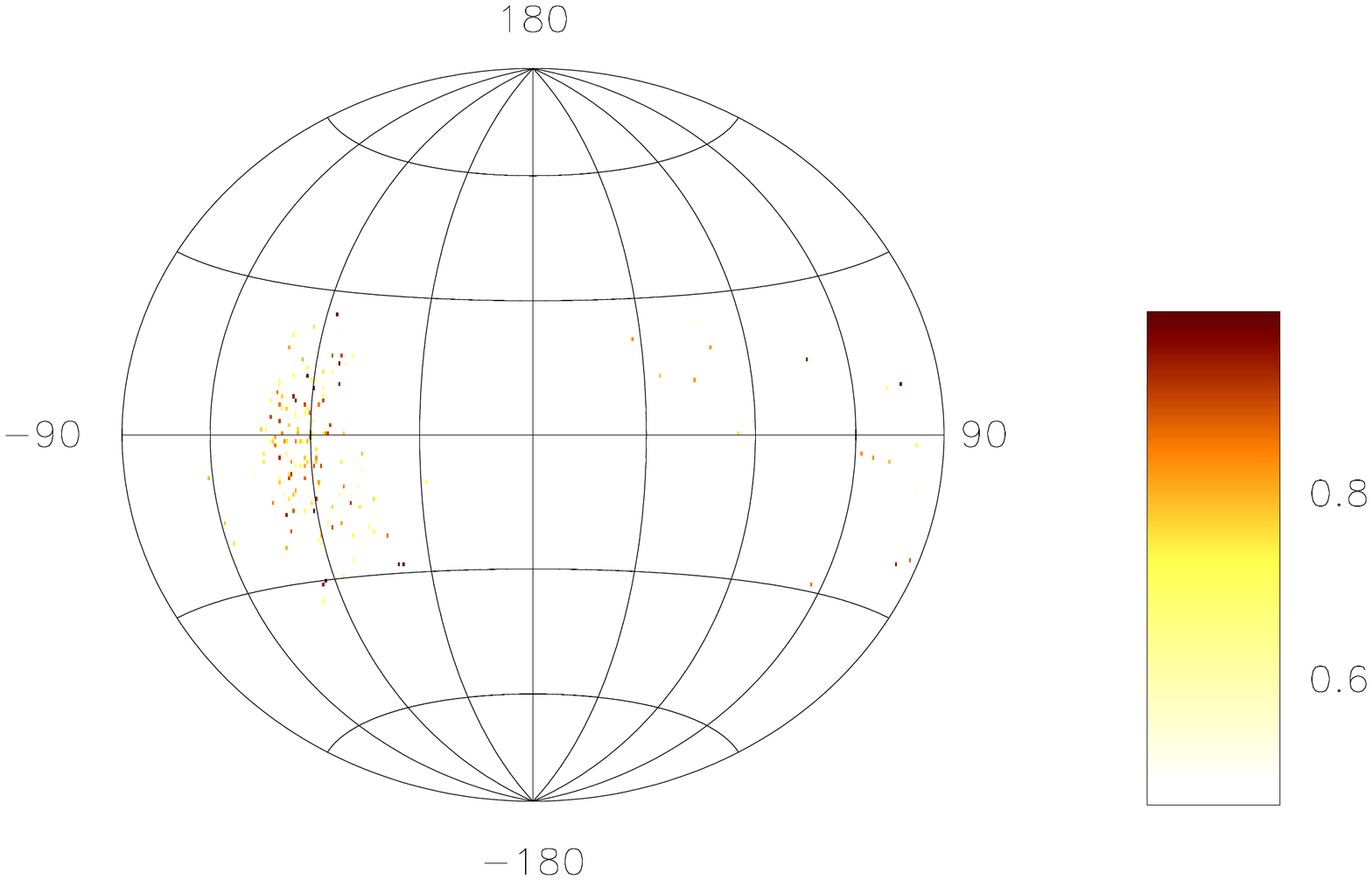}}}&
{\mbox{\includegraphics[width=9cm,height=6cm,angle=0.]{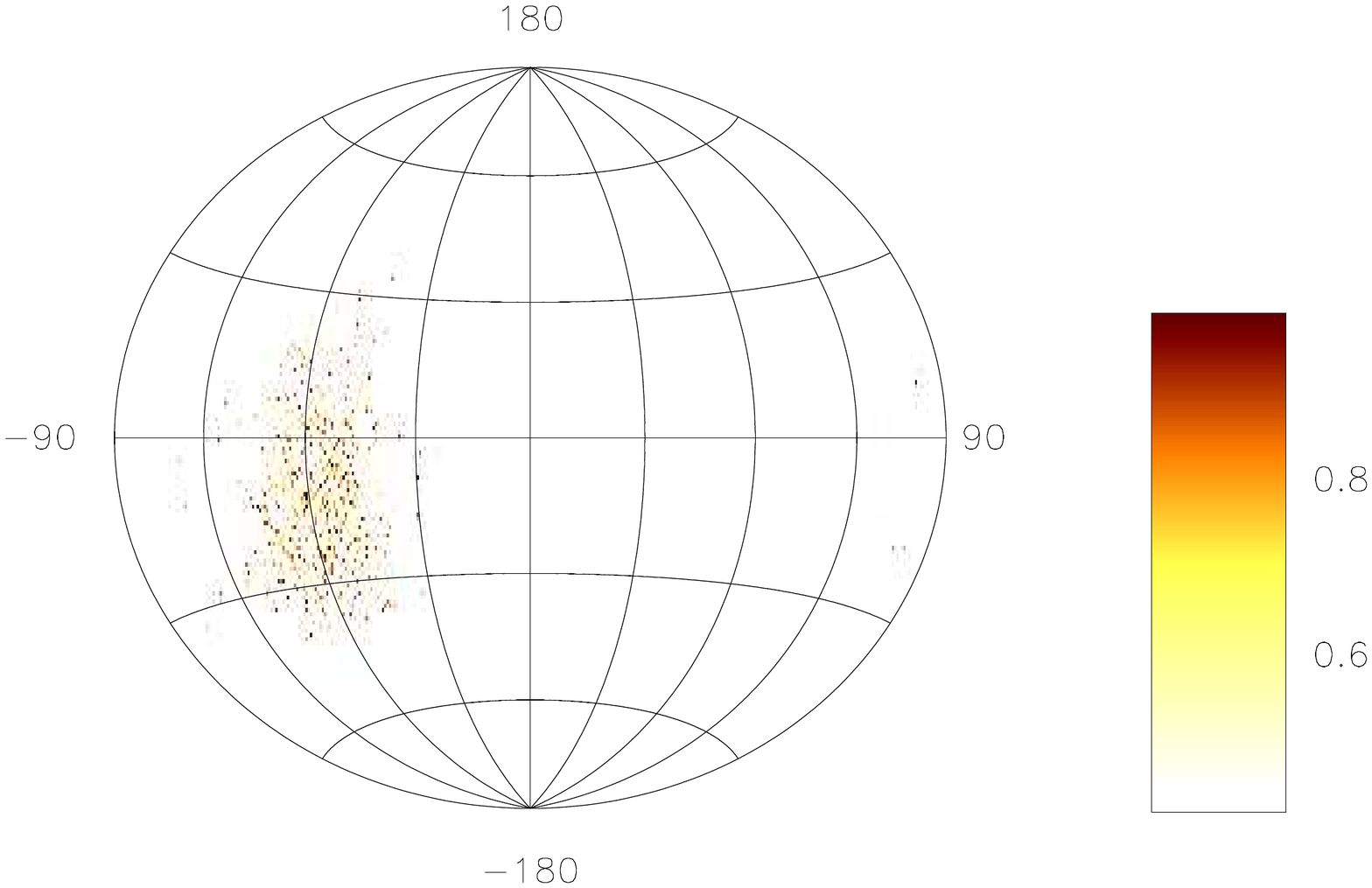}}}\\
\end{tabular}
\caption{Top panel (upper window) shows the average profile with total
intensity (Stokes I; solid black lines), total linear polarization (dashed red
line) and circular polarization (Stokes V; dotted blue line). Top panel (lower
window) also shows the single pulse PPA distribution (colour scale) along with
the average PPA (red error bars). The RVM fits to the average PPA (dashed pink
line) is also shown in this plot. Middle panel show
the $\chi^2$ contours for the parameters $\alpha$ and $\beta$ obtained from RVM
fits. Bottom panel shows the Hammer-Aitoff projection of the polarized time
samples with the colour scheme representing the fractional polarization level.}
\label{a2}
\end{center}
\end{figure*}


\begin{figure*}
\begin{center}
\begin{tabular}{cc}
{\mbox{\includegraphics[width=9cm,height=6cm,angle=0.]{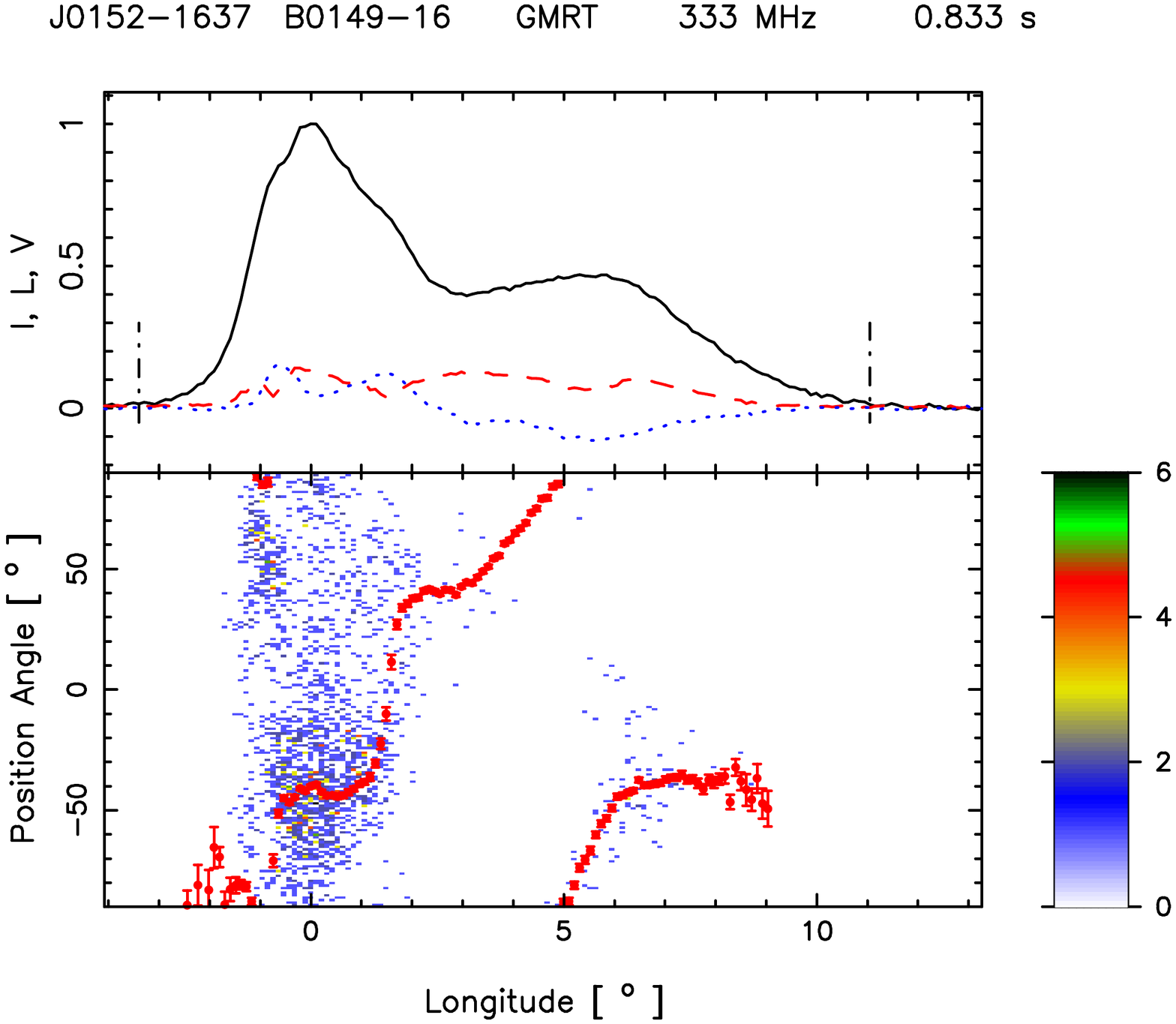}}}&
{\mbox{\includegraphics[width=9cm,height=6cm,angle=0.]{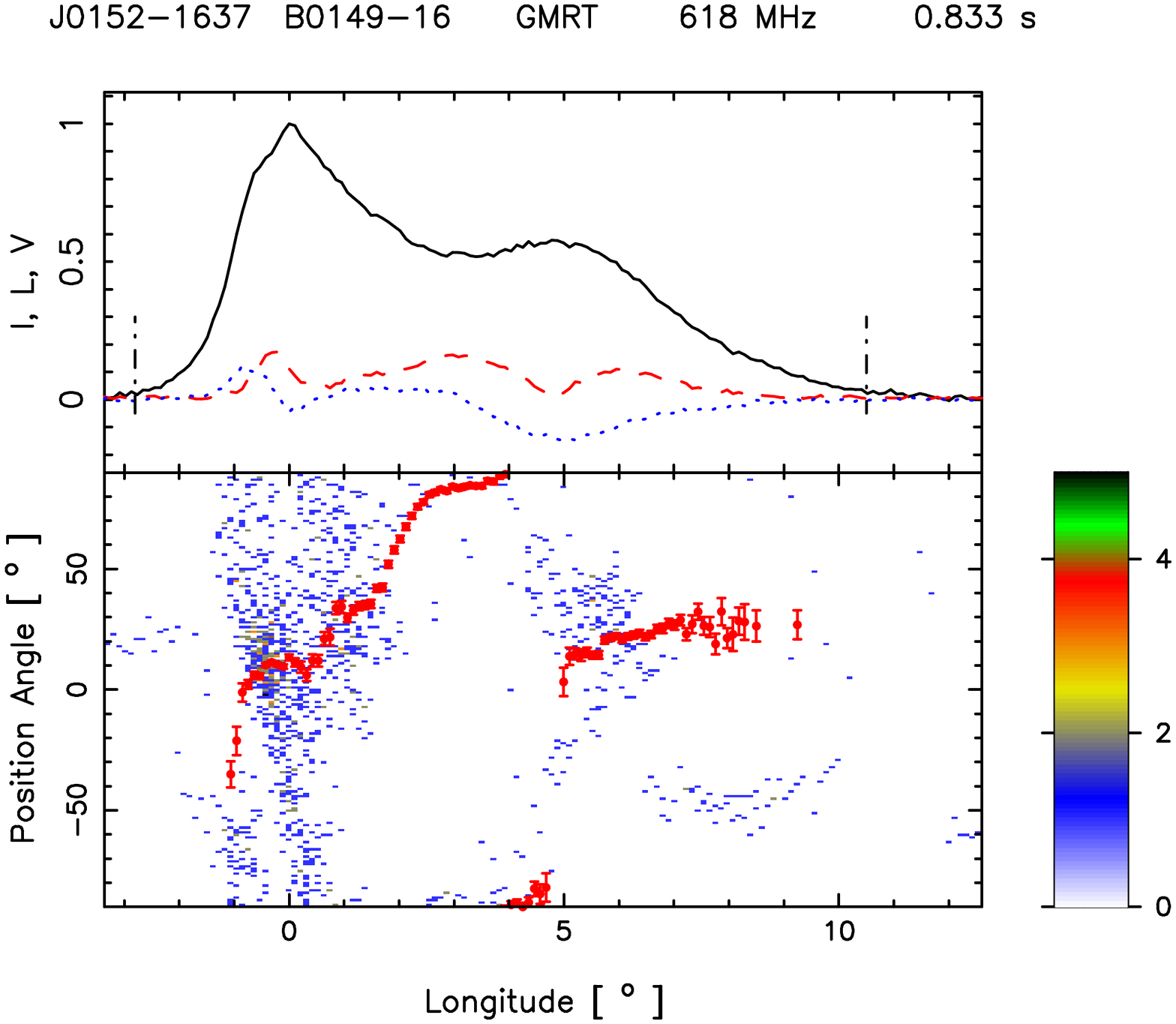}}}\\
&
\\
{\mbox{\includegraphics[width=9cm,height=6cm,angle=0.]{J0151-0635_317MHz_16Feb14.dat.epn2a.71.pdf}}}&
{\mbox{\includegraphics[width=9cm,height=6cm,angle=0.]{J0151-0635_602MHz_23Mar14.dat.epn2a.71.pdf}}}\\
\end{tabular}
\caption{Top panel (upper window) shows the average profile with total
intensity (Stokes I; solid black lines), total linear polarization (dashed red
line) and circular polarization (Stokes V; dotted blue line). Top panel (lower
window) also shows the single pulse PPA distribution (colour scale) along with
the average PPA (red error bars).
Bottom panel shows the Hammer-Aitoff projection of the polarized time
samples with the colour scheme representing the fractional polarization level.}
\label{a3}
\end{center}
\end{figure*}

\begin{figure*}
\begin{center}
\begin{tabular}{cc}
{\mbox{\includegraphics[width=9cm,height=6cm,angle=0.]{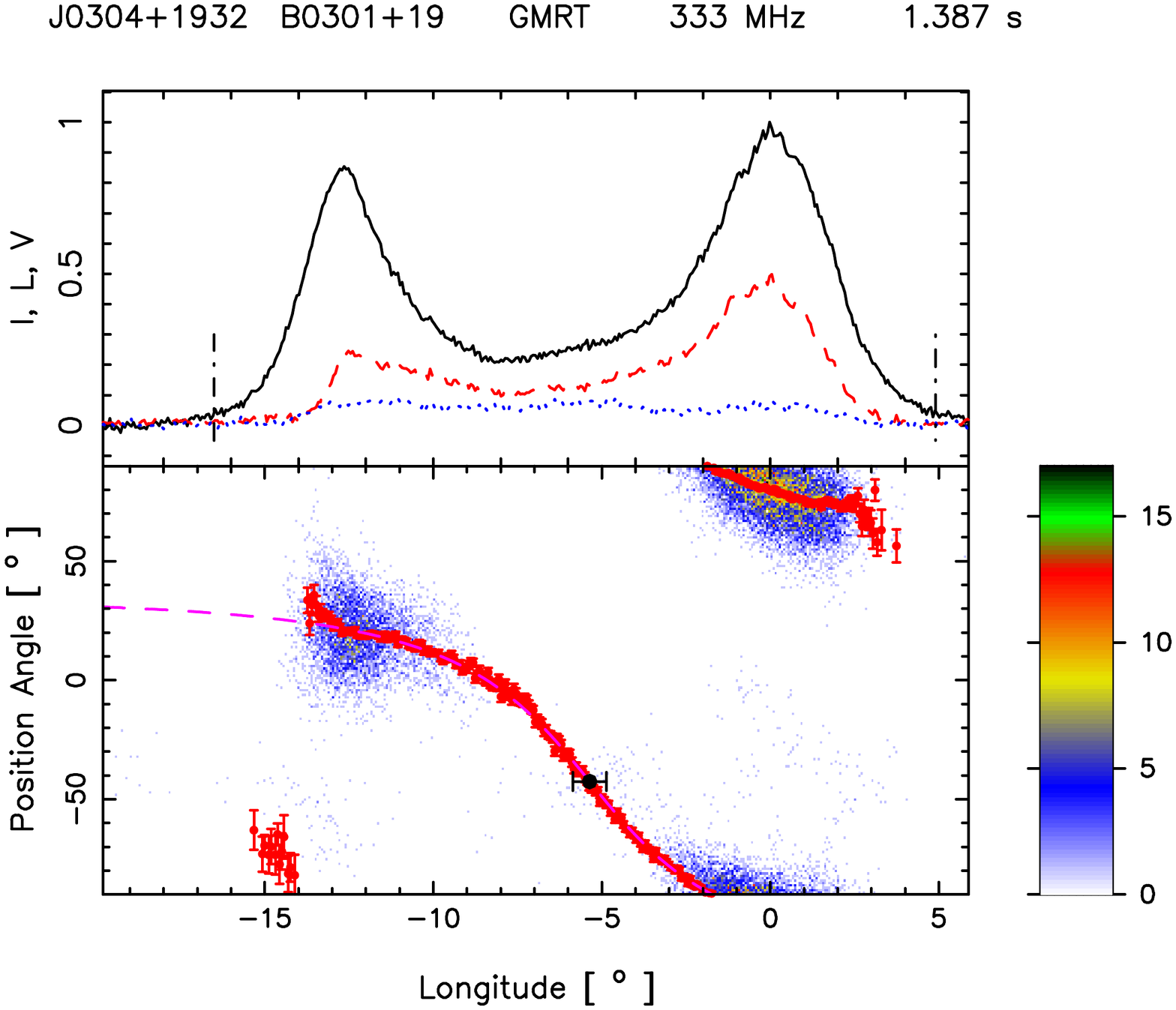}}}&
{\mbox{\includegraphics[width=9cm,height=6cm,angle=0.]{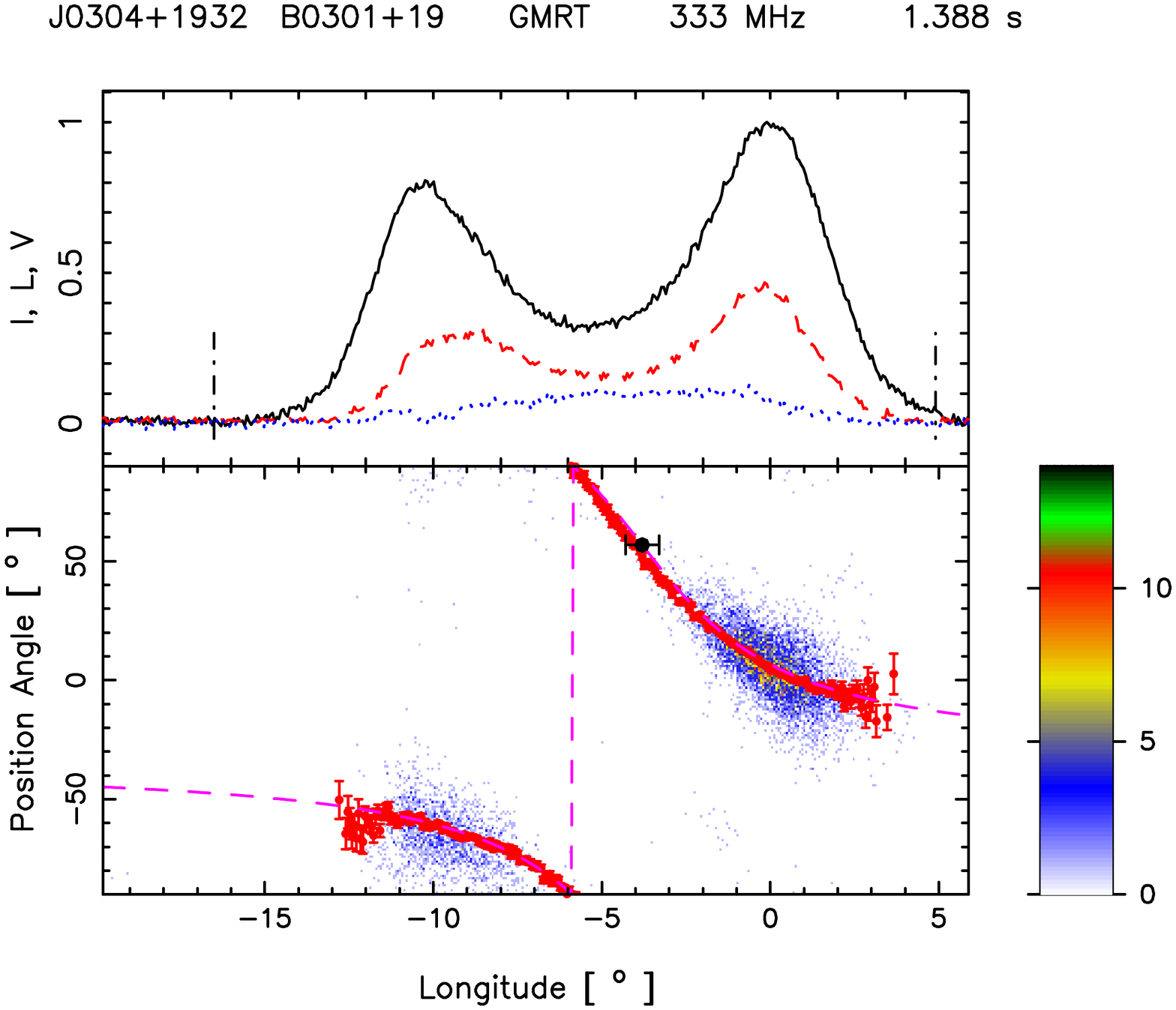}}}\\
{\mbox{\includegraphics[width=9cm,height=6cm,angle=0.]{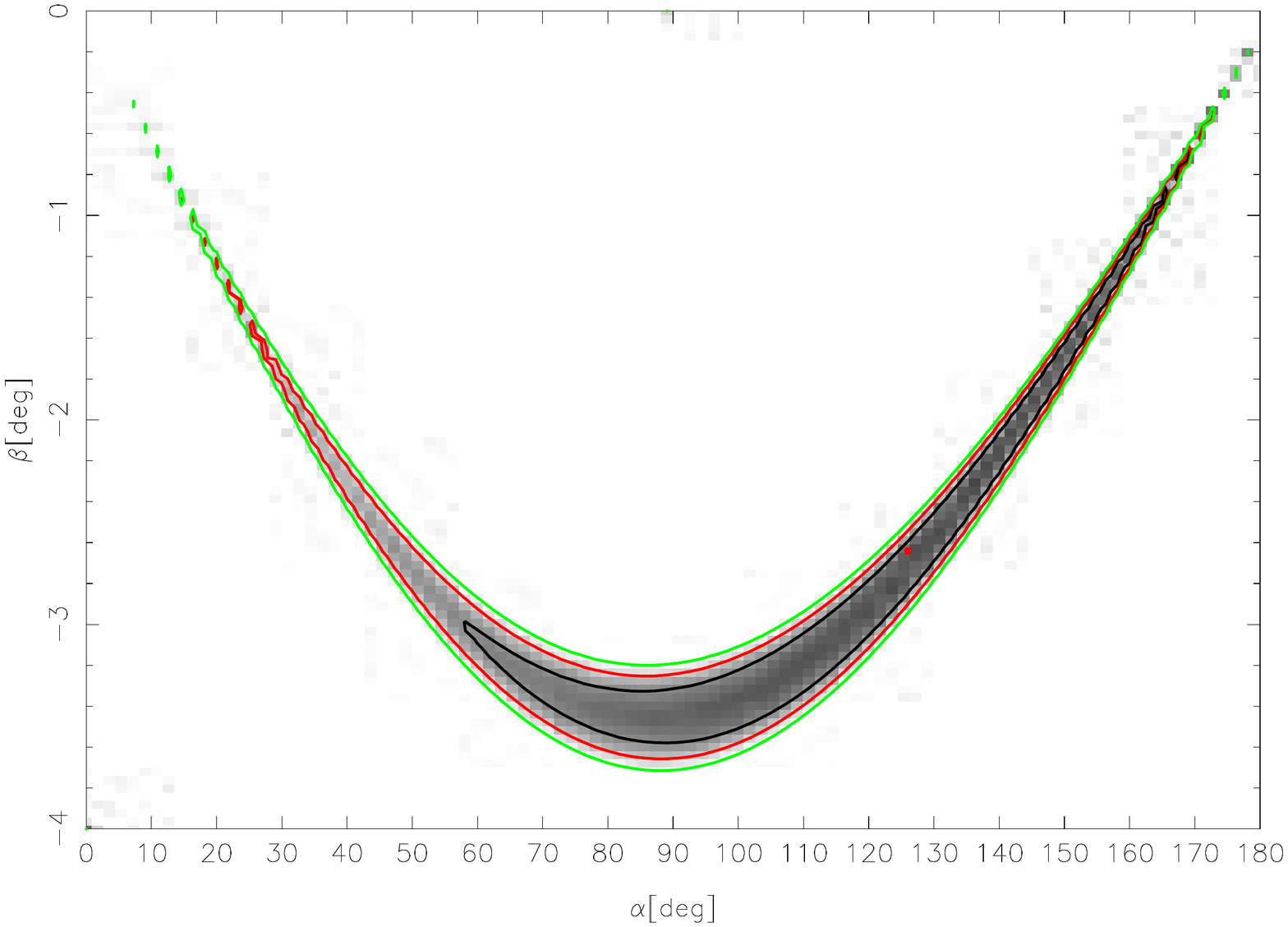}}}&
{\mbox{\includegraphics[width=9cm,height=6cm,angle=0.]{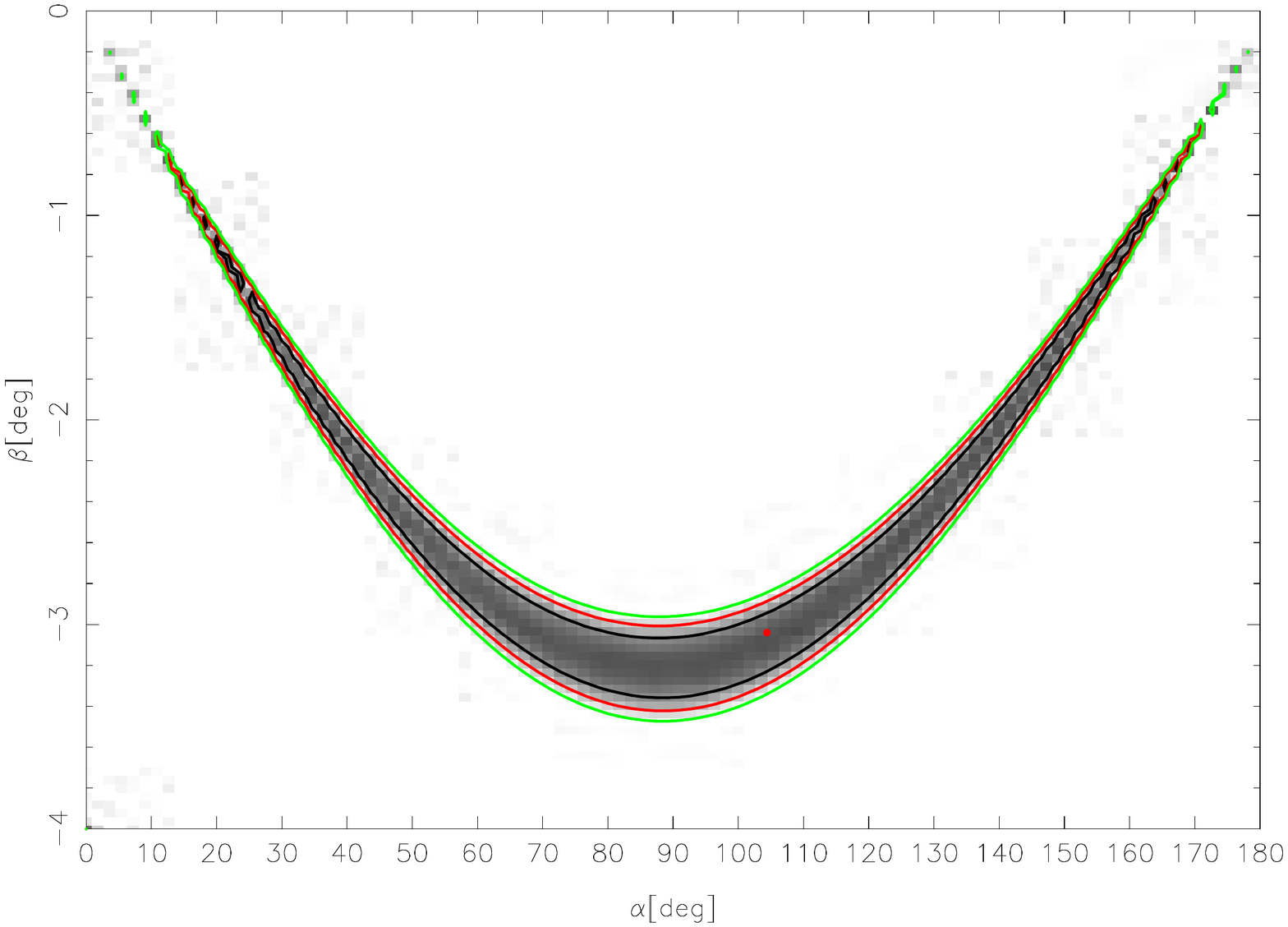}}}\\
{\mbox{\includegraphics[width=9cm,height=6cm,angle=0.]{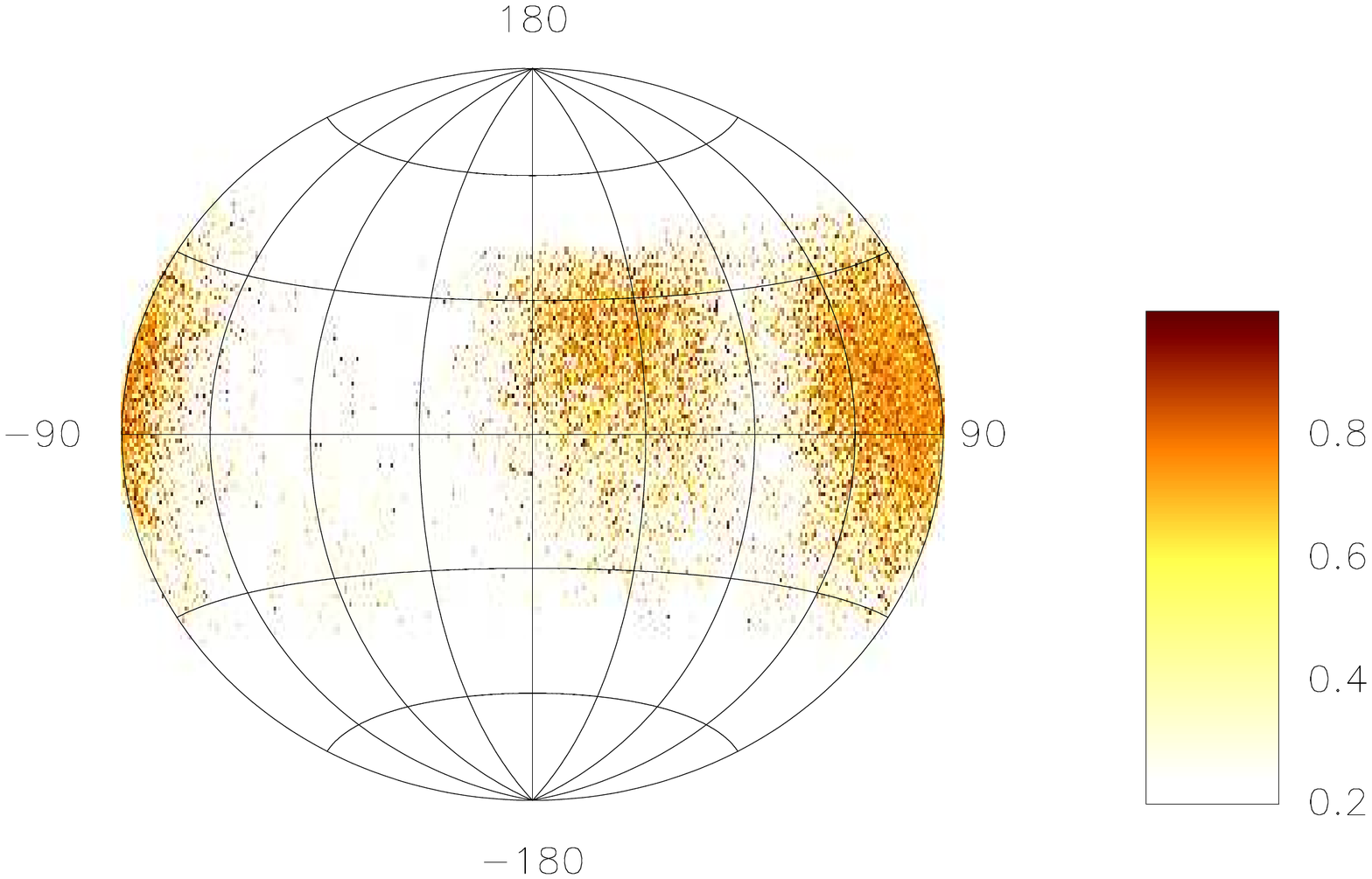}}}&
{\mbox{\includegraphics[width=9cm,height=6cm,angle=0.]{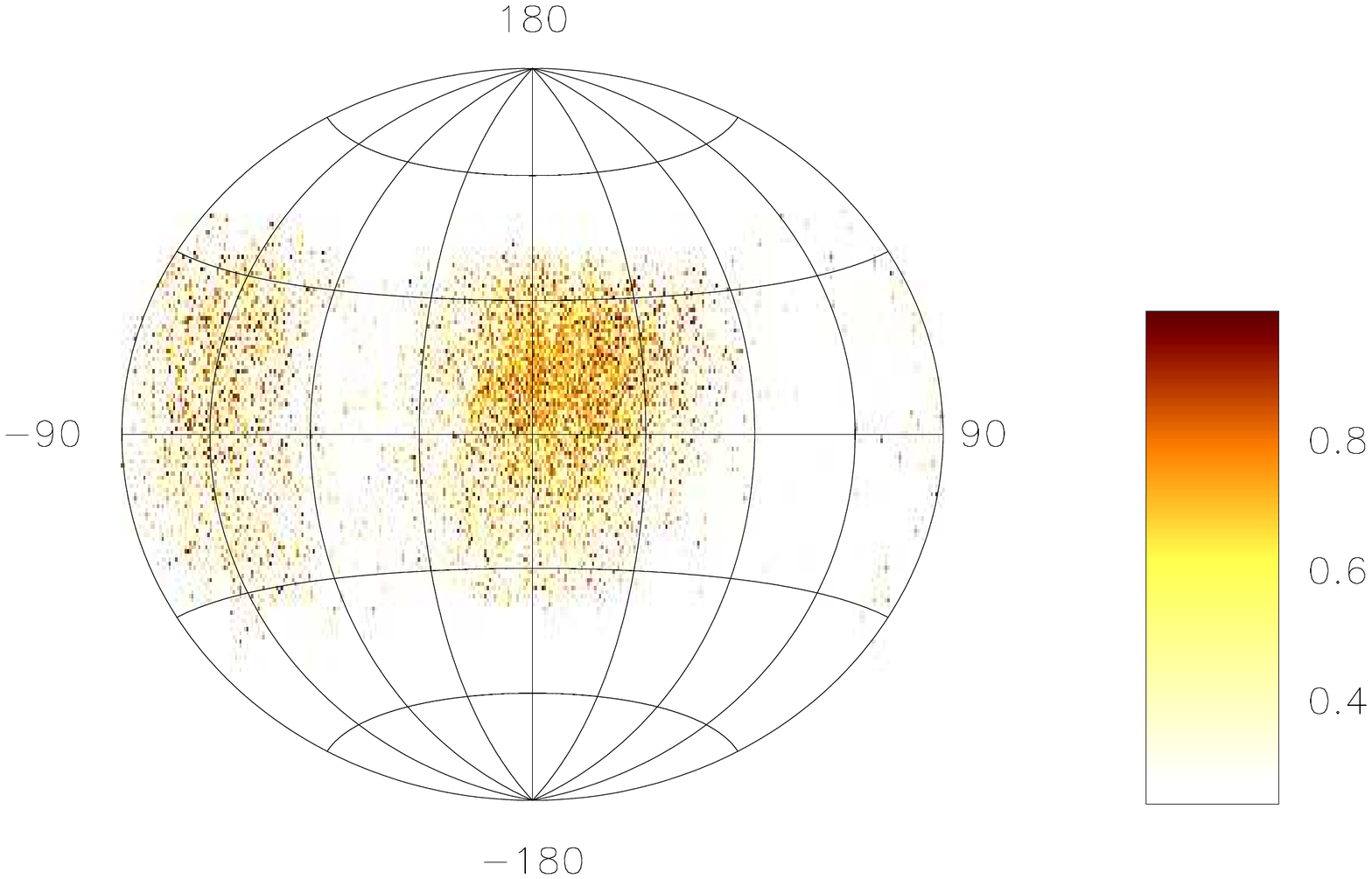}}}\\
\end{tabular}
\caption{Top panel (upper window) shows the average profile with total
intensity (Stokes I; solid black lines), total linear polarization (dashed red
line) and circular polarization (Stokes V; dotted blue line). Top panel (lower
window) also shows the single pulse PPA distribution (colour scale) along with
the average PPA (red error bars).
The RVM fits to the average PPA (dashed pink
line) is also shown in this plot. Middle panel show
the $\chi^2$ contours for the parameters $\alpha$ and $\beta$ obtained from RVM
fits.
Bottom panel shows the Hammer-Aitoff projection of the polarized time
samples with the colour scheme representing the fractional polarization level.}
\label{a4}
\end{center}
\end{figure*}


\begin{figure*}
\begin{center}
\begin{tabular}{cc}
{\mbox{\includegraphics[width=9cm,height=6cm,angle=0.]{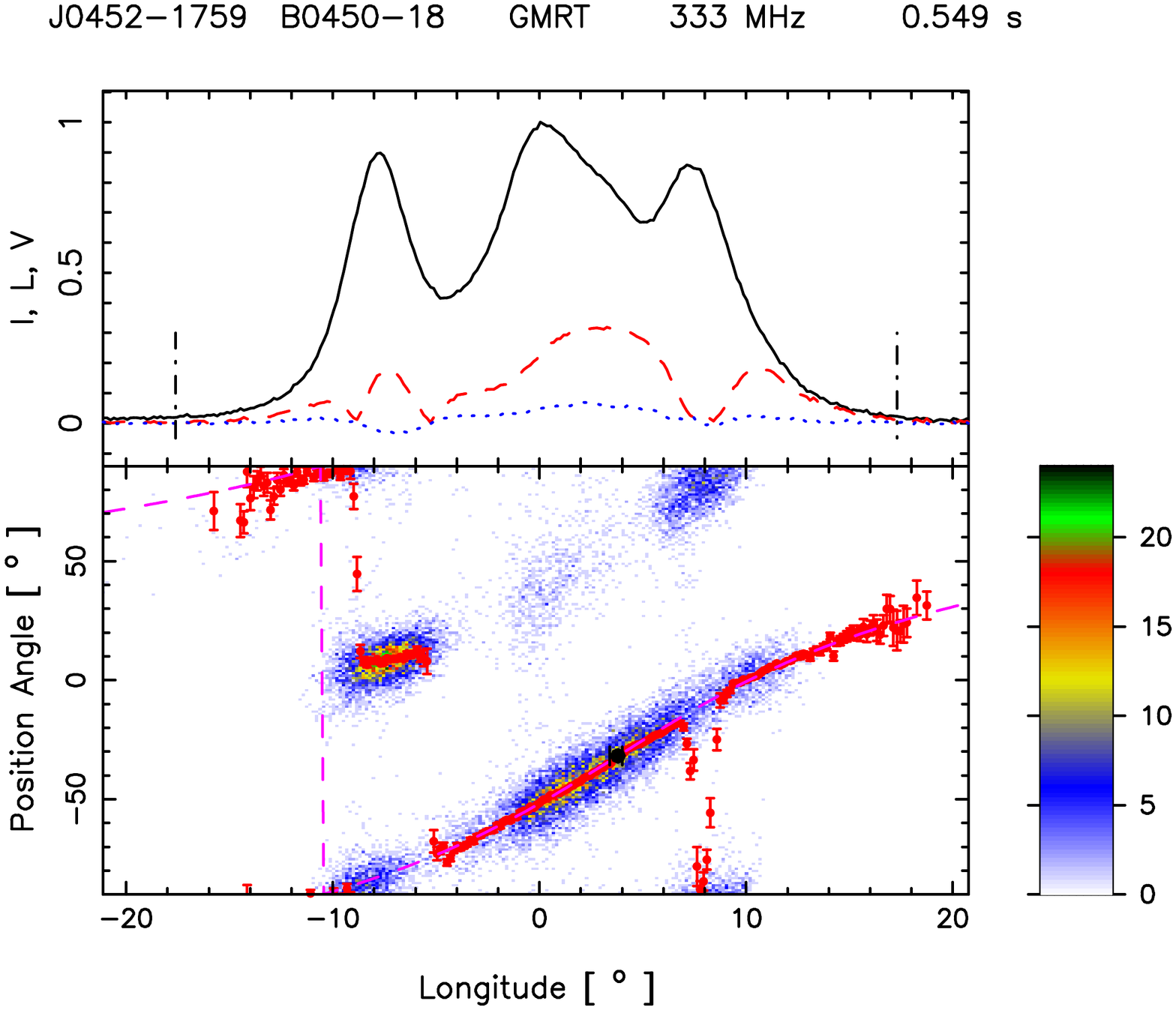}}}&
{\mbox{\includegraphics[width=9cm,height=6cm,angle=0.]{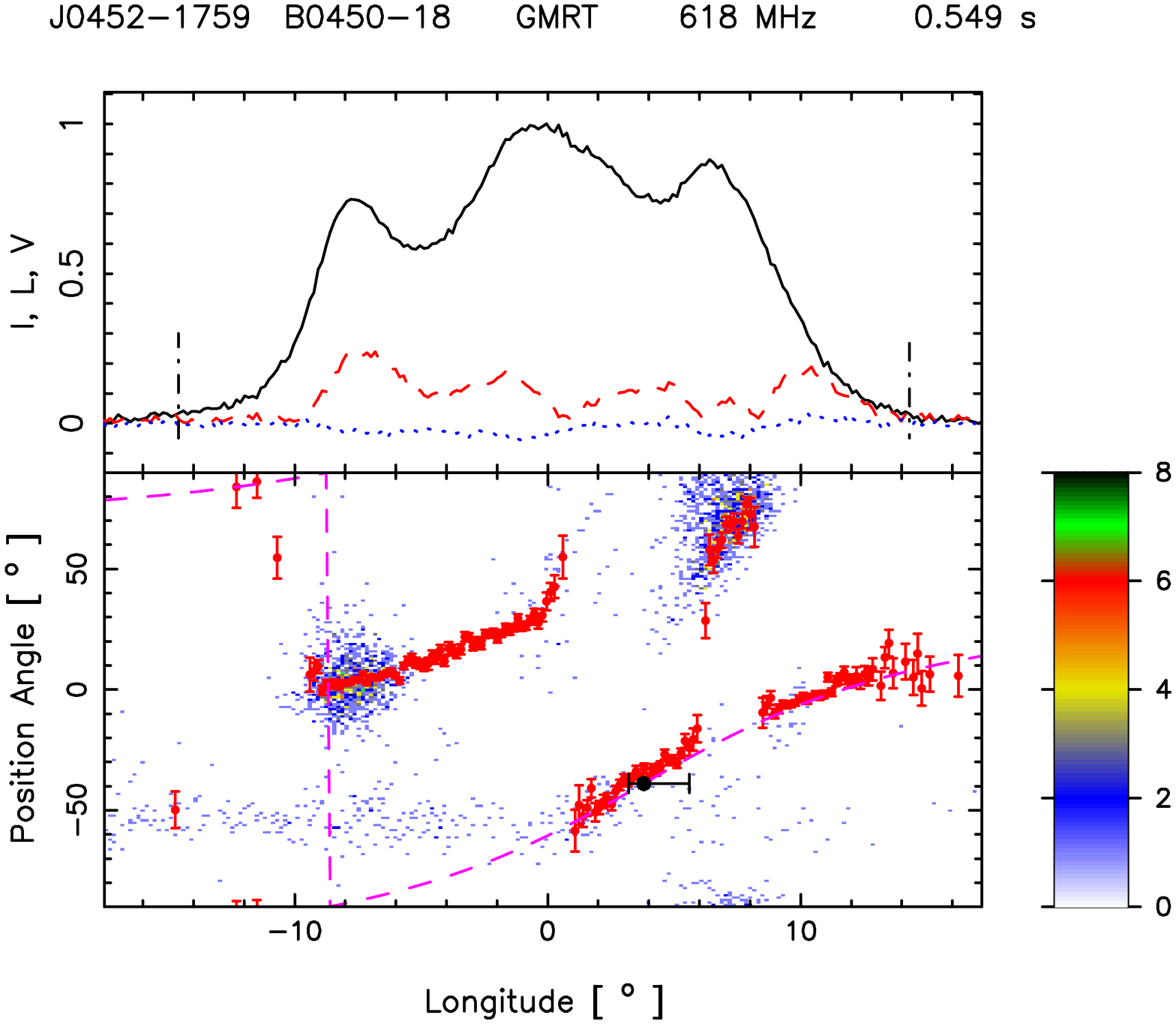}}}\\
{\mbox{\includegraphics[width=9cm,height=6cm,angle=0.]{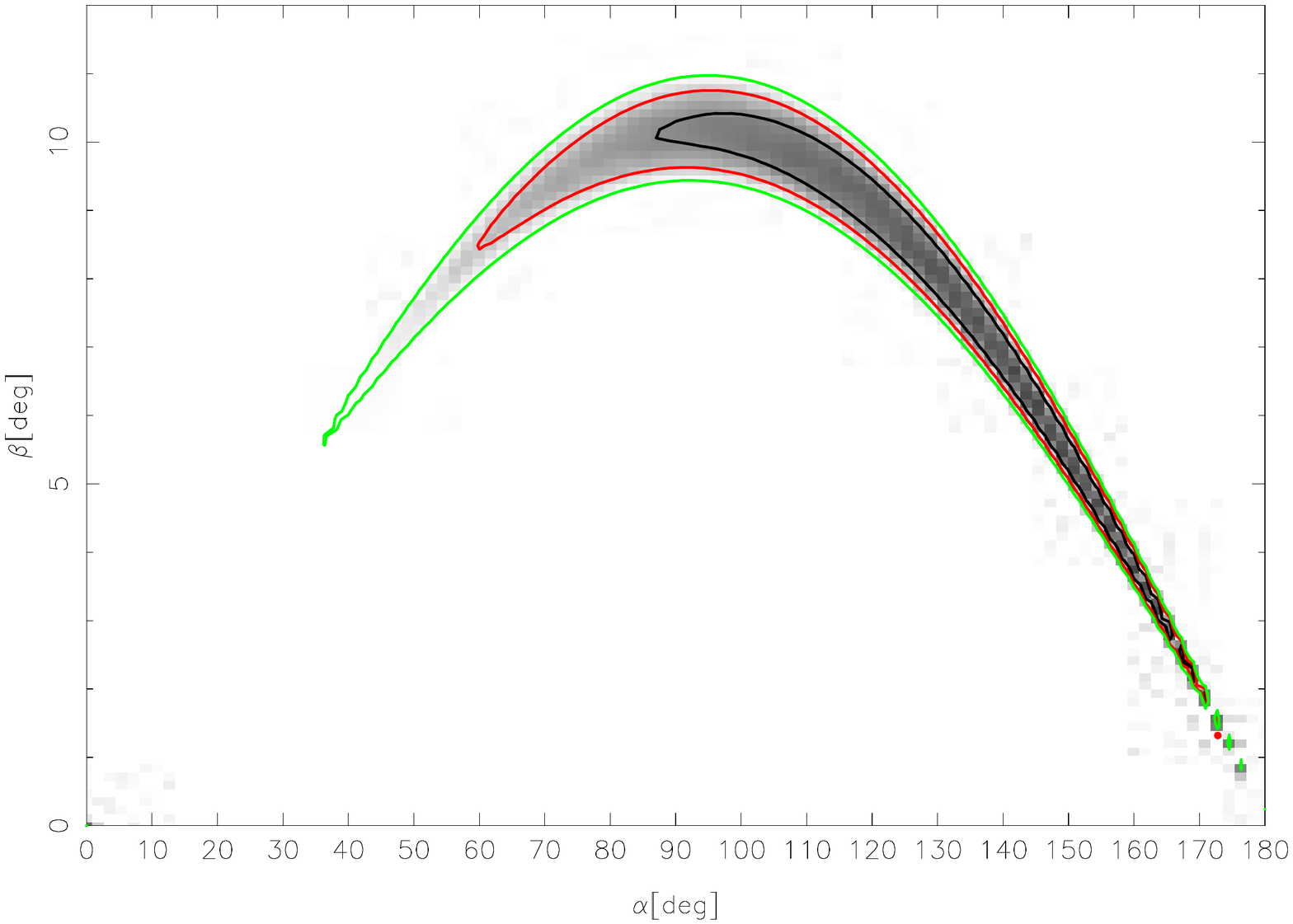}}}&
{\mbox{\includegraphics[width=9cm,height=6cm,angle=0.]{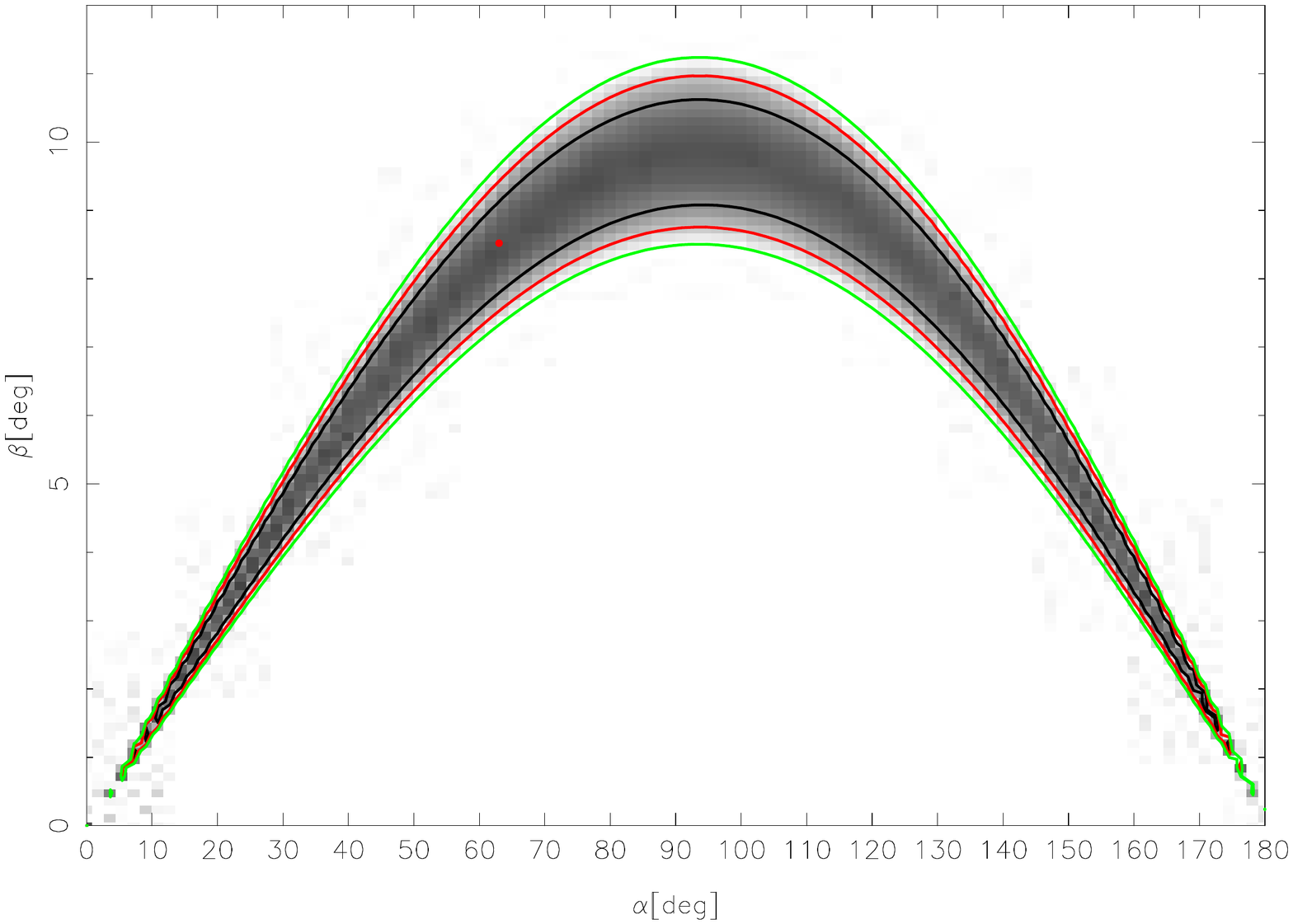}}}\\
{\mbox{\includegraphics[width=9cm,height=6cm,angle=0.]{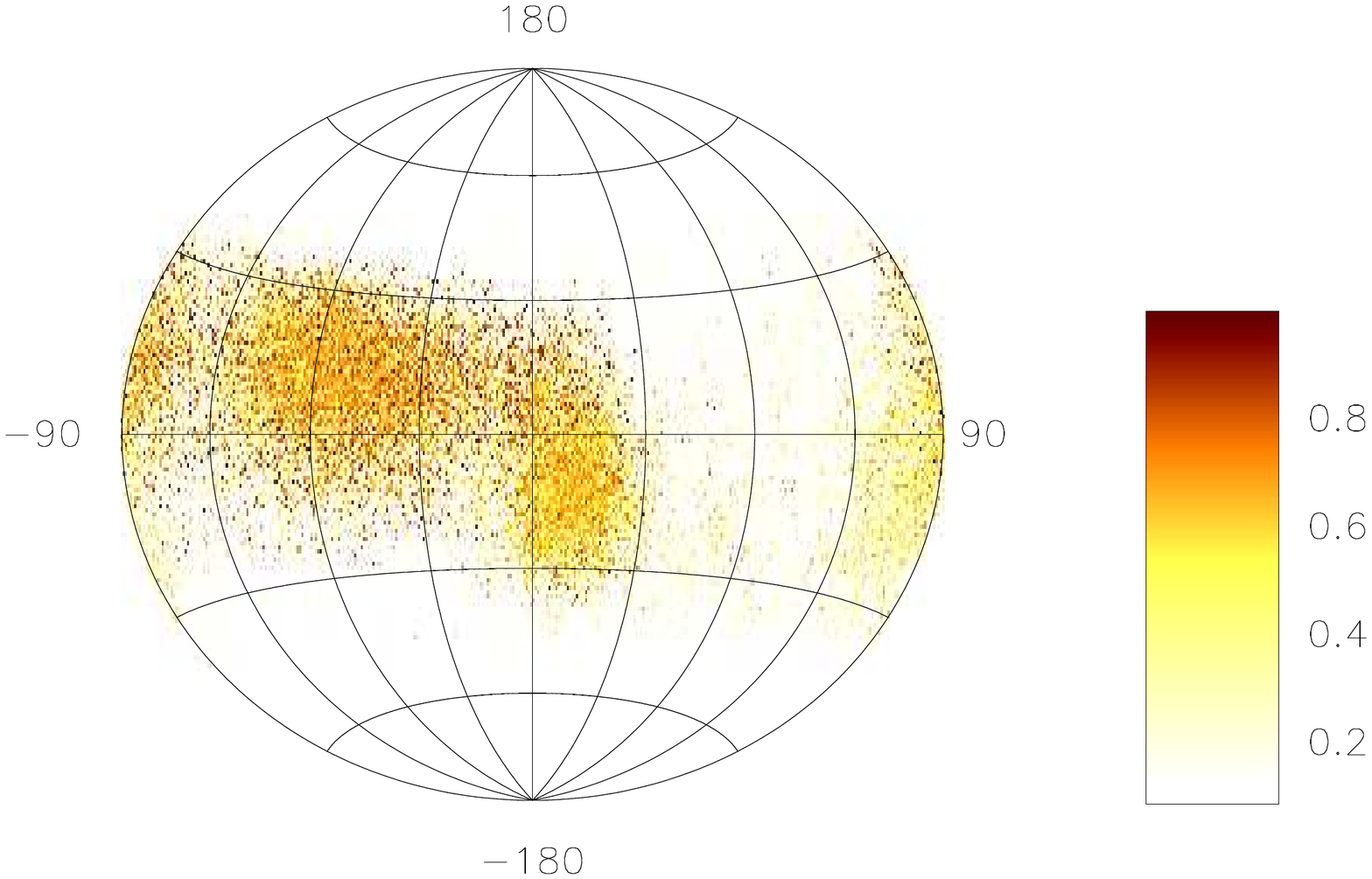}}}&
{\mbox{\includegraphics[width=9cm,height=6cm,angle=0.]{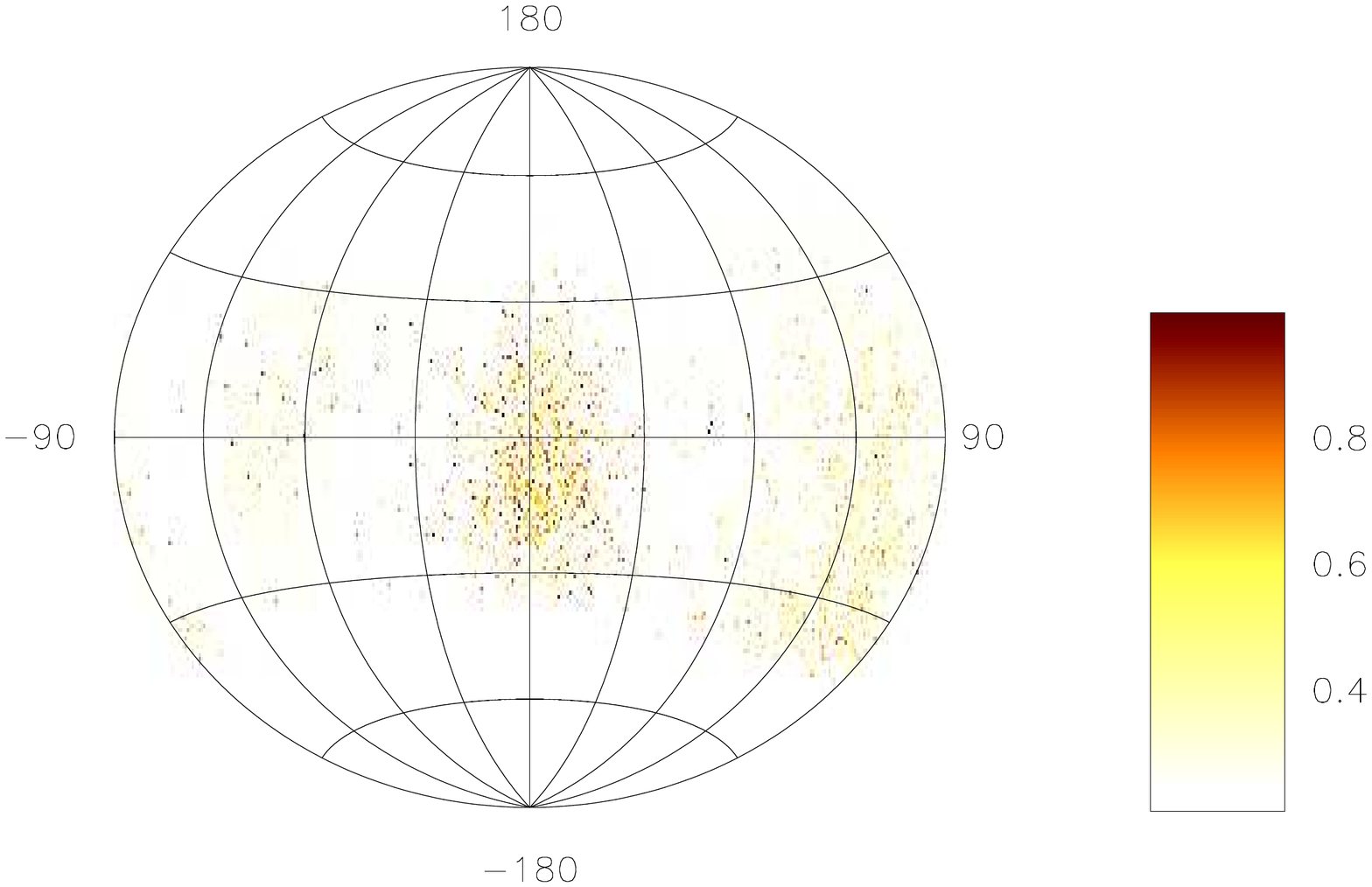}}}\\
\end{tabular}
\caption{Top panel (upper window) shows the average profile with total
intensity (Stokes I; solid black lines), total linear polarization (dashed red
line) and circular polarization (Stokes V; dotted blue line). Top panel (lower
window) also shows the single pulse PPA distribution (colour scale) along with
the average PPA (red error bars).
The RVM fits to the average PPA (dashed pink
line) is also shown in this plot. Middle panel show
the $\chi^2$ contours for the parameters $\alpha$ and $\beta$ obtained from RVM
fits.
Bottom panel shows the Hammer-Aitoff projection of the polarized time
samples with the colour scheme representing the fractional polarization level.}
\label{a5}
\end{center}
\end{figure*}

\begin{figure*}
\begin{center}
\begin{tabular}{cc}
{\mbox{\includegraphics[width=9cm,height=6cm,angle=0.]{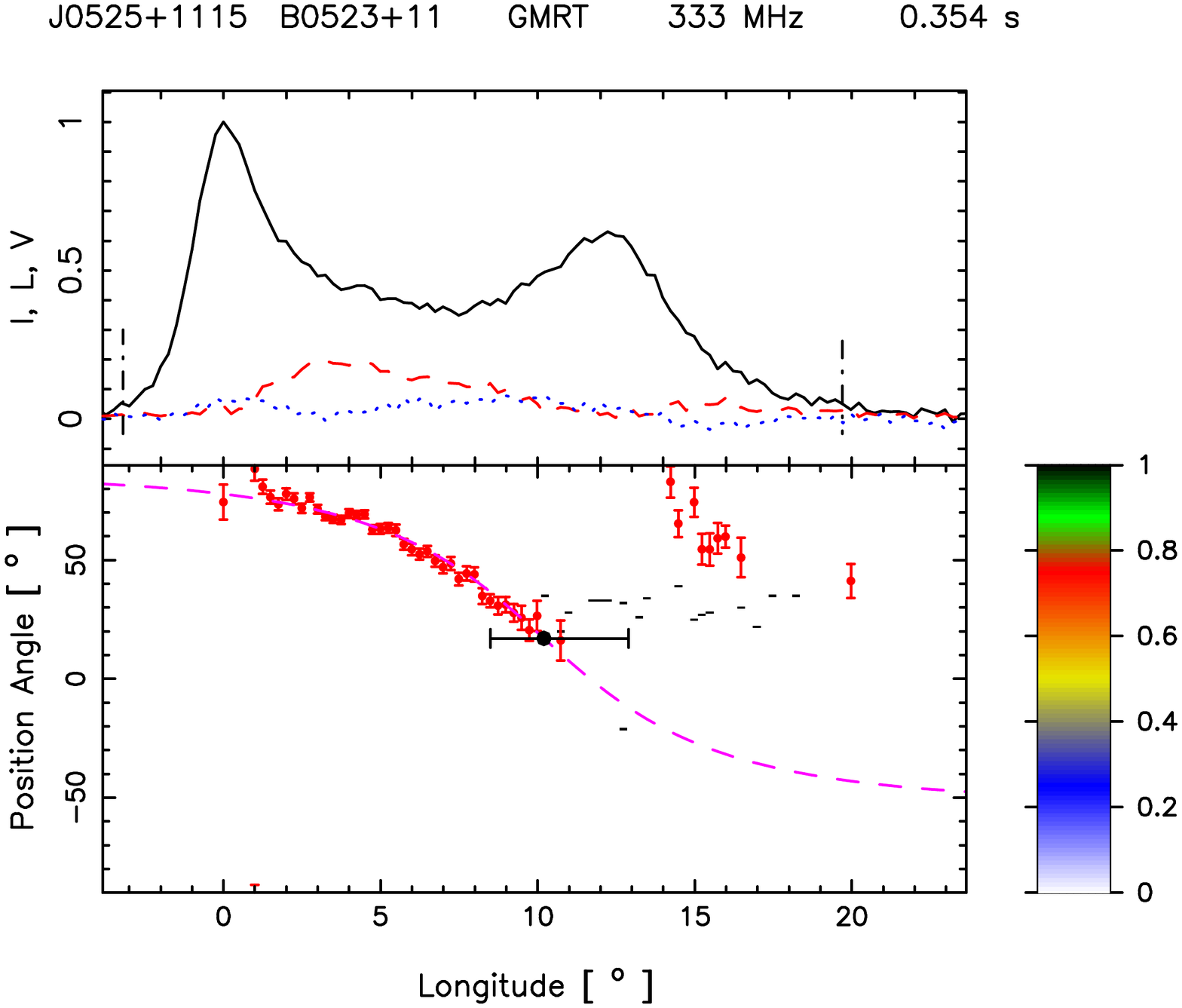}}}&
{\mbox{\includegraphics[width=9cm,height=6cm,angle=0.]{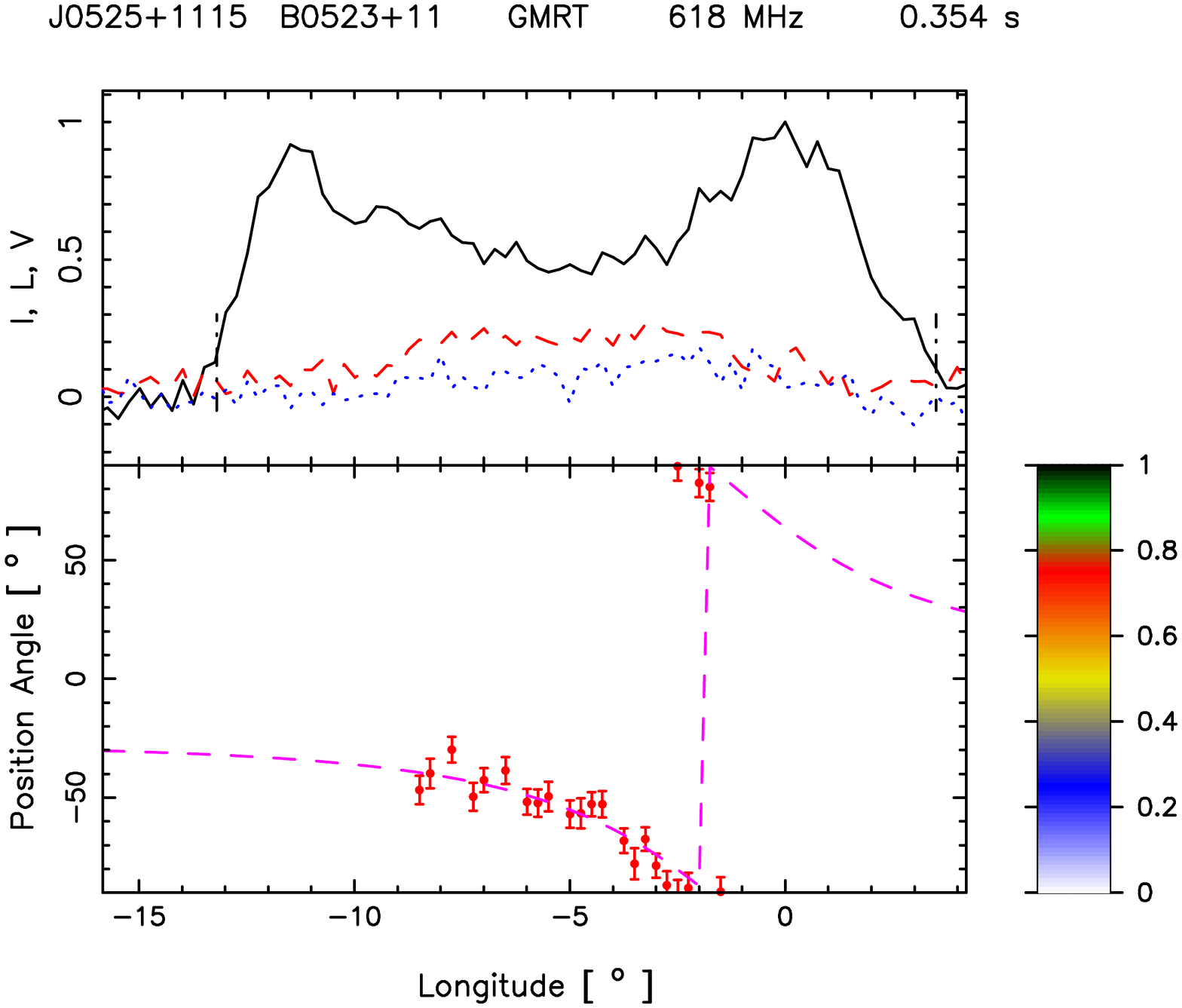}}}\\
{\mbox{\includegraphics[width=9cm,height=6cm,angle=0.]{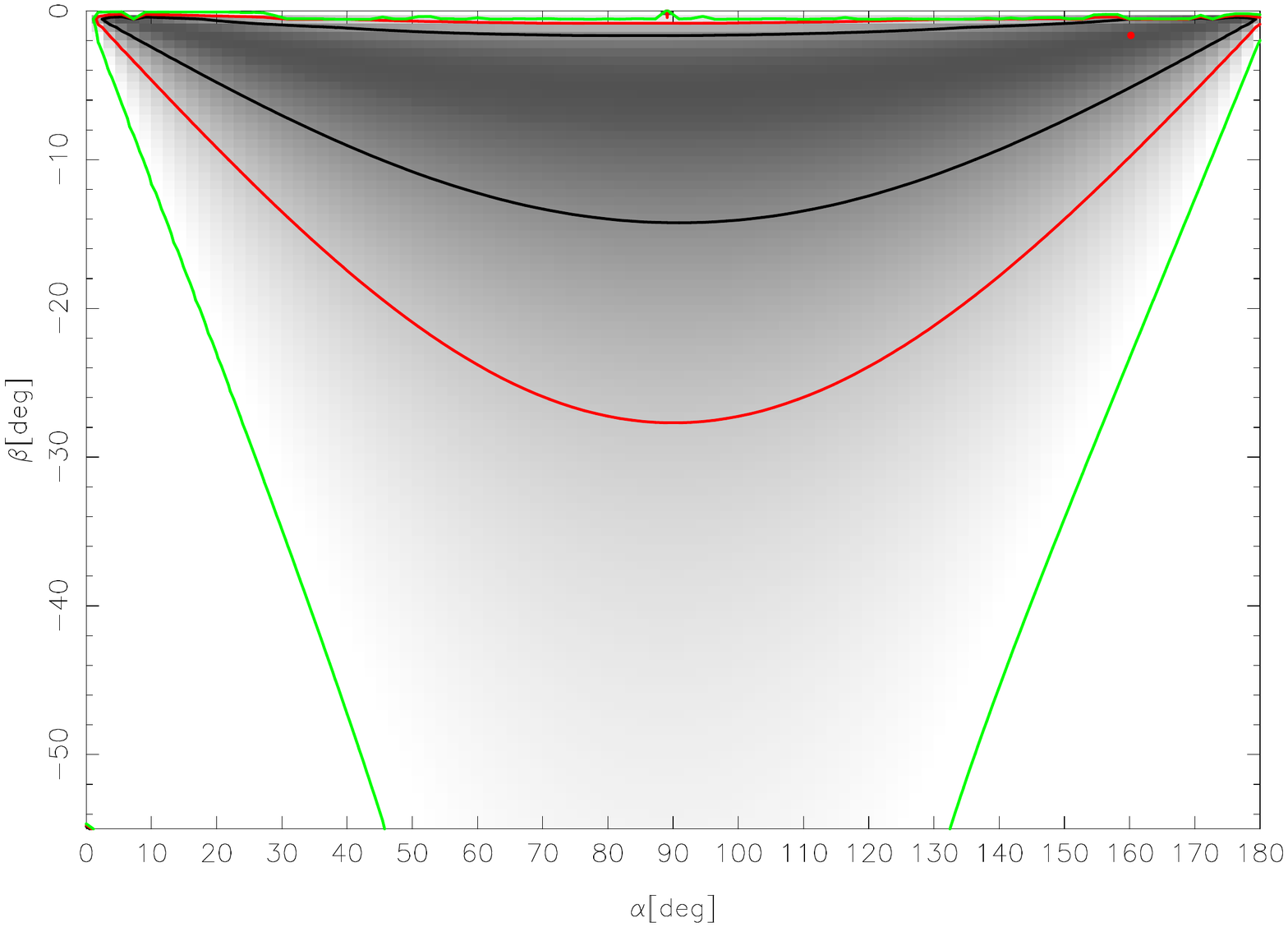}}}&
{\mbox{\includegraphics[width=9cm,height=6cm,angle=0.]{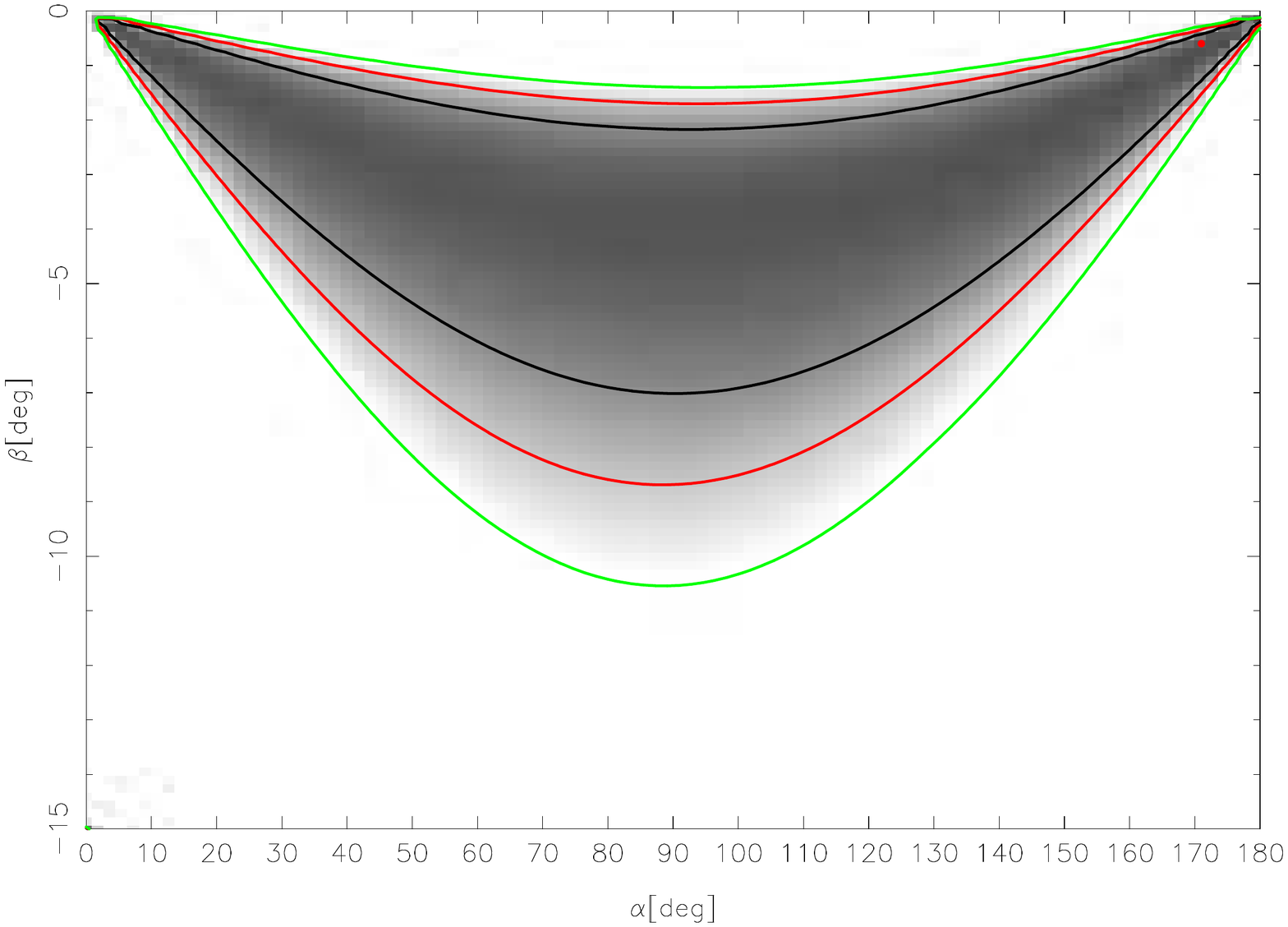}}}\\
&
\\
\end{tabular}
\caption{Top panel (upper window) shows the average profile with total
intensity (Stokes I; solid black lines), total linear polarization (dashed red
line) and circular polarization (Stokes V; dotted blue line). Top panel (lower
window) also shows the single pulse PPA distribution (colour scale) along with
the average PPA (red error bars).
The RVM fits to the average PPA (dashed pink
line) is also shown in this plot. Bottom panel show
the $\chi^2$ contours for the parameters $\alpha$ and $\beta$ obtained from RVM
fits.}
\label{a6}
\end{center}
\end{figure*}


\begin{figure*}
\begin{center}
\begin{tabular}{cc}
{\mbox{\includegraphics[width=9cm,height=6cm,angle=0.]{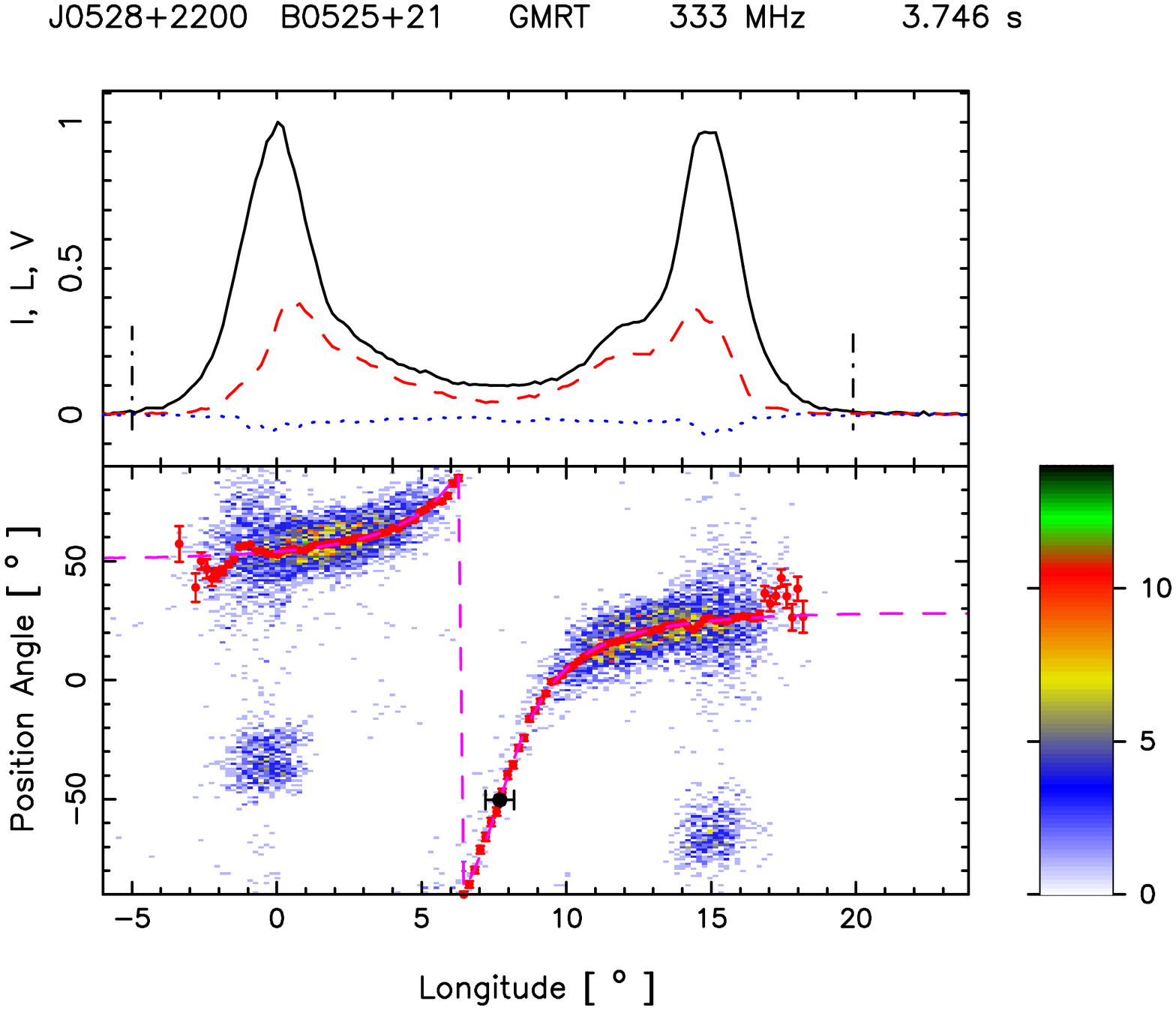}}}&
\\
{\mbox{\includegraphics[width=9cm,height=6cm,angle=0.]{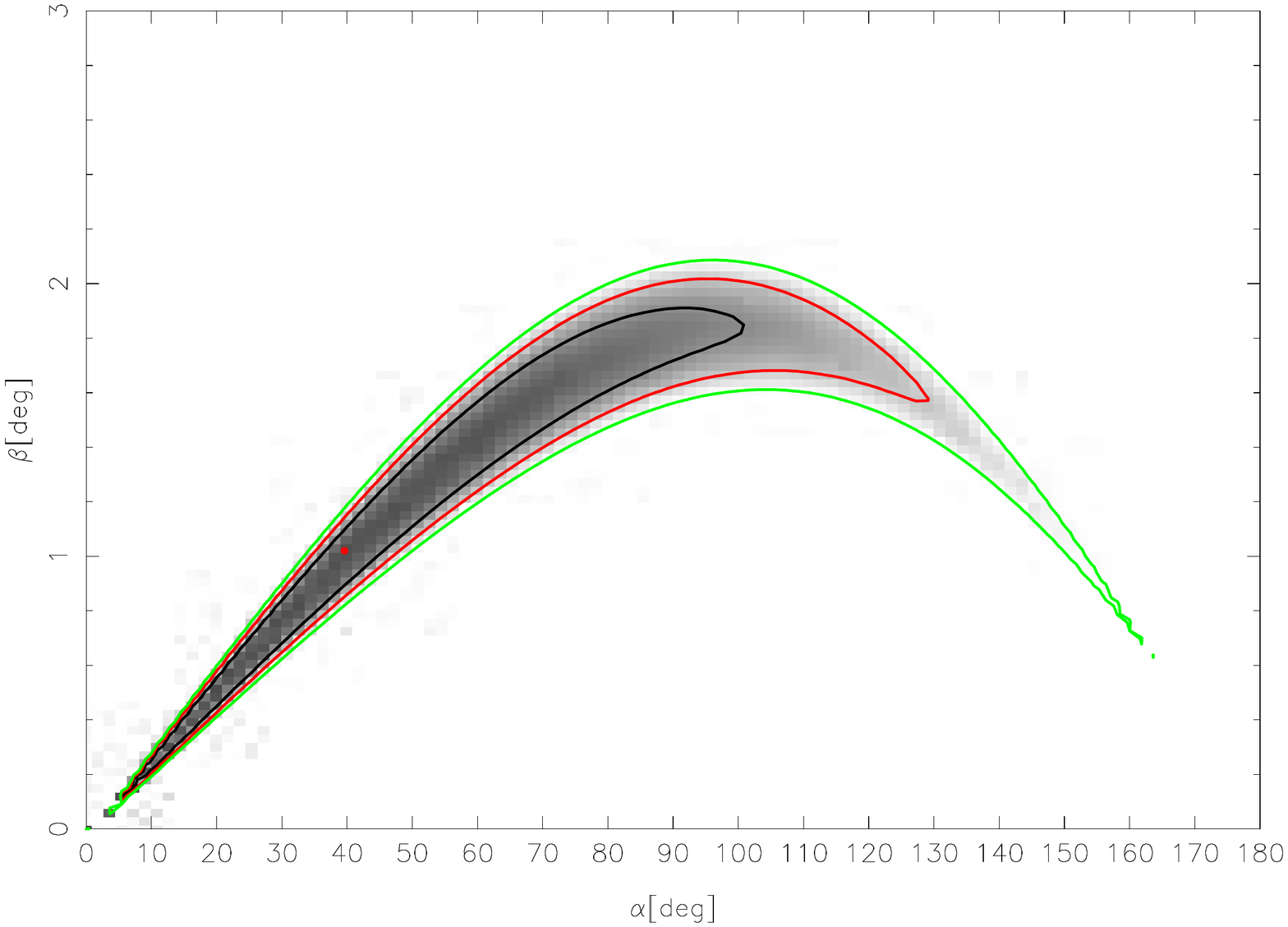}}}&
\\
{\mbox{\includegraphics[width=9cm,height=6cm,angle=0.]{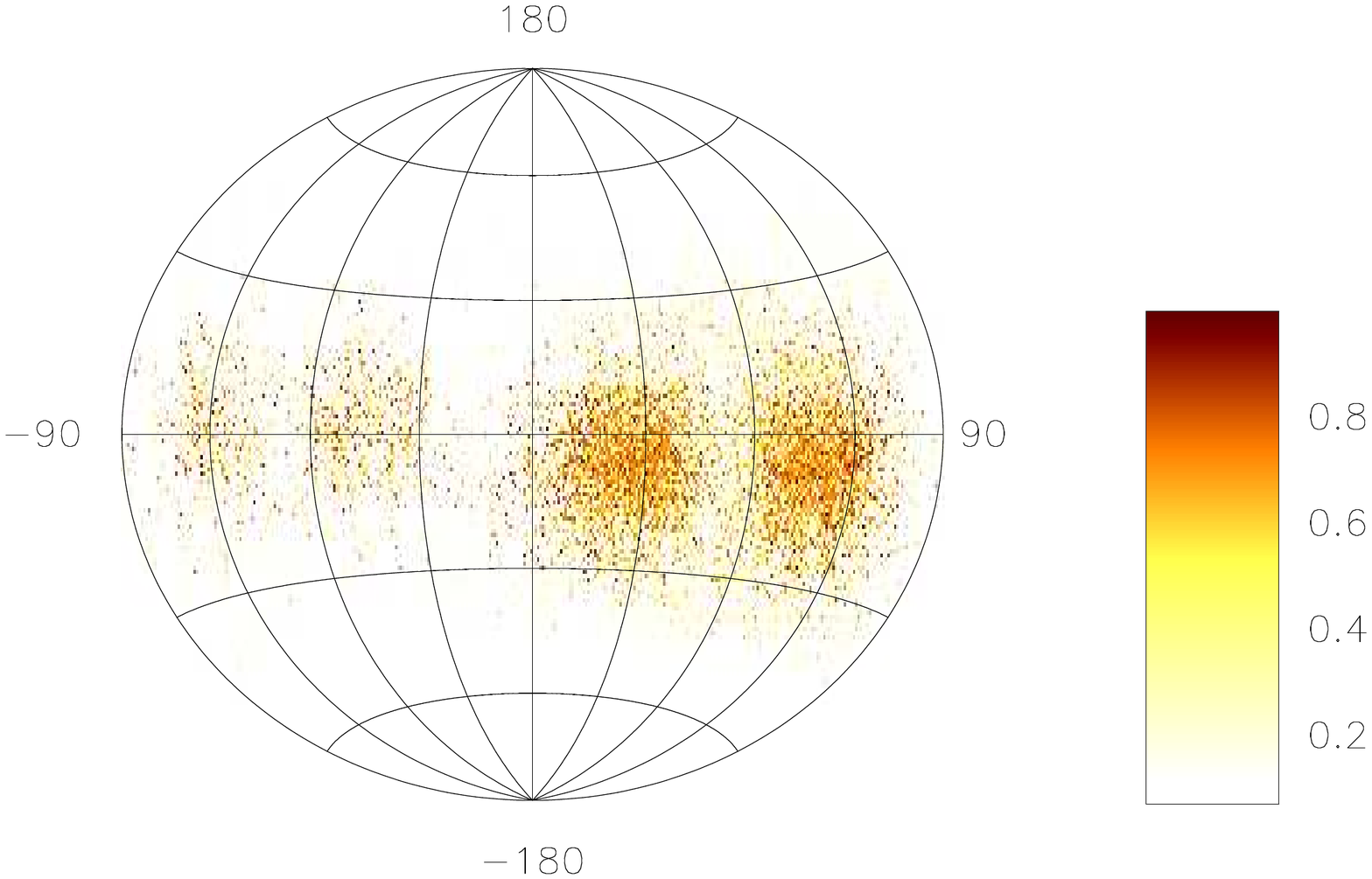}}}&
\\
\end{tabular}
\caption{Top panel (upper window) shows the average profile with total
intensity (Stokes I; solid black lines), total linear polarization (dashed red
line) and circular polarization (Stokes V; dotted blue line). Top panel (lower
window) also shows the single pulse PPA distribution (colour scale) along with
the average PPA (red error bars).
The RVM fits to the average PPA (dashed pink
line) is also shown in this plot. Middle panel show
the $\chi^2$ contours for the parameters $\alpha$ and $\beta$ obtained from RVM
fits.
Bottom panel shows the Hammer-Aitoff projection of the polarized time
samples with the colour scheme representing the fractional polarization level.}
\label{a7}
\end{center}
\end{figure*}

\begin{figure*}
\begin{center}
\begin{tabular}{cc}
{\mbox{\includegraphics[width=9cm,height=6cm,angle=0.]{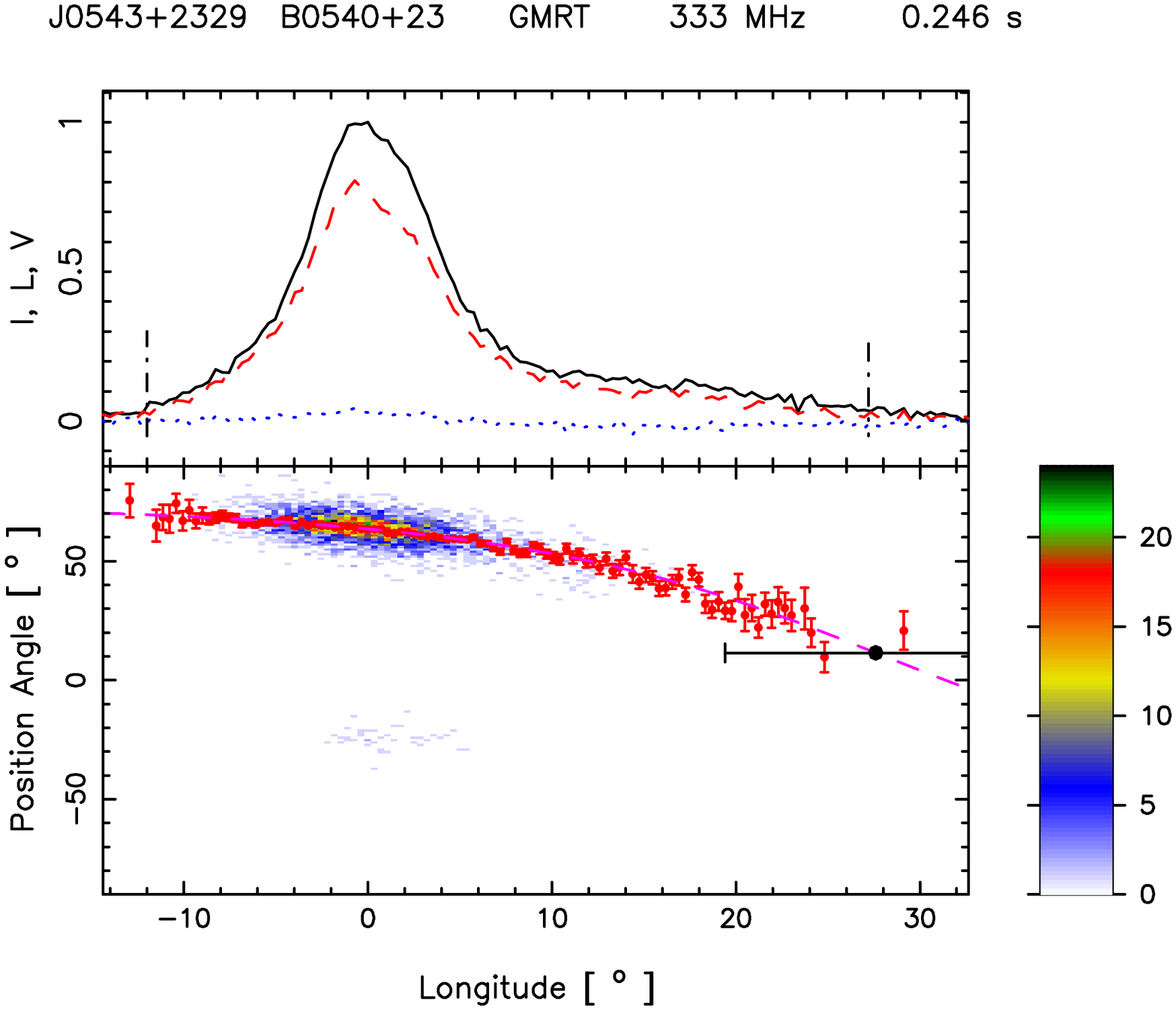}}}&
{\mbox{\includegraphics[width=9cm,height=6cm,angle=0.]{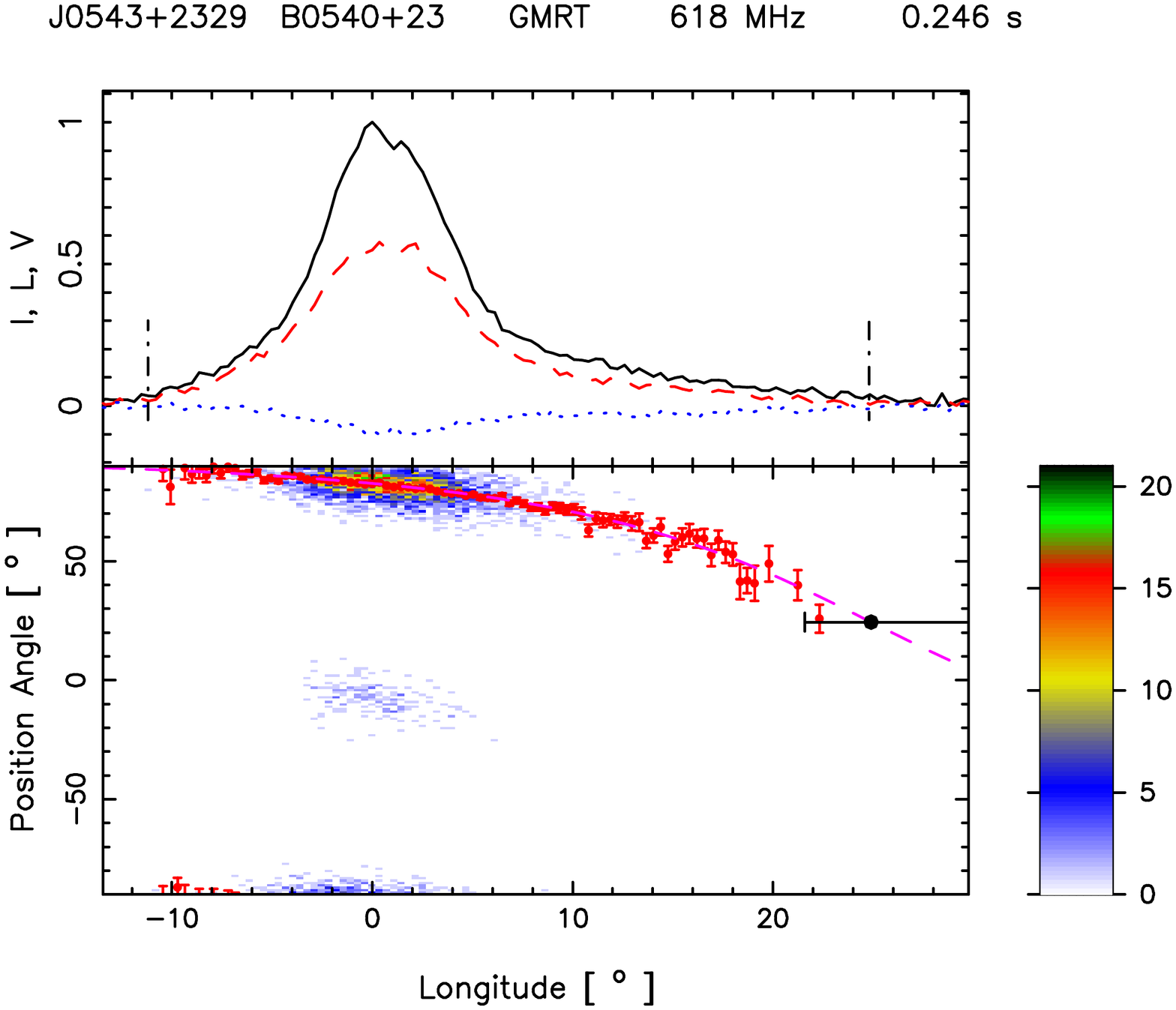}}}\\
{\mbox{\includegraphics[width=9cm,height=6cm,angle=0.]{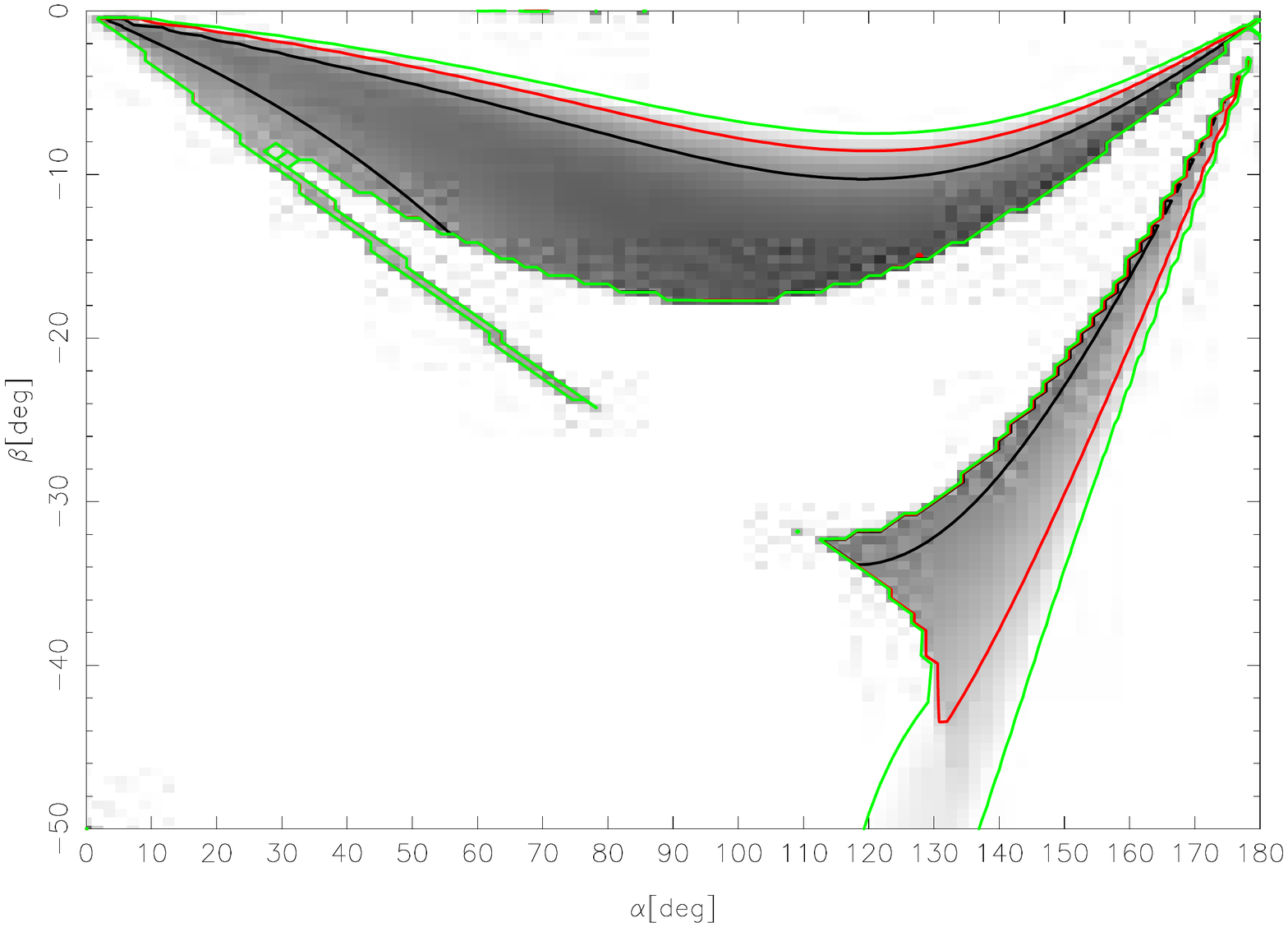}}}&
{\mbox{\includegraphics[width=9cm,height=6cm,angle=0.]{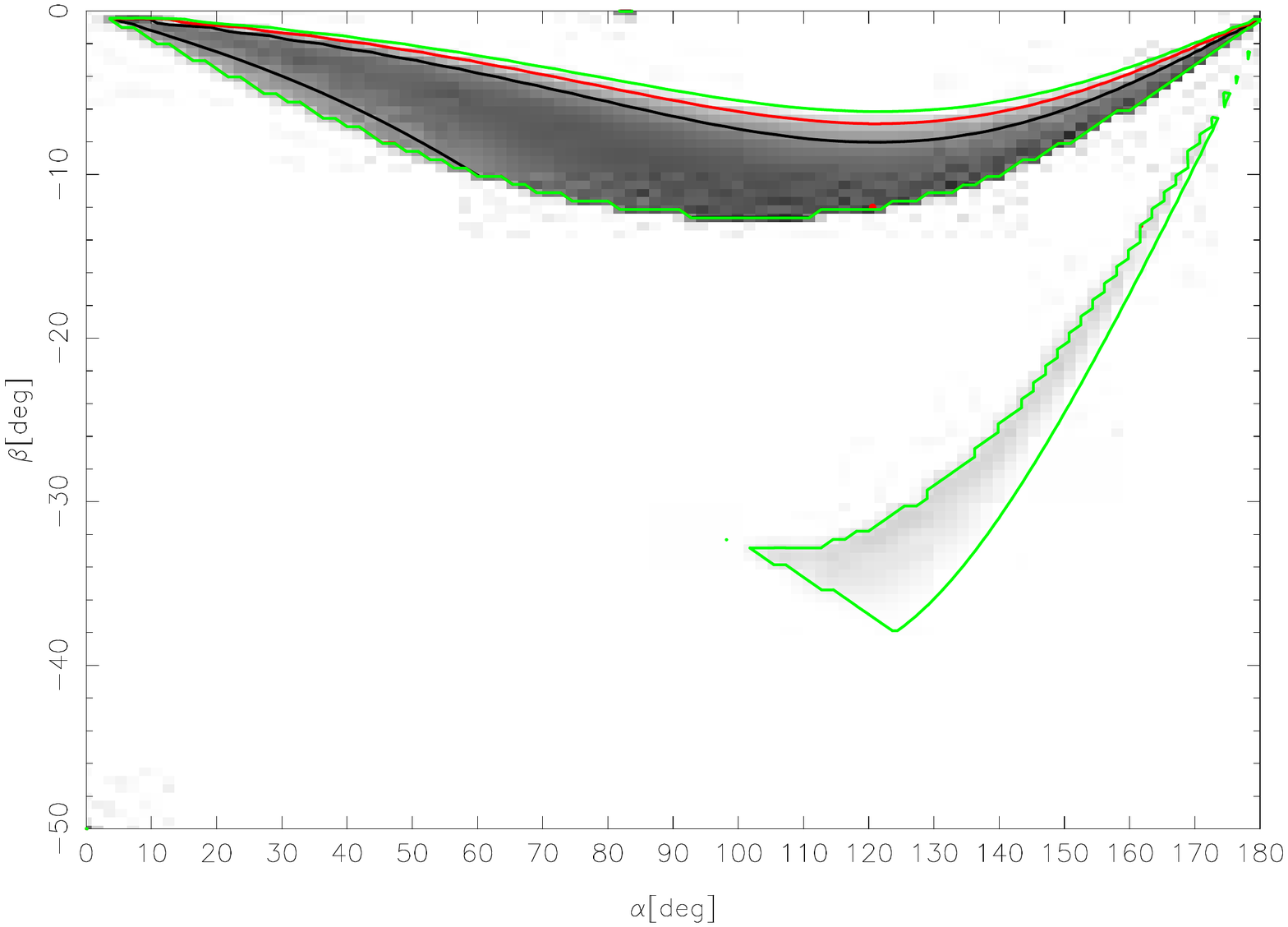}}}\\
{\mbox{\includegraphics[width=9cm,height=6cm,angle=0.]{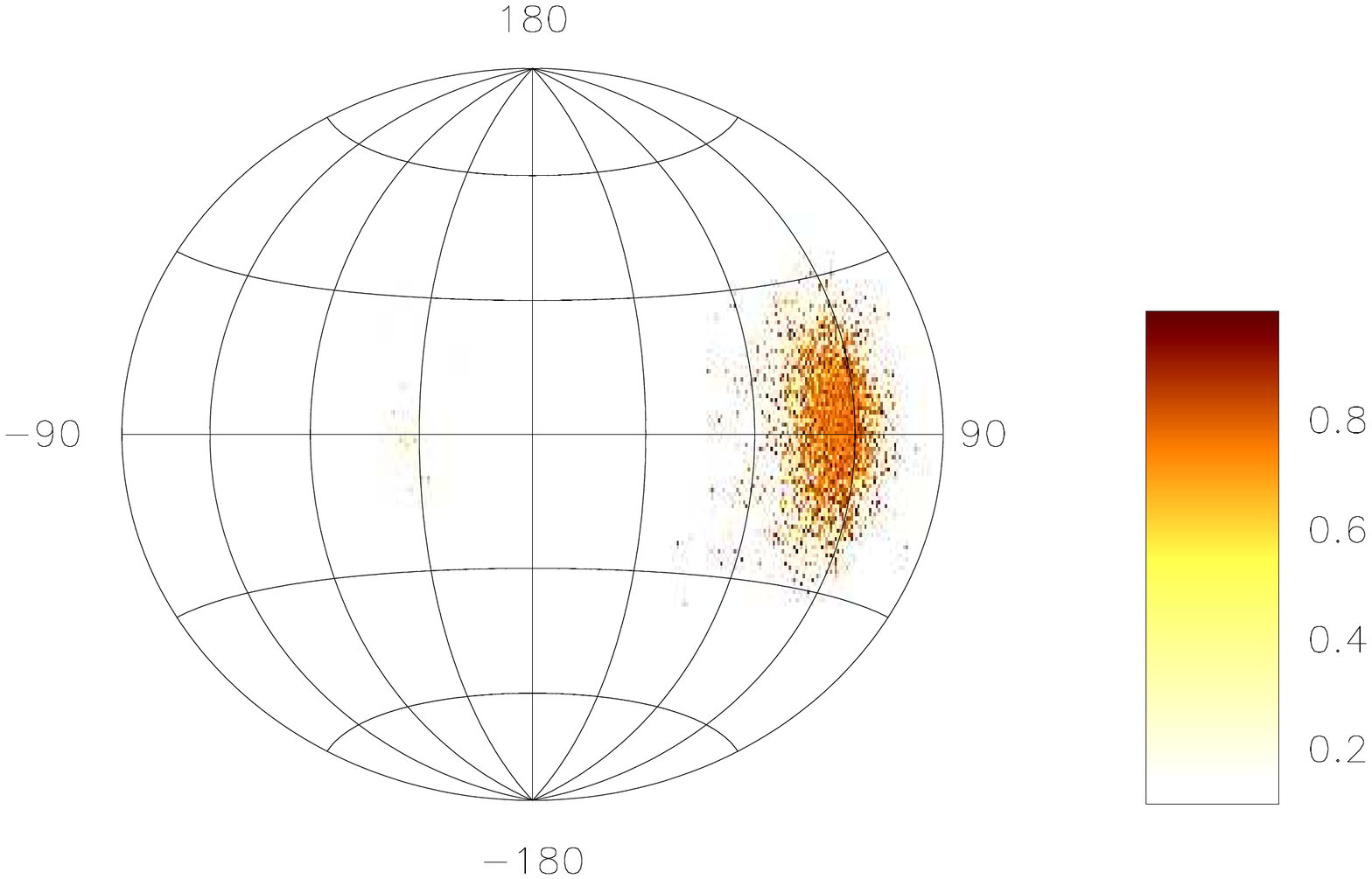}}}&
{\mbox{\includegraphics[width=9cm,height=6cm,angle=0.]{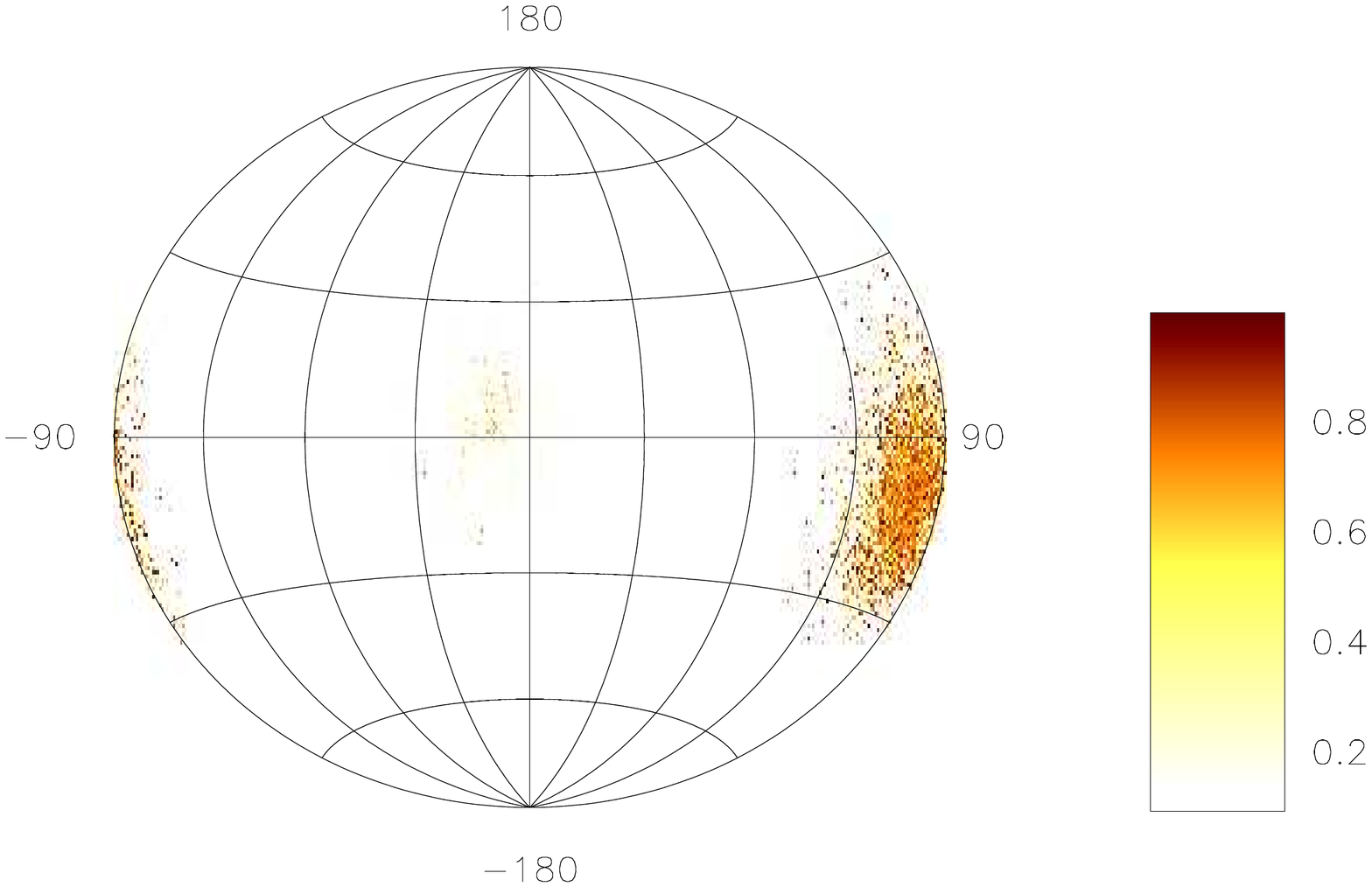}}}\\
\end{tabular}
\caption{Top panel (upper window) shows the average profile with total
intensity (Stokes I; solid black lines), total linear polarization (dashed red
line) and circular polarization (Stokes V; dotted blue line). Top panel (lower
window) also shows the single pulse PPA distribution (colour scale) along with
the average PPA (red error bars).
The RVM fits to the average PPA (dashed pink
line) is also shown in this plot. Middle panel show
the $\chi^2$ contours for the parameters $\alpha$ and $\beta$ obtained from RVM
fits.
Bottom panel shows the Hammer-Aitoff projection of the polarized time
samples with the colour scheme representing the fractional polarization level.}
\label{a8}
\end{center}
\end{figure*}

\begin{figure*}
\begin{center}
\begin{tabular}{cc}
{\mbox{\includegraphics[width=9cm,height=6cm,angle=0.]{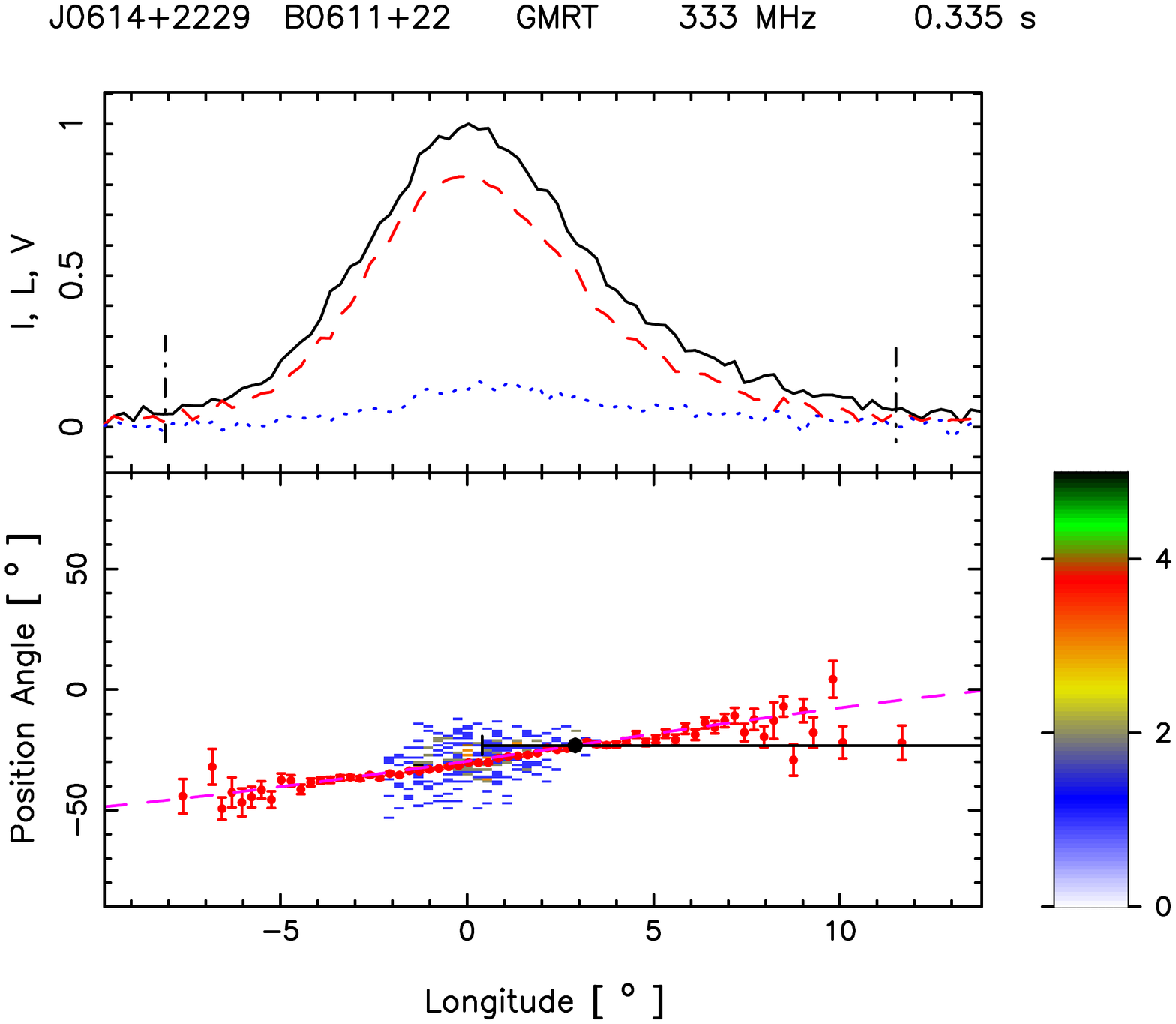}}}&
{\mbox{\includegraphics[width=9cm,height=6cm,angle=0.]{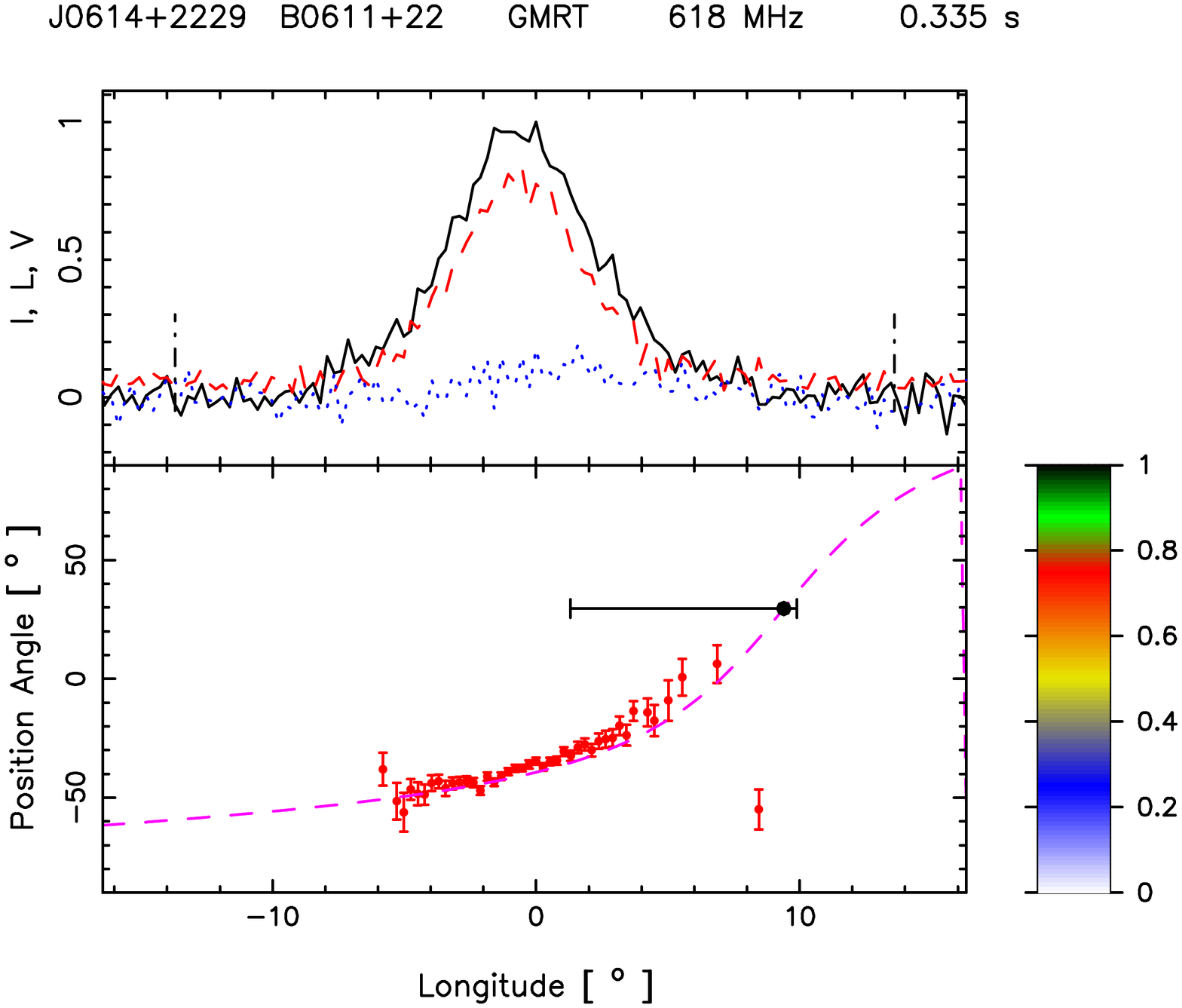}}}\\
{\mbox{\includegraphics[width=9cm,height=6cm,angle=0.]{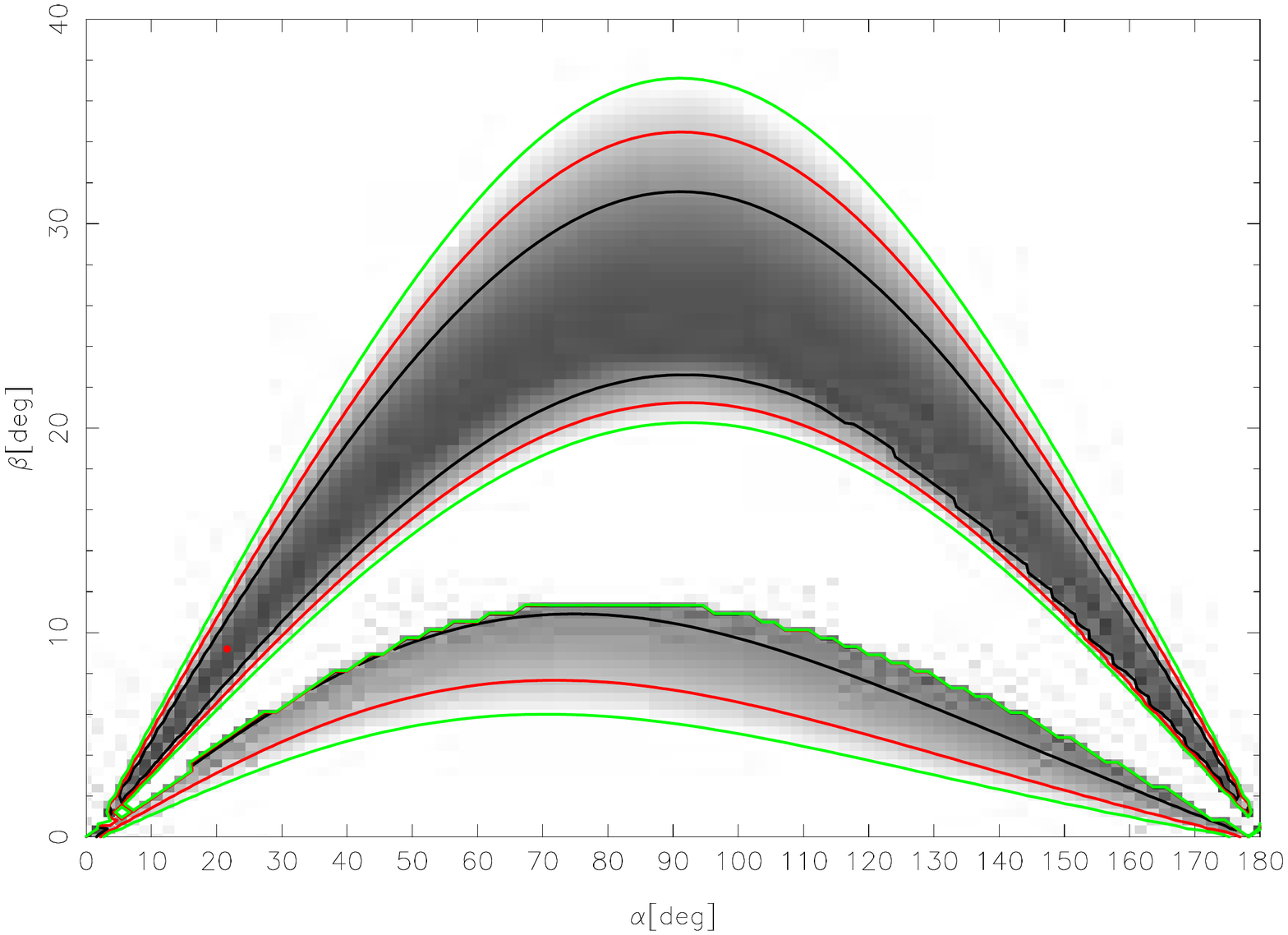}}}&
{\mbox{\includegraphics[width=9cm,height=6cm,angle=0.]{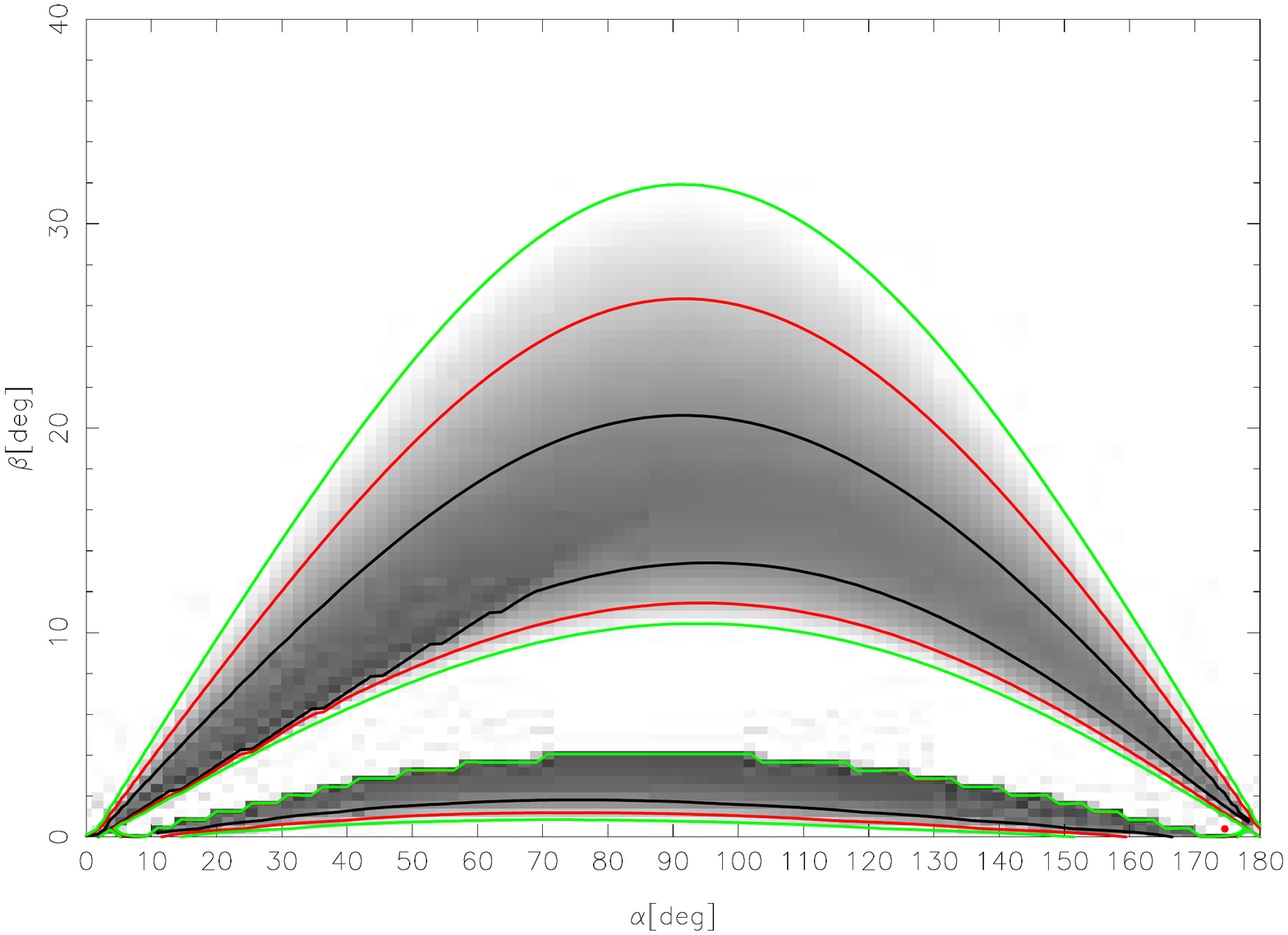}}}\\
{\mbox{\includegraphics[width=9cm,height=6cm,angle=0.]{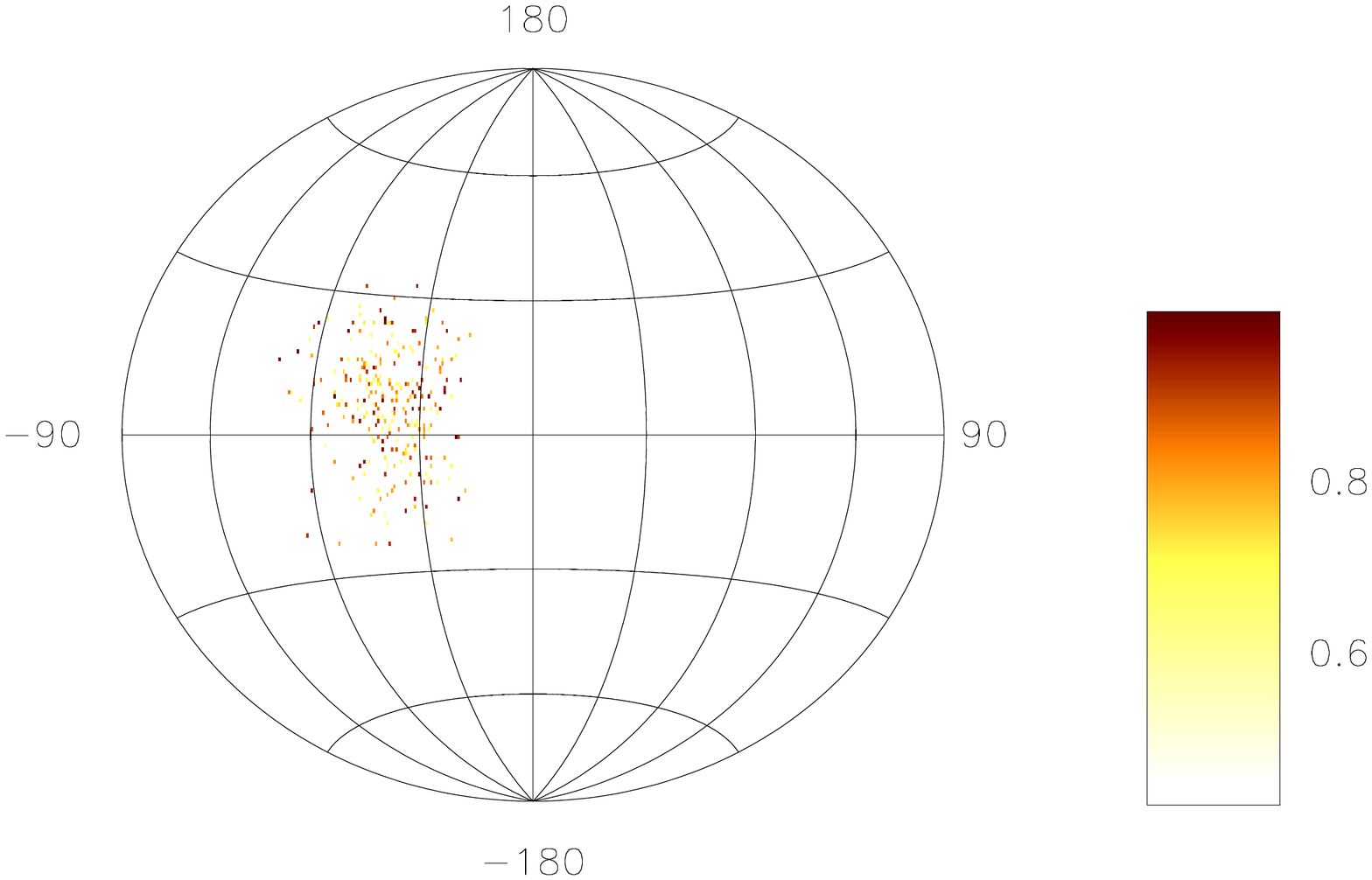}}}&
\\
\end{tabular}
\caption{Top panel (upper window) shows the average profile with total
intensity (Stokes I; solid black lines), total linear polarization (dashed red
line) and circular polarization (Stokes V; dotted blue line). Top panel (lower
window) also shows the single pulse PPA distribution (colour scale) along with
the average PPA (red error bars).
The RVM fits to the average PPA (dashed pink
line) is also shown in this plot. Middle panel show
the $\chi^2$ contours for the parameters $\alpha$ and $\beta$ obtained from RVM
fits.
Bottom panel only for 333 MHz shows the Hammer-Aitoff projection of the polarized time
samples with the colour scheme representing the fractional polarization level.}
\label{a9}
\end{center}
\end{figure*}


\begin{figure*}
\begin{center}
\begin{tabular}{cc}
{\mbox{\includegraphics[width=9cm,height=6cm,angle=0.]{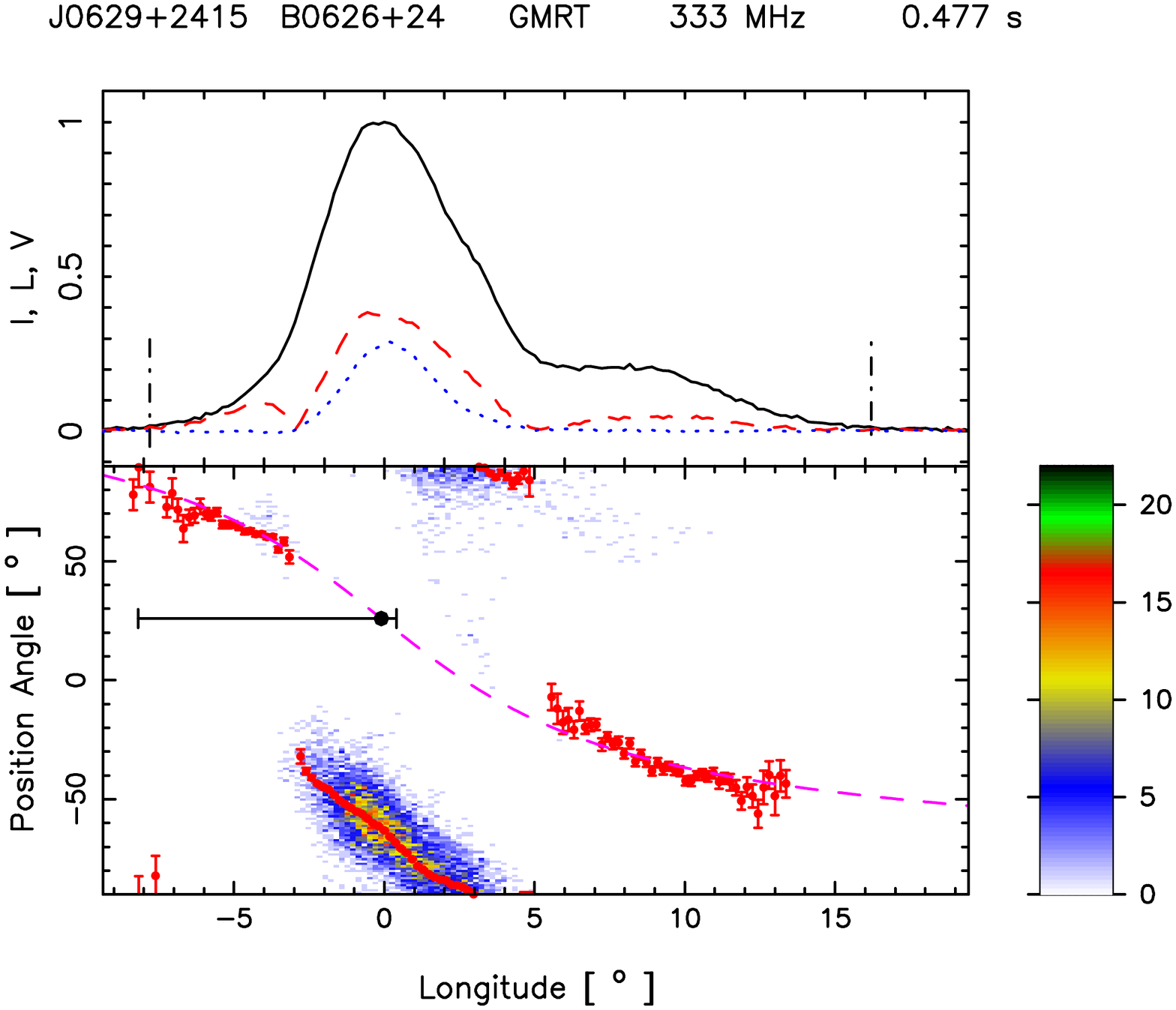}}}&
{\mbox{\includegraphics[width=9cm,height=6cm,angle=0.]{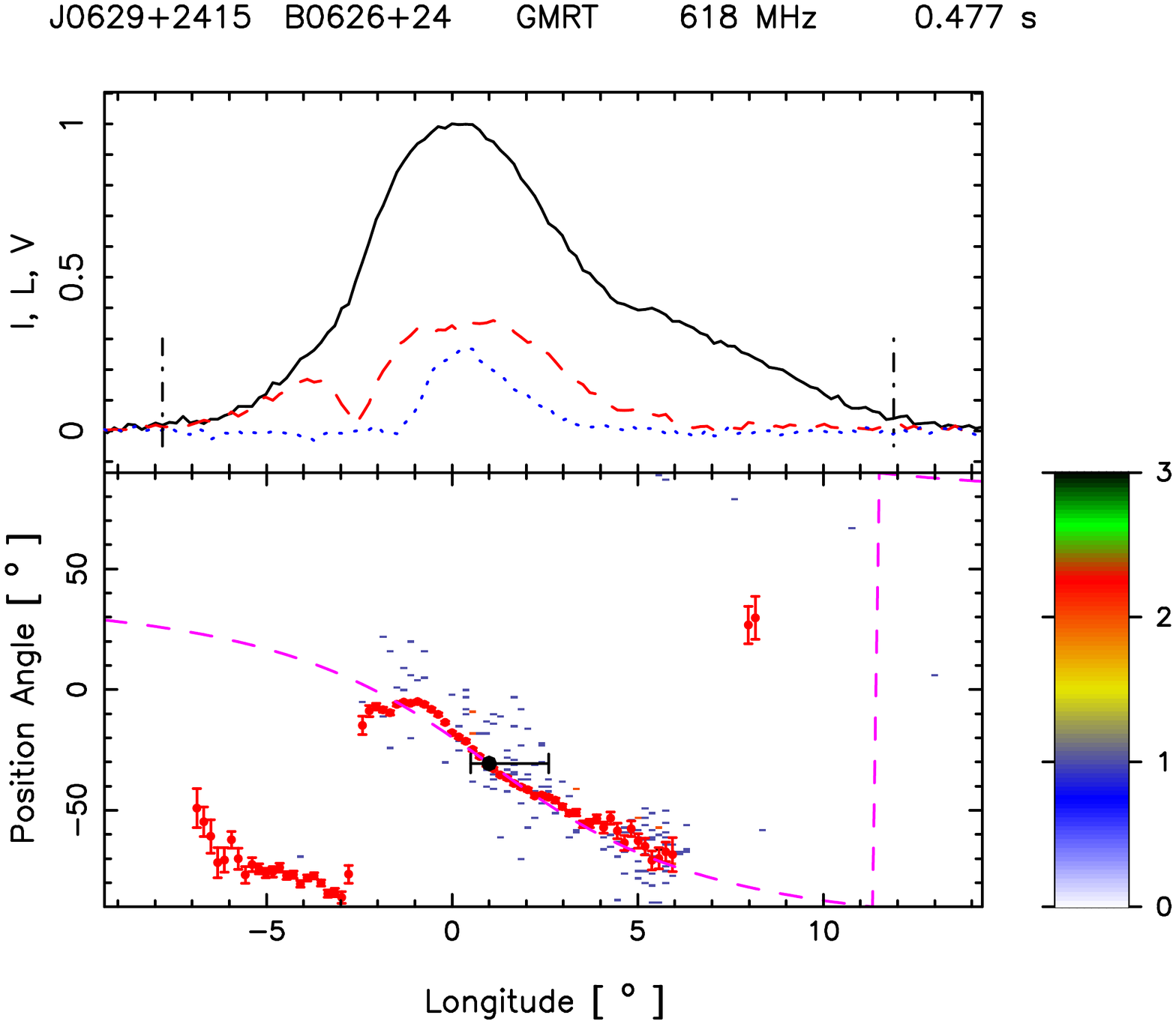}}}\\
{\mbox{\includegraphics[width=9cm,height=6cm,angle=0.]{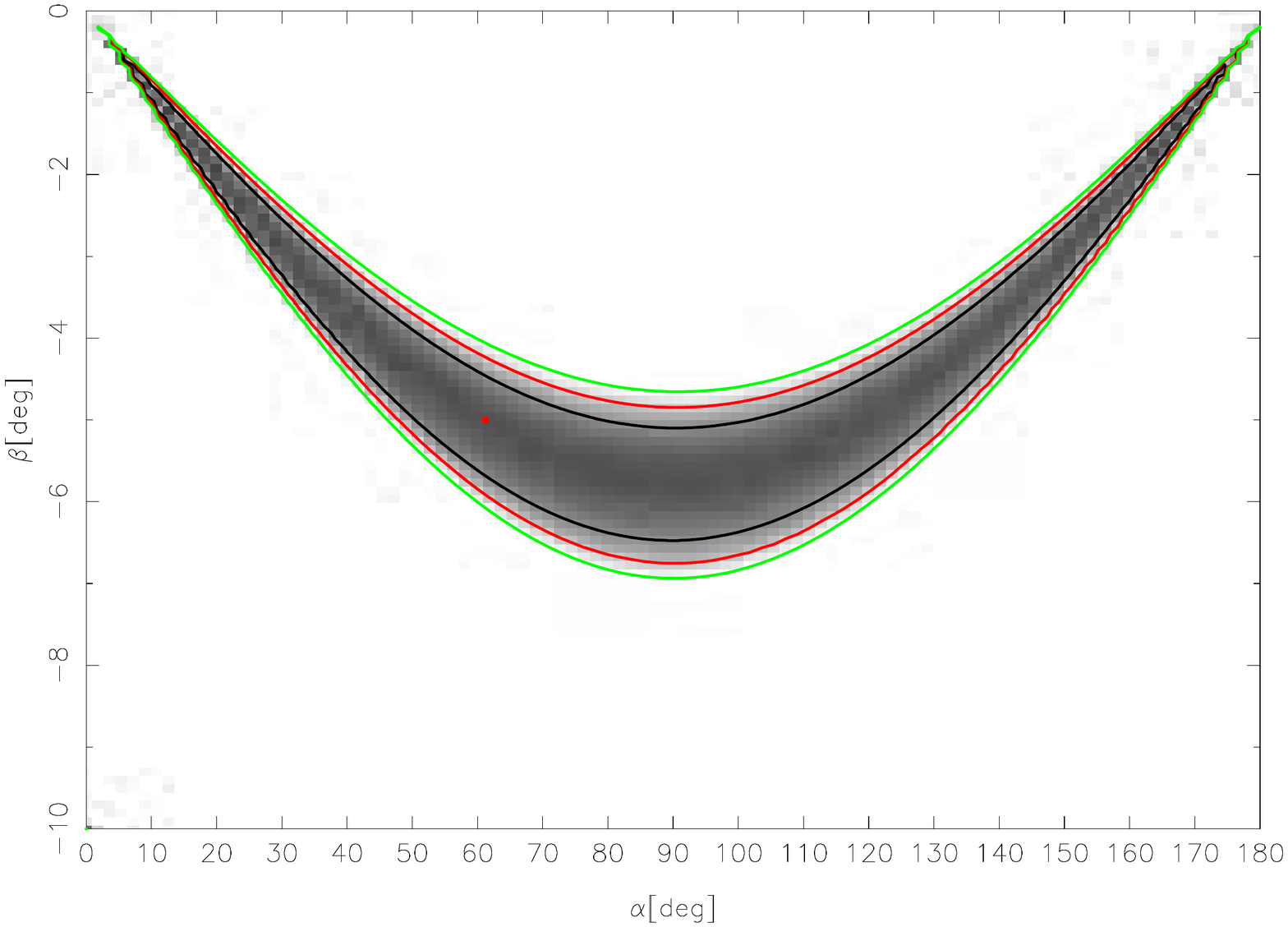}}}&
{\mbox{\includegraphics[width=9cm,height=6cm,angle=0.]{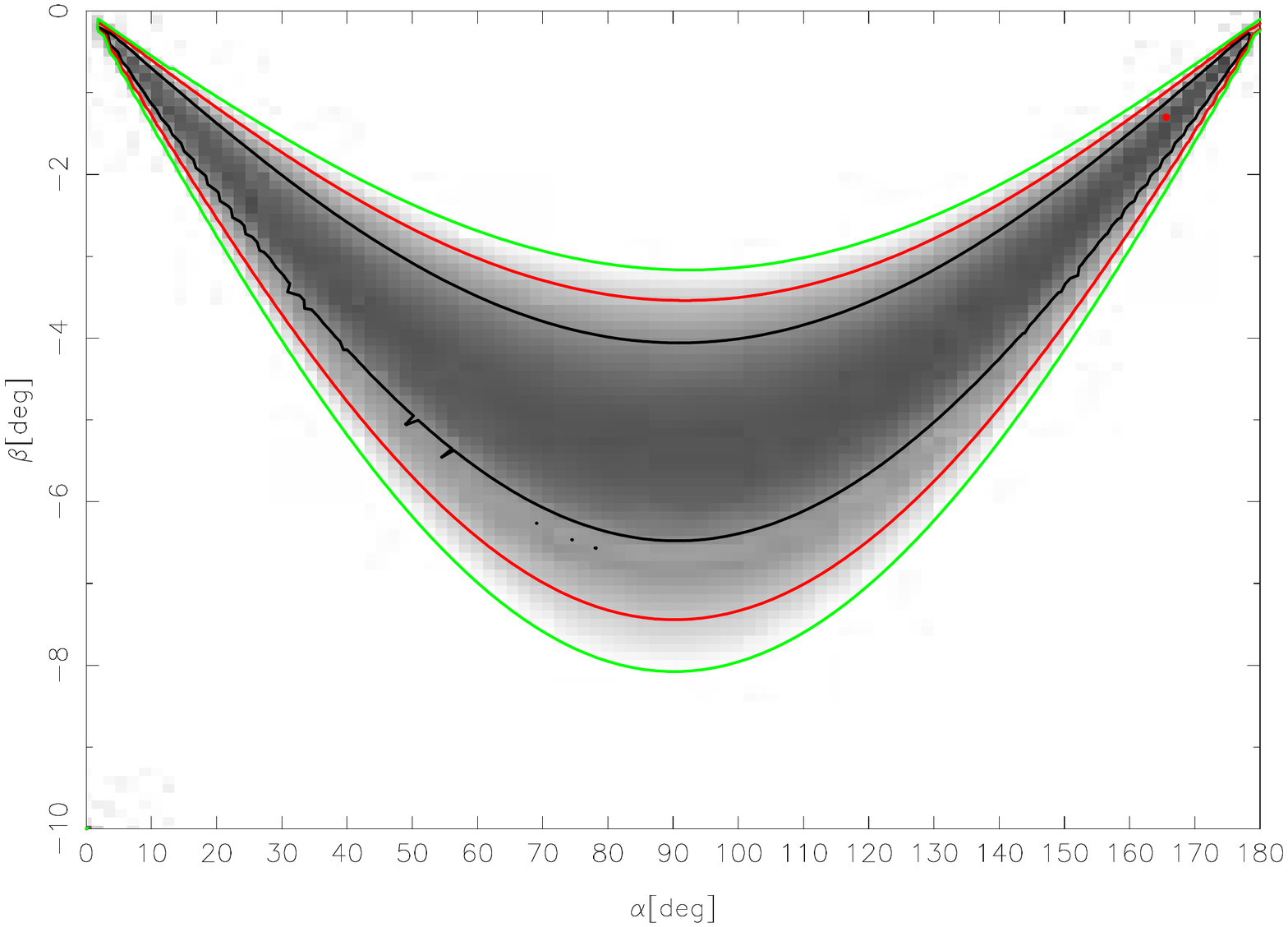}}}\\
{\mbox{\includegraphics[width=9cm,height=6cm,angle=0.]{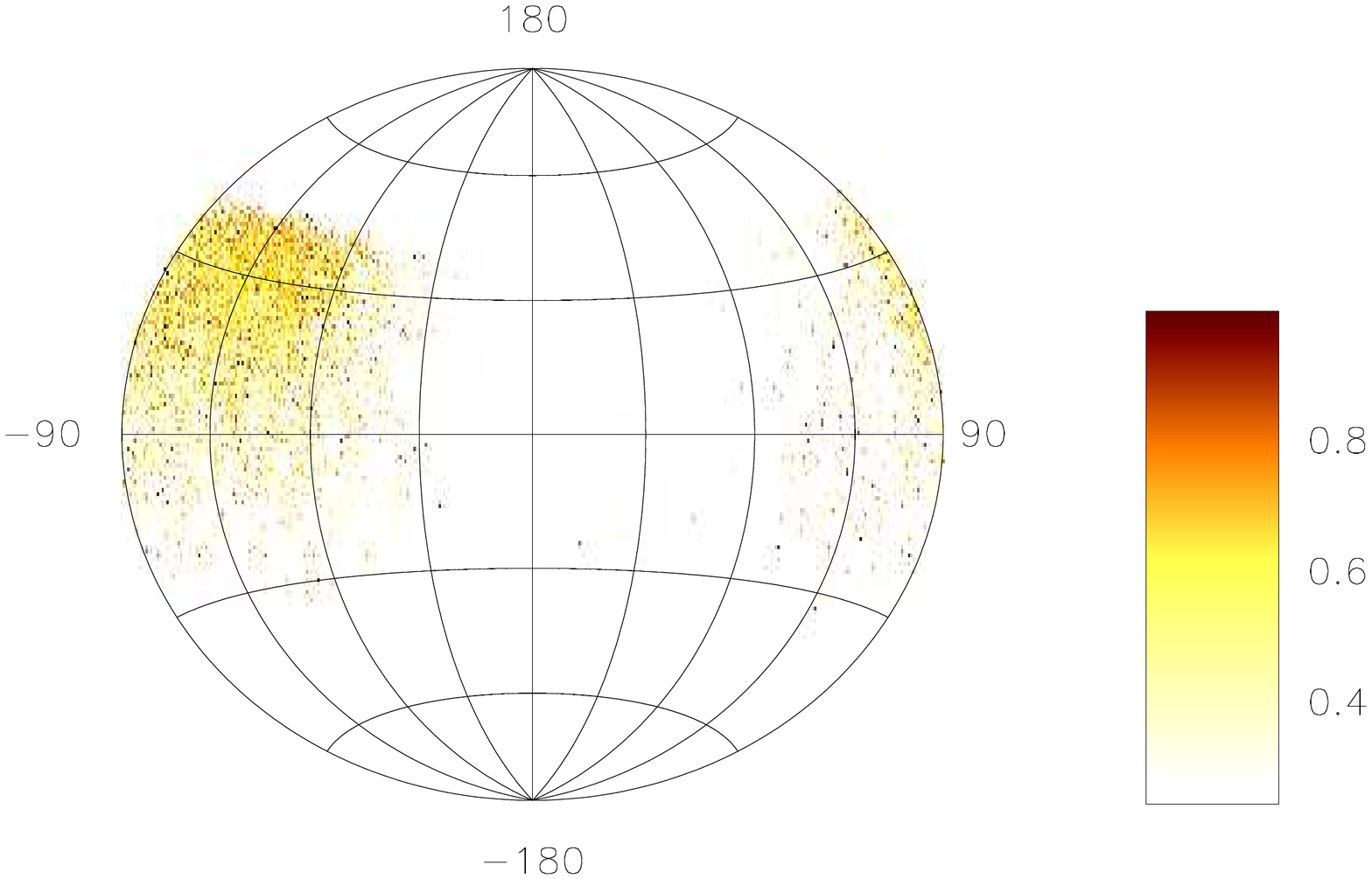}}}&
{\mbox{\includegraphics[width=9cm,height=6cm,angle=0.]{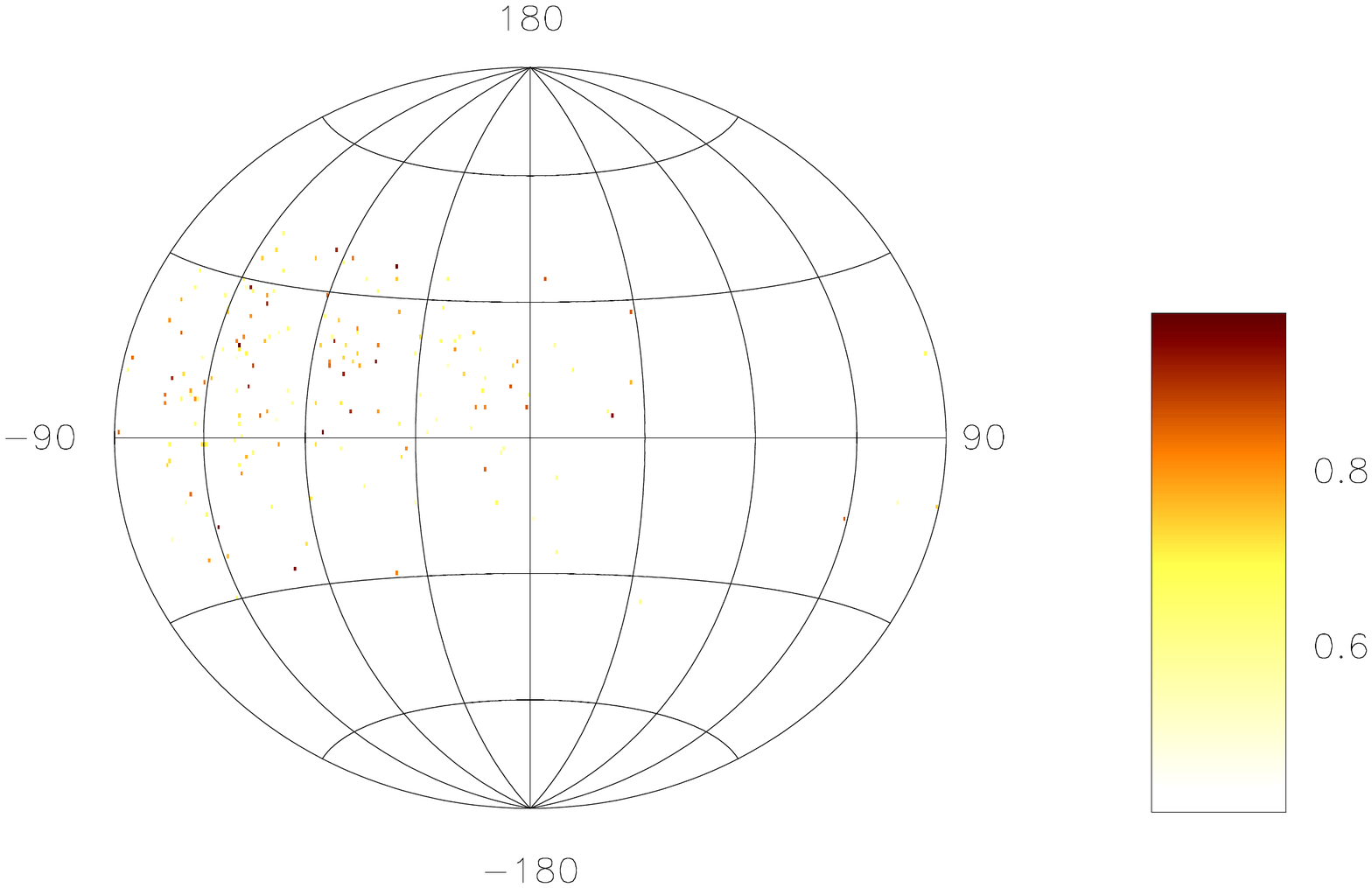}}}\\
\end{tabular}
\caption{Top panel (upper window) shows the average profile with total
intensity (Stokes I; solid black lines), total linear polarization (dashed red
line) and circular polarization (Stokes V; dotted blue line). Top panel (lower
window) also shows the single pulse PPA distribution (colour scale) along with
the average PPA (red error bars).
The RVM fits to the average PPA (dashed pink
line) is also shown in this plot. Middle panel show
the $\chi^2$ contours for the parameters $\alpha$ and $\beta$ obtained from RVM
fits.
Bottom panel shows the Hammer-Aitoff projection of the polarized time
samples with the colour scheme representing the fractional polarization level.}
\label{a10}
\end{center}
\end{figure*}

\begin{figure*}
\begin{center}
\begin{tabular}{cc}
{\mbox{\includegraphics[width=9cm,height=6cm,angle=0.]{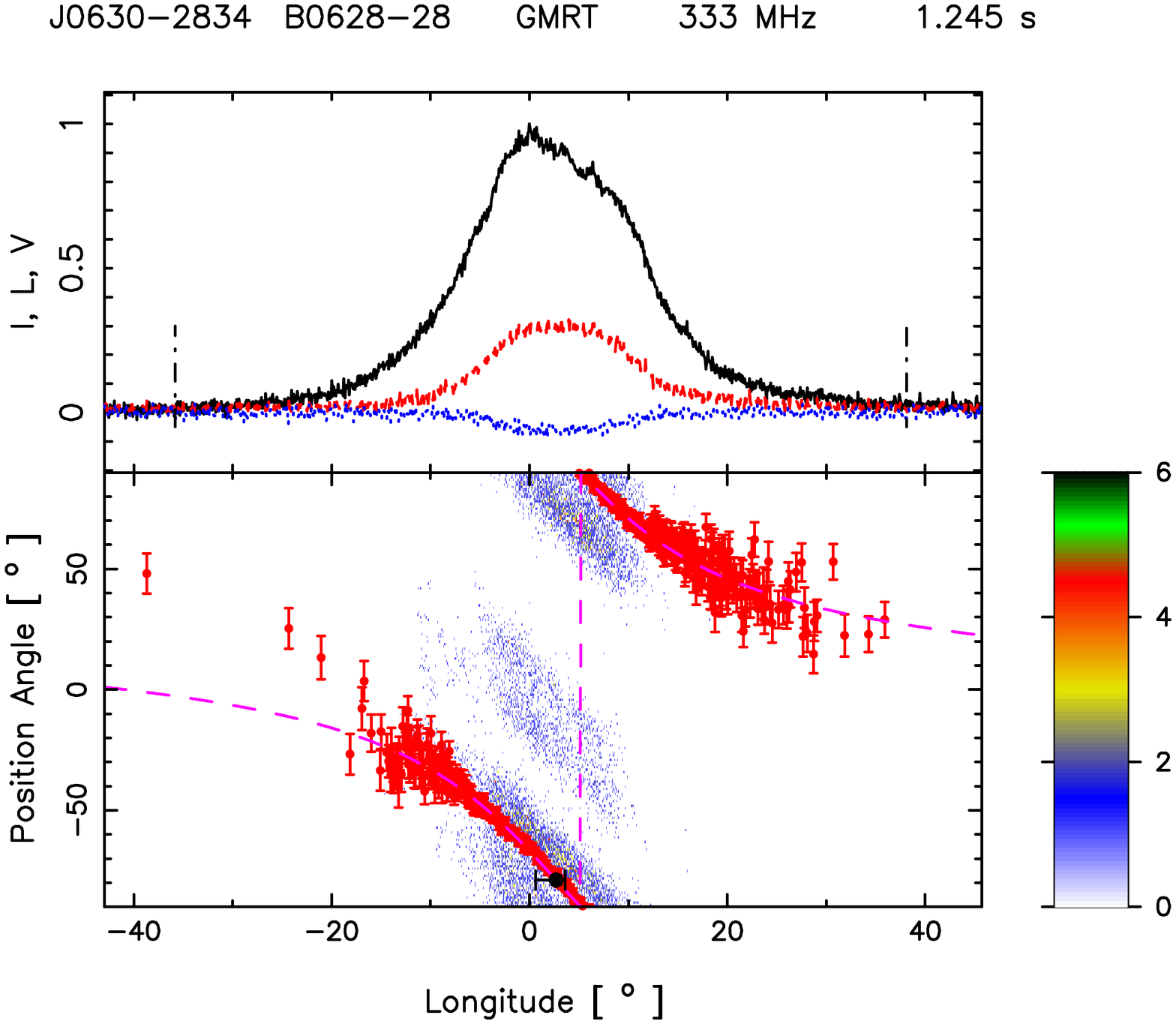}}}&
{\mbox{\includegraphics[width=9cm,height=6cm,angle=0.]{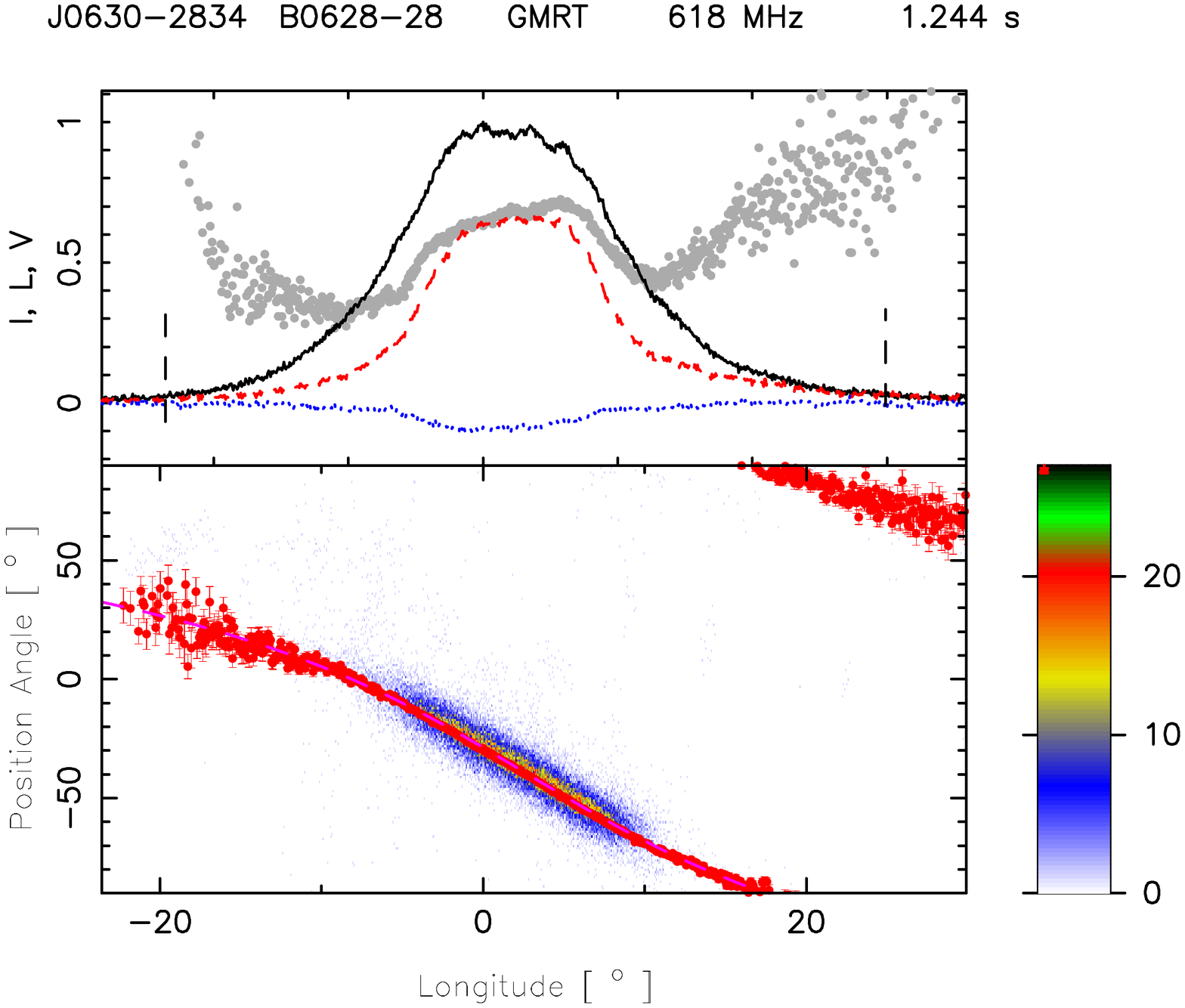}}}\\
{\mbox{\includegraphics[width=9cm,height=6cm,angle=0.]{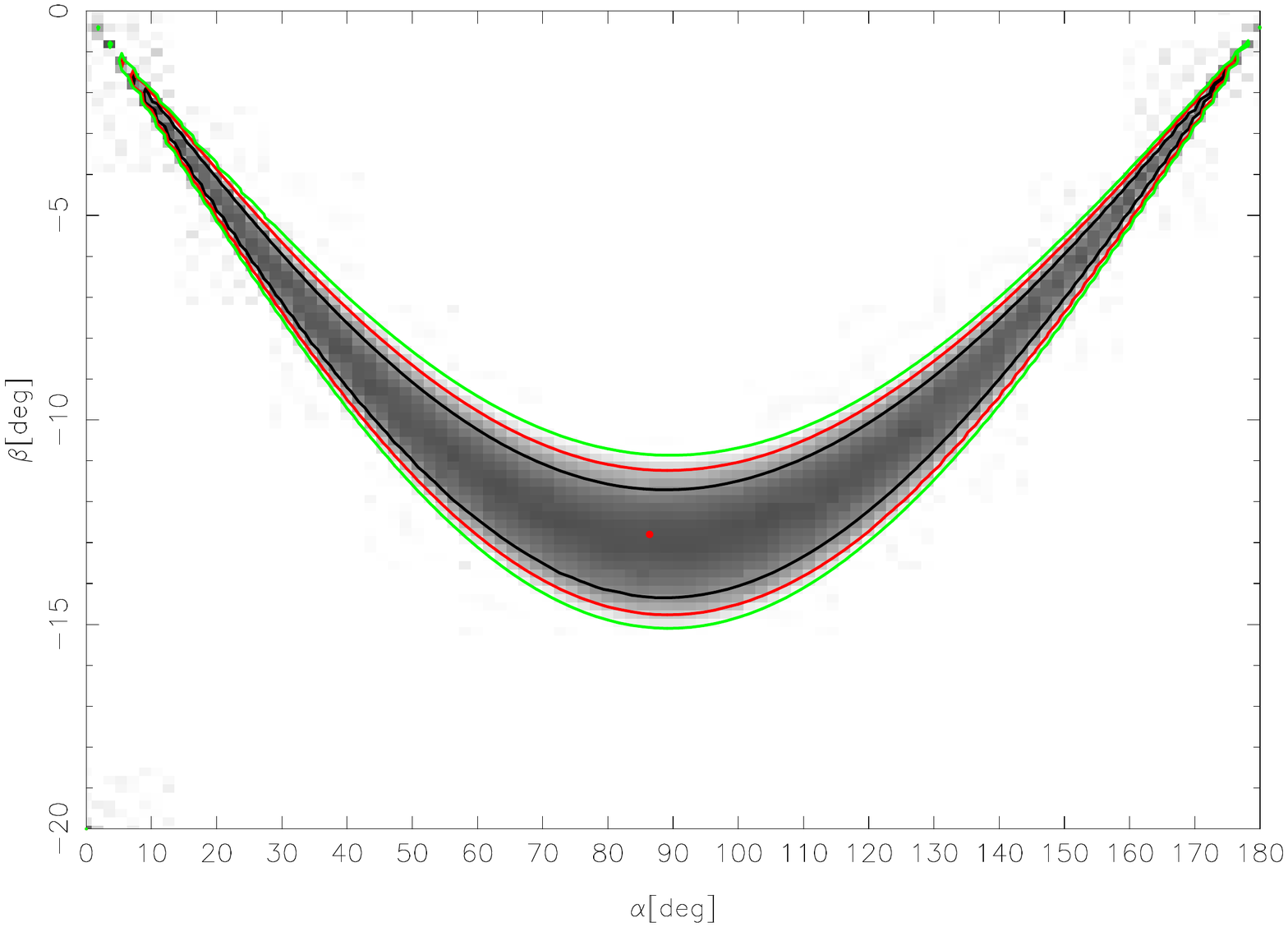}}}&
{\mbox{\includegraphics[width=9cm,height=6cm,angle=0.]{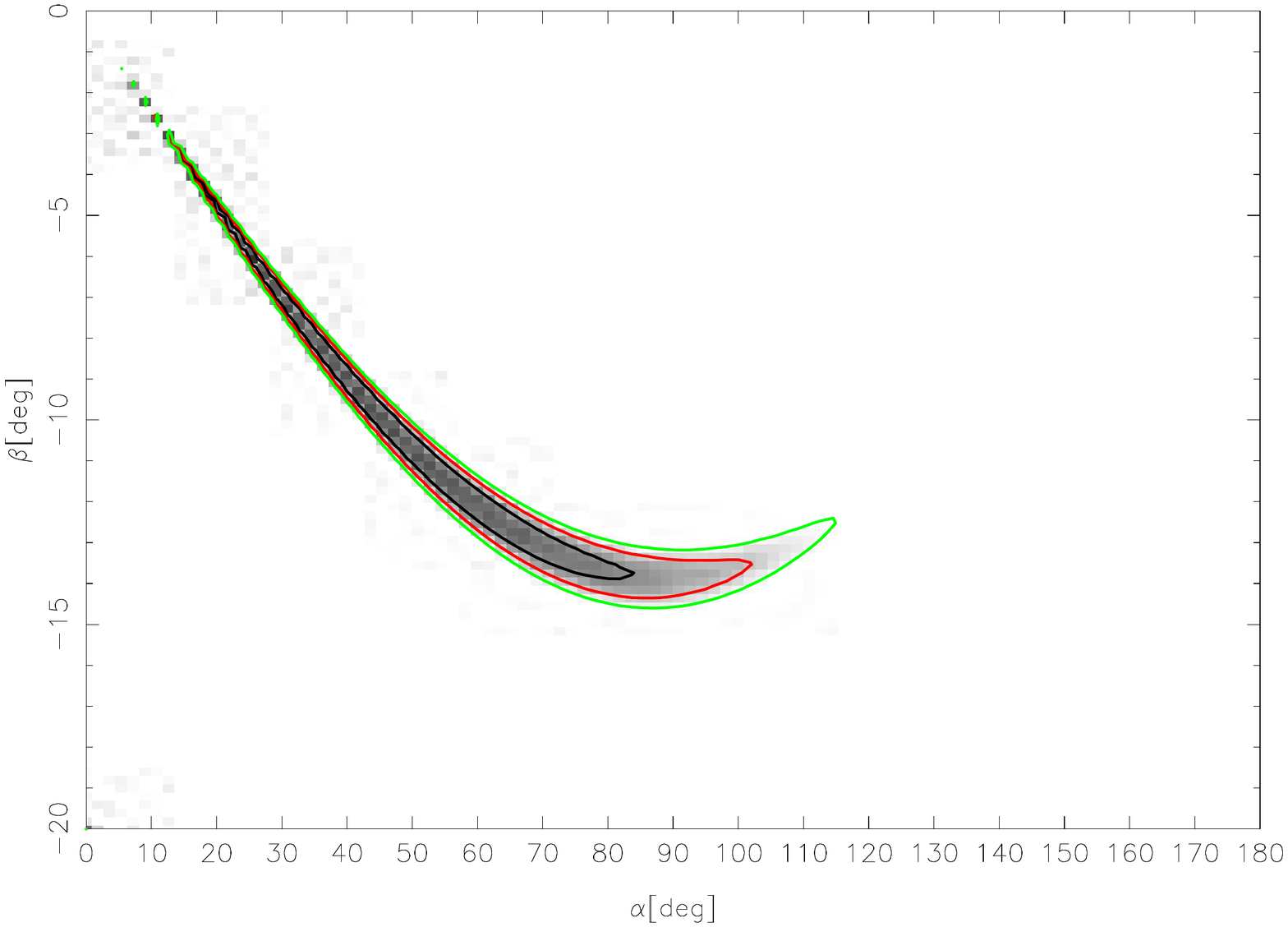}}}\\
{\mbox{\includegraphics[width=9cm,height=6cm,angle=0.]{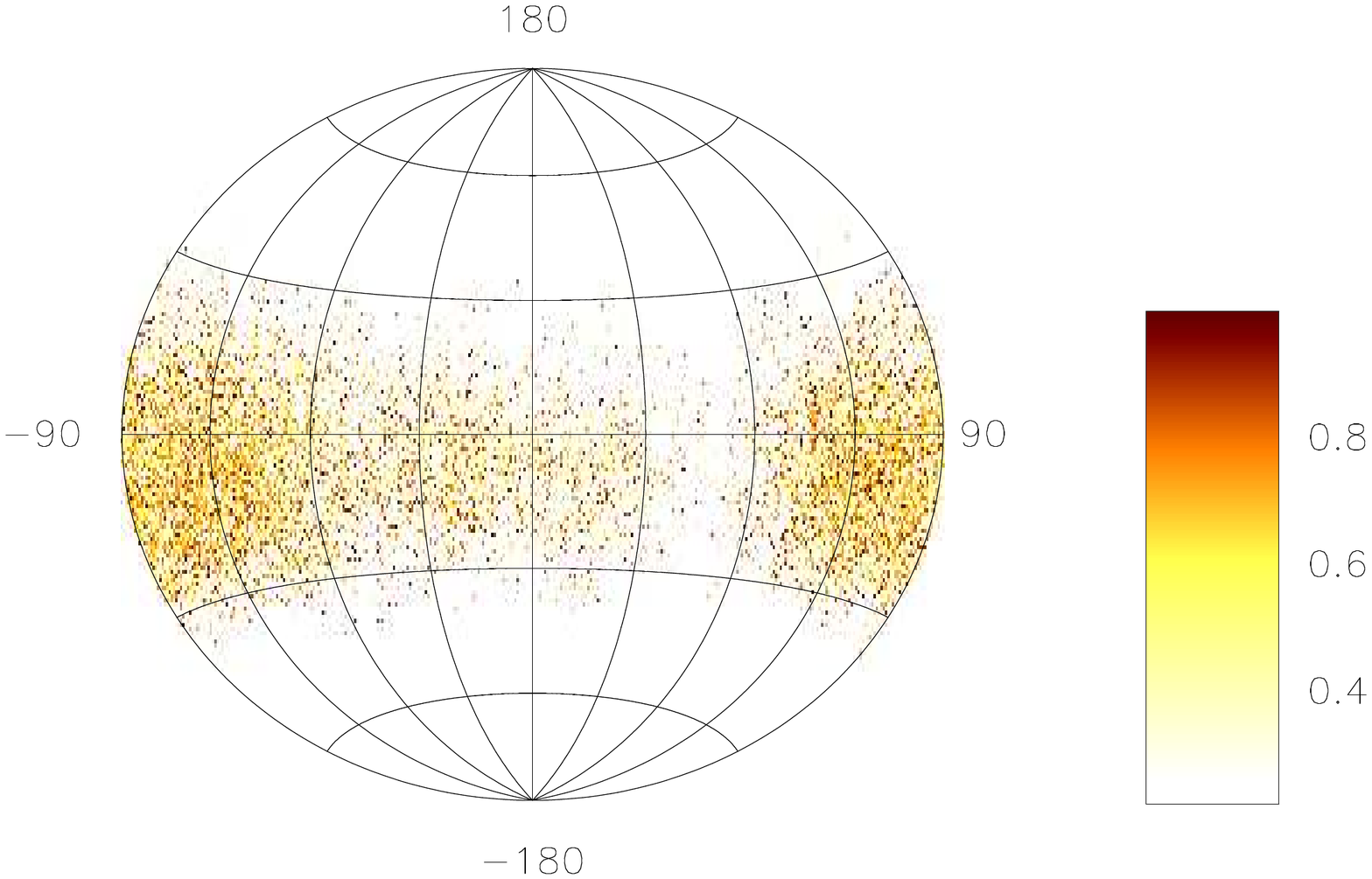}}}&
{\mbox{\includegraphics[width=9cm,height=6cm,angle=0.]{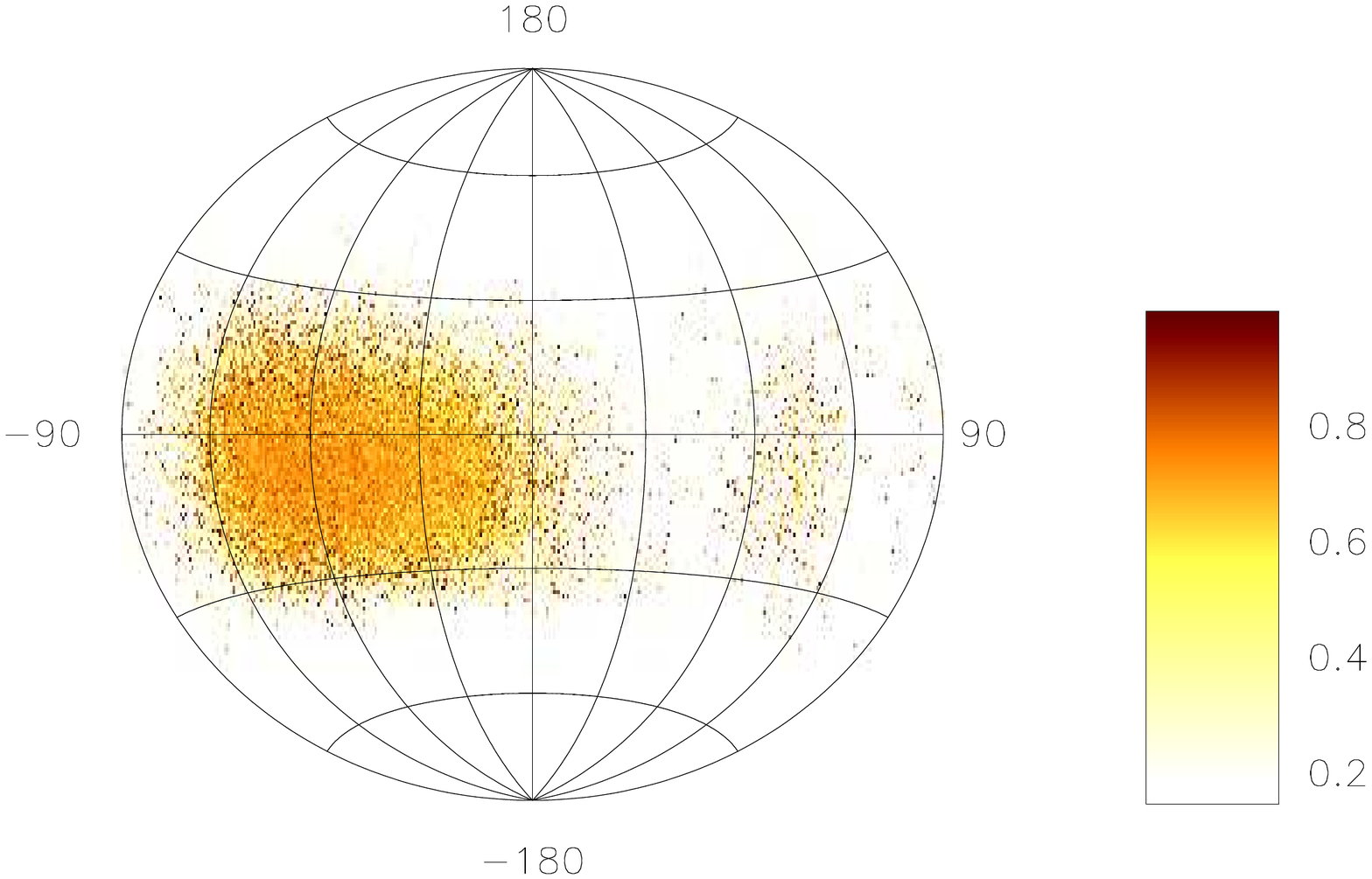}}}\\
\end{tabular}
\caption{Top panel (upper window) shows the average profile with total
intensity (Stokes I; solid black lines), total linear polarization (dashed red
line) and circular polarization (Stokes V; dotted blue line). Top panel (lower
window) also shows the single pulse PPA distribution (colour scale) along with
the average PPA (red error bars).
The RVM fits to the average PPA (dashed pink
line) is also shown in this plot. Middle panel show
the $\chi^2$ contours for the parameters $\alpha$ and $\beta$ obtained from RVM
fits.
Bottom panel shows the Hammer-Aitoff projection of the polarized time
samples with the colour scheme representing the fractional polarization level.}
\label{a11}
\end{center}
\end{figure*}


\begin{figure*}
\begin{center}
\begin{tabular}{cc}
{\mbox{\includegraphics[width=9cm,height=6cm,angle=0.]{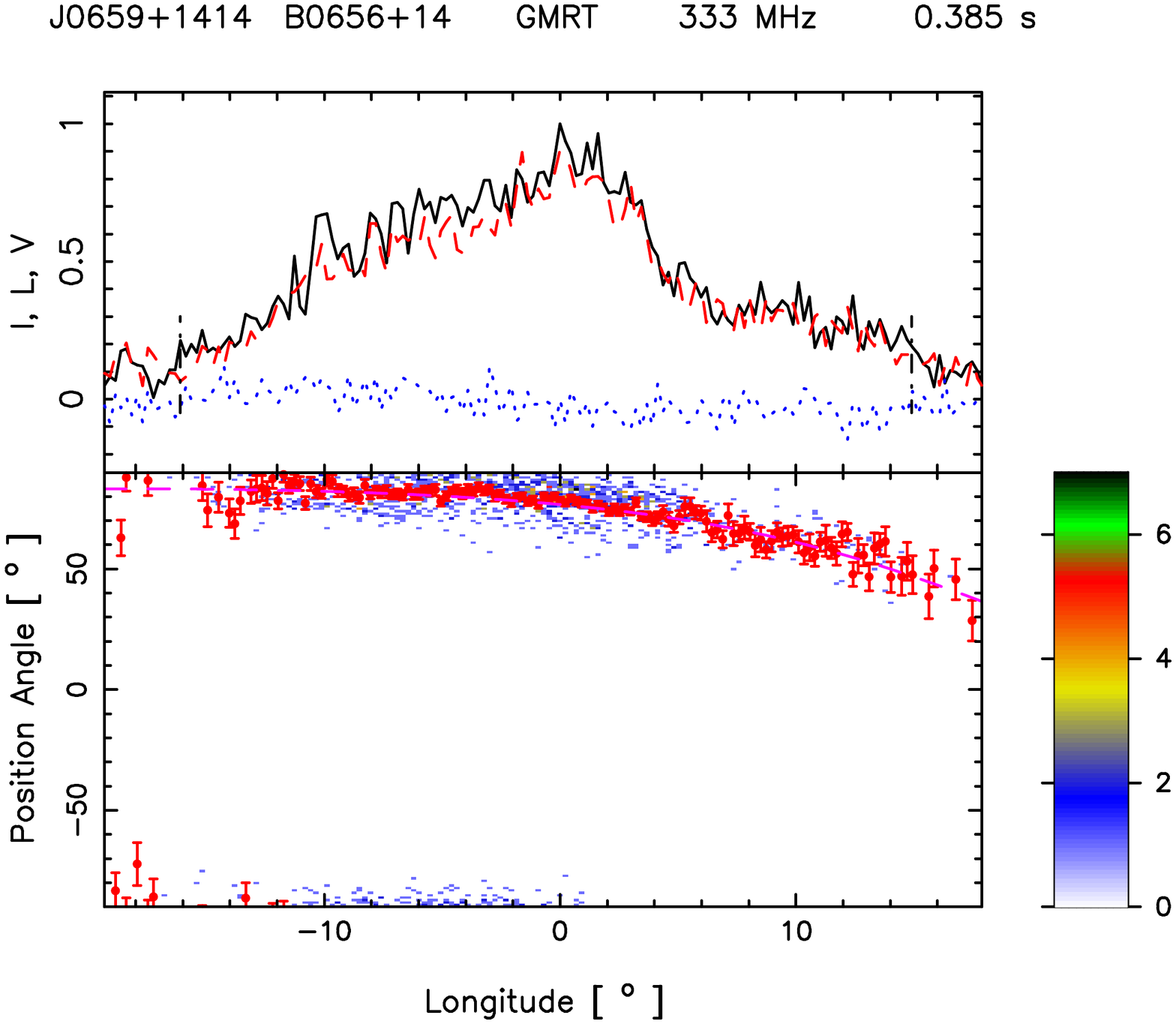}}}&
{\mbox{\includegraphics[width=9cm,height=6cm,angle=0.]{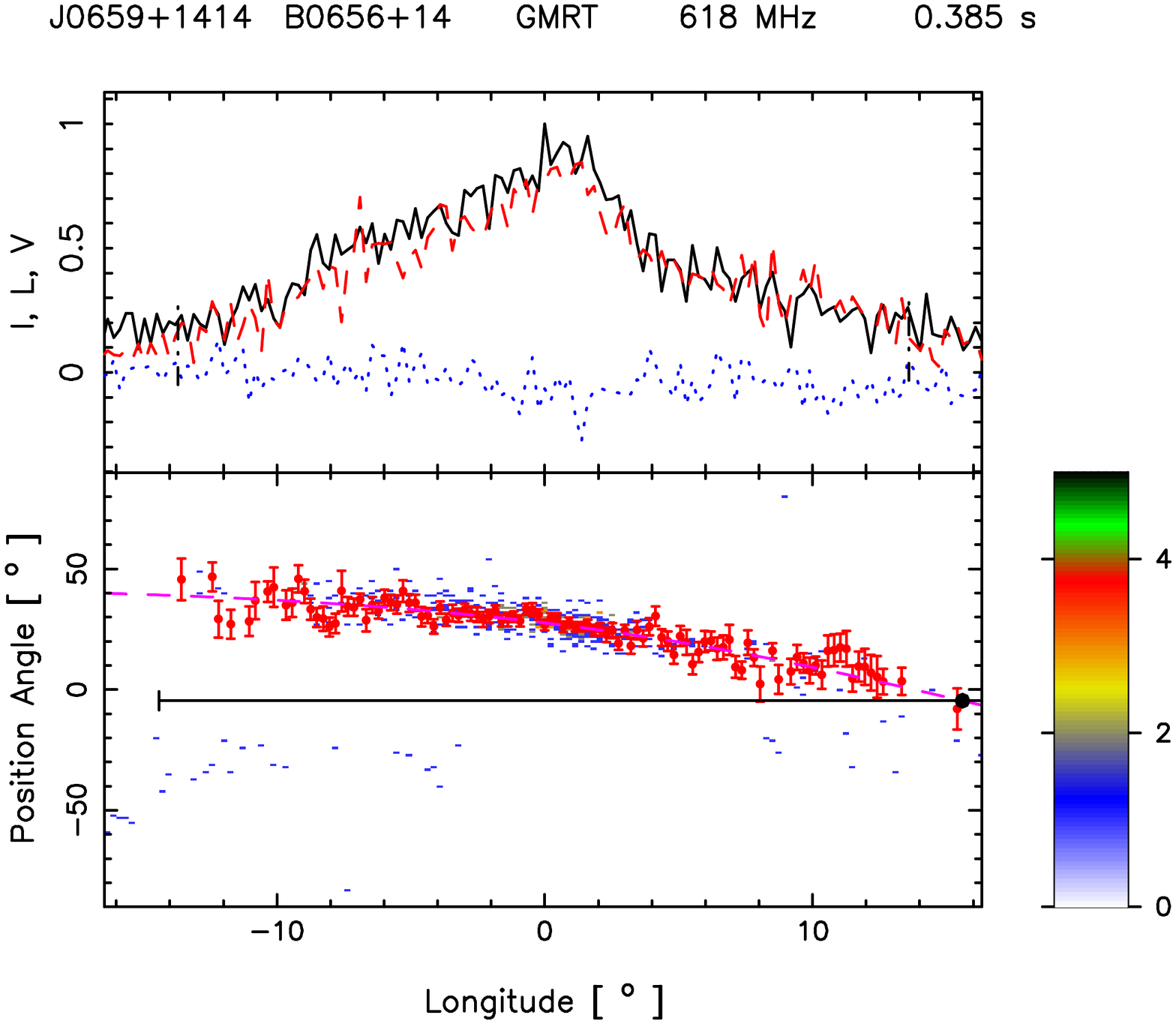}}}\\
{\mbox{\includegraphics[width=9cm,height=6cm,angle=0.]{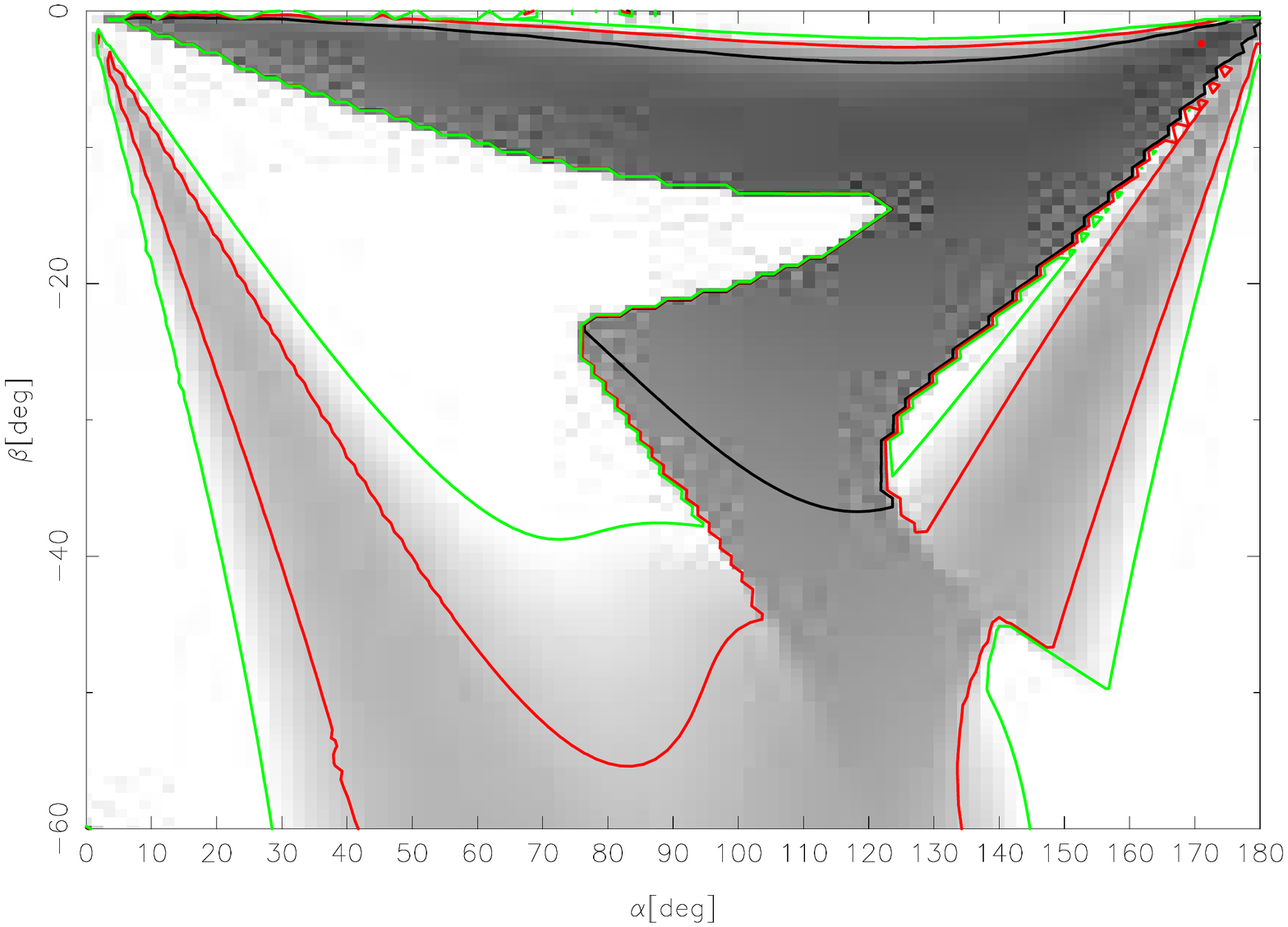}}}&
{\mbox{\includegraphics[width=9cm,height=6cm,angle=0.]{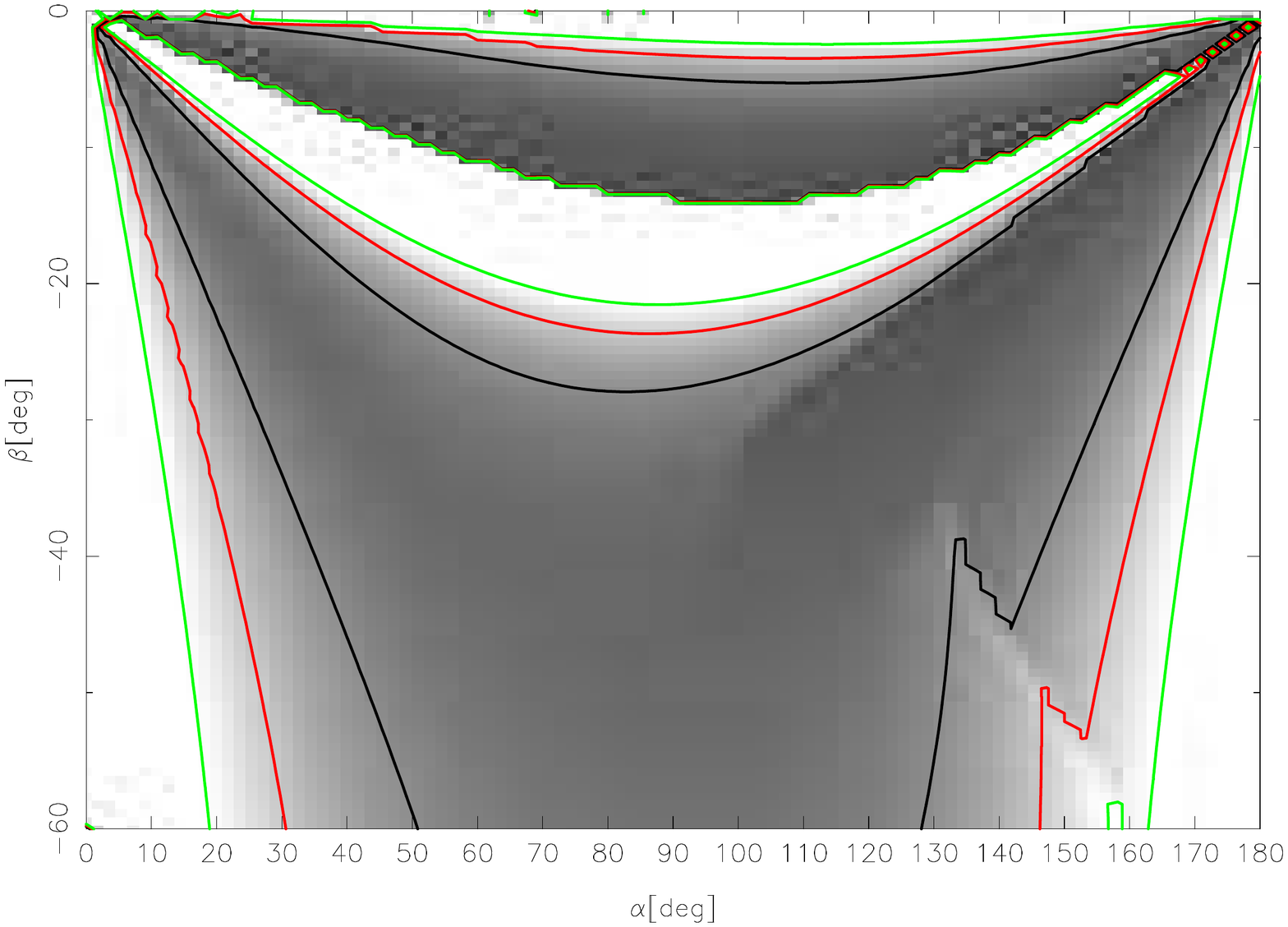}}}\\
{\mbox{\includegraphics[width=9cm,height=6cm,angle=0.]{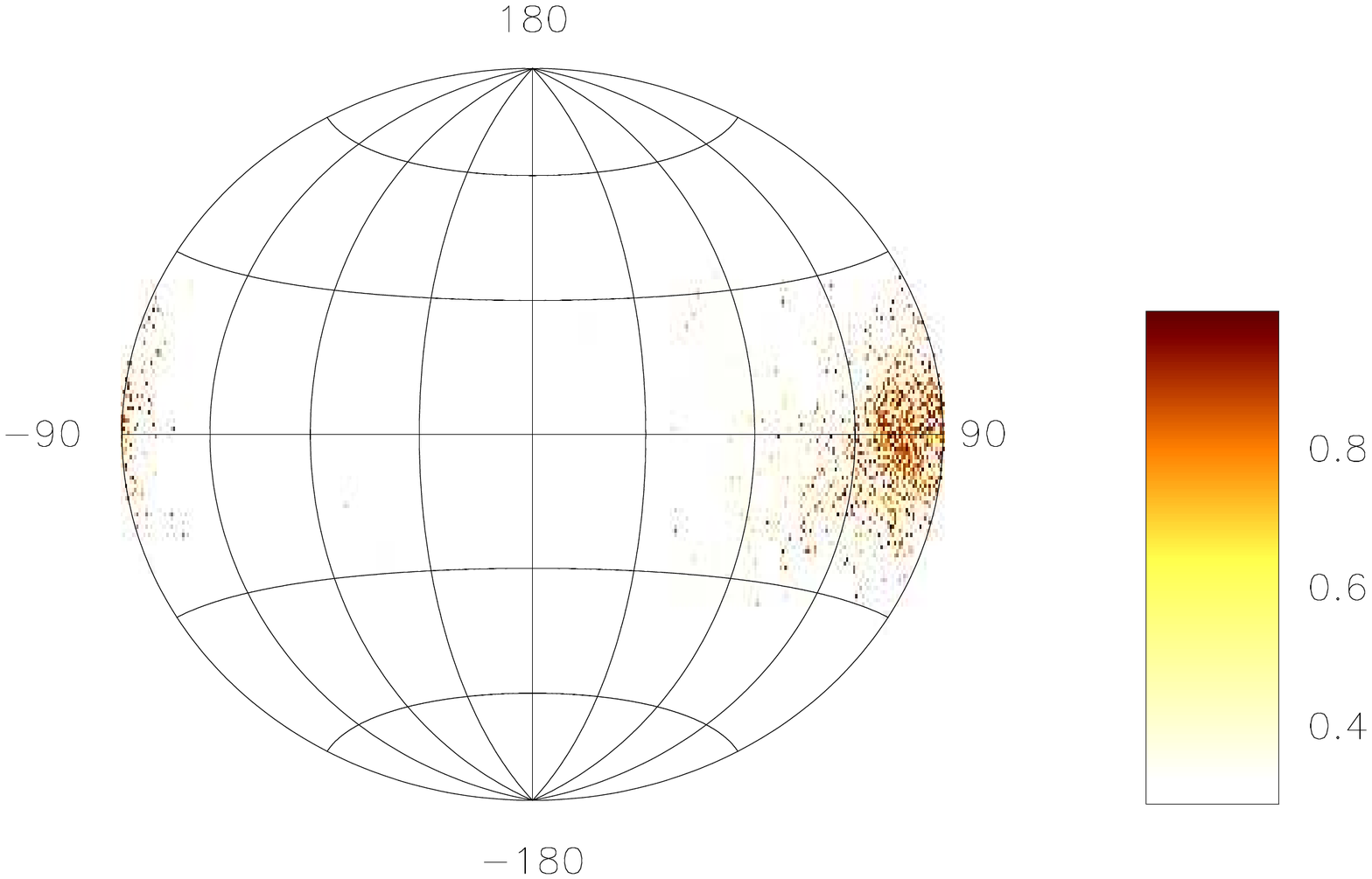}}}&
{\mbox{\includegraphics[width=9cm,height=6cm,angle=0.]{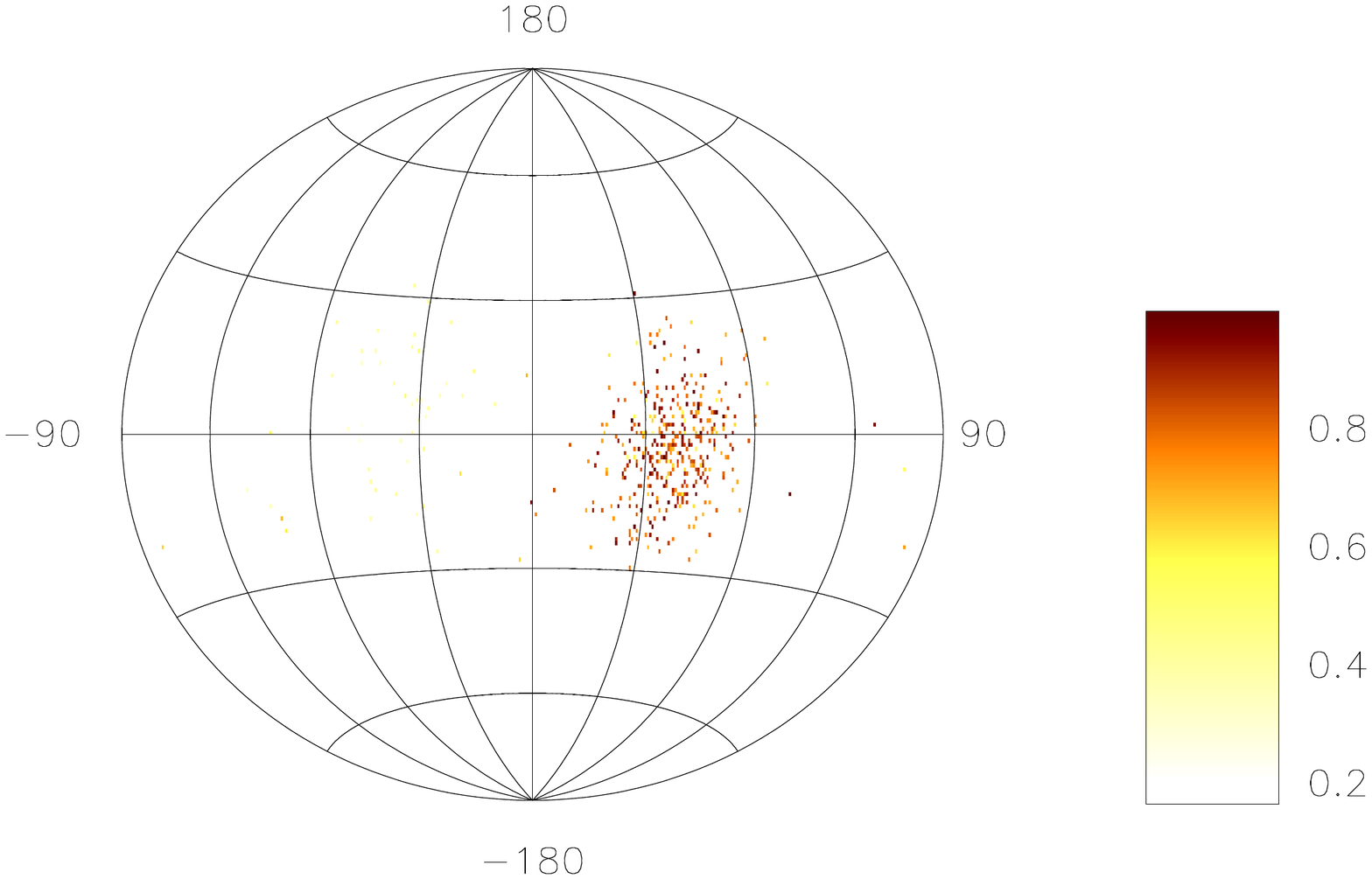}}}\\
\end{tabular}
\caption{Top panel (upper window) shows the average profile with total
intensity (Stokes I; solid black lines), total linear polarization (dashed red
line) and circular polarization (Stokes V; dotted blue line). Top panel (lower
window) also shows the single pulse PPA distribution (colour scale) along with
the average PPA (red error bars).
The RVM fits to the average PPA (dashed pink
line) is also shown in this plot. Middle panel show
the $\chi^2$ contours for the parameters $\alpha$ and $\beta$ obtained from RVM
fits.
Bottom panel shows the Hammer-Aitoff projection of the polarized time
samples with the colour scheme representing the fractional polarization level.}
\label{a12}
\end{center}
\end{figure*}


\begin{figure*}
\begin{center}
\begin{tabular}{cc}
{\mbox{\includegraphics[width=9cm,height=6cm,angle=0.]{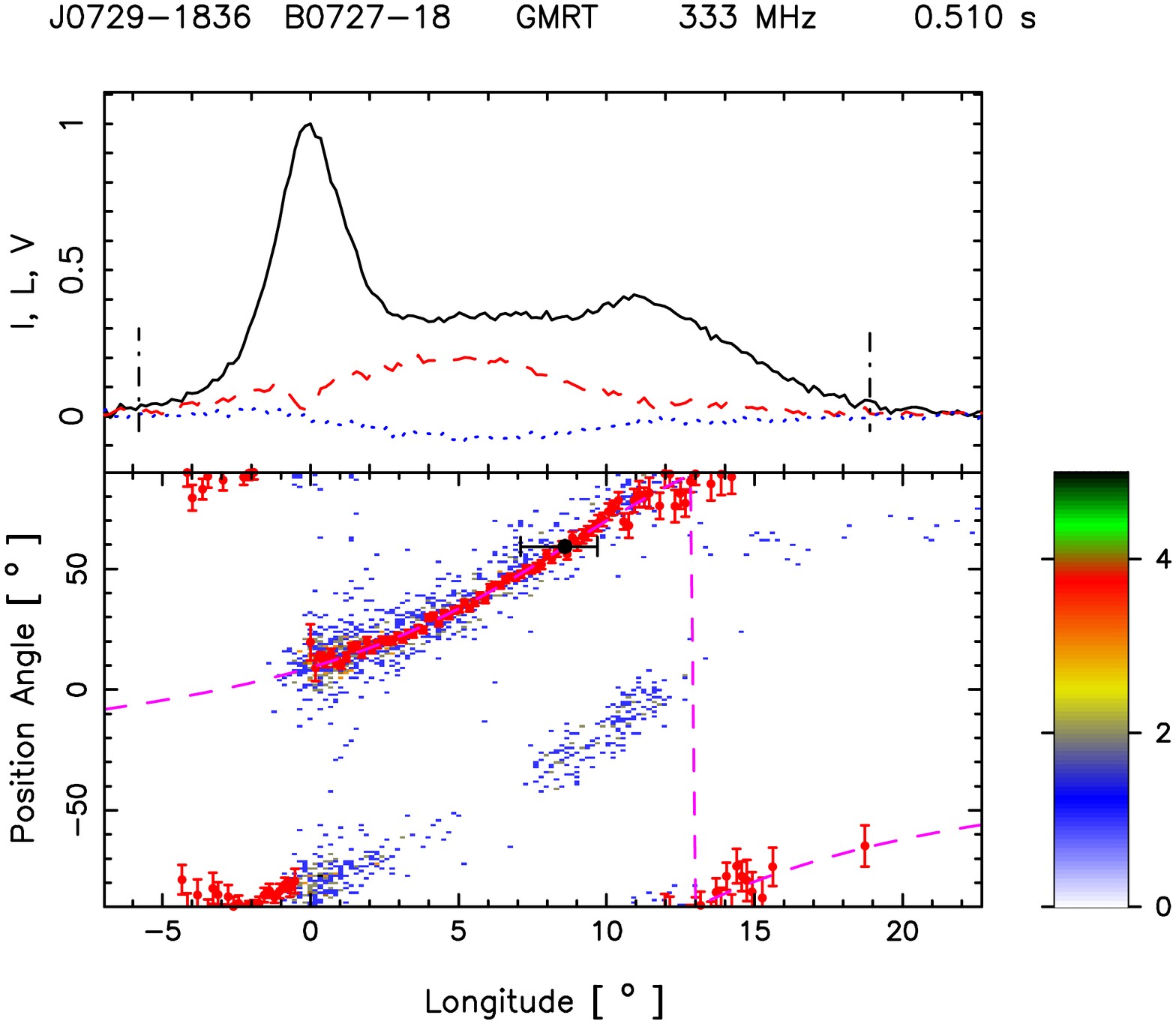}}}&
{\mbox{\includegraphics[width=9cm,height=6cm,angle=0.]{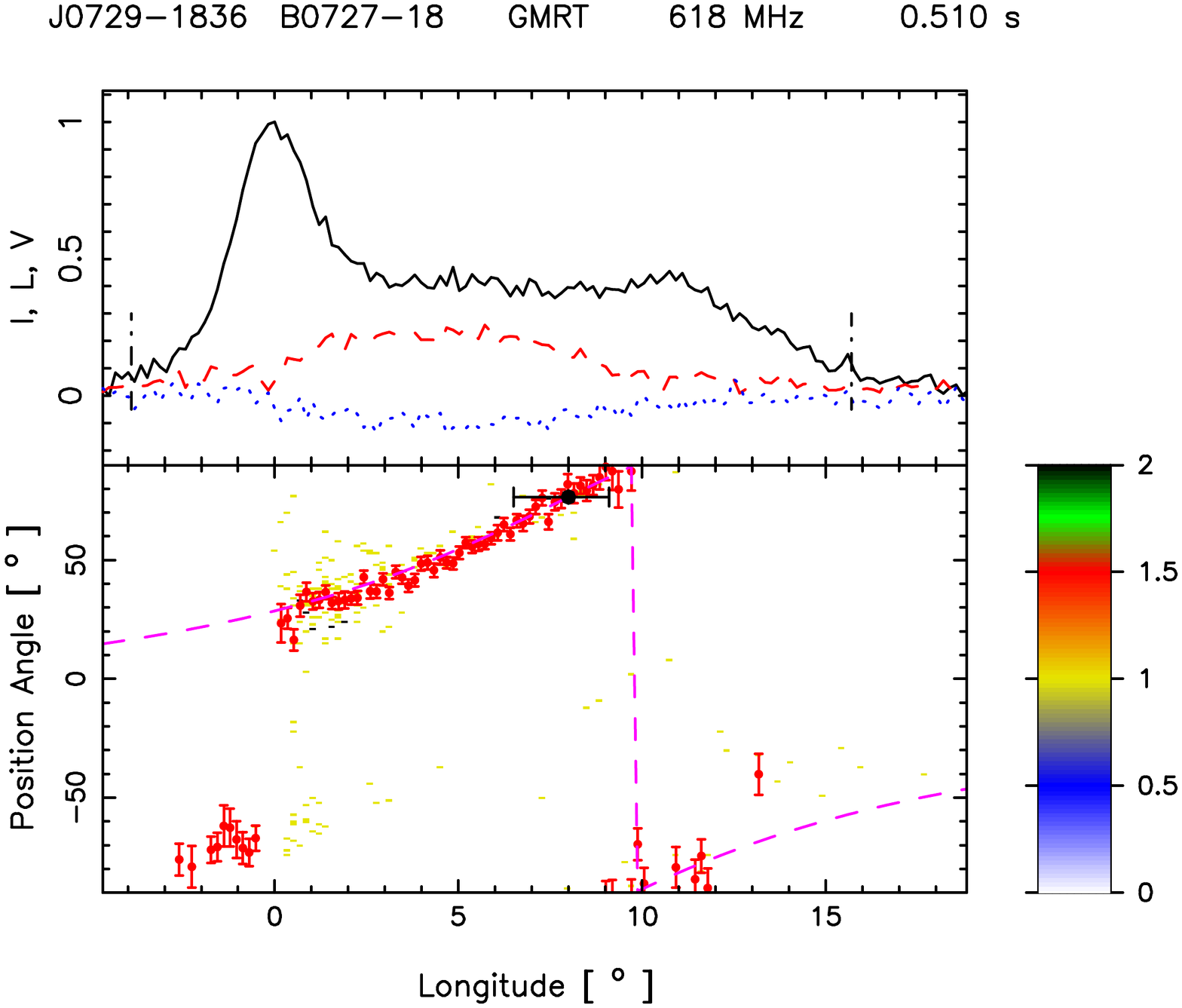}}}\\
{\mbox{\includegraphics[width=9cm,height=6cm,angle=0.]{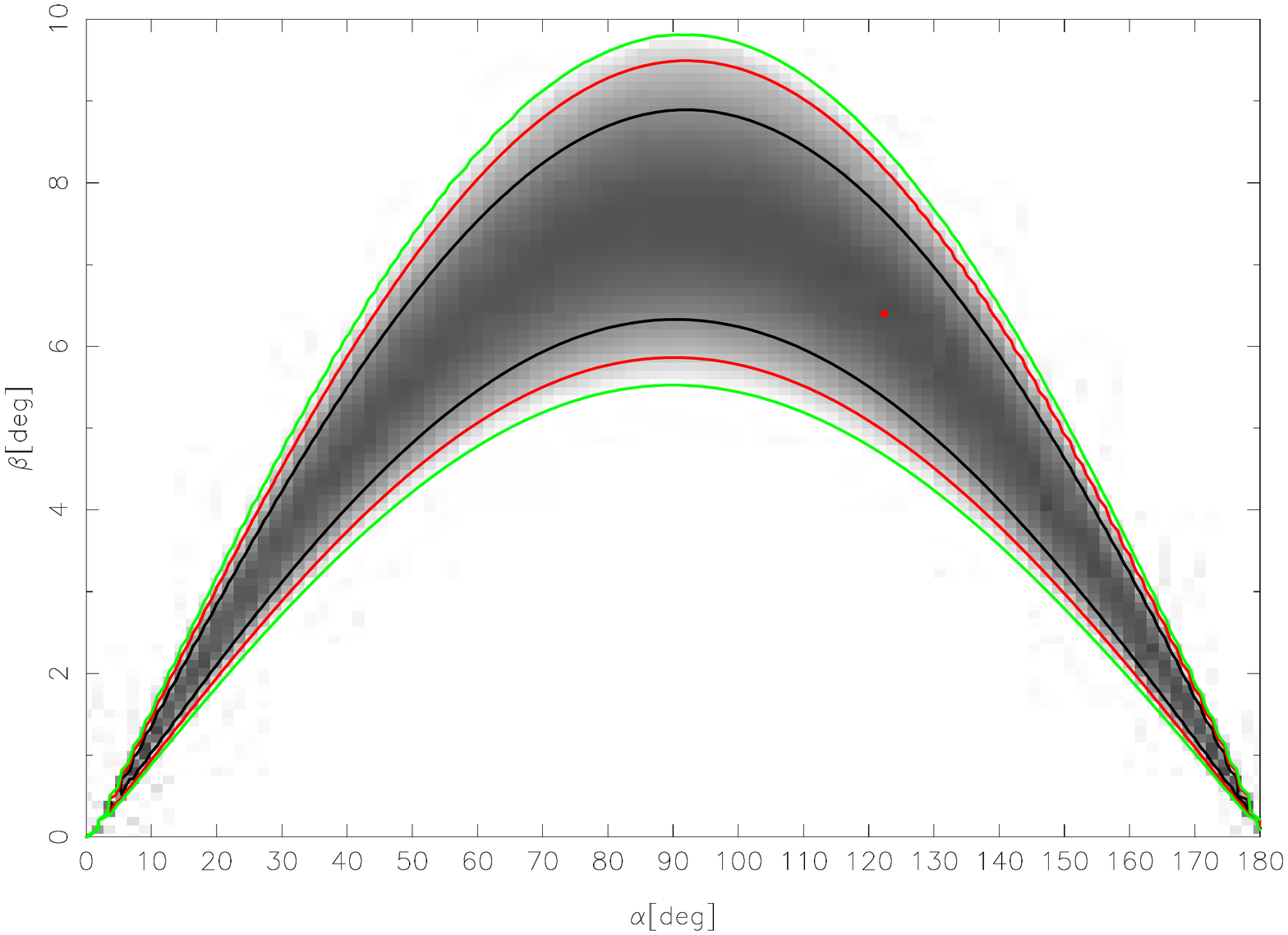}}}&
{\mbox{\includegraphics[width=9cm,height=6cm,angle=0.]{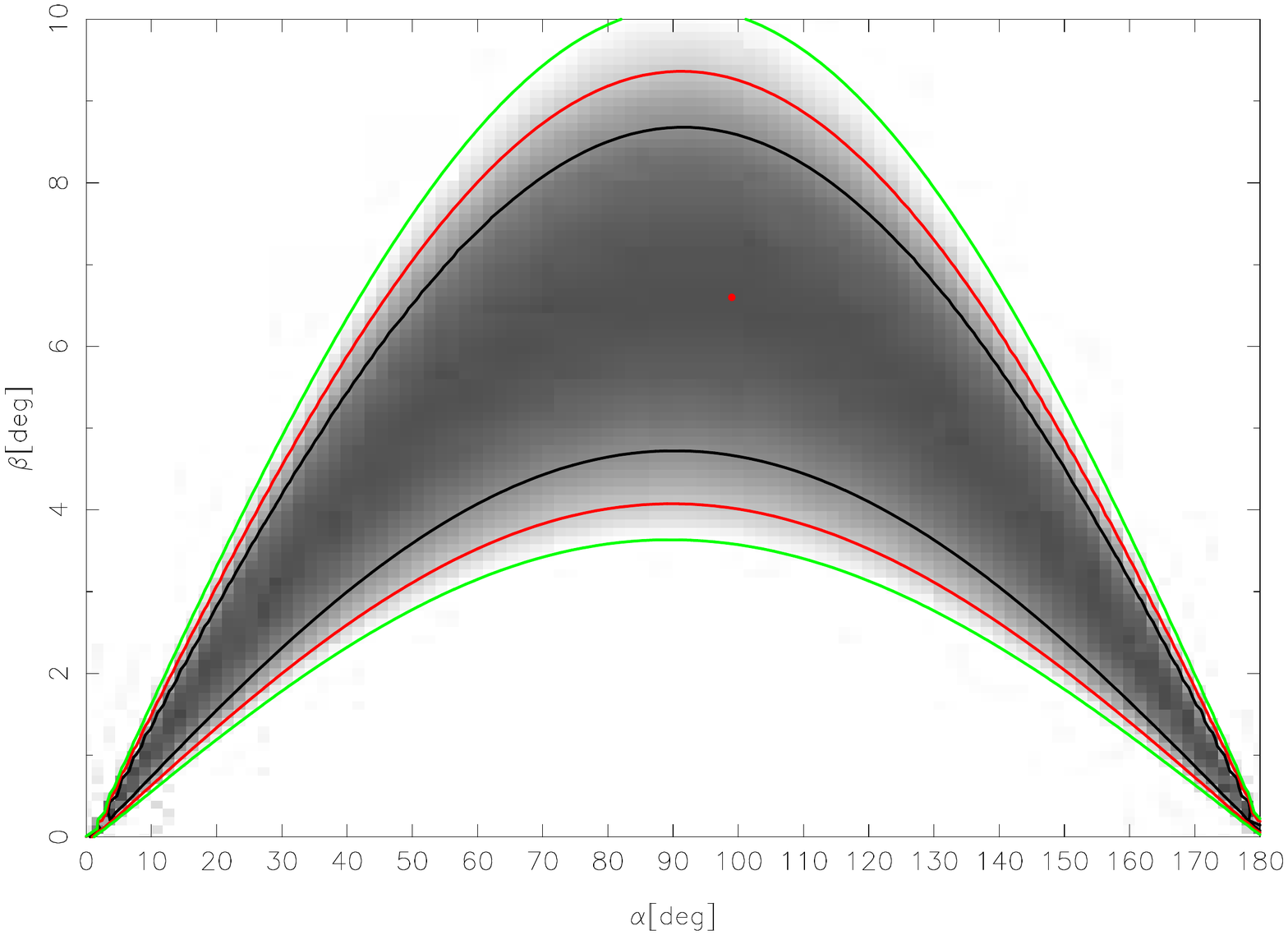}}}\\
{\mbox{\includegraphics[width=9cm,height=6cm,angle=0.]{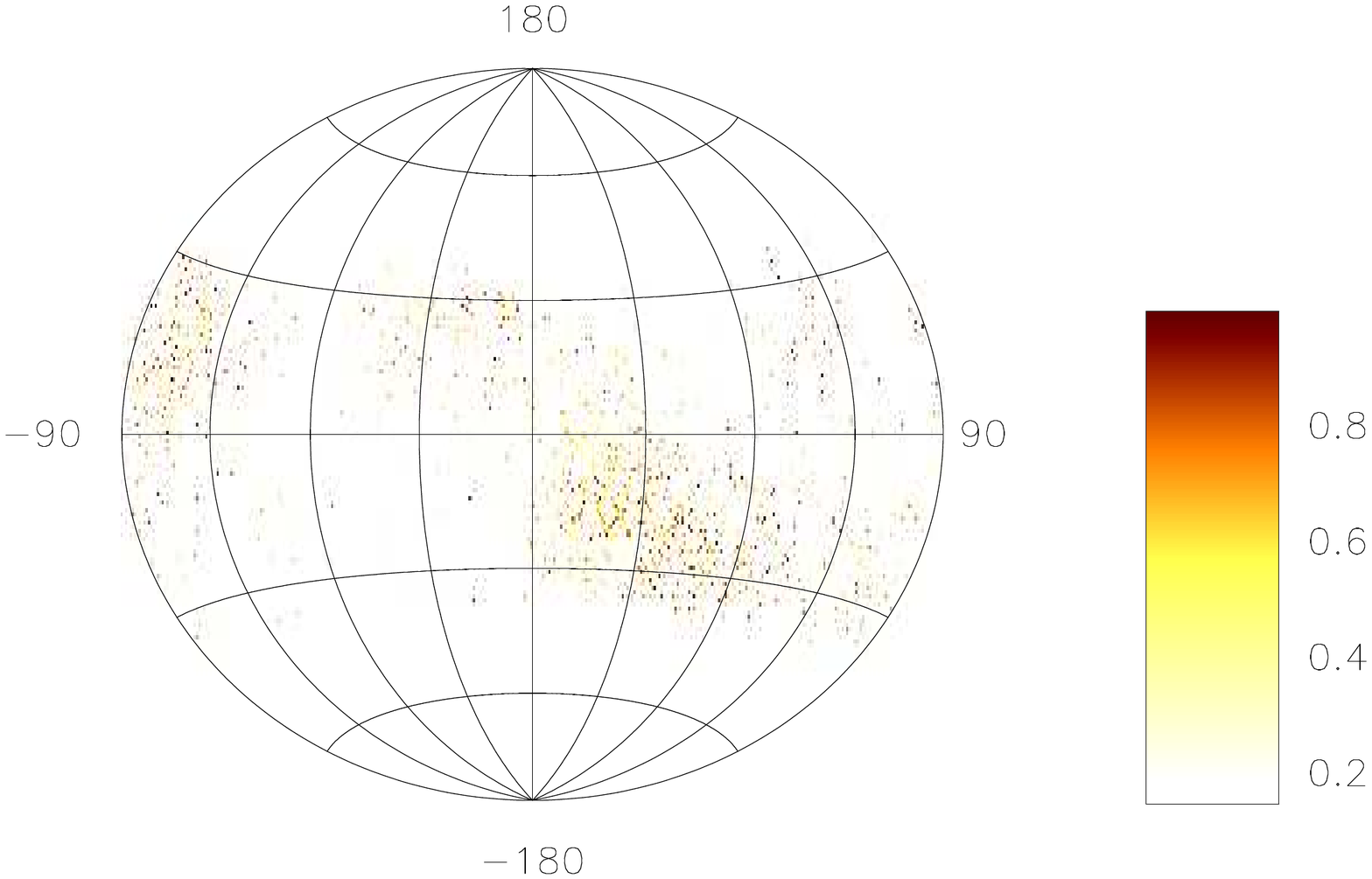}}}&
{\mbox{\includegraphics[width=9cm,height=6cm,angle=0.]{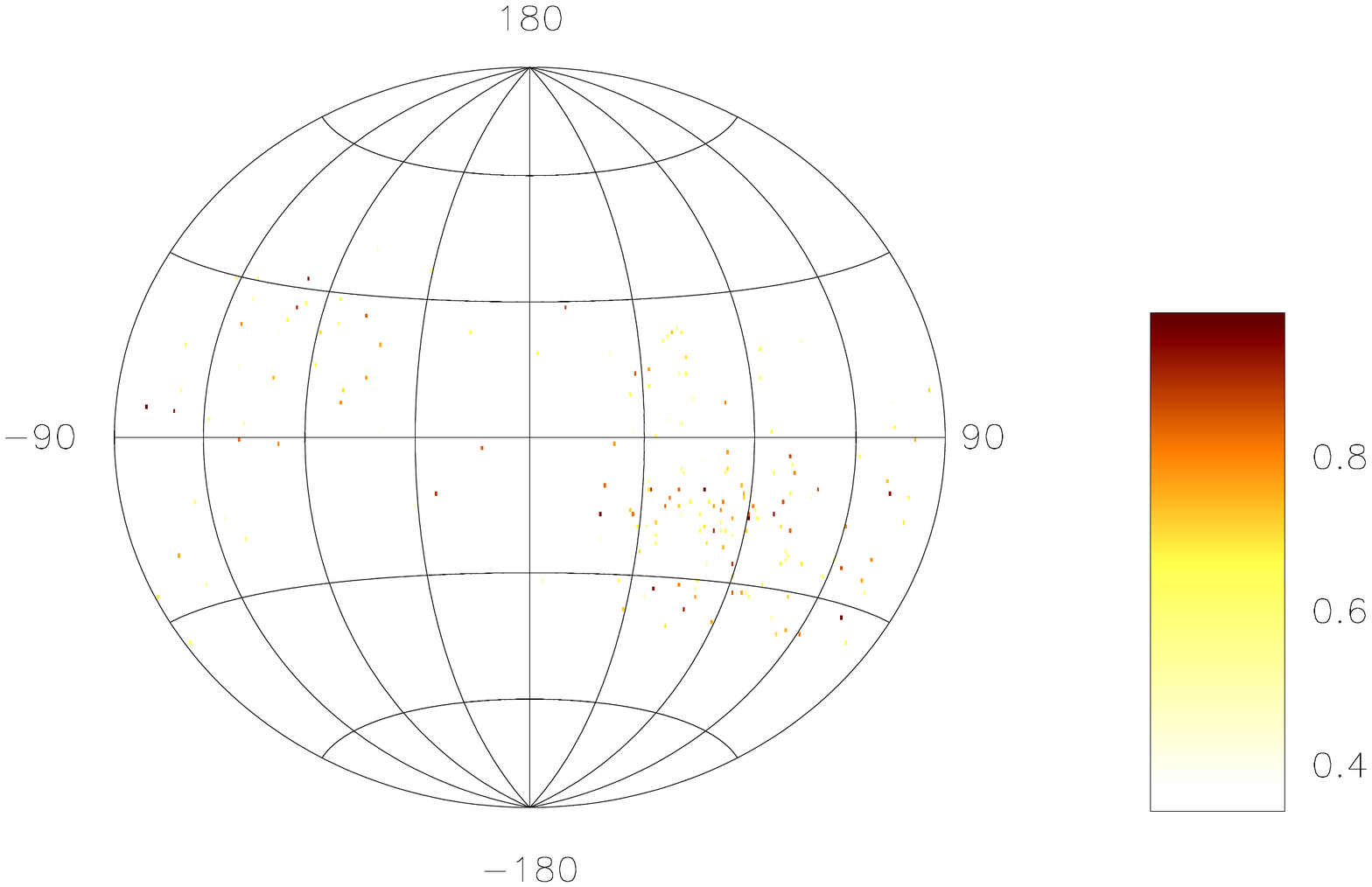}}}\\
\end{tabular}
\caption{Top panel (upper window) shows the average profile with total
intensity (Stokes I; solid black lines), total linear polarization (dashed red
line) and circular polarization (Stokes V; dotted blue line). Top panel (lower
window) also shows the single pulse PPA distribution (colour scale) along with
the average PPA (red error bars).
The RVM fits to the average PPA (dashed pink
line) is also shown in this plot. Middle panel show
the $\chi^2$ contours for the parameters $\alpha$ and $\beta$ obtained from RVM
fits.
Bottom panel shows the Hammer-Aitoff projection of the polarized time
samples with the colour scheme representing the fractional polarization level.}
\label{a13}
\end{center}
\end{figure*}

\begin{figure*}
\begin{center}
\begin{tabular}{cc}
&
{\mbox{\includegraphics[width=9cm,height=6cm,angle=0.]{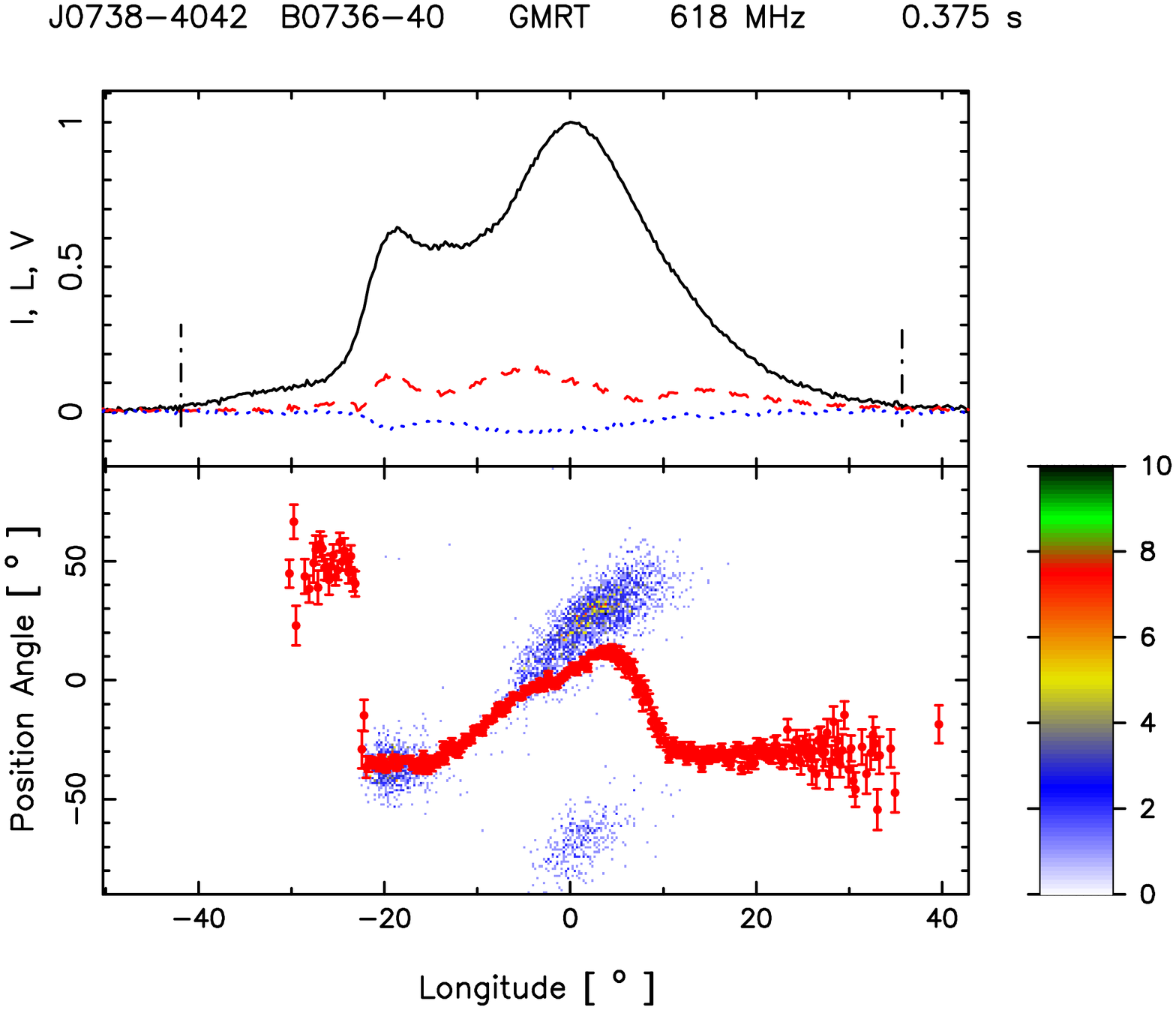}}}\\
&
\\
&
{\mbox{\includegraphics[width=9cm,height=6cm,angle=0.]{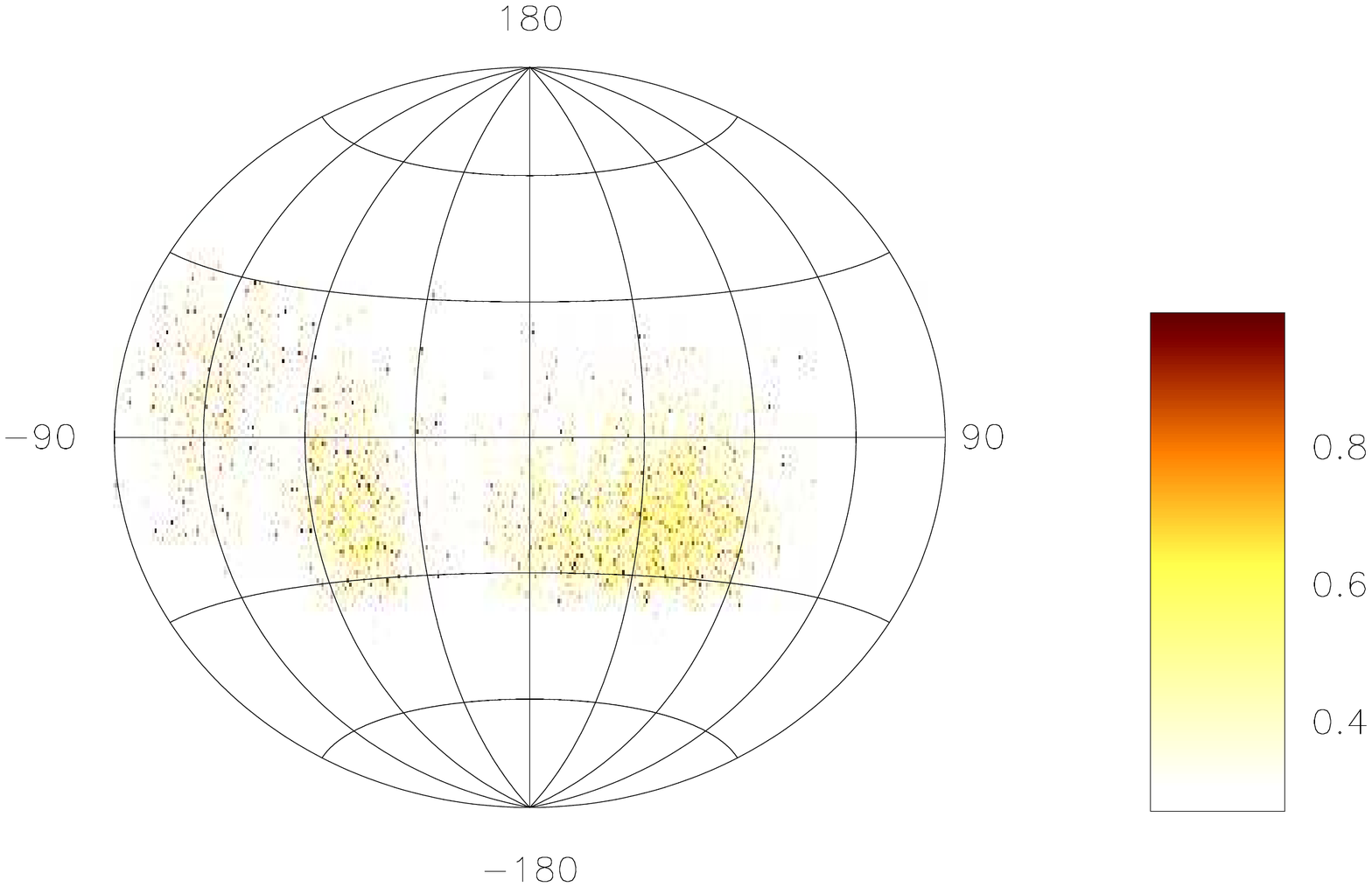}}}\\
\end{tabular}
\caption{Top panel only for 618 MHz (upper window) shows the average profile with total
intensity (Stokes I; solid black lines), total linear polarization (dashed red
line) and circular polarization (Stokes V; dotted blue line). Top panel (lower
window) also shows the single pulse PPA distribution (colour scale) along with
the average PPA (red error bars).
Bottom panel only for 618 MHz shows the Hammer-Aitoff projection of the polarized time
samples with the colour scheme representing the fractional polarization level.}
\label{a14}
\end{center}
\end{figure*}

\begin{figure*}
\begin{center}
\begin{tabular}{cc}
{\mbox{\includegraphics[width=9cm,height=6cm,angle=0.]{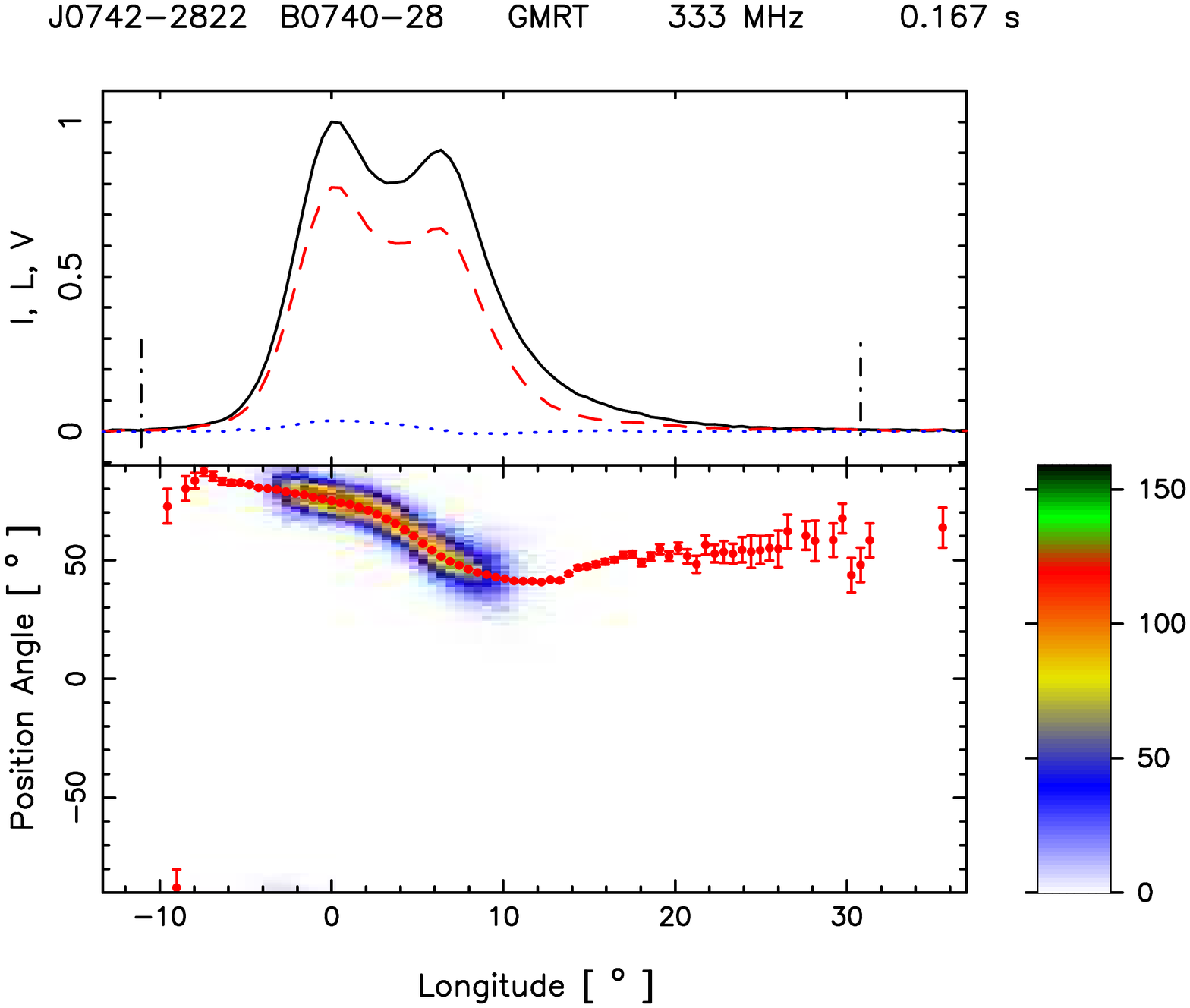}}}&
{\mbox{\includegraphics[width=9cm,height=6cm,angle=0.]{J0742-2822_602MHz_23Mar14.dat.epn2a.pdf}}}\\
&
{\mbox{\includegraphics[width=9cm,height=6cm,angle=0.]{J0742-2822_602MHz_23Mar14.dat.orderf.dat_chisq.pdf}}}\\
{\mbox{\includegraphics[width=9cm,height=6cm,angle=0.]{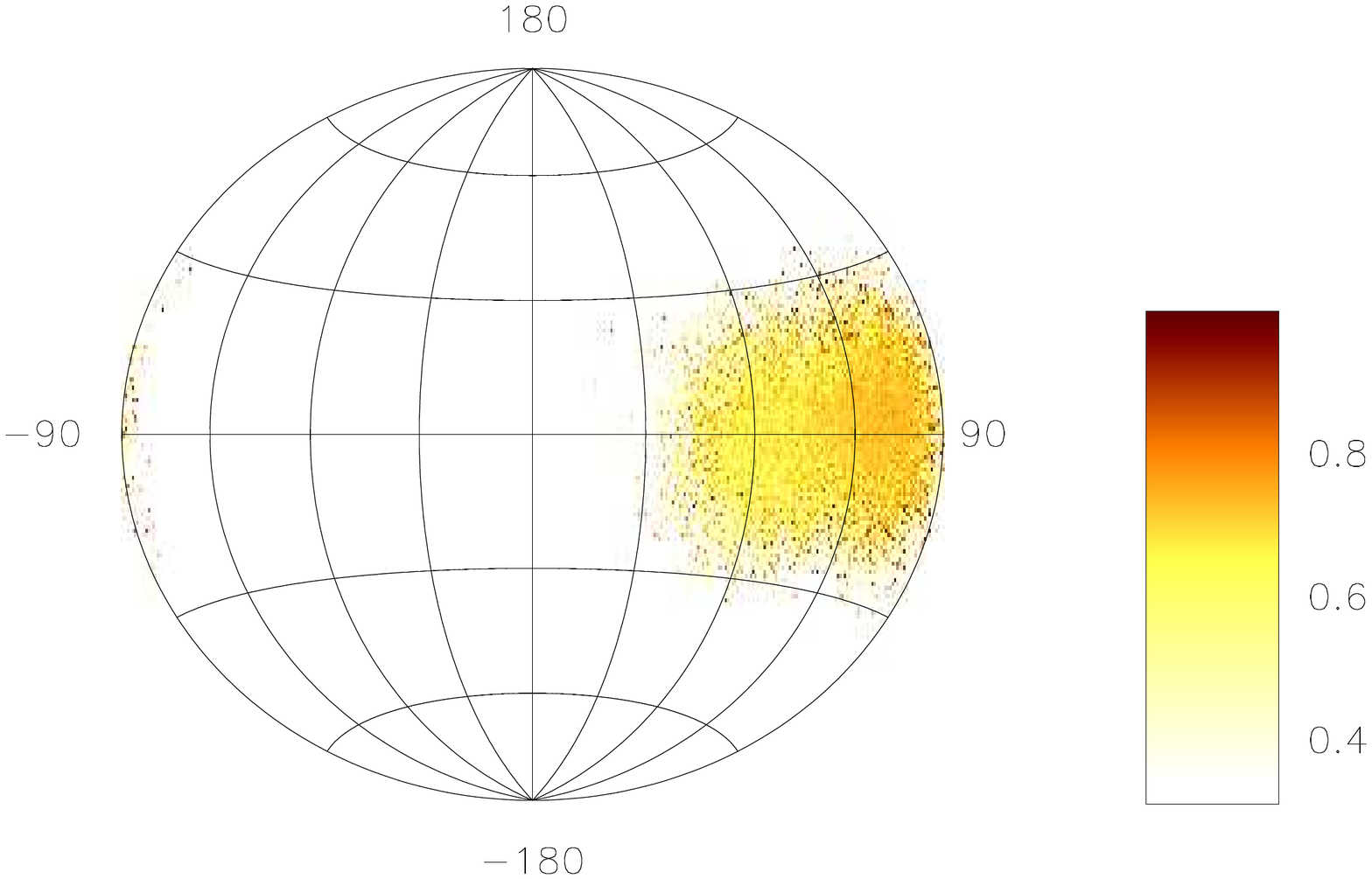}}}&
{\mbox{\includegraphics[width=9cm,height=6cm,angle=0.]{J0742-2822_602MHz_23Mar14.dat.epn2a.71.pdf}}}\\
\end{tabular}
\caption{Top panel (upper window) shows the average profile with total
intensity (Stokes I; solid black lines), total linear polarization (dashed red
line) and circular polarization (Stokes V; dotted blue line). Top panel (lower
window) also shows the single pulse PPA distribution (colour scale) along with
the average PPA (red error bars).
The RVM fits to the average PPA (dashed pink
line) is also shown in this plot. Middle panel only for 618 MHz show
the $\chi^2$ contours for the parameters $\alpha$ and $\beta$ obtained from RVM
fits.
Bottom panel shows the Hammer-Aitoff projection of the polarized time
samples with the colour scheme representing the fractional polarization level.}
\label{a15}
\end{center}
\end{figure*}


\begin{figure*}
\begin{center}
\begin{tabular}{cc}
&
{\mbox{\includegraphics[width=9cm,height=6cm,angle=0.]{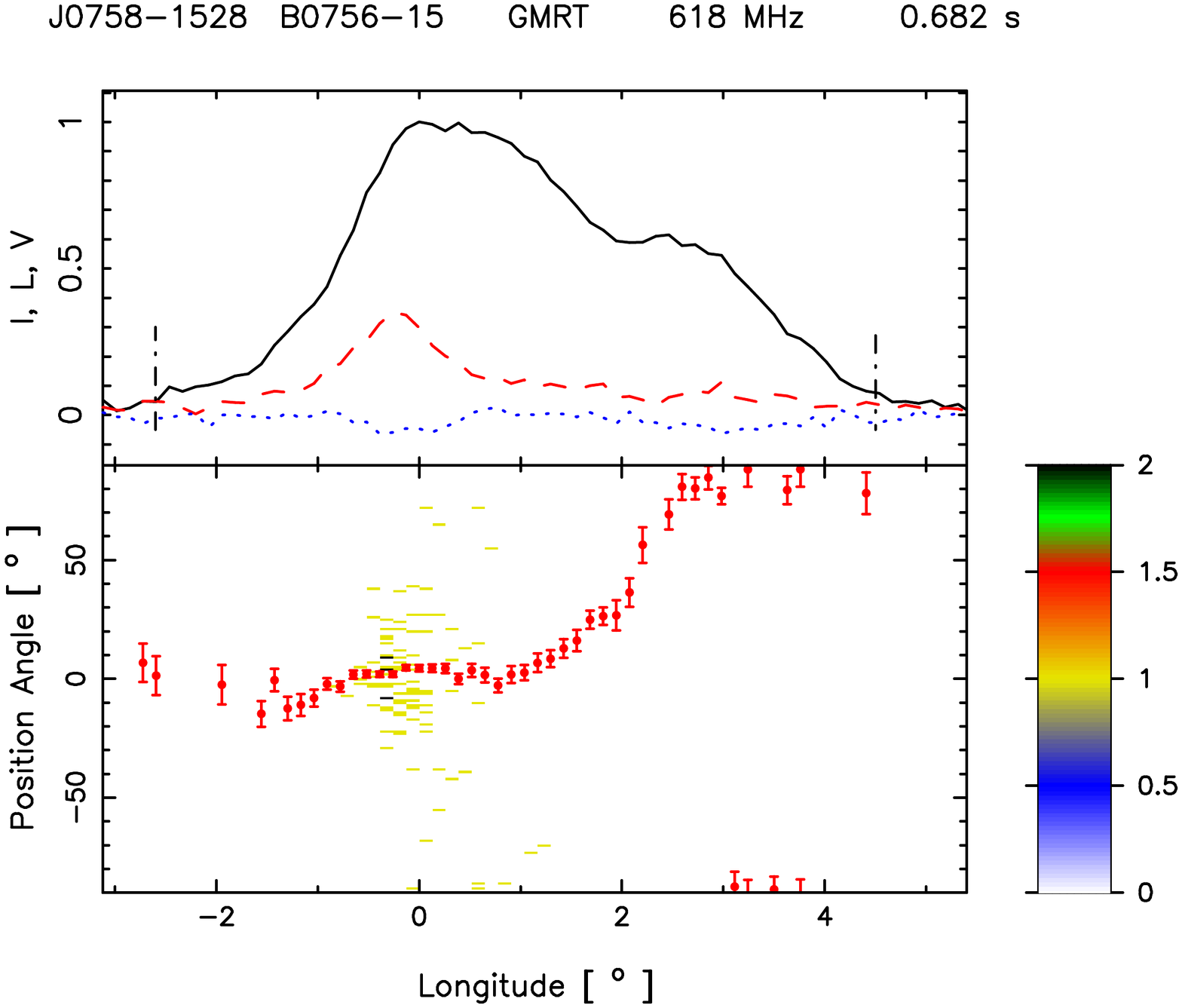}}}\\
&
\\
&
{\mbox{\includegraphics[width=9cm,height=6cm,angle=0.]{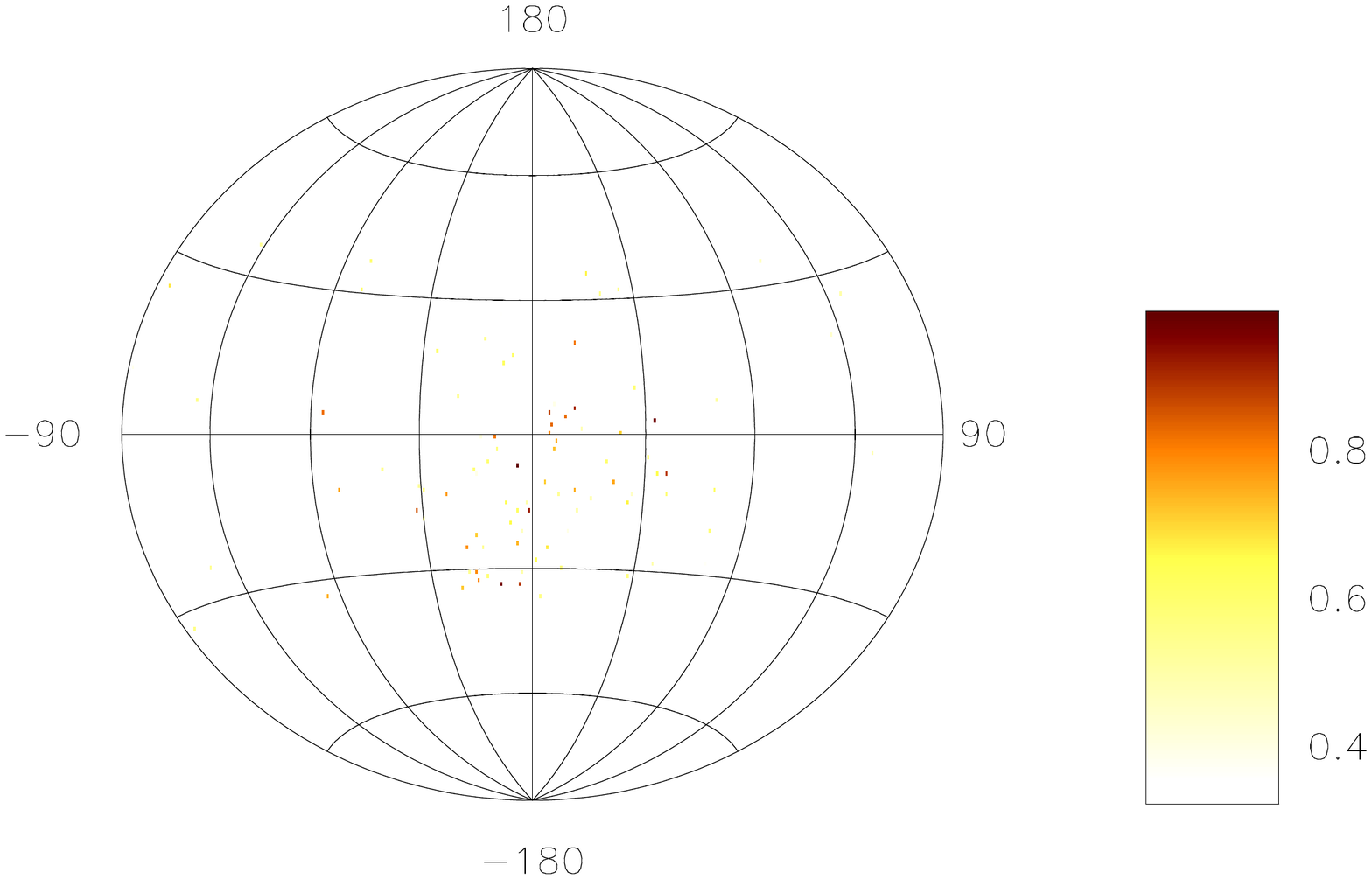}}}\\
\end{tabular}
\caption{Top panel only for 618 MHz (upper window) shows the average profile with total
intensity (Stokes I; solid black lines), total linear polarization (dashed red
line) and circular polarization (Stokes V; dotted blue line). Top panel (lower
window) also shows the single pulse PPA distribution (colour scale) along with
the average PPA (red error bars).
Bottom panel shows the Hammer-Aitoff projection of the polarized time
samples with the colour scheme representing the fractional polarization level.}
\label{a16}
\end{center}
\end{figure*}


\begin{figure*}
\begin{center}
\begin{tabular}{cc}
&
{\mbox{\includegraphics[width=9cm,height=6cm,angle=0.]{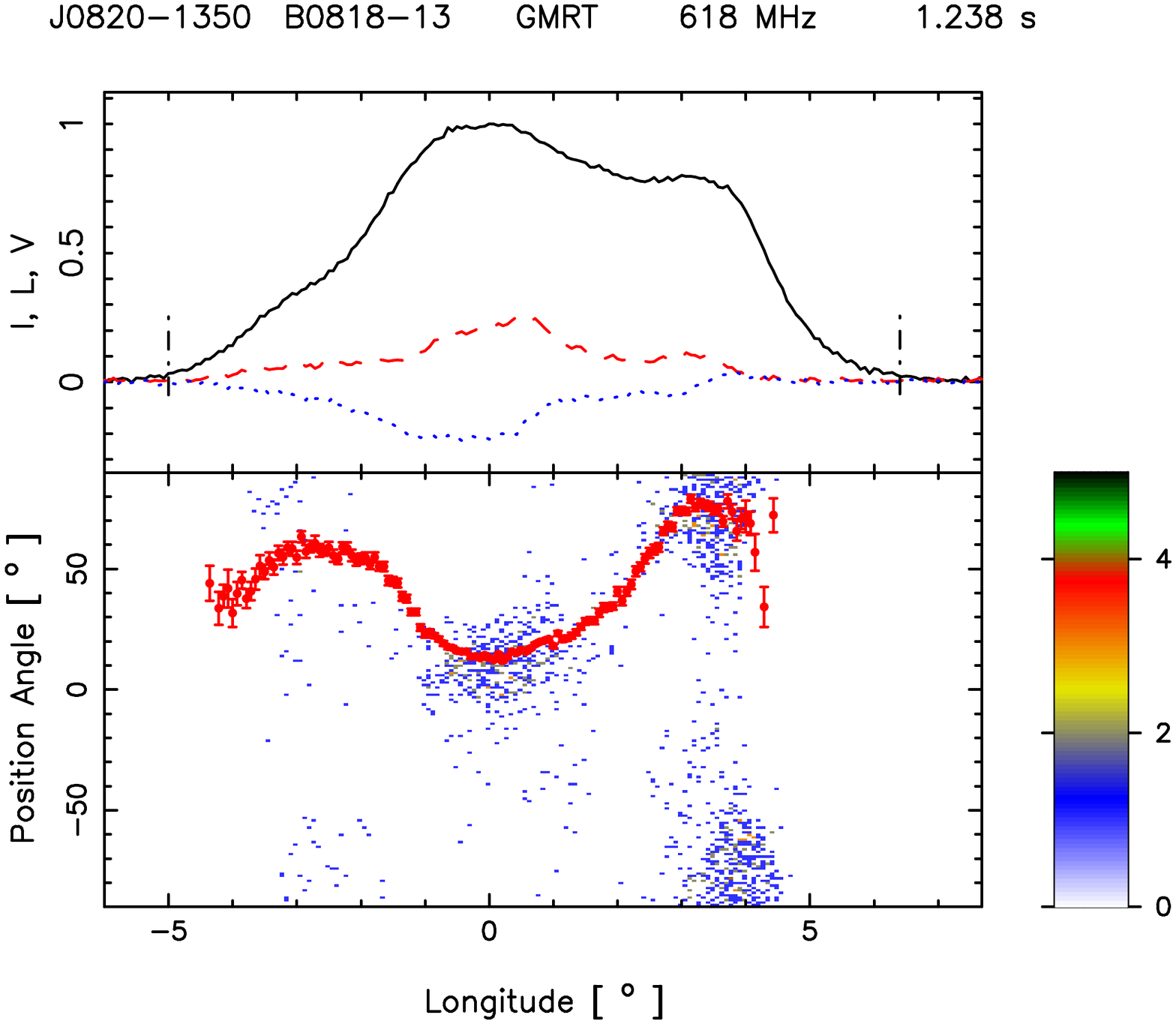}}}\\
&
\\
&
{\mbox{\includegraphics[width=9cm,height=6cm,angle=0.]{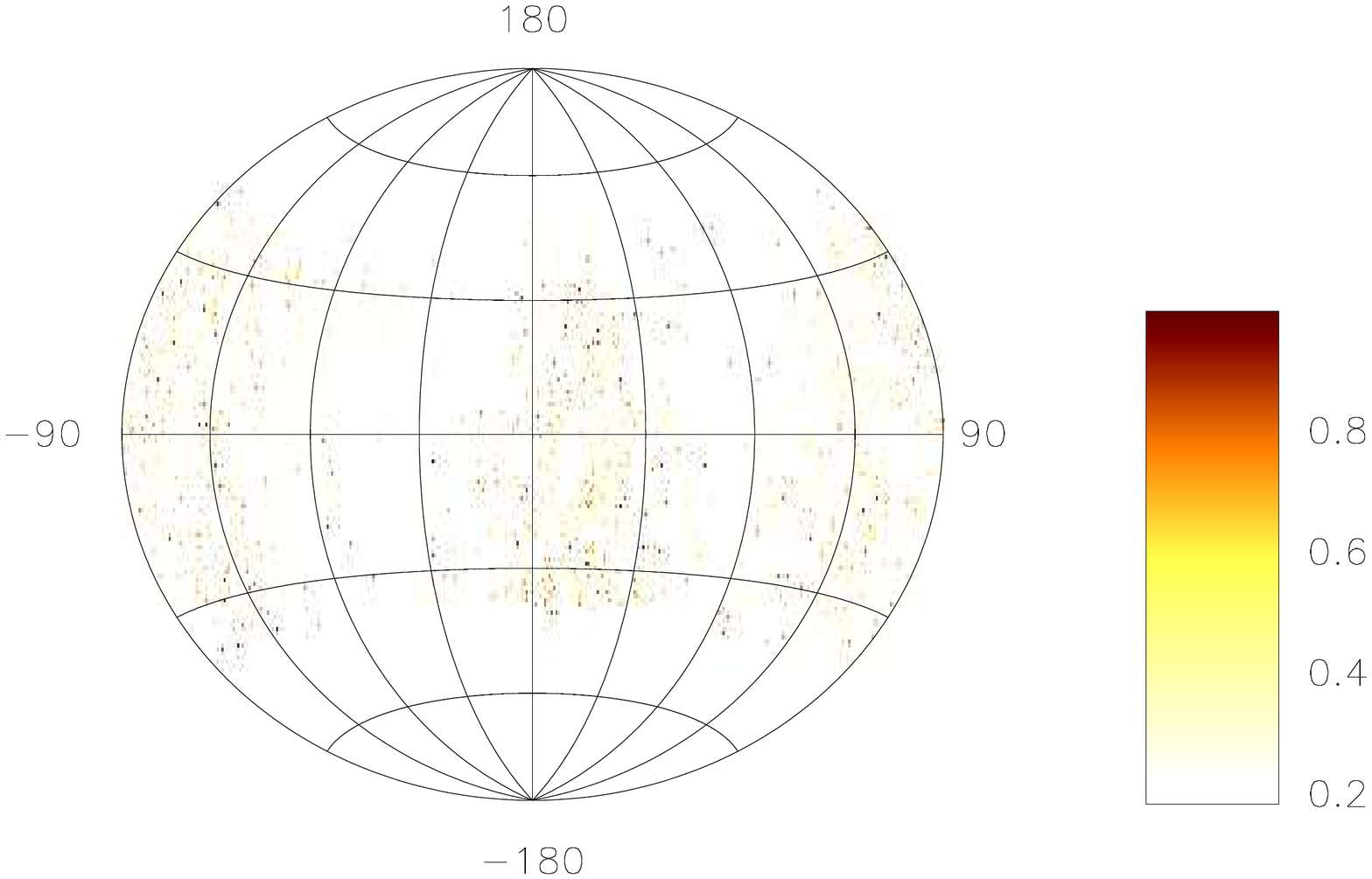}}}\\
\end{tabular}
\caption{Top panel only for 618 MHz (upper window) shows the average profile with total
intensity (Stokes I; solid black lines), total linear polarization (dashed red
line) and circular polarization (Stokes V; dotted blue line). Top panel (lower
window) also shows the single pulse PPA distribution (colour scale) along with
the average PPA (red error bars).
Bottom panel shows the Hammer-Aitoff projection of the polarized time
samples with the colour scheme representing the fractional polarization level.}
\label{a17}
\end{center}
\end{figure*}


\begin{figure*}
\begin{center}
\begin{tabular}{cc}
{\mbox{\includegraphics[width=9cm,height=6cm,angle=0.]{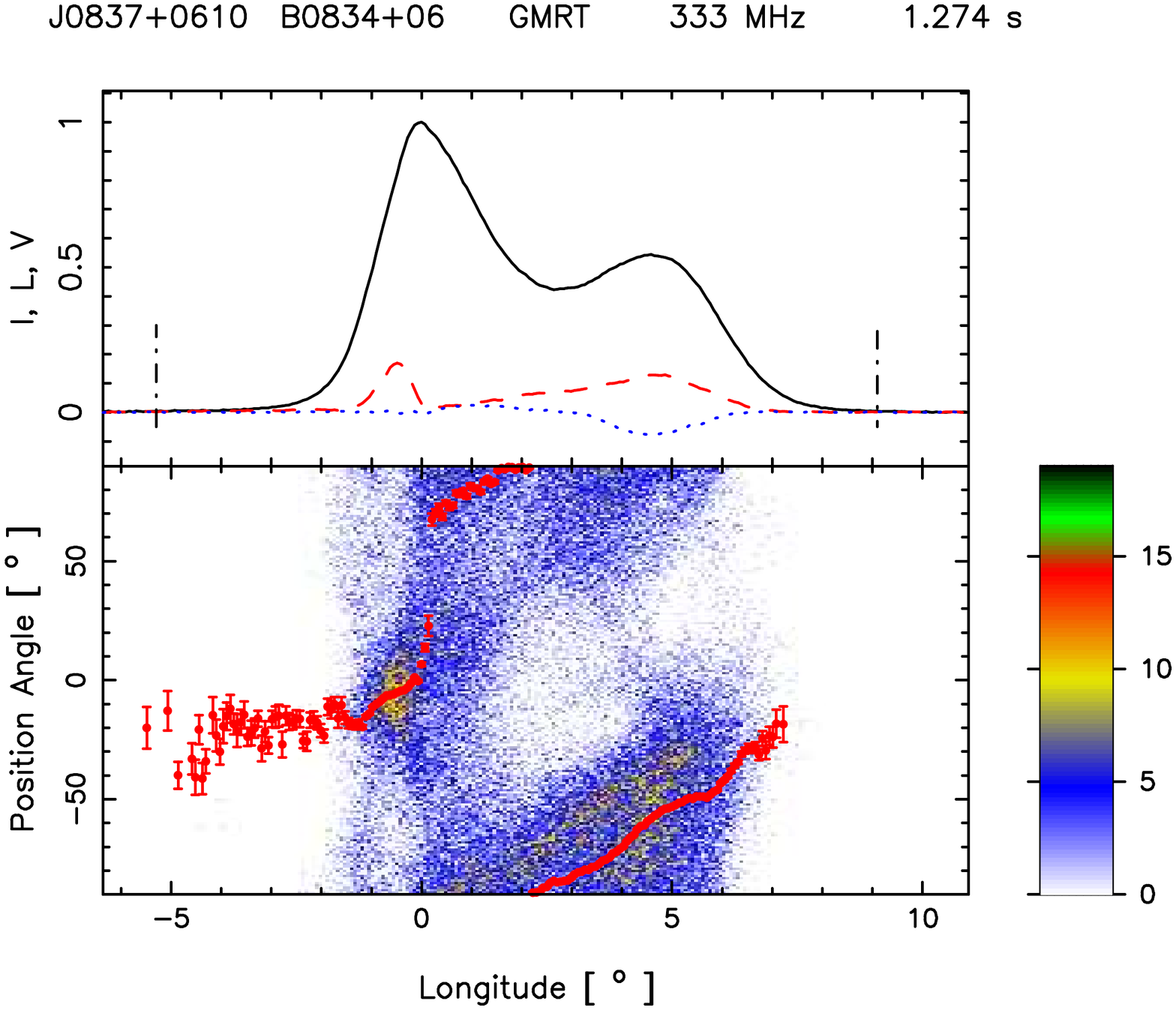}}}&
{\mbox{\includegraphics[width=9cm,height=6cm,angle=0.]{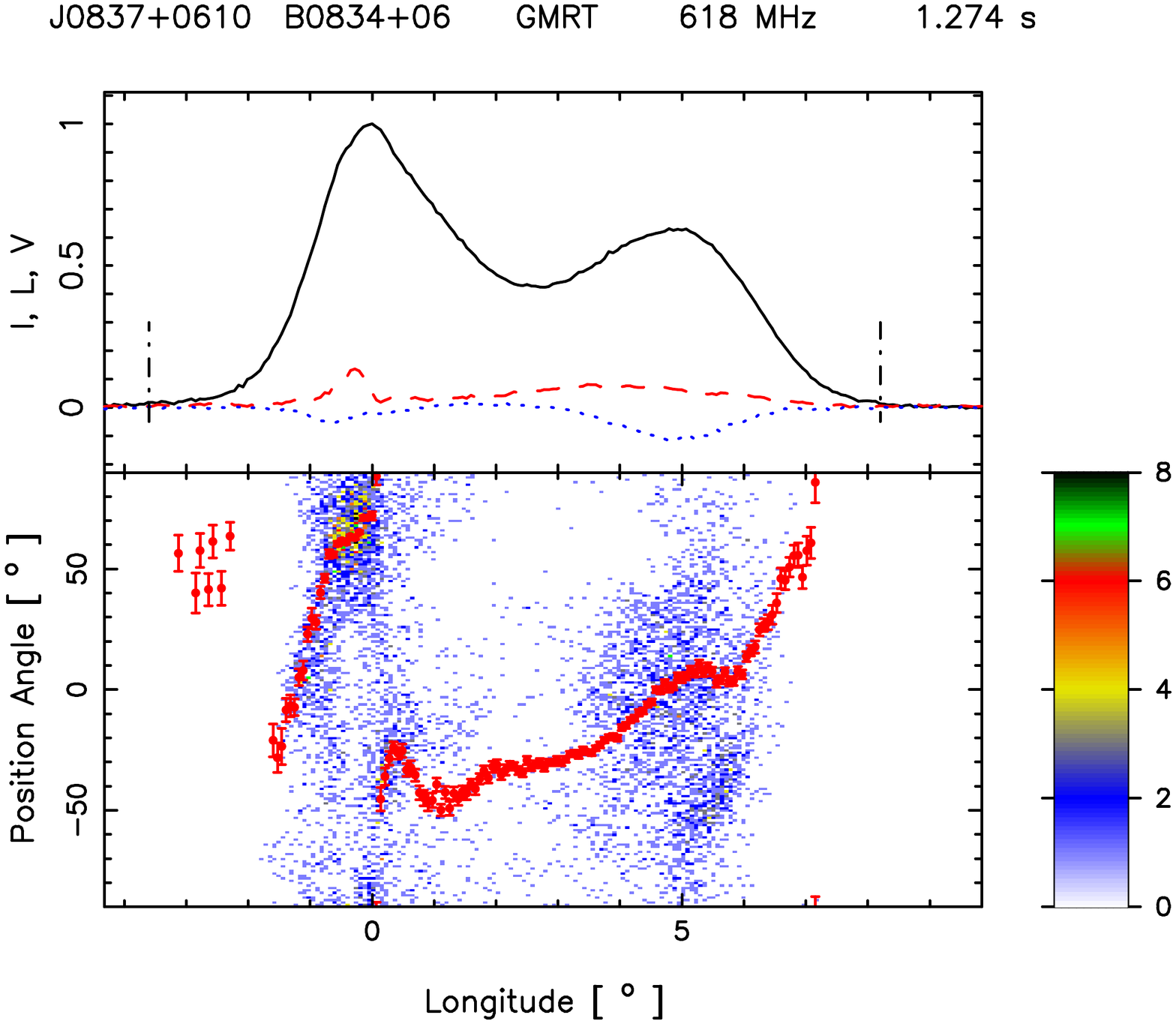}}}\\
&
\\
{\mbox{\includegraphics[width=9cm,height=6cm,angle=0.]{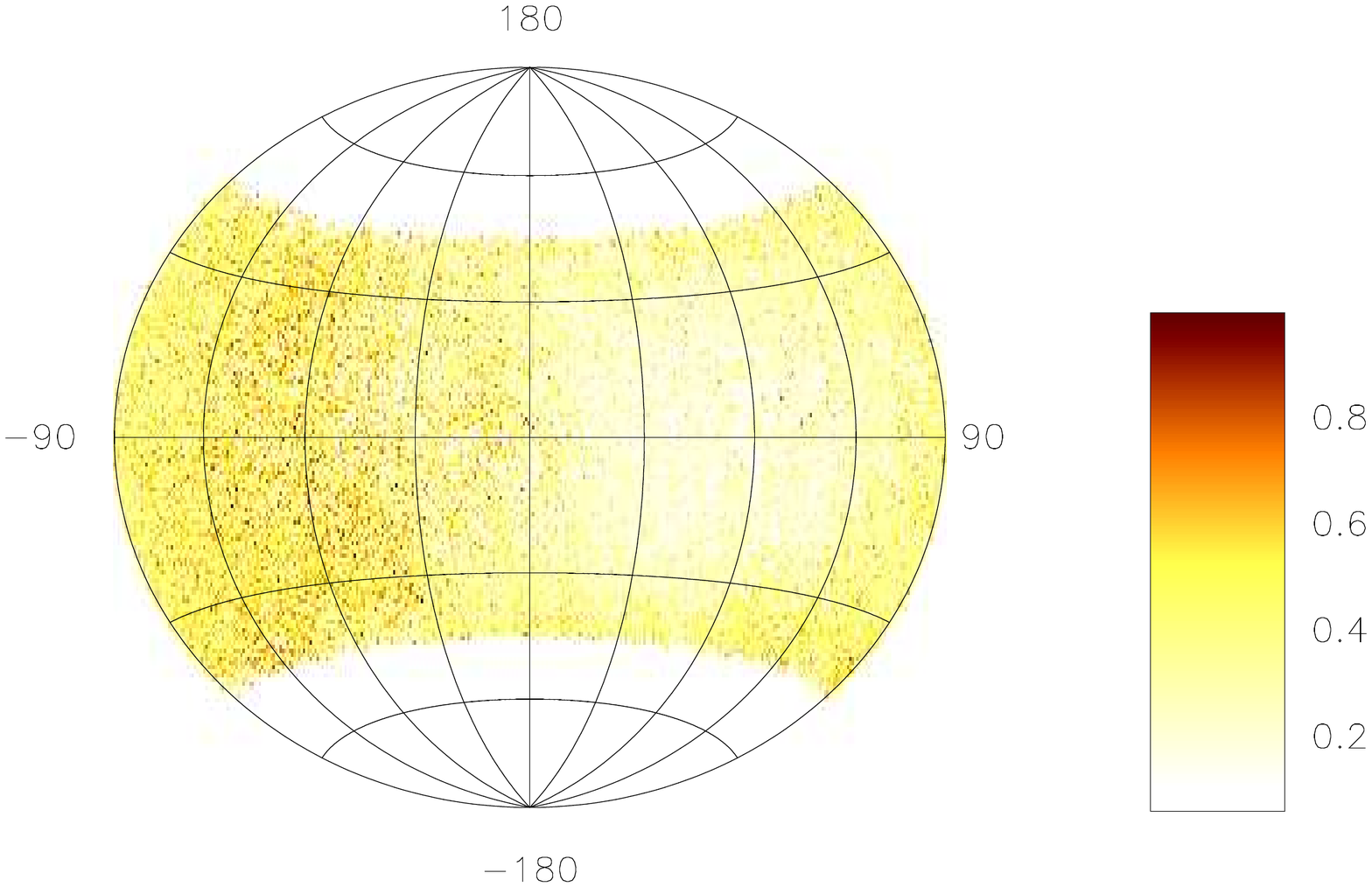}}}&
{\mbox{\includegraphics[width=9cm,height=6cm,angle=0.]{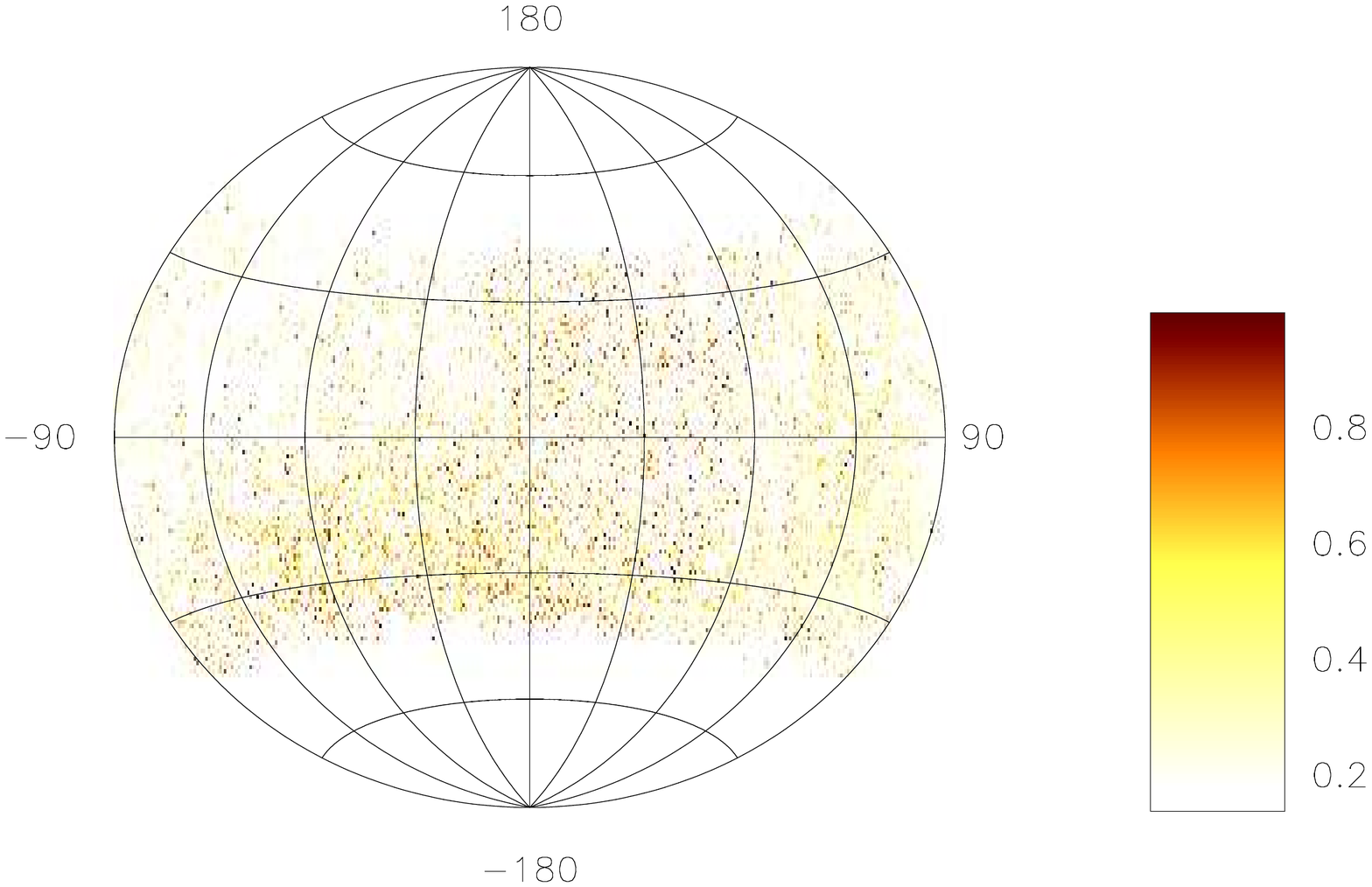}}}\\
\end{tabular}
\caption{Top panel (upper window) shows the average profile with total
intensity (Stokes I; solid black lines), total linear polarization (dashed red
line) and circular polarization (Stokes V; dotted blue line). Top panel (lower
window) also shows the single pulse PPA distribution (colour scale) along with
the average PPA (red error bars).
Bottom panel shows the Hammer-Aitoff projection of the polarized time
samples with the colour scheme representing the fractional polarization level.}
\label{a18}
\end{center}
\end{figure*}

\begin{figure*}
\begin{center}
\begin{tabular}{cc}
{\mbox{\includegraphics[width=9cm,height=6cm,angle=0.]{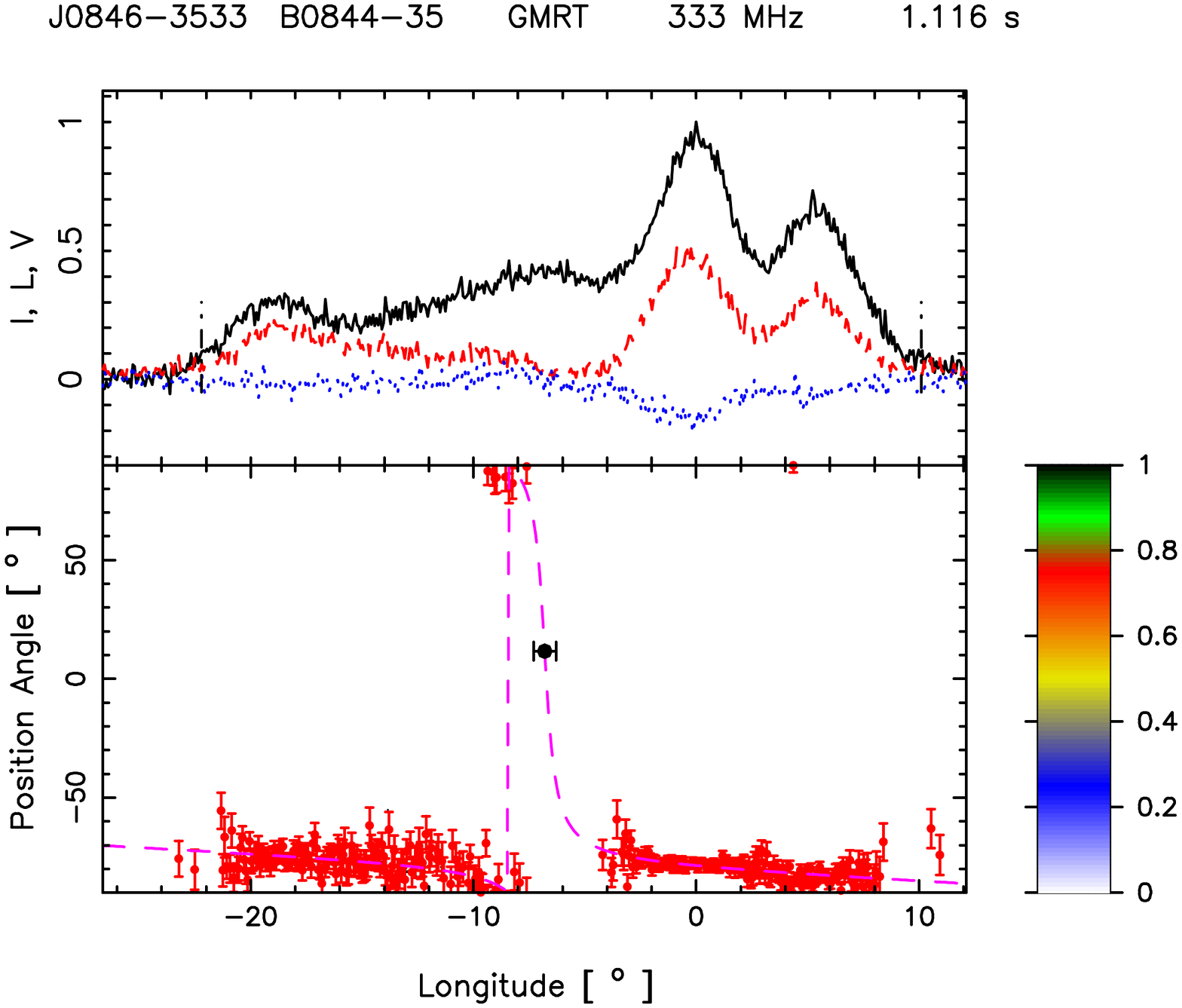}}}&
{\mbox{\includegraphics[width=9cm,height=6cm,angle=0.]{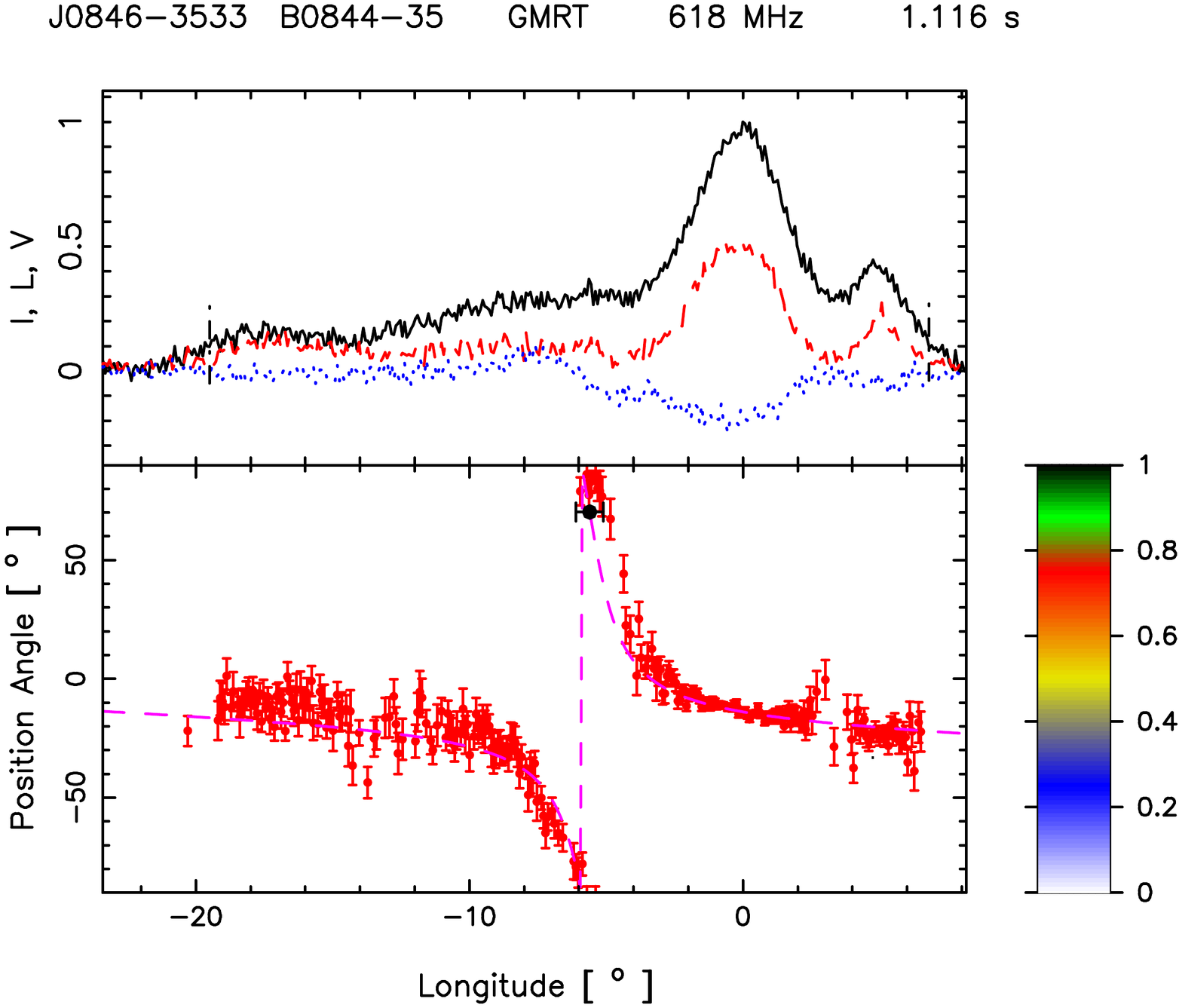}}}\\
{\mbox{\includegraphics[width=9cm,height=6cm,angle=0.]{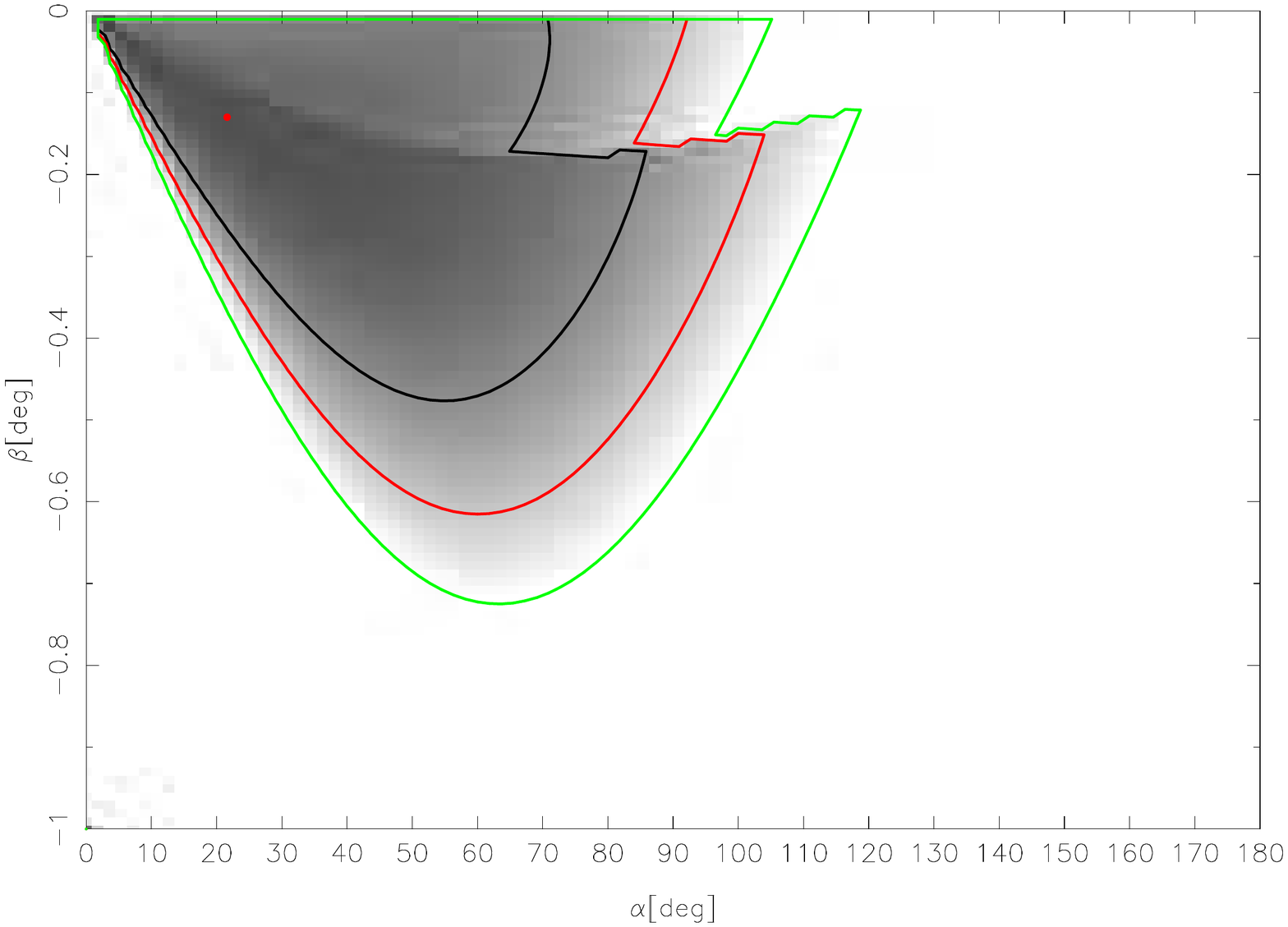}}}&
{\mbox{\includegraphics[width=9cm,height=6cm,angle=0.]{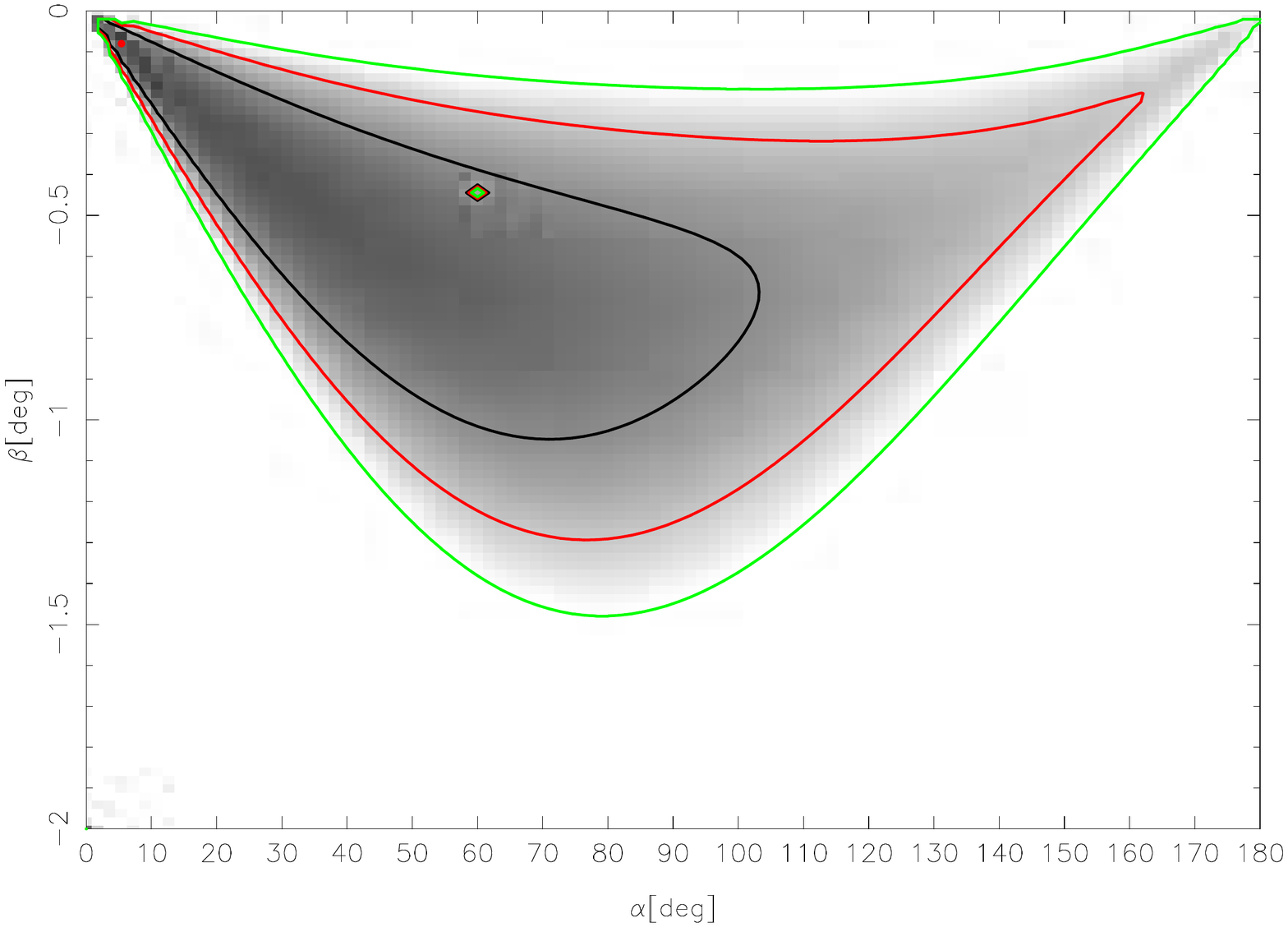}}}\\
&
\\
\end{tabular}
\caption{Top panel (upper window) shows the average profile with total
intensity (Stokes I; solid black lines), total linear polarization (dashed red
line) and circular polarization (Stokes V; dotted blue line). Top panel (lower
window) also shows the single pulse PPA distribution (colour scale) along with
the average PPA (red error bars).
The RVM fits to the average PPA (dashed pink
line) is also shown in this plot. Bottom panel show
the $\chi^2$ contours for the parameters $\alpha$ and $\beta$ obtained from RVM
fits.}
\label{a19}
\end{center}
\end{figure*}


\begin{figure*}
\begin{center}
\begin{tabular}{cc}
&
{\mbox{\includegraphics[width=9cm,height=6cm,angle=0.]{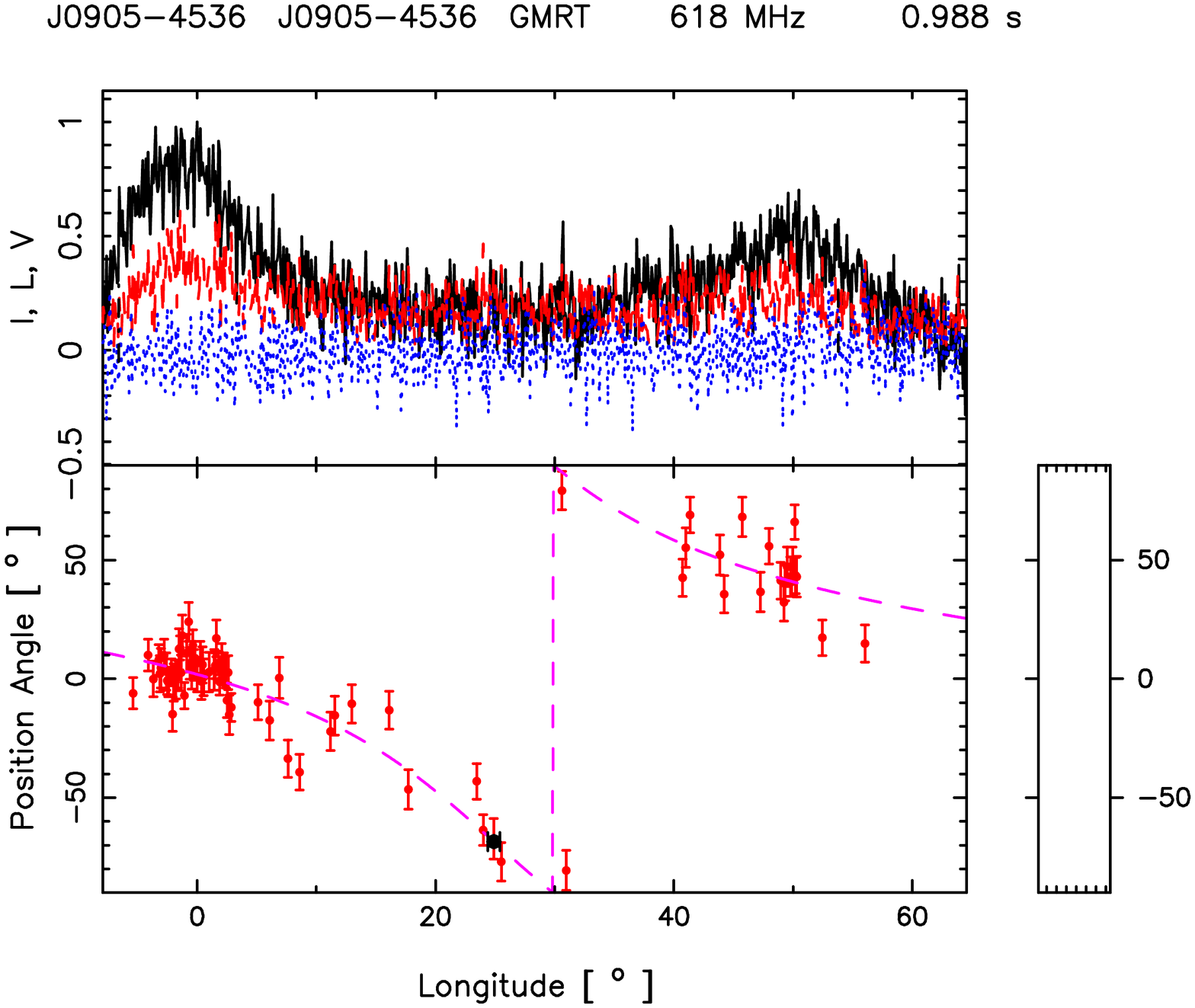}}}\\
&
{\mbox{\includegraphics[width=9cm,height=6cm,angle=0.]{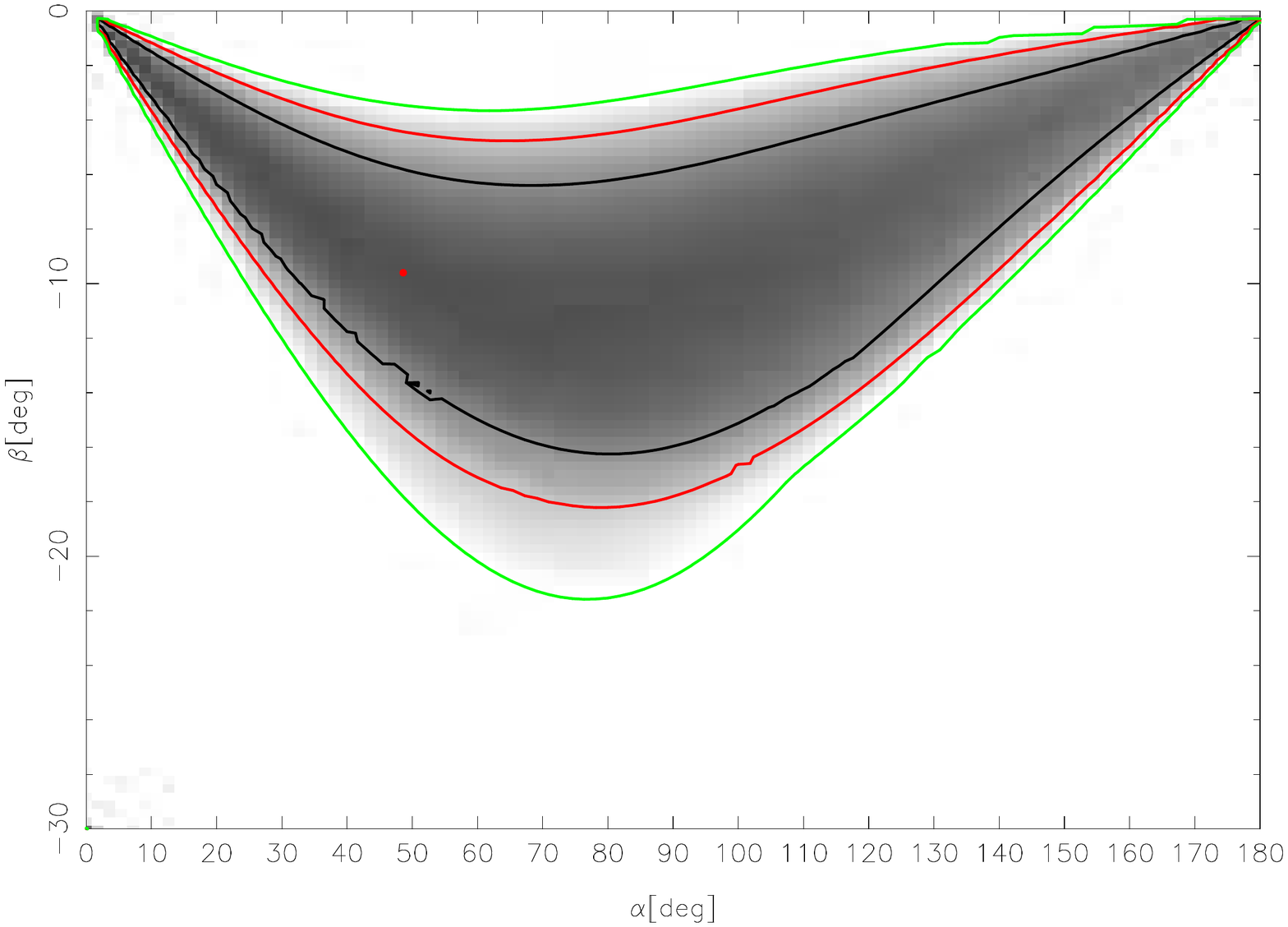}}}
\\
&
\\
\end{tabular}
\caption{Top panel only for 618 MHz (upper window) shows the average profile with total
intensity (Stokes I; solid black lines), total linear polarization (dashed red
line) and circular polarization (Stokes V; dotted blue line). Top panel (lower
window) also shows the single pulse PPA distribution (colour scale) along with
the average PPA (red error bars).
The RVM fits to the average PPA (dashed pink
line) is also shown in this plot. Bottom panel show
the $\chi^2$ contours for the parameters $\alpha$ and $\beta$ obtained from RVM
fits.}
\label{a20}
\end{center}
\end{figure*}


\begin{figure*}
\begin{center}
\begin{tabular}{cc}
{\mbox{\includegraphics[width=9cm,height=6cm,angle=0.]{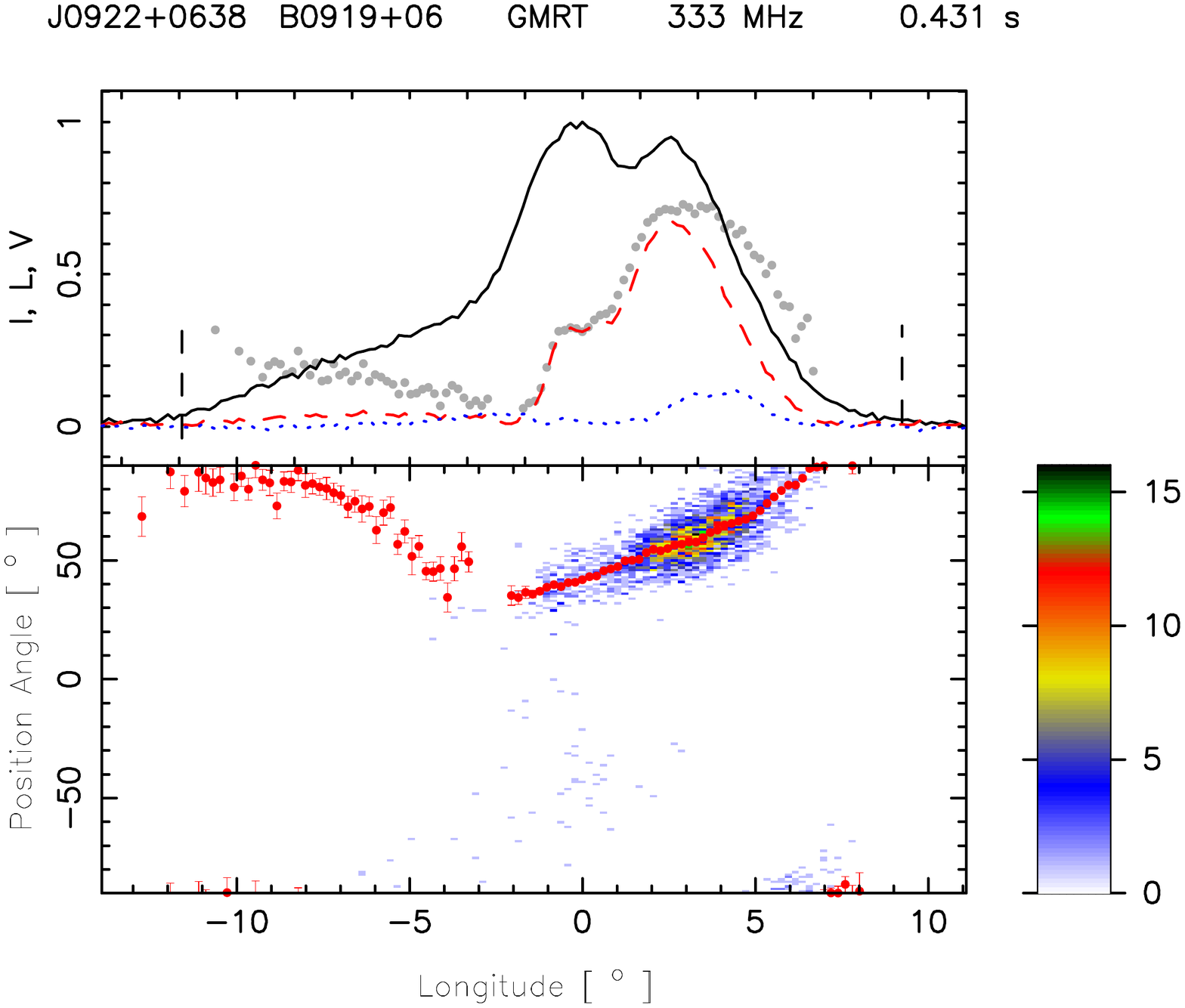}}}&
{\mbox{\includegraphics[width=9cm,height=6cm,angle=0.]{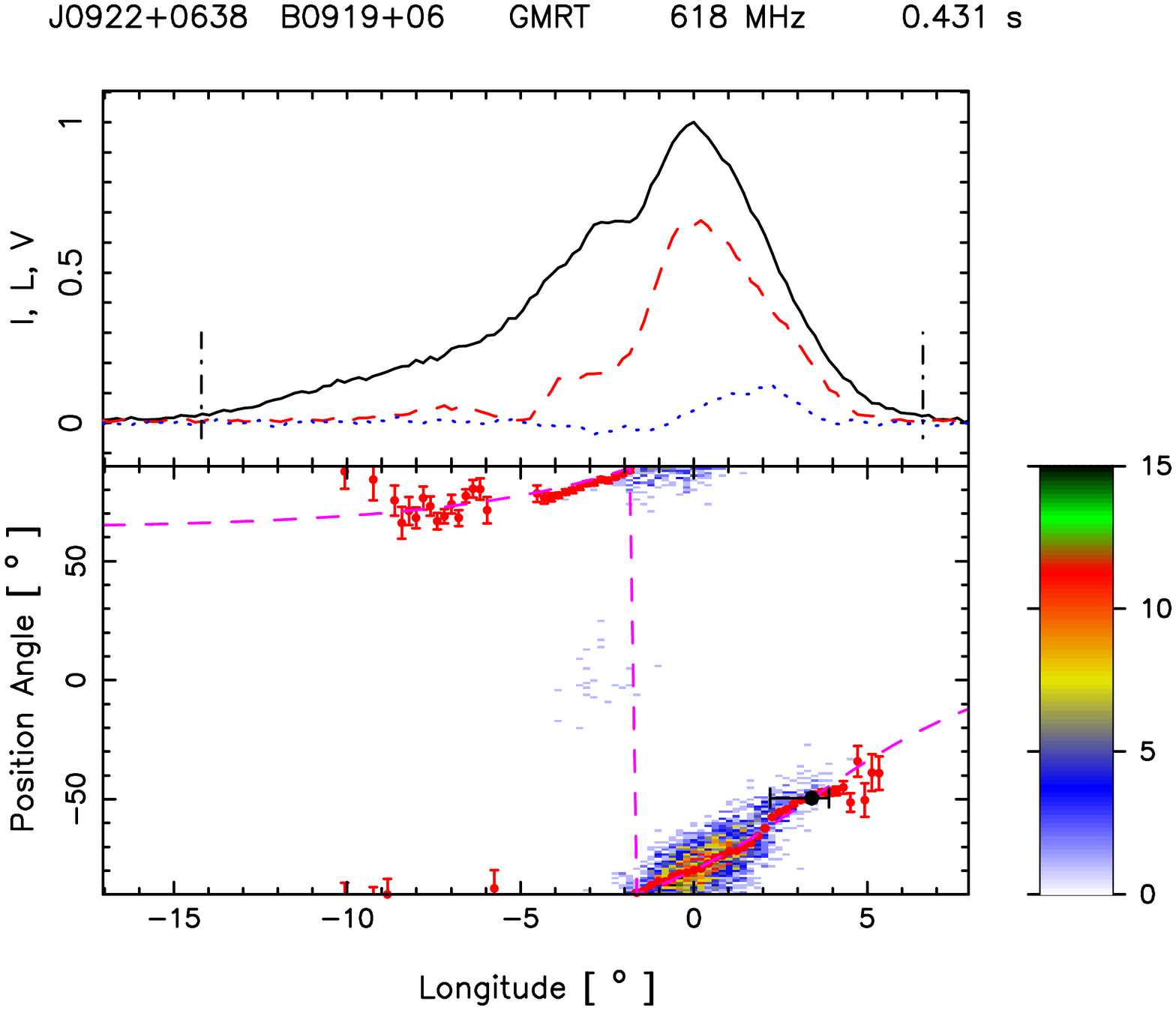}}}\\
&
{\mbox{\includegraphics[width=9cm,height=6cm,angle=0.]{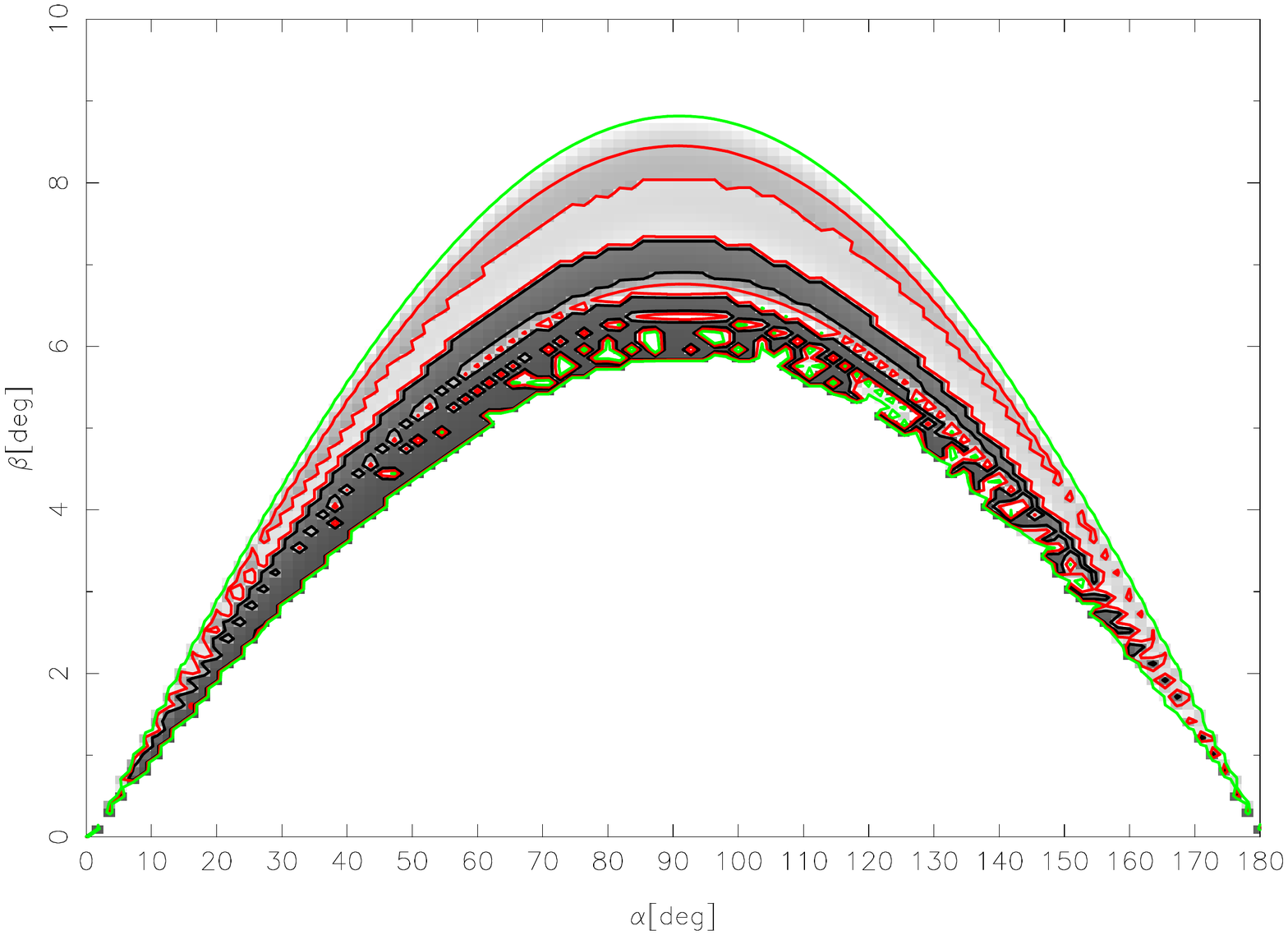}}}\\
{\mbox{\includegraphics[width=9cm,height=6cm,angle=0.]{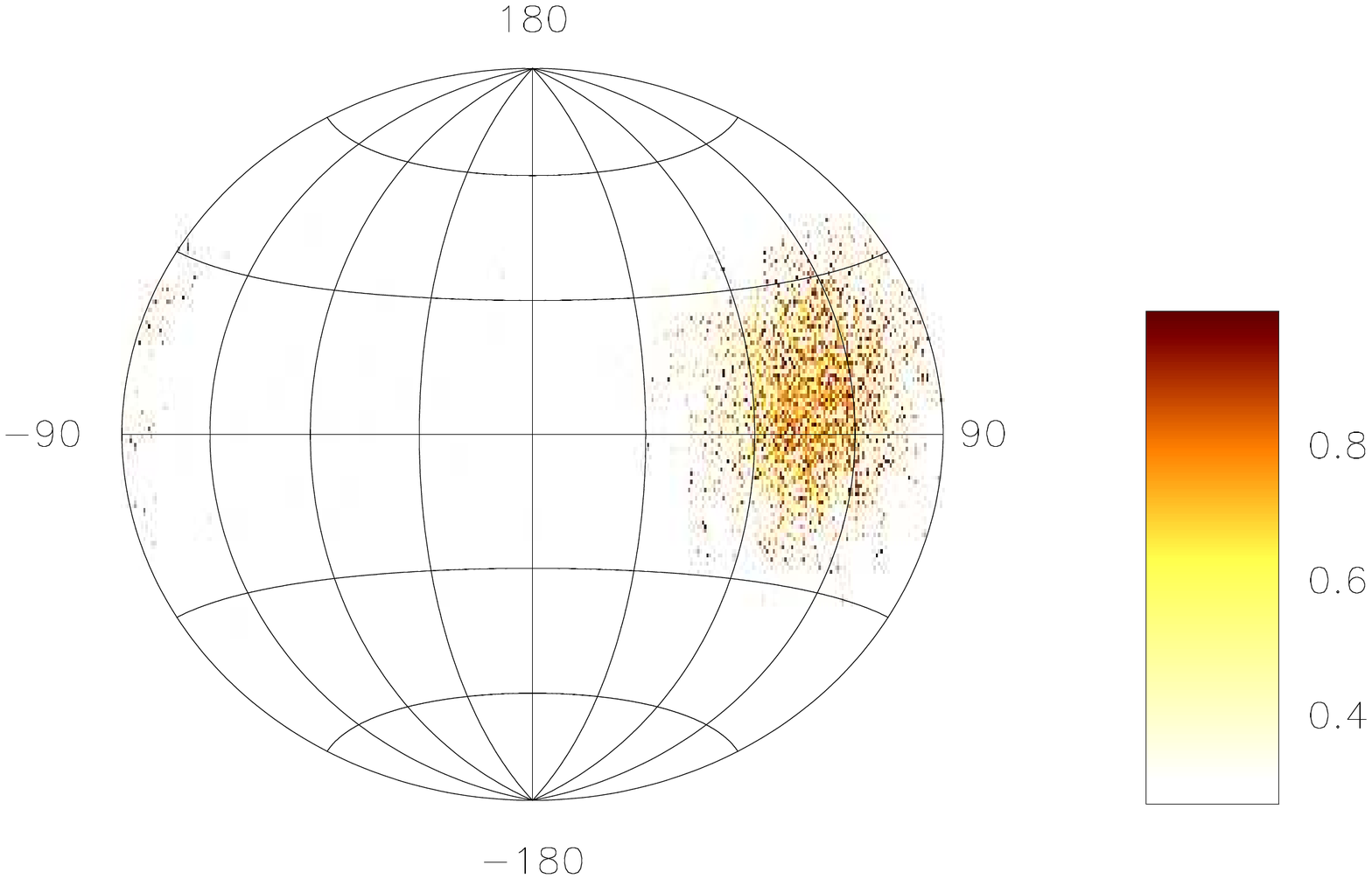}}}&
{\mbox{\includegraphics[width=9cm,height=6cm,angle=0.]{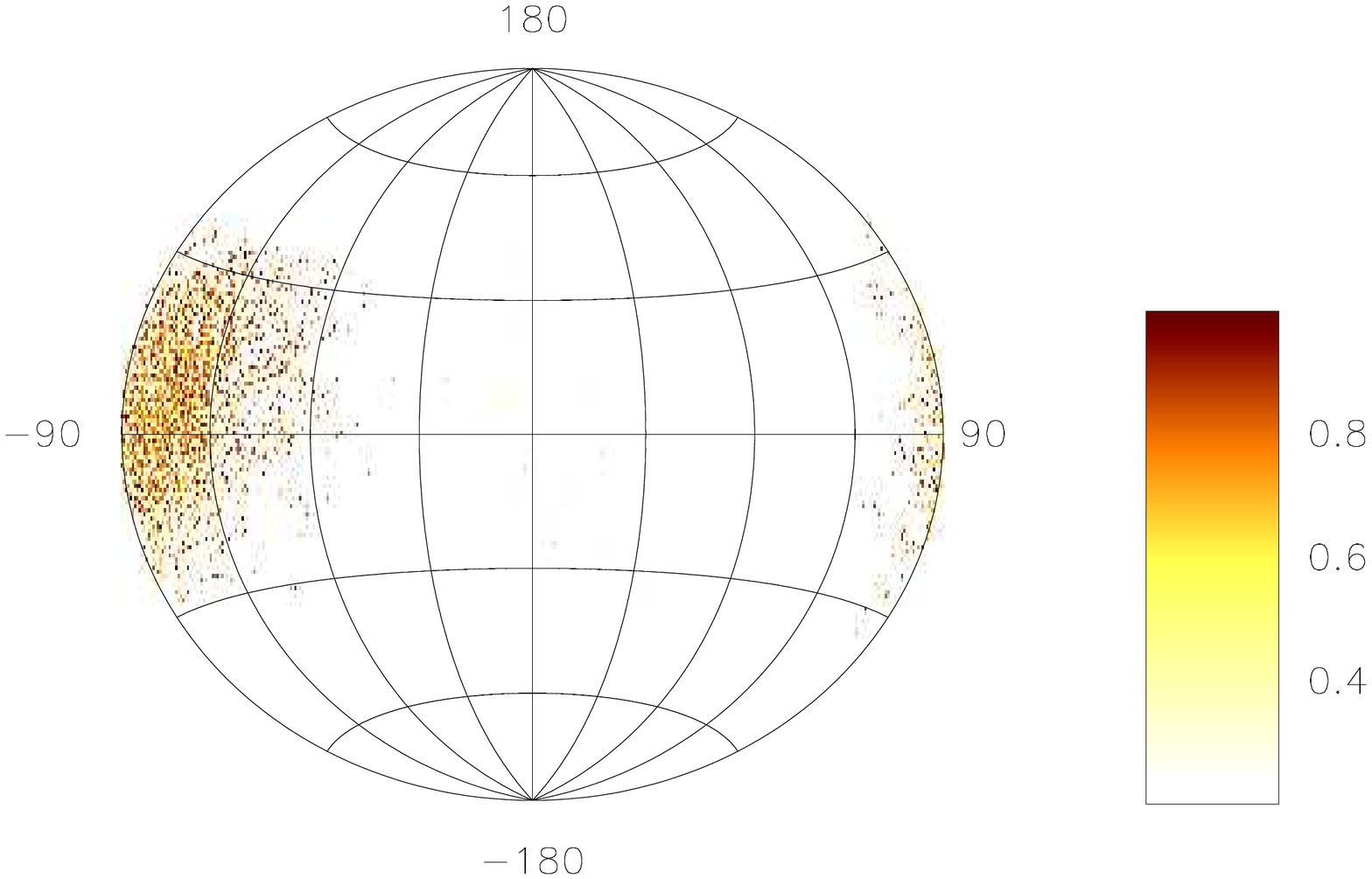}}}\\
\end{tabular}
\caption{Top panel (upper window) shows the average profile with total
intensity (Stokes I; solid black lines), total linear polarization (dashed red
line) and circular polarization (Stokes V; dotted blue line). Top panel (lower
window) also shows the single pulse PPA distribution (colour scale) along with
the average PPA (red error bars).
The RVM fits to the average PPA (dashed pink
line) is also shown in this plot. Middle panel only for 618 MHz show
the $\chi^2$ contours for the parameters $\alpha$ and $\beta$ obtained from RVM
fits.
Bottom panel shows the Hammer-Aitoff projection of the polarized time
samples with the colour scheme representing the fractional polarization level.}
\label{a21}
\end{center}
\end{figure*}


\begin{figure*}
\begin{center}
\begin{tabular}{cc}
{\mbox{\includegraphics[width=9cm,height=6cm,angle=0.]{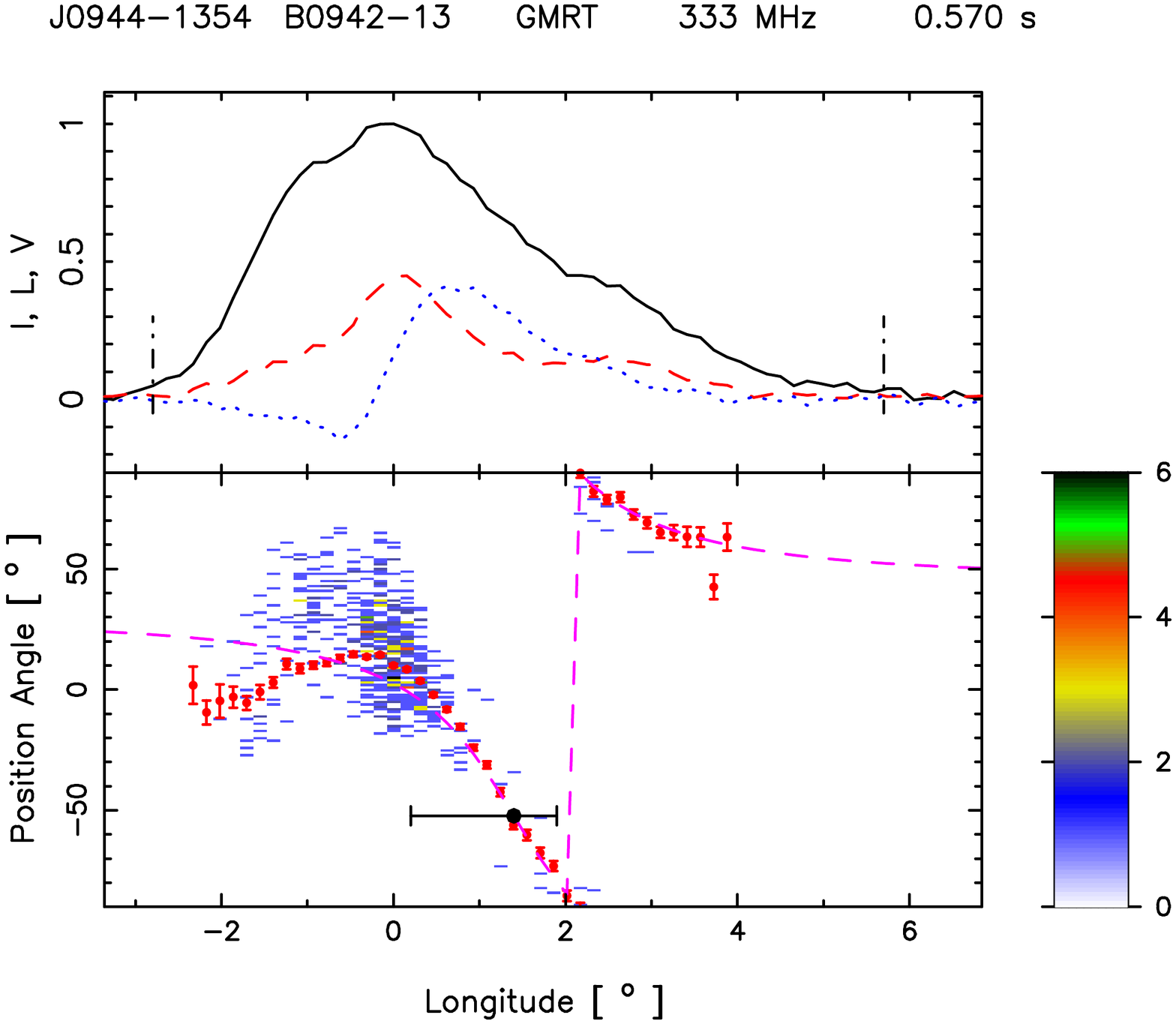}}}&
{\mbox{\includegraphics[width=9cm,height=6cm,angle=0.]{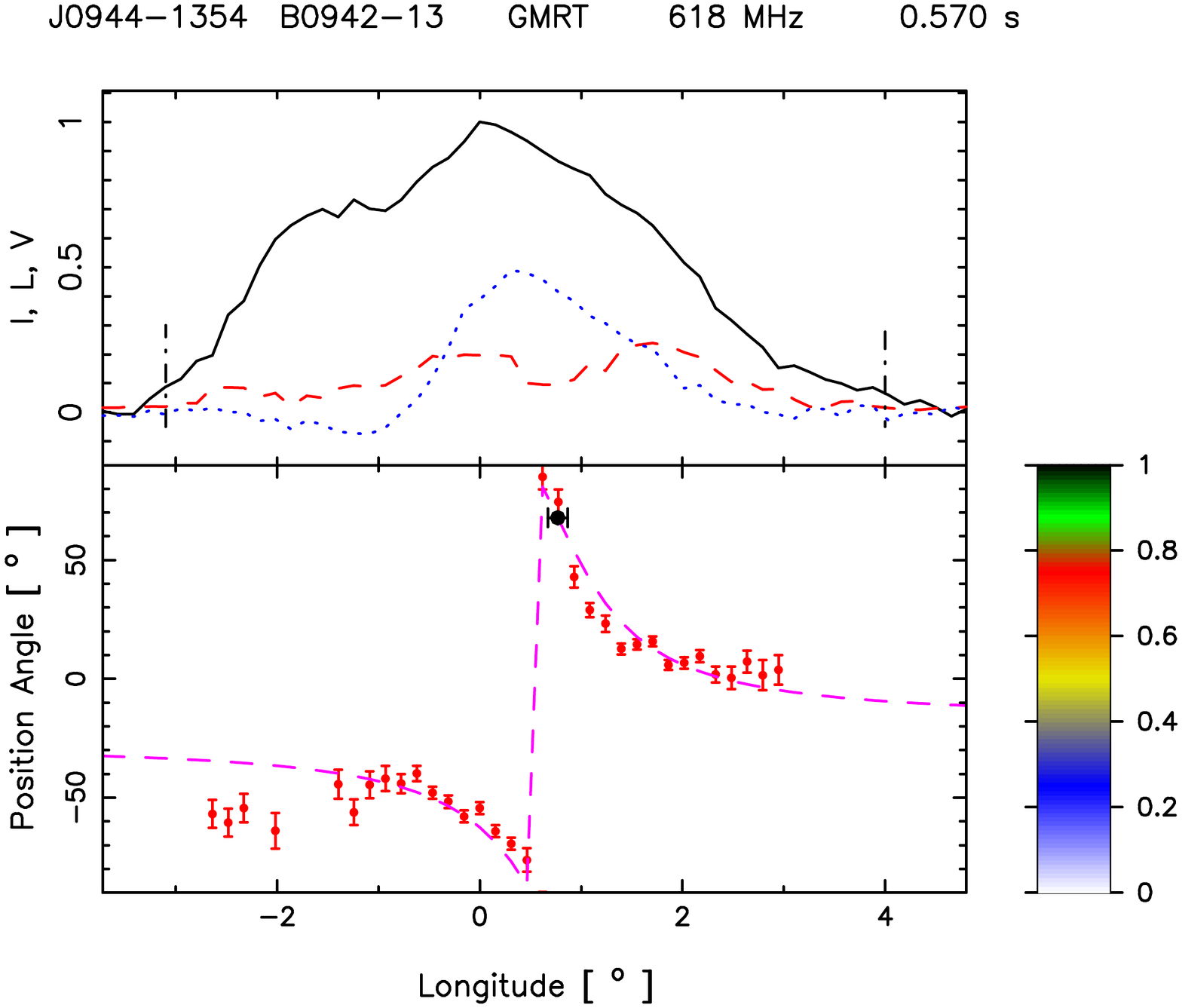}}}\\
{\mbox{\includegraphics[width=9cm,height=6cm,angle=0.]{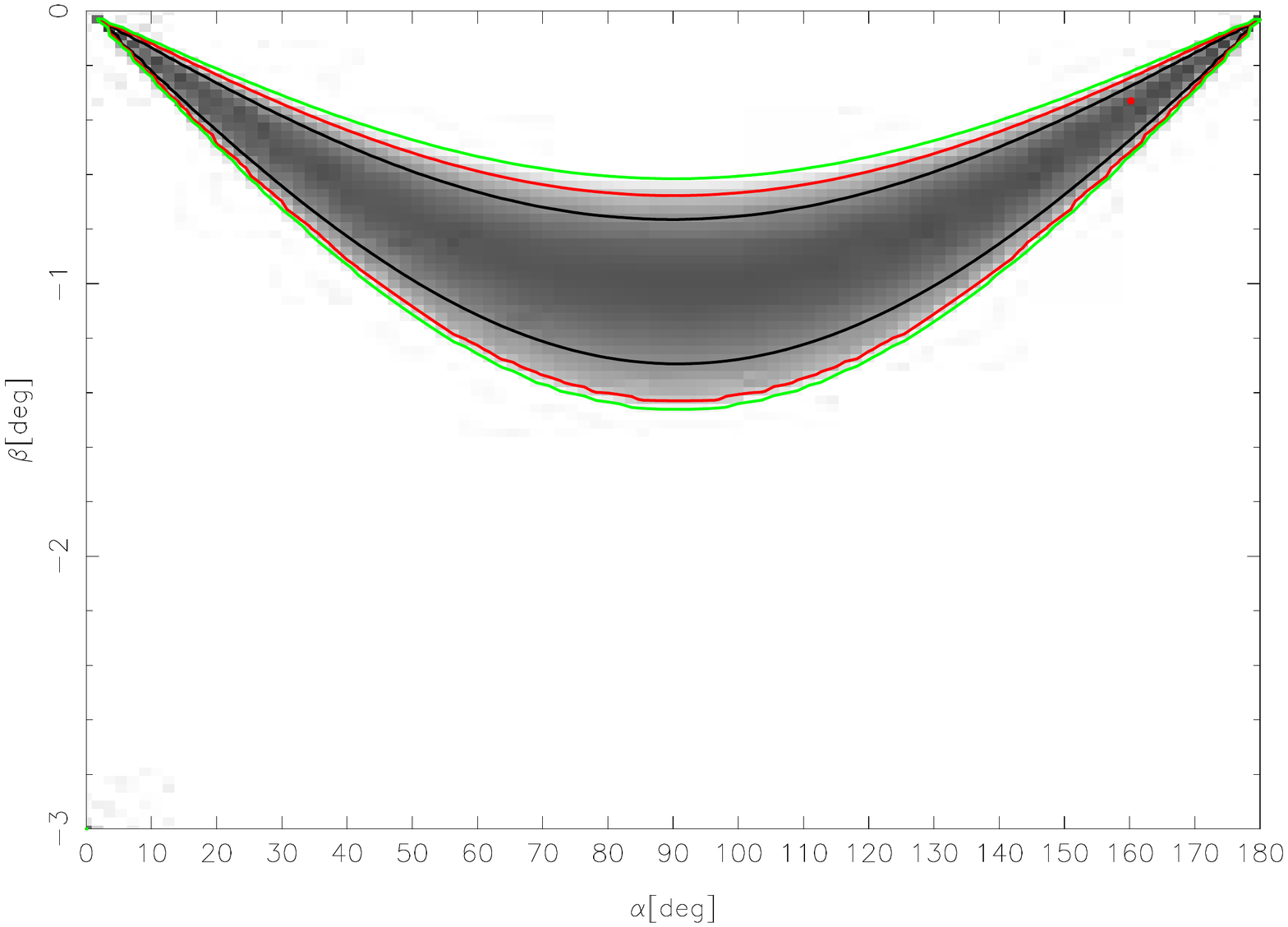}}}&
{\mbox{\includegraphics[width=9cm,height=6cm,angle=0.]{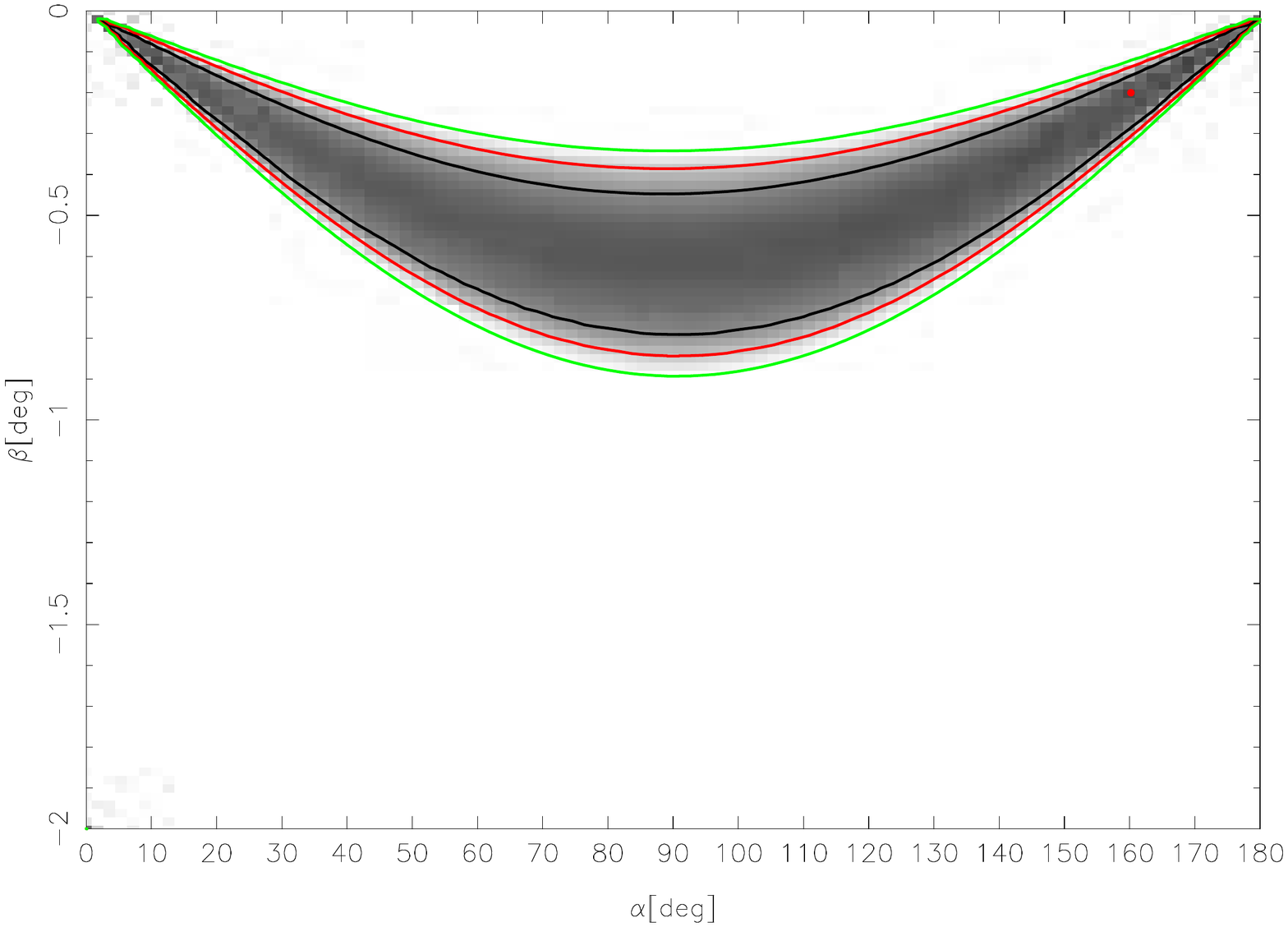}}}\\
{\mbox{\includegraphics[width=9cm,height=6cm,angle=0.]{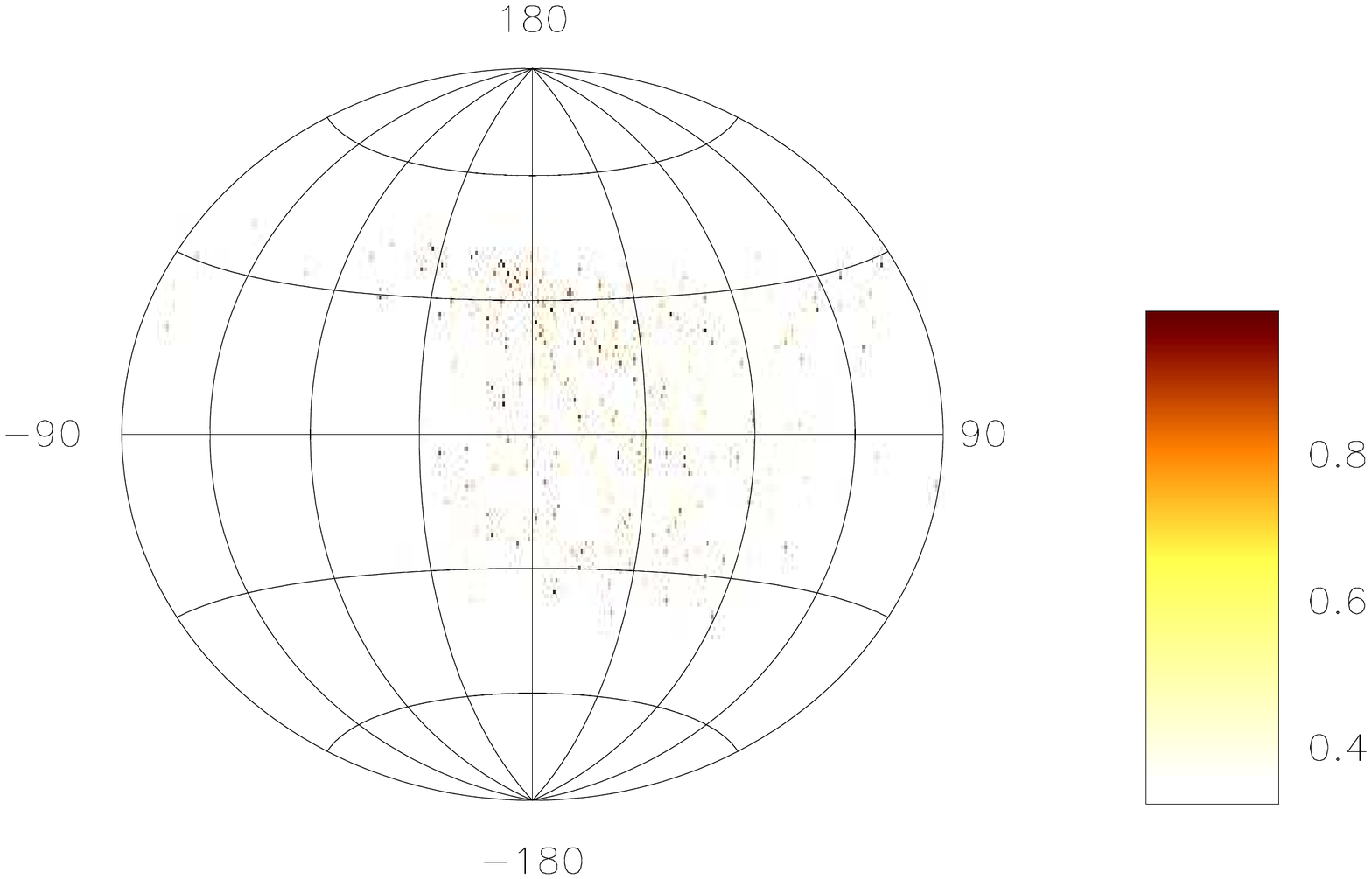}}}&
\\
\end{tabular}
\caption{Top panel (upper window) shows the average profile with total
intensity (Stokes I; solid black lines), total linear polarization (dashed red
line) and circular polarization (Stokes V; dotted blue line). Top panel (lower
window) also shows the single pulse PPA distribution (colour scale) along with
the average PPA (red error bars).
The RVM fits to the average PPA (dashed pink
line) is also shown in this plot. Middle panel show
the $\chi^2$ contours for the parameters $\alpha$ and $\beta$ obtained from RVM
fits.
Bottom panel only for 333 MHz shows the Hammer-Aitoff projection of the polarized time
samples with the colour scheme representing the fractional polarization level.}
\label{a22}
\end{center}
\end{figure*}


\begin{figure*}
\begin{center}
\begin{tabular}{cc}
{\mbox{\includegraphics[width=9cm,height=6cm,angle=0.]{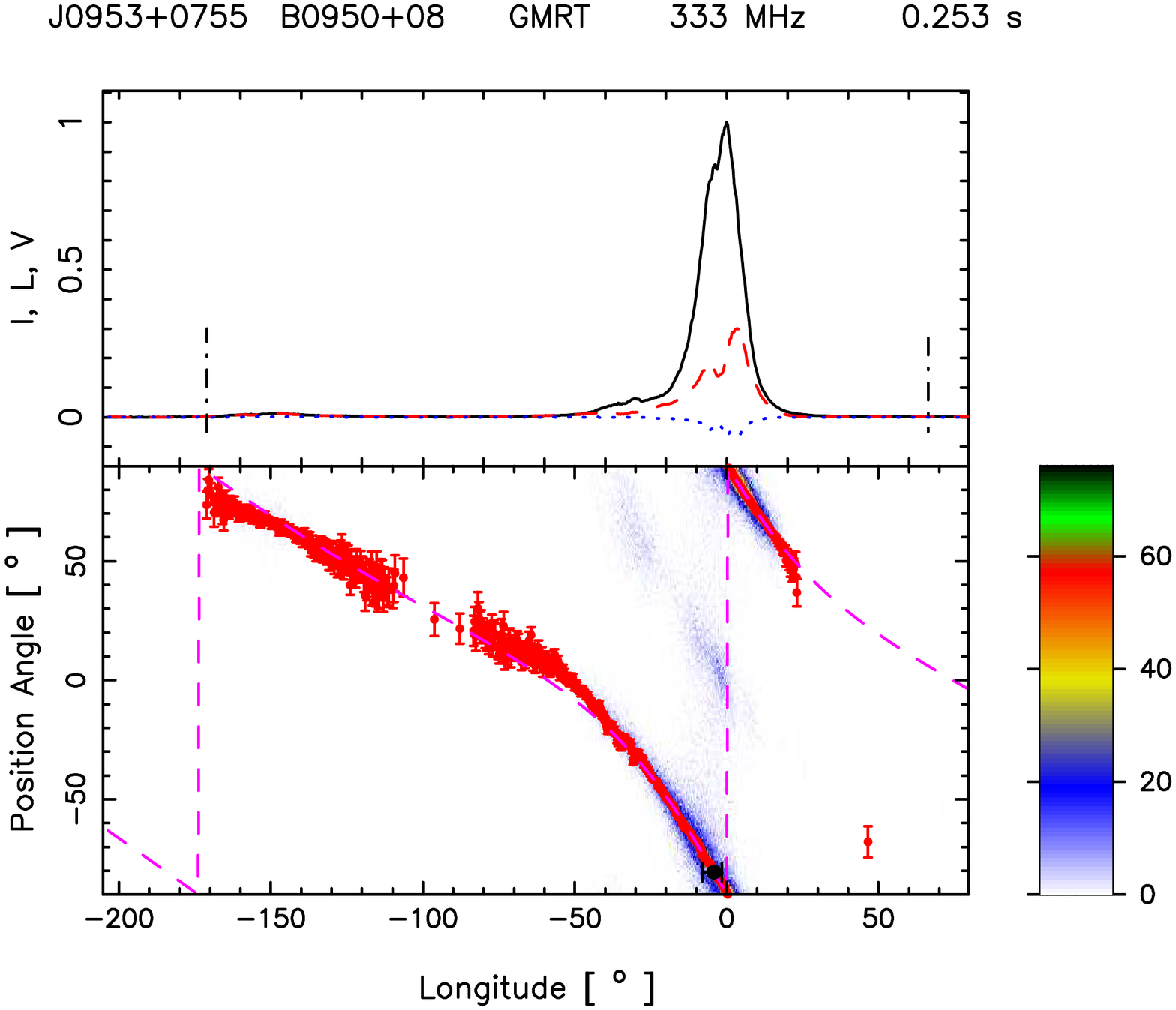}}}&
{\mbox{\includegraphics[width=9cm,height=6cm,angle=0.]{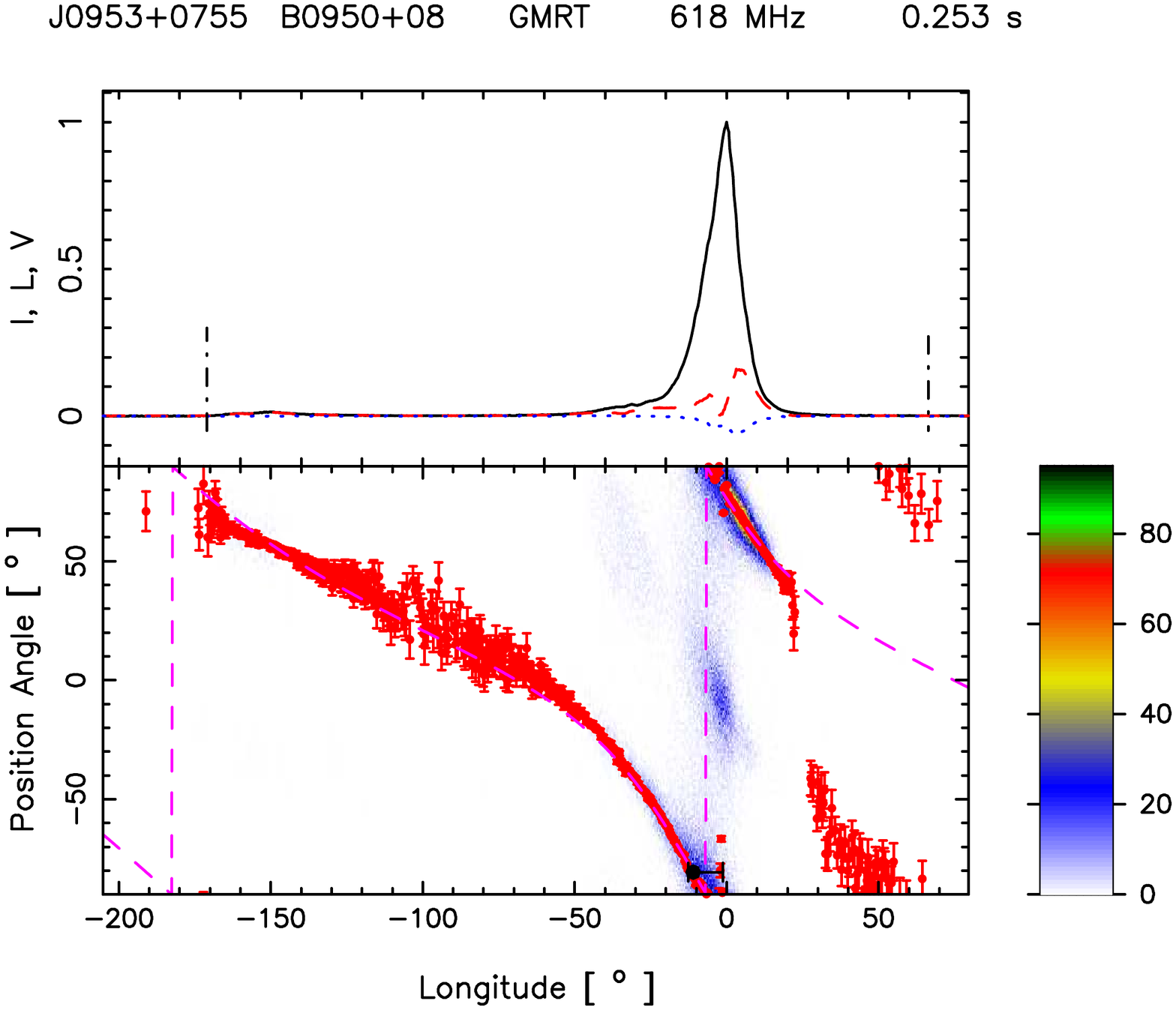}}}\\
{\mbox{\includegraphics[width=9cm,height=6cm,angle=0.]{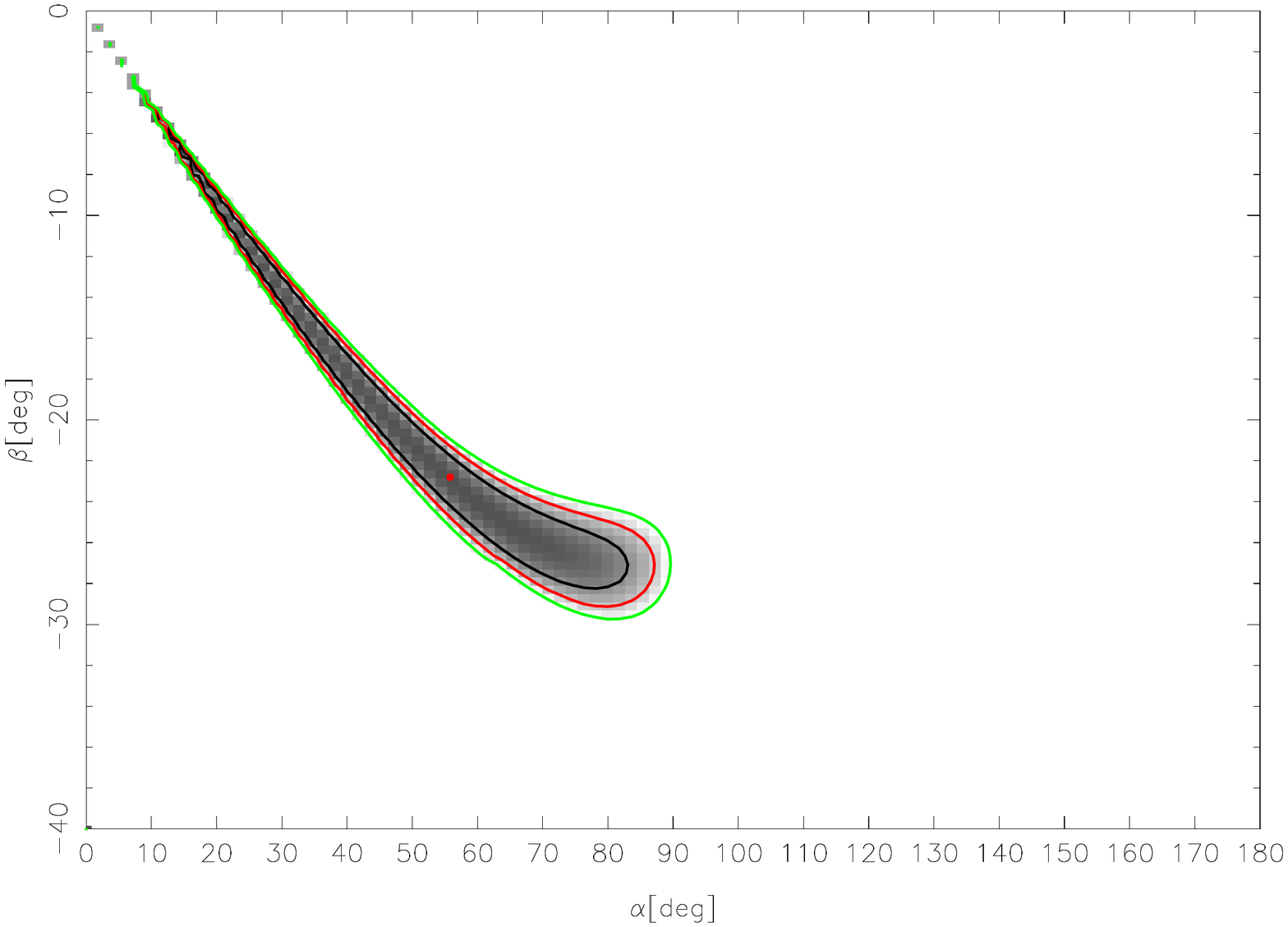}}}&
{\mbox{\includegraphics[width=9cm,height=6cm,angle=0.]{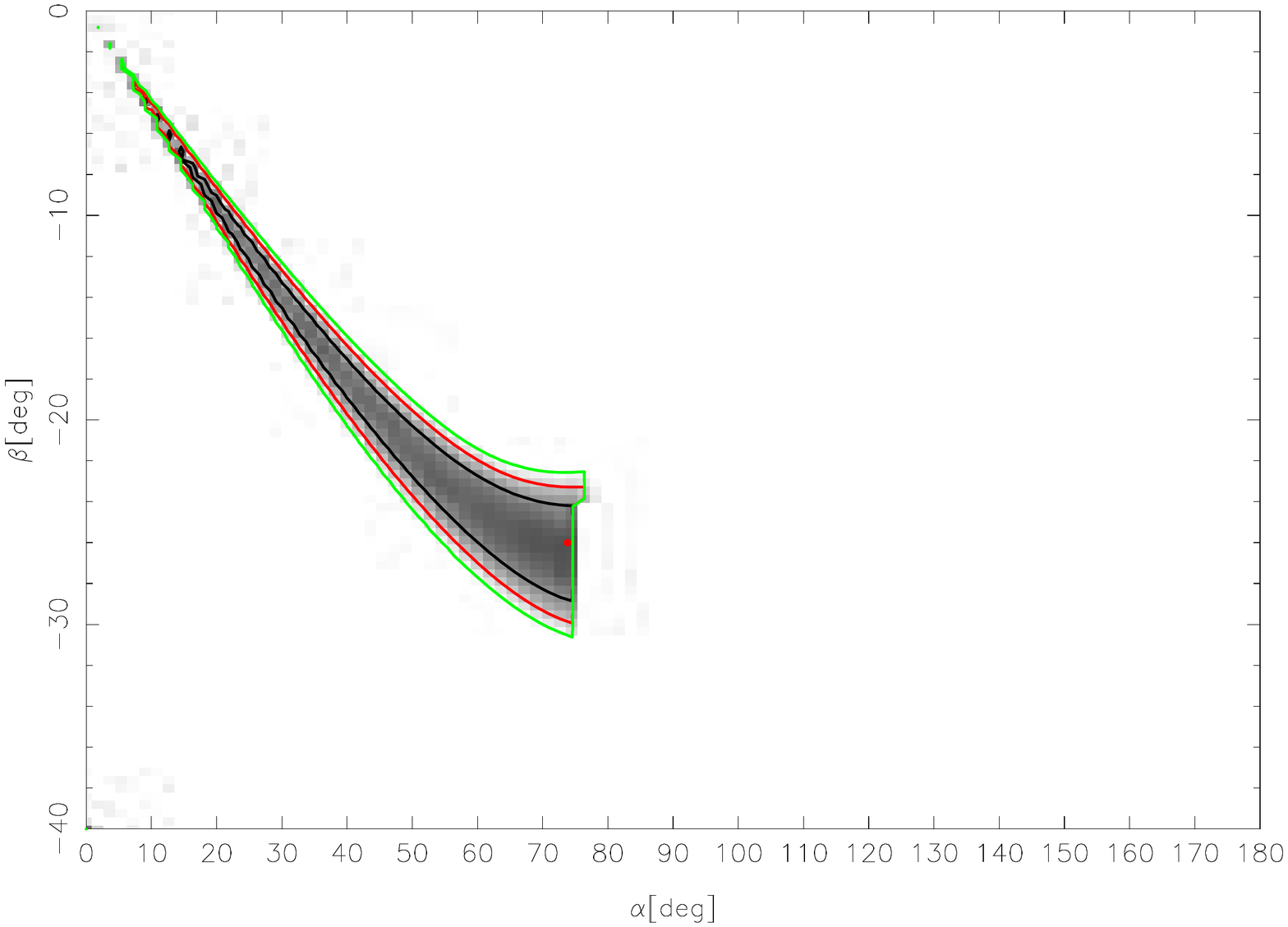}}}\\
{\mbox{\includegraphics[width=9cm,height=6cm,angle=0.]{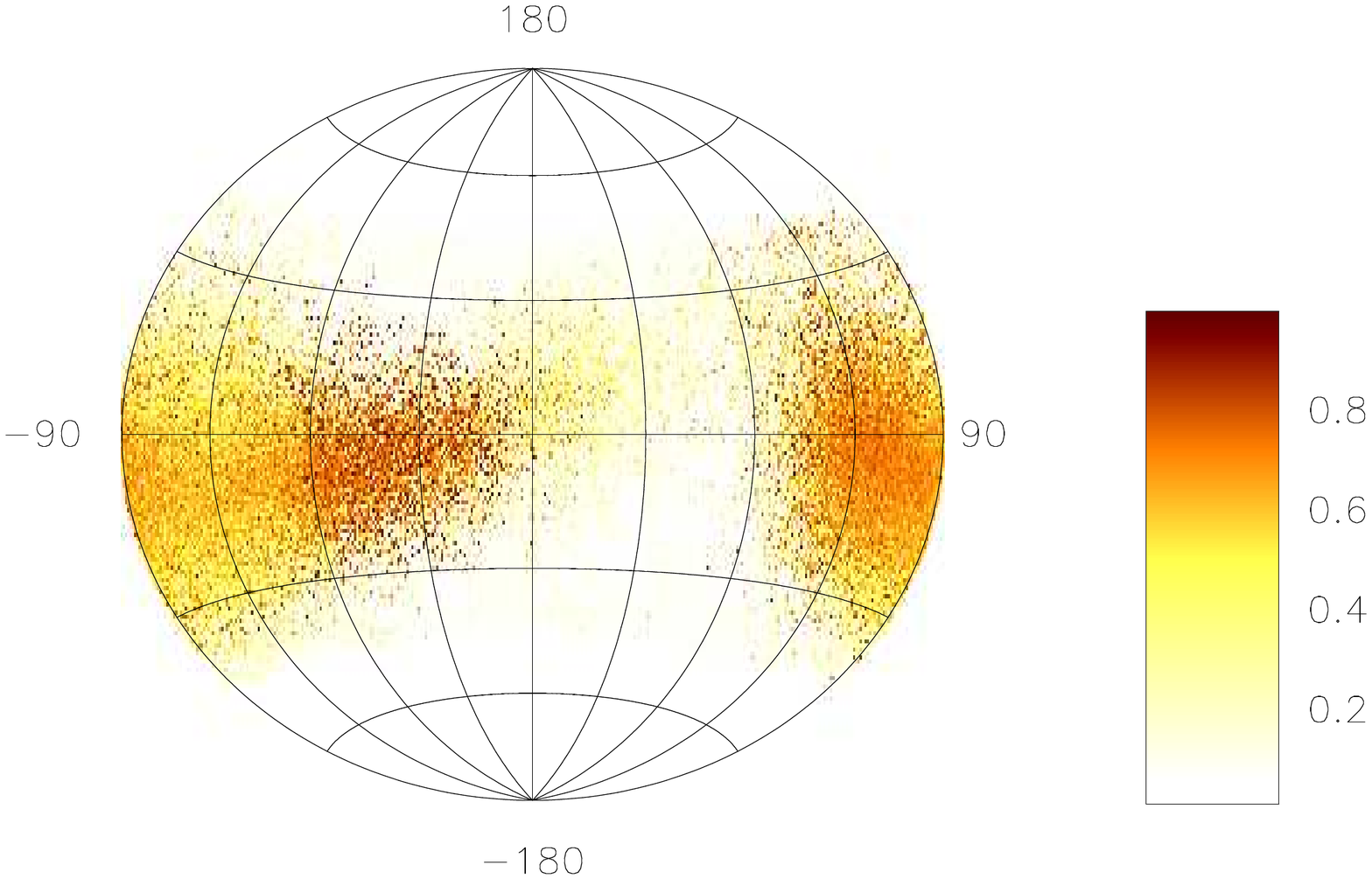}}}&
{\mbox{\includegraphics[width=9cm,height=6cm,angle=0.]{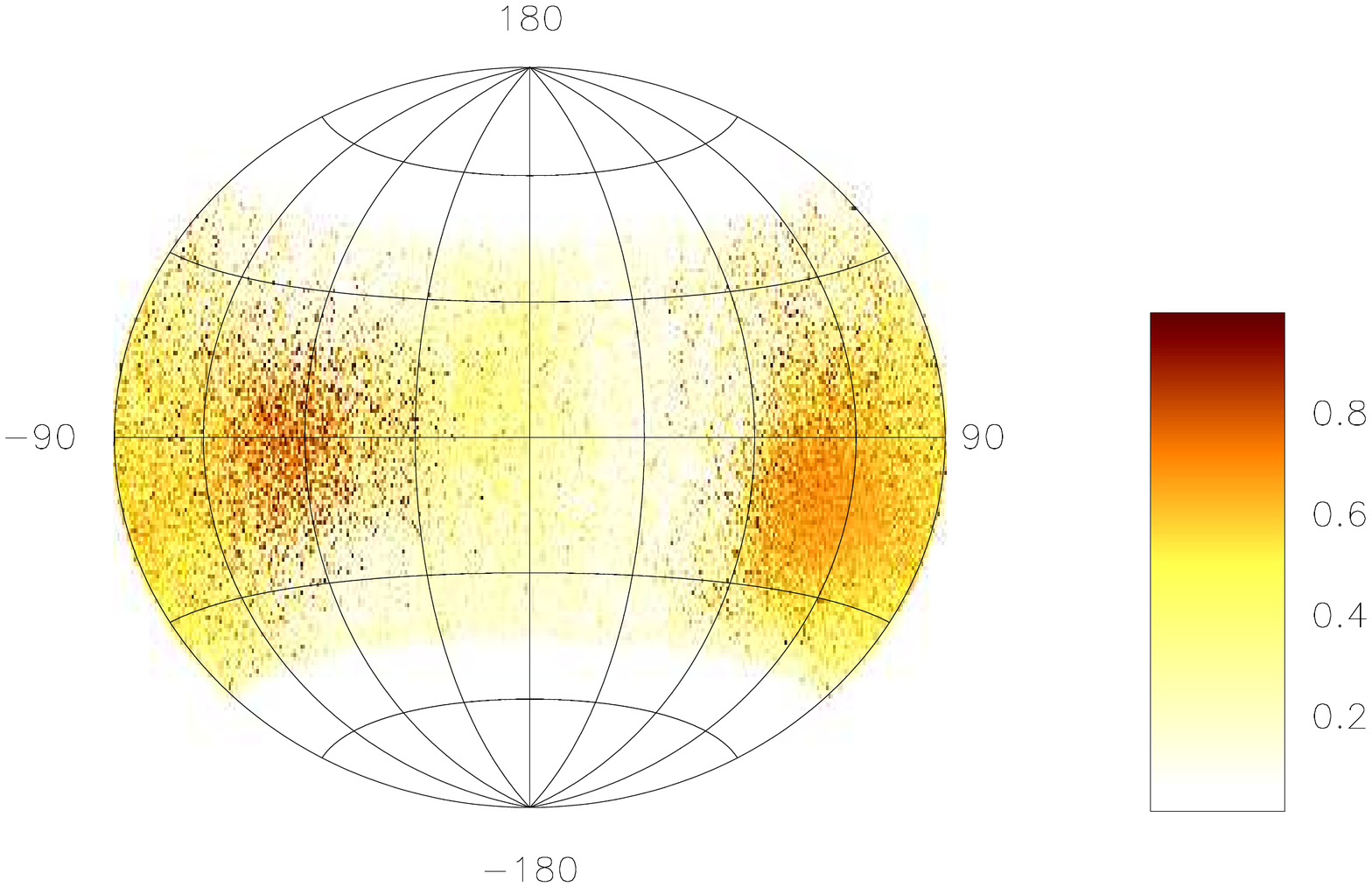}}}\\
\end{tabular}
\caption{Top panel (upper window) shows the average profile with total
intensity (Stokes I; solid black lines), total linear polarization (dashed red
line) and circular polarization (Stokes V; dotted blue line). Top panel (lower
window) also shows the single pulse PPA distribution (colour scale) along with
the average PPA (red error bars).
The RVM fits to the average PPA (dashed pink
line) is also shown in this plot. Middle panel show
the $\chi^2$ contours for the parameters $\alpha$ and $\beta$ obtained from RVM
fits.
Bottom panel shows the Hammer-Aitoff projection of the polarized time
samples with the colour scheme representing the fractional polarization level.}
\label{a23}
\end{center}
\end{figure*}


\begin{figure*}
\begin{center}
\begin{tabular}{cc}
&
{\mbox{\includegraphics[width=9cm,height=6cm,angle=0.]{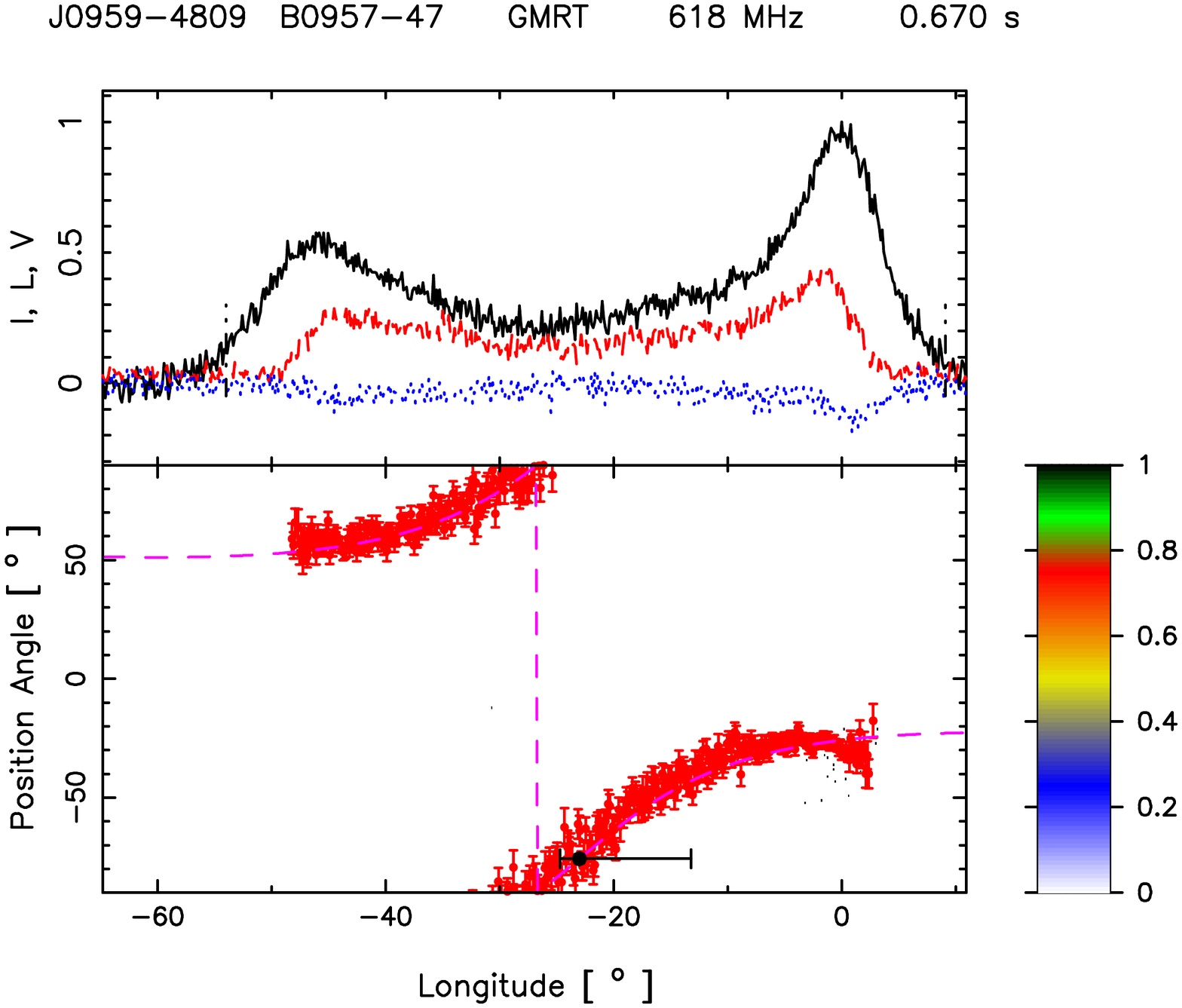}}}\\
&
{\mbox{\includegraphics[width=9cm,height=6cm,angle=0.]{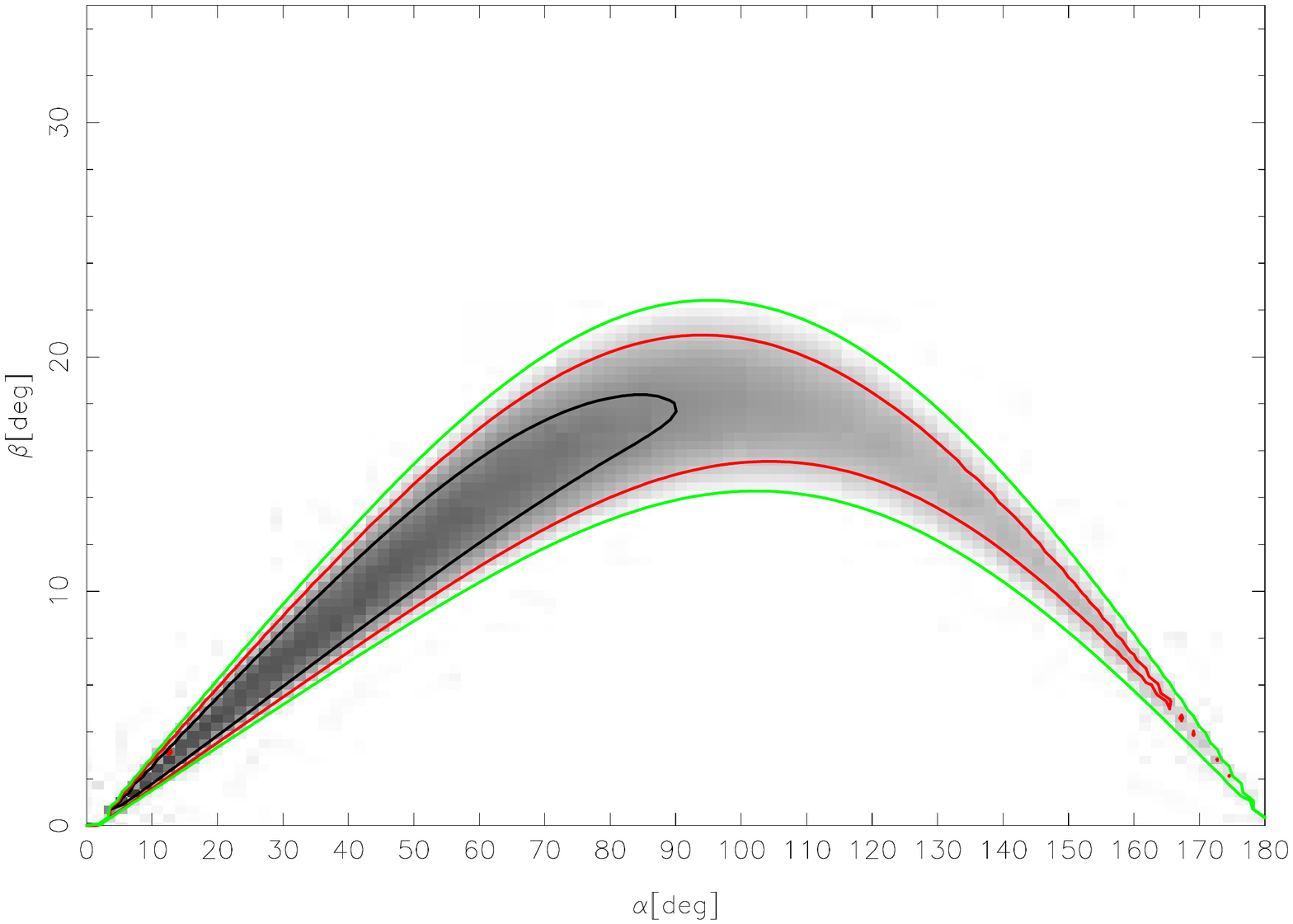}}}\\
&
\\
\end{tabular}
\caption{op panel only for 618 MHz (upper window) shows the average profile with total
intensity (Stokes I; solid black lines), total linear polarization (dashed red
line) and circular polarization (Stokes V; dotted blue line). Top panel (lower
window) also shows the single pulse PPA distribution (colour scale) along with
the average PPA (red error bars).
The RVM fits to the average PPA (dashed pink
line) is also shown in this plot. Bottom panel show
the $\chi^2$ contours for the parameters $\alpha$ and $\beta$ obtained from RVM
fits.}
\label{a24}
\end{center}
\end{figure*}


\begin{figure*}
\begin{center}
\begin{tabular}{cc}
{\mbox{\includegraphics[width=9cm,height=6cm,angle=0.]{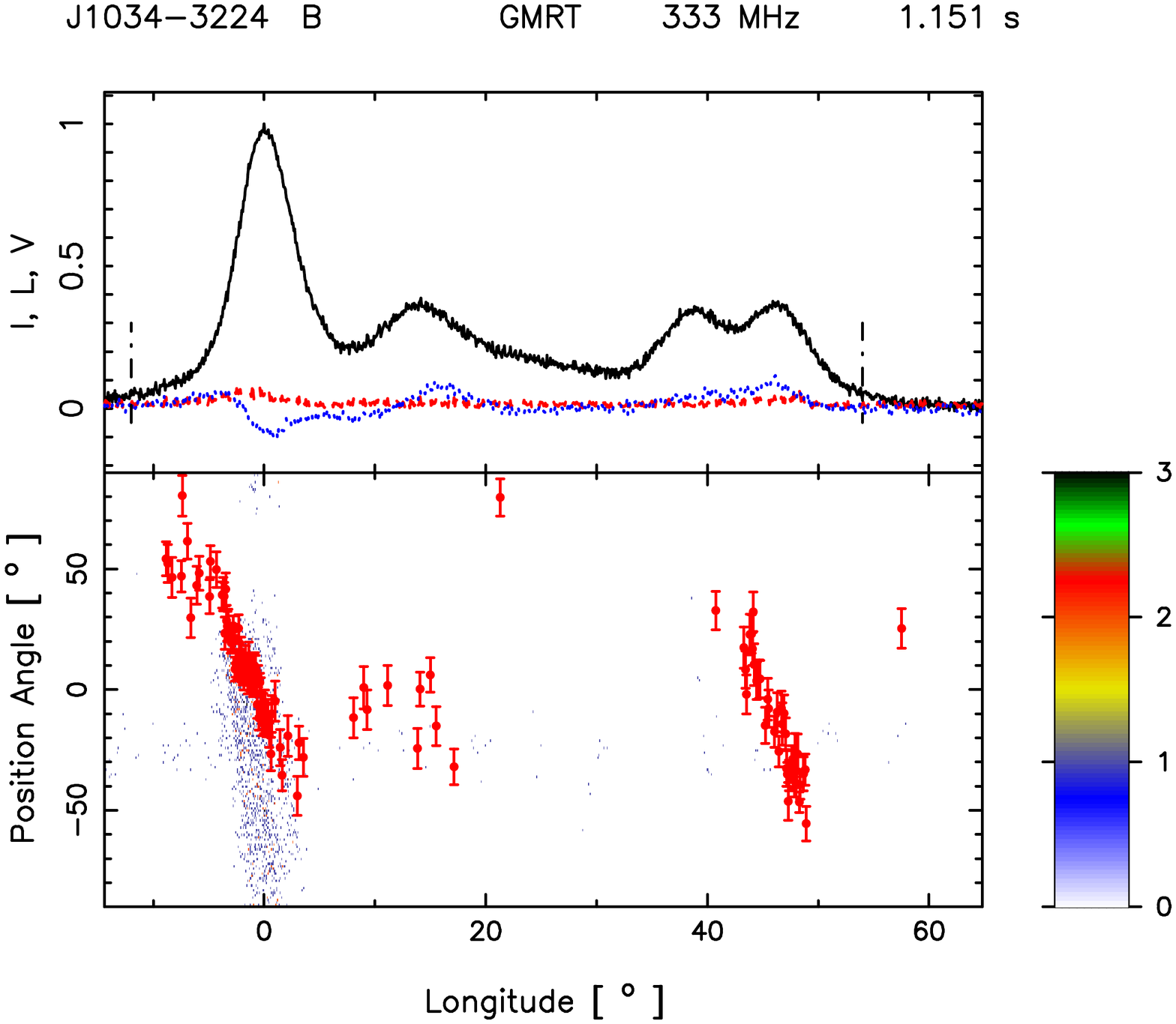}}}&
{\mbox{\includegraphics[width=9cm,height=6cm,angle=0.]{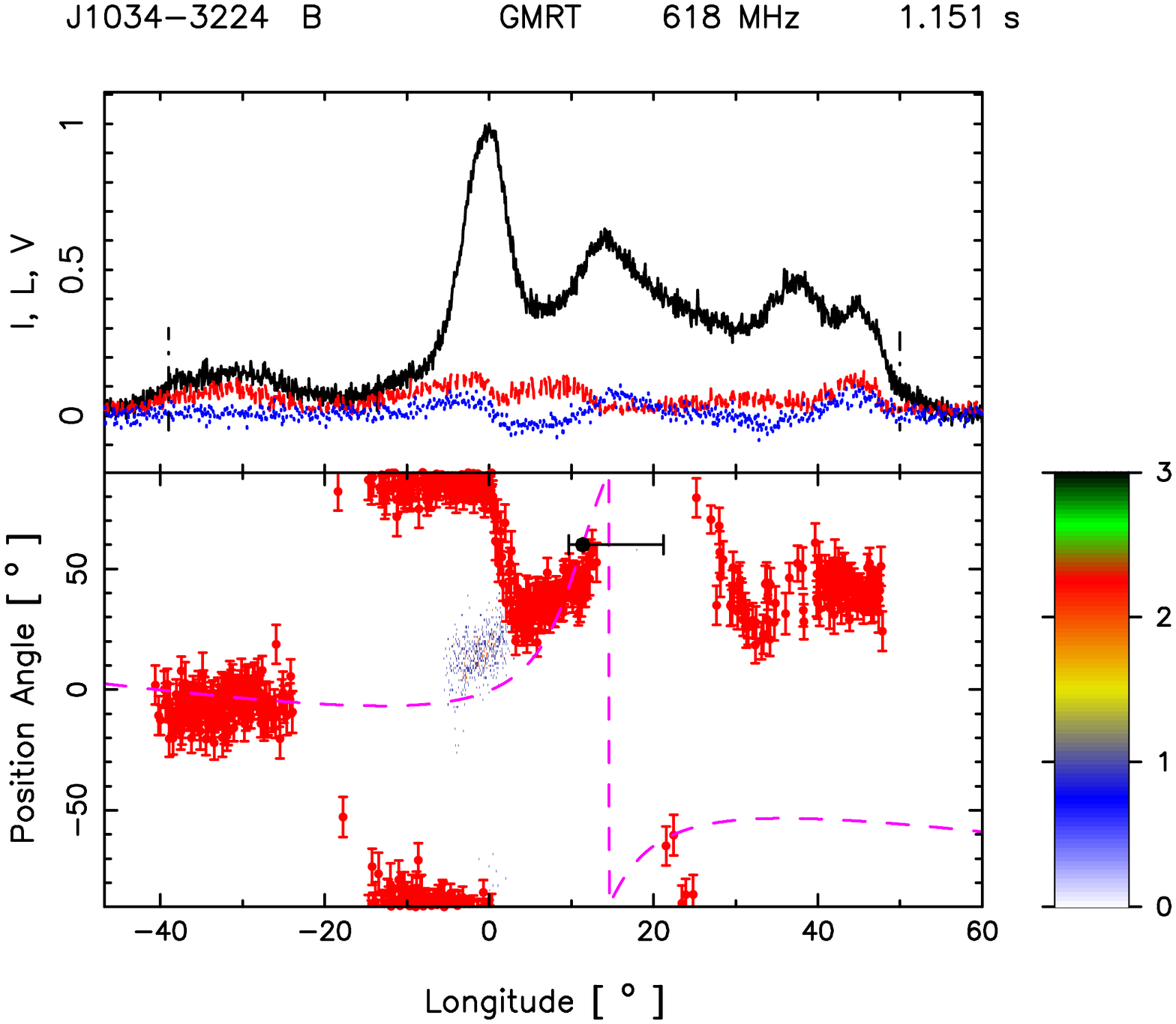}}}\\
&
\\
{\mbox{\includegraphics[width=9cm,height=6cm,angle=0.]{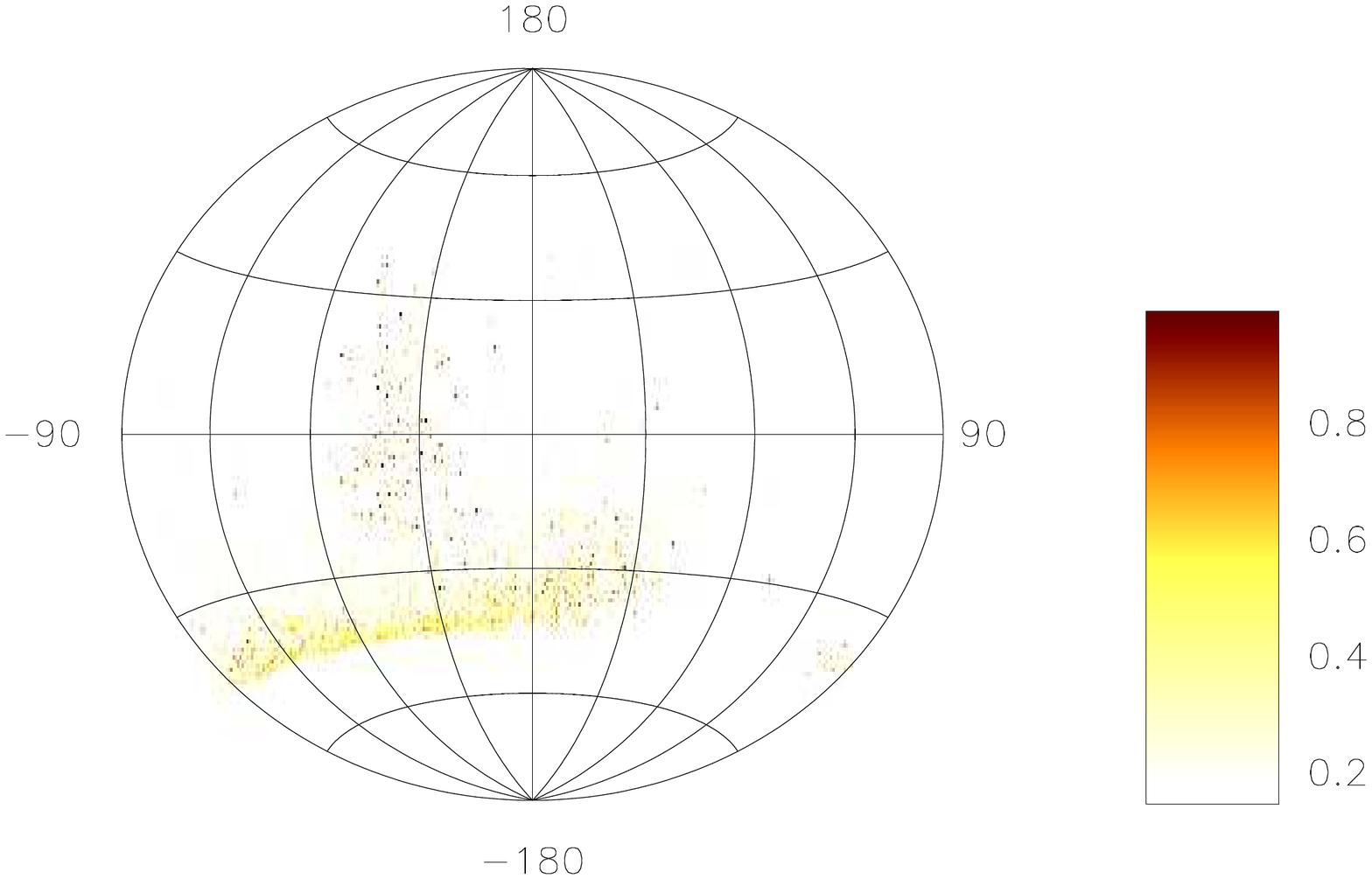}}}&
{\mbox{\includegraphics[width=9cm,height=6cm,angle=0.]{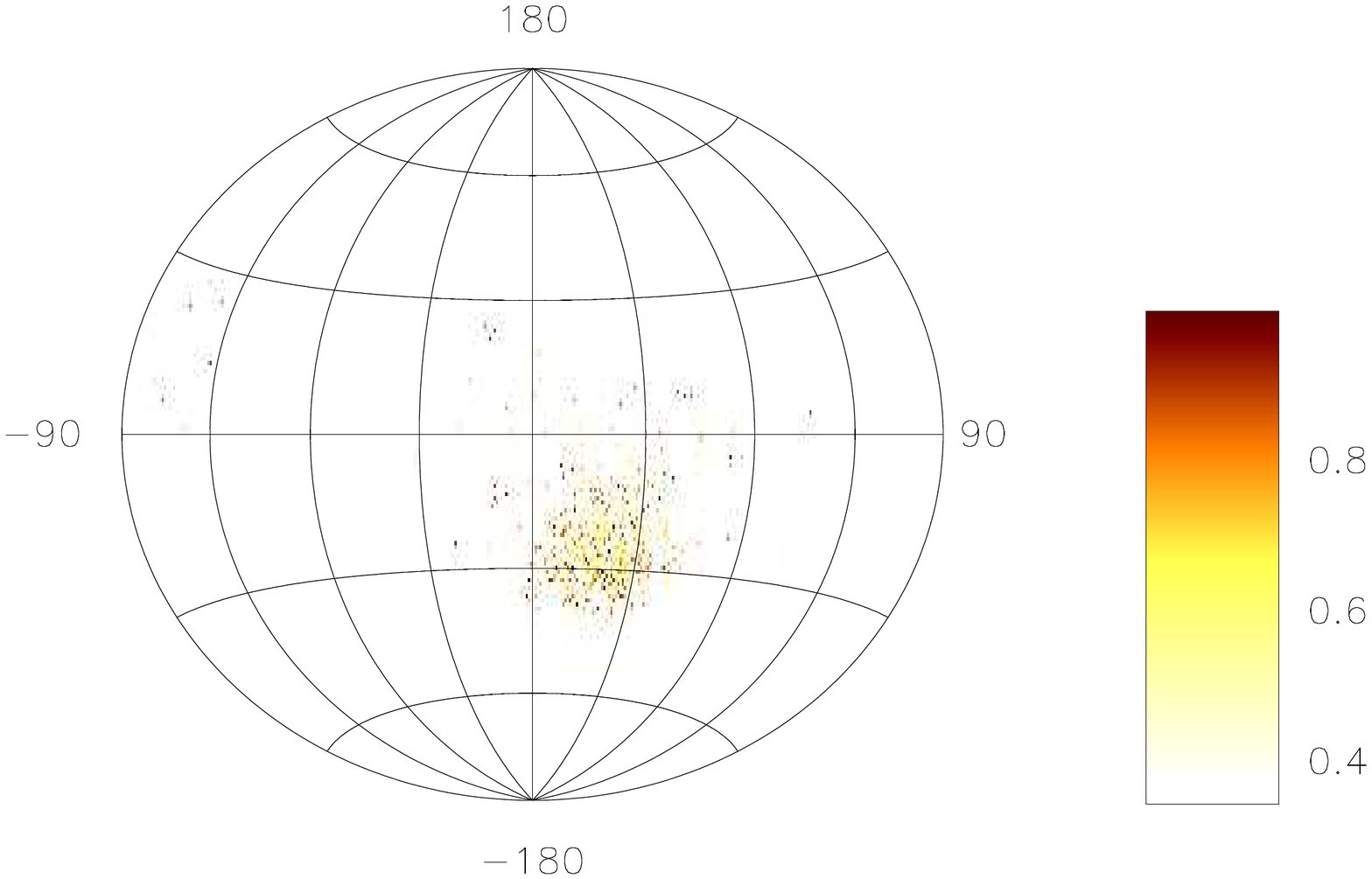}}}\\
\end{tabular}
\caption{Top panel (upper window) shows the average profile with total
intensity (Stokes I; solid black lines), total linear polarization (dashed red
line) and circular polarization (Stokes V; dotted blue line). Top panel (lower
window) also shows the single pulse PPA distribution (colour scale) along with
the average PPA (red error bars).
The RVM fits to the average PPA (dashed pink
line) is also shown in this plot. Middle panel only for 618 MHz show
the $\chi^2$ contours for the parameters $\alpha$ and $\beta$ obtained from RVM
fits.
Bottom panel shows the Hammer-Aitoff projection of the polarized time
samples with the colour scheme representing the fractional polarization level.}
\label{a25}
\end{center}
\end{figure*}


\begin{figure*}
\begin{center}
\begin{tabular}{cc}
&
{\mbox{\includegraphics[width=9cm,height=6cm,angle=0.]{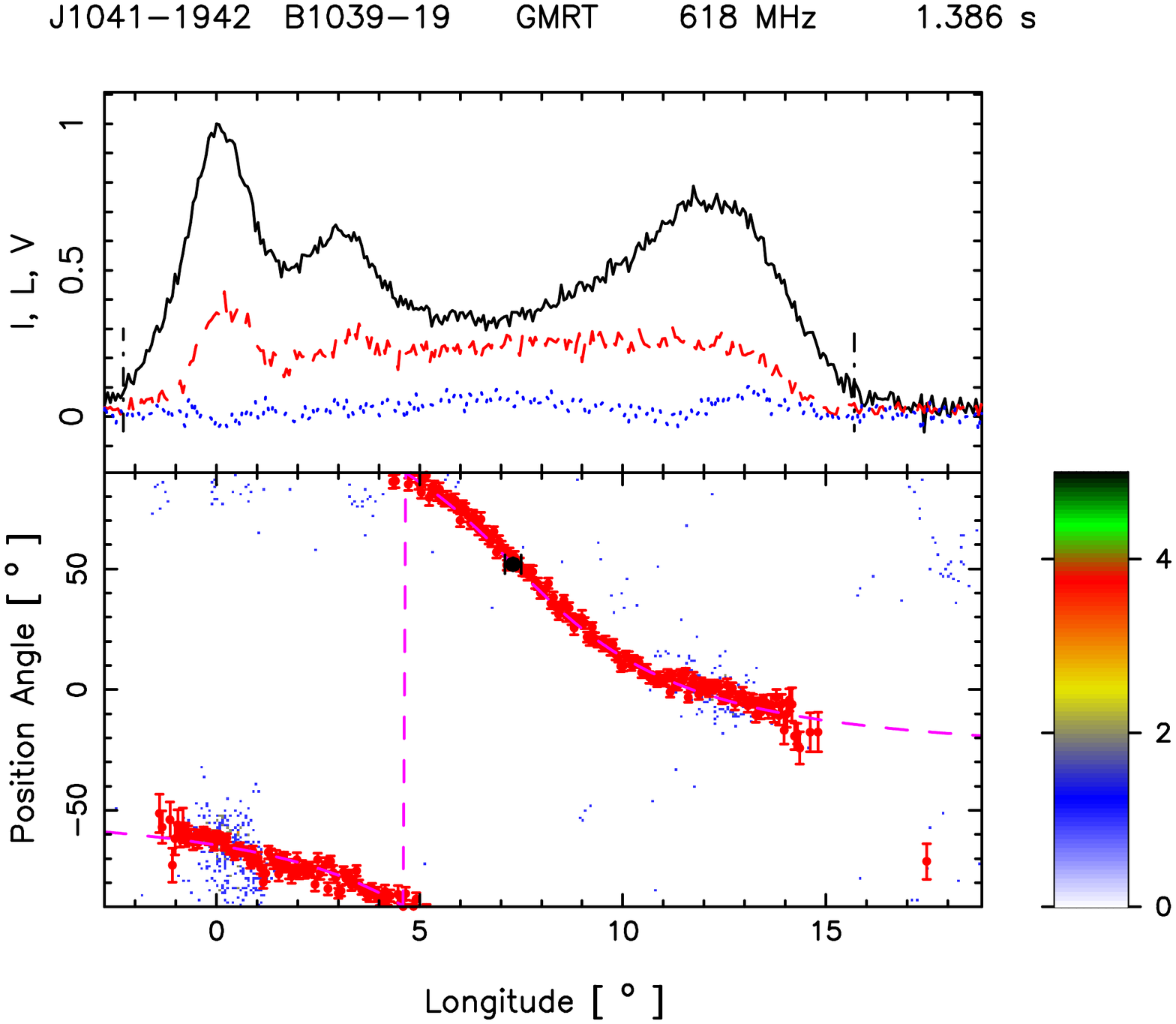}}}\\
&
{\mbox{\includegraphics[width=9cm,height=6cm,angle=0.]{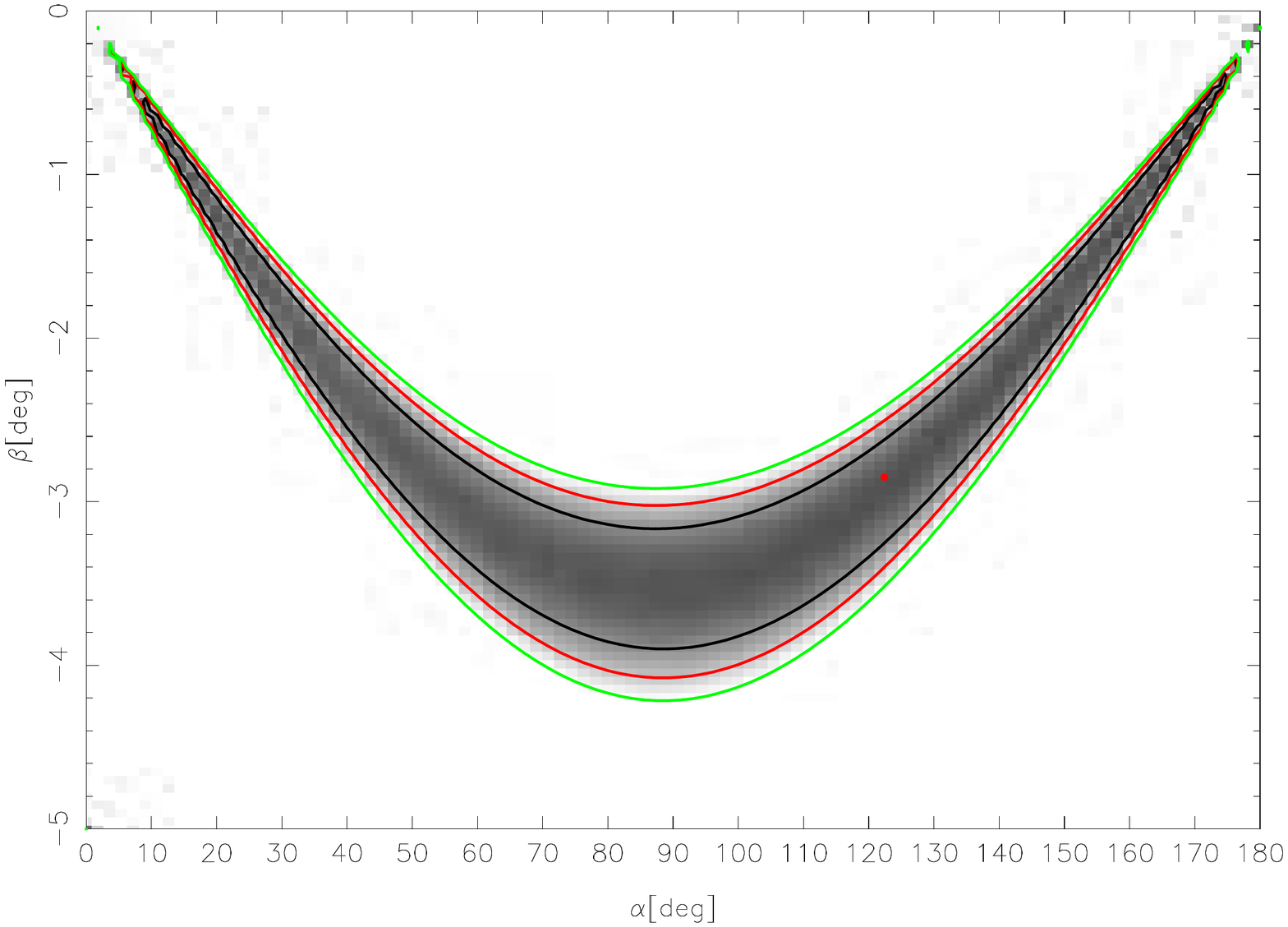}}}\\
&
{\mbox{\includegraphics[width=9cm,height=6cm,angle=0.]{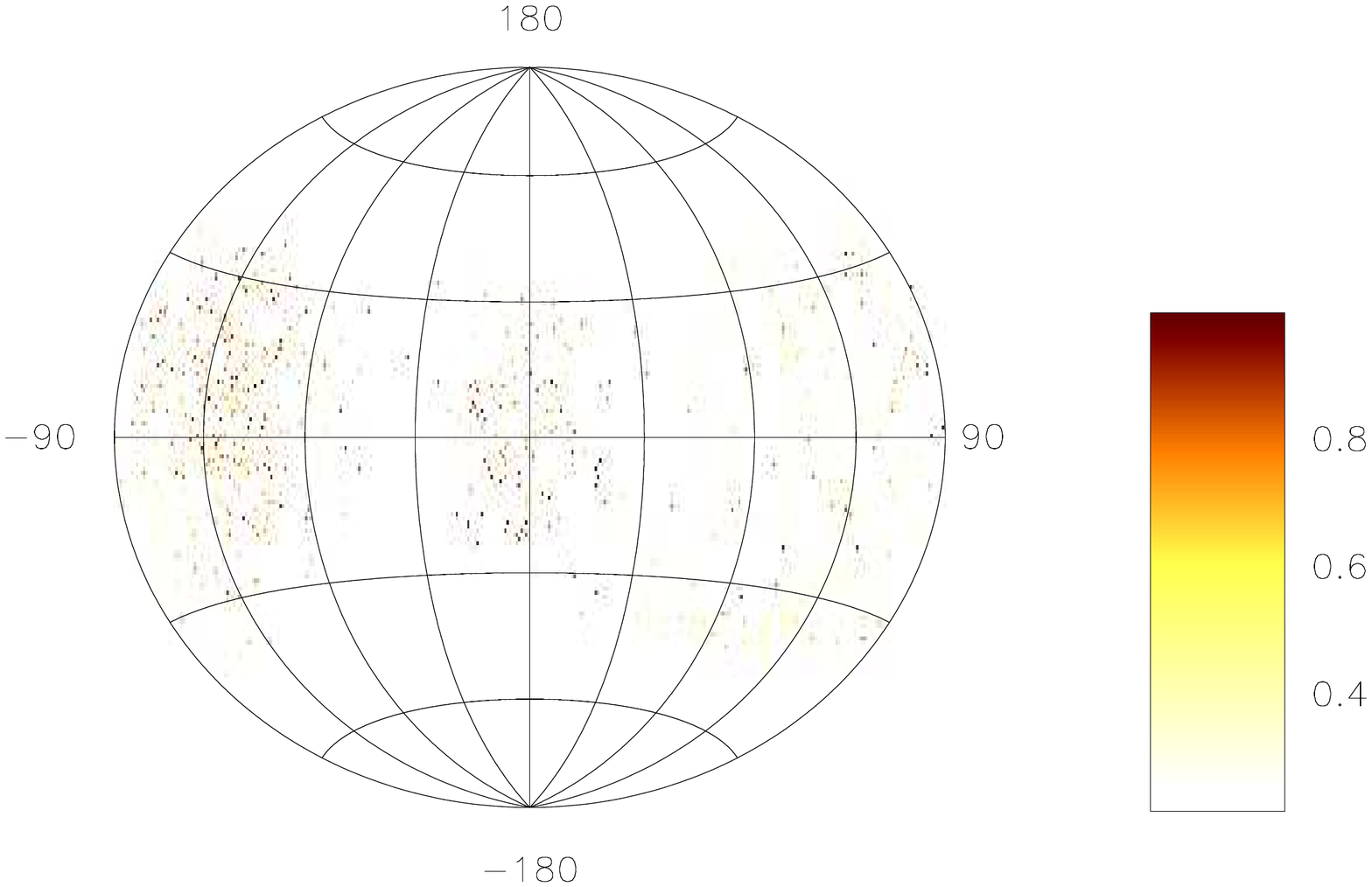}}}\\
\end{tabular}
\caption{Top panel only for 618 MHz (upper window) shows the average profile with total
intensity (Stokes I; solid black lines), total linear polarization (dashed red
line) and circular polarization (Stokes V; dotted blue line). Top panel (lower
window) also shows the single pulse PPA distribution (colour scale) along with
the average PPA (red error bars).
The RVM fits to the average PPA (dashed pink
line) is also shown in this plot. Middle panel only for 618 MHz show
the $\chi^2$ contours for the parameters $\alpha$ and $\beta$ obtained from RVM
fits.
Bottom panel only for 618 MHz shows the Hammer-Aitoff projection of the polarized time
samples with the colour scheme representing the fractional polarization level.}
\label{a26}
\end{center}
\end{figure*}


\begin{figure*}
\begin{center}
\begin{tabular}{cc}
{\mbox{\includegraphics[width=9cm,height=6cm,angle=0.]{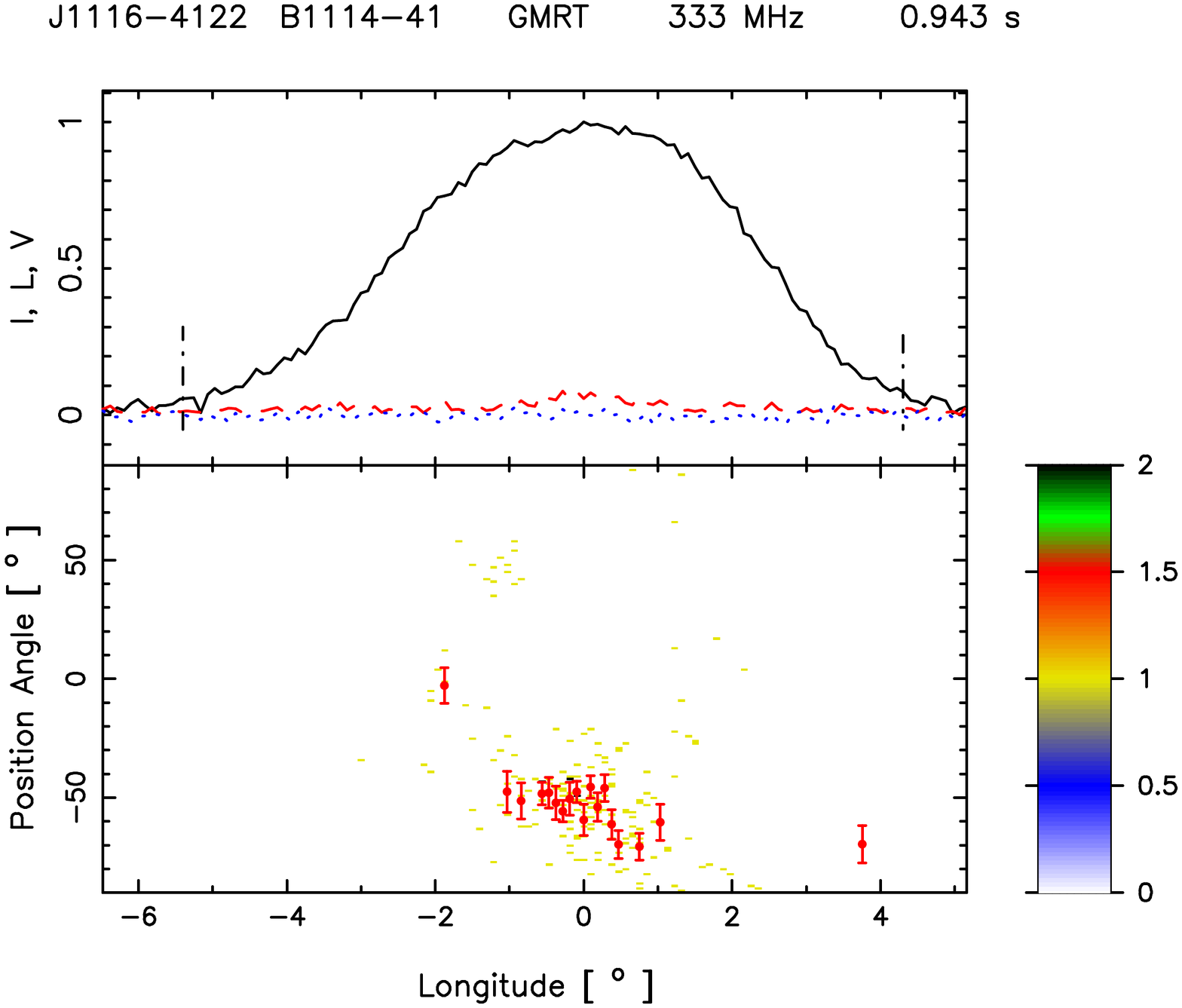}}}&
\\
&
\\
{\mbox{\includegraphics[width=9cm,height=6cm,angle=0.]{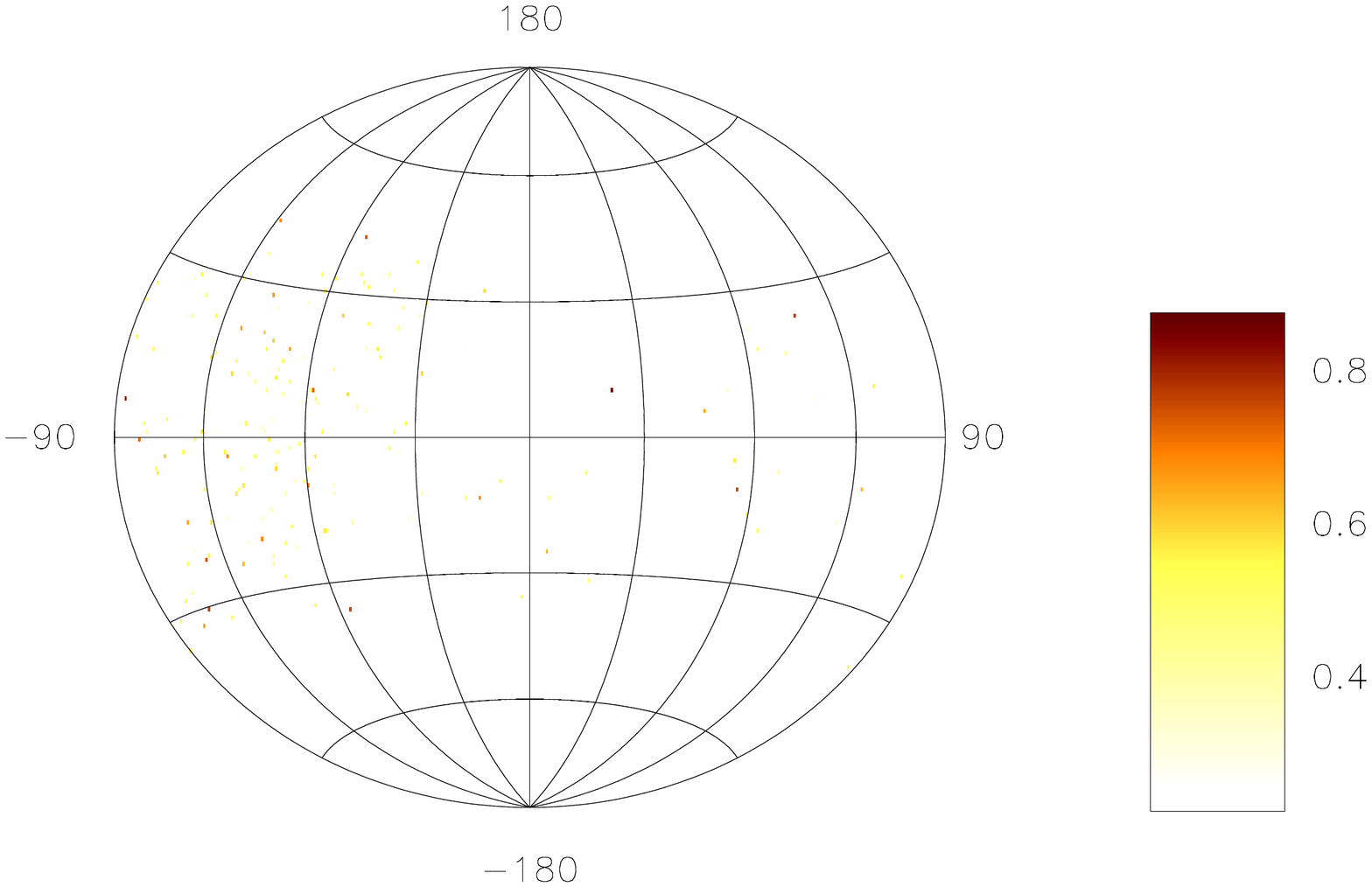}}}&
\\
\end{tabular}
\caption{Top panel only for 333 MHz (upper window) shows the average profile with total
intensity (Stokes I; solid black lines), total linear polarization (dashed red
line) and circular polarization (Stokes V; dotted blue line). Top panel (lower
window) also shows the single pulse PPA distribution (colour scale) along with
the average PPA (red error bars).
Bottom panel only for 333 MHz shows the Hammer-Aitoff projection of the polarized time
samples with the colour scheme representing the fractional polarization level.}
\label{a27}
\end{center}
\end{figure*}


\begin{figure*}
\begin{center}
\begin{tabular}{cc}
{\mbox{\includegraphics[width=9cm,height=6cm,angle=0.]{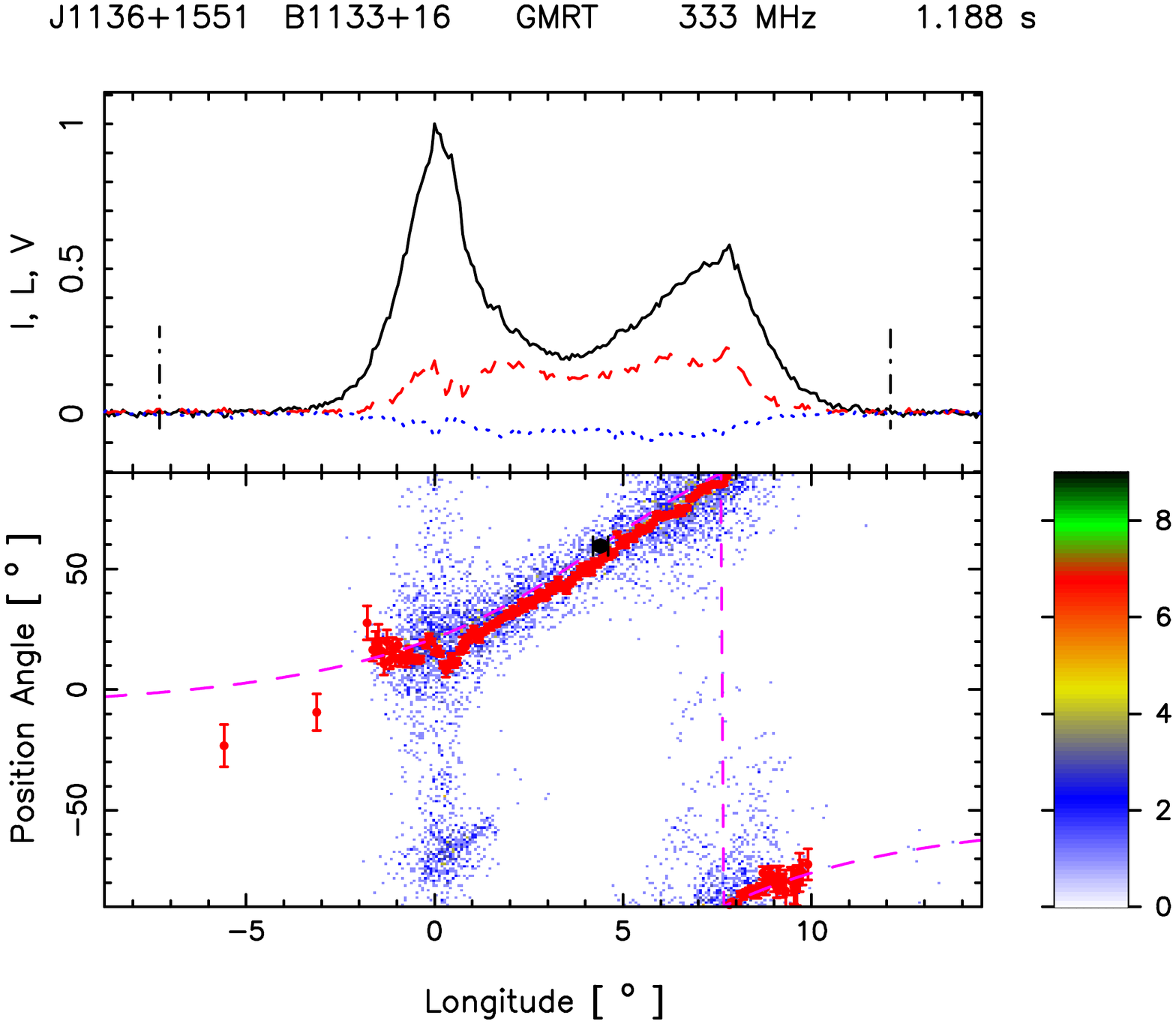}}}&
{\mbox{\includegraphics[width=9cm,height=6cm,angle=0.]{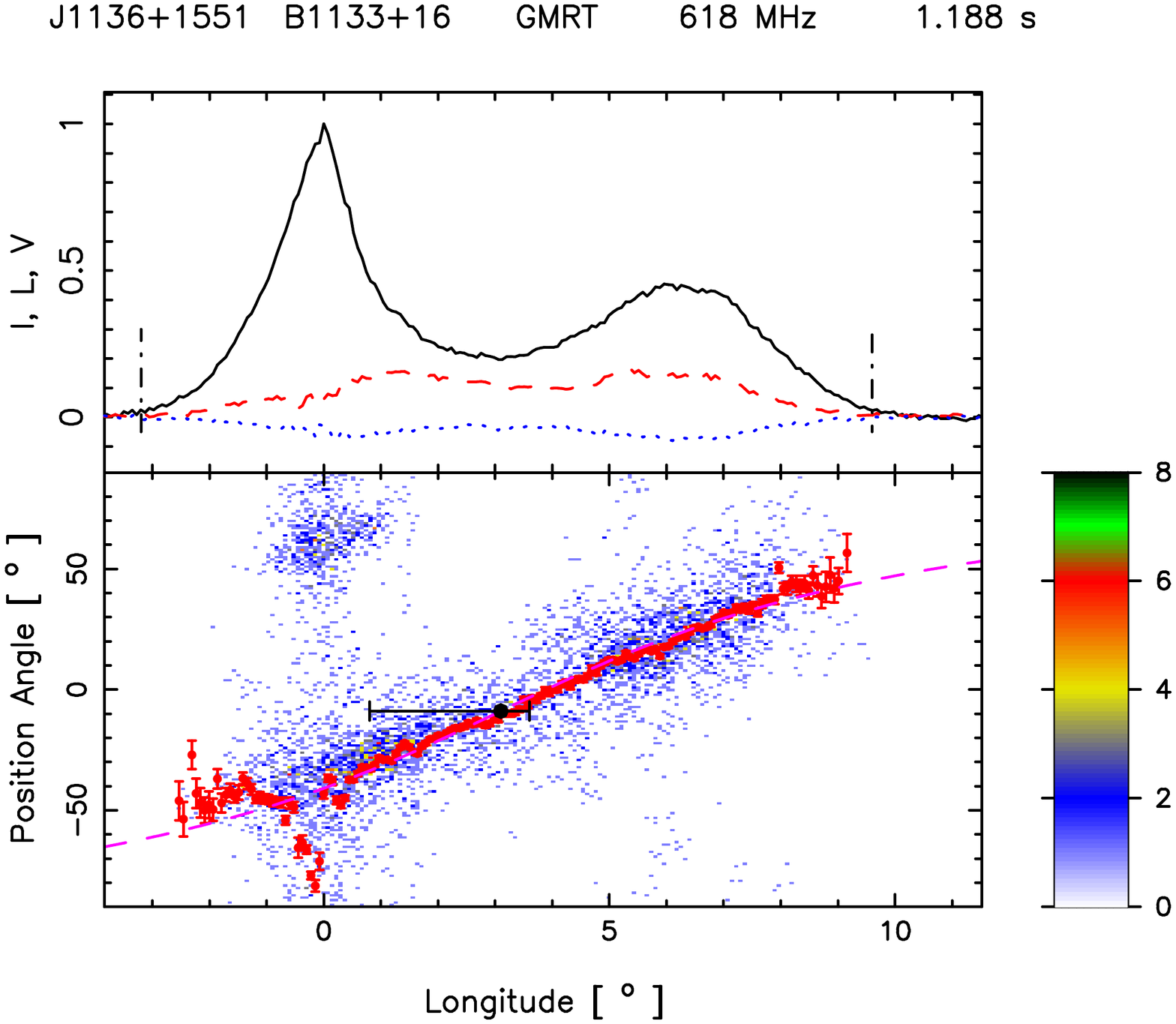}}}\\
{\mbox{\includegraphics[width=9cm,height=6cm,angle=0.]{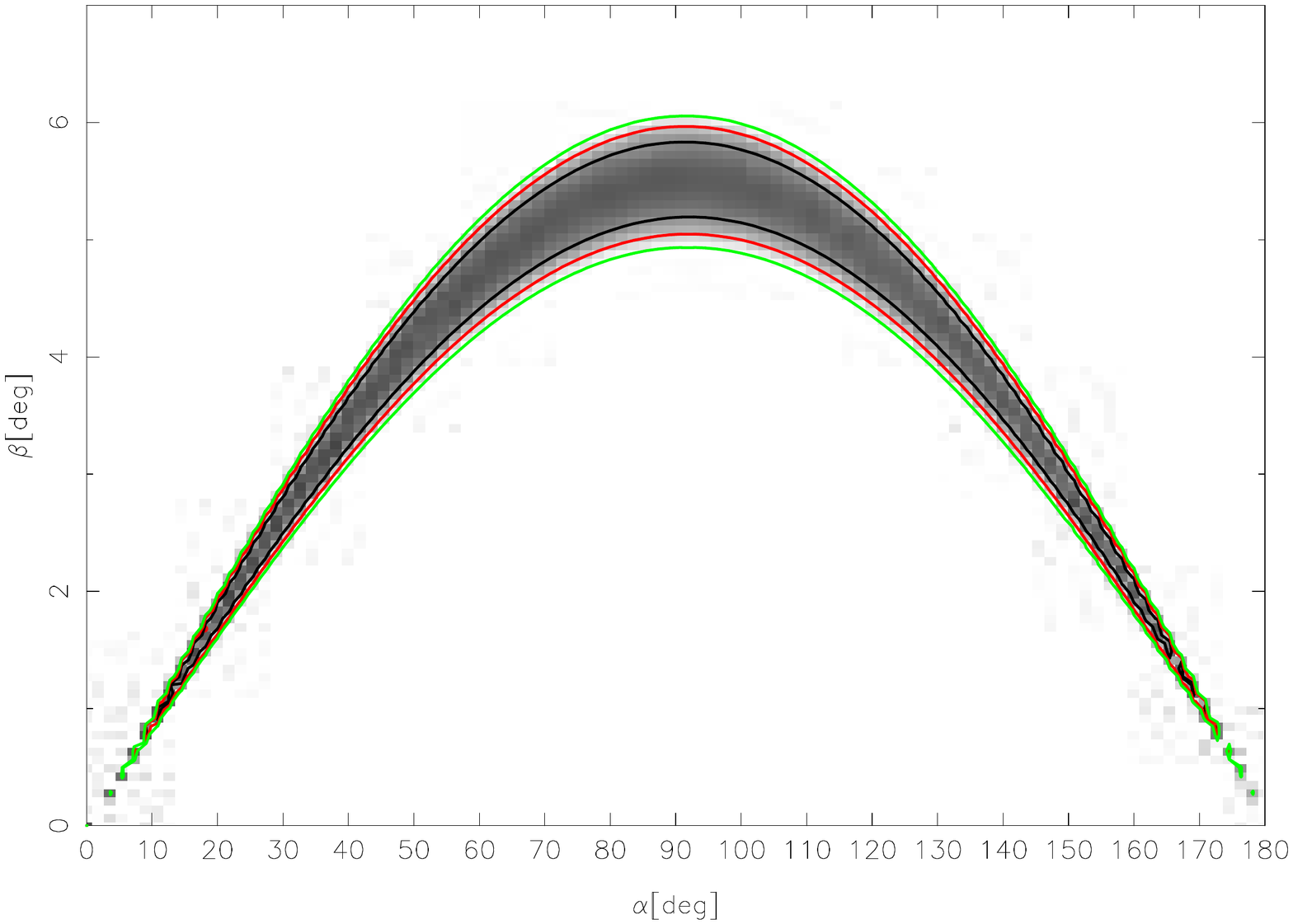}}}&
{\mbox{\includegraphics[width=9cm,height=6cm,angle=0.]{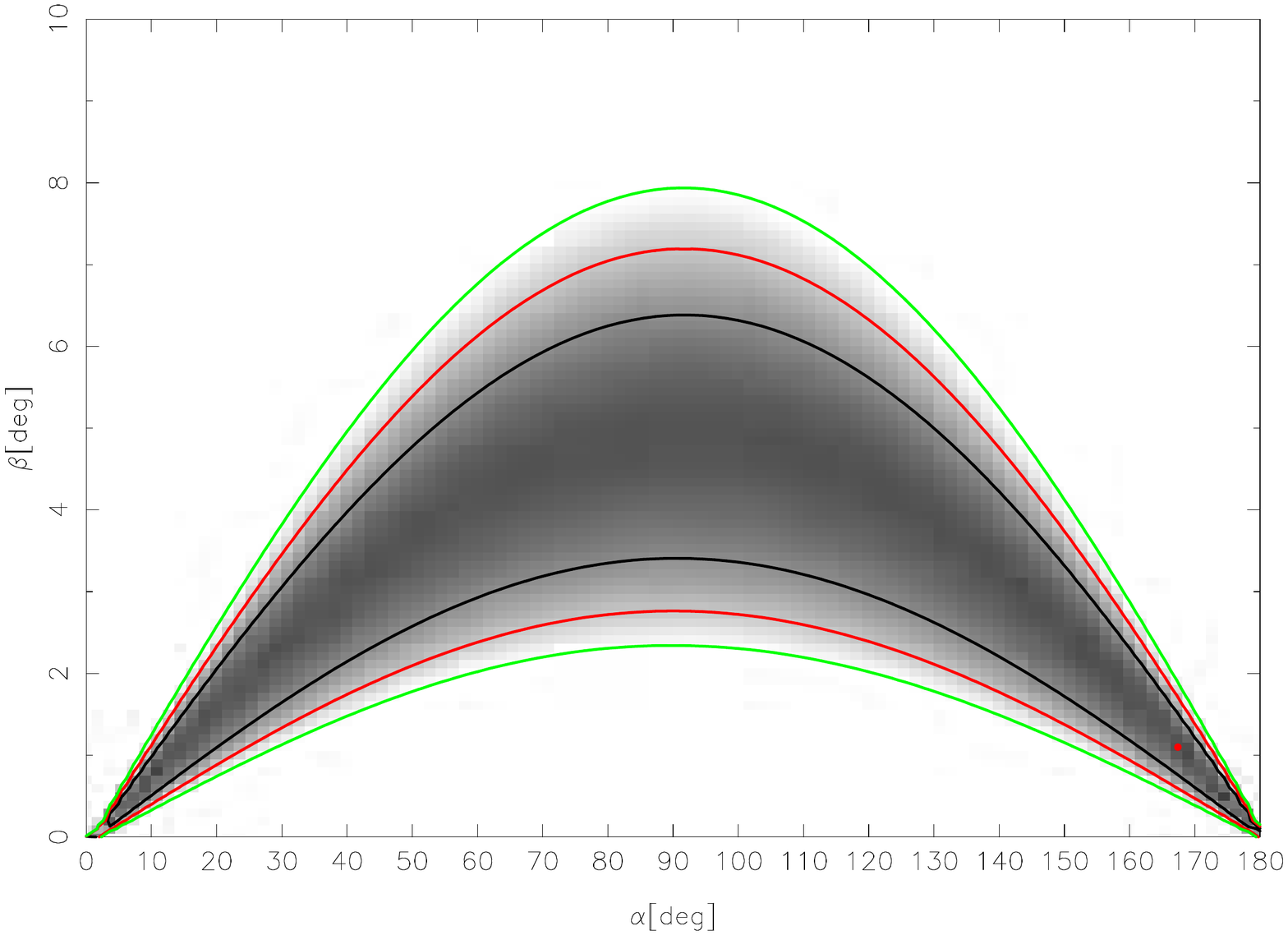}}}\\
{\mbox{\includegraphics[width=9cm,height=6cm,angle=0.]{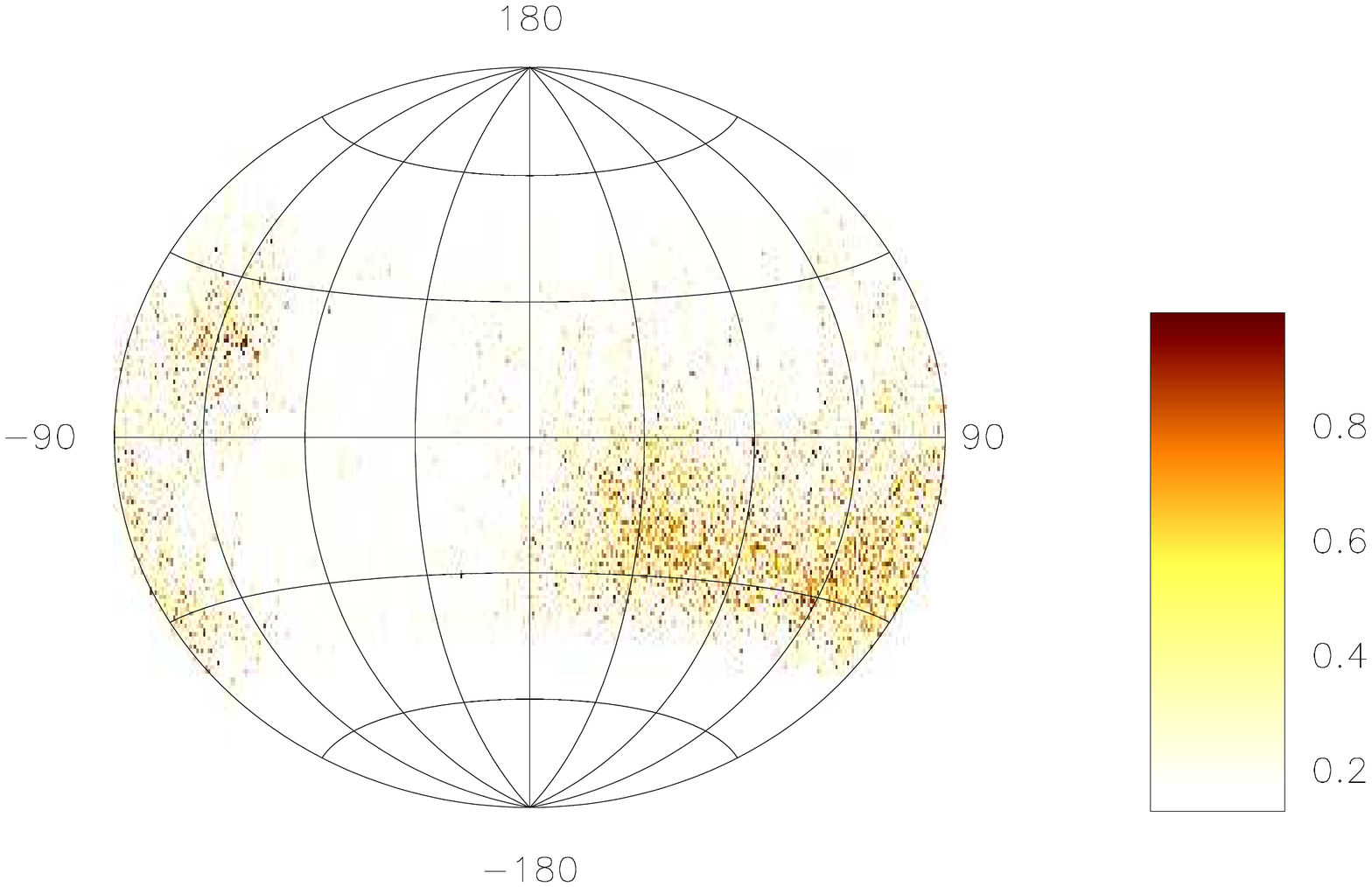}}}&
{\mbox{\includegraphics[width=9cm,height=6cm,angle=0.]{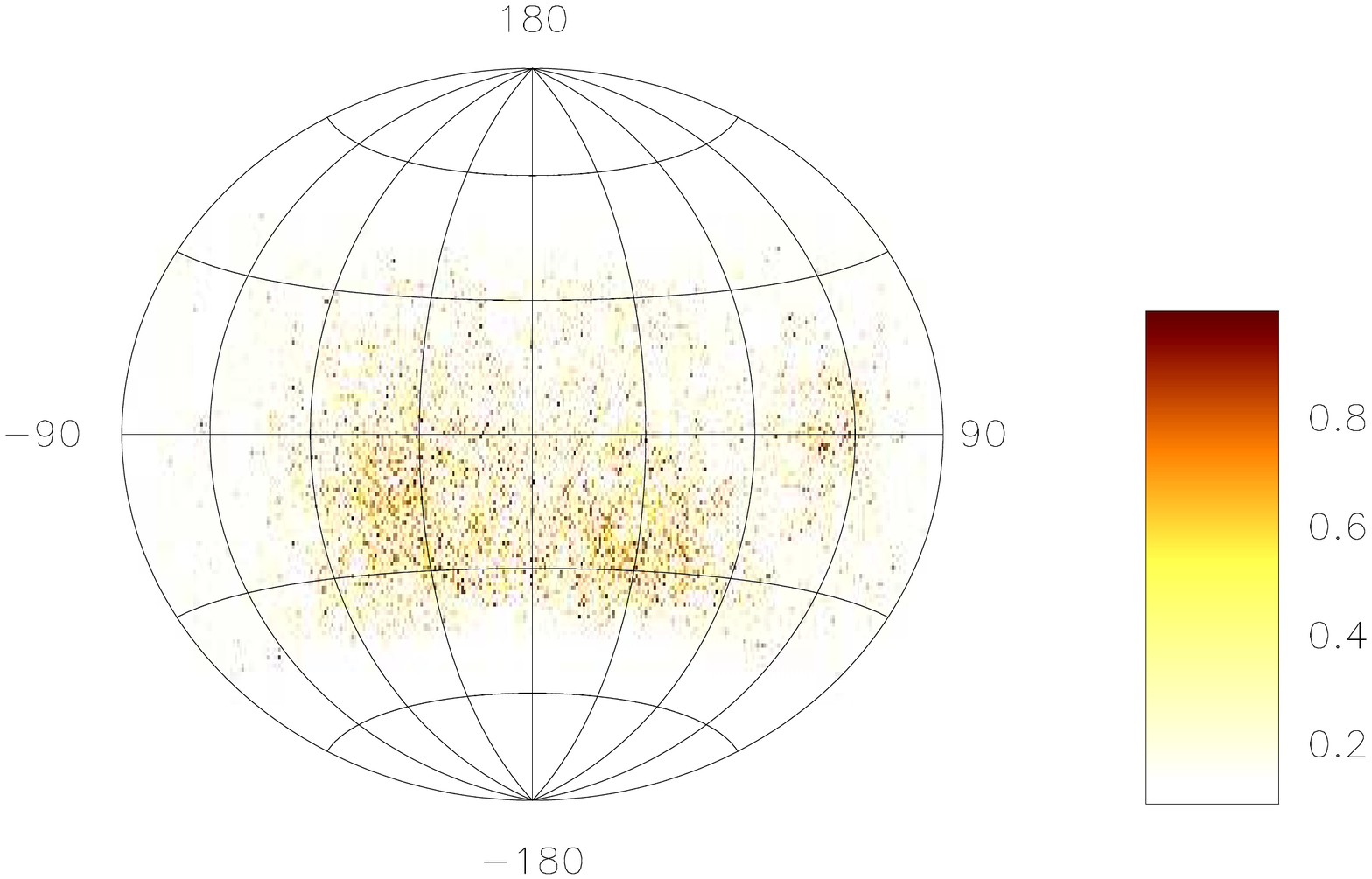}}}\\
\end{tabular}
\caption{Top panel (upper window) shows the average profile with total
intensity (Stokes I; solid black lines), total linear polarization (dashed red
line) and circular polarization (Stokes V; dotted blue line). Top panel (lower
window) also shows the single pulse PPA distribution (colour scale) along with
the average PPA (red error bars).
The RVM fits to the average PPA (dashed pink
line) is also shown in this plot. Middle panel show
the $\chi^2$ contours for the parameters $\alpha$ and $\beta$ obtained from RVM
fits.
Bottom panel shows the Hammer-Aitoff projection of the polarized time
samples with the colour scheme representing the fractional polarization level.}
\label{a28}
\end{center}
\end{figure*}


\begin{figure*}
\begin{center}
\begin{tabular}{cc}
{\mbox{\includegraphics[width=9cm,height=6cm,angle=0.]{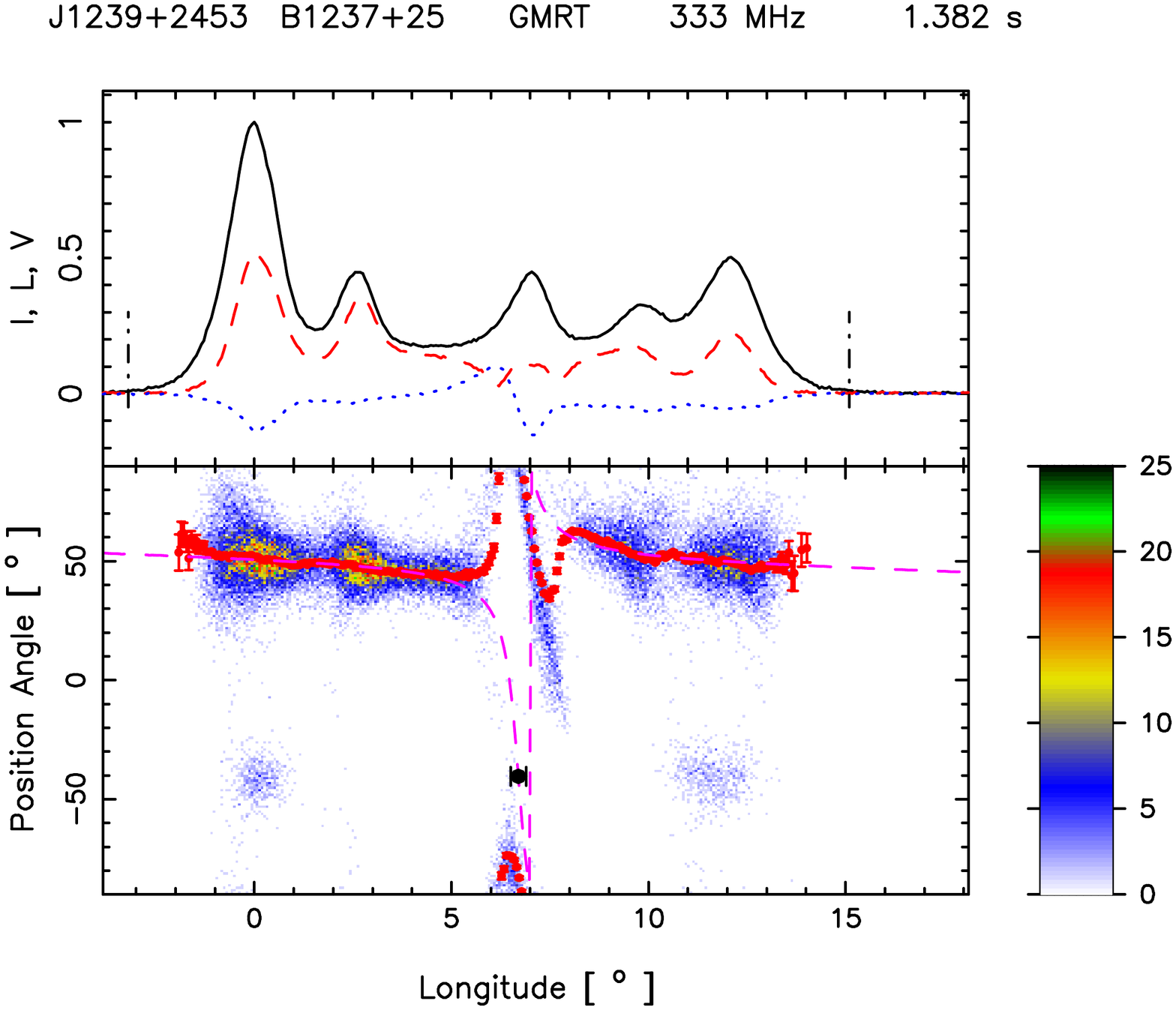}}}&
{\mbox{\includegraphics[width=9cm,height=6cm,angle=0.]{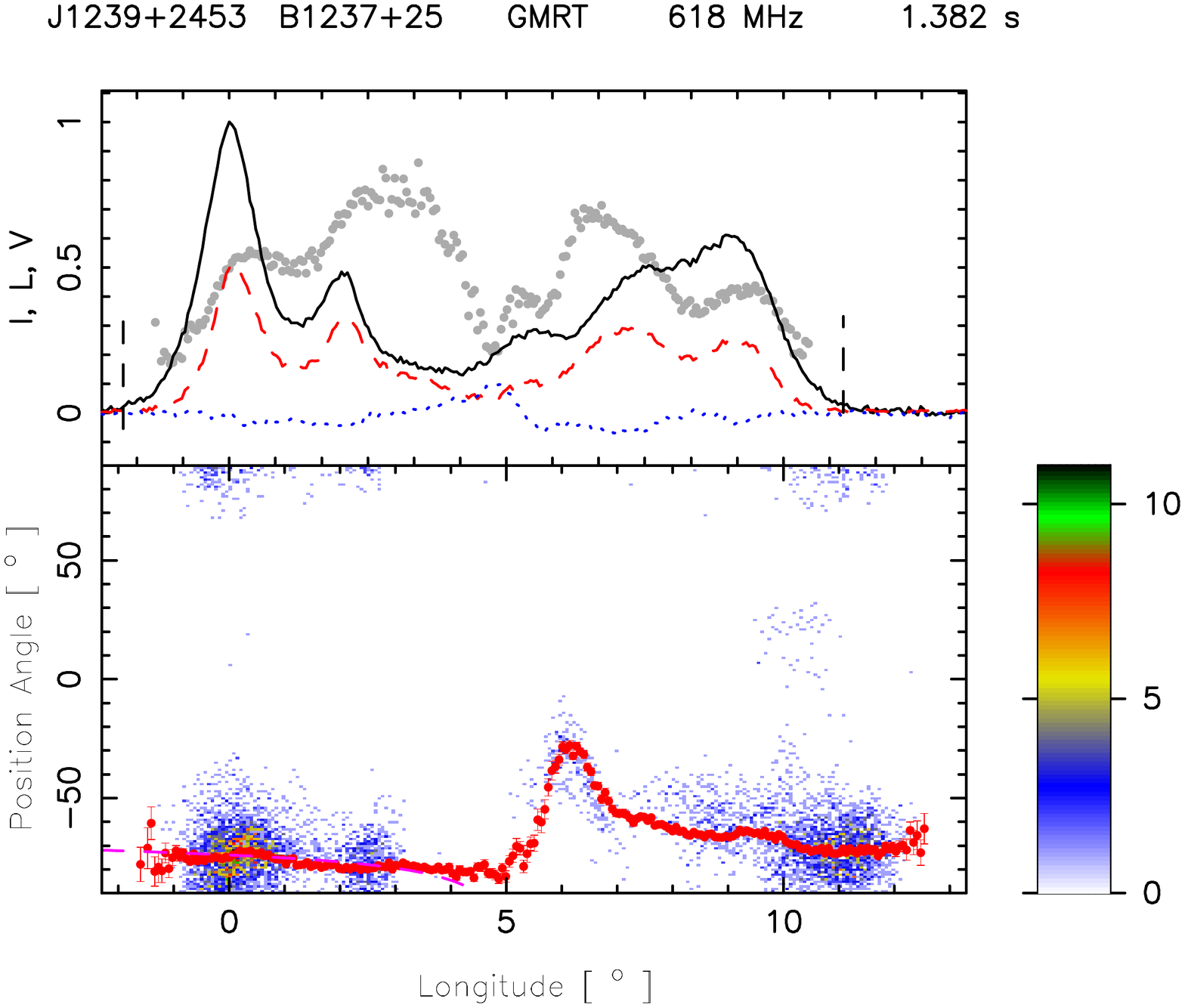}}}\\
{\mbox{\includegraphics[width=9cm,height=6cm,angle=0.]{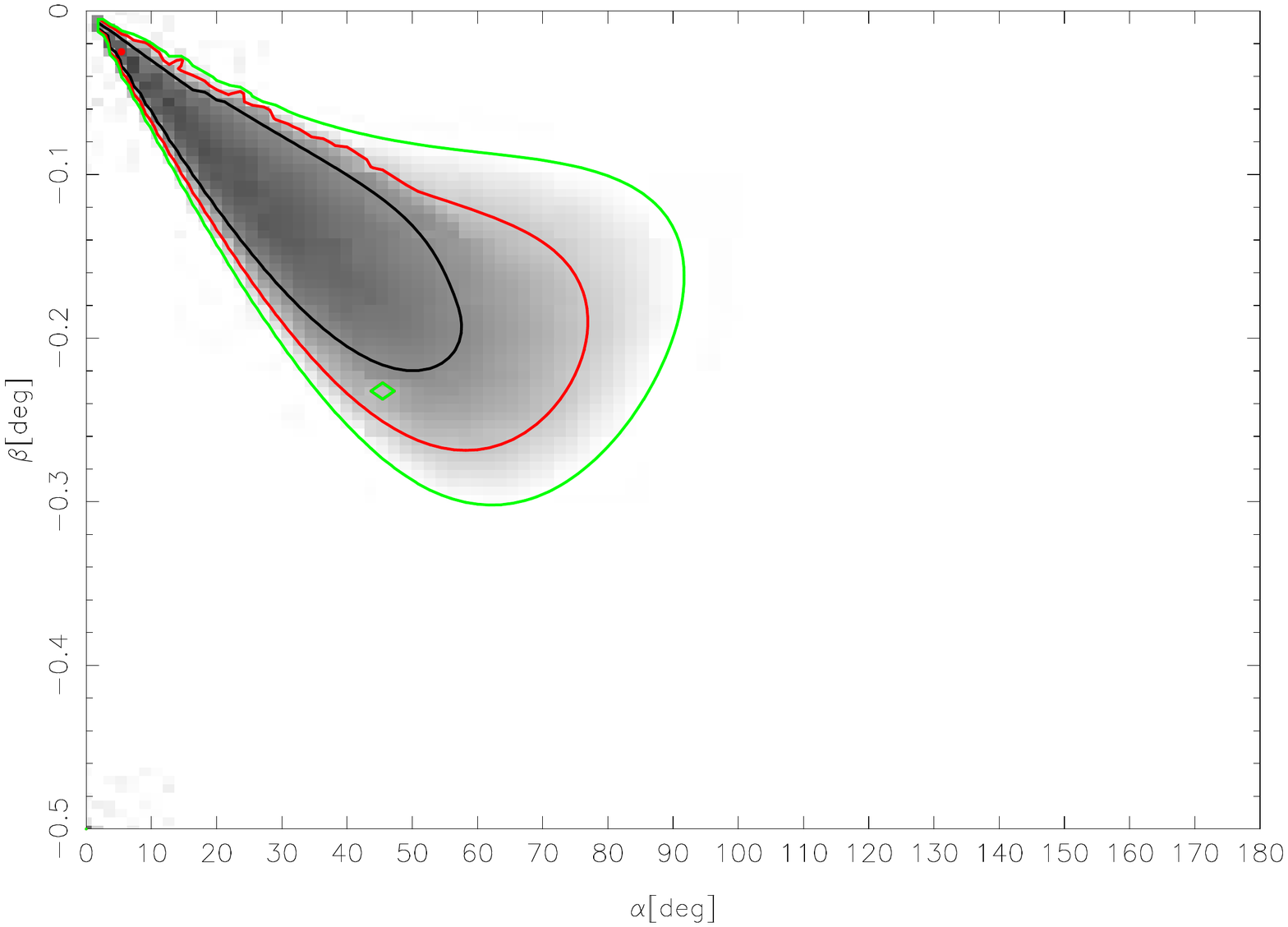}}}&
{\mbox{\includegraphics[width=9cm,height=6cm,angle=0.]{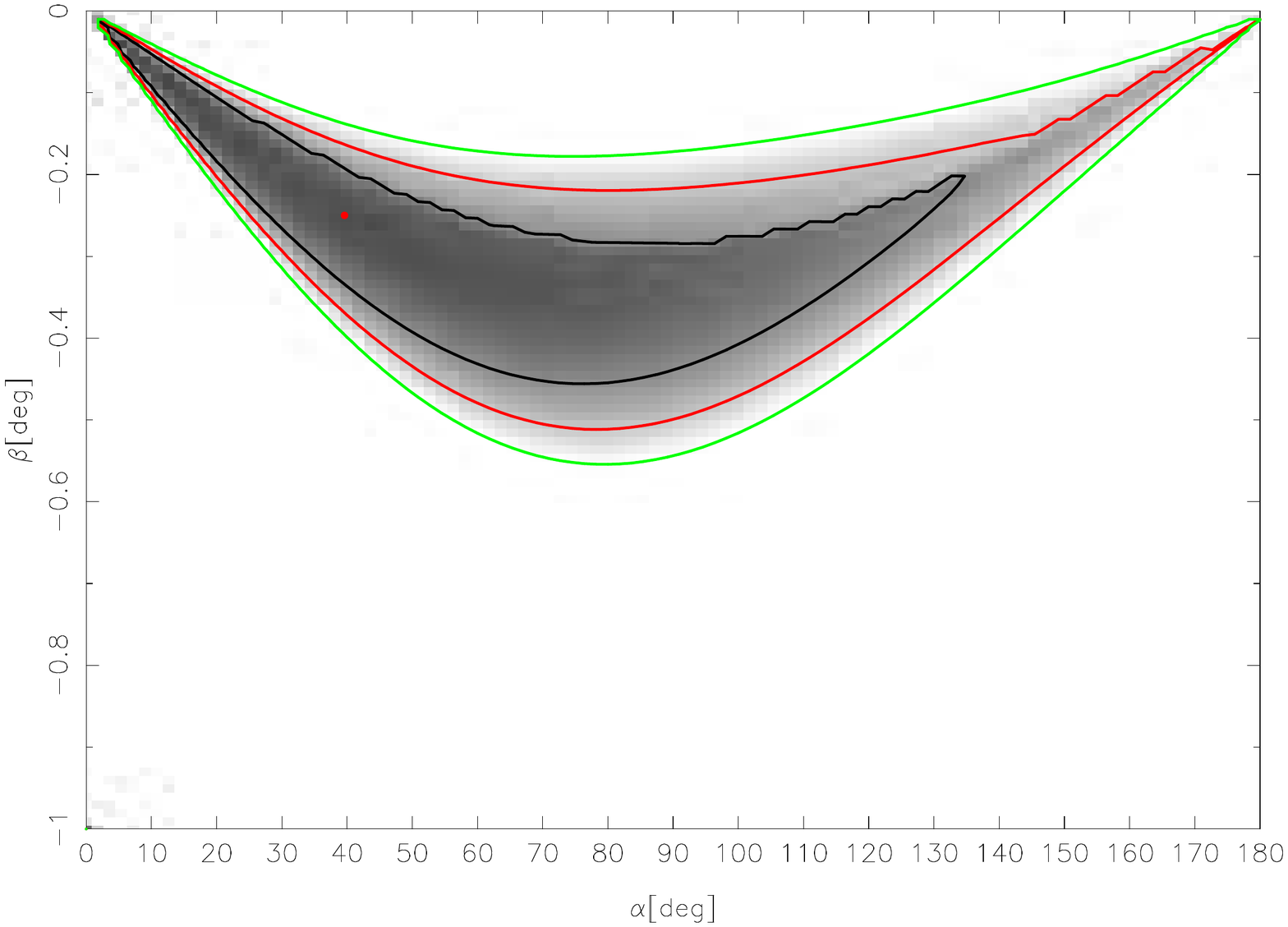}}}\\
{\mbox{\includegraphics[width=9cm,height=6cm,angle=0.]{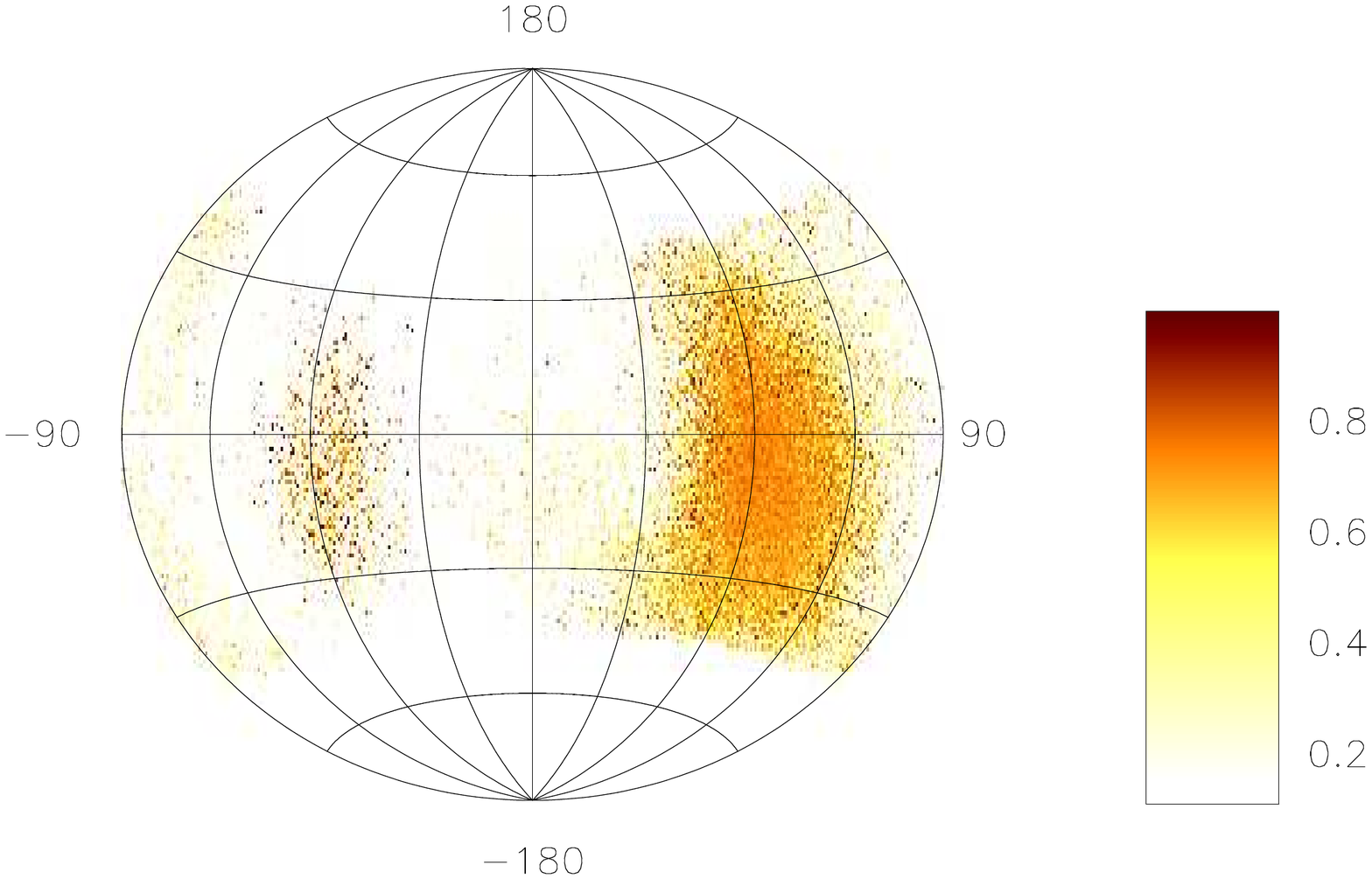}}}&
{\mbox{\includegraphics[width=9cm,height=6cm,angle=0.]{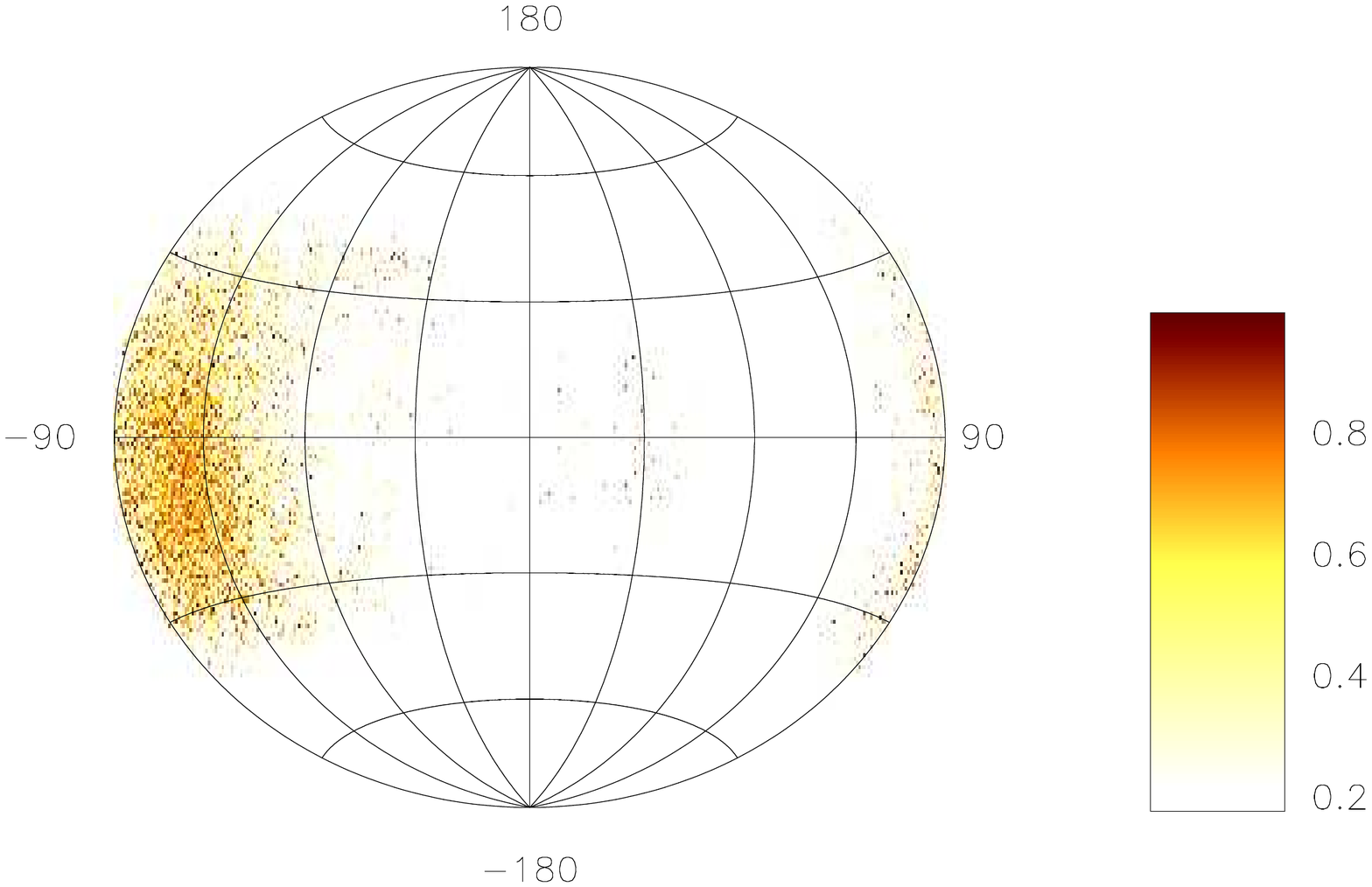}}}\\
\end{tabular}
\caption{Top panel (upper window) shows the average profile with total
intensity (Stokes I; solid black lines), total linear polarization (dashed red
line) and circular polarization (Stokes V; dotted blue line). Top panel (lower
window) also shows the single pulse PPA distribution (colour scale) along with
the average PPA (red error bars).
The RVM fits to the average PPA (dashed pink
line) is also shown in this plot. Middle panel show
the $\chi^2$ contours for the parameters $\alpha$ and $\beta$ obtained from RVM
fits.
Bottom panel shows the Hammer-Aitoff projection of the polarized time
samples with the colour scheme representing the fractional polarization level.}
\label{a29}
\end{center}
\end{figure*}


\begin{figure*}
\begin{center}
\begin{tabular}{cc}
&
{\mbox{\includegraphics[width=9cm,height=6cm,angle=0.]{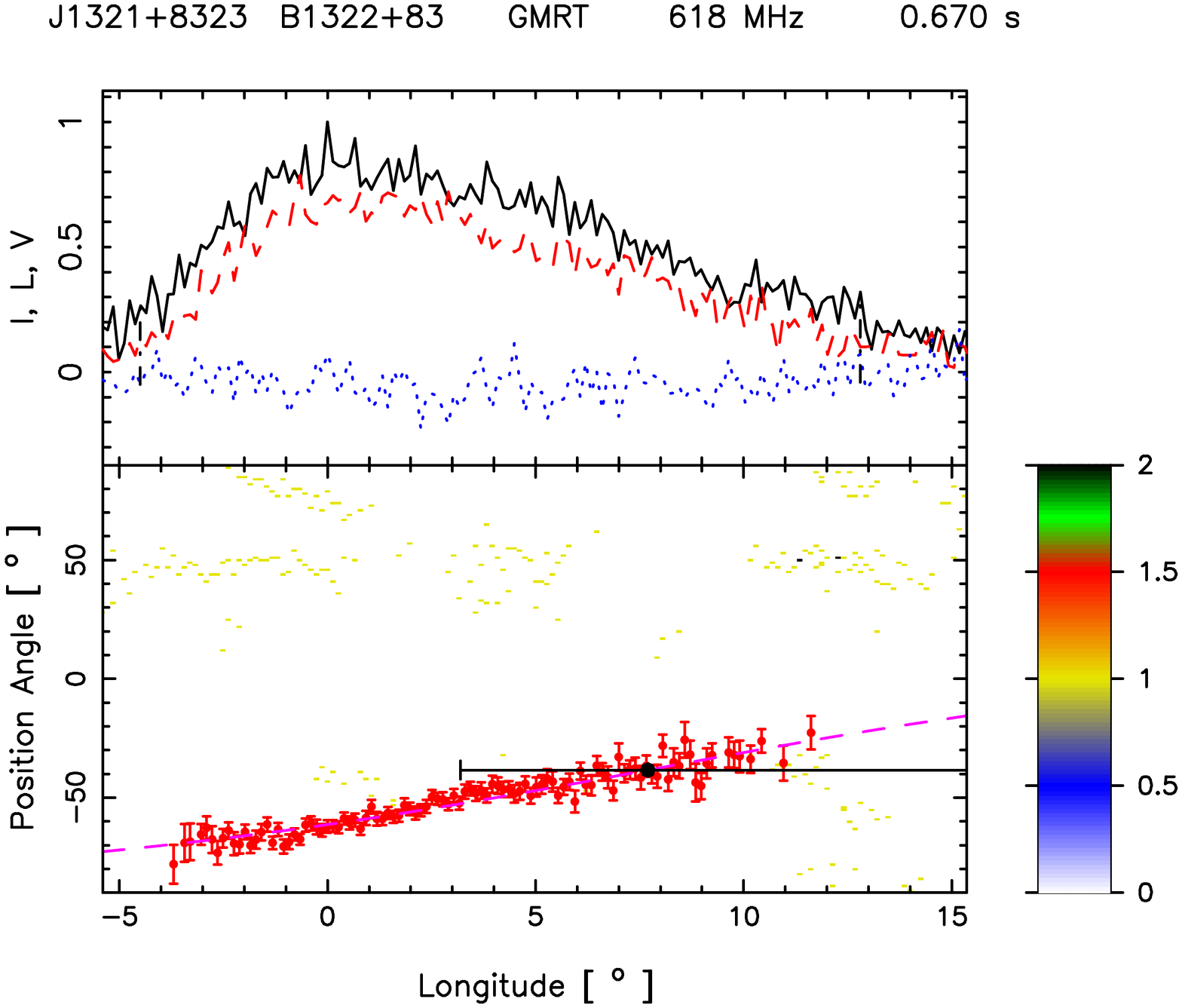}}}\\
&
{\mbox{\includegraphics[width=9cm,height=6cm,angle=0.]{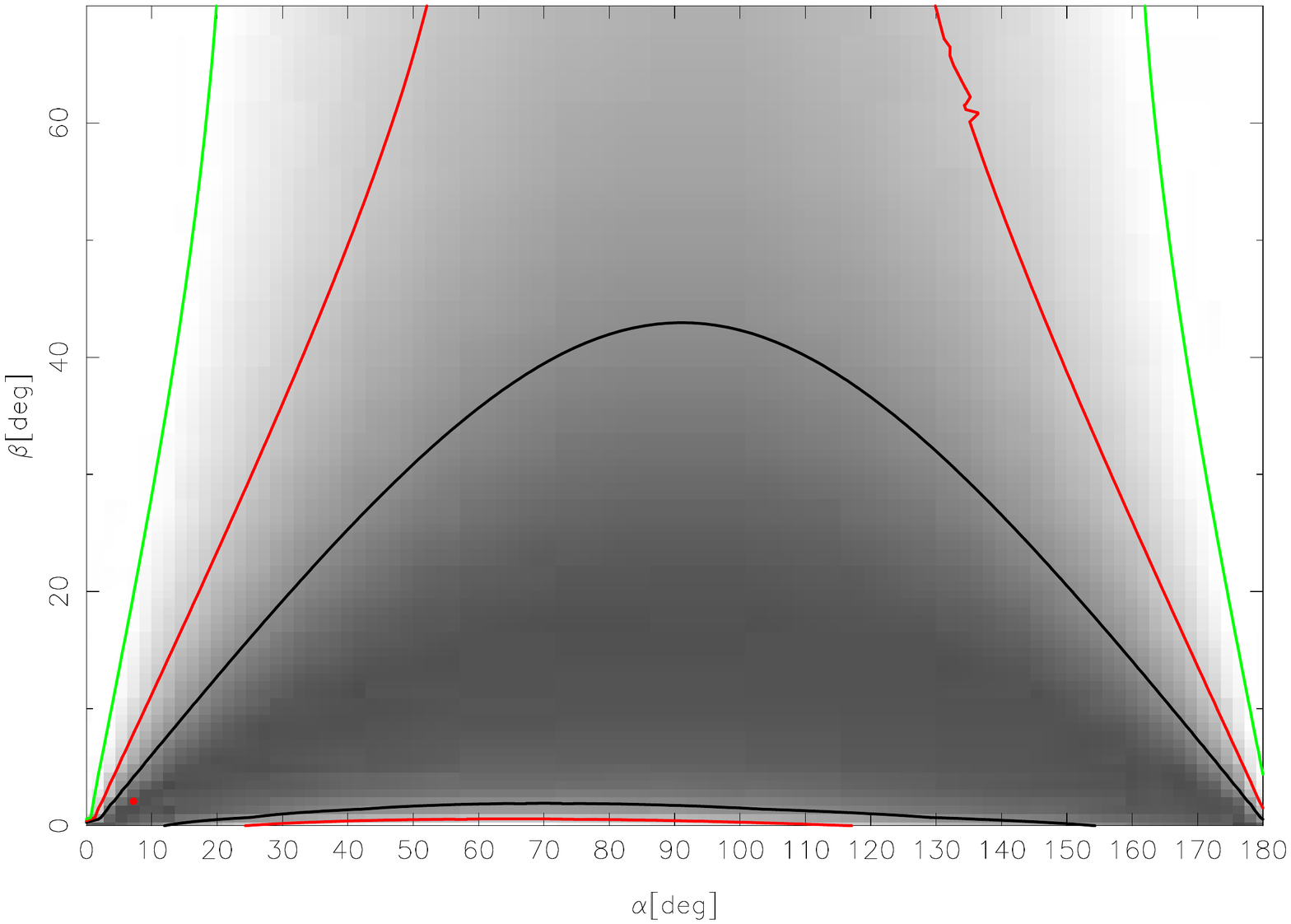}}}\\
&
\\
\end{tabular}
\caption{Top panel only for 618 MHz (upper window) shows the average profile with total
intensity (Stokes I; solid black lines), total linear polarization (dashed red
line) and circular polarization (Stokes V; dotted blue line). Top panel (lower
window) also shows the single pulse PPA distribution (colour scale) along with
the average PPA (red error bars).
The RVM fits to the average PPA (dashed pink
line) is also shown in this plot. Bottom panel only for 618 MHz show
the $\chi^2$ contours for the parameters $\alpha$ and $\beta$ obtained from RVM
fits.}
\label{a30}
\end{center}
\end{figure*}

\begin{figure*}
\begin{center}
\begin{tabular}{cc}
{\mbox{\includegraphics[width=9cm,height=6cm,angle=0.]{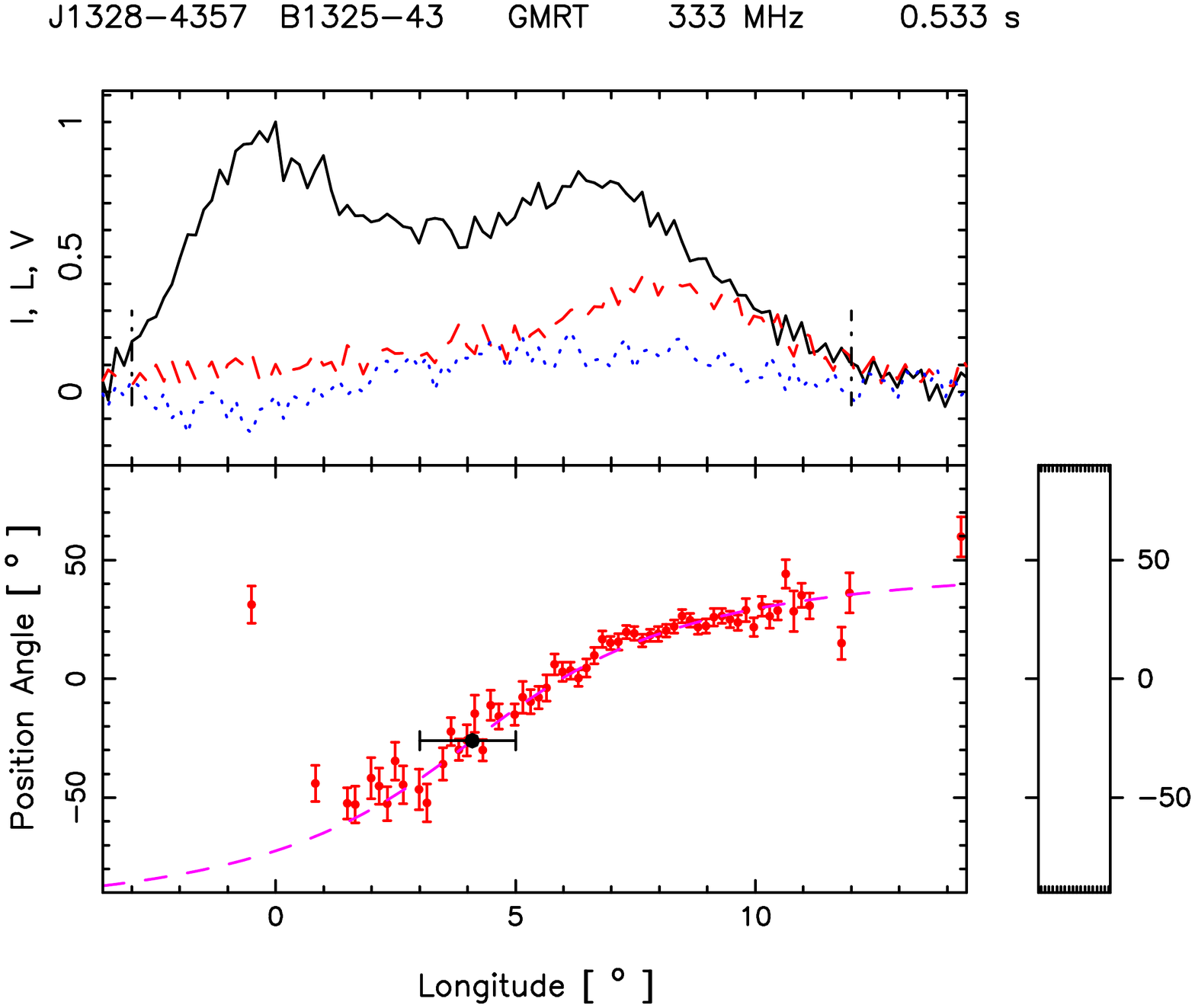}}}&
{\mbox{\includegraphics[width=9cm,height=6cm,angle=0.]{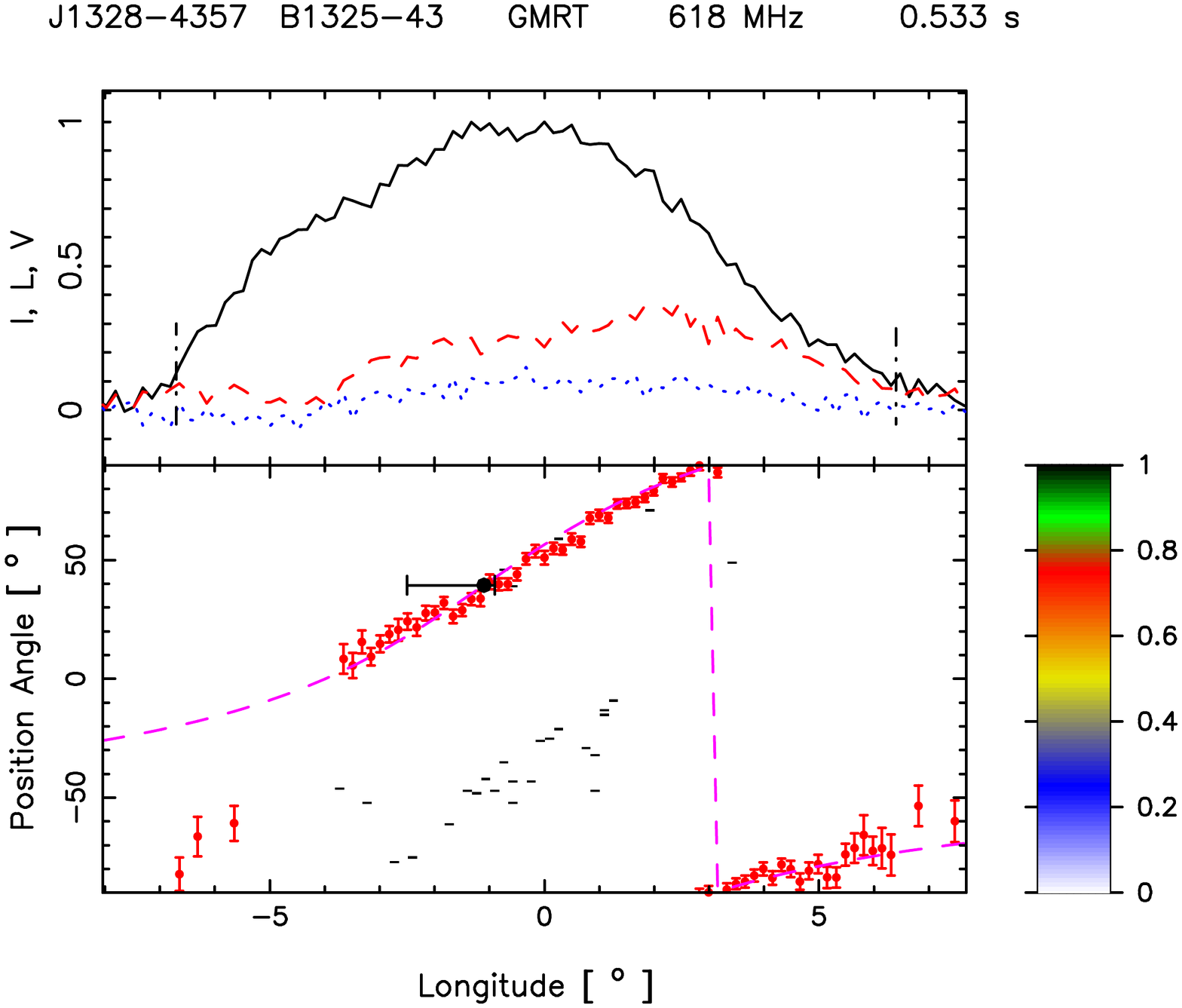}}}\\
{\mbox{\includegraphics[width=9cm,height=6cm,angle=0.]{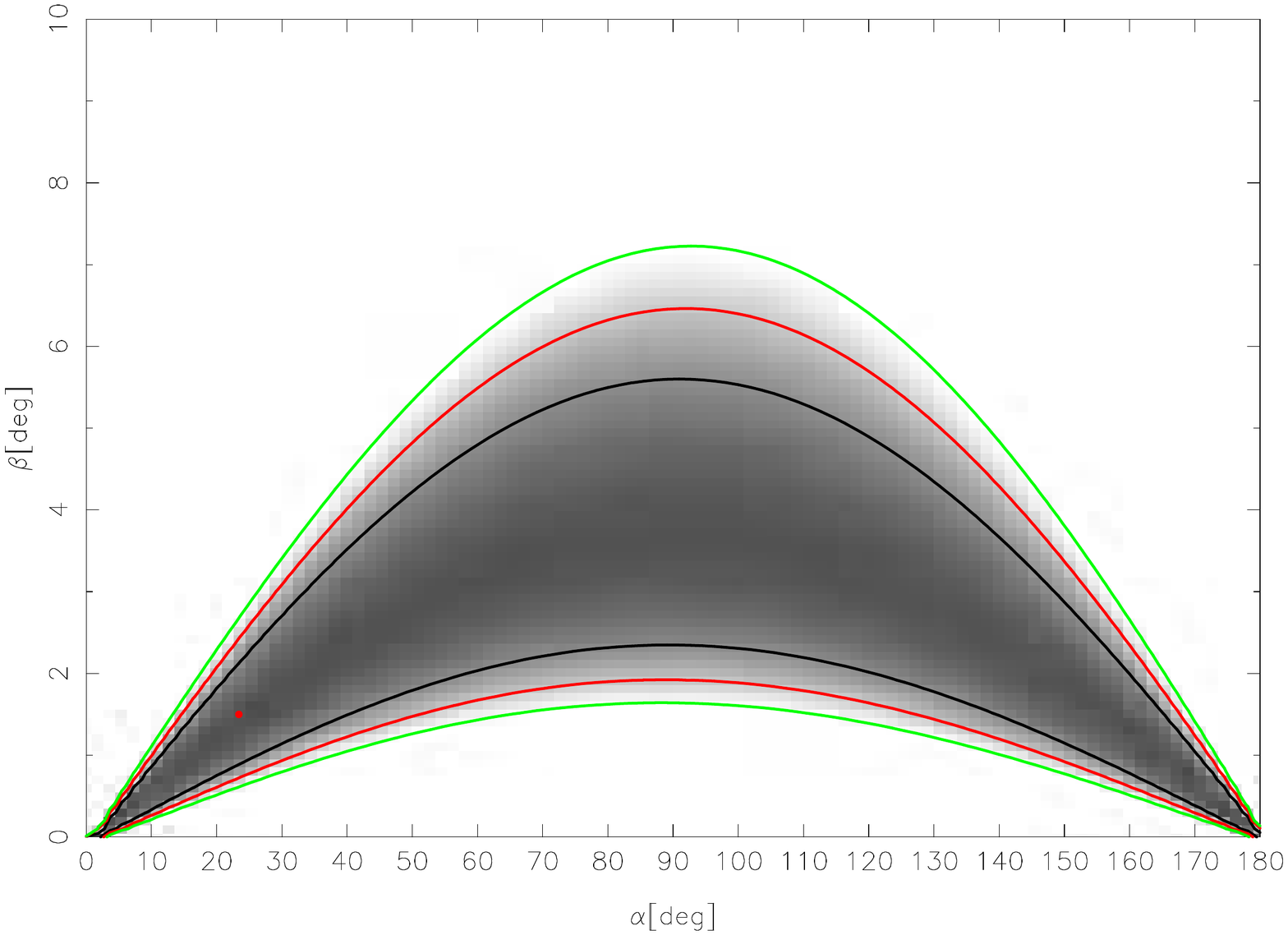}}}&
{\mbox{\includegraphics[width=9cm,height=6cm,angle=0.]{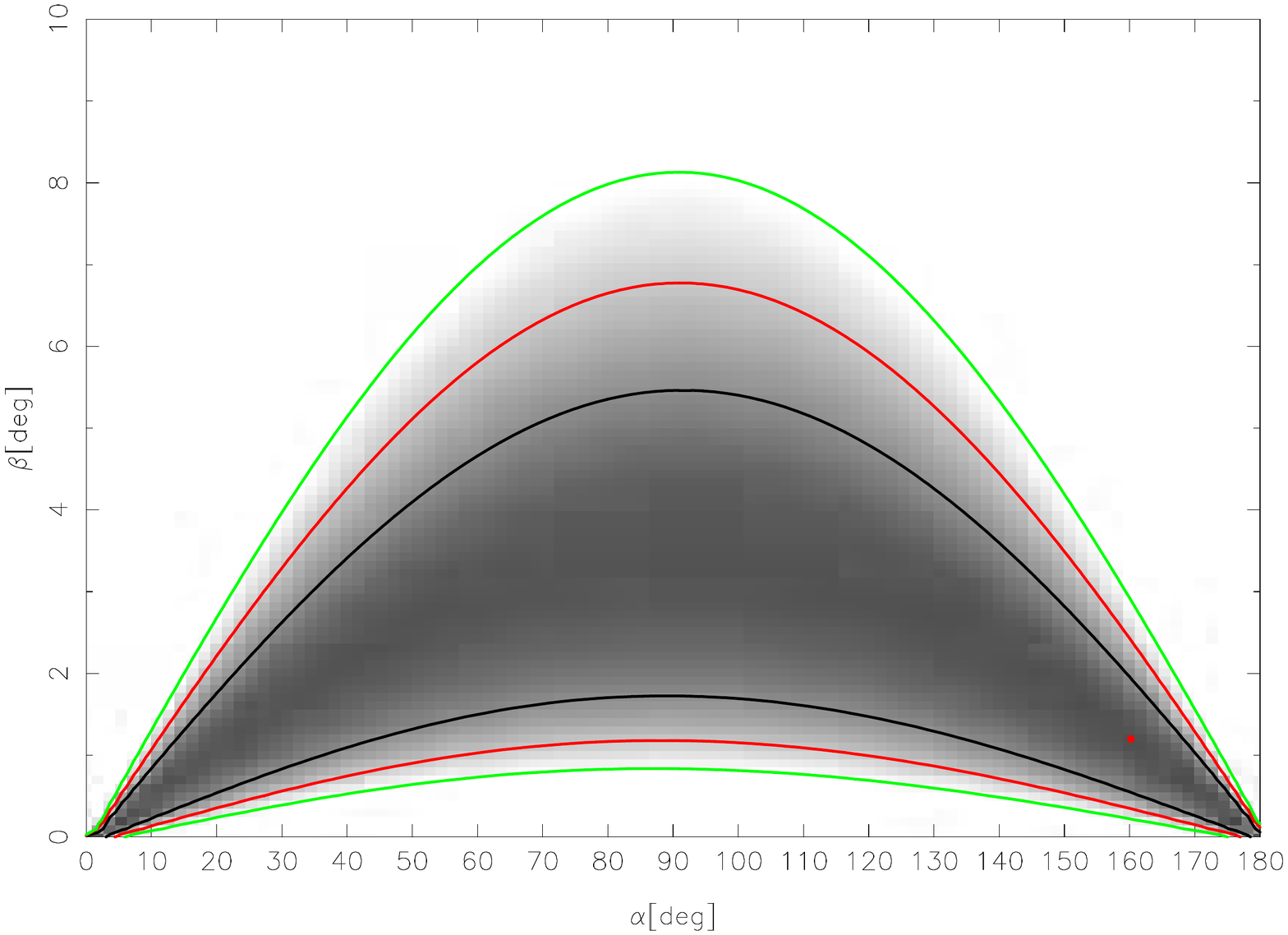}}}\\
&
\\
\end{tabular}
\caption{Top panel (upper window) shows the average profile with total
intensity (Stokes I; solid black lines), total linear polarization (dashed red
line) and circular polarization (Stokes V; dotted blue line). Top panel (lower
window) also shows the single pulse PPA distribution (colour scale) along with
the average PPA (red error bars).
The RVM fits to the average PPA (dashed pink
line) is also shown in this plot. Bottom panel show
the $\chi^2$ contours for the parameters $\alpha$ and $\beta$ obtained from RVM
fits.}
\label{a31}
\end{center}
\end{figure*}


\begin{figure*}
\begin{center}
\begin{tabular}{cc}
{\mbox{\includegraphics[width=9cm,height=6cm,angle=0.]{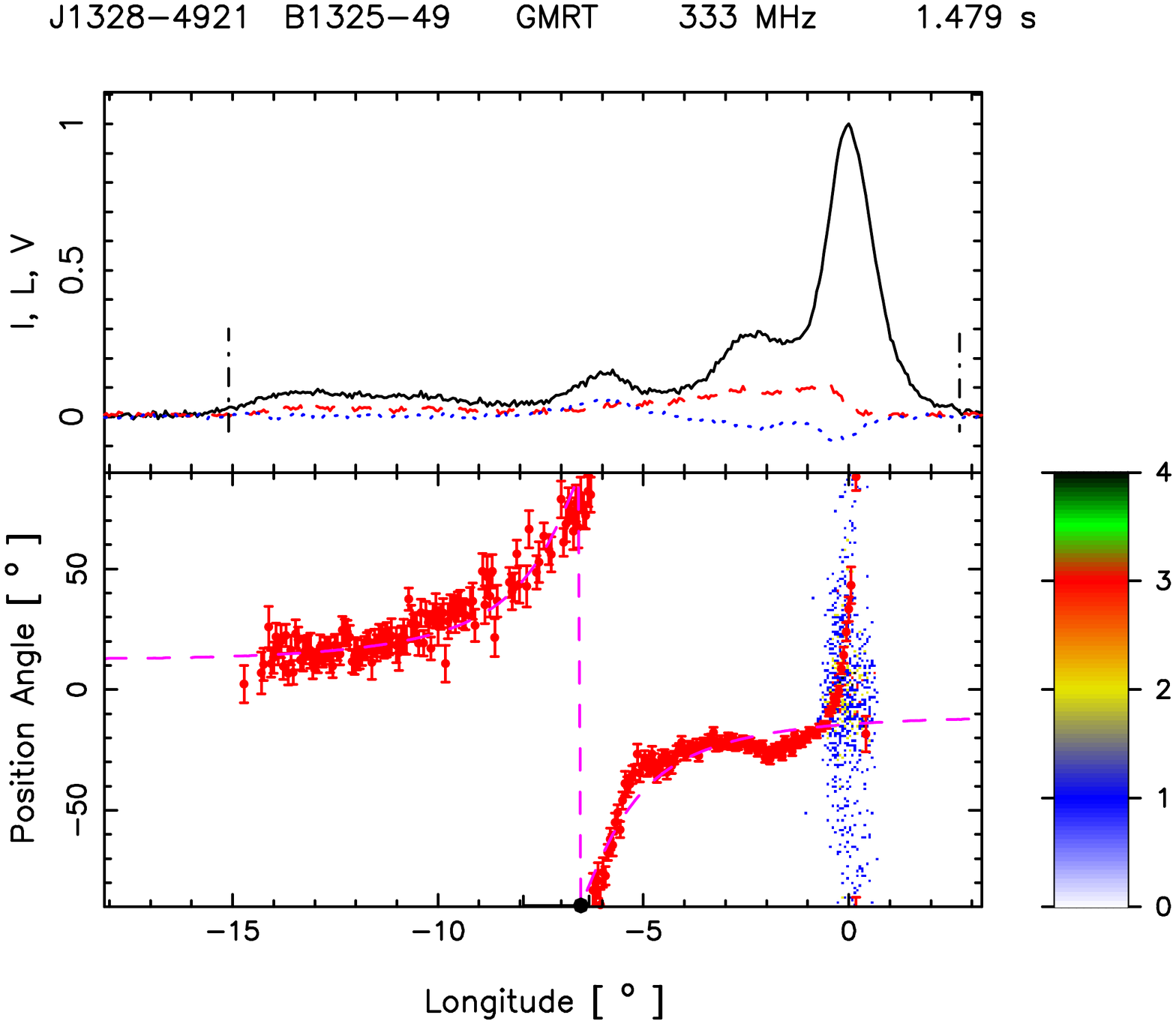}}}&
{\mbox{\includegraphics[width=9cm,height=6cm,angle=0.]{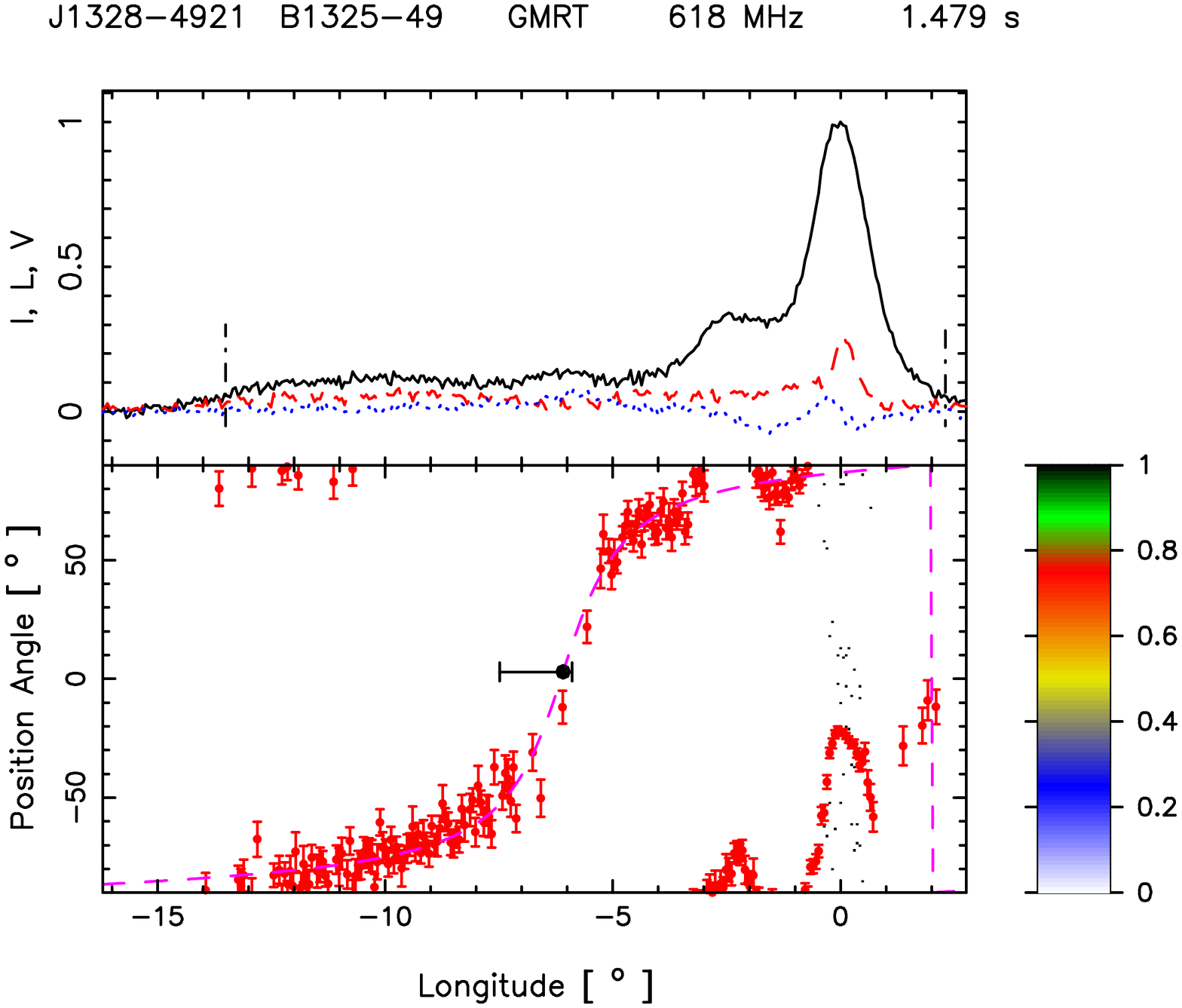}}}\\
{\mbox{\includegraphics[width=9cm,height=6cm,angle=0.]{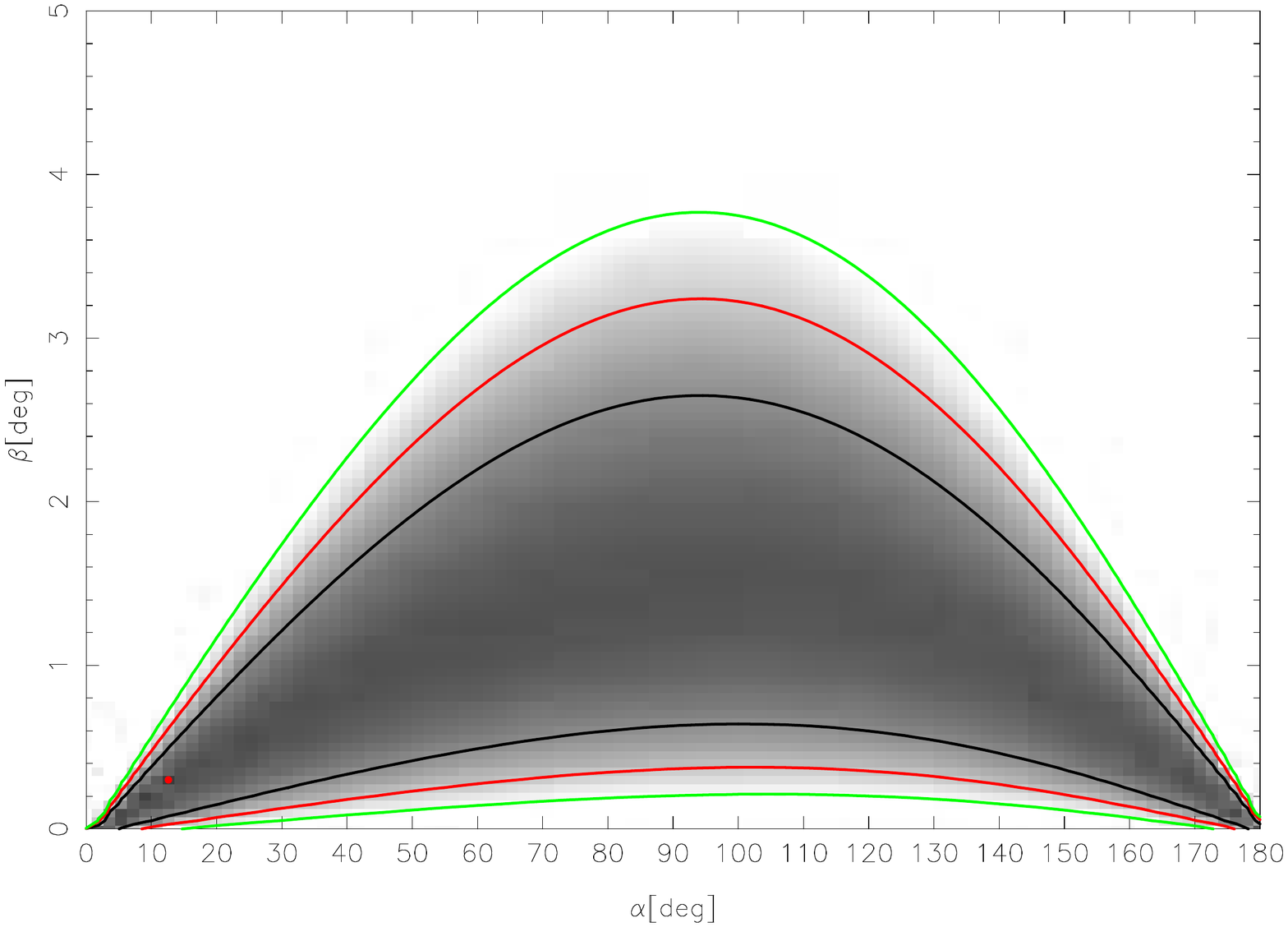}}}&
{\mbox{\includegraphics[width=9cm,height=6cm,angle=0.]{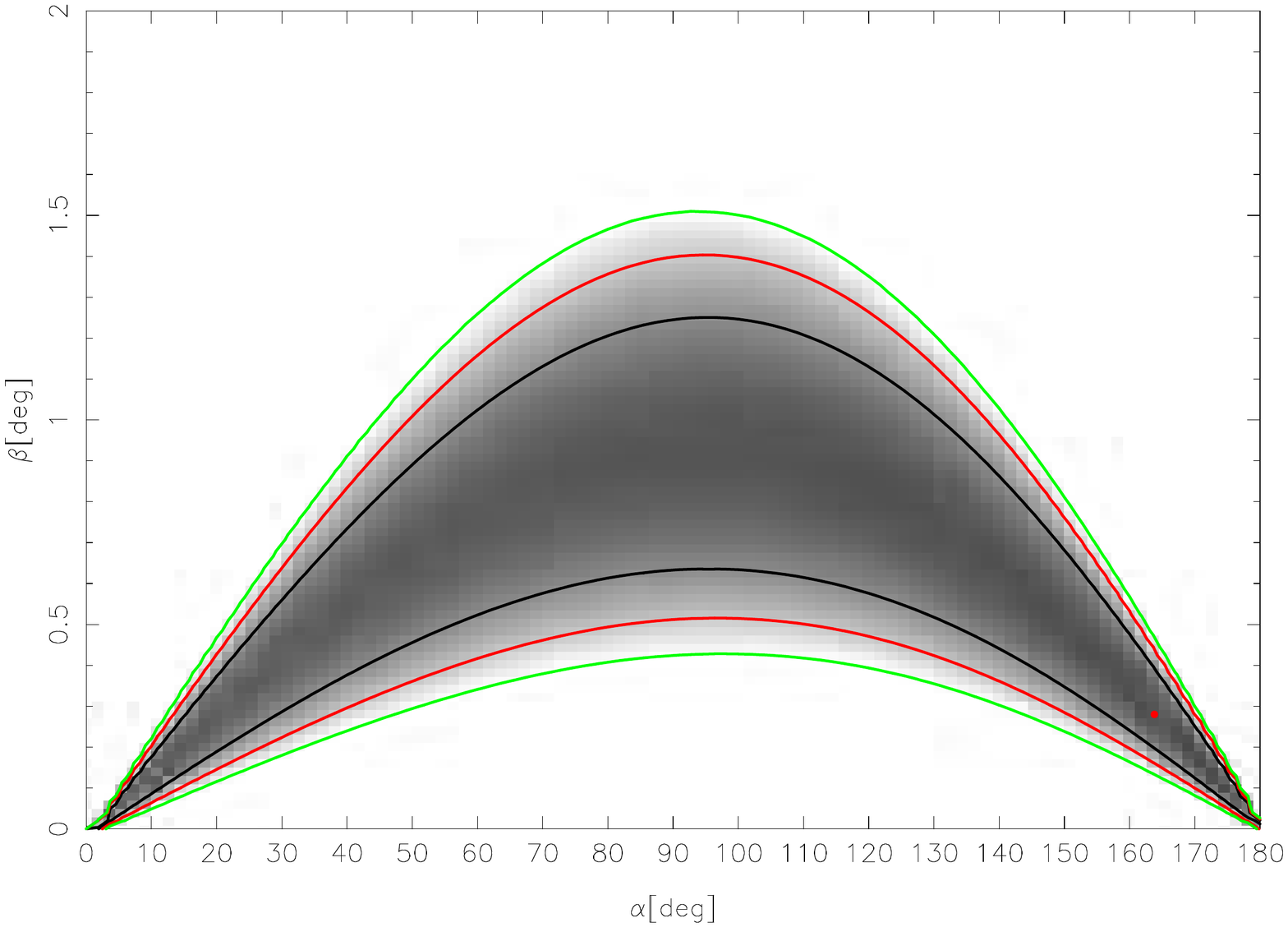}}}\\
{\mbox{\includegraphics[width=9cm,height=6cm,angle=0.]{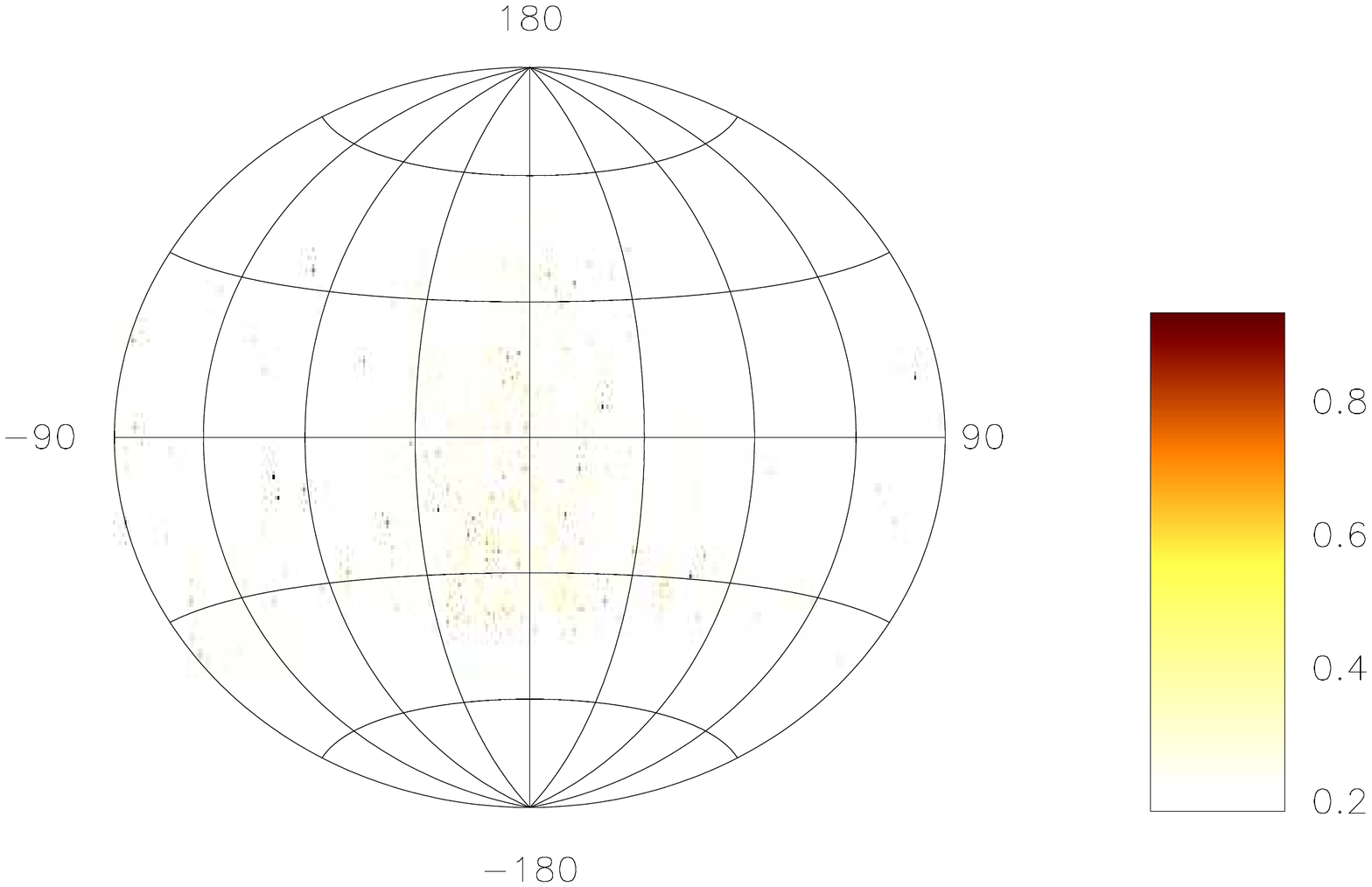}}}&
\\
\end{tabular}
\caption{Top panel (upper window) shows the average profile with total
intensity (Stokes I; solid black lines), total linear polarization (dashed red
line) and circular polarization (Stokes V; dotted blue line). Top panel (lower
window) also shows the single pulse PPA distribution (colour scale) along with
the average PPA (red error bars).
The RVM fits to the average PPA (dashed pink
line) is also shown in this plot. Middle panel show
the $\chi^2$ contours for the parameters $\alpha$ and $\beta$ obtained from RVM
fits.
Bottom panel only for 333 MHz shows the Hammer-Aitoff projection of the polarized time
samples with the colour scheme representing the fractional polarization level.}
\label{a32}
\end{center}
\end{figure*}


\begin{figure*}
\begin{center}
\begin{tabular}{cc}
{\mbox{\includegraphics[width=9cm,height=6cm,angle=0.]{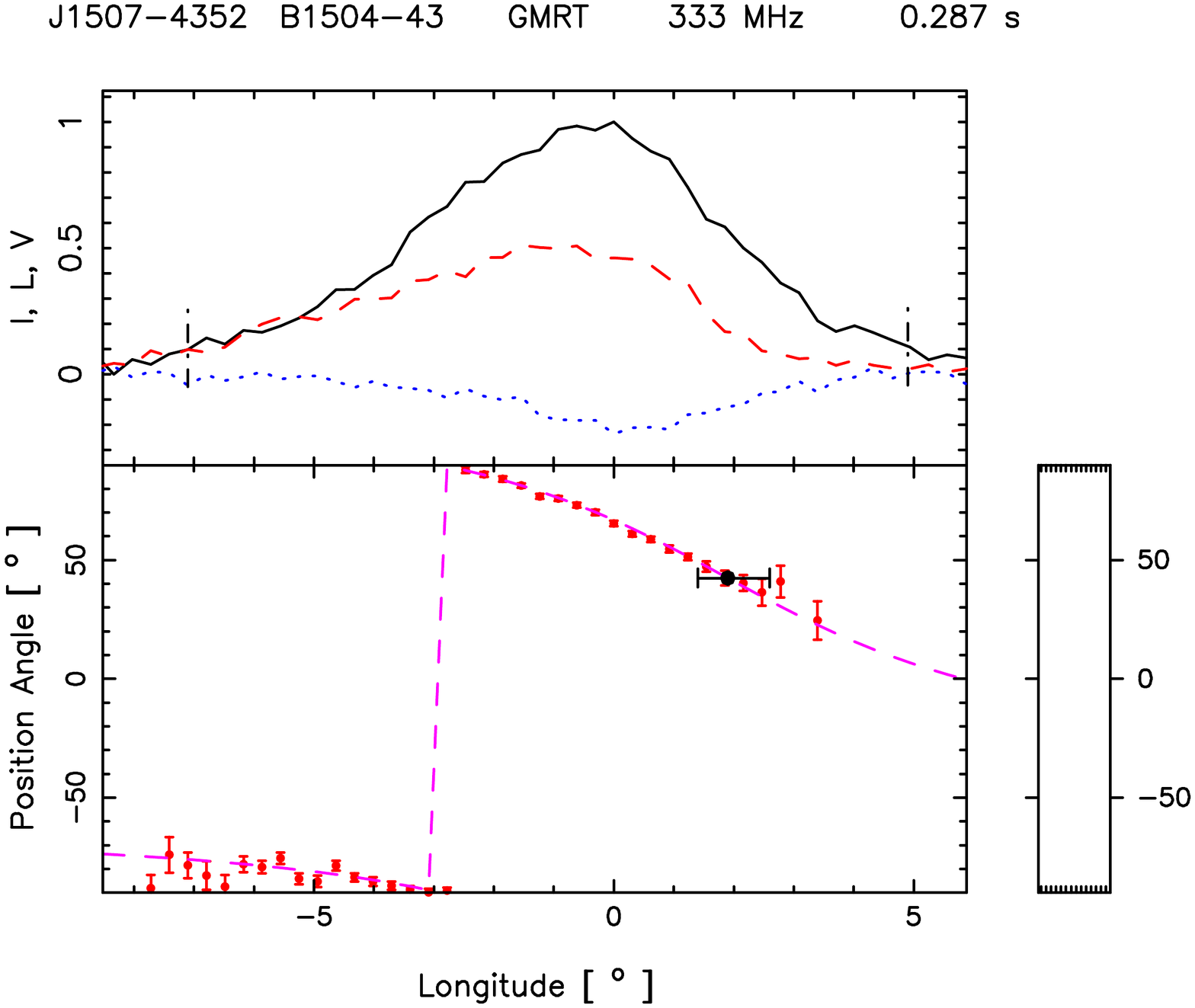}}}&
{\mbox{\includegraphics[width=9cm,height=6cm,angle=0.]{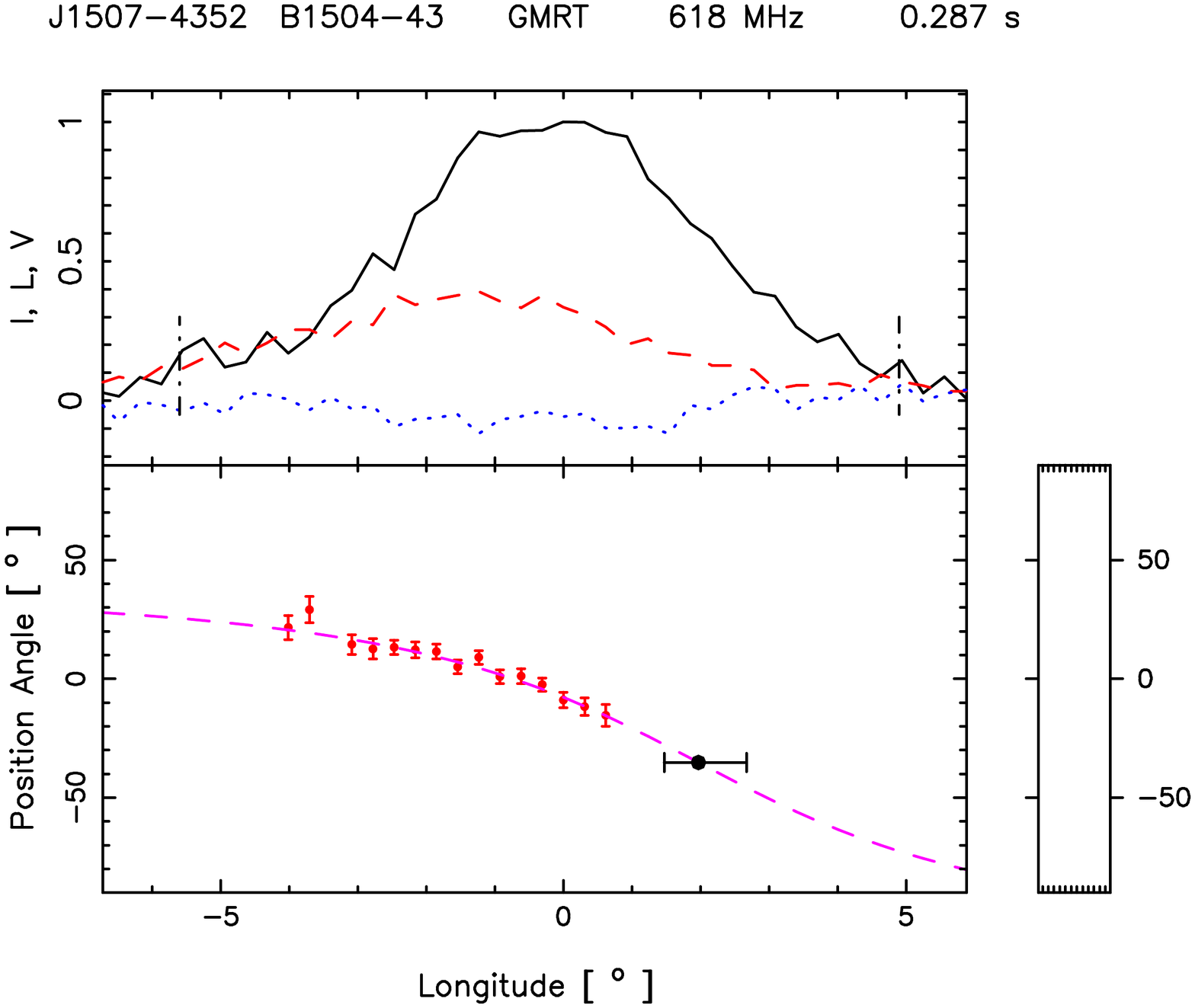}}}\\
{\mbox{\includegraphics[width=9cm,height=6cm,angle=0.]{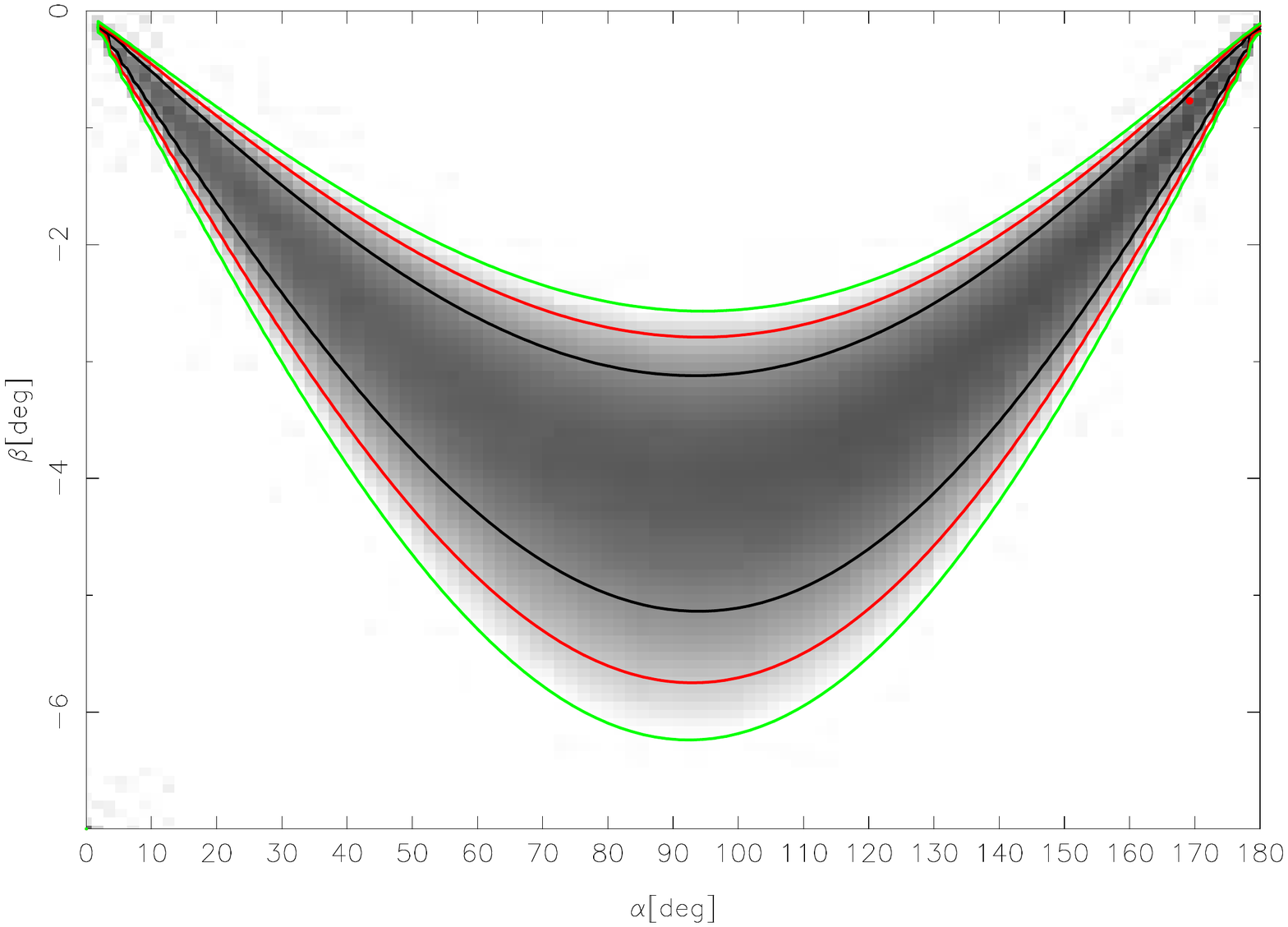}}}&
{\mbox{\includegraphics[width=9cm,height=6cm,angle=0.]{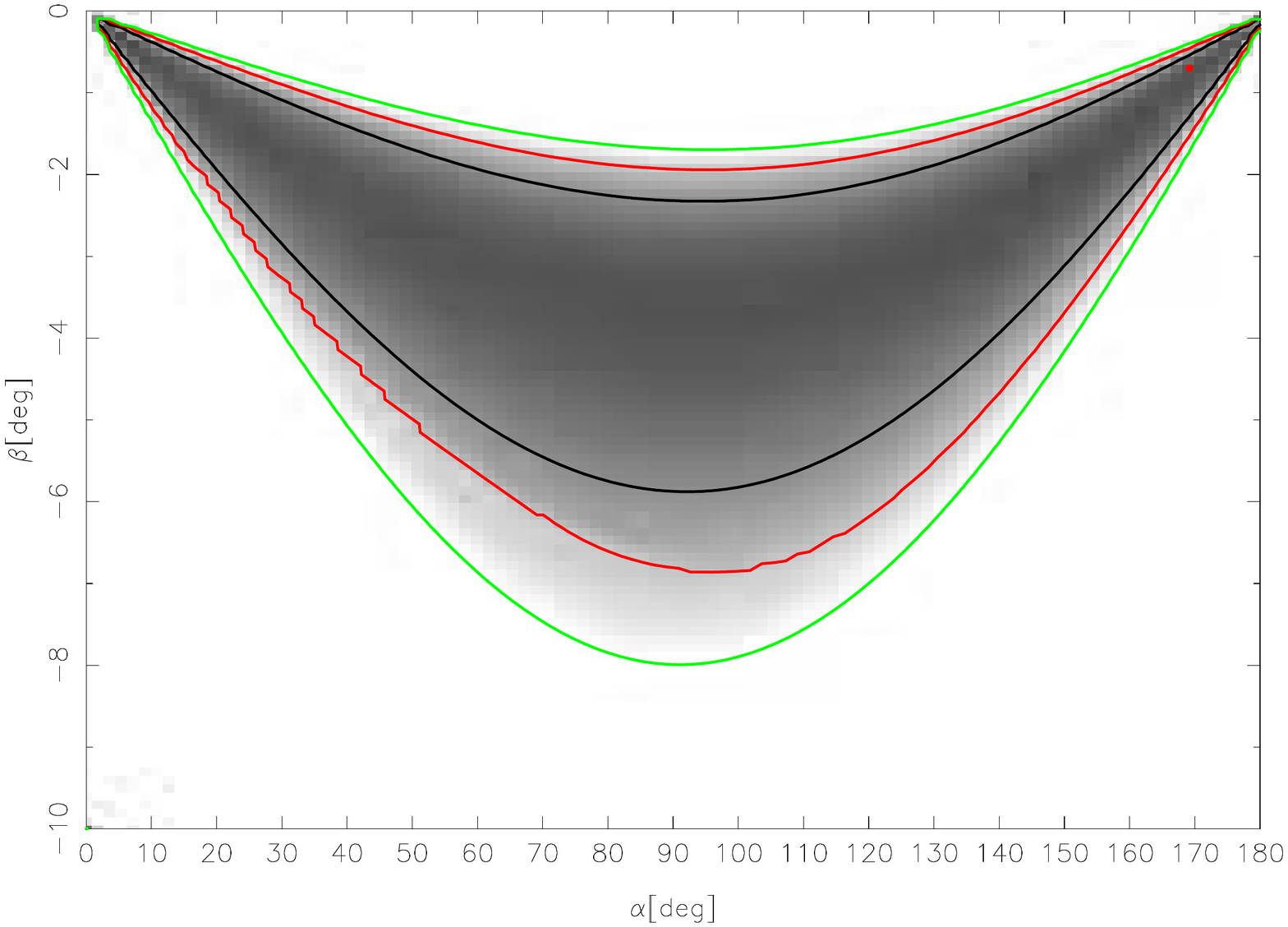}}}\\
&
\\
\end{tabular}
\caption{Top panel (upper window) shows the average profile with total
intensity (Stokes I; solid black lines), total linear polarization (dashed red
line) and circular polarization (Stokes V; dotted blue line). Top panel (lower
window) also shows the single pulse PPA distribution (colour scale) along with
the average PPA (red error bars).
The RVM fits to the average PPA (dashed pink
line) is also shown in this plot. Bottom panel show
the $\chi^2$ contours for the parameters $\alpha$ and $\beta$ obtained from RVM
fits.}
\label{a33}
\end{center}
\end{figure*}


\begin{figure*}
\begin{center}
\begin{tabular}{cc}
{\mbox{\includegraphics[width=9cm,height=6cm,angle=0.]{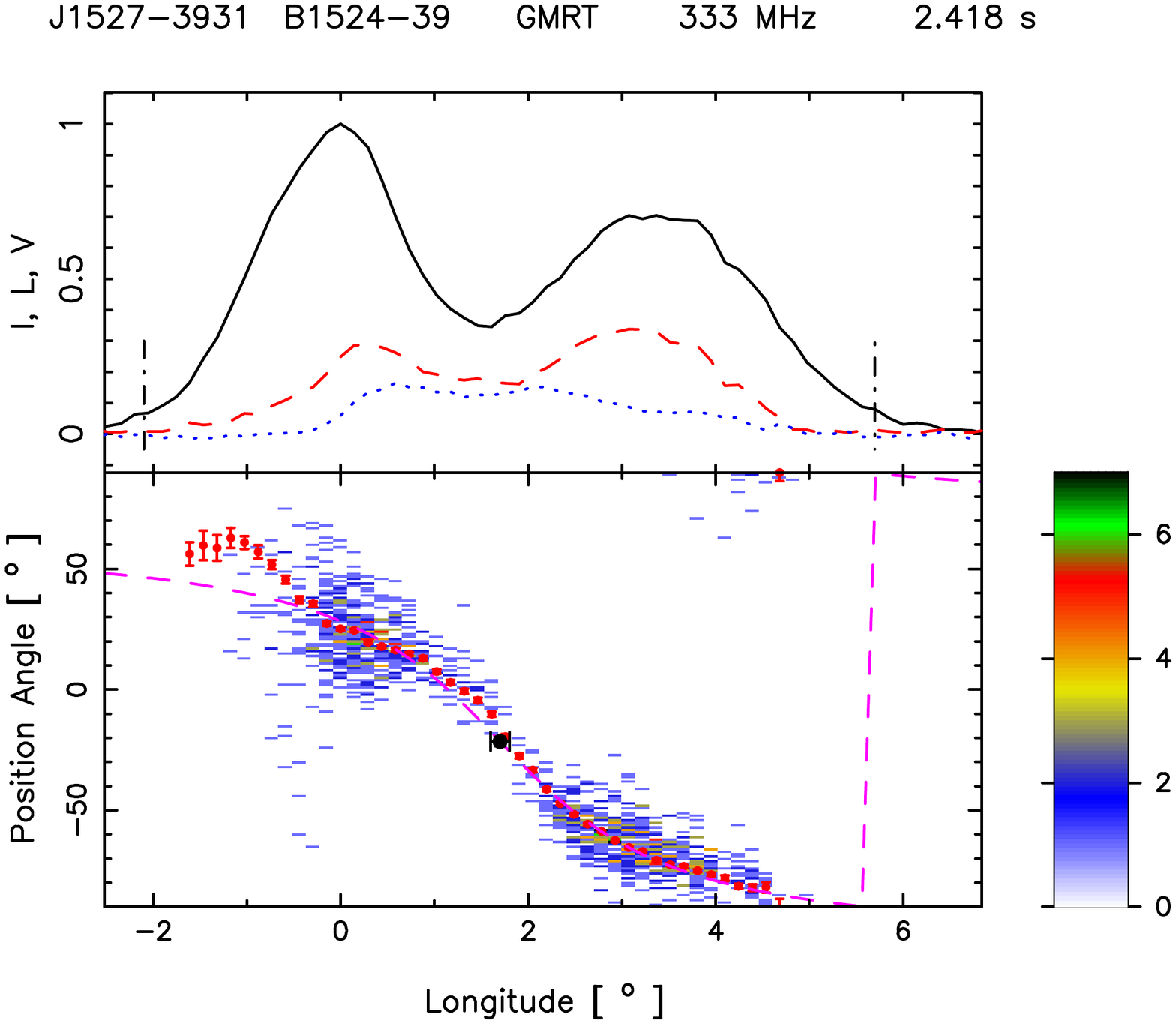}}}&
{\mbox{\includegraphics[width=9cm,height=6cm,angle=0.]{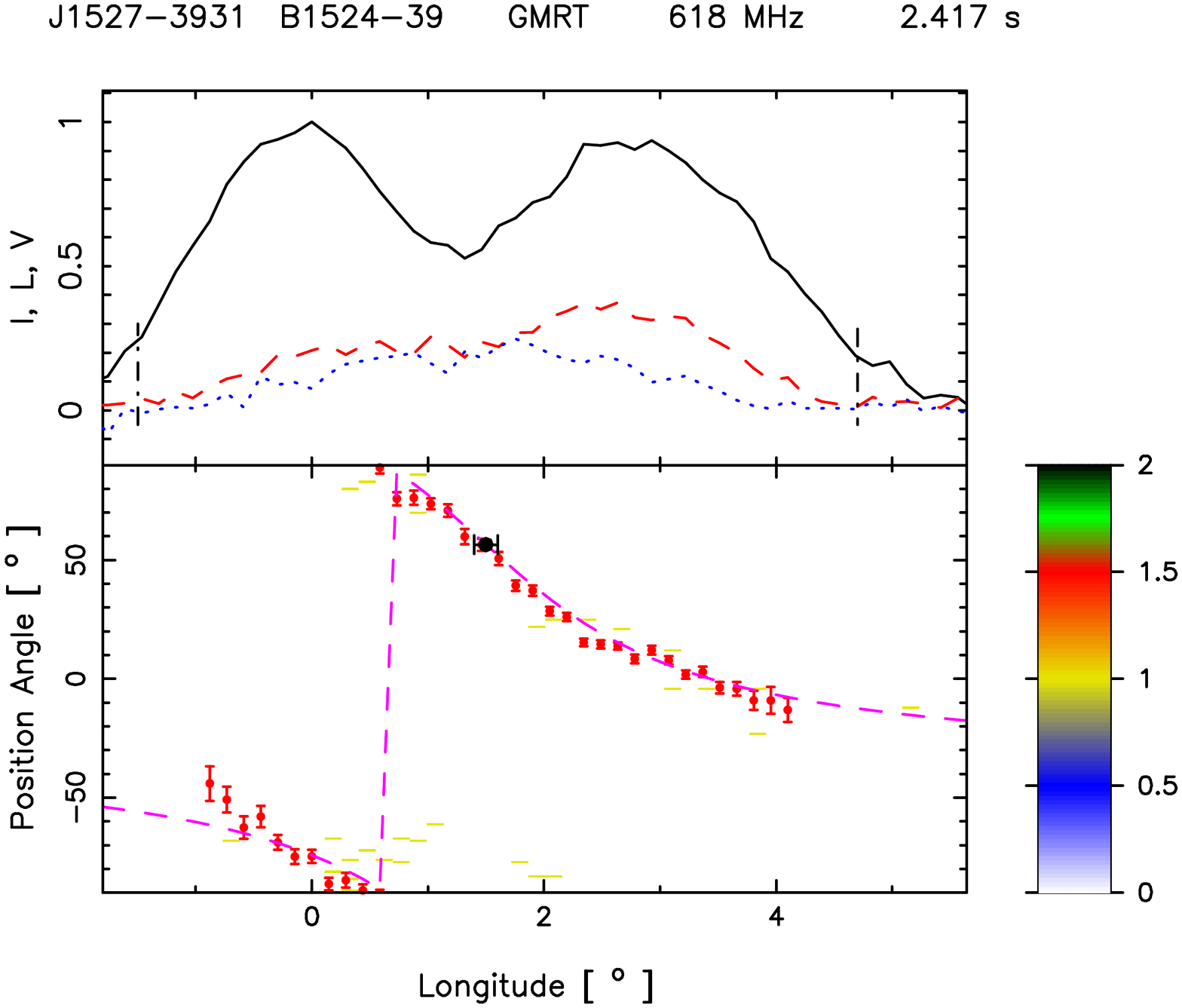}}}\\
{\mbox{\includegraphics[width=9cm,height=6cm,angle=0.]{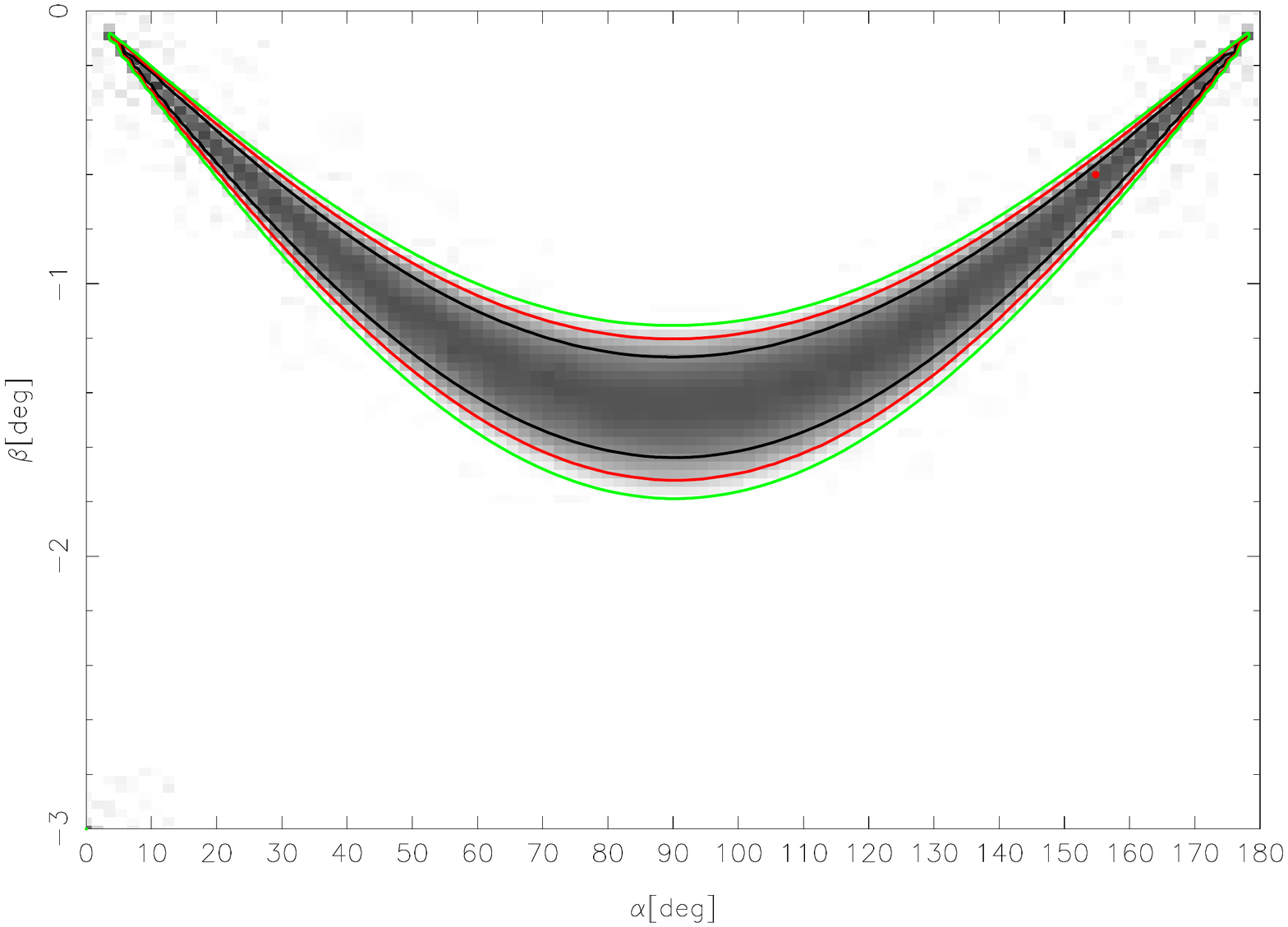}}}&
{\mbox{\includegraphics[width=9cm,height=6cm,angle=0.]{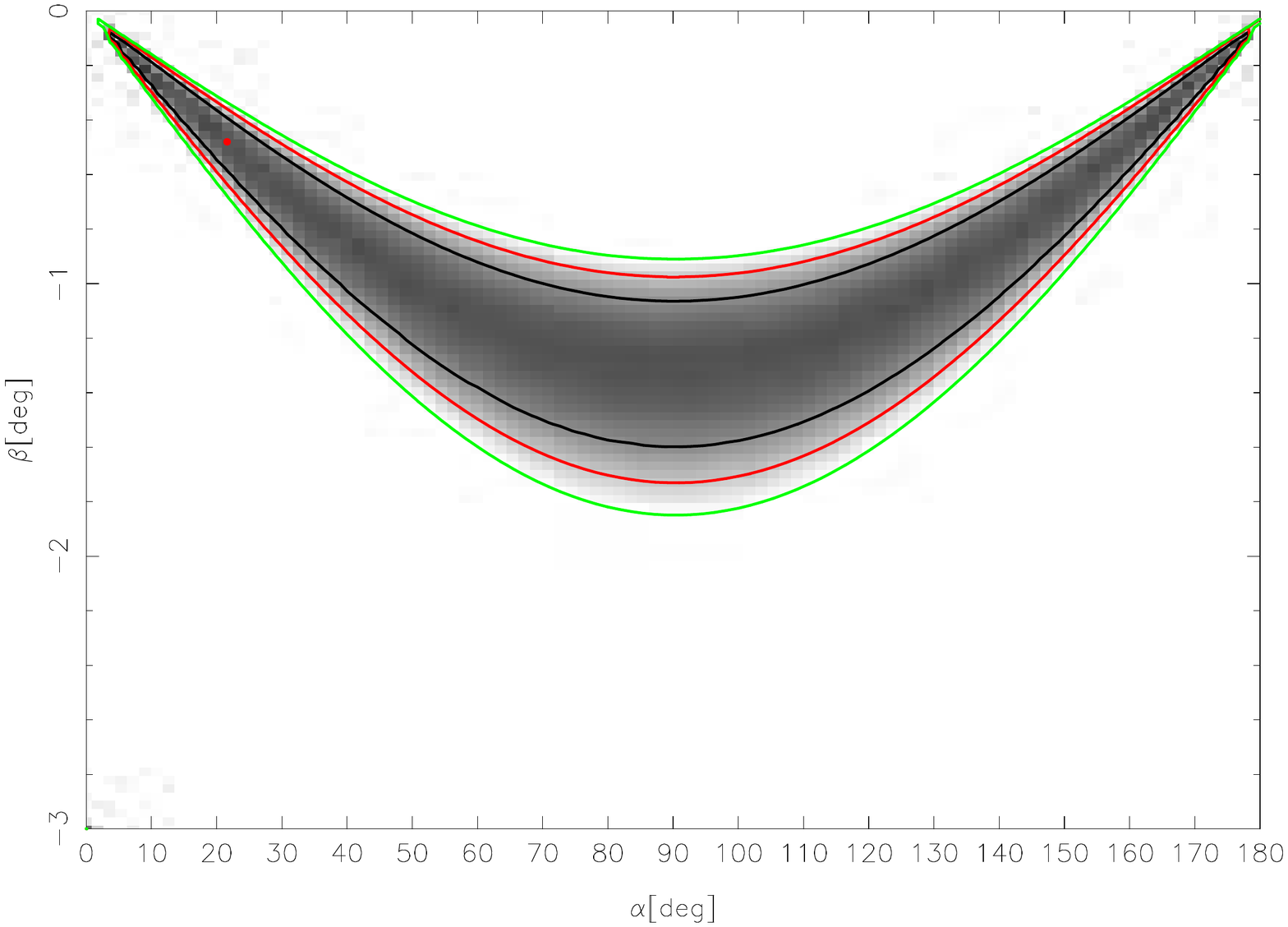}}}\\
{\mbox{\includegraphics[width=9cm,height=6cm,angle=0.]{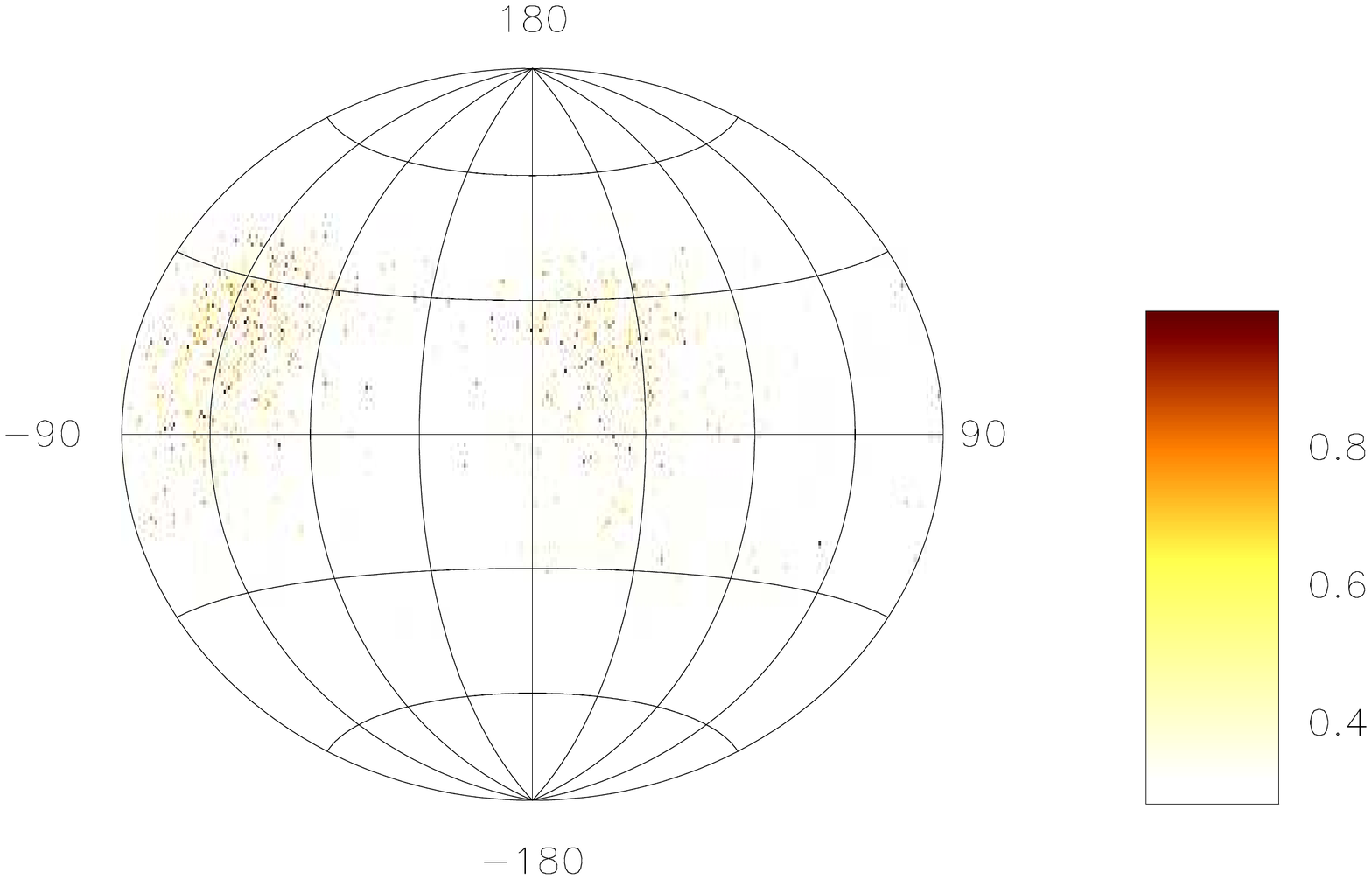}}}&
{\mbox{\includegraphics[width=9cm,height=6cm,angle=0.]{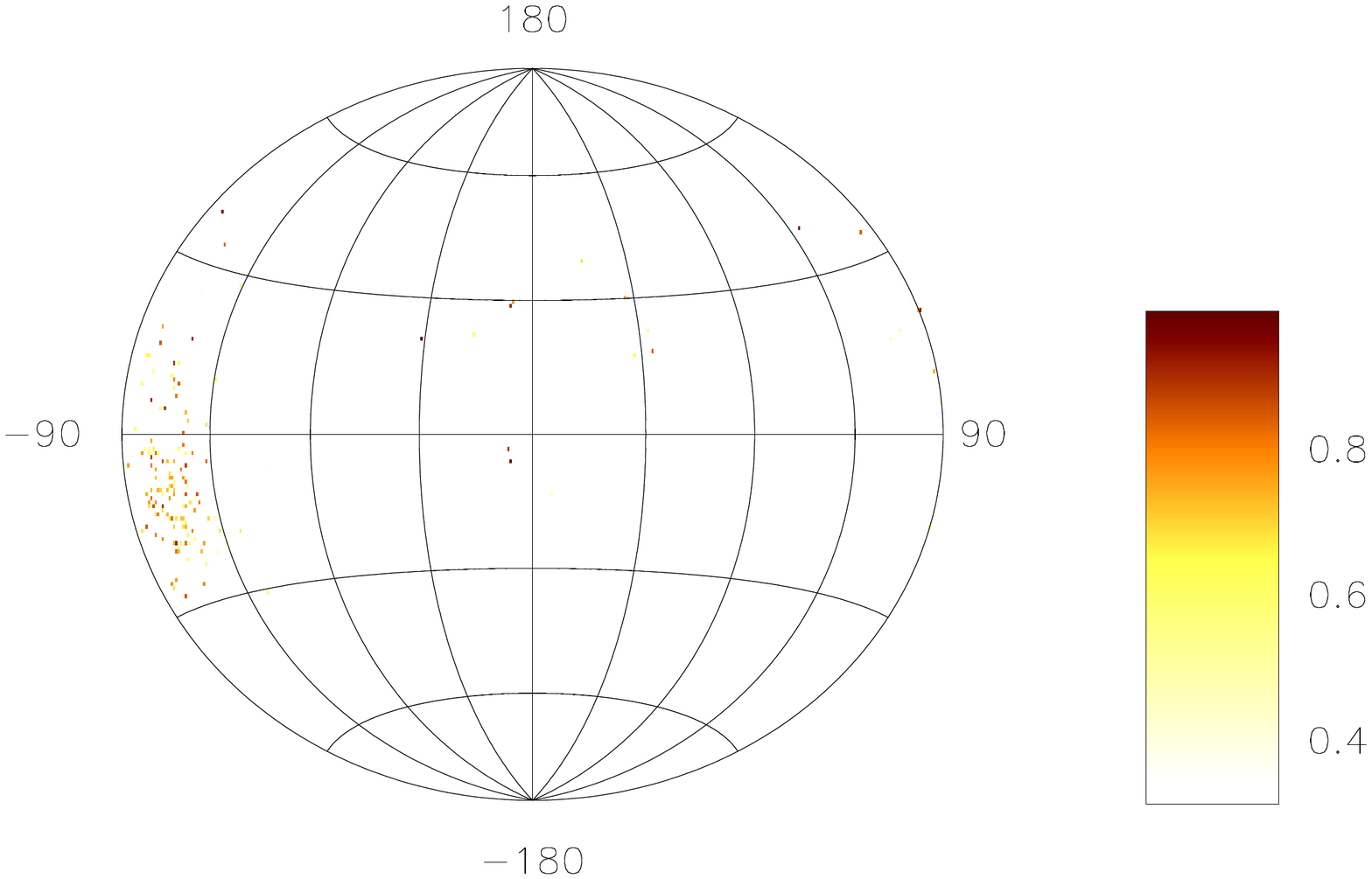}}}\\
\end{tabular}
\caption{Top panel (upper window) shows the average profile with total
intensity (Stokes I; solid black lines), total linear polarization (dashed red
line) and circular polarization (Stokes V; dotted blue line). Top panel (lower
window) also shows the single pulse PPA distribution (colour scale) along with
the average PPA (red error bars).
The RVM fits to the average PPA (dashed pink
line) is also shown in this plot. Middle panel show
the $\chi^2$ contours for the parameters $\alpha$ and $\beta$ obtained from RVM
fits.
Bottom panel shows the Hammer-Aitoff projection of the polarized time
samples with the colour scheme representing the fractional polarization level.}
\label{a34}
\end{center}
\end{figure*}


\begin{figure*}
\begin{center}
\begin{tabular}{cc}
{\mbox{\includegraphics[width=9cm,height=6cm,angle=0.]{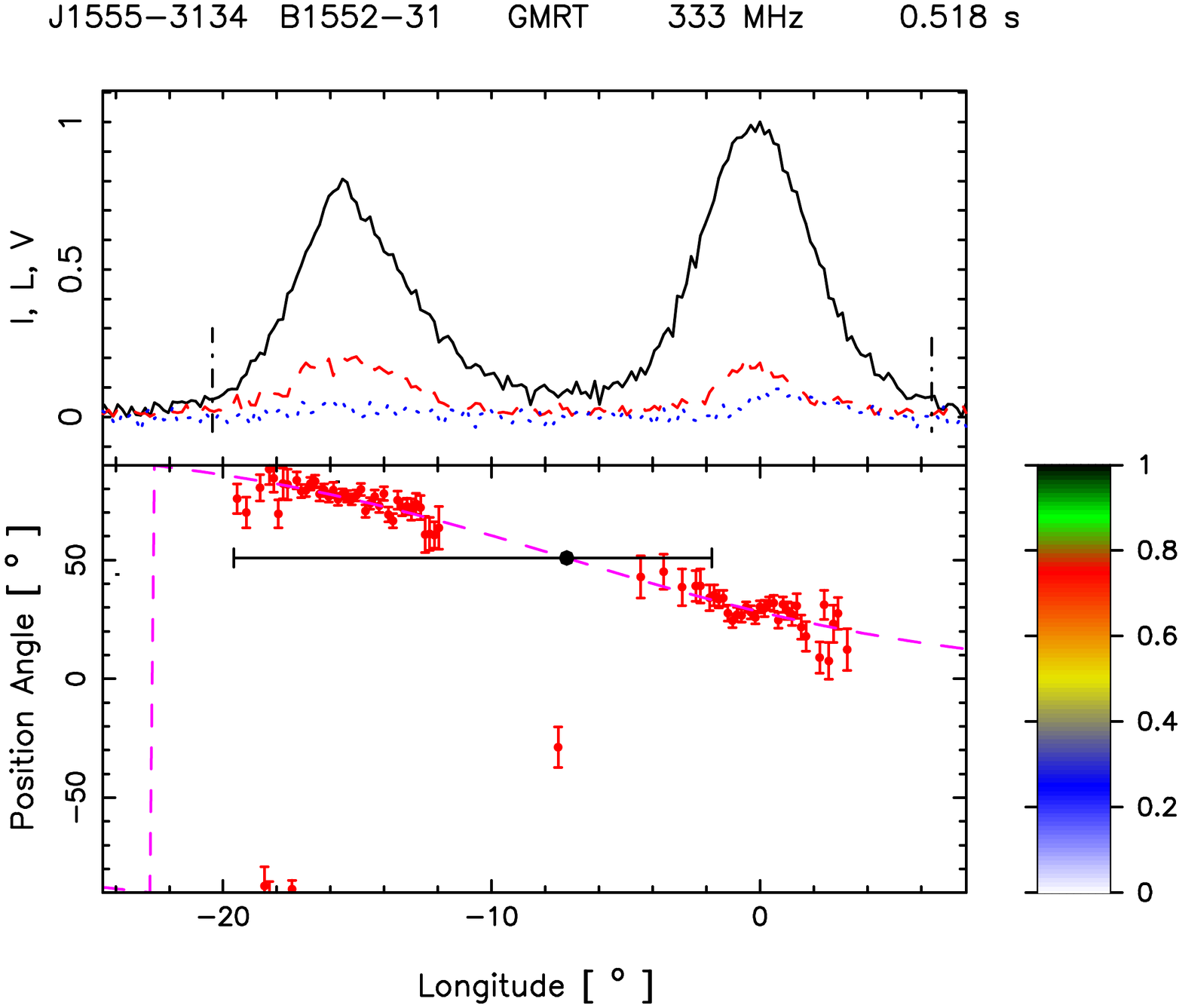}}}&
{\mbox{\includegraphics[width=9cm,height=6cm,angle=0.]{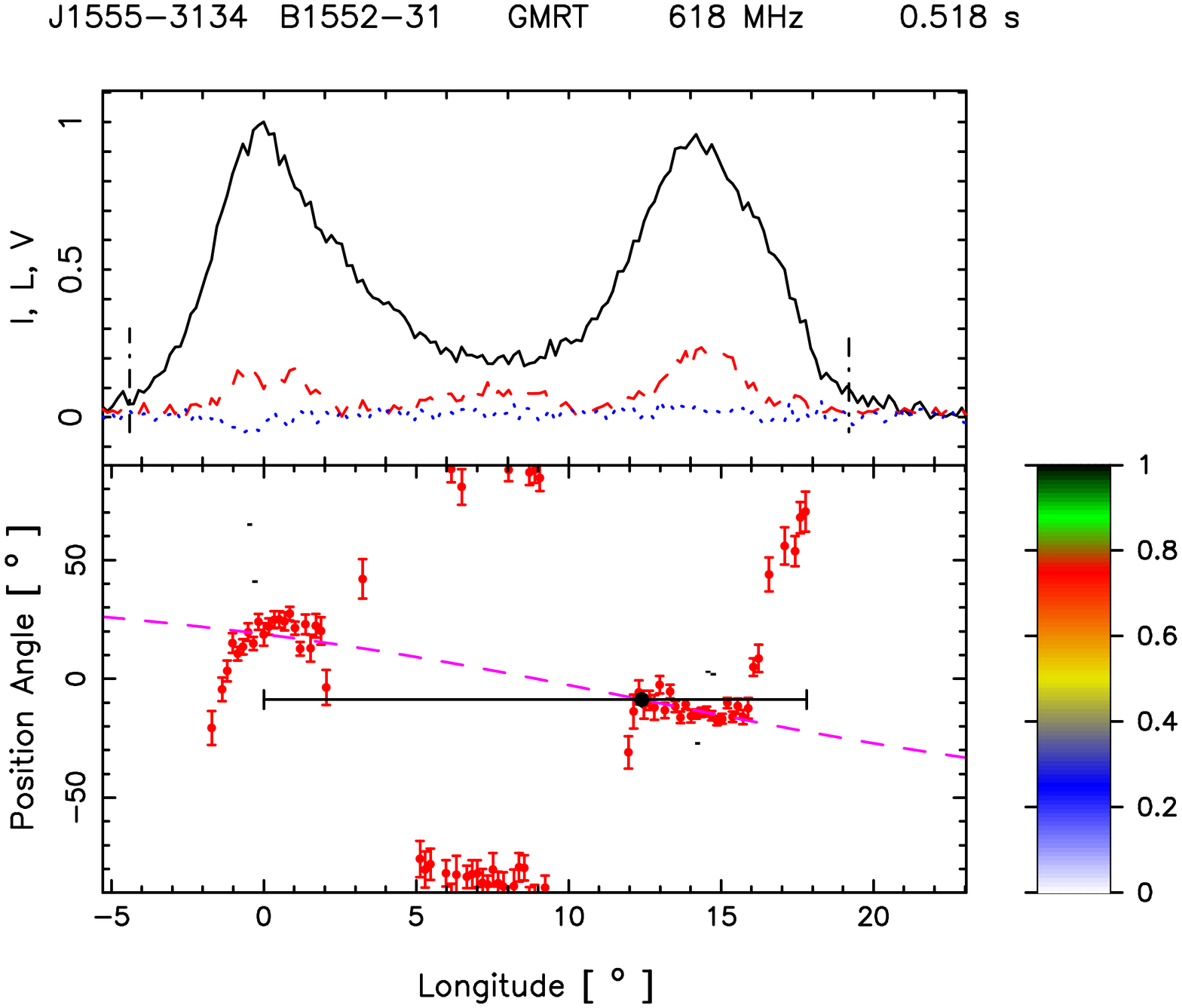}}}\\
{\mbox{\includegraphics[width=9cm,height=6cm,angle=0.]{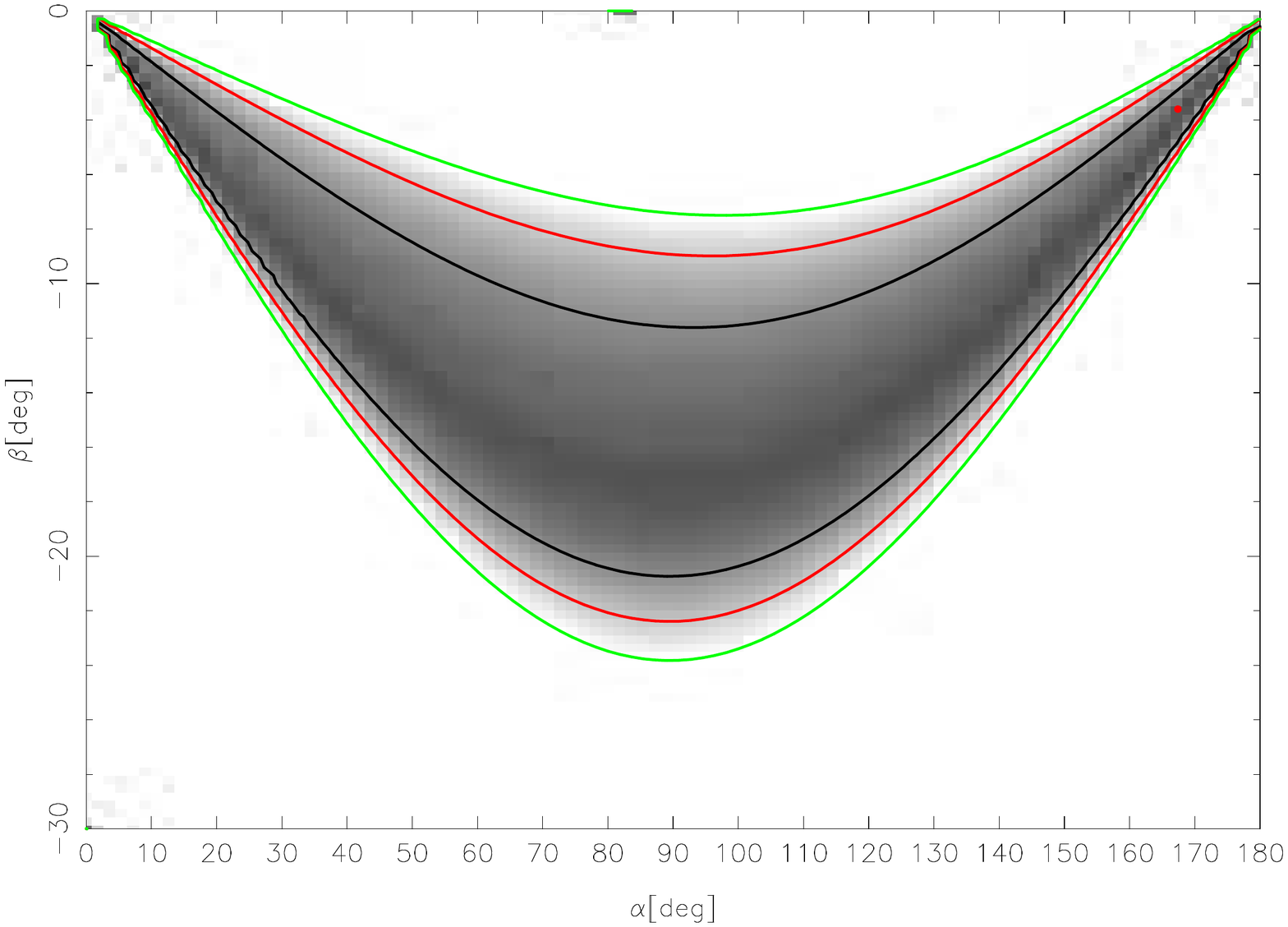}}}&
{\mbox{\includegraphics[width=9cm,height=6cm,angle=0.]{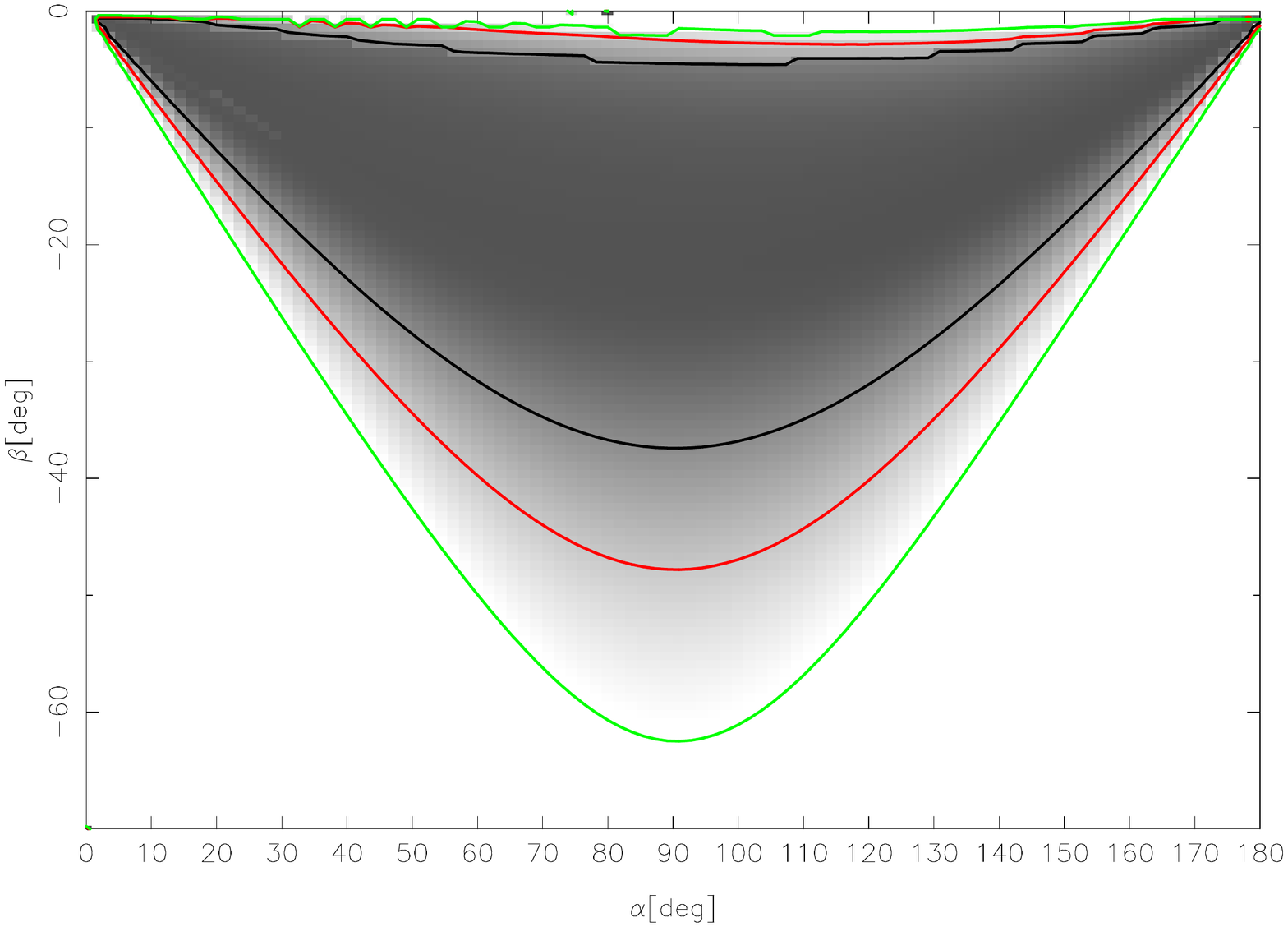}}}\\
&
\\
\end{tabular}
\caption{Top panel (upper window) shows the average profile with total
intensity (Stokes I; solid black lines), total linear polarization (dashed red
line) and circular polarization (Stokes V; dotted blue line). Top panel (lower
window) also shows the single pulse PPA distribution (colour scale) along with
the average PPA (red error bars).
The RVM fits to the average PPA (dashed pink
line) is also shown in this plot. Bottom panel show
the $\chi^2$ contours for the parameters $\alpha$ and $\beta$ obtained from RVM
fits.}
\label{a35}
\end{center}
\end{figure*}


\begin{figure*}
\begin{center}
\begin{tabular}{cc}
{\mbox{\includegraphics[width=9cm,height=6cm,angle=0.]{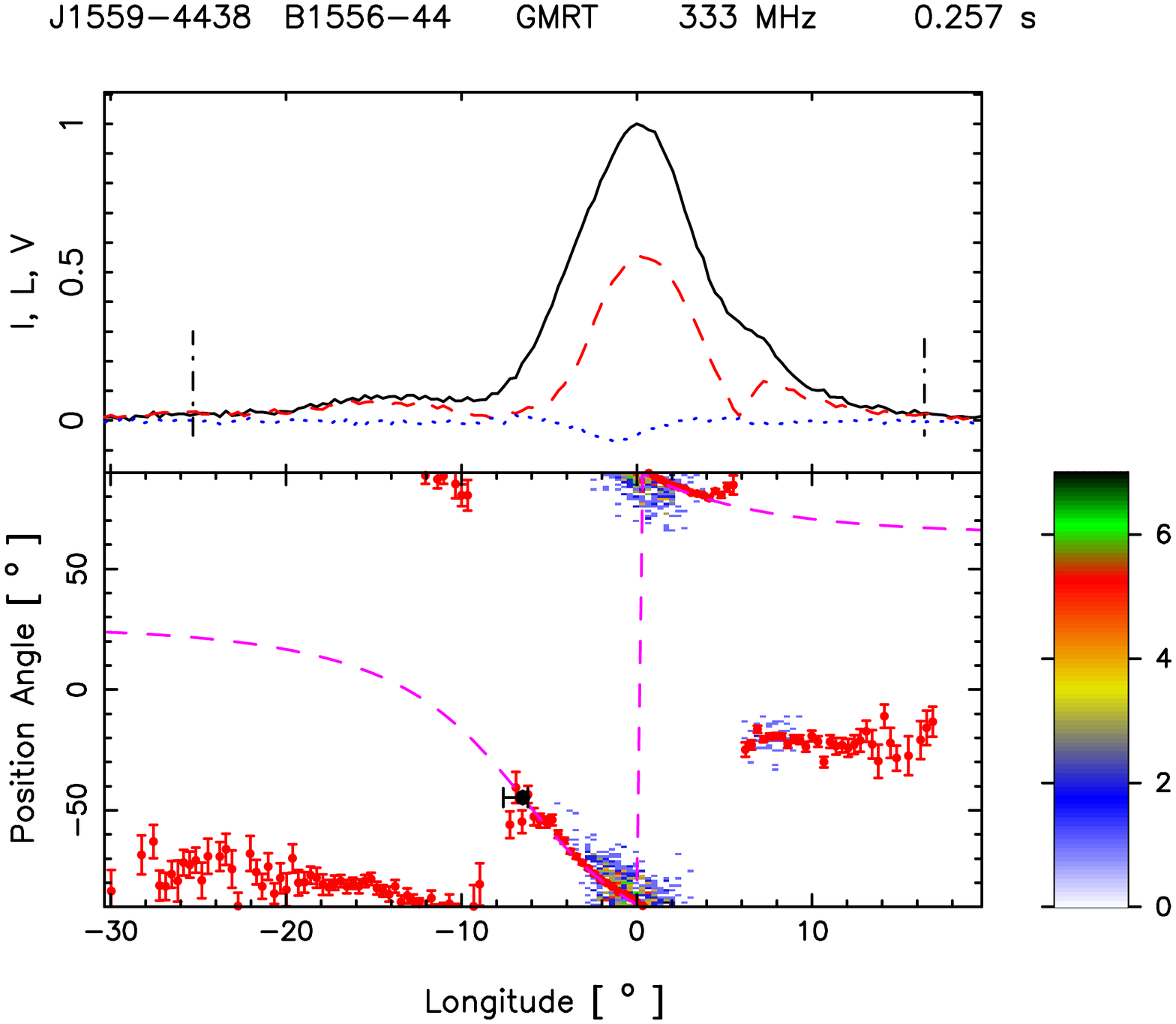}}}&
{\mbox{\includegraphics[width=9cm,height=6cm,angle=0.]{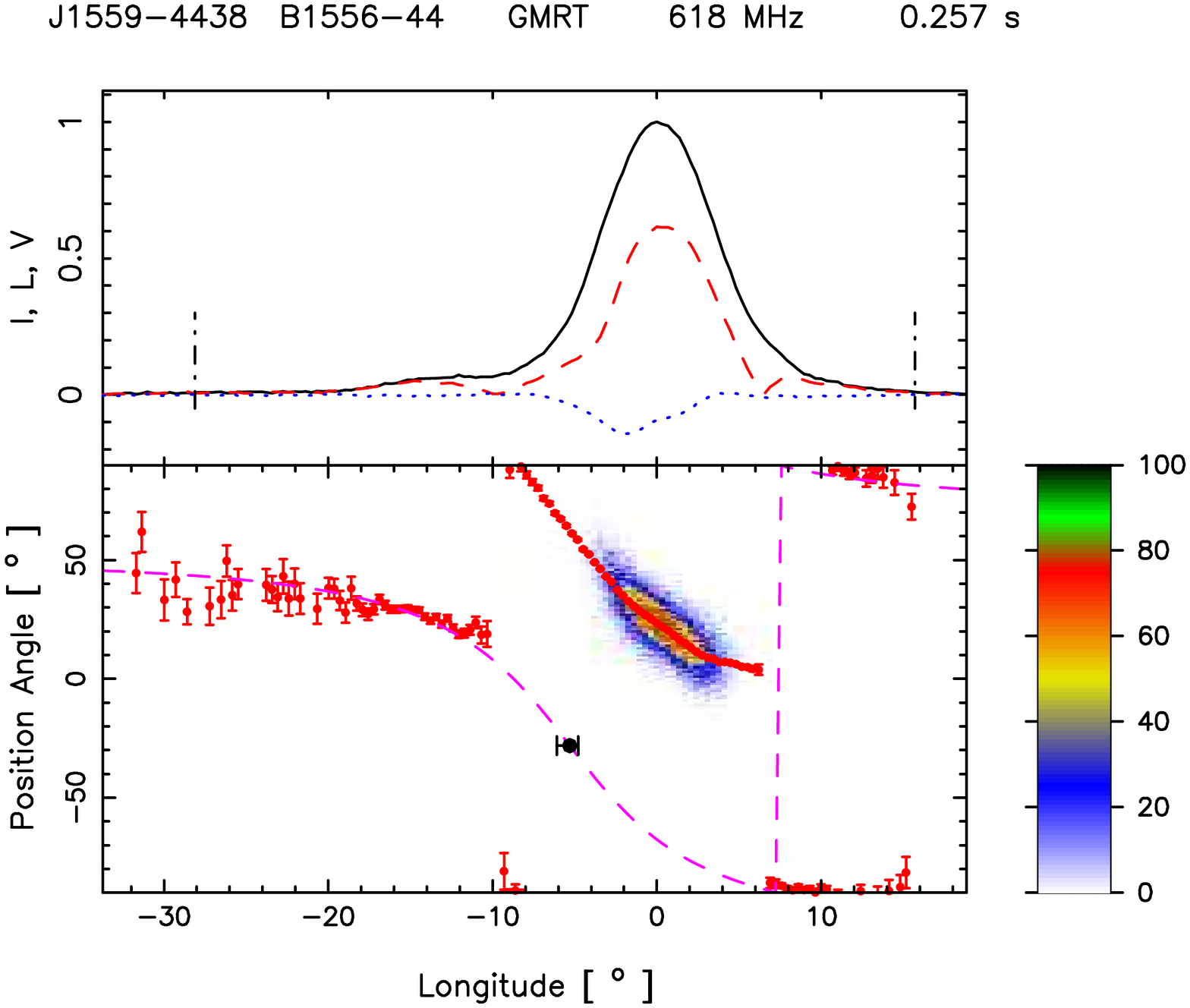}}}\\
{\mbox{\includegraphics[width=9cm,height=6cm,angle=0.]{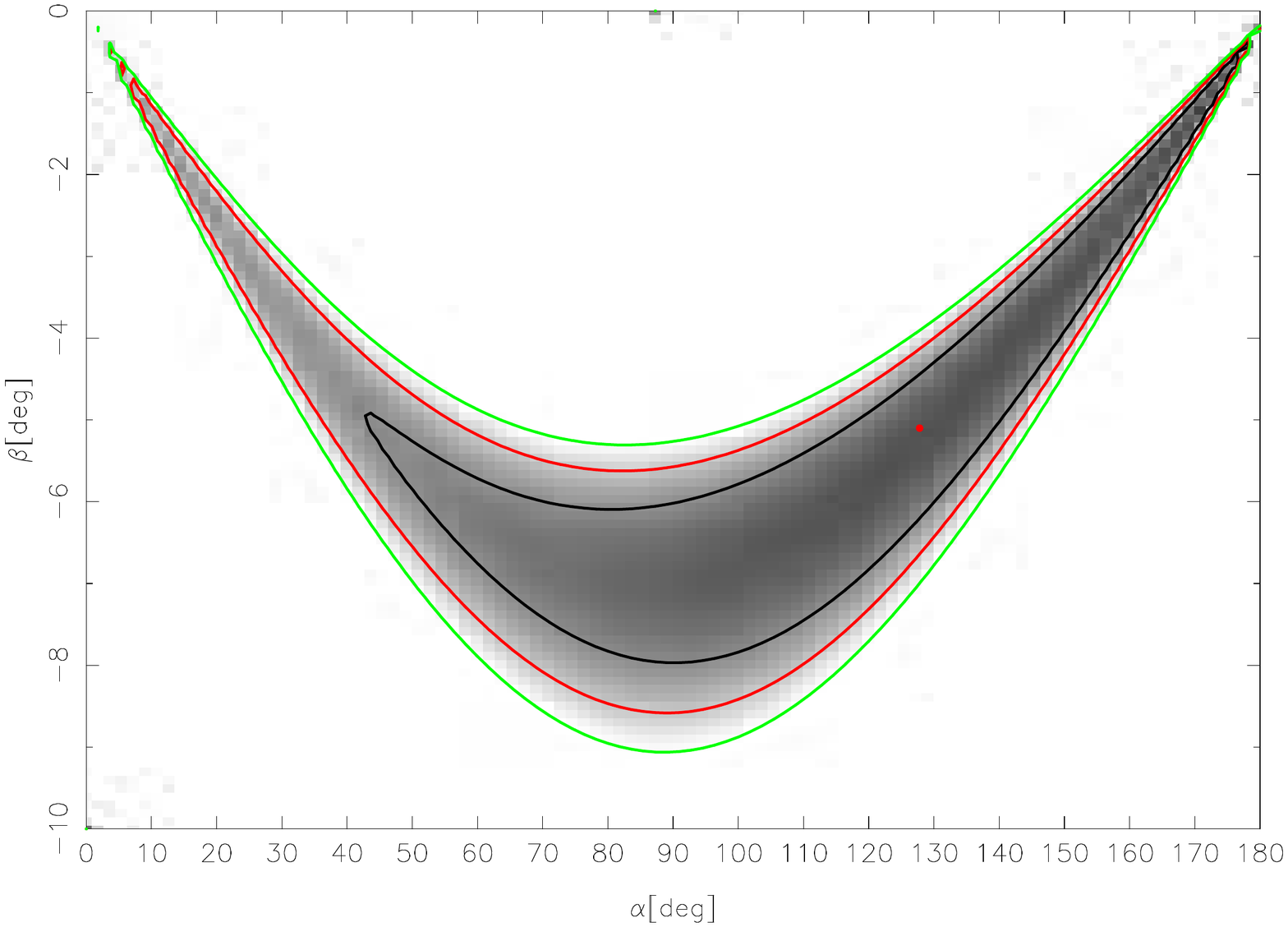}}}&
{\mbox{\includegraphics[width=9cm,height=6cm,angle=0.]{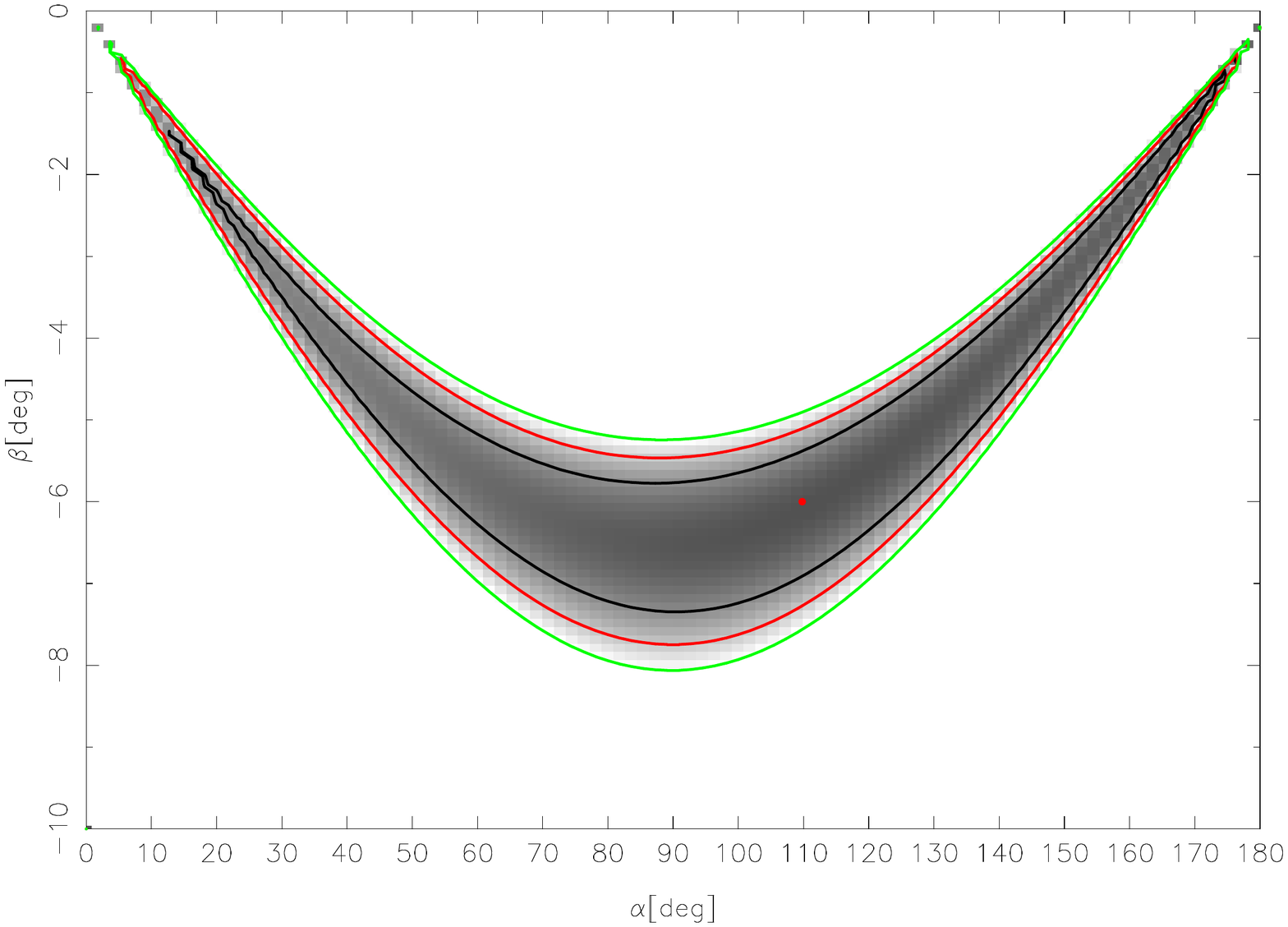}}}\\
{\mbox{\includegraphics[width=9cm,height=6cm,angle=0.]{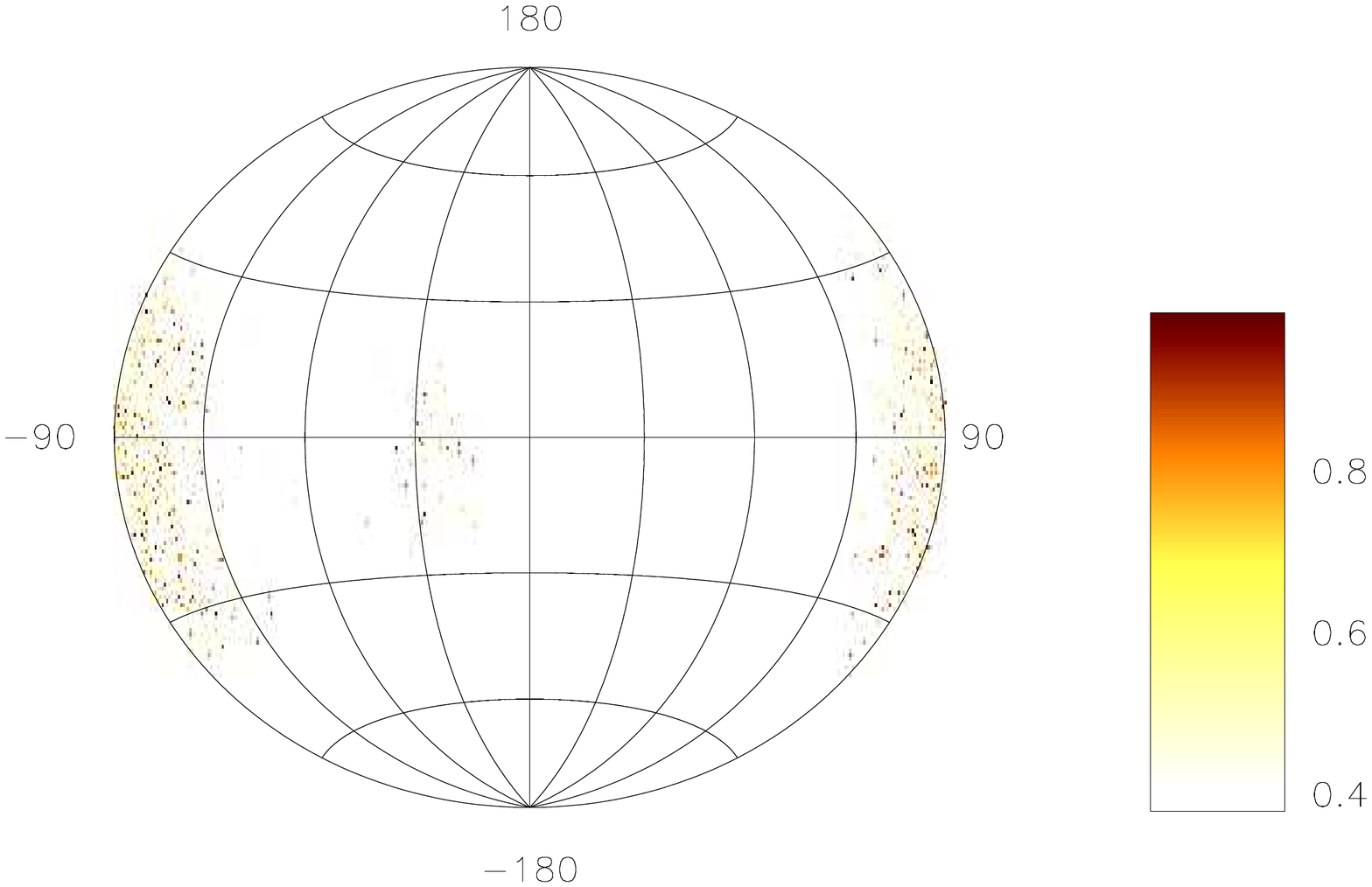}}}&
{\mbox{\includegraphics[width=9cm,height=6cm,angle=0.]{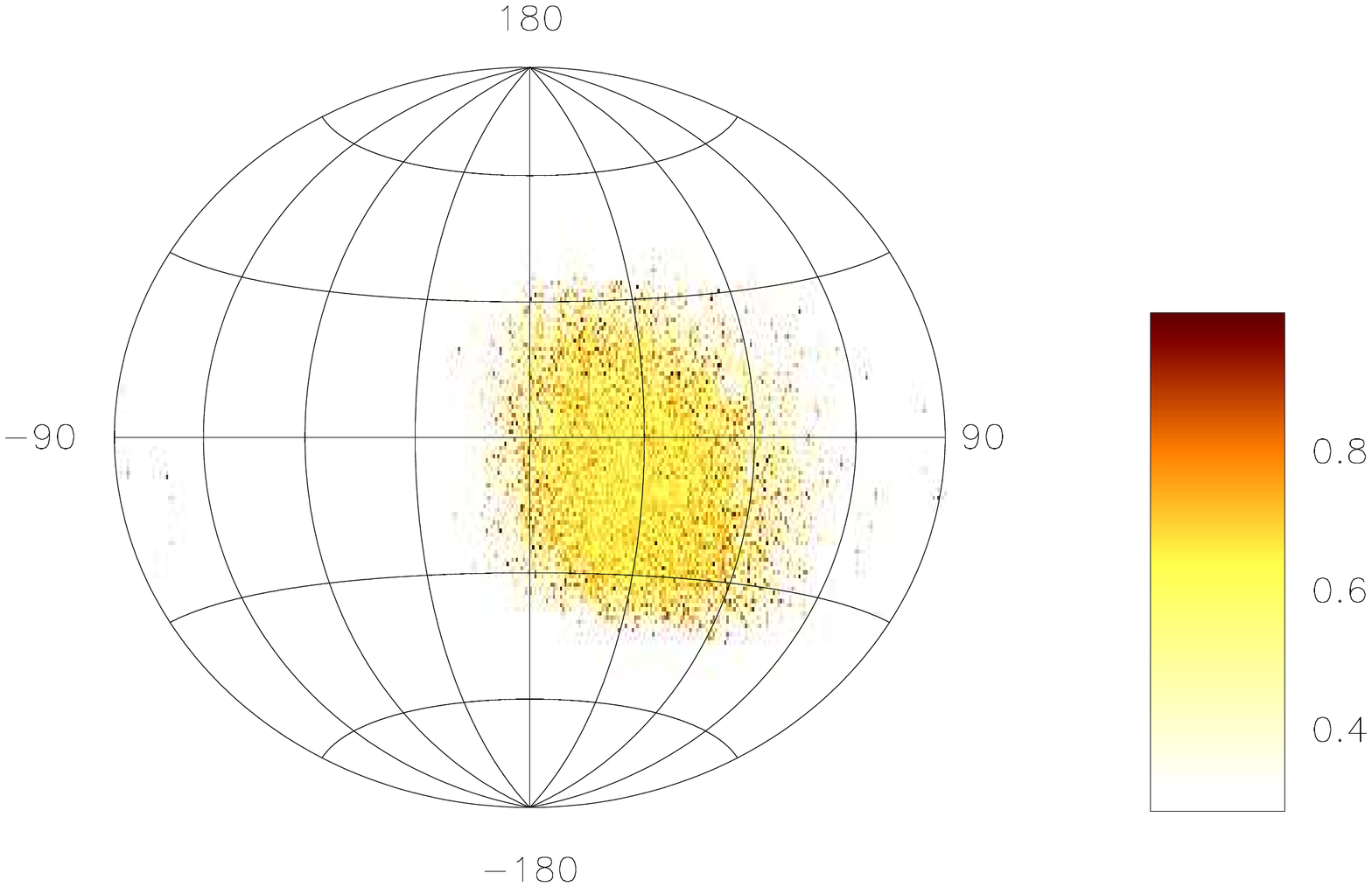}}}\\
\end{tabular}
\caption{Top panel (upper window) shows the average profile with total
intensity (Stokes I; solid black lines), total linear polarization (dashed red
line) and circular polarization (Stokes V; dotted blue line). Top panel (lower
window) also shows the single pulse PPA distribution (colour scale) along with
the average PPA (red error bars).
The RVM fits to the average PPA (dashed pink
line) is also shown in this plot. Middle panel show
the $\chi^2$ contours for the parameters $\alpha$ and $\beta$ obtained from RVM
fits.
Bottom panel shows the Hammer-Aitoff projection of the polarized time
samples with the colour scheme representing the fractional polarization level.}
\label{a36}
\end{center}
\end{figure*}


\begin{figure*}
\begin{center}
\begin{tabular}{cc}
{\mbox{\includegraphics[width=9cm,height=6cm,angle=0.]{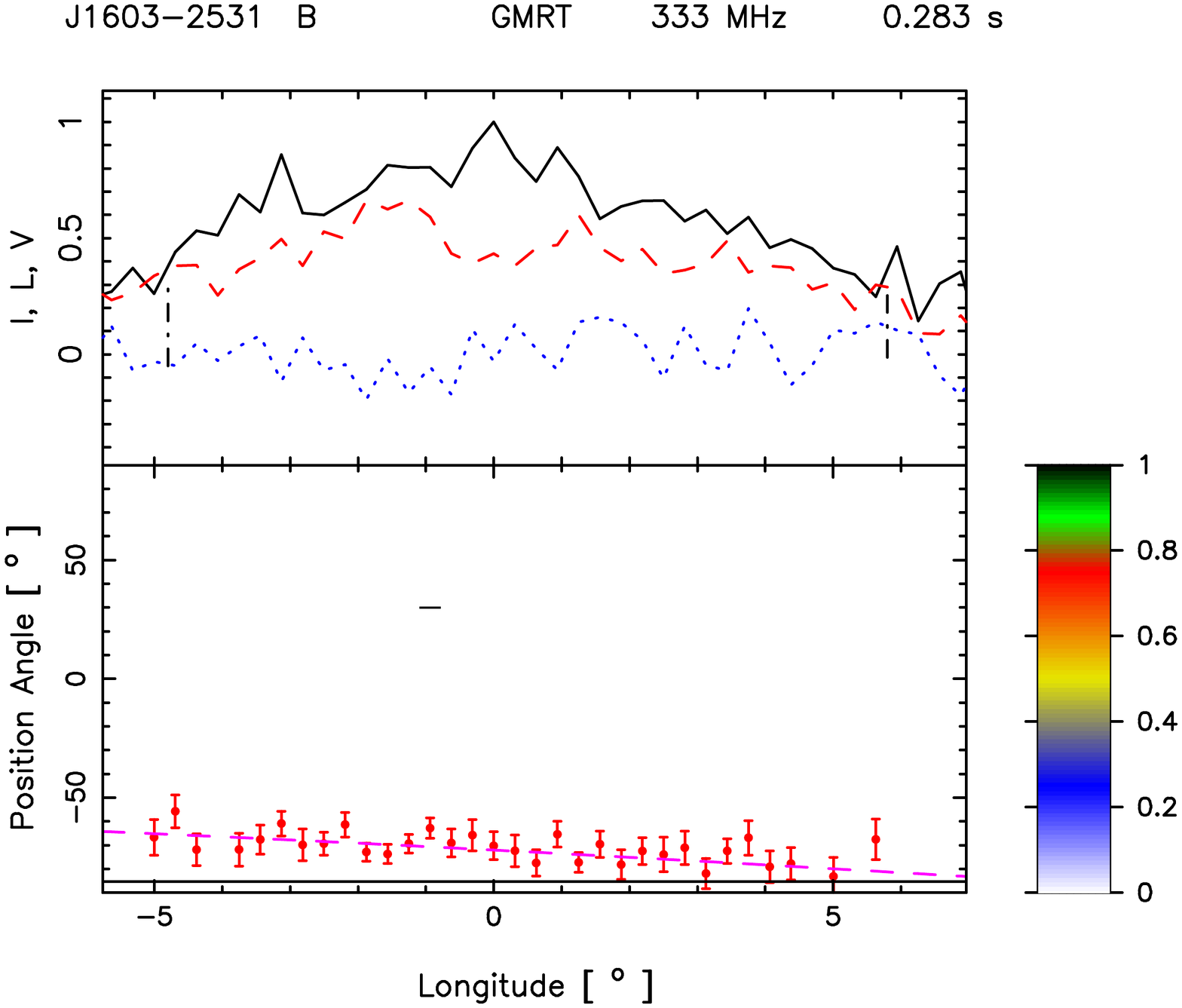}}}&
{\mbox{\includegraphics[width=9cm,height=6cm,angle=0.]{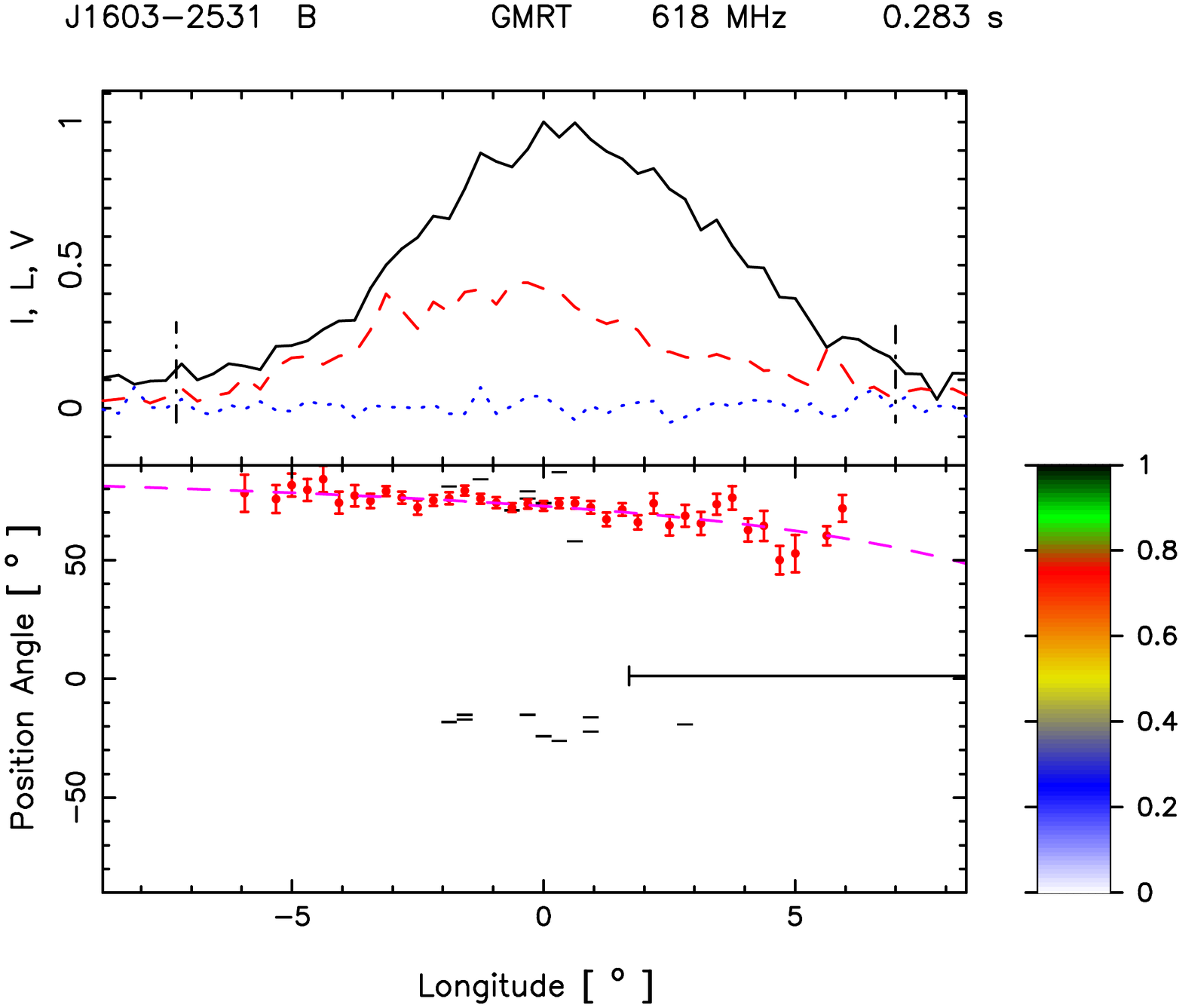}}}\\
{\mbox{\includegraphics[width=9cm,height=6cm,angle=0.]{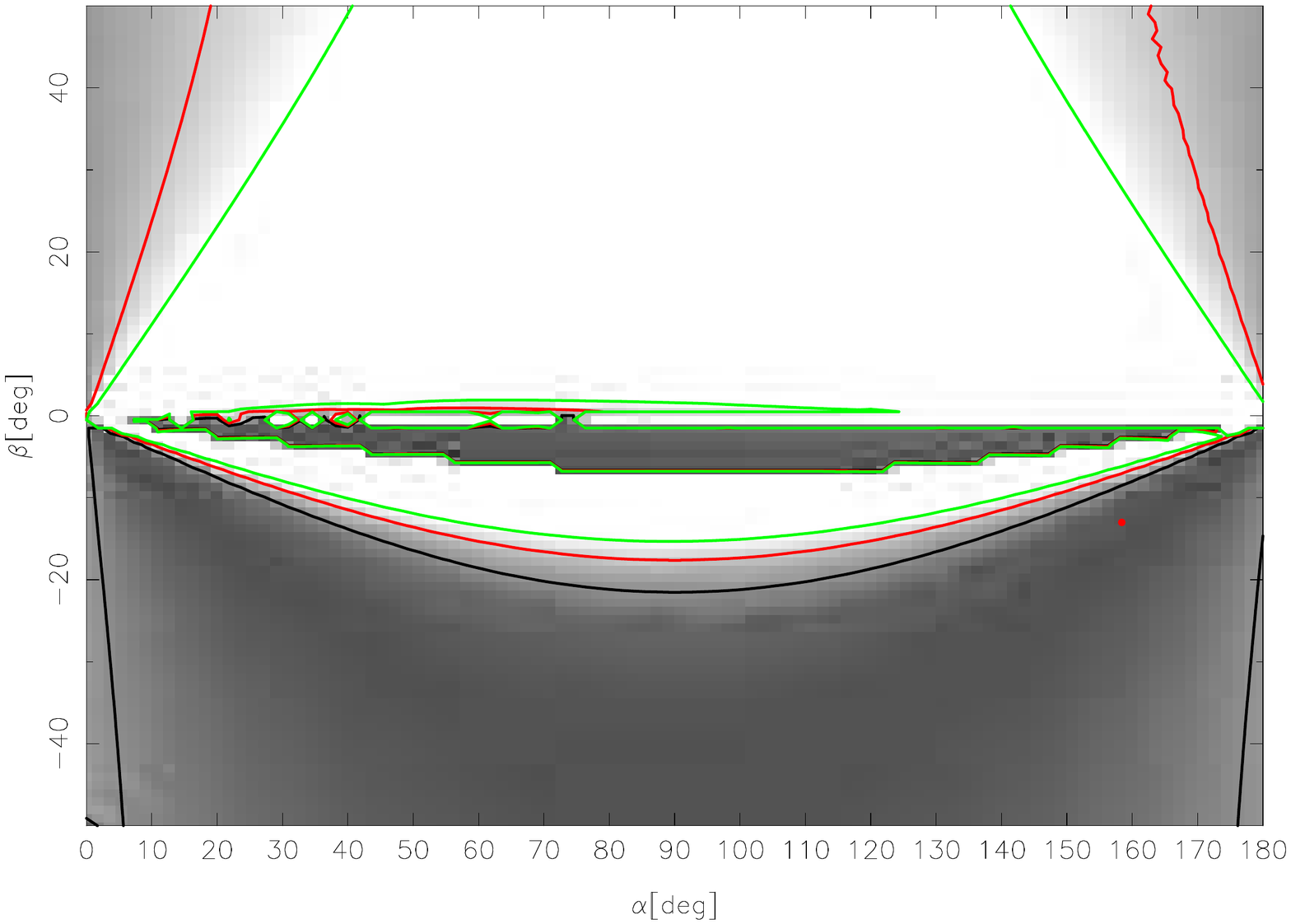}}}&
{\mbox{\includegraphics[width=9cm,height=6cm,angle=0.]{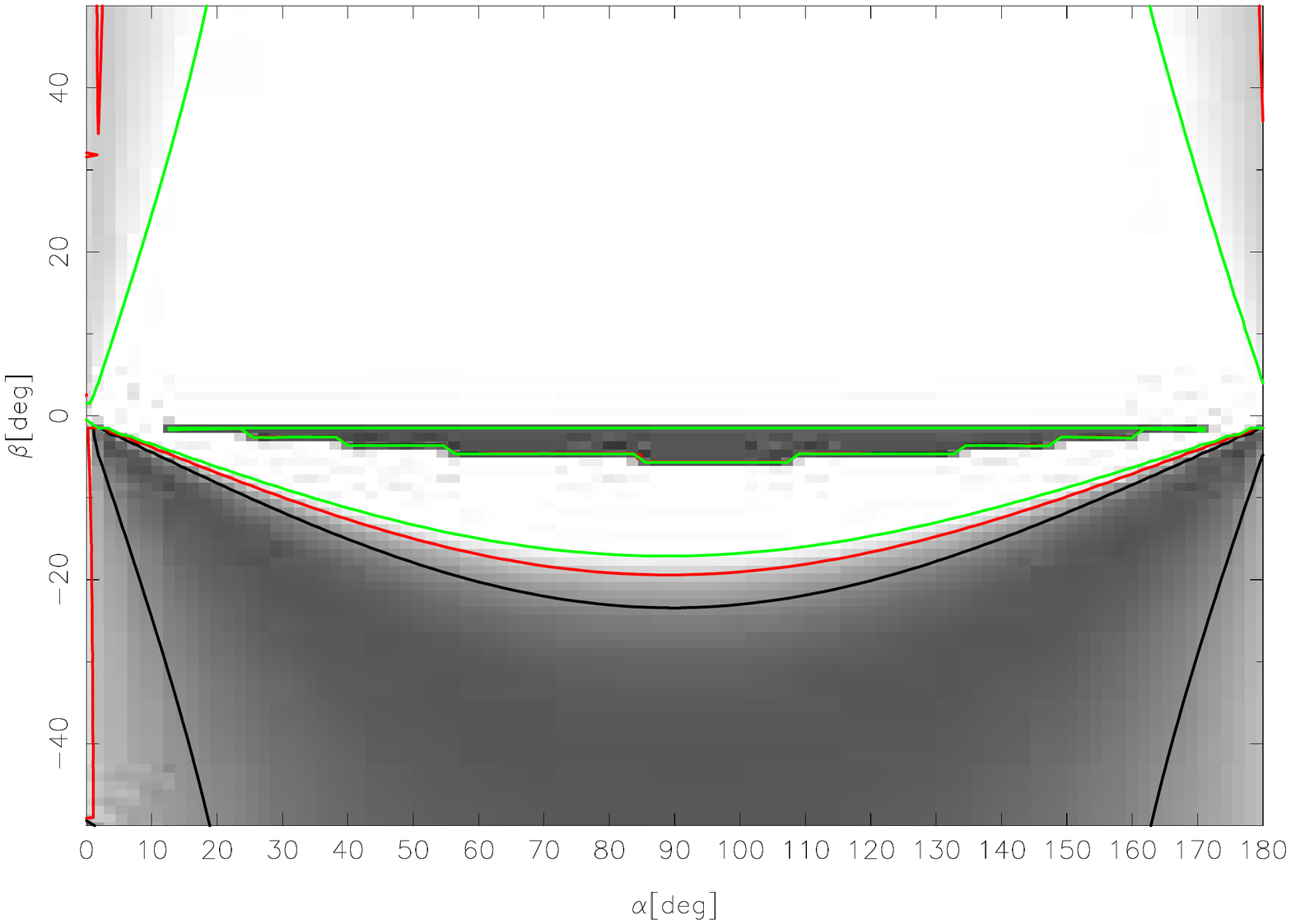}}}\\
&
\\
\end{tabular}
\caption{Top panel (upper window) shows the average profile with total
intensity (Stokes I; solid black lines), total linear polarization (dashed red
line) and circular polarization (Stokes V; dotted blue line). Top panel (lower
window) also shows the single pulse PPA distribution (colour scale) along with
the average PPA (red error bars).
The RVM fits to the average PPA (dashed pink
line) is also shown in this plot. Bottom panel show
the $\chi^2$ contours for the parameters $\alpha$ and $\beta$ obtained from RVM
fits.}
\label{a37}
\end{center}
\end{figure*}


\begin{figure*}
\begin{center}
\begin{tabular}{cc}
&
{\mbox{\includegraphics[width=9cm,height=6cm,angle=0.]{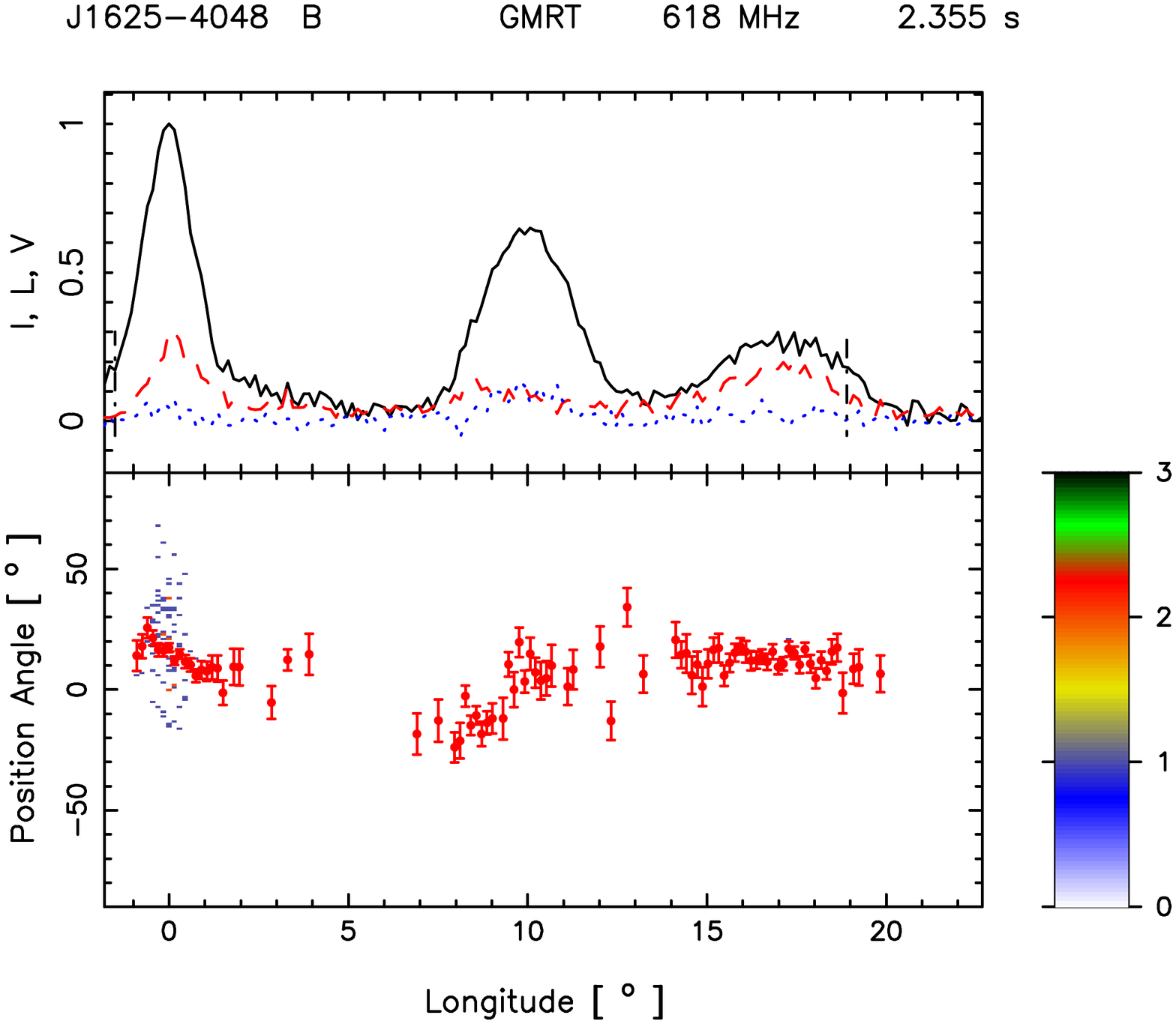}}}\\
&
\\
&
{\mbox{\includegraphics[width=9cm,height=6cm,angle=0.]{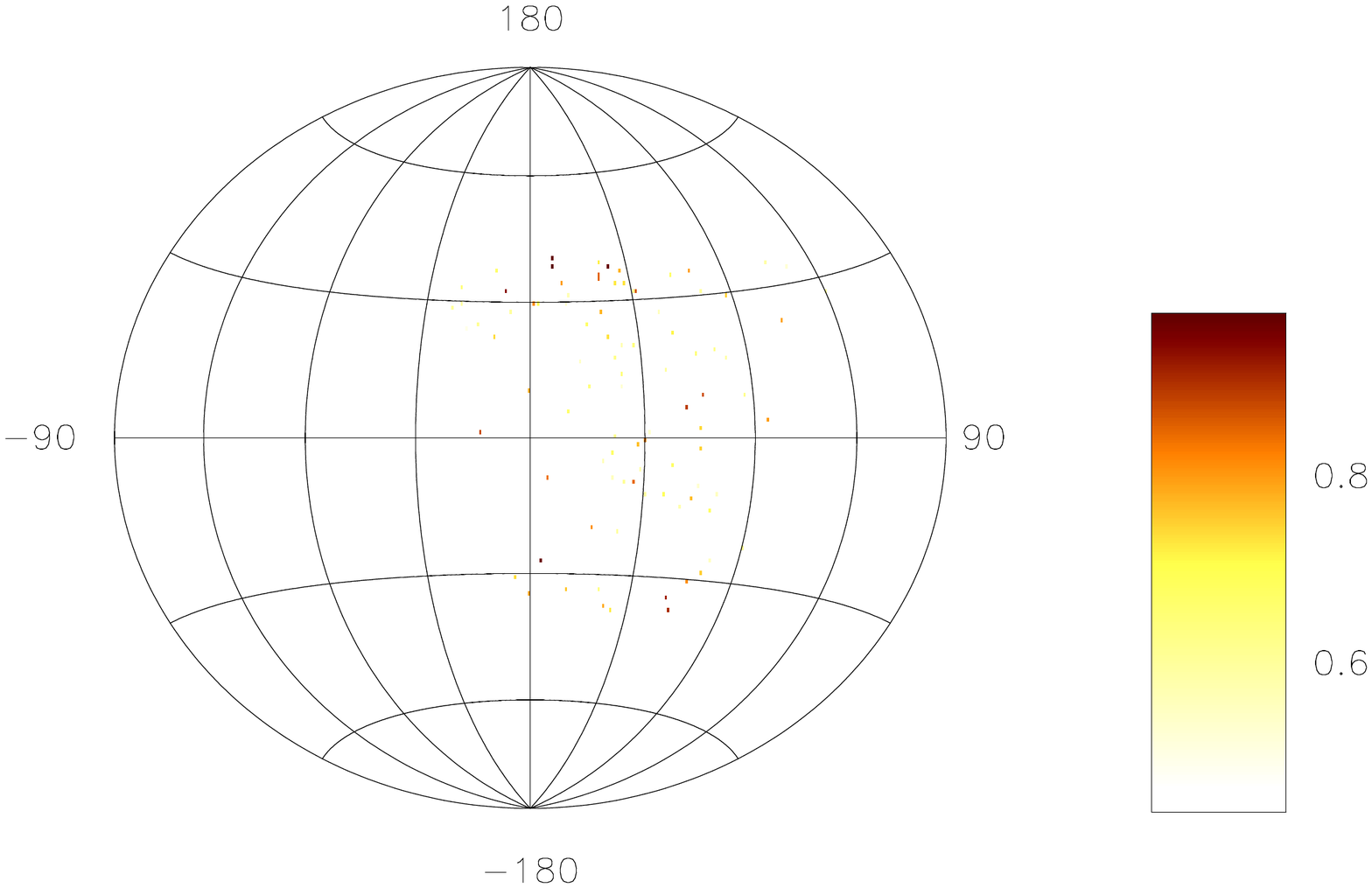}}}\\
\end{tabular}
\caption{Top panel only for 618 MHz (upper window) shows the average profile with total
intensity (Stokes I; solid black lines), total linear polarization (dashed red
line) and circular polarization (Stokes V; dotted blue line). Top panel (lower
window) also shows the single pulse PPA distribution (colour scale) along with
the average PPA (red error bars).
Bottom panel only for 618 MHz shows the Hammer-Aitoff projection of the polarized time
samples with the colour scheme representing the fractional polarization level.}
\label{a38}
\end{center}
\end{figure*}


\begin{figure*}
\begin{center}
\begin{tabular}{cc}
&
{\mbox{\includegraphics[width=9cm,height=6cm,angle=0.]{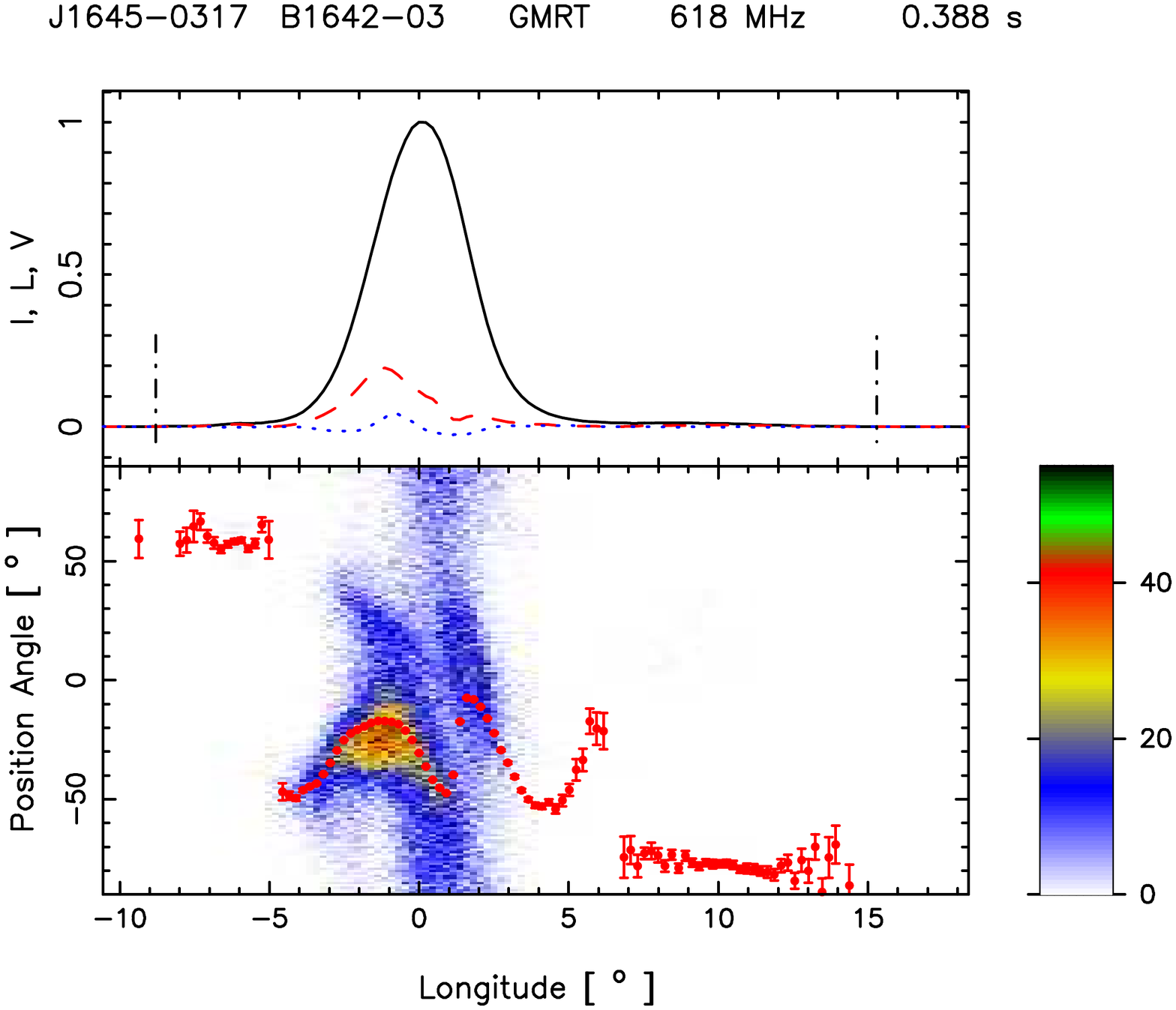}}}\\
&
\\
&
{\mbox{\includegraphics[width=9cm,height=6cm,angle=0.]{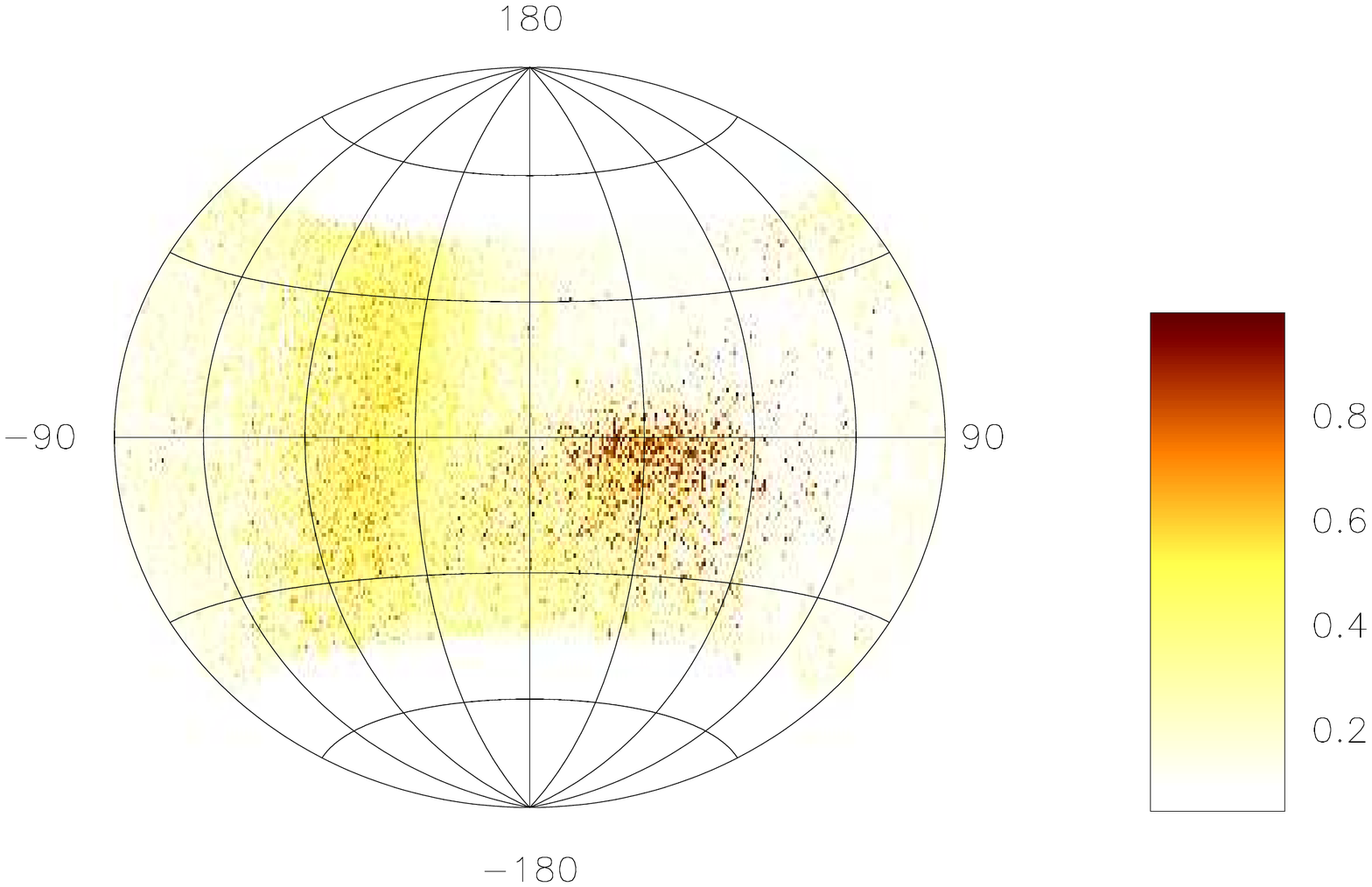}}}\\
\end{tabular}
\caption{Top panel only for 618 MHz (upper window) shows the average profile with total
intensity (Stokes I; solid black lines), total linear polarization (dashed red
line) and circular polarization (Stokes V; dotted blue line). Top panel (lower
window) also shows the single pulse PPA distribution (colour scale) along with
the average PPA (red error bars).
Bottom panel only for 618 MHz shows the Hammer-Aitoff projection of the polarized time
samples with the colour scheme representing the fractional polarization level.}
\label{a39}
\end{center}
\end{figure*}


\begin{figure*}
\begin{center}
\begin{tabular}{cc}
&
{\mbox{\includegraphics[width=9cm,height=6cm,angle=0.]{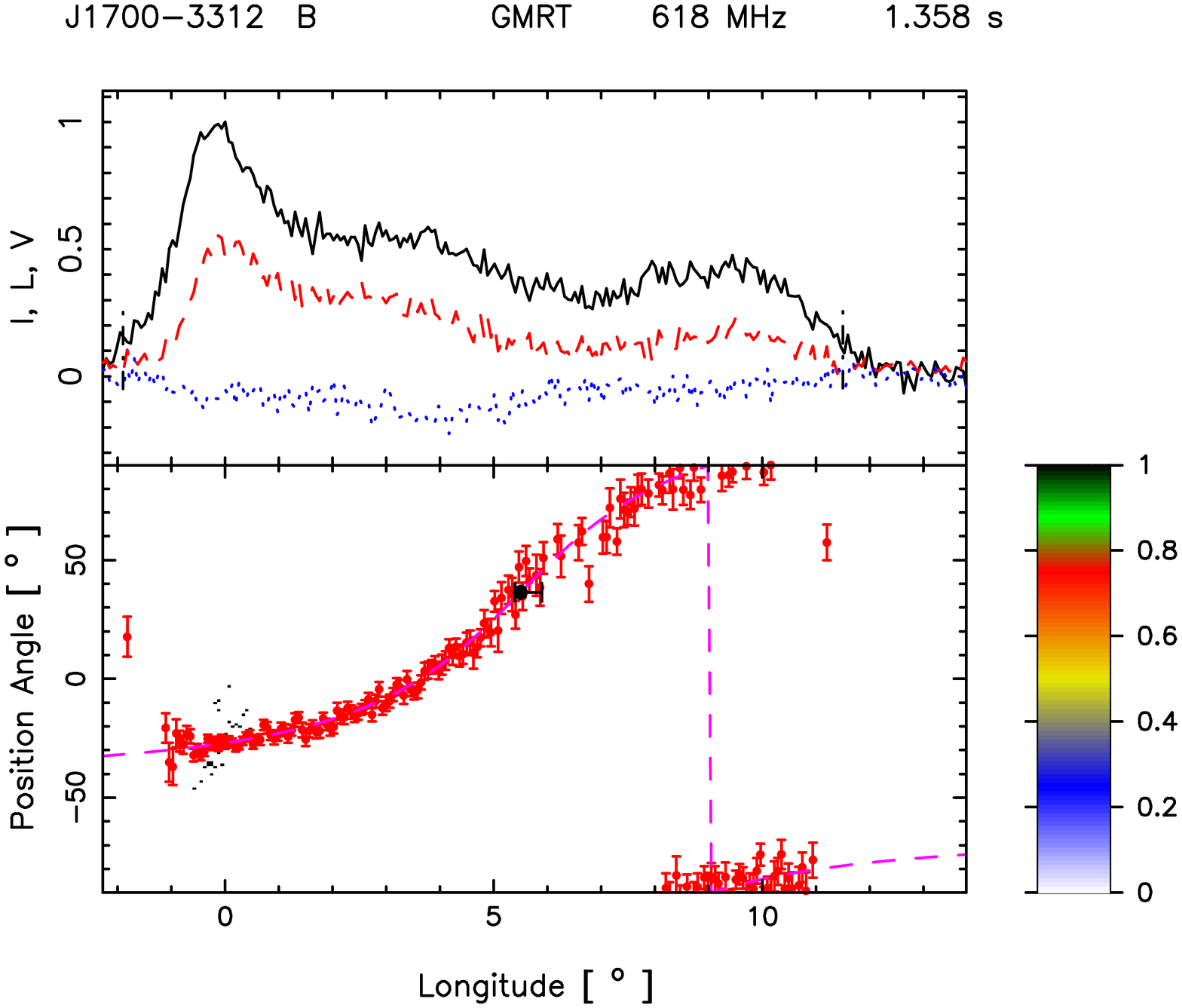}}}\\
&
{\mbox{\includegraphics[width=9cm,height=6cm,angle=0.]{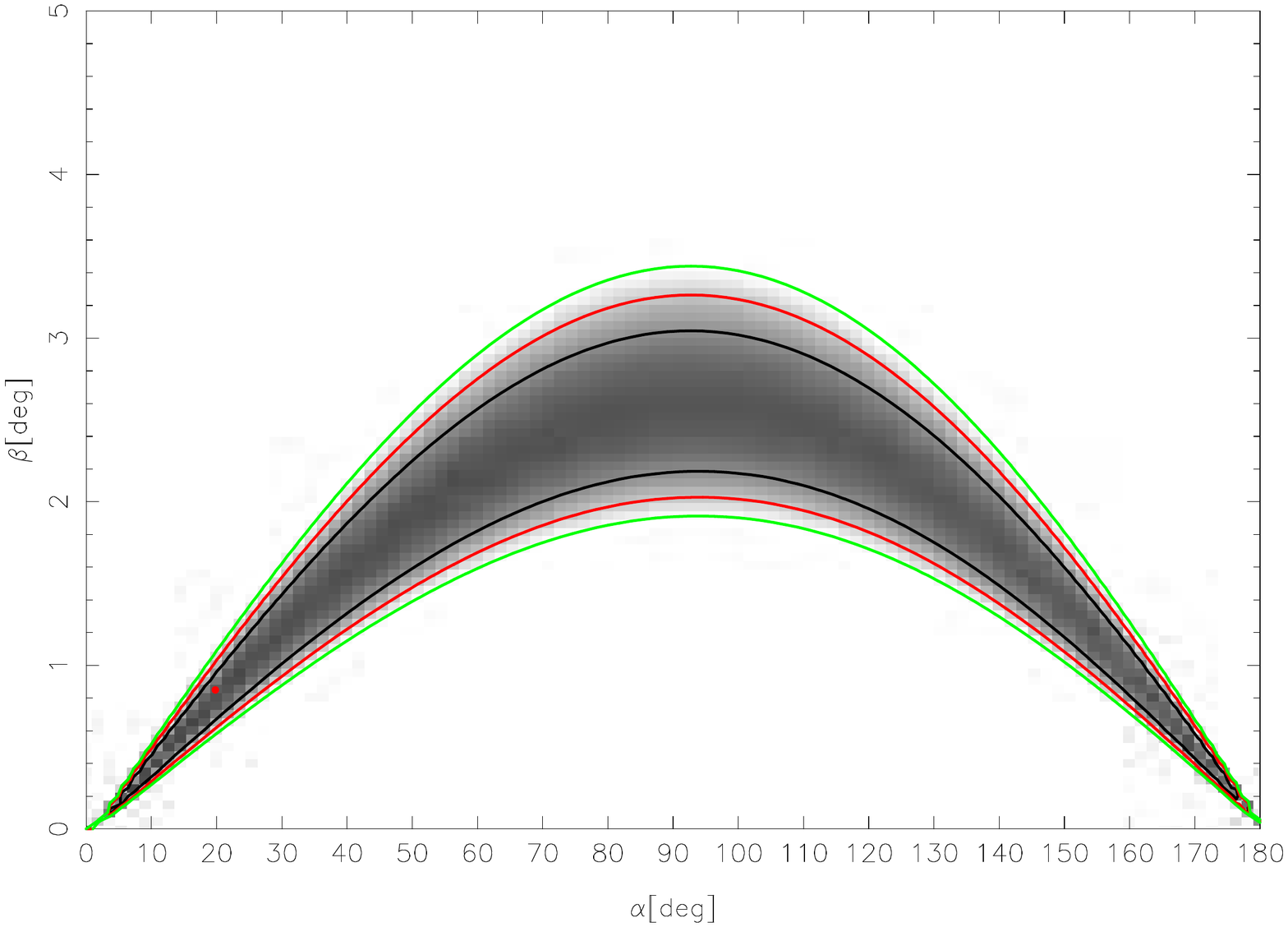}}}\\
&
\\
\end{tabular}
\caption{Top panel only for 618 MHz (upper window) shows the average profile with total
intensity (Stokes I; solid black lines), total linear polarization (dashed red
line) and circular polarization (Stokes V; dotted blue line). Top panel (lower
window) also shows the single pulse PPA distribution (colour scale) along with
the average PPA (red error bars).
The RVM fits to the average PPA (dashed pink
line) is also shown in this plot. Bottom panel only for 618 MHz show
the $\chi^2$ contours for the parameters $\alpha$ and $\beta$ obtained from RVM
fits.}
\label{a40}
\end{center}
\end{figure*}


\begin{figure*}
\begin{center}
\begin{tabular}{cc}
{\mbox{\includegraphics[width=9cm,height=6cm,angle=0.]{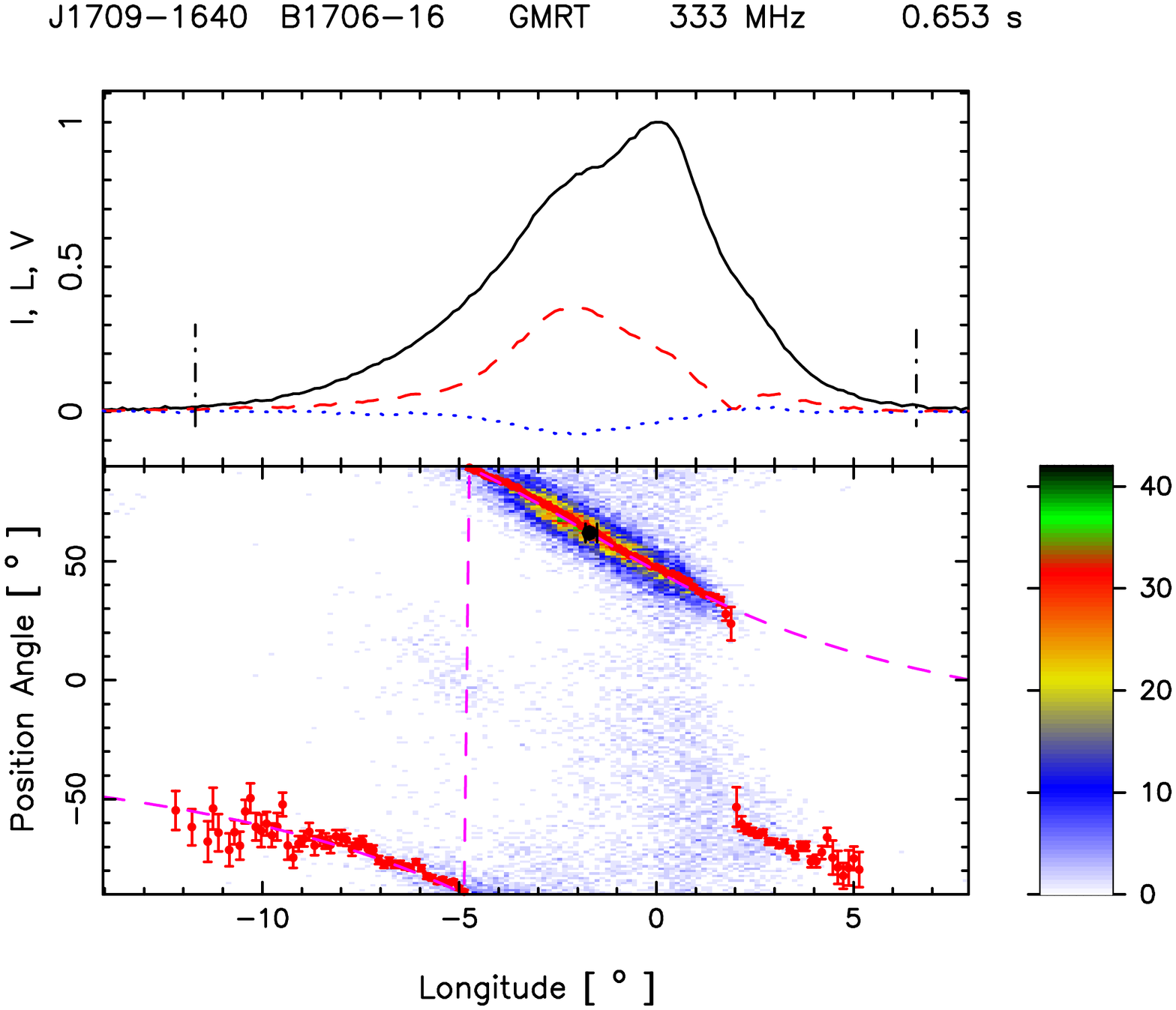}}}&
{\mbox{\includegraphics[width=9cm,height=6cm,angle=0.]{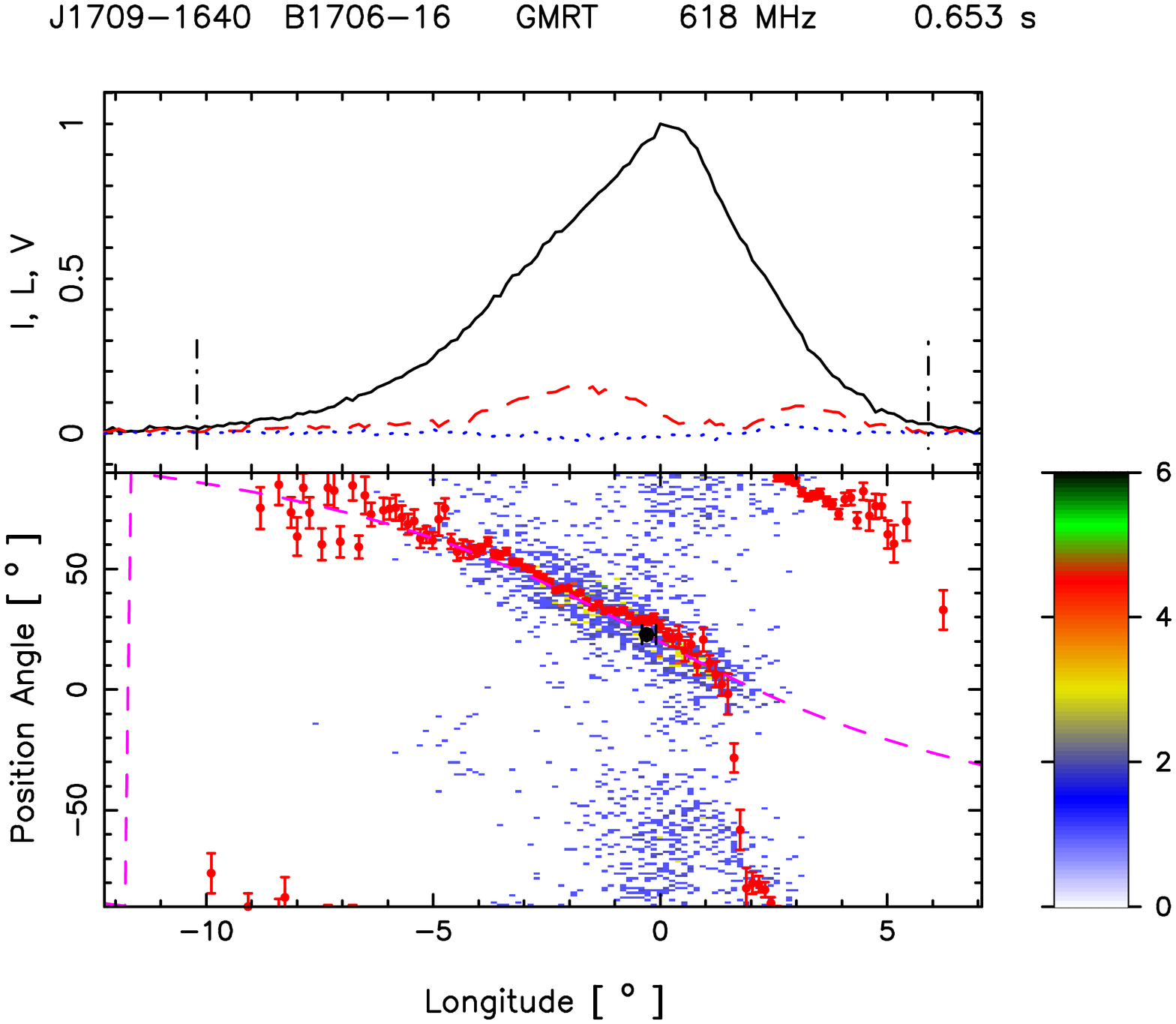}}}\\
{\mbox{\includegraphics[width=9cm,height=6cm,angle=0.]{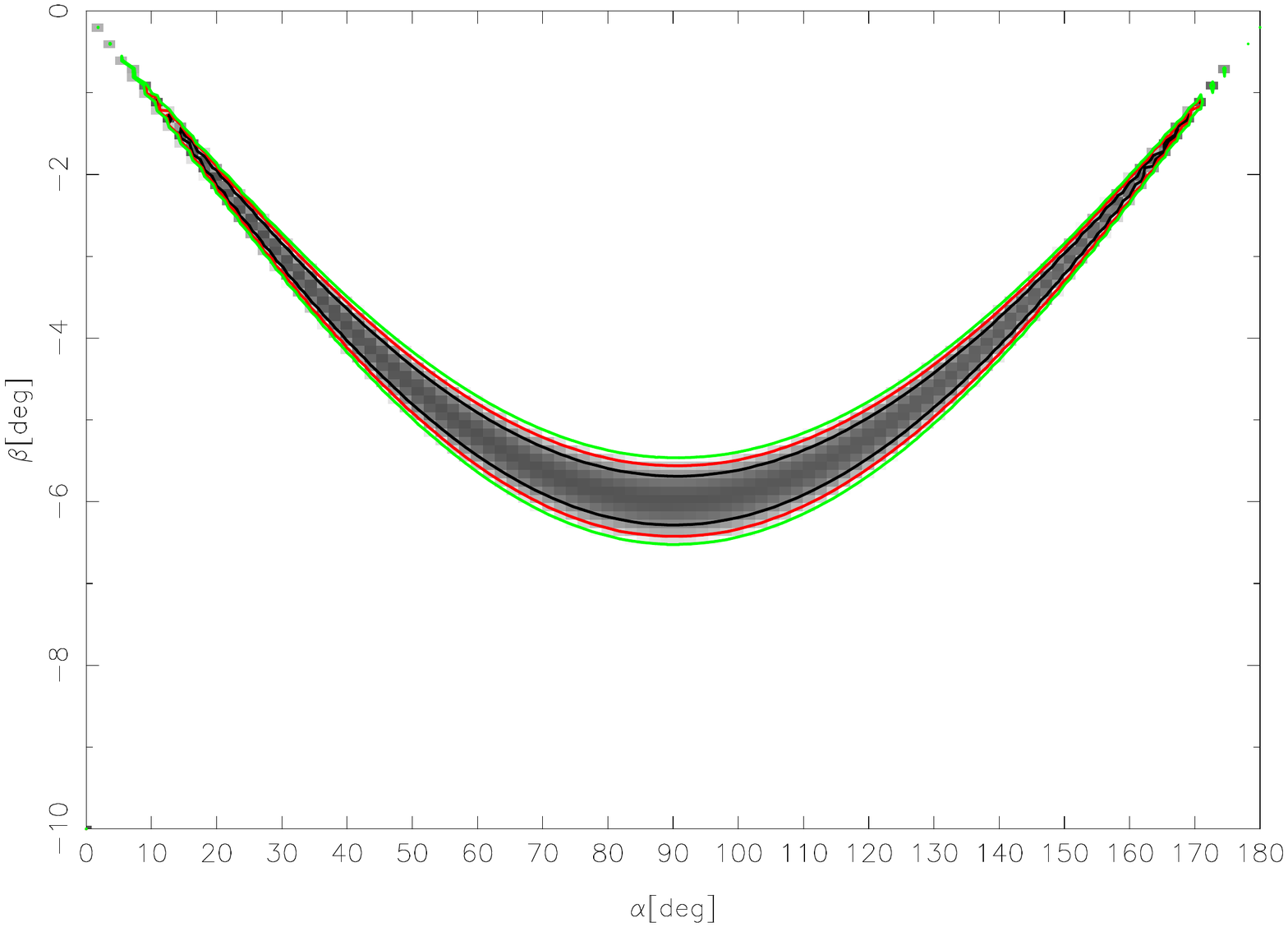}}}&
{\mbox{\includegraphics[width=9cm,height=6cm,angle=0.]{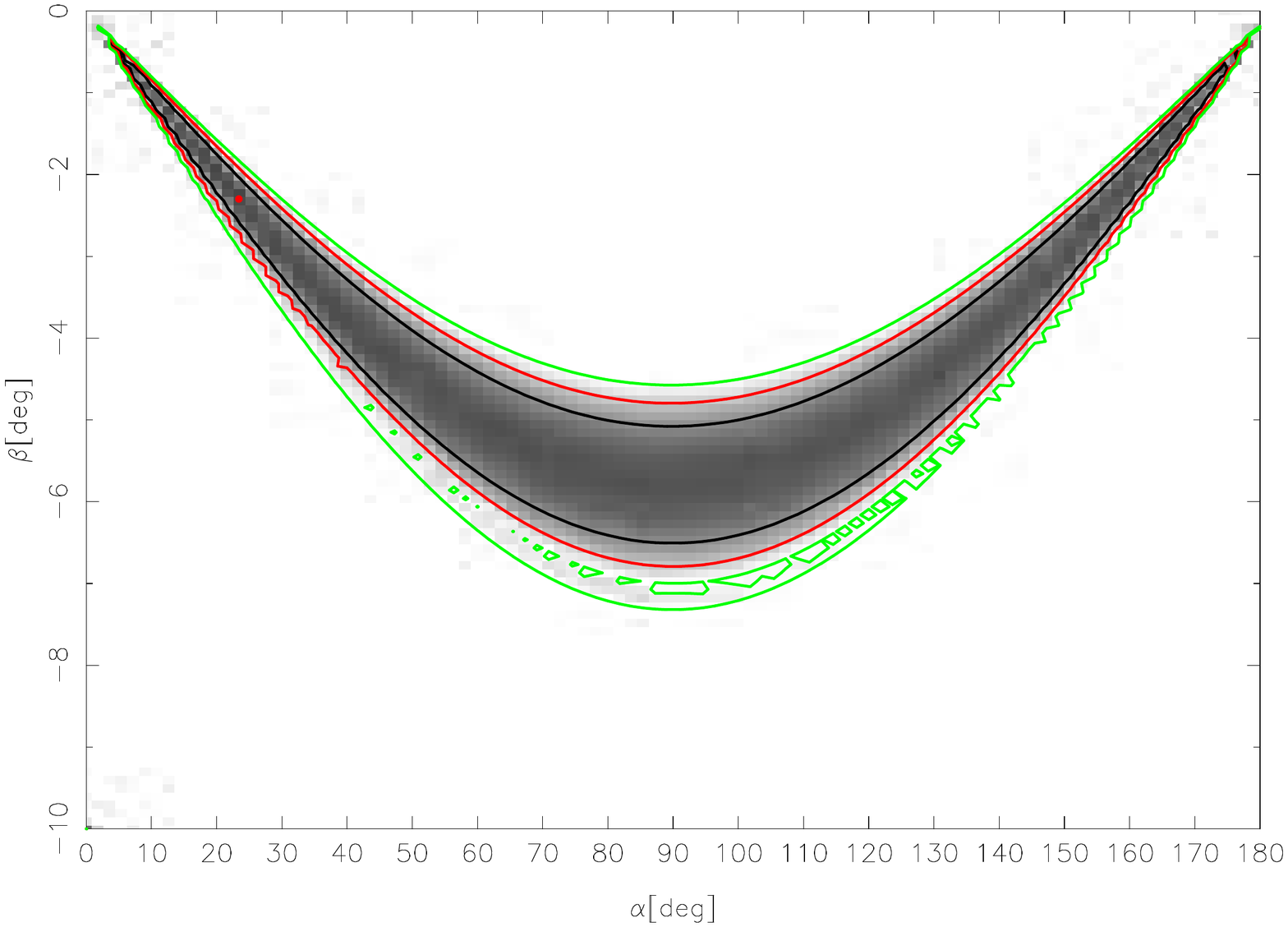}}}\\
{\mbox{\includegraphics[width=9cm,height=6cm,angle=0.]{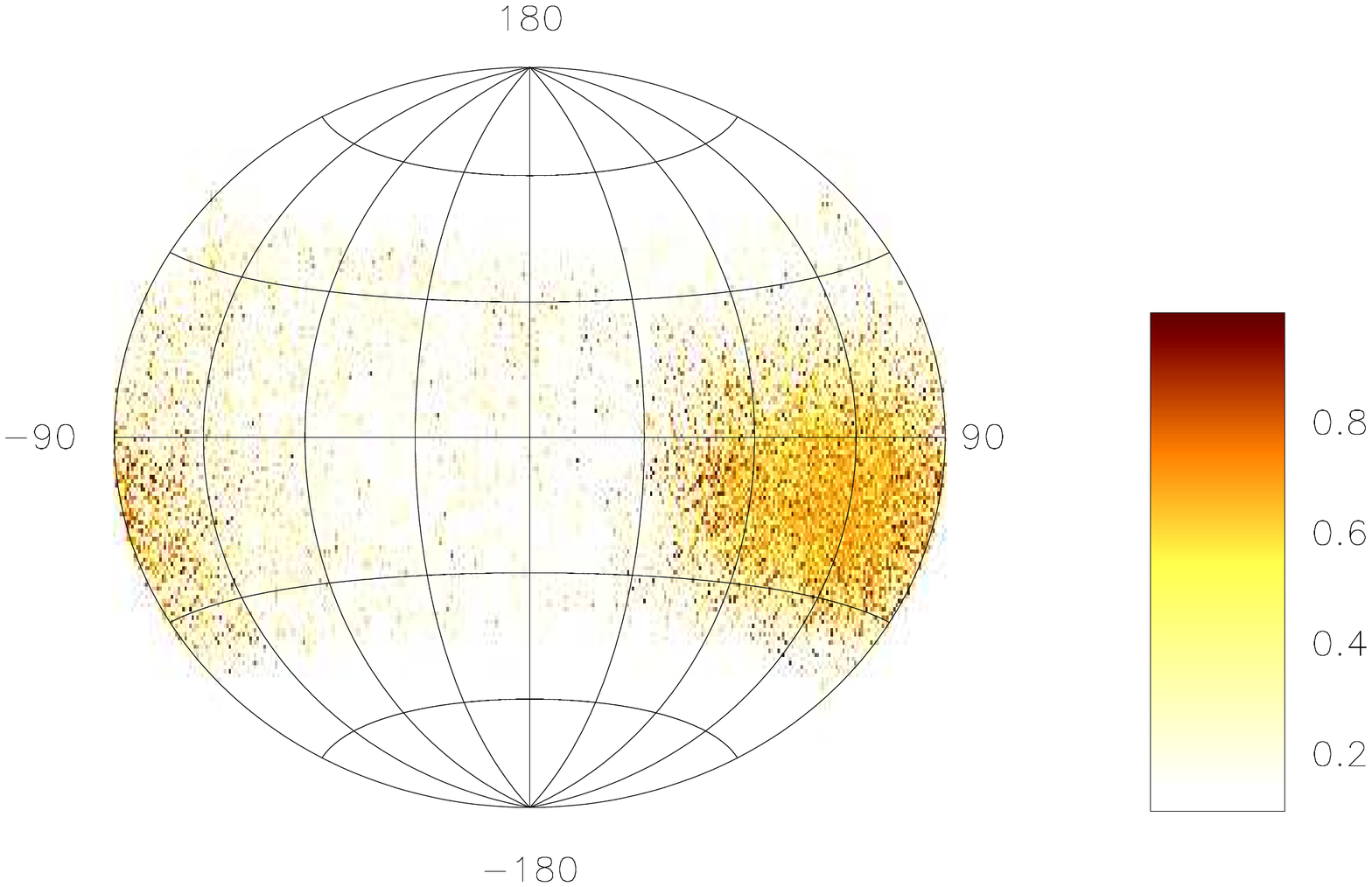}}}&
{\mbox{\includegraphics[width=9cm,height=6cm,angle=0.]{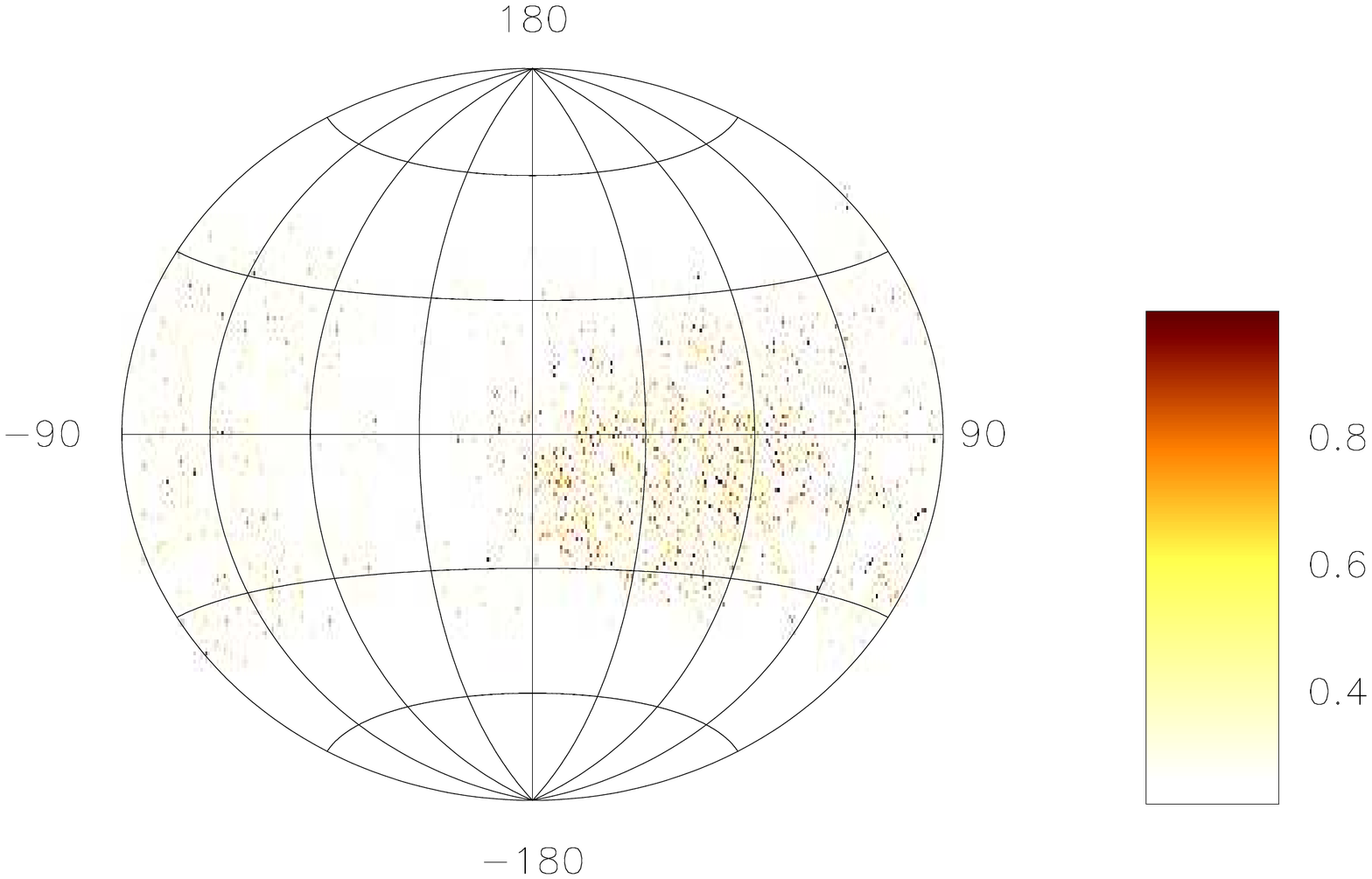}}}\\
\end{tabular}
\caption{Top panel (upper window) shows the average profile with total
intensity (Stokes I; solid black lines), total linear polarization (dashed red
line) and circular polarization (Stokes V; dotted blue line). Top panel (lower
window) also shows the single pulse PPA distribution (colour scale) along with
the average PPA (red error bars).
The RVM fits to the average PPA (dashed pink
line) is also shown in this plot. Middle panel show
the $\chi^2$ contours for the parameters $\alpha$ and $\beta$ obtained from RVM
fits.
Bottom panel shows the Hammer-Aitoff projection of the polarized time
samples with the colour scheme representing the fractional polarization level.}
\label{a41}
\end{center}
\end{figure*}


\begin{figure*}
\begin{center}
\begin{tabular}{cc}
&
{\mbox{\includegraphics[width=9cm,height=6cm,angle=0.]{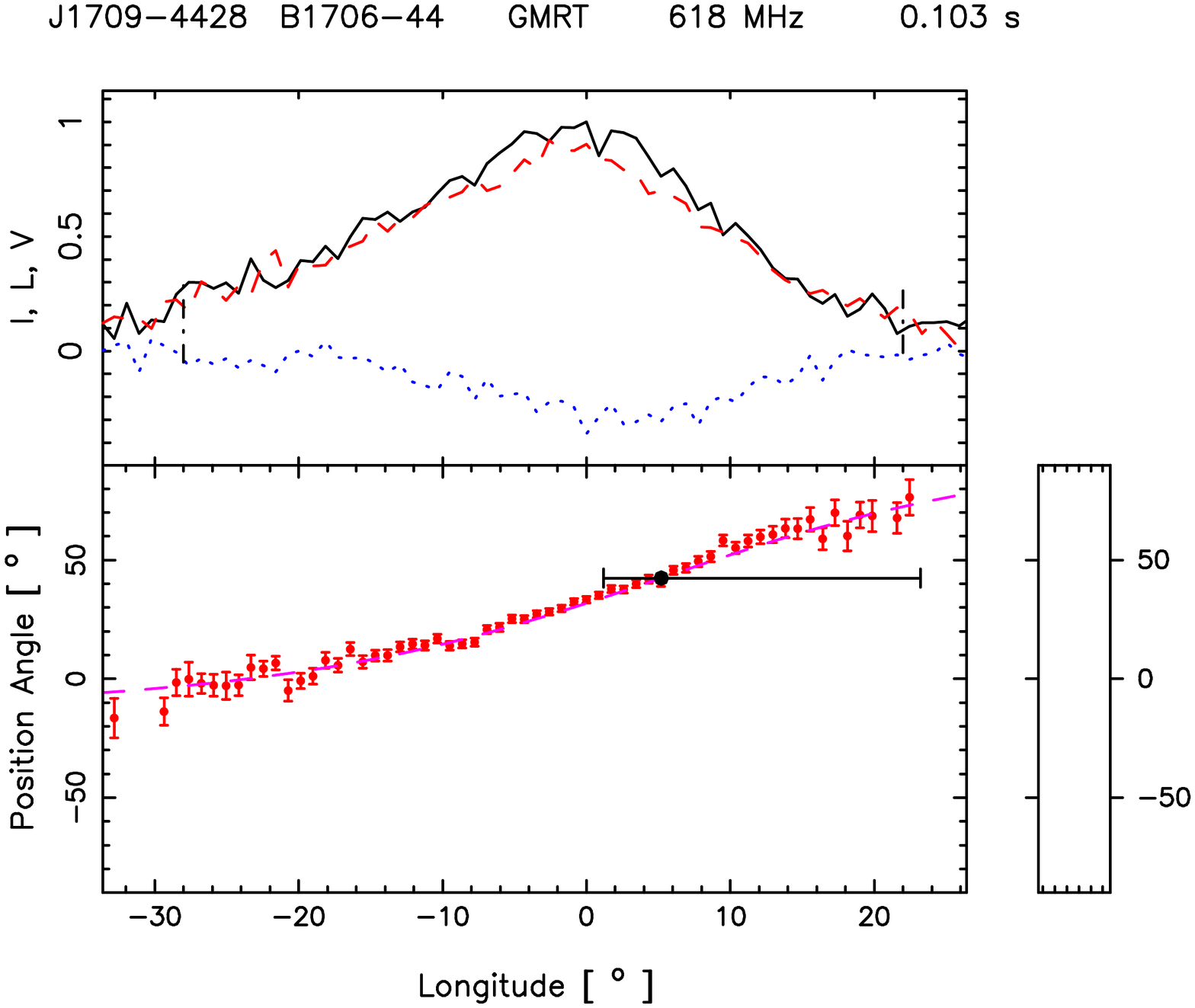}}}\\
&
{\mbox{\includegraphics[width=9cm,height=6cm,angle=0.]{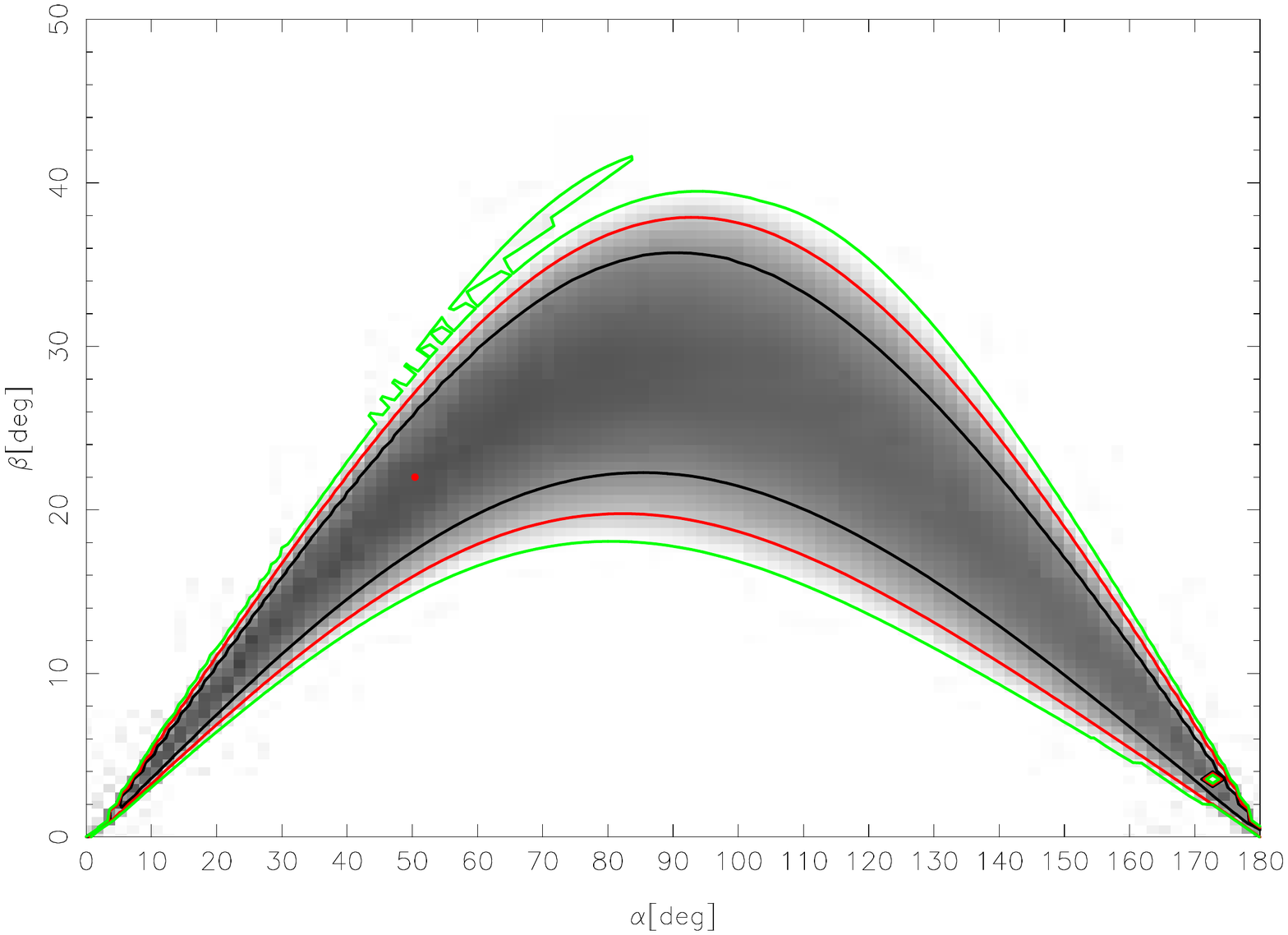}}}\\
&
\\
\end{tabular}
\caption{Top panel only for 618 MHz (upper window) shows the average profile with total
intensity (Stokes I; solid black lines), total linear polarization (dashed red
line) and circular polarization (Stokes V; dotted blue line). Top panel (lower
window) also shows the single pulse PPA distribution (colour scale) along with
the average PPA (red error bars).
The RVM fits to the average PPA (dashed pink
line) is also shown in this plot. Bottom panel only for 618 MHz show
the $\chi^2$ contours for the parameters $\alpha$ and $\beta$ obtained from RVM
fits.}
\label{a42}
\end{center}
\end{figure*}


\begin{figure*}
\begin{center}
\begin{tabular}{cc}
{\mbox{\includegraphics[width=9cm,height=6cm,angle=0.]{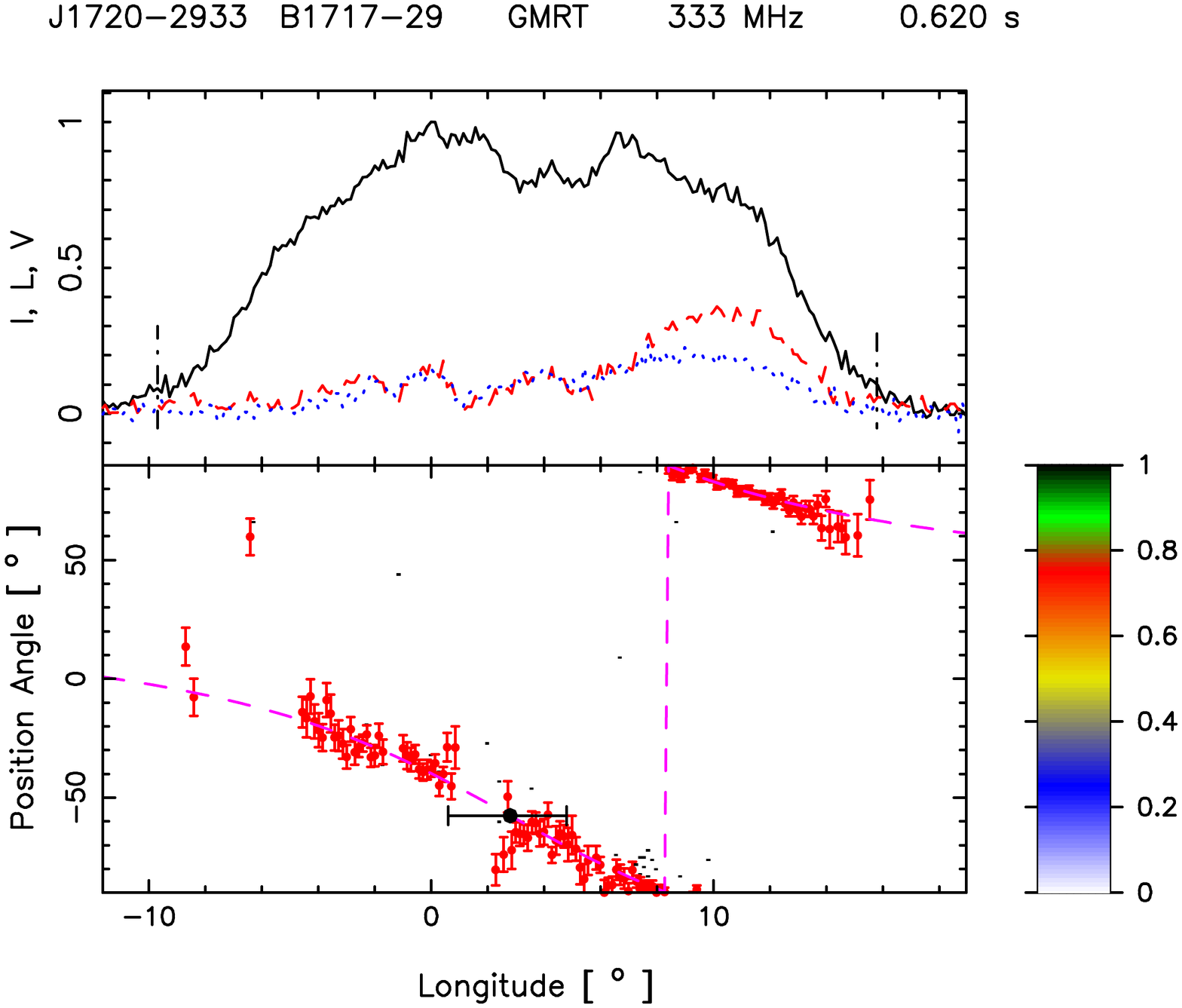}}}&
{\mbox{\includegraphics[width=9cm,height=6cm,angle=0.]{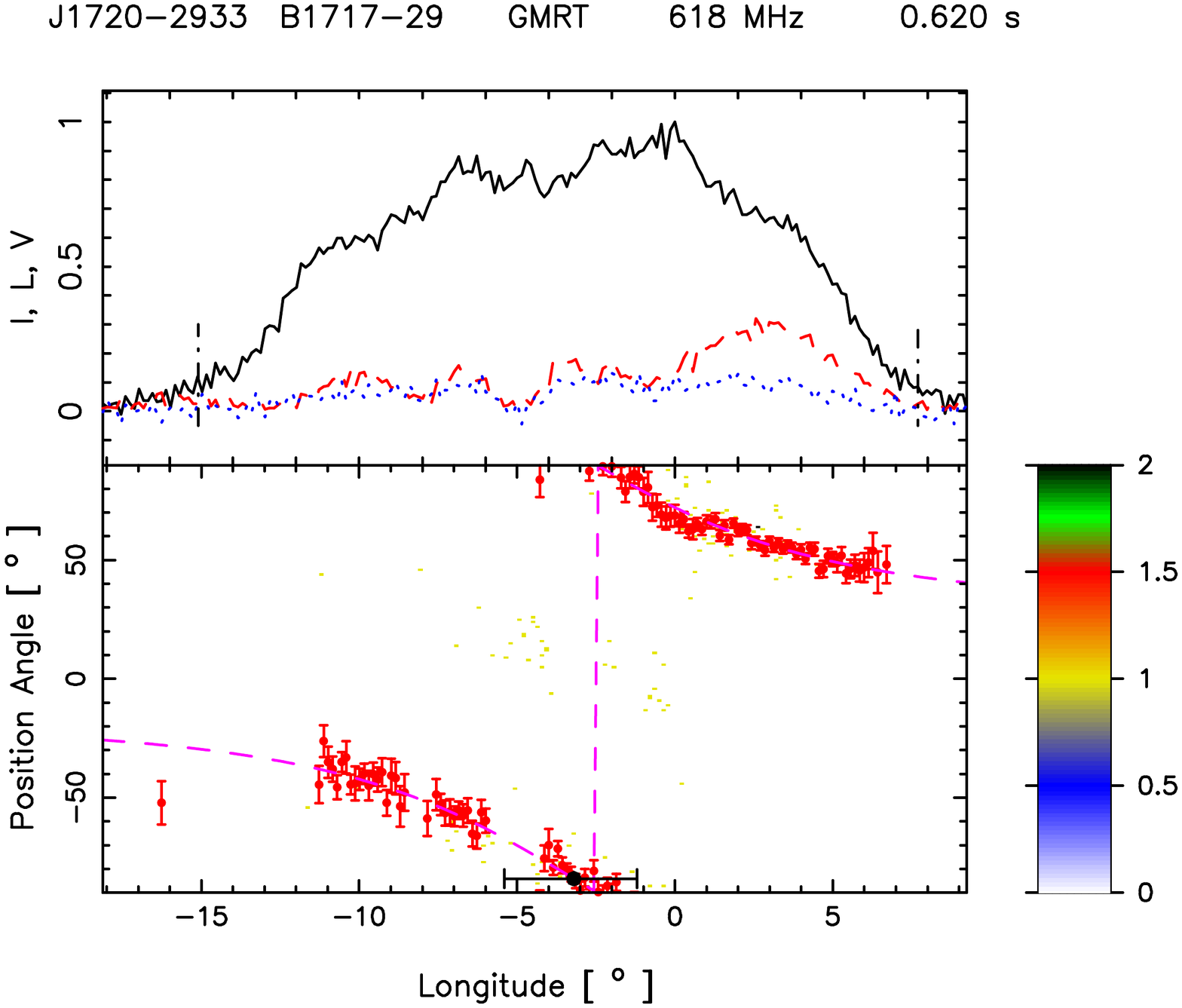}}}\\
{\mbox{\includegraphics[width=9cm,height=6cm,angle=0.]{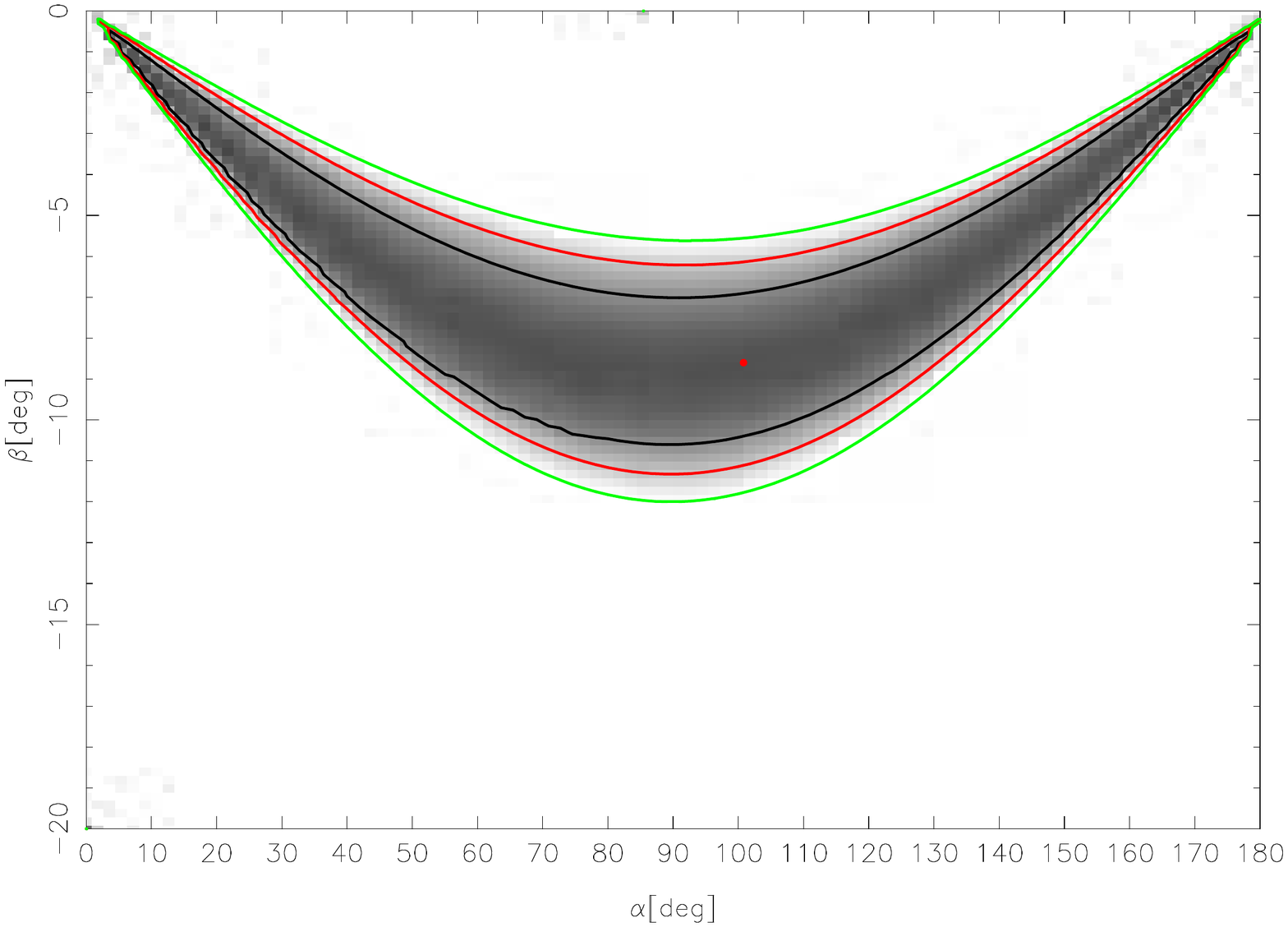}}}&
{\mbox{\includegraphics[width=9cm,height=6cm,angle=0.]{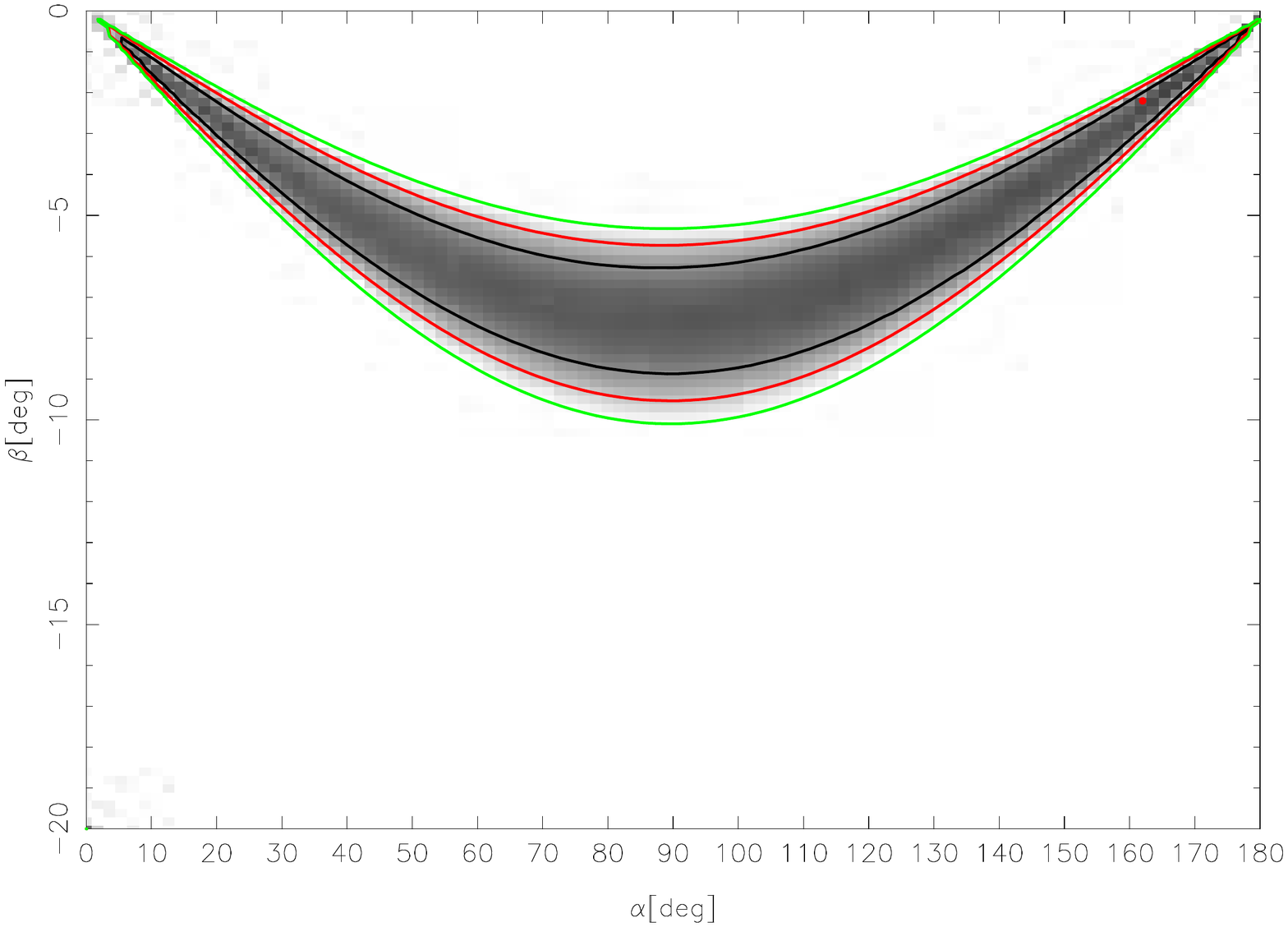}}}\\
&
{\mbox{\includegraphics[width=9cm,height=6cm,angle=0.]{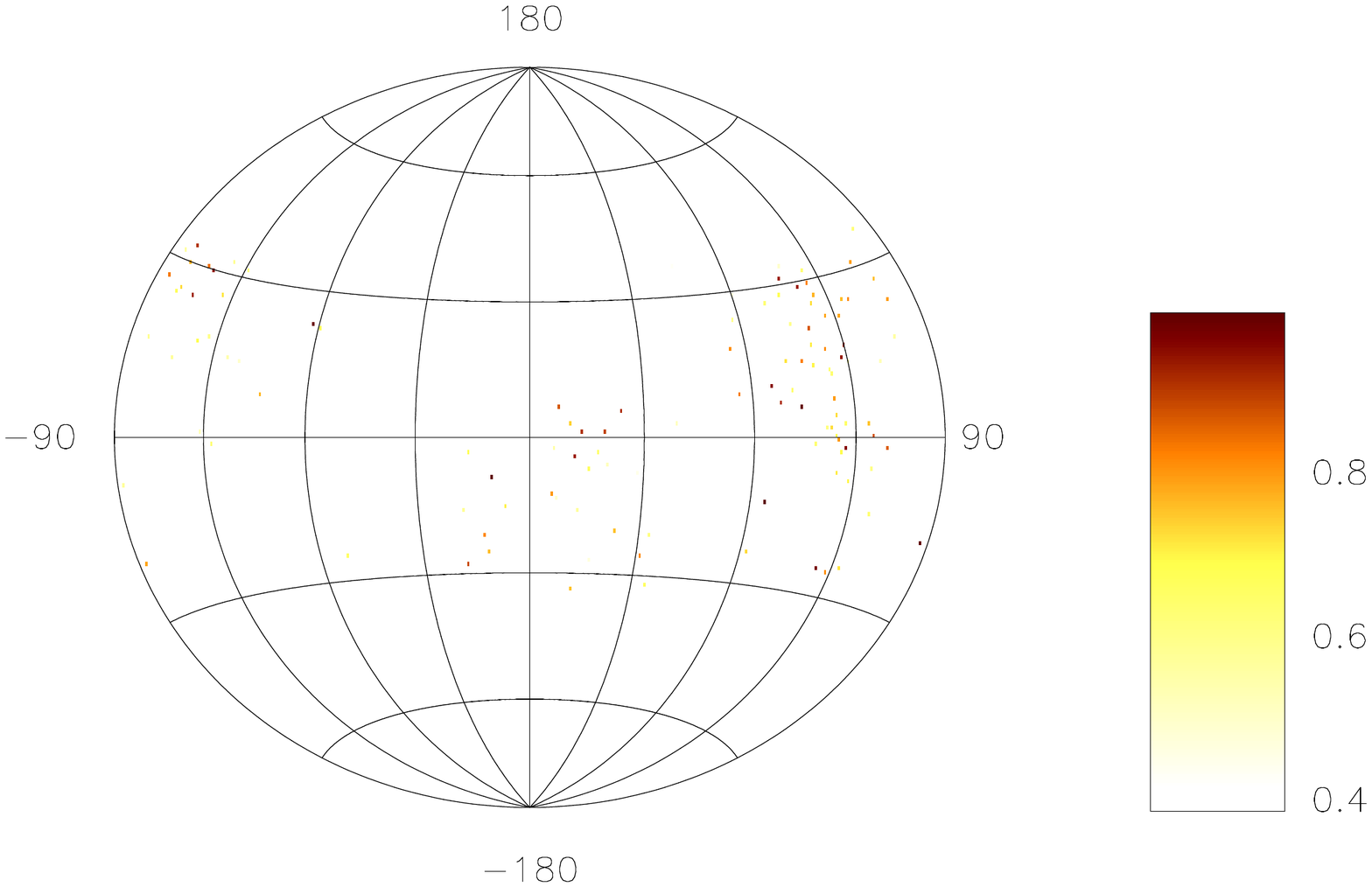}}}\\
\end{tabular}
\caption{Top panel (upper window) shows the average profile with total
intensity (Stokes I; solid black lines), total linear polarization (dashed red
line) and circular polarization (Stokes V; dotted blue line). Top panel (lower
window) also shows the single pulse PPA distribution (colour scale) along with
the average PPA (red error bars).
The RVM fits to the average PPA (dashed pink
line) is also shown in this plot. Middle panel show
the $\chi^2$ contours for the parameters $\alpha$ and $\beta$ obtained from RVM
fits.
Bottom panel only for 618 MHz shows the Hammer-Aitoff projection of the polarized time
samples with the colour scheme representing the fractional polarization level.}
\label{a43}
\end{center}
\end{figure*}

\begin{figure*}
\begin{center}
\begin{tabular}{cc}
{\mbox{\includegraphics[width=9cm,height=6cm,angle=0.]{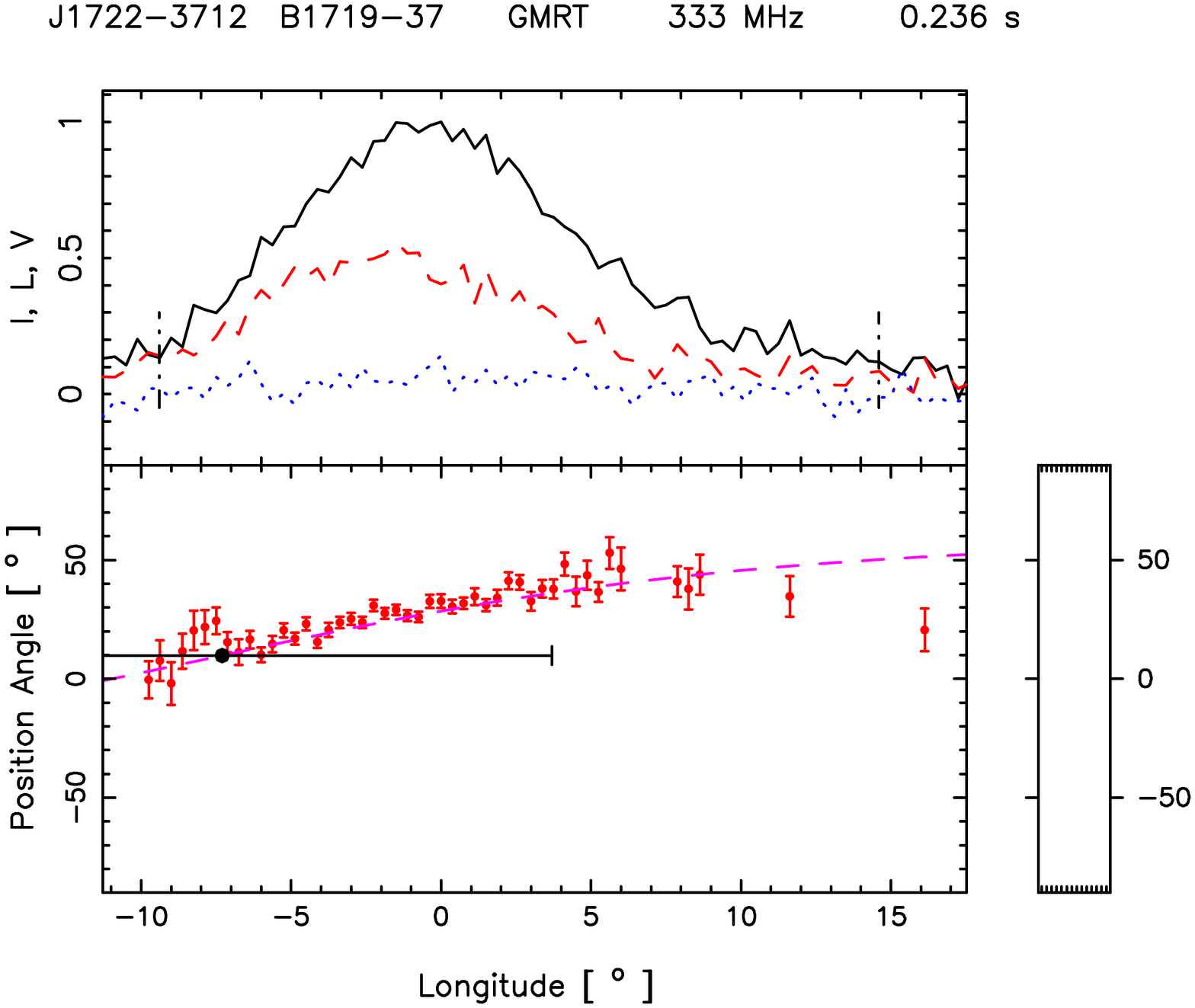}}}&
{\mbox{\includegraphics[width=9cm,height=6cm,angle=0.]{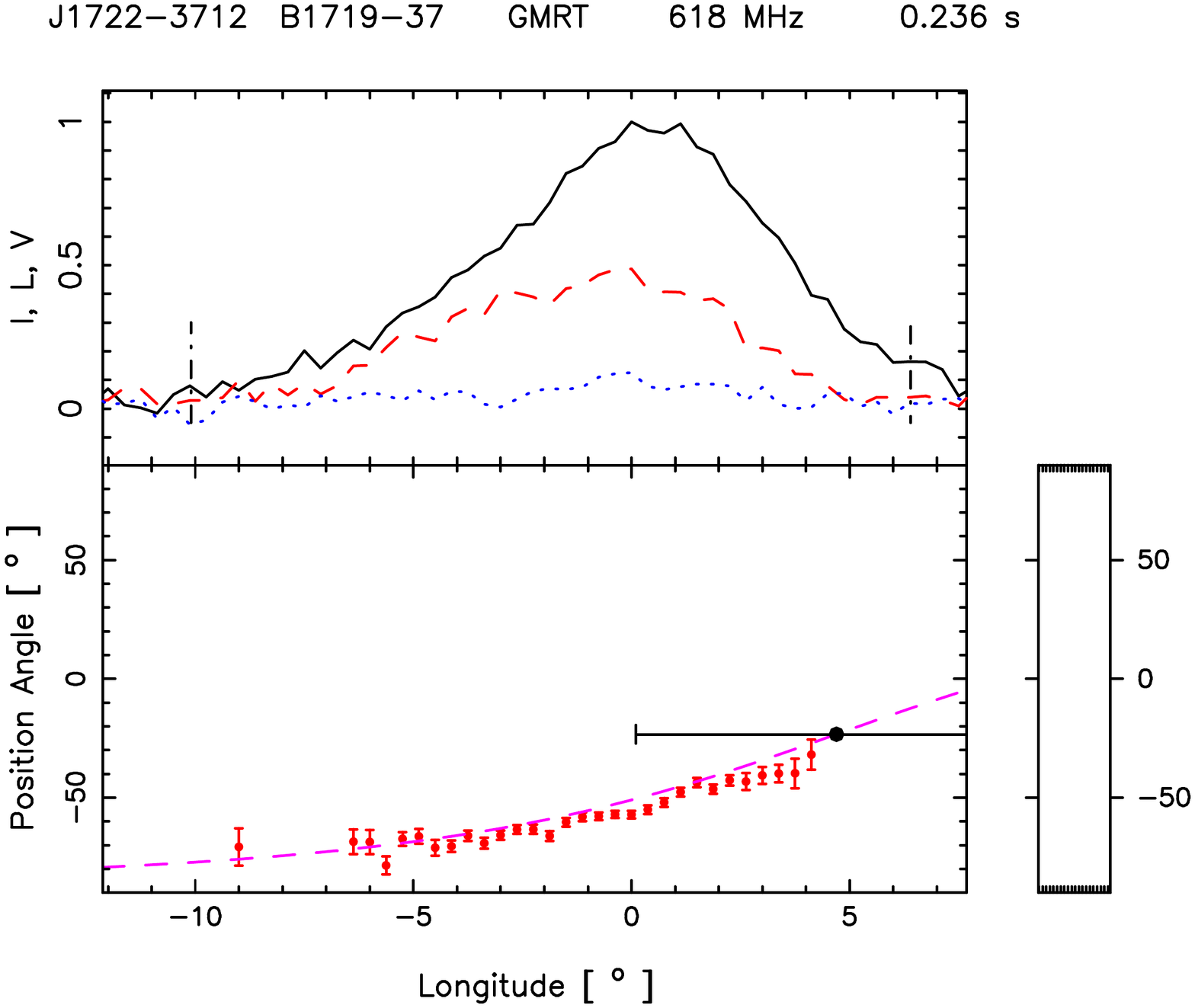}}}\\
{\mbox{\includegraphics[width=9cm,height=6cm,angle=0.]{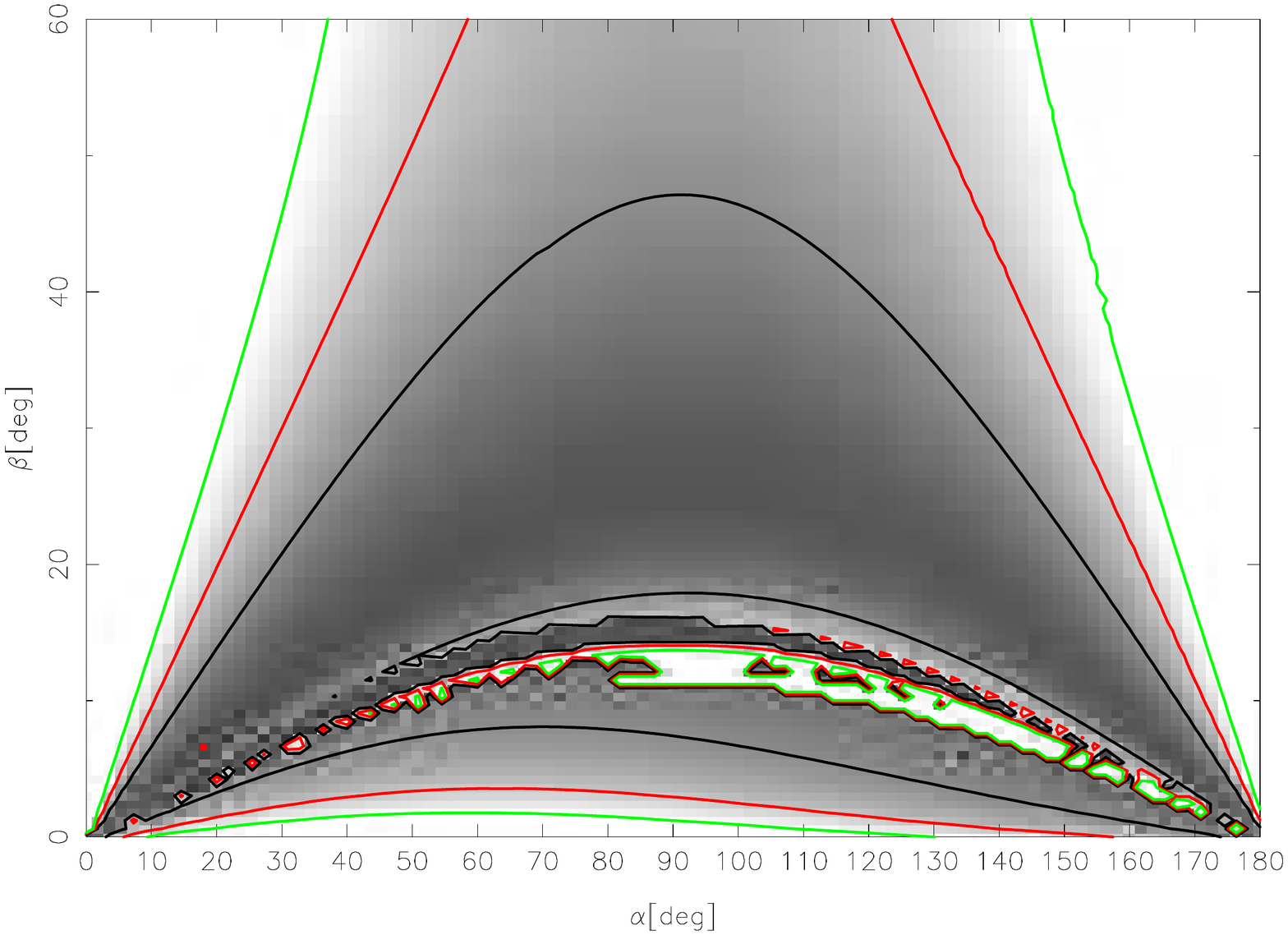}}}&
{\mbox{\includegraphics[width=9cm,height=6cm,angle=0.]{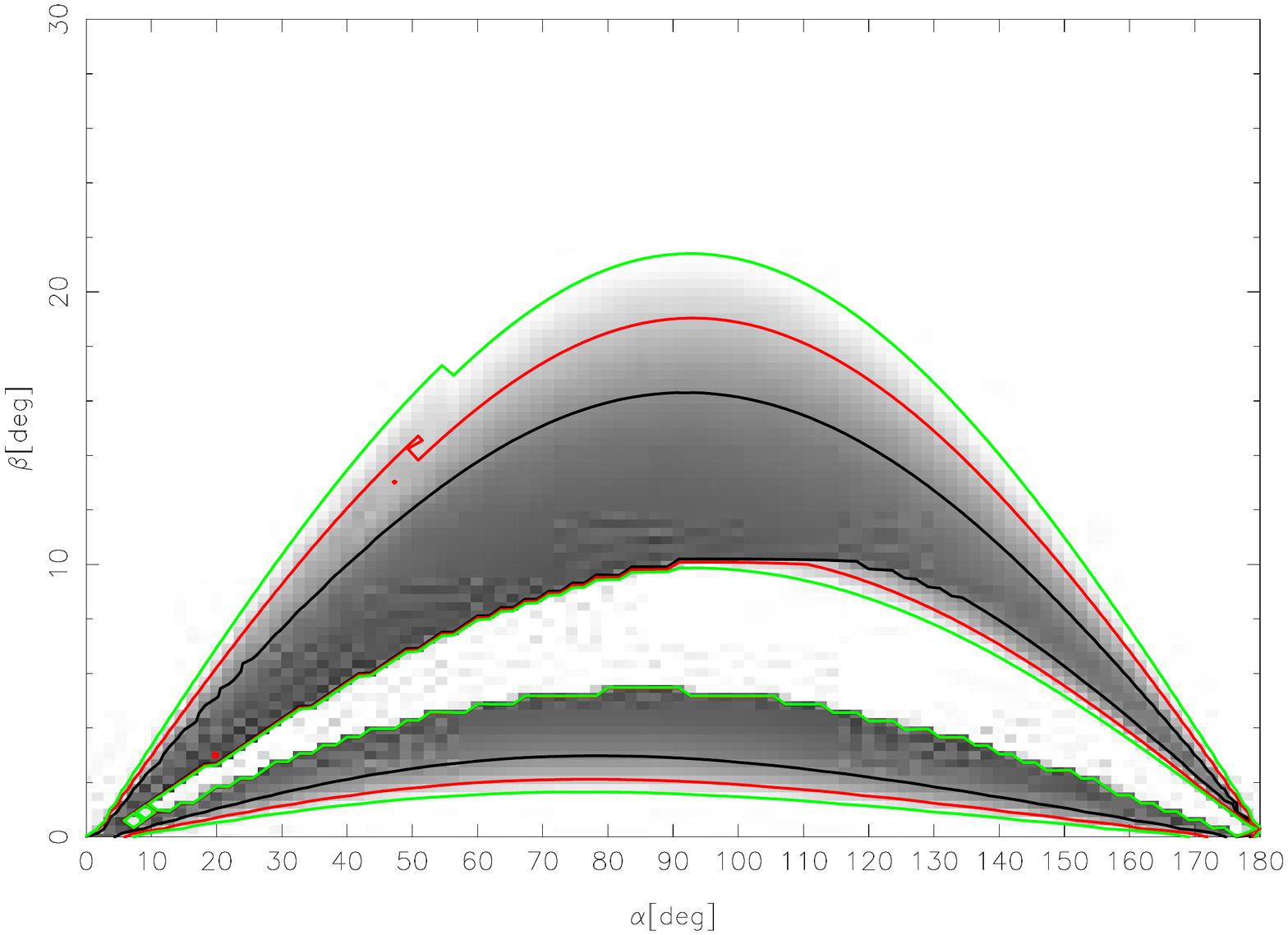}}}\\
&
\\
\end{tabular}
\caption{Top panel (upper window) shows the average profile with total
intensity (Stokes I; solid black lines), total linear polarization (dashed red
line) and circular polarization (Stokes V; dotted blue line). Top panel (lower
window) also shows the single pulse PPA distribution (colour scale) along with
the average PPA (red error bars).
The RVM fits to the average PPA (dashed pink
line) is also shown in this plot. Bottom panel show
the $\chi^2$ contours for the parameters $\alpha$ and $\beta$ obtained from RVM
fits.}
\label{a44}
\end{center}
\end{figure*}


\begin{figure*}
\begin{center}
\begin{tabular}{cc}
{\mbox{\includegraphics[width=9cm,height=6cm,angle=0.]{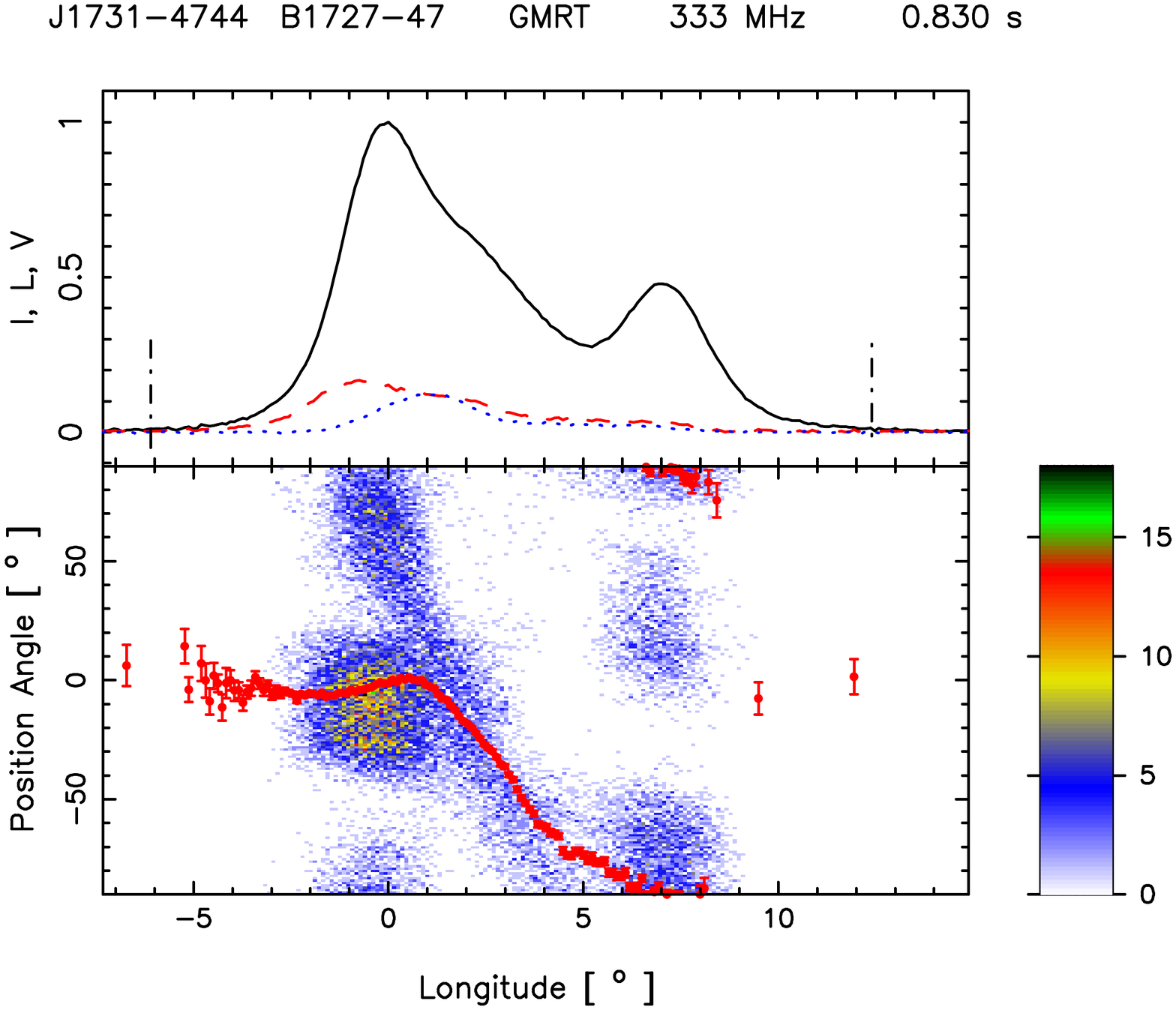}}}&
{\mbox{\includegraphics[width=9cm,height=6cm,angle=0.]{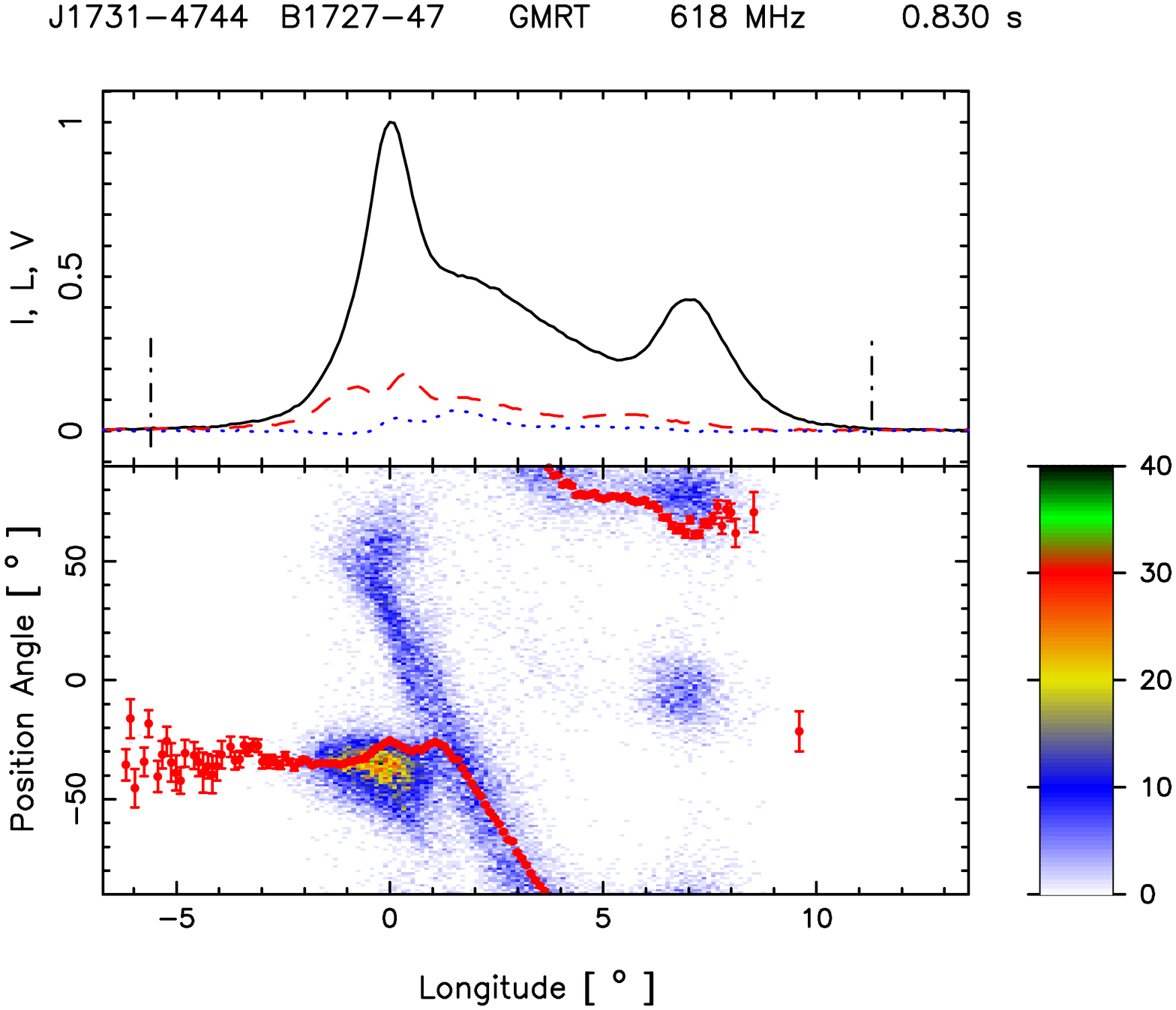}}}\\
&
\\
{\mbox{\includegraphics[width=9cm,height=6cm,angle=0.]{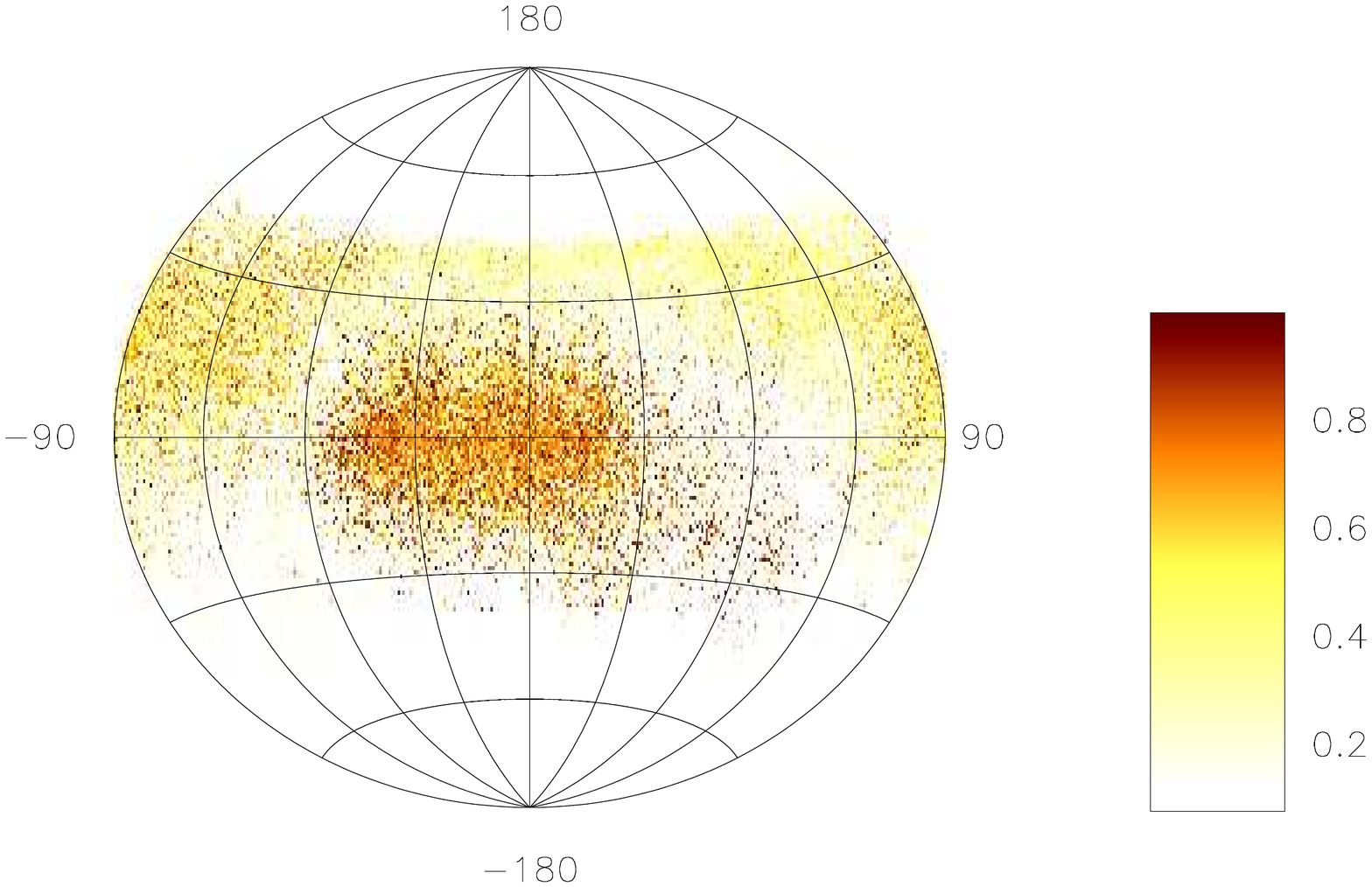}}}&
{\mbox{\includegraphics[width=9cm,height=6cm,angle=0.]{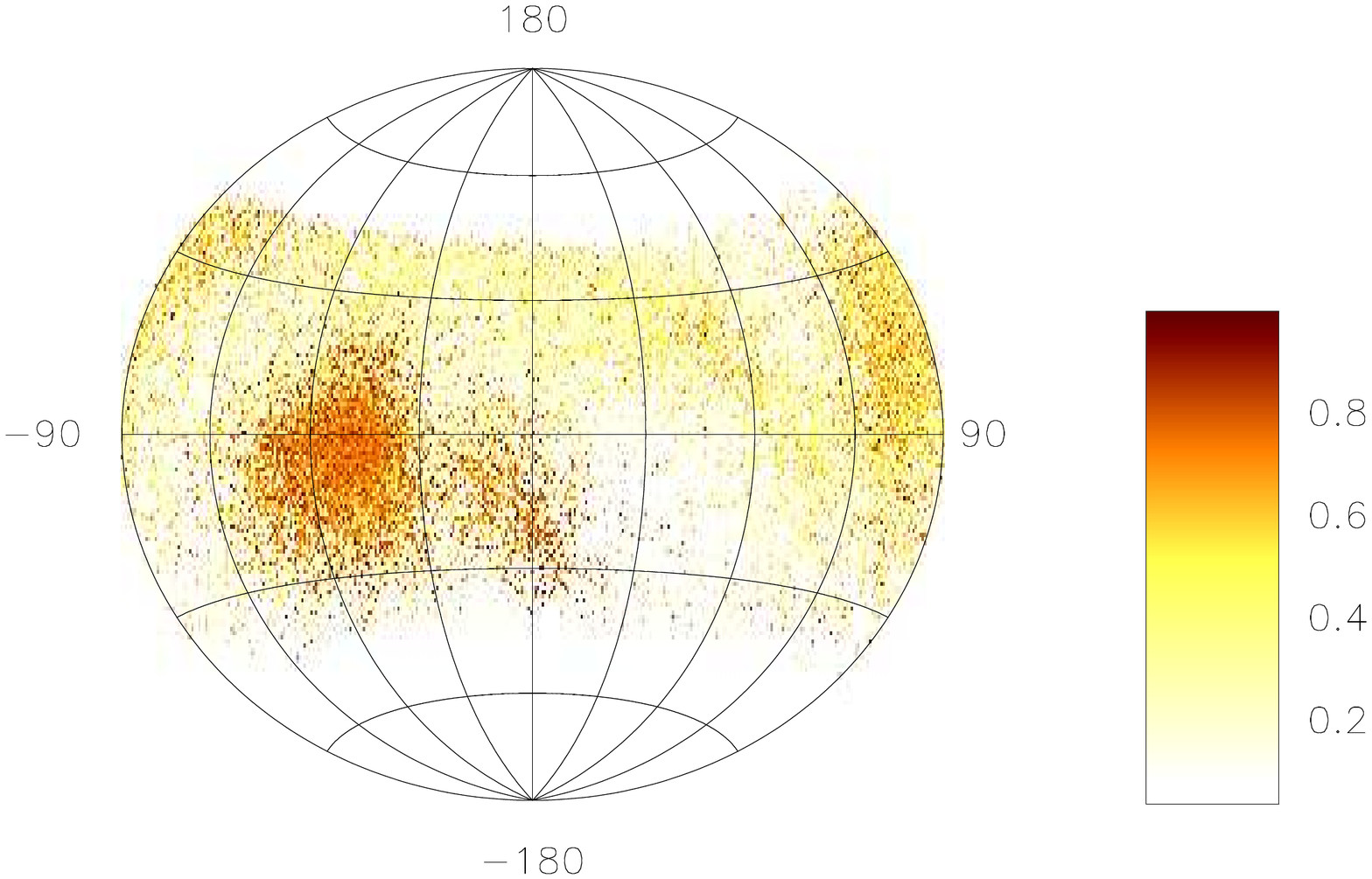}}}\\
\end{tabular}
\caption{Top panel (upper window) shows the average profile with total
intensity (Stokes I; solid black lines), total linear polarization (dashed red
line) and circular polarization (Stokes V; dotted blue line). Top panel (lower
window) also shows the single pulse PPA distribution (colour scale) along with
the average PPA (red error bars).
Bottom panel shows the Hammer-Aitoff projection of the polarized time
samples with the colour scheme representing the fractional polarization level.}
\label{a45}
\end{center}
\end{figure*}


\begin{figure*}
\begin{center}
\begin{tabular}{cc}
&
{\mbox{\includegraphics[width=9cm,height=6cm,angle=0.]{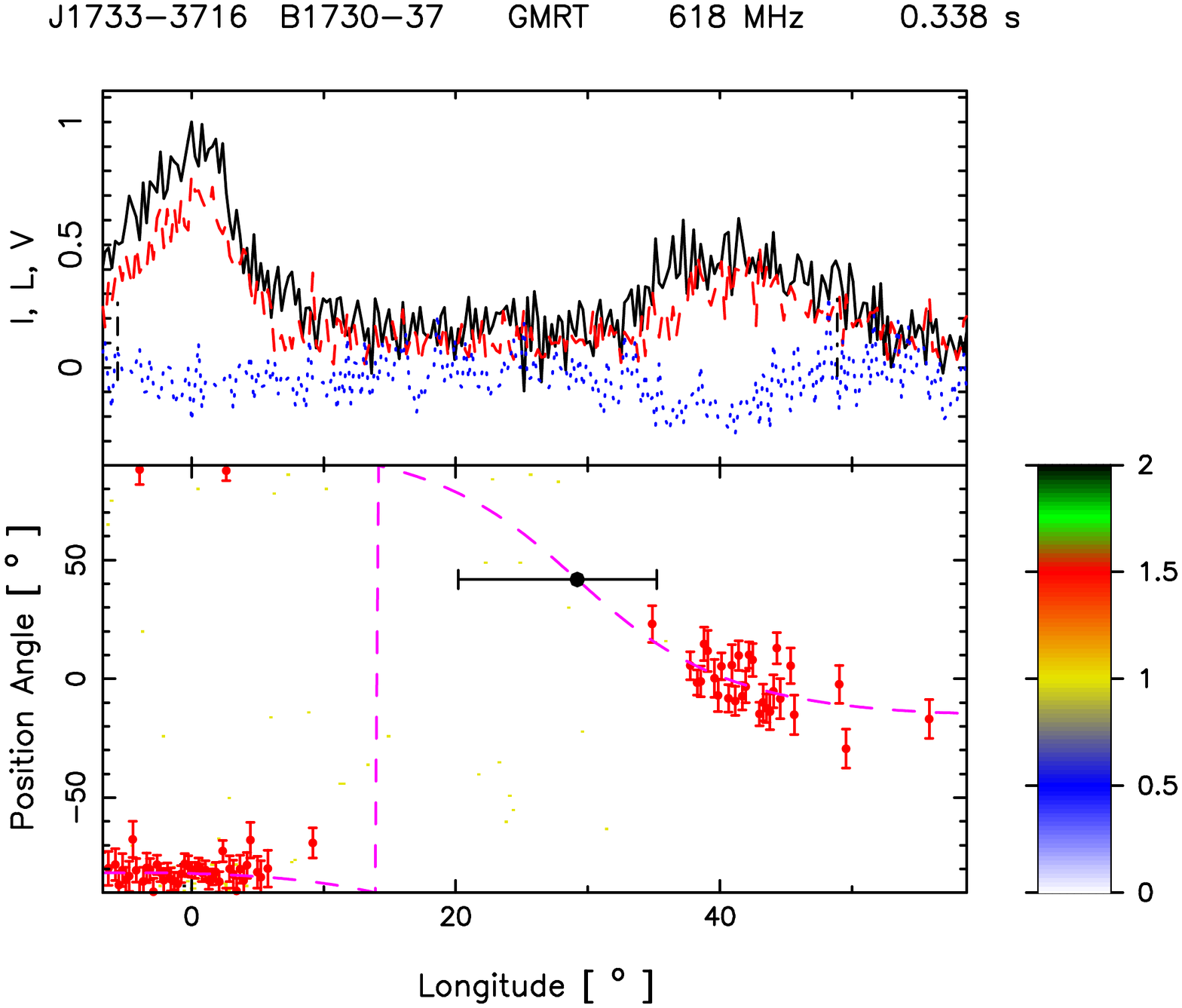}}}\\
&
{\mbox{\includegraphics[width=9cm,height=6cm,angle=0.]{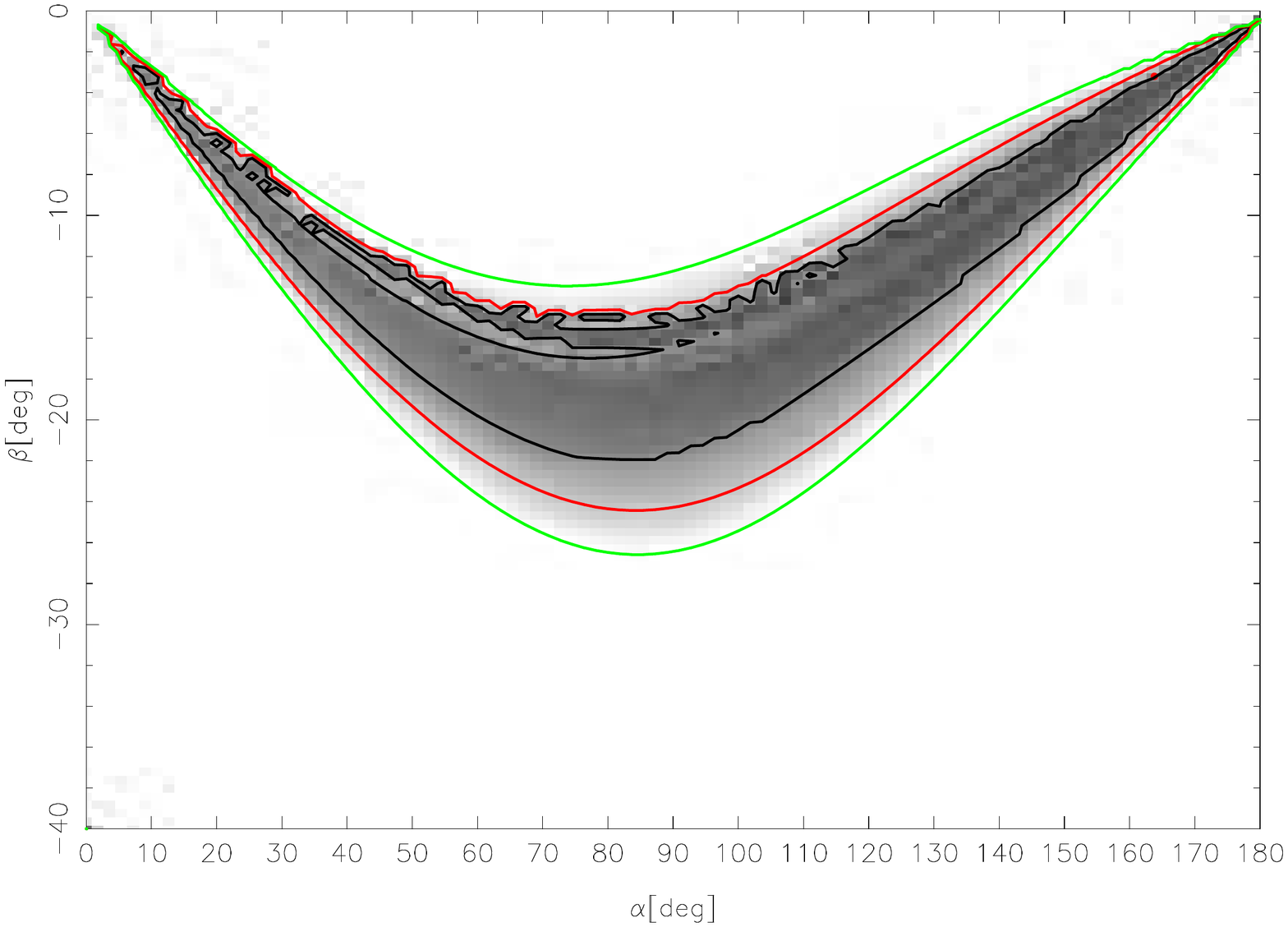}}}\\
&
{\mbox{\includegraphics[width=9cm,height=6cm,angle=0.]{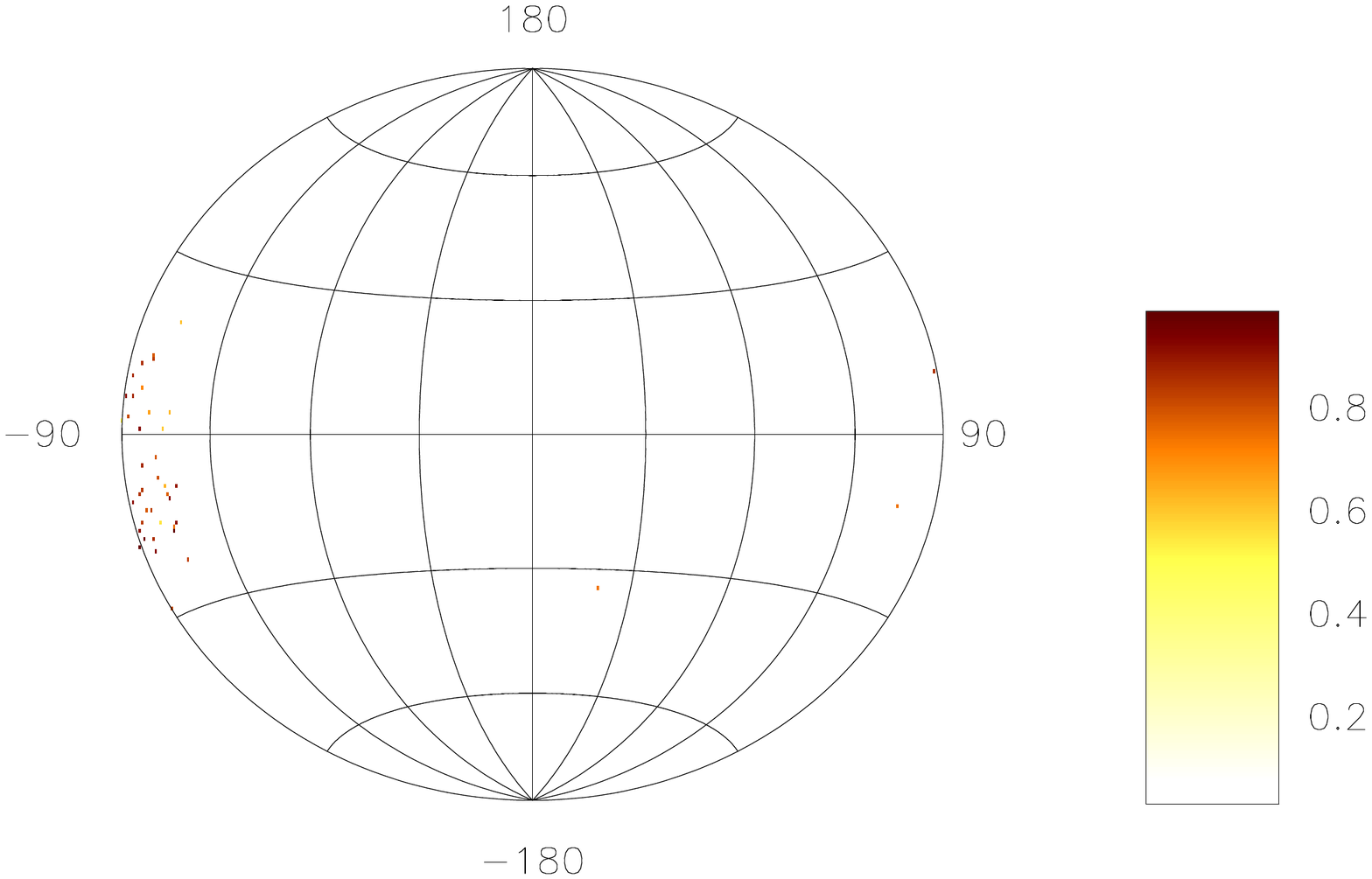}}}\\
\end{tabular}
\caption{Top panel only for 618 MHz (upper window) shows the average profile with total
intensity (Stokes I; solid black lines), total linear polarization (dashed red
line) and circular polarization (Stokes V; dotted blue line). Top panel (lower
window) also shows the single pulse PPA distribution (colour scale) along with
the average PPA (red error bars).
The RVM fits to the average PPA (dashed pink
line) is also shown in this plot. Middle panel only for 618 MHz show
the $\chi^2$ contours for the parameters $\alpha$ and $\beta$ obtained from RVM
fits.
Bottom panel only for 618 MHz shows the Hammer-Aitoff projection of the polarized time
samples with the colour scheme representing the fractional polarization level.}
\label{a46}
\end{center}
\end{figure*}


\begin{figure*}
\begin{center}
\begin{tabular}{cc}
{\mbox{\includegraphics[width=9cm,height=6cm,angle=0.]{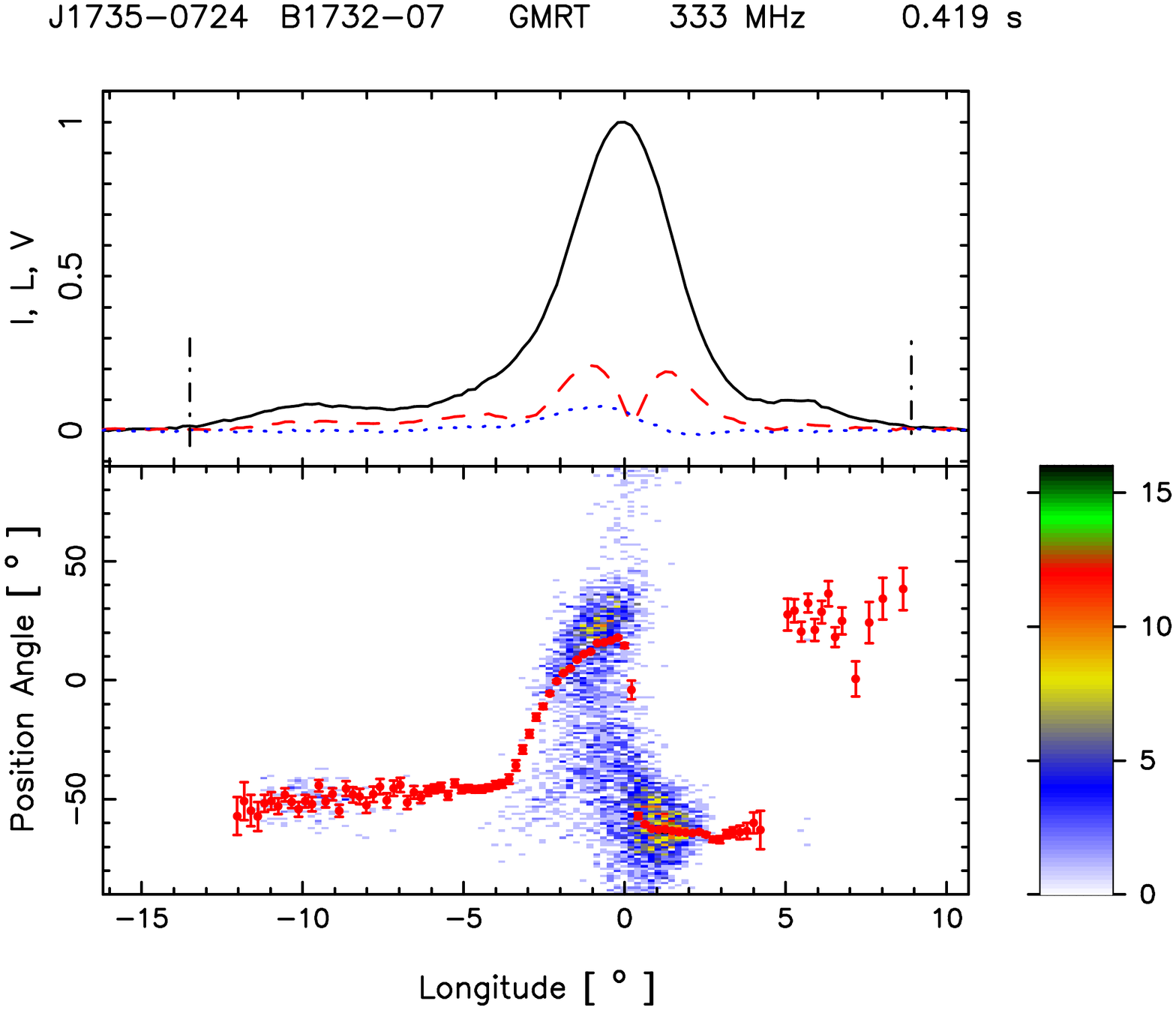}}}&
\\
&
\\
{\mbox{\includegraphics[width=9cm,height=6cm,angle=0.]{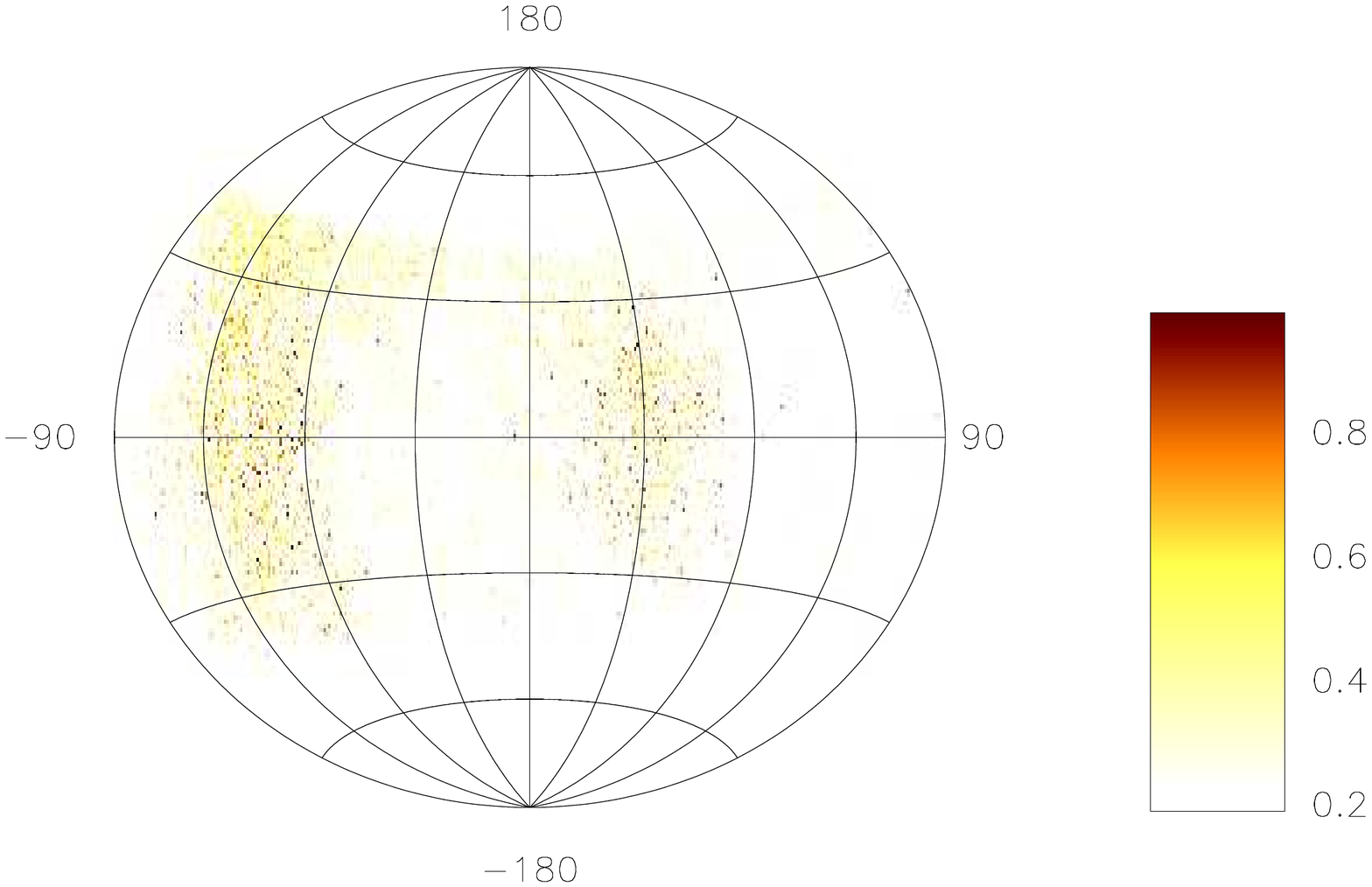}}}&
\\
\end{tabular}
\caption{Top panel only for 333 MHz (upper window) shows the average profile with total
intensity (Stokes I; solid black lines), total linear polarization (dashed red
line) and circular polarization (Stokes V; dotted blue line). Top panel (lower
window) also shows the single pulse PPA distribution (colour scale) along with
the average PPA (red error bars).
Bottom panel only for 333 MHz shows the Hammer-Aitoff projection of the polarized time
samples with the colour scheme representing the fractional polarization level.}
\label{a47}
\end{center}
\end{figure*}


\begin{figure*}
\begin{center}
\begin{tabular}{cc}
{\mbox{\includegraphics[width=9cm,height=6cm,angle=0.]{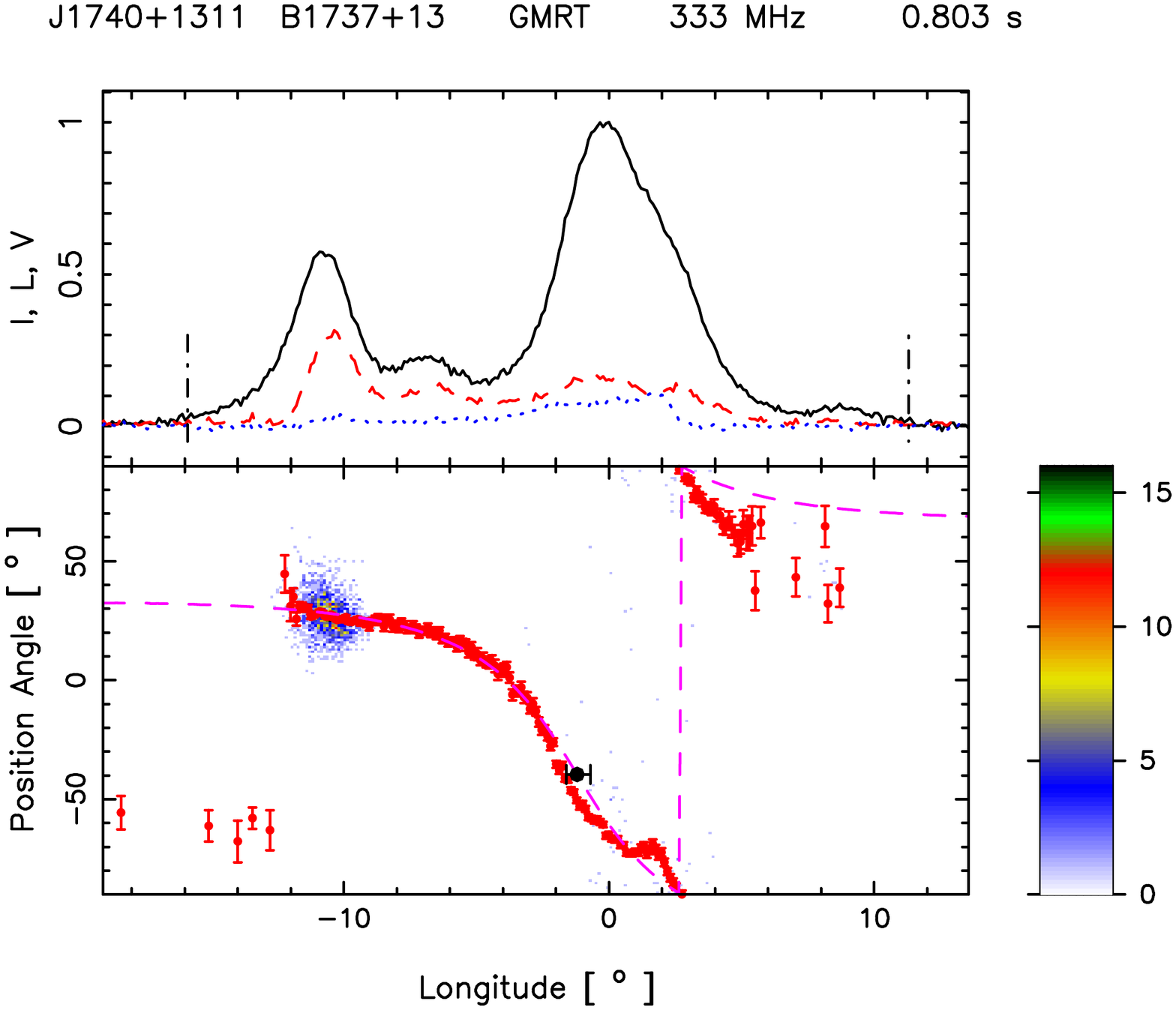}}}&
{\mbox{\includegraphics[width=9cm,height=6cm,angle=0.]{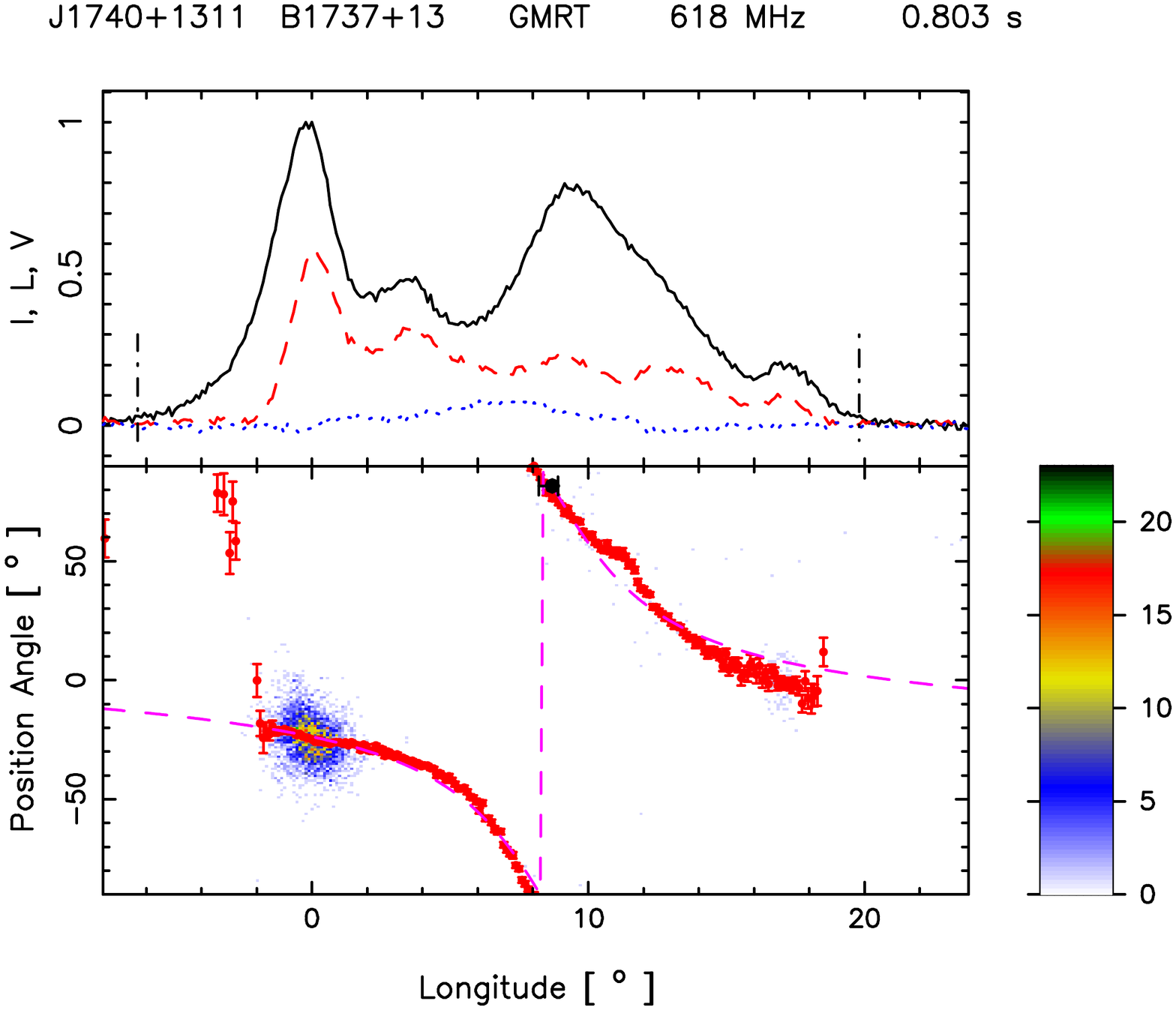}}}\\
{\mbox{\includegraphics[width=9cm,height=6cm,angle=0.]{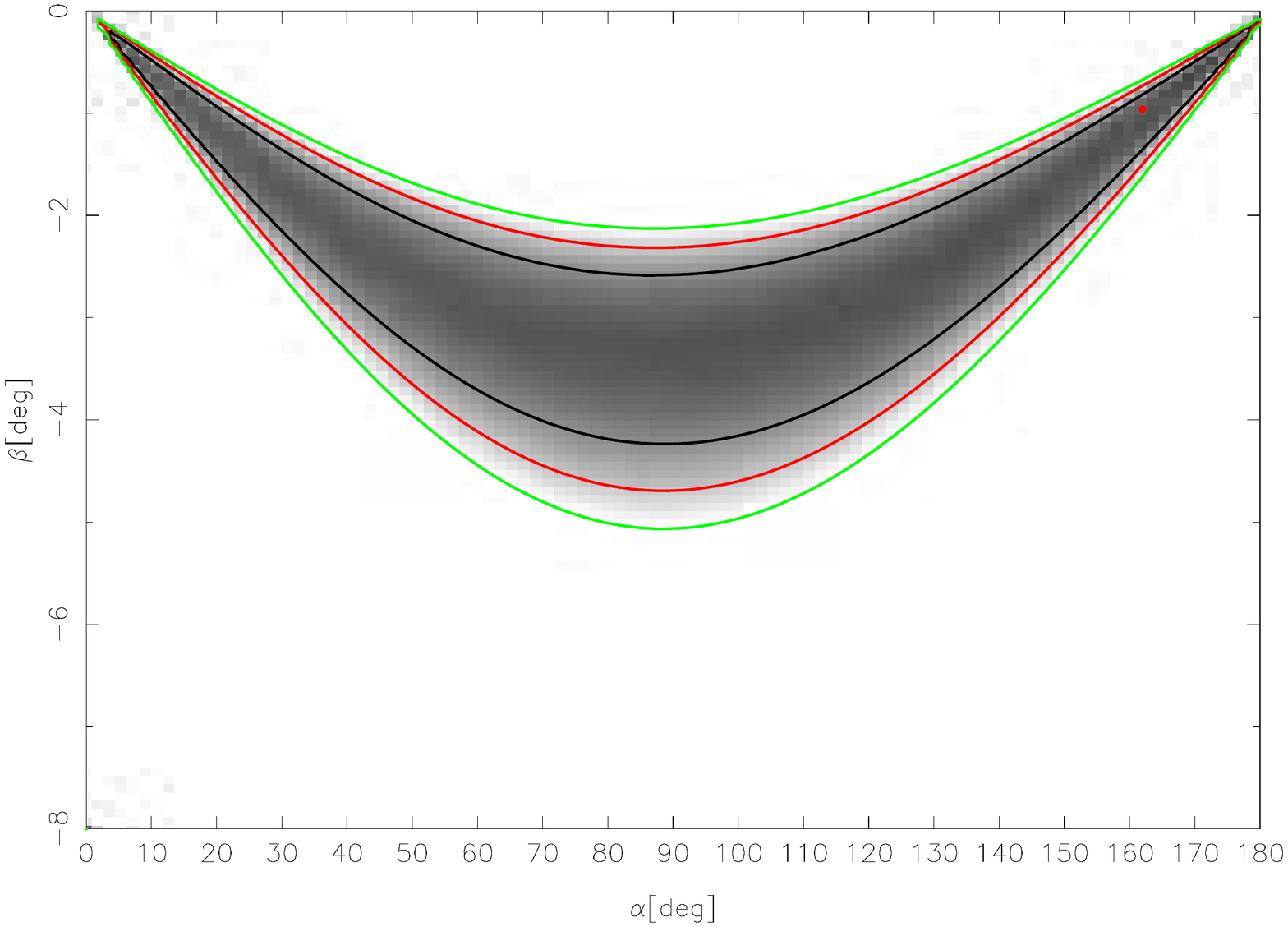}}}&
{\mbox{\includegraphics[width=9cm,height=6cm,angle=0.]{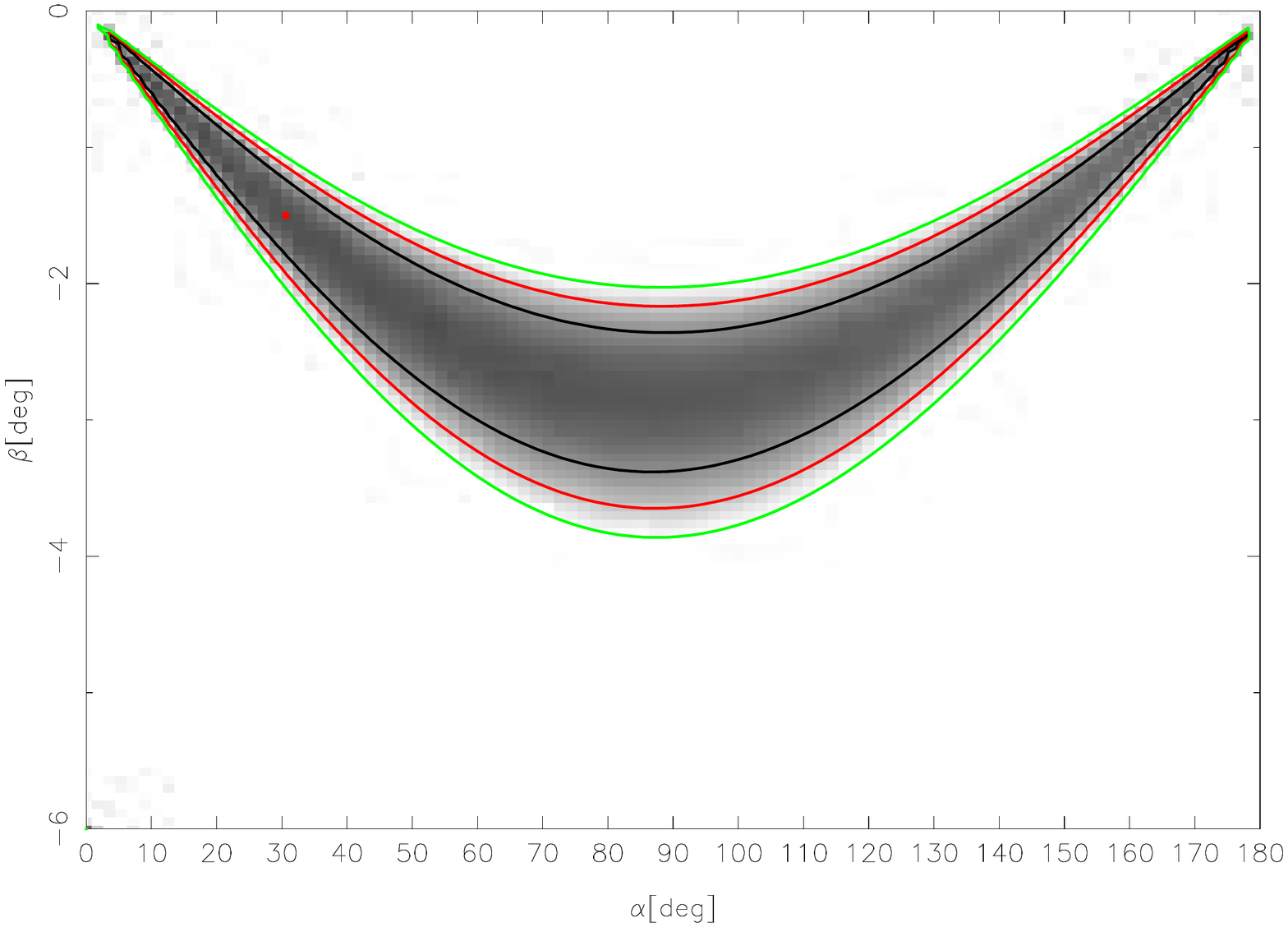}}}\\
{\mbox{\includegraphics[width=9cm,height=6cm,angle=0.]{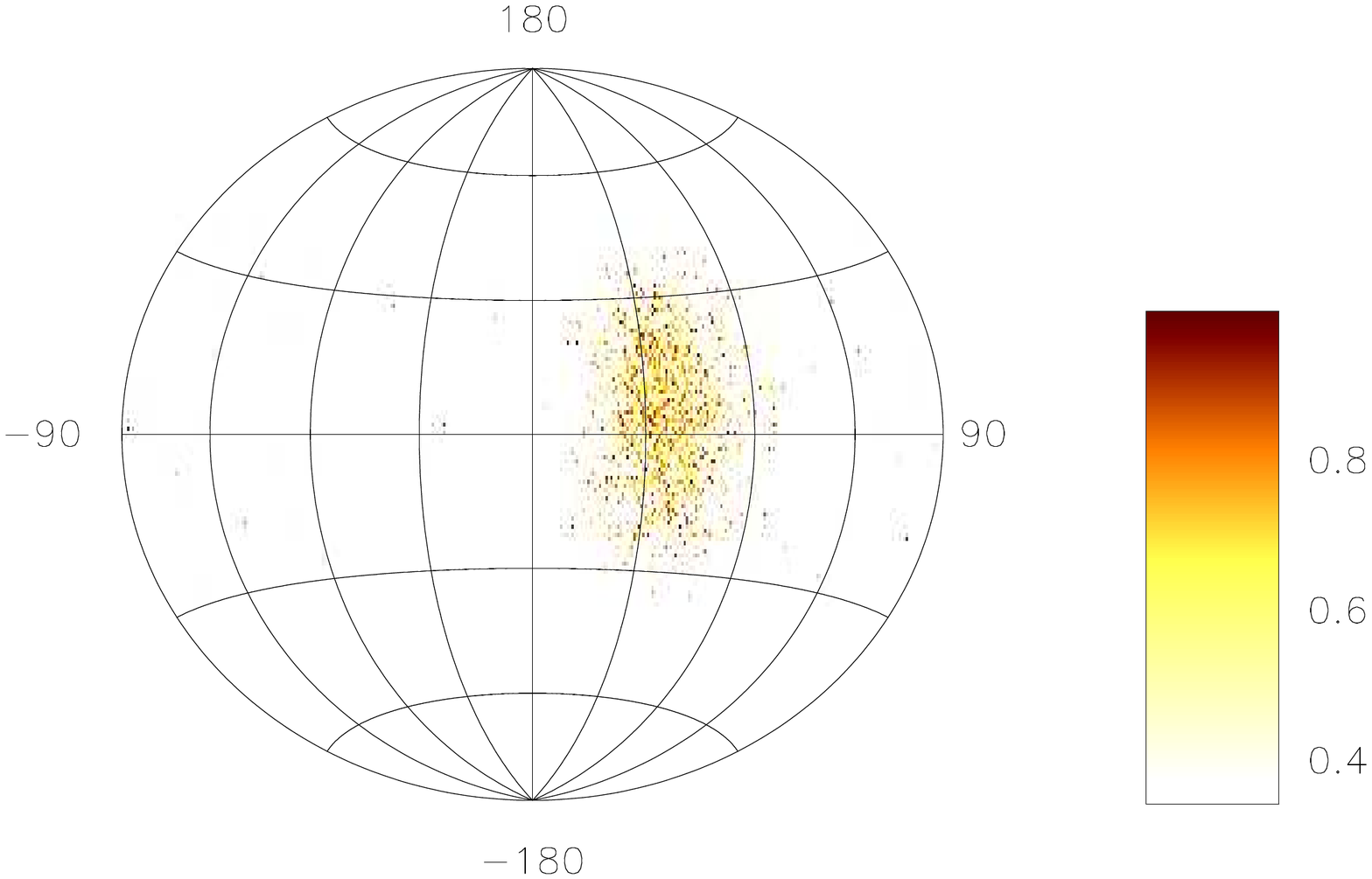}}}&
{\mbox{\includegraphics[width=9cm,height=6cm,angle=0.]{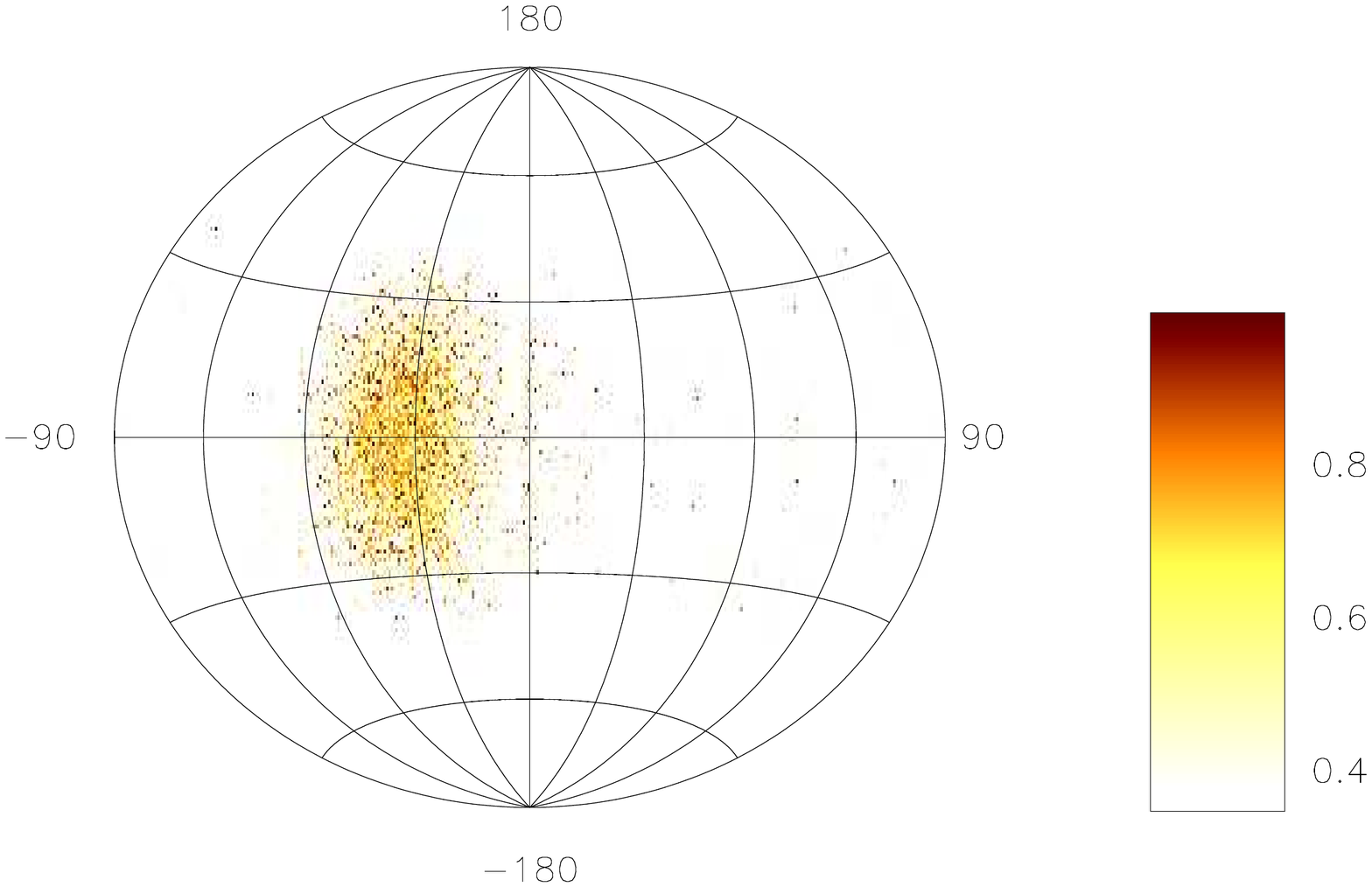}}}\\
\end{tabular}
\caption{Top panel (upper window) shows the average profile with total
intensity (Stokes I; solid black lines), total linear polarization (dashed red
line) and circular polarization (Stokes V; dotted blue line). Top panel (lower
window) also shows the single pulse PPA distribution (colour scale) along with
the average PPA (red error bars).
The RVM fits to the average PPA (dashed pink
line) is also shown in this plot. Middle panel show
the $\chi^2$ contours for the parameters $\alpha$ and $\beta$ obtained from RVM
fits.
Bottom panel shows the Hammer-Aitoff projection of the polarized time
samples with the colour scheme representing the fractional polarization level.}
\label{a48}
\end{center}
\end{figure*}


\begin{figure*}
\begin{center}
\begin{tabular}{cc}
{\mbox{\includegraphics[width=9cm,height=6cm,angle=0.]{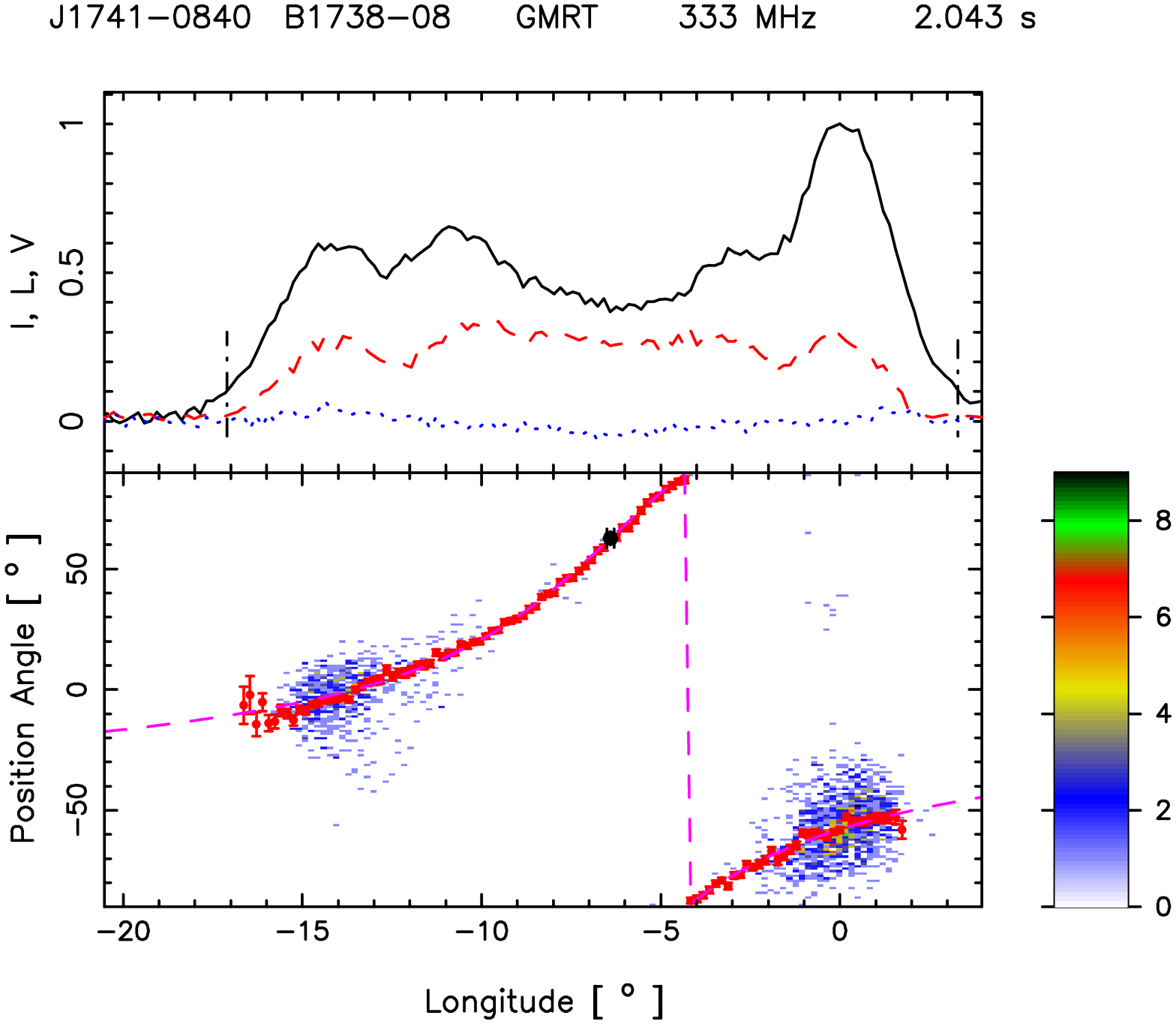}}}&
{\mbox{\includegraphics[width=9cm,height=6cm,angle=0.]{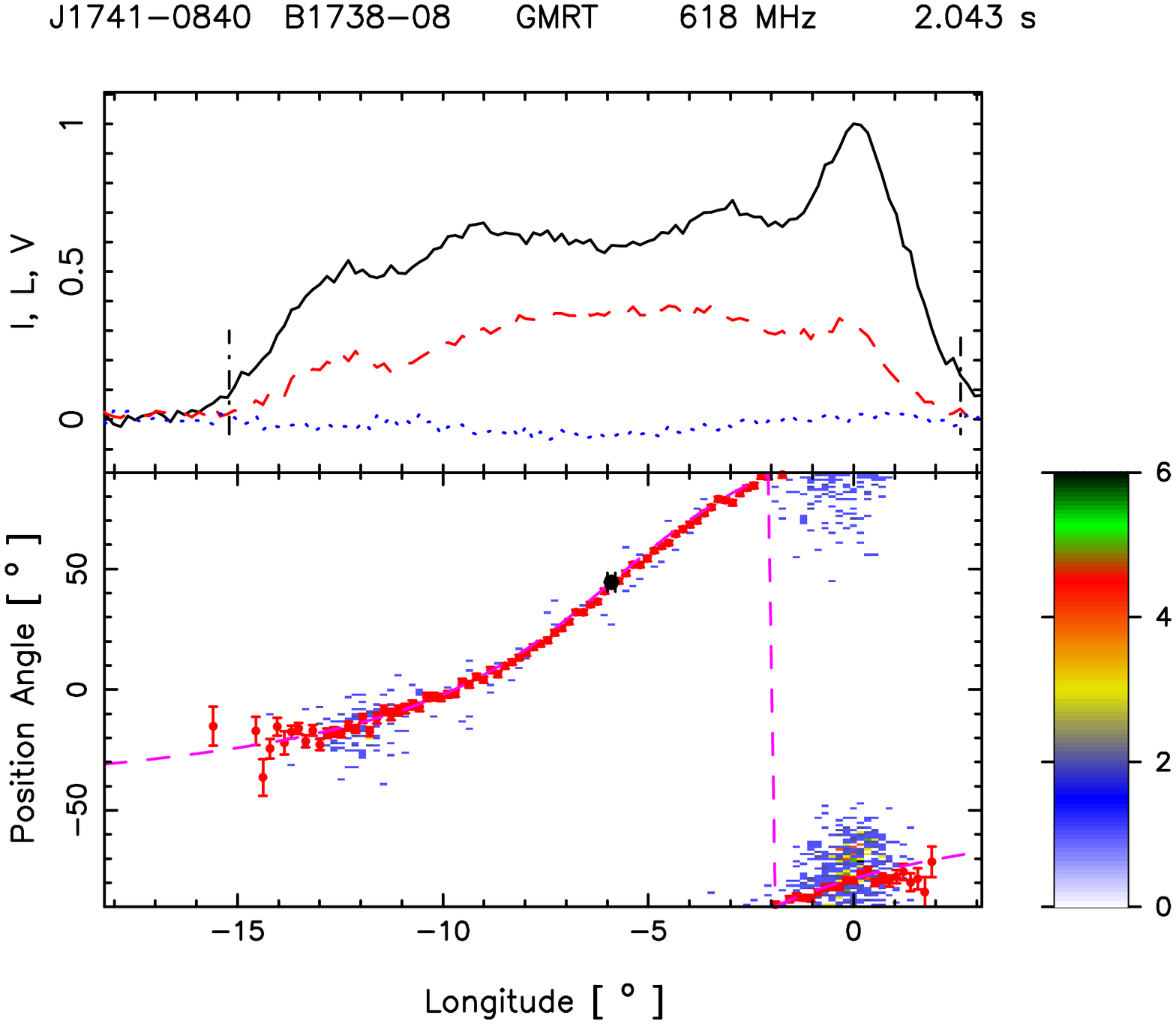}}}\\
{\mbox{\includegraphics[width=9cm,height=6cm,angle=0.]{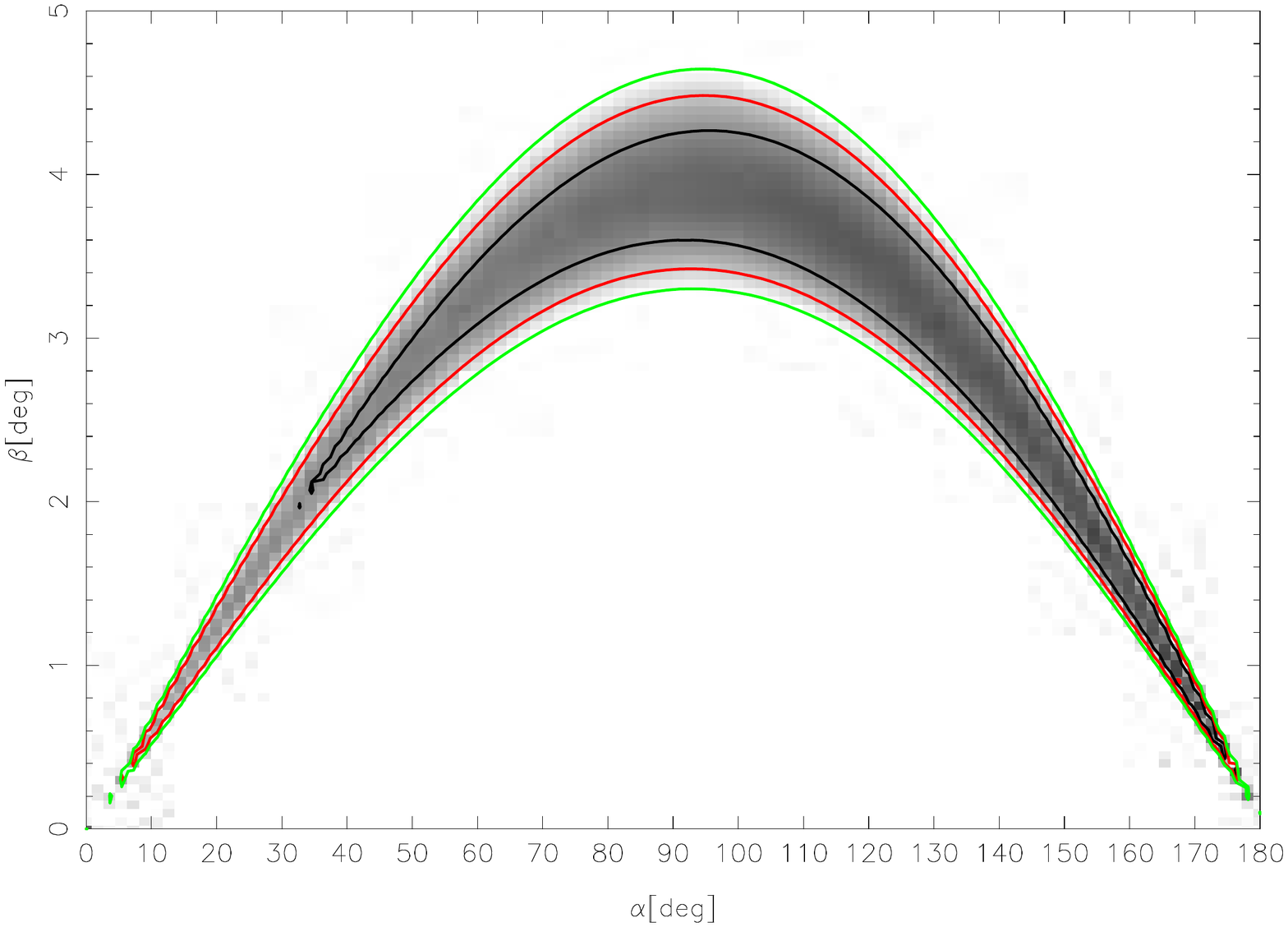}}}&
{\mbox{\includegraphics[width=9cm,height=6cm,angle=0.]{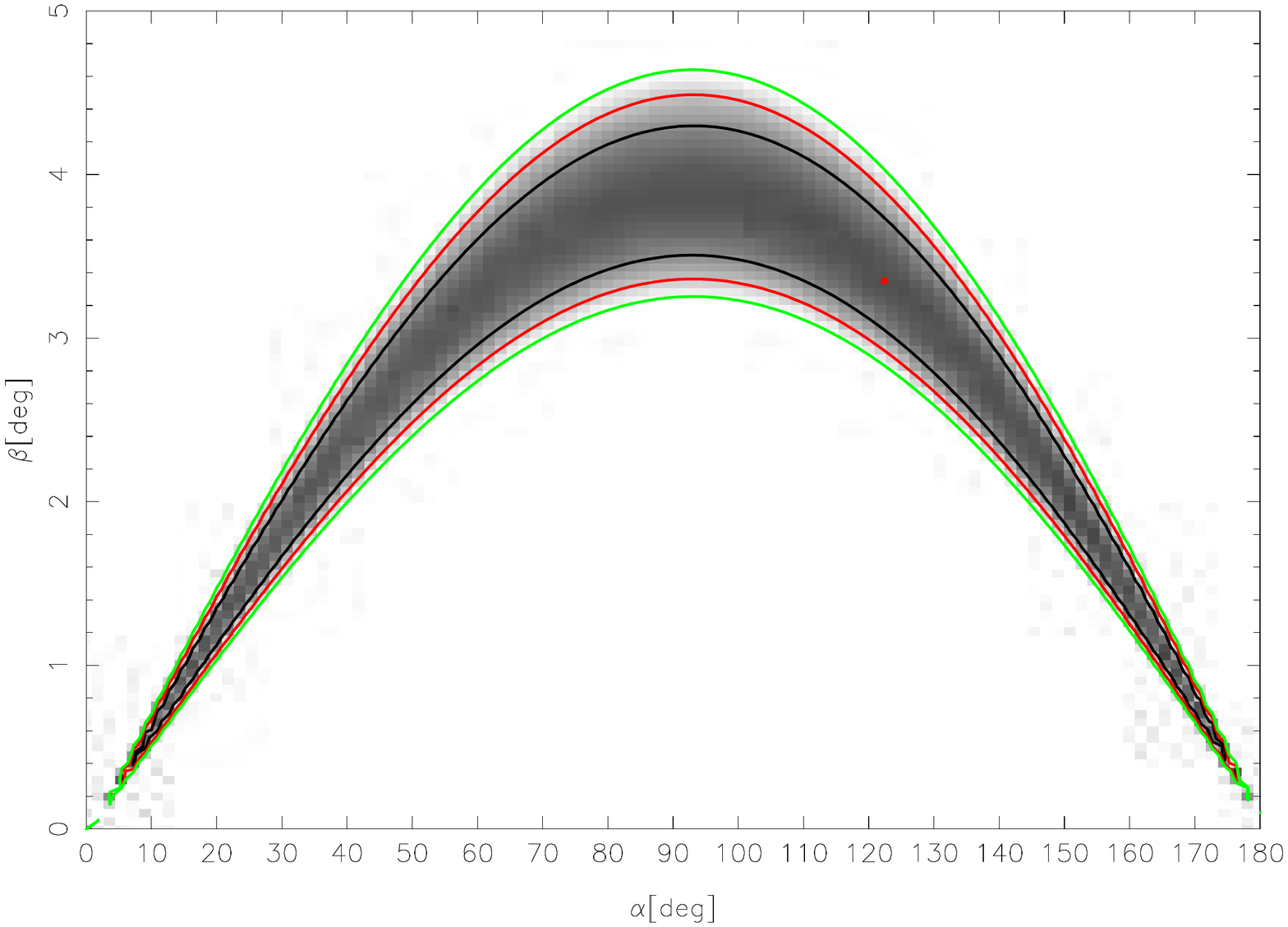}}}\\
{\mbox{\includegraphics[width=9cm,height=6cm,angle=0.]{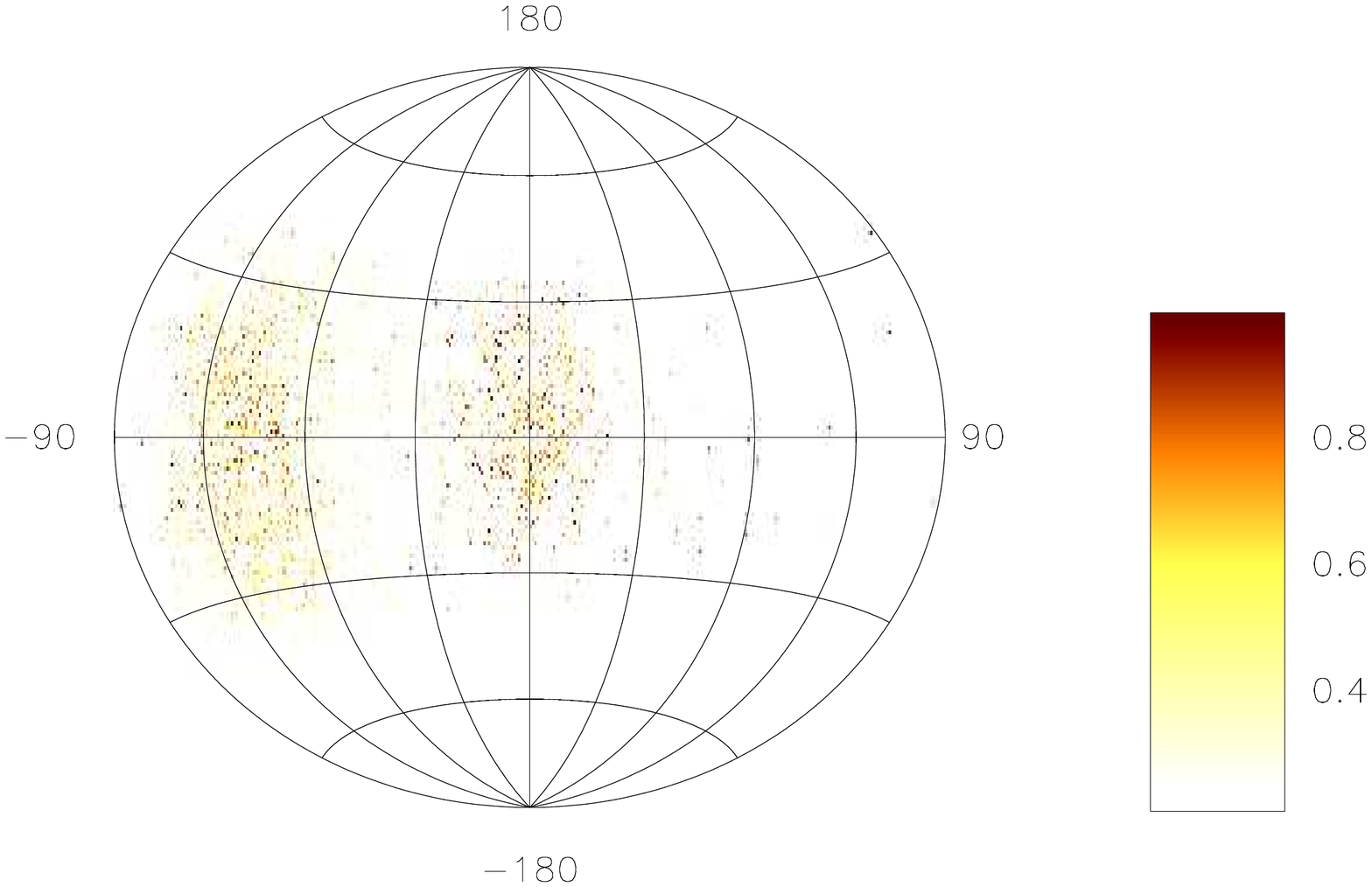}}}&
{\mbox{\includegraphics[width=9cm,height=6cm,angle=0.]{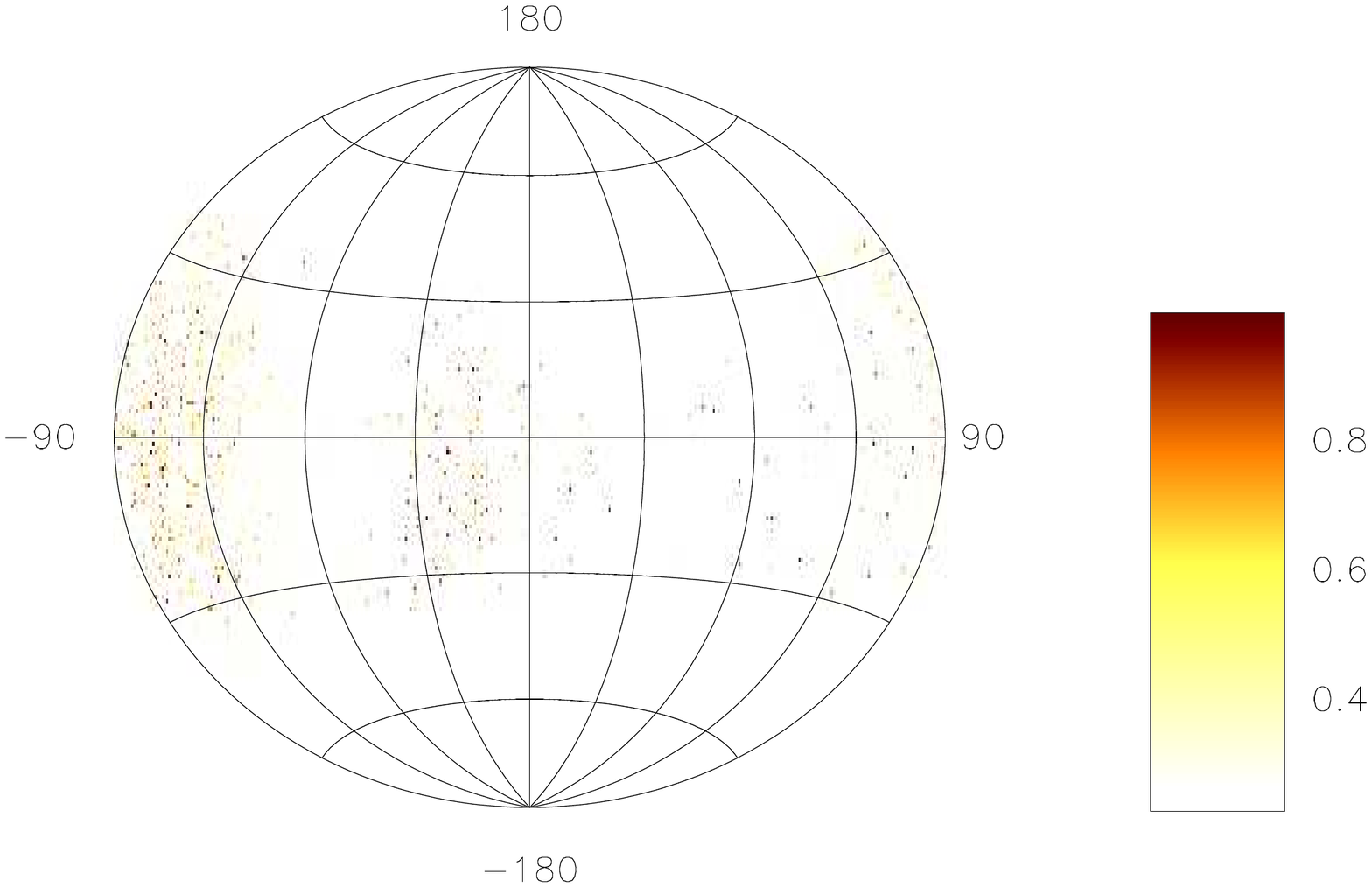}}}\\
\end{tabular}
\caption{Top panel (upper window) shows the average profile with total
intensity (Stokes I; solid black lines), total linear polarization (dashed red
line) and circular polarization (Stokes V; dotted blue line). Top panel (lower
window) also shows the single pulse PPA distribution (colour scale) along with
the average PPA (red error bars).
The RVM fits to the average PPA (dashed pink
line) is also shown in this plot. Middle panel show
the $\chi^2$ contours for the parameters $\alpha$ and $\beta$ obtained from RVM
fits.
Bottom panel shows the Hammer-Aitoff projection of the polarized time
samples with the colour scheme representing the fractional polarization level.}
\label{a49}
\end{center}
\end{figure*}


\begin{figure*}
\begin{center}
\begin{tabular}{cc}
&
{\mbox{\includegraphics[width=9cm,height=6cm,angle=0.]{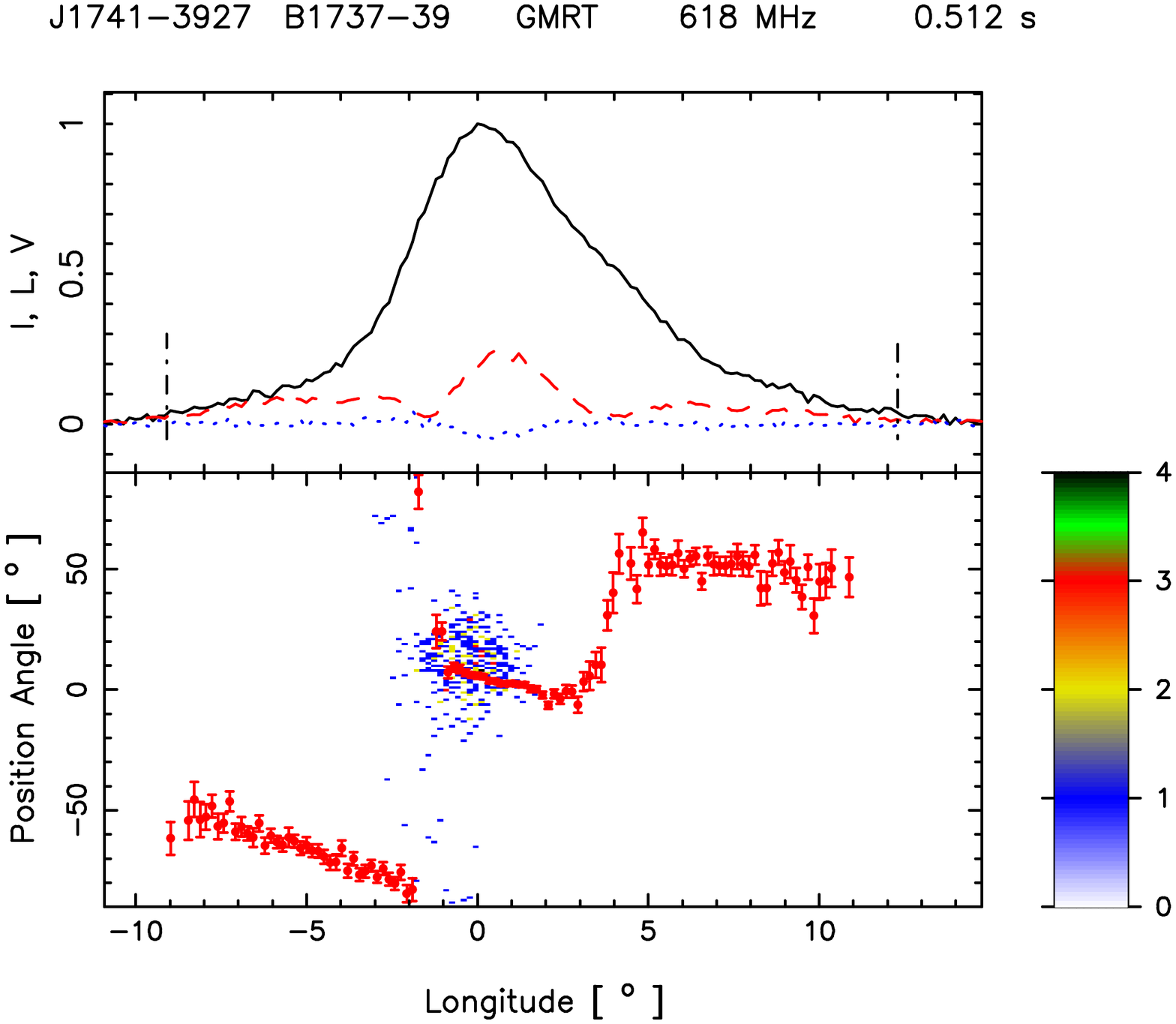}}}\\
&
{\mbox{\includegraphics[width=9cm,height=6cm,angle=0.]{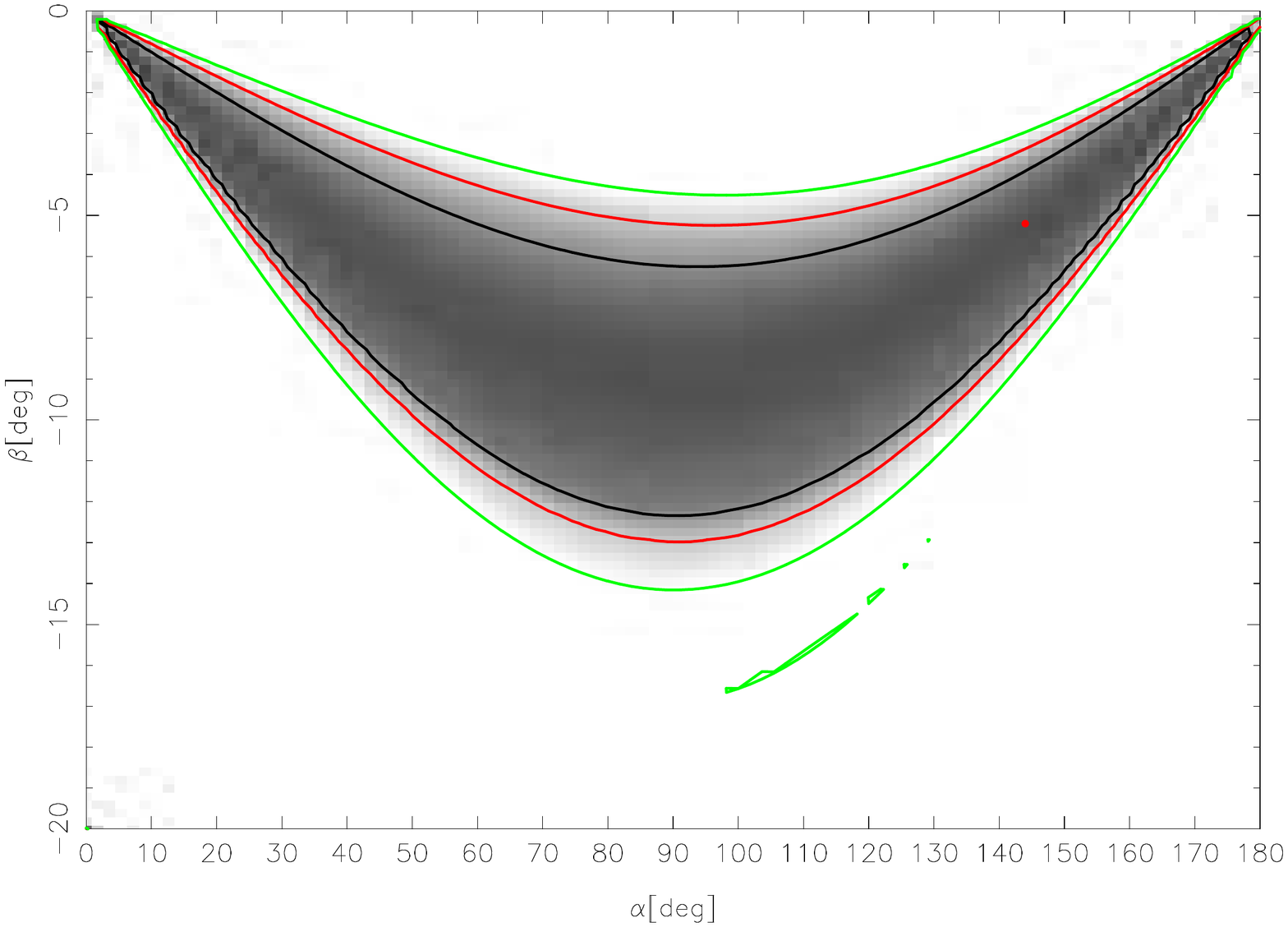}}}\\
&
{\mbox{\includegraphics[width=9cm,height=6cm,angle=0.]{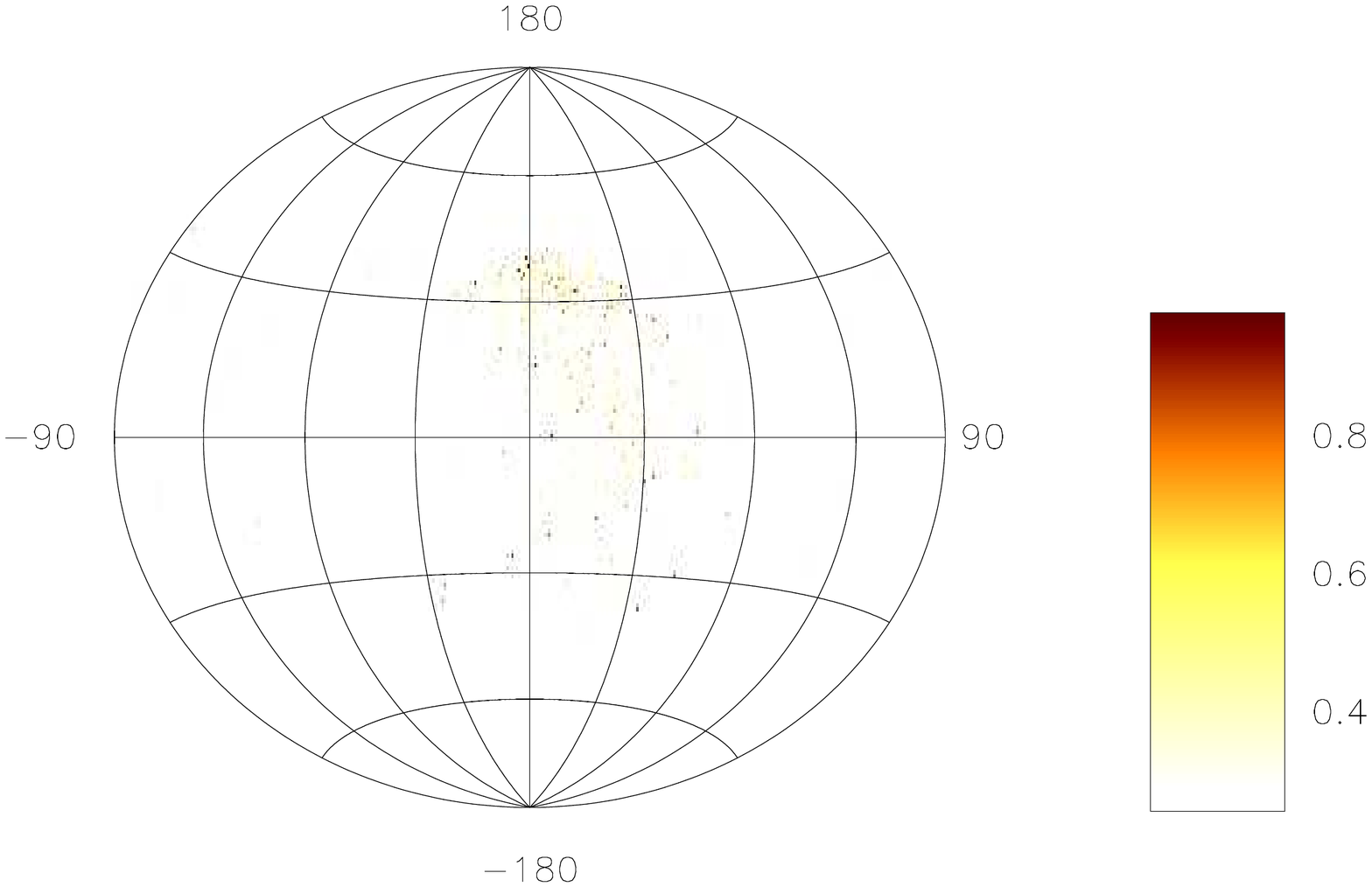}}}\\
\end{tabular}
\caption{Top panel only for 618 MHz (upper window) shows the average profile with total
intensity (Stokes I; solid black lines), total linear polarization (dashed red
line) and circular polarization (Stokes V; dotted blue line). Top panel (lower
window) also shows the single pulse PPA distribution (colour scale) along with
the average PPA (red error bars).
The RVM fits to the average PPA (dashed pink
line) is also shown in this plot. Middle panel only for 618 MHz show
the $\chi^2$ contours for the parameters $\alpha$ and $\beta$ obtained from RVM
fits.
Bottom panel only for 618 MHz shows the Hammer-Aitoff projection of the polarized time
samples with the colour scheme representing the fractional polarization level.}
\label{a50}
\end{center}
\end{figure*}


\begin{figure*}
\begin{center}
\begin{tabular}{cc}
{\mbox{\includegraphics[width=9cm,height=6cm,angle=0.]{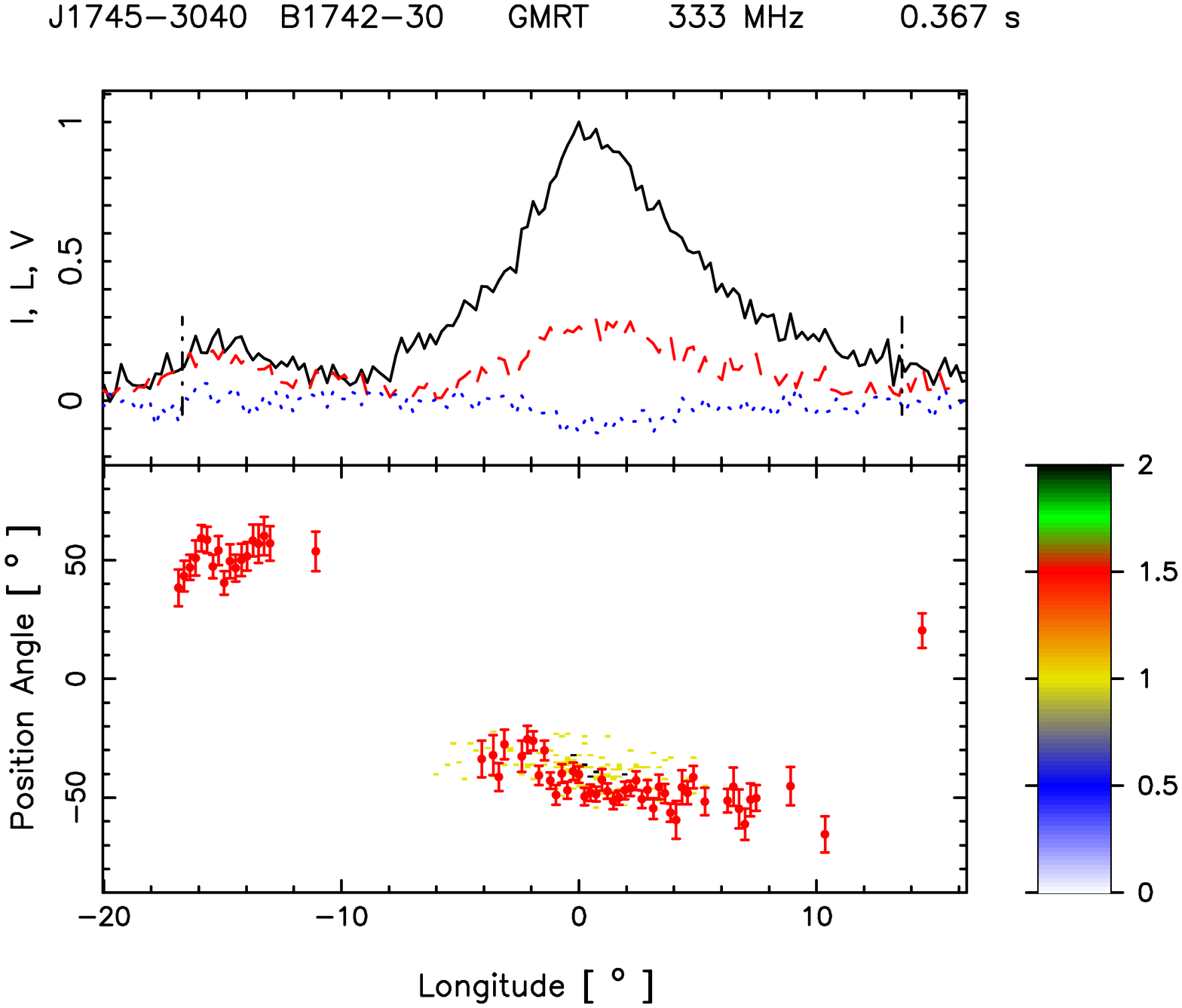}}}&
{\mbox{\includegraphics[width=9cm,height=6cm,angle=0.]{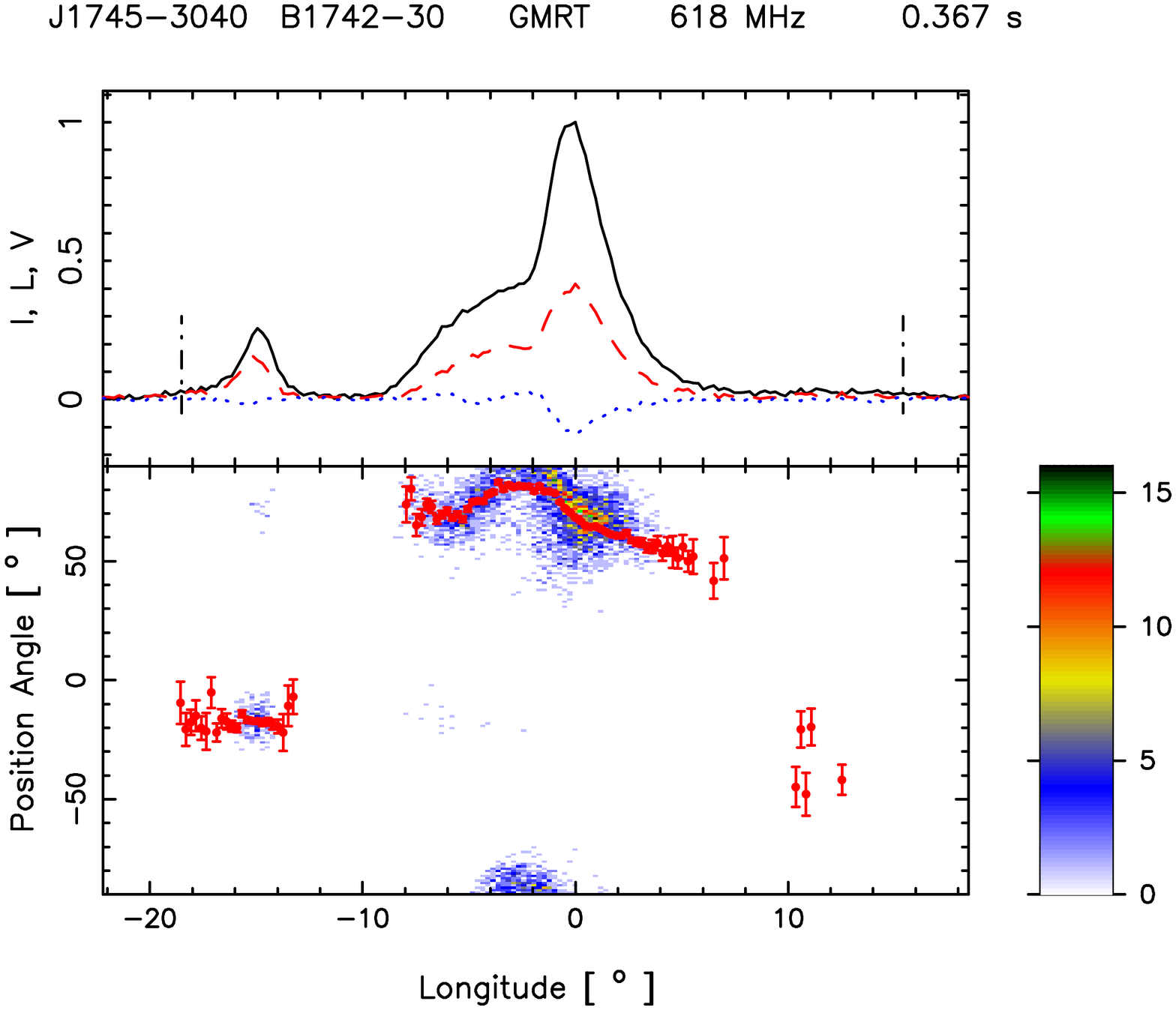}}}\\
&
\\
{\mbox{\includegraphics[width=9cm,height=6cm,angle=0.]{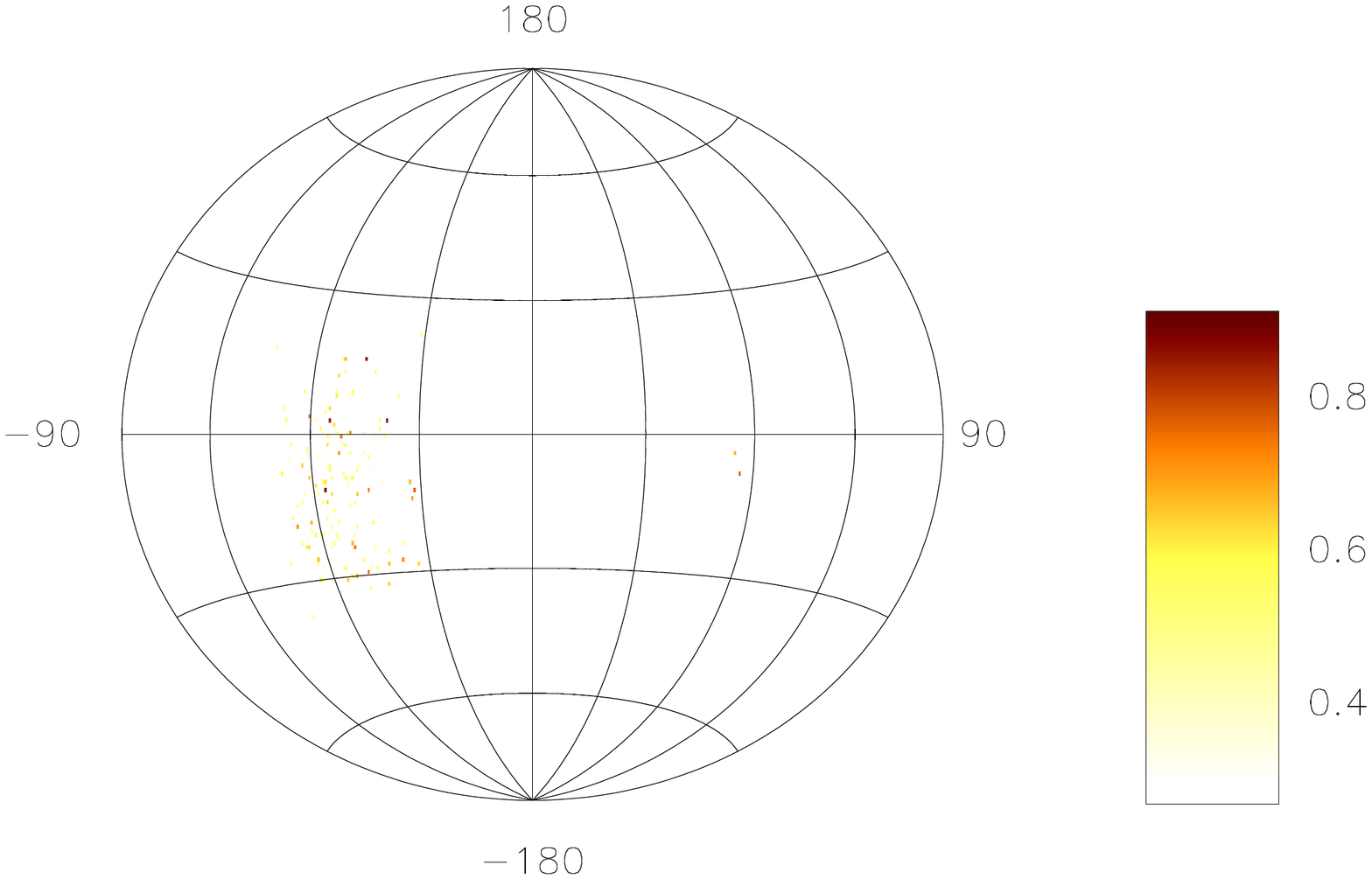}}}&
{\mbox{\includegraphics[width=9cm,height=6cm,angle=0.]{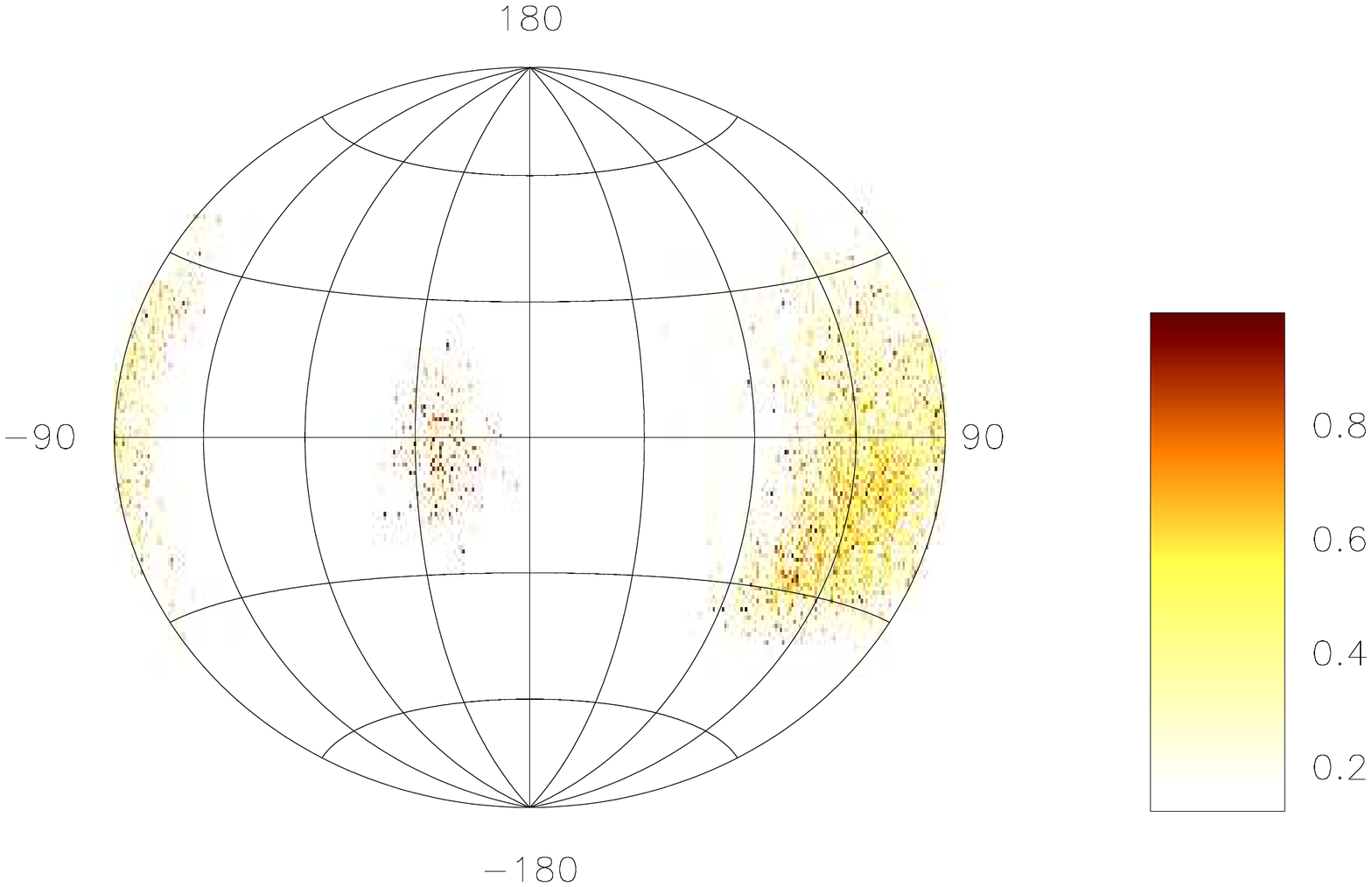}}}\\
\end{tabular}
\caption{Top panel (upper window) shows the average profile with total
intensity (Stokes I; solid black lines), total linear polarization (dashed red
line) and circular polarization (Stokes V; dotted blue line). Top panel (lower
window) also shows the single pulse PPA distribution (colour scale) along with
the average PPA (red error bars).
Bottom panel shows the Hammer-Aitoff projection of the polarized time
samples with the colour scheme representing the fractional polarization level.}
\label{a51}
\end{center}
\end{figure*}


\begin{figure*}
\begin{center}
\begin{tabular}{cc}
{\mbox{\includegraphics[width=9cm,height=6cm,angle=0.]{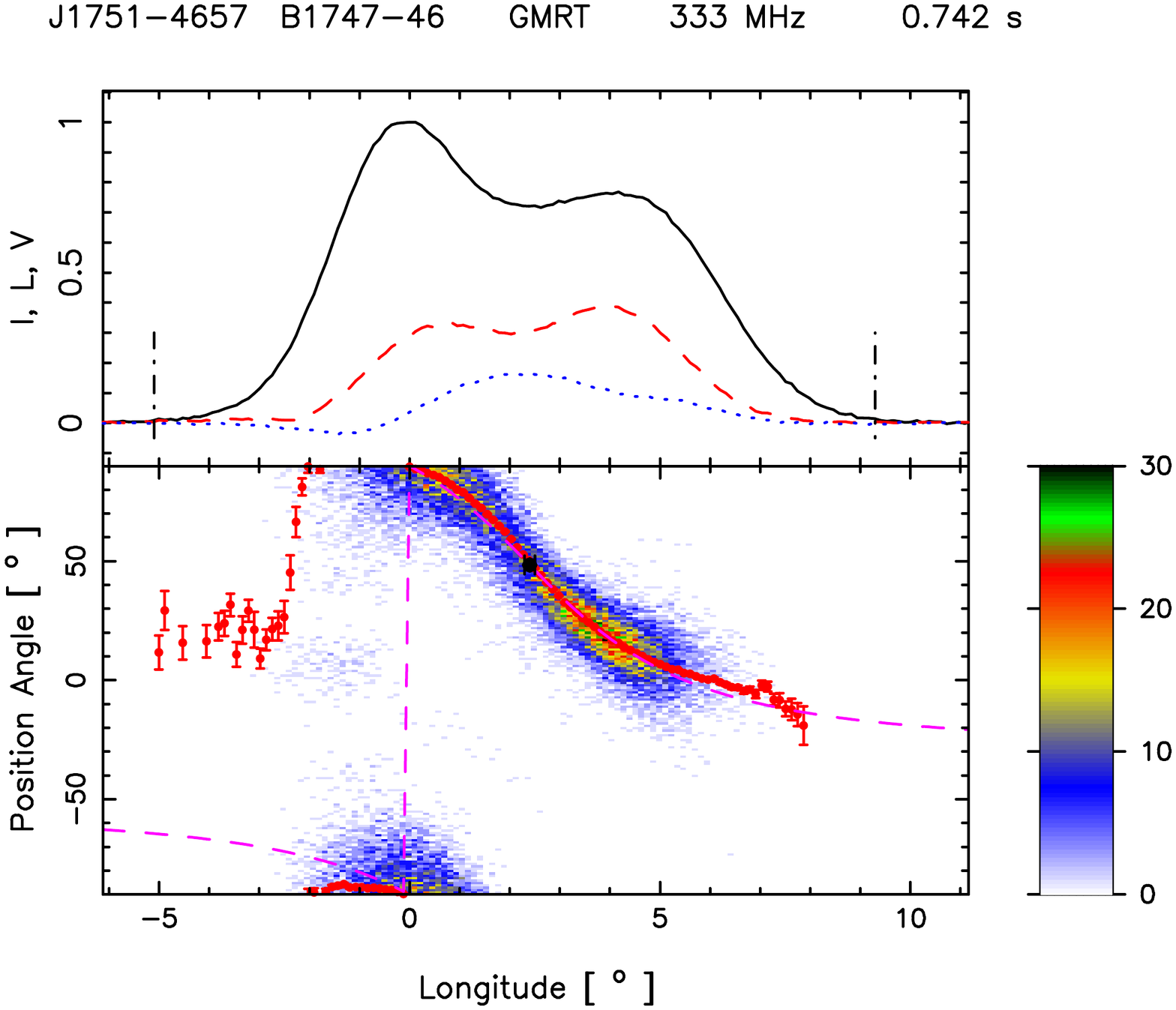}}}&
{\mbox{\includegraphics[width=9cm,height=6cm,angle=0.]{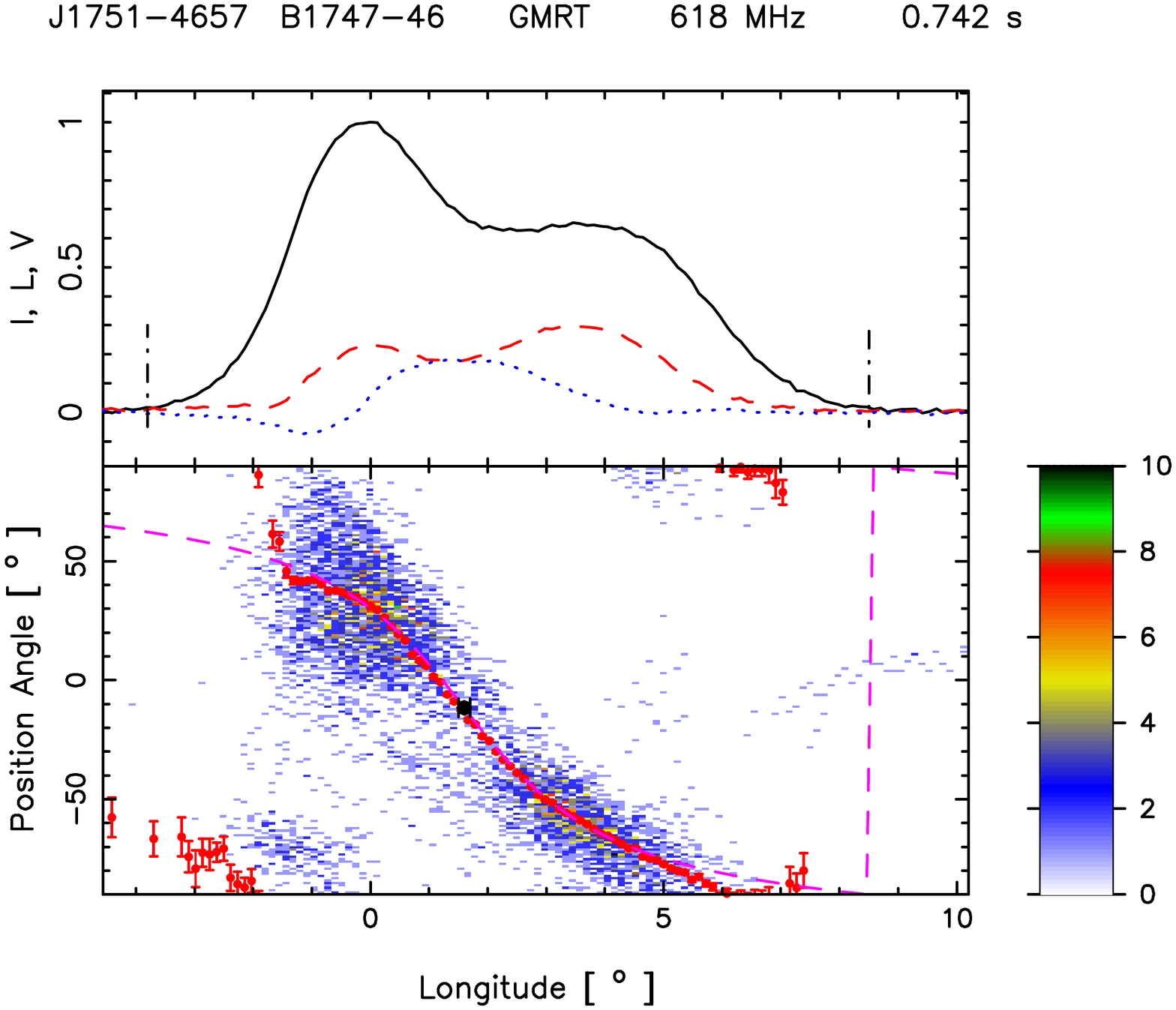}}}\\
{\mbox{\includegraphics[width=9cm,height=6cm,angle=0.]{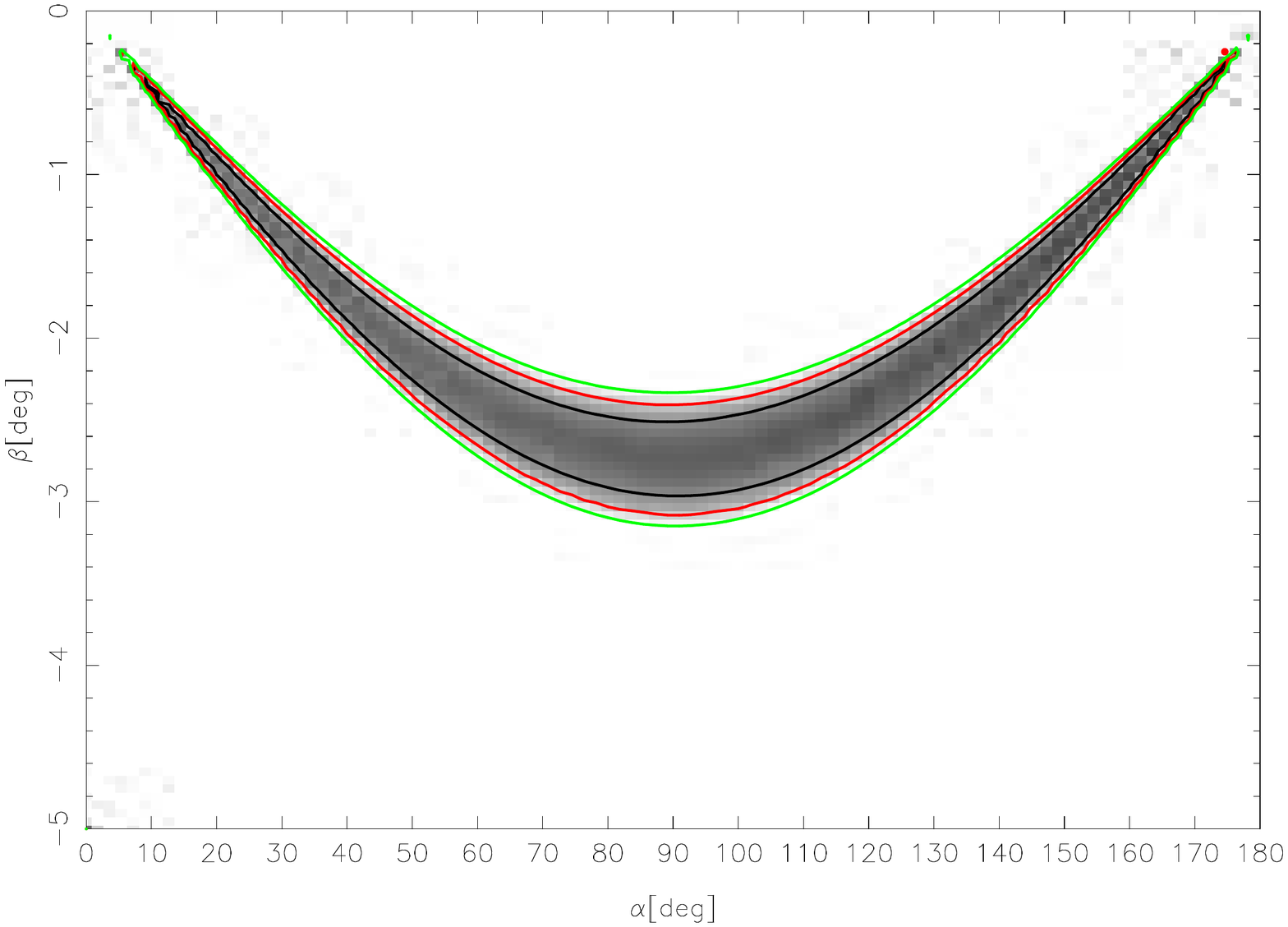}}}&
{\mbox{\includegraphics[width=9cm,height=6cm,angle=0.]{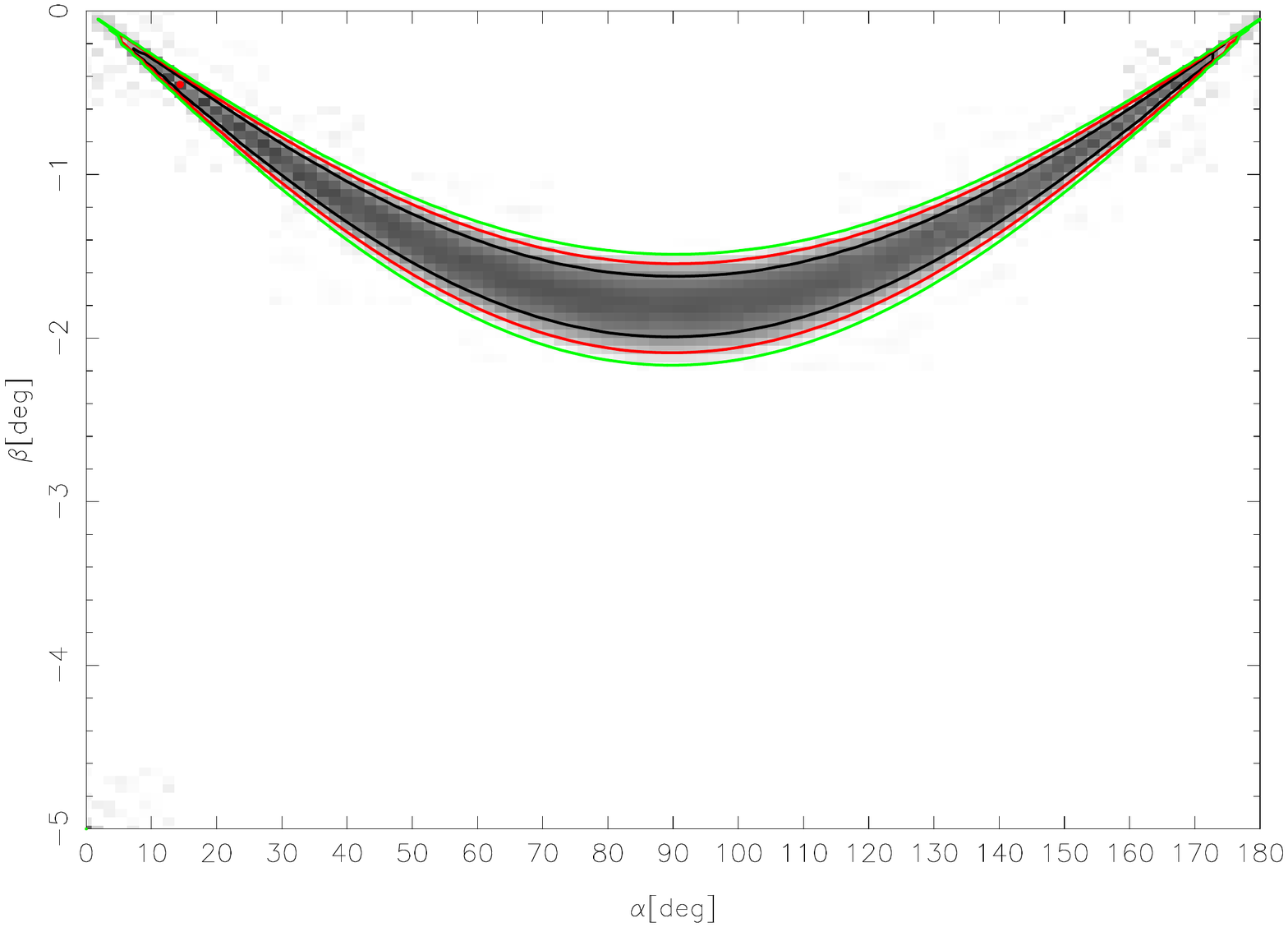}}}\\
{\mbox{\includegraphics[width=9cm,height=6cm,angle=0.]{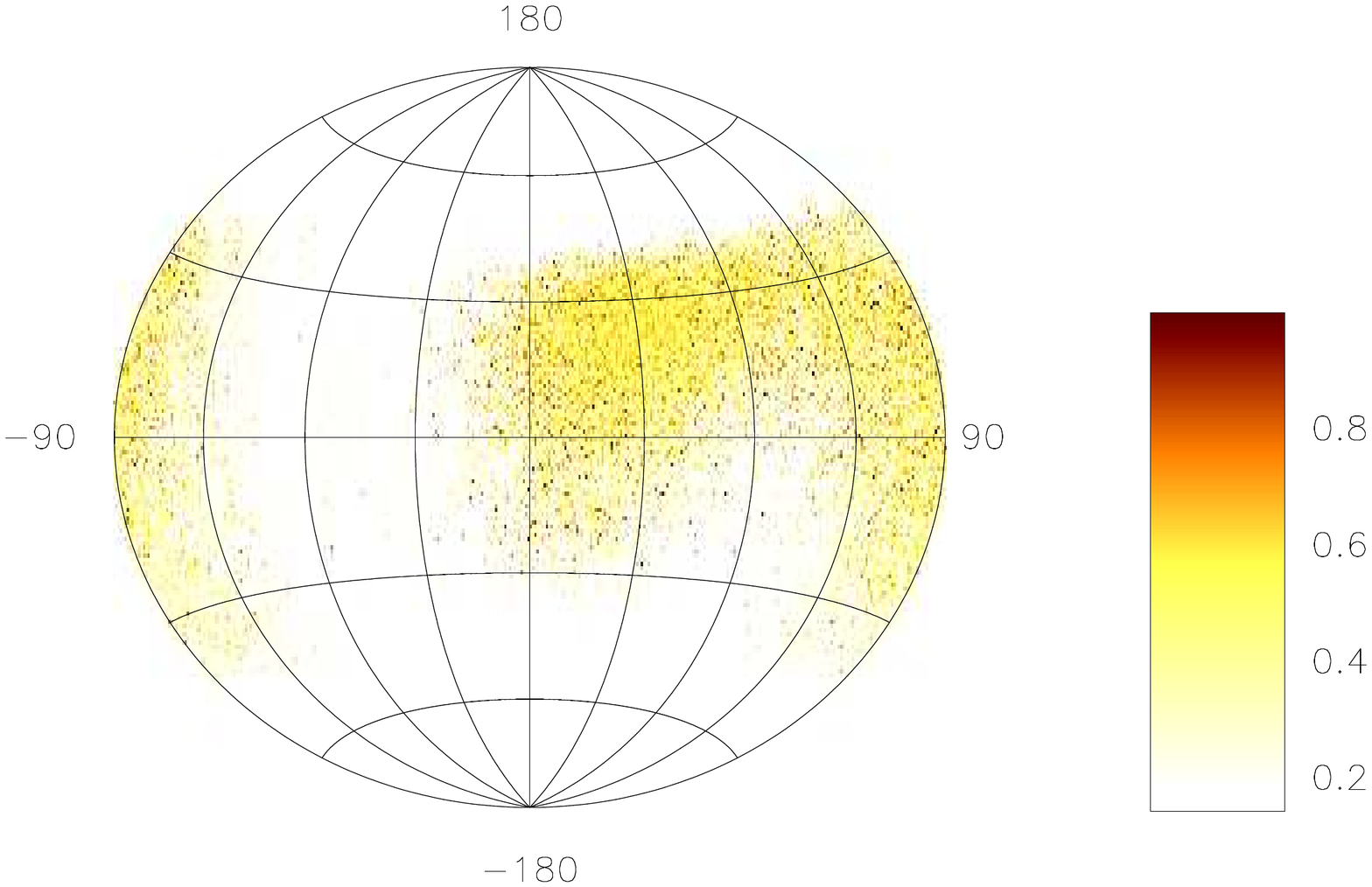}}}&
{\mbox{\includegraphics[width=9cm,height=6cm,angle=0.]{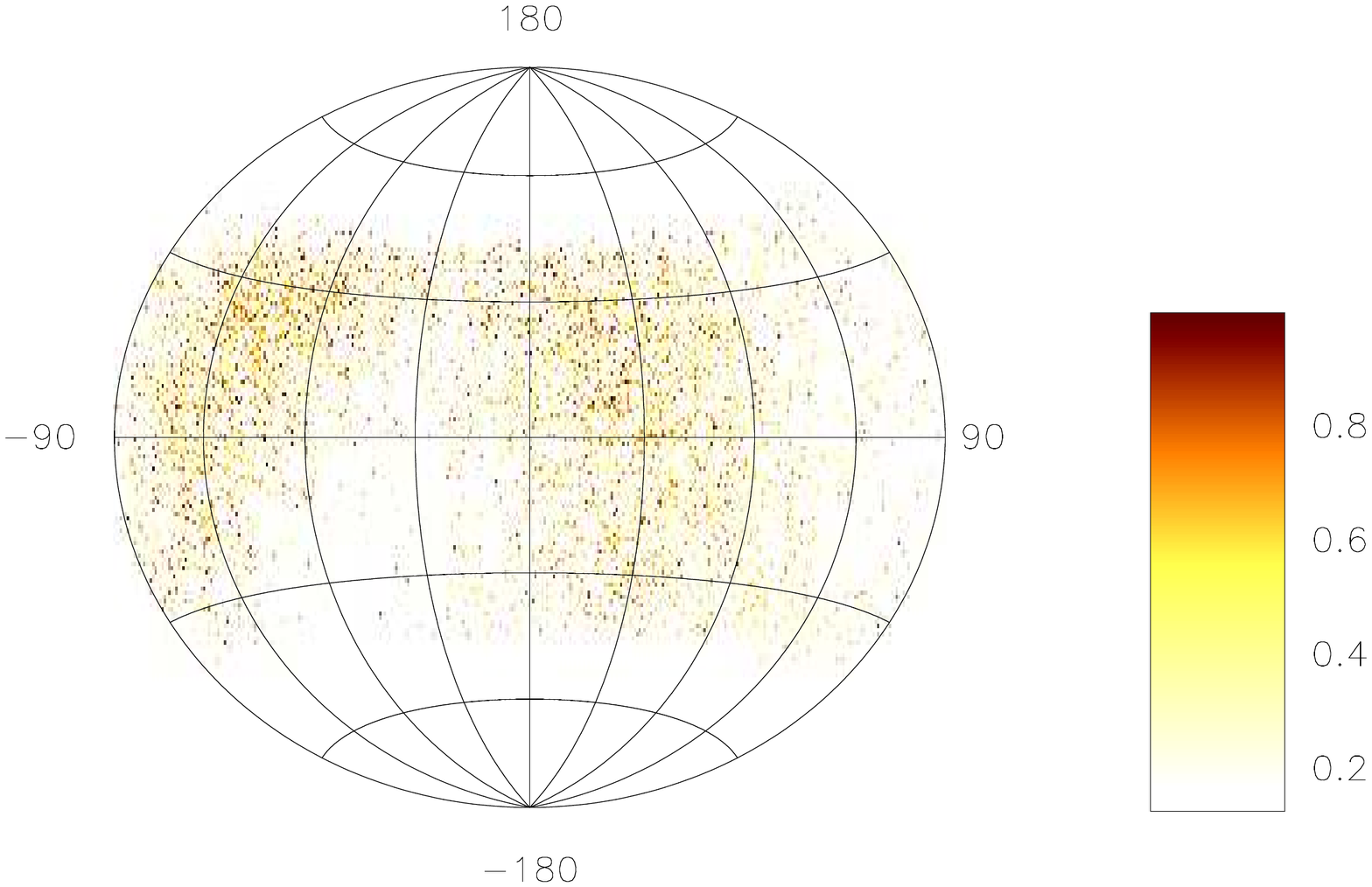}}}\\
\end{tabular}
\caption{Top panel (upper window) shows the average profile with total
intensity (Stokes I; solid black lines), total linear polarization (dashed red
line) and circular polarization (Stokes V; dotted blue line). Top panel (lower
window) also shows the single pulse PPA distribution (colour scale) along with
the average PPA (red error bars).
The RVM fits to the average PPA (dashed pink
line) is also shown in this plot. Middle panel show
the $\chi^2$ contours for the parameters $\alpha$ and $\beta$ obtained from RVM
fits.
Bottom panel shows the Hammer-Aitoff projection of the polarized time
samples with the colour scheme representing the fractional polarization level.}
\label{a52}
\end{center}
\end{figure*}


\begin{figure*}
\begin{center}
\begin{tabular}{cc}
{\mbox{\includegraphics[width=9cm,height=6cm,angle=0.]{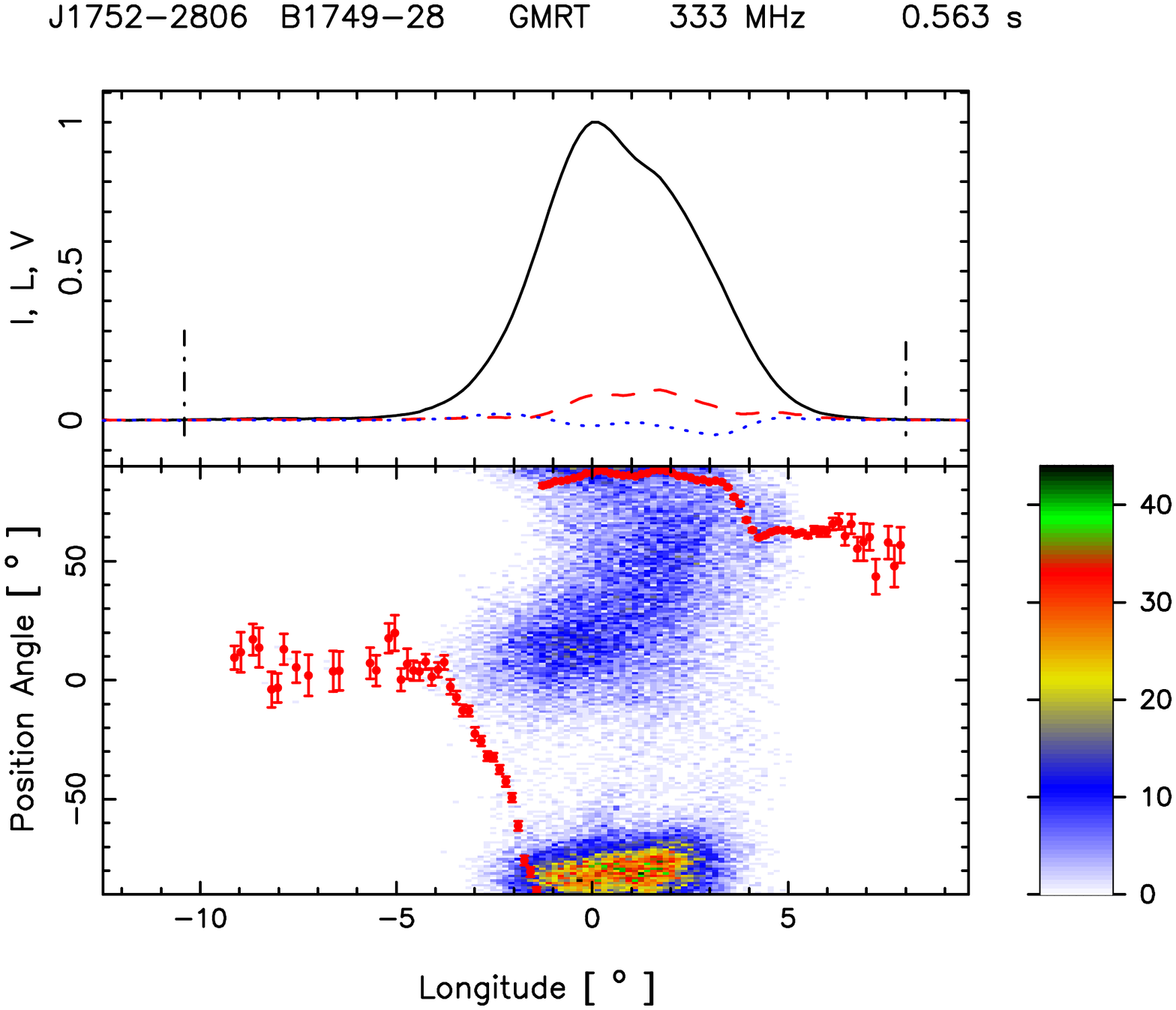}}}&
{\mbox{\includegraphics[width=9cm,height=6cm,angle=0.]{J1752-2806_602MHz_27Mar2014.dat.epn2a.pdf}}}\\
&
\\
{\mbox{\includegraphics[width=9cm,height=6cm,angle=0.]{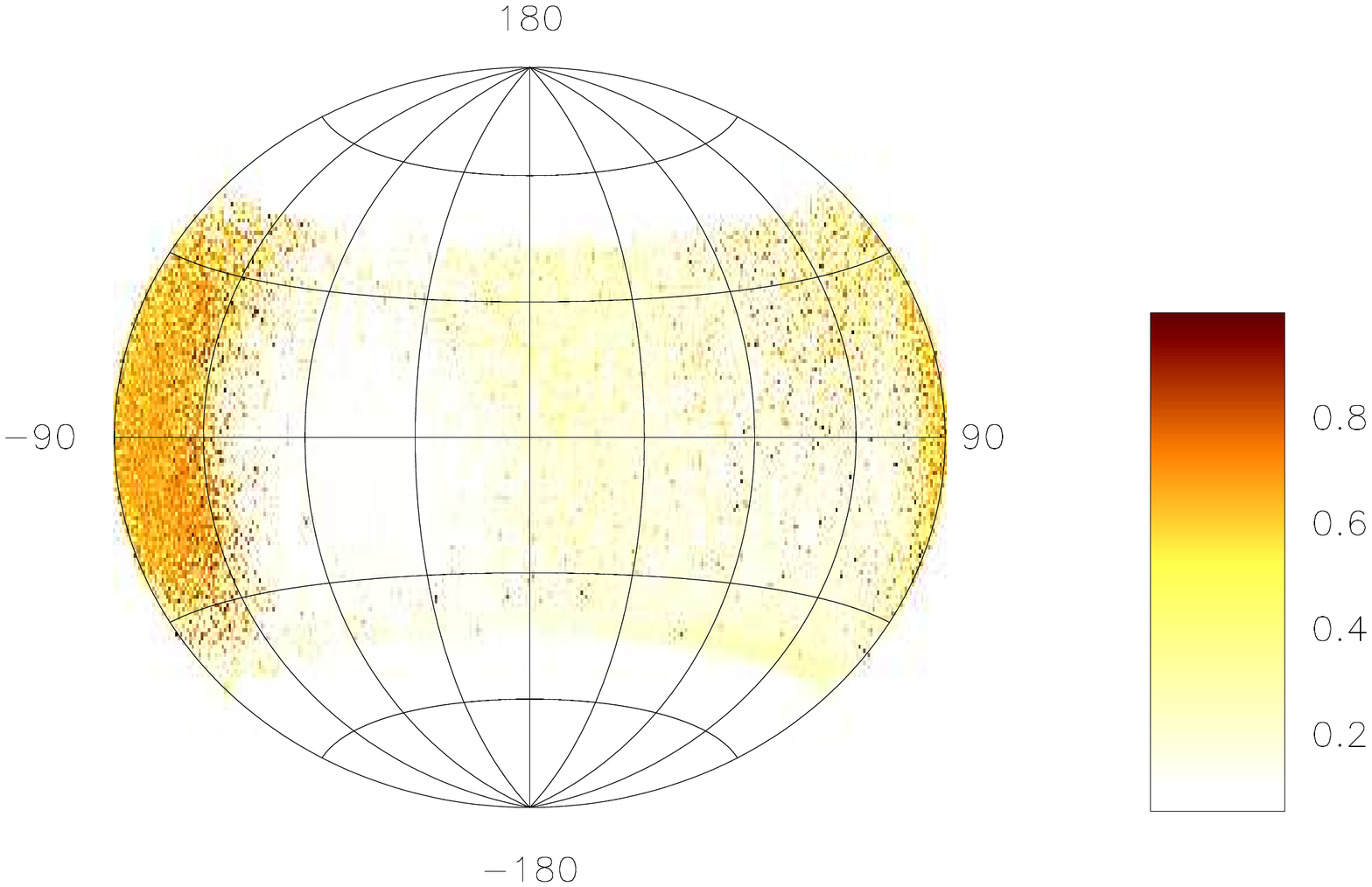}}}&
{\mbox{\includegraphics[width=9cm,height=6cm,angle=0.]{J1752-2806_602MHz_27Mar2014.dat.epn2a.71.pdf}}}\\
\end{tabular}
\caption{Top panel (upper window) shows the average profile with total
intensity (Stokes I; solid black lines), total linear polarization (dashed red
line) and circular polarization (Stokes V; dotted blue line). Top panel (lower
window) also shows the single pulse PPA distribution (colour scale) along with
the average PPA (red error bars).
Bottom panel shows the Hammer-Aitoff projection of the polarized time
samples with the colour scheme representing the fractional polarization level.}
\label{a53}
\end{center}
\end{figure*}


\begin{figure*}
\begin{center}
\begin{tabular}{cc}
{\mbox{\includegraphics[width=9cm,height=6cm,angle=0.]{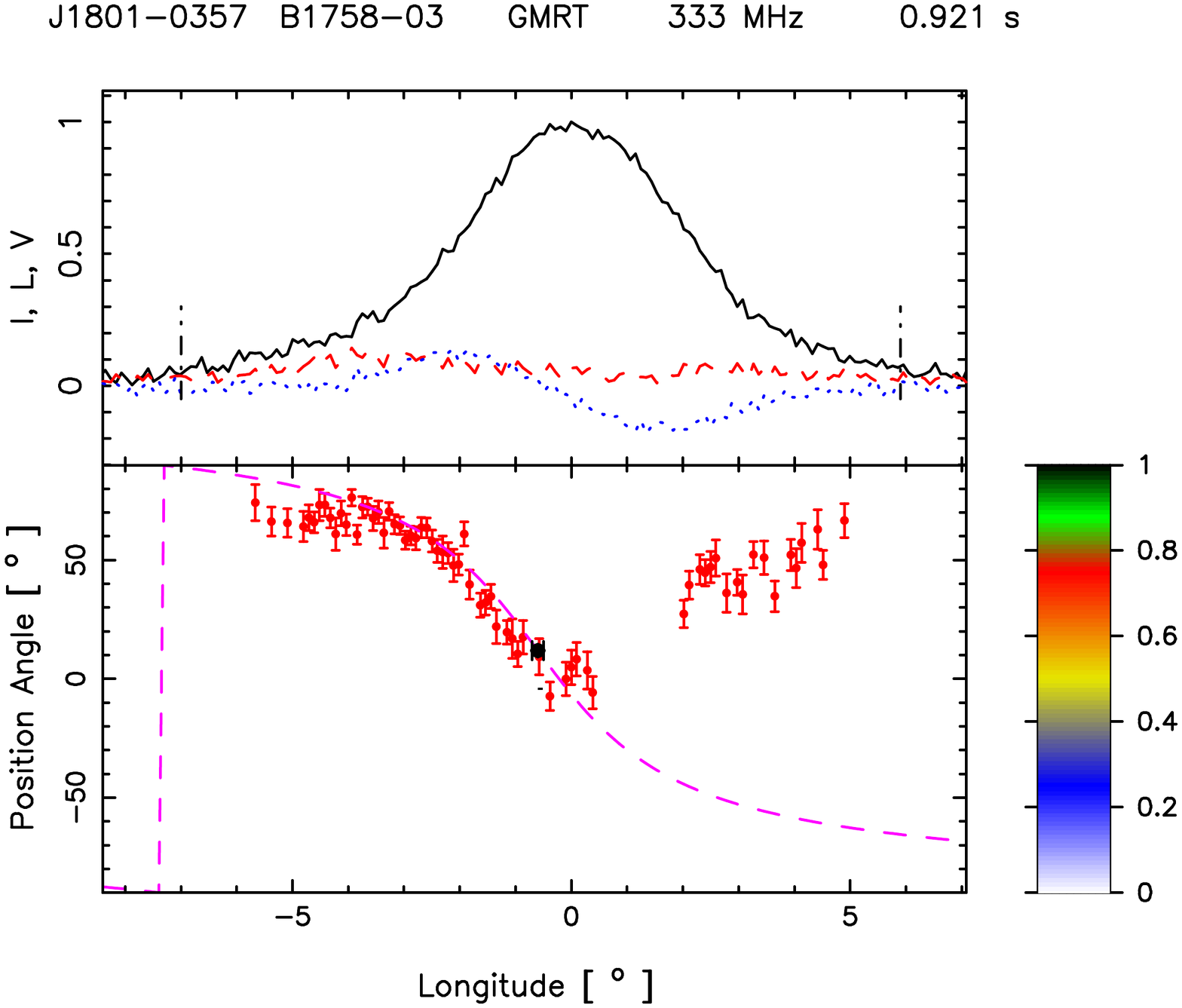}}}&
{\mbox{\includegraphics[width=9cm,height=6cm,angle=0.]{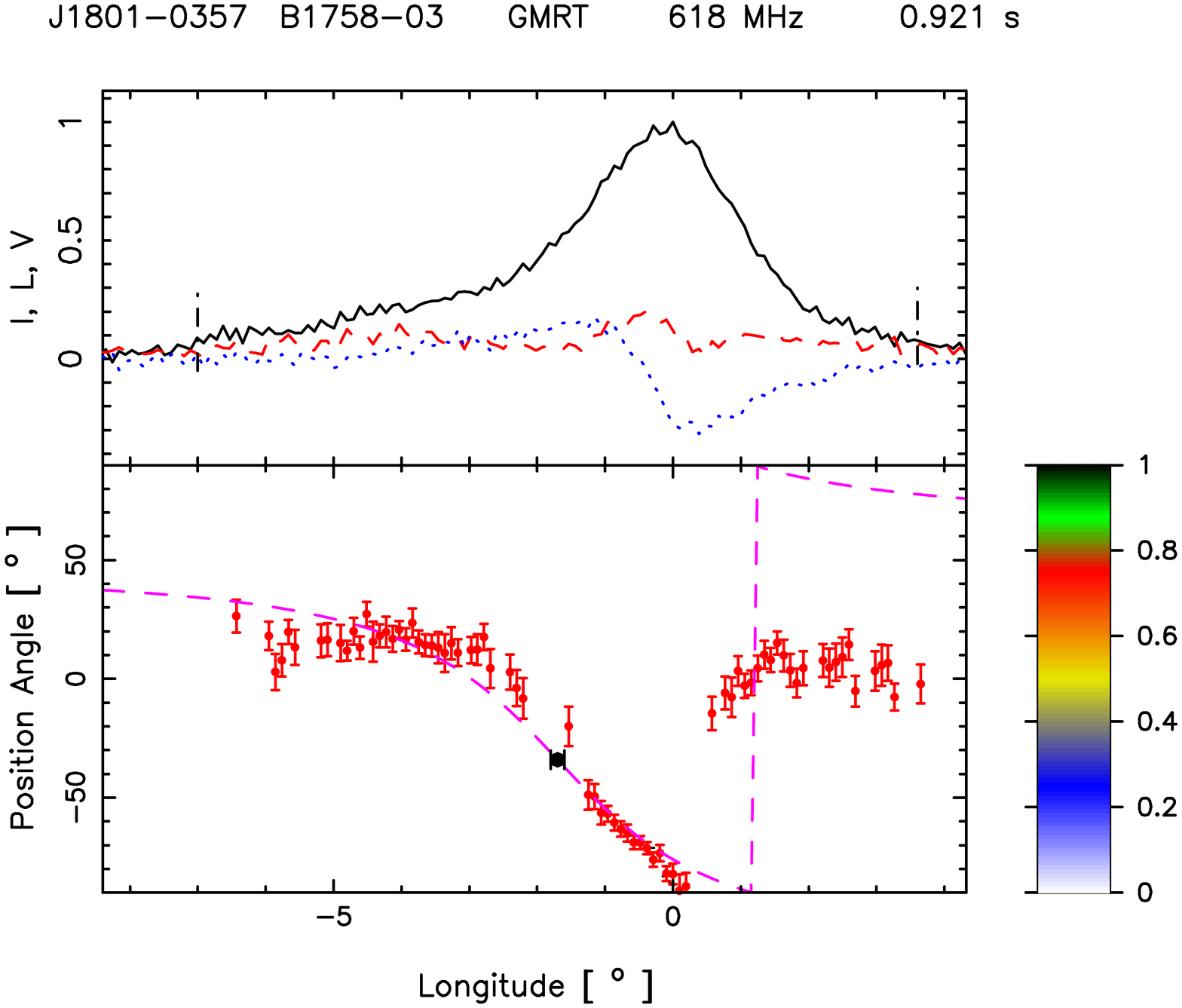}}}\\
{\mbox{\includegraphics[width=9cm,height=6cm,angle=0.]{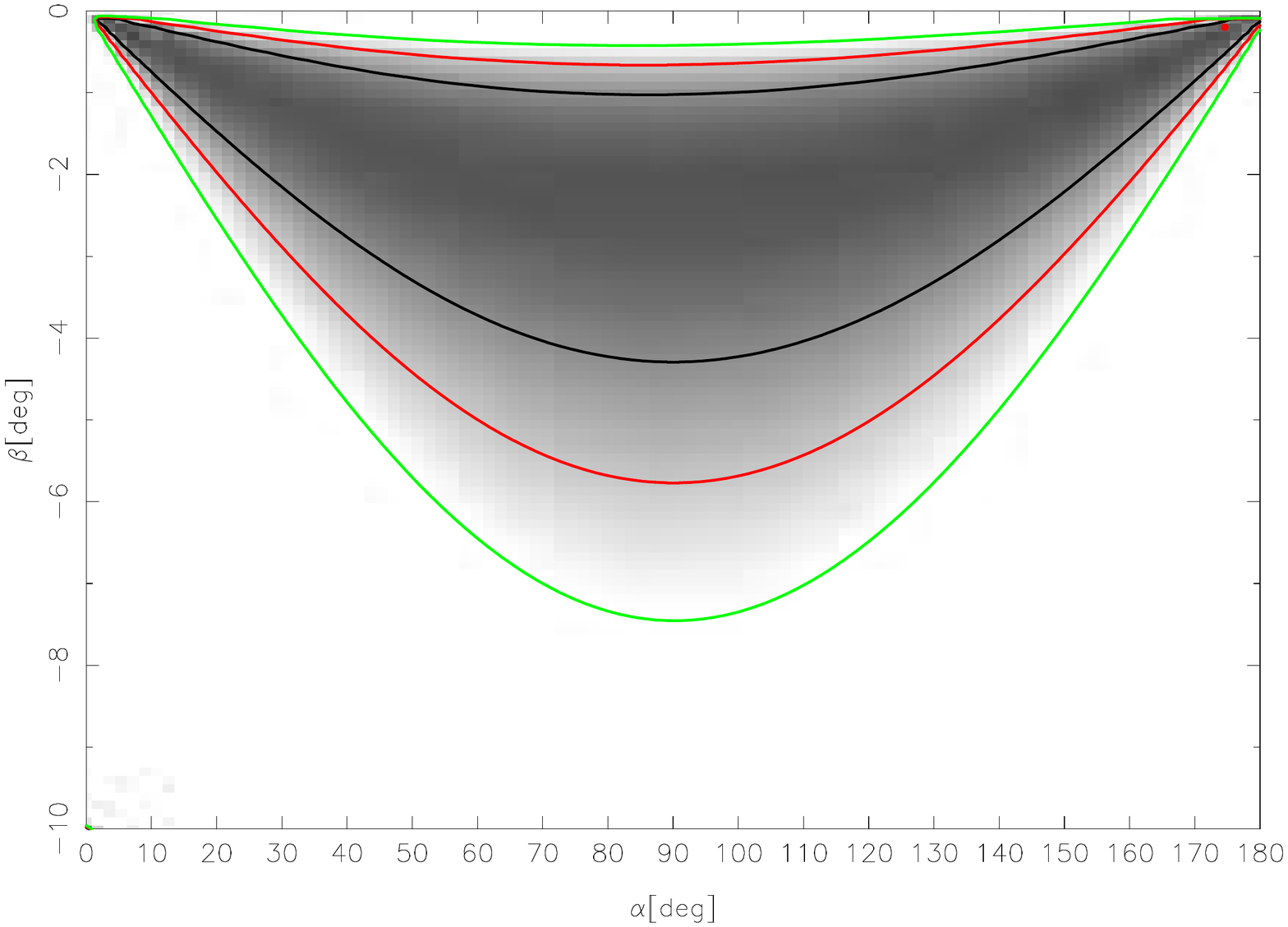}}}&
{\mbox{\includegraphics[width=9cm,height=6cm,angle=0.]{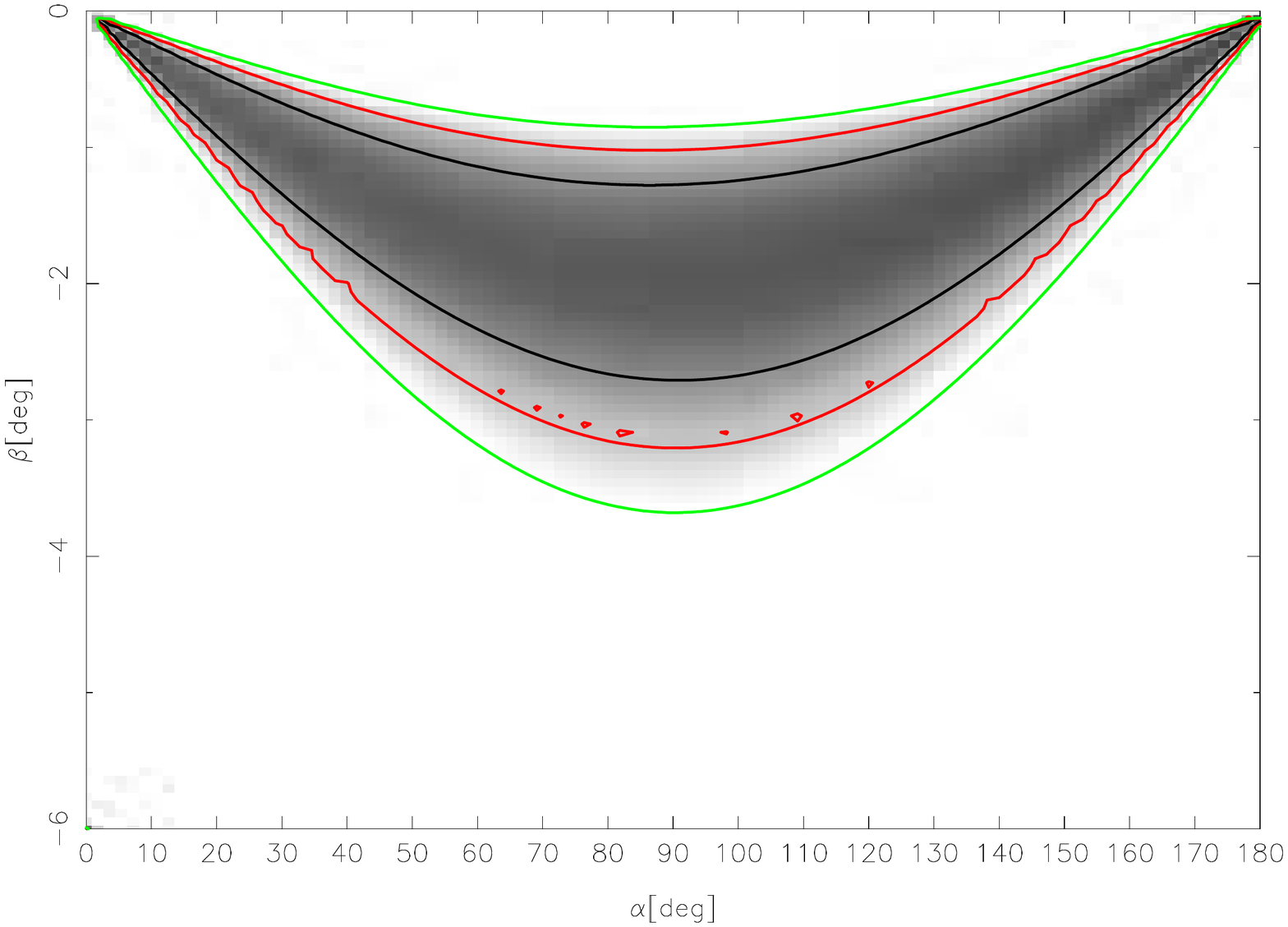}}}\\
&
\\
\end{tabular}
\caption{Top panel (upper window) shows the average profile with total
intensity (Stokes I; solid black lines), total linear polarization (dashed red
line) and circular polarization (Stokes V; dotted blue line). Top panel (lower
window) also shows the single pulse PPA distribution (colour scale) along with
the average PPA (red error bars).
The RVM fits to the average PPA (dashed pink
line) is also shown in this plot. Bottom panel show
the $\chi^2$ contours for the parameters $\alpha$ and $\beta$ obtained from RVM
fits.}
\label{a54}
\end{center}
\end{figure*}
\clearpage


\begin{figure*}
\begin{center}
\begin{tabular}{cc}
{\mbox{\includegraphics[width=9cm,height=6cm,angle=0.]{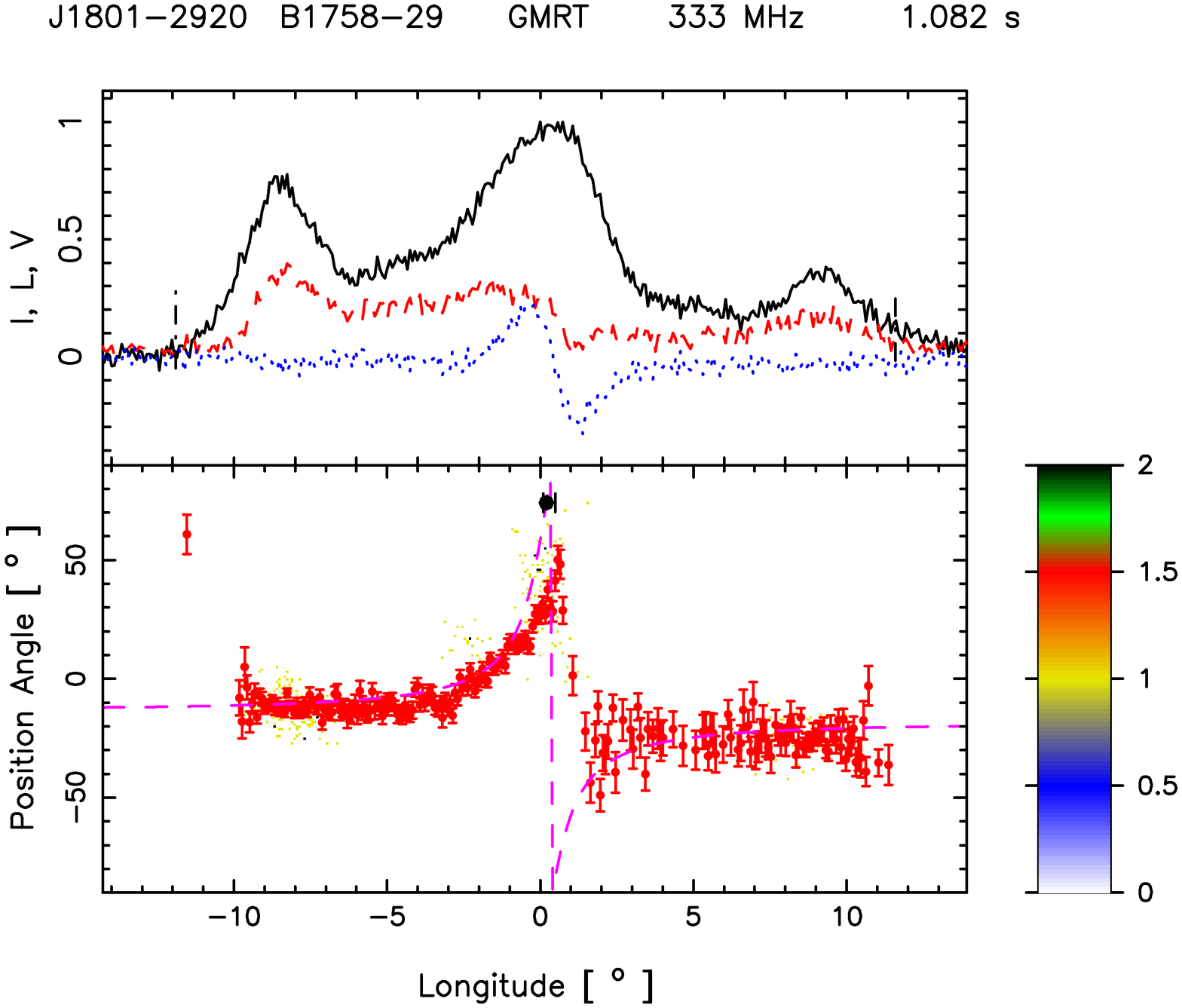}}}&
{\mbox{\includegraphics[width=9cm,height=6cm,angle=0.]{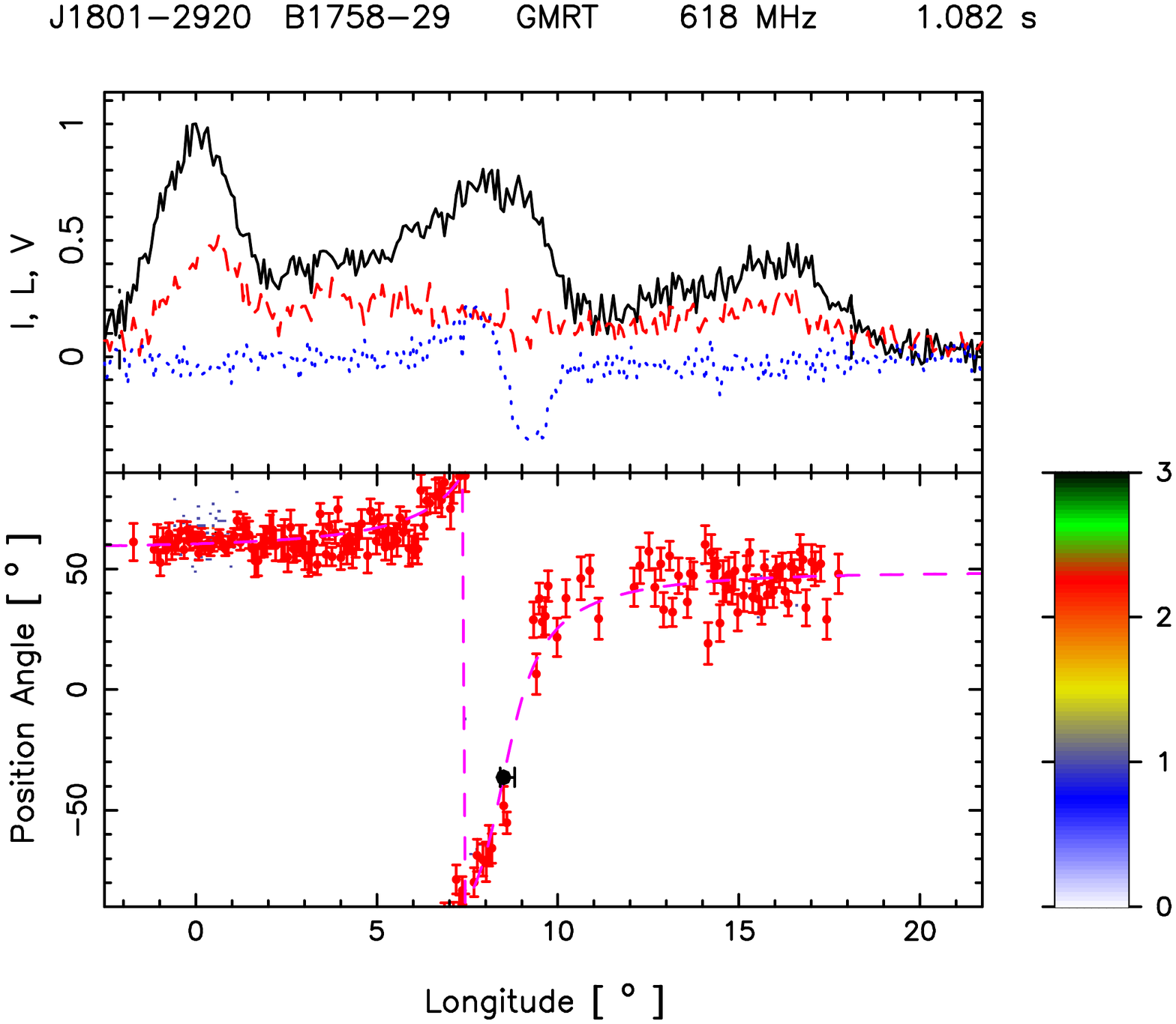}}}\\
{\mbox{\includegraphics[width=9cm,height=6cm,angle=0.]{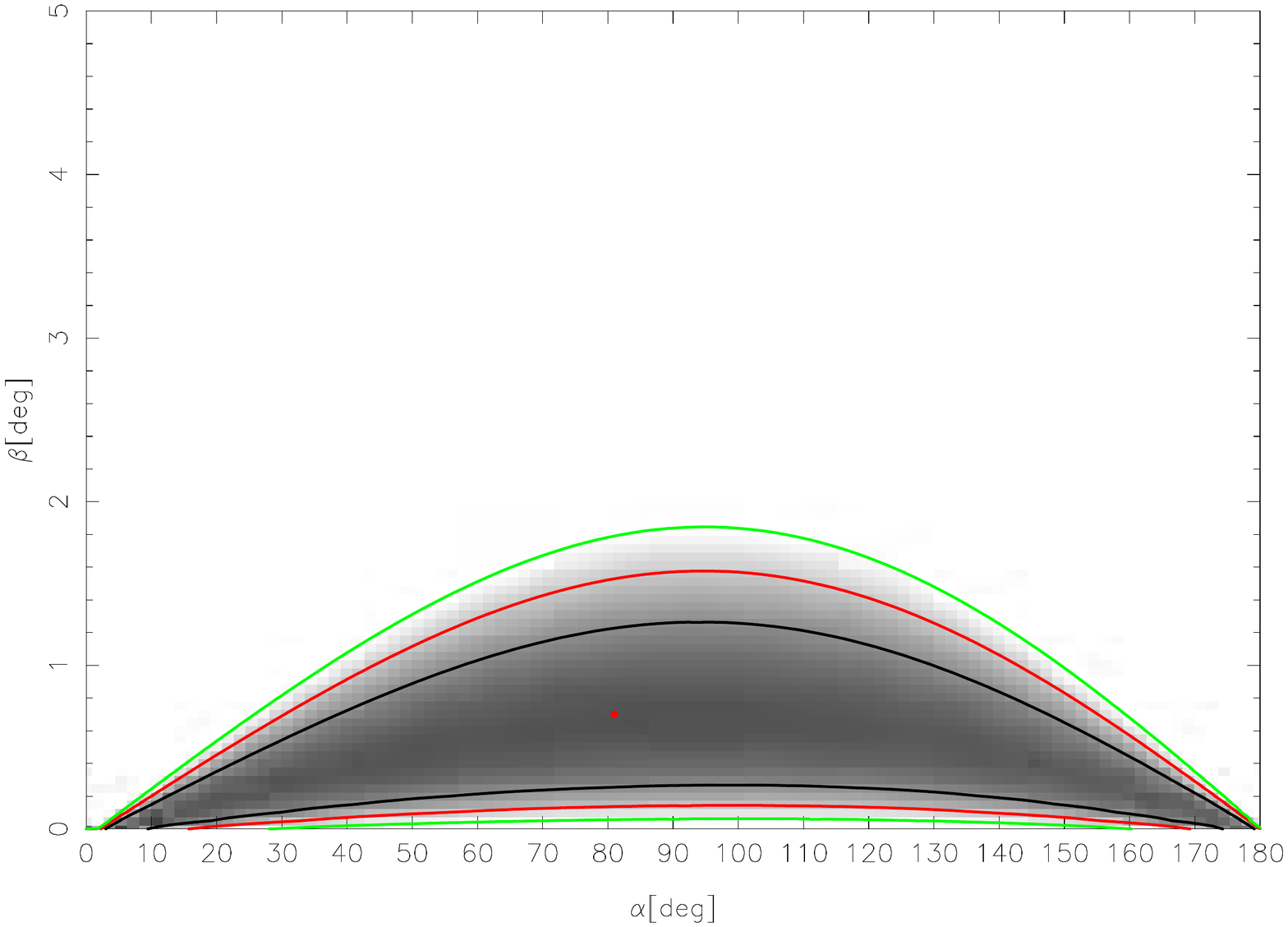}}}&
{\mbox{\includegraphics[width=9cm,height=6cm,angle=0.]{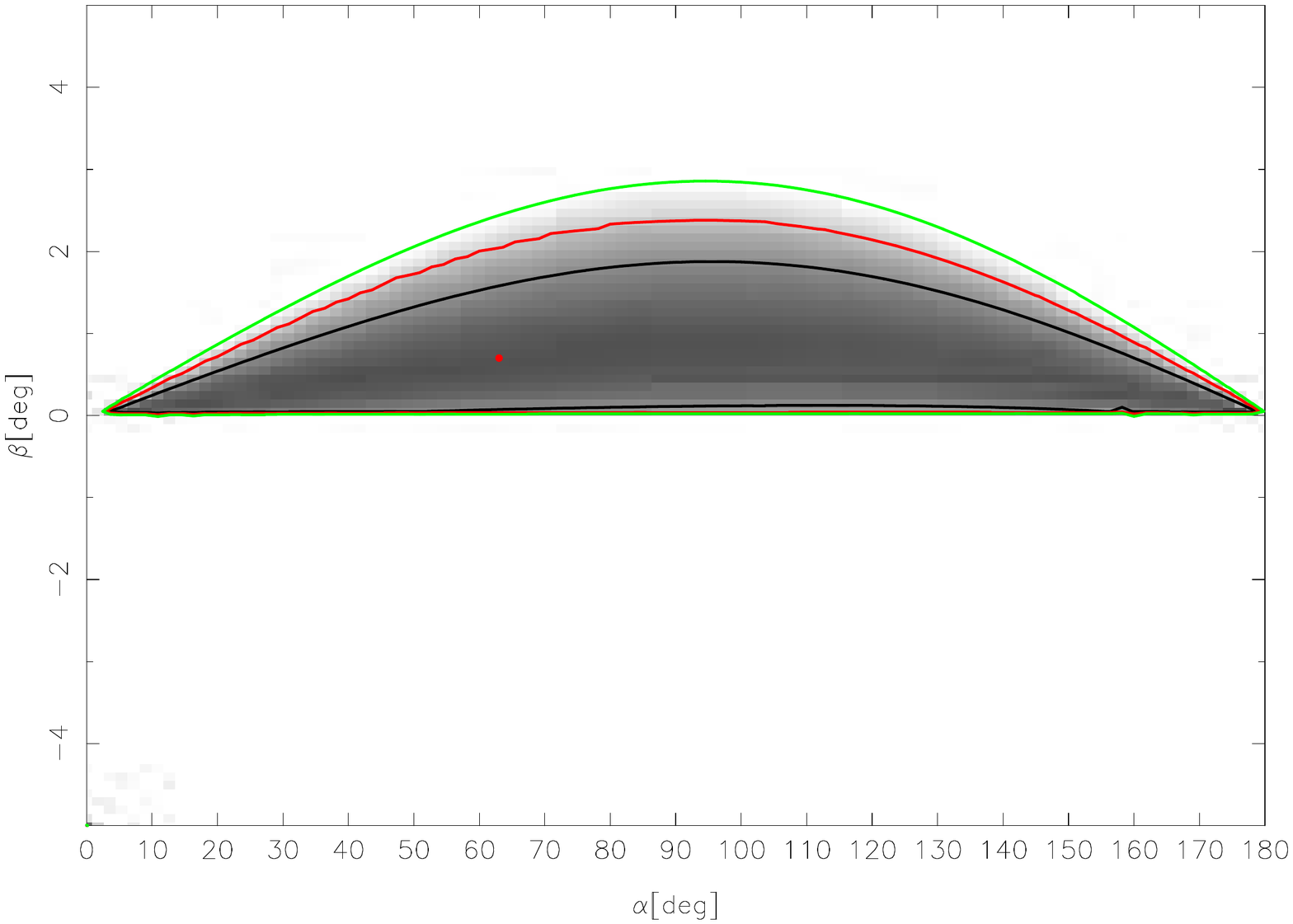}}}\\
{\mbox{\includegraphics[width=9cm,height=6cm,angle=0.]{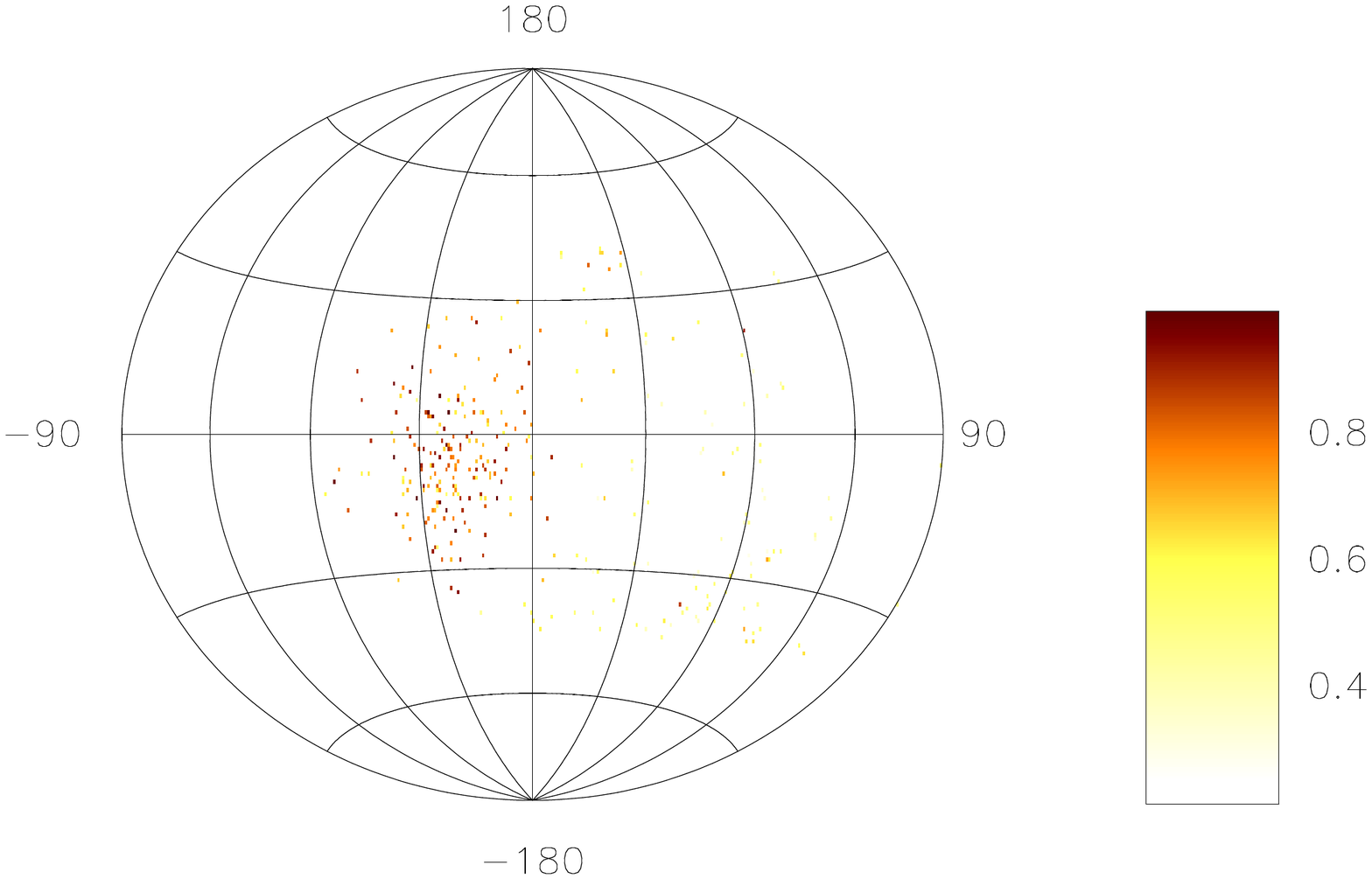}}}&
\\
\end{tabular}
\caption{Top panel (upper window) shows the average profile with total
intensity (Stokes I; solid black lines), total linear polarization (dashed red
line) and circular polarization (Stokes V; dotted blue line). Top panel (lower
window) also shows the single pulse PPA distribution (colour scale) along with
the average PPA (red error bars).
The RVM fits to the average PPA (dashed pink
line) is also shown in this plot. Middle panel show
the $\chi^2$ contours for the parameters $\alpha$ and $\beta$ obtained from RVM
fits.
Bottom panel only for 333 MHz shows the Hammer-Aitoff projection of the polarized time
samples with the colour scheme representing the fractional polarization level.}
\label{a55}
\end{center}
\end{figure*}


\begin{figure*}
\begin{center}
\begin{tabular}{cc}
{\mbox{\includegraphics[width=9cm,height=6cm,angle=0.]{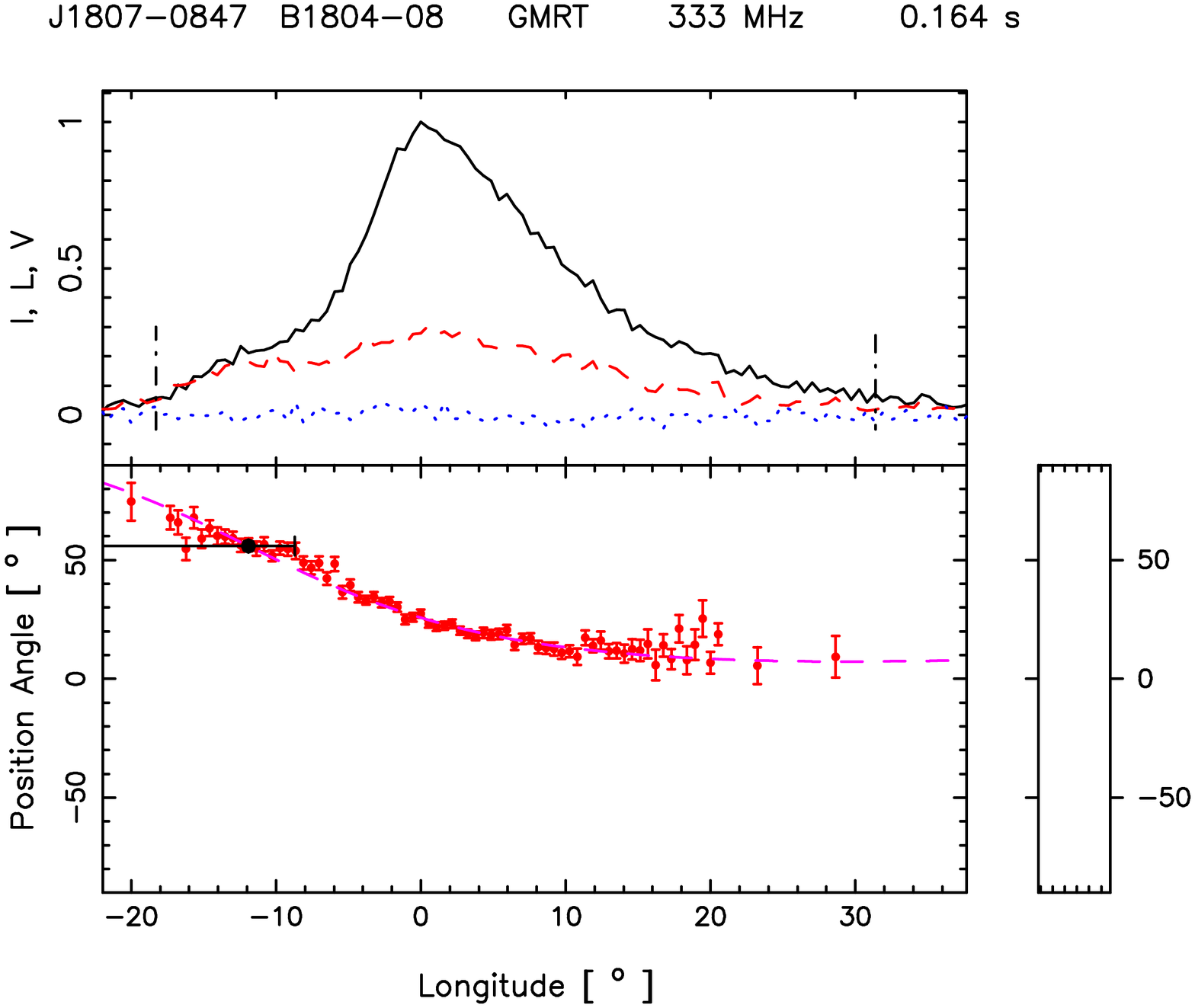}}}&
{\mbox{\includegraphics[width=9cm,height=6cm,angle=0.]{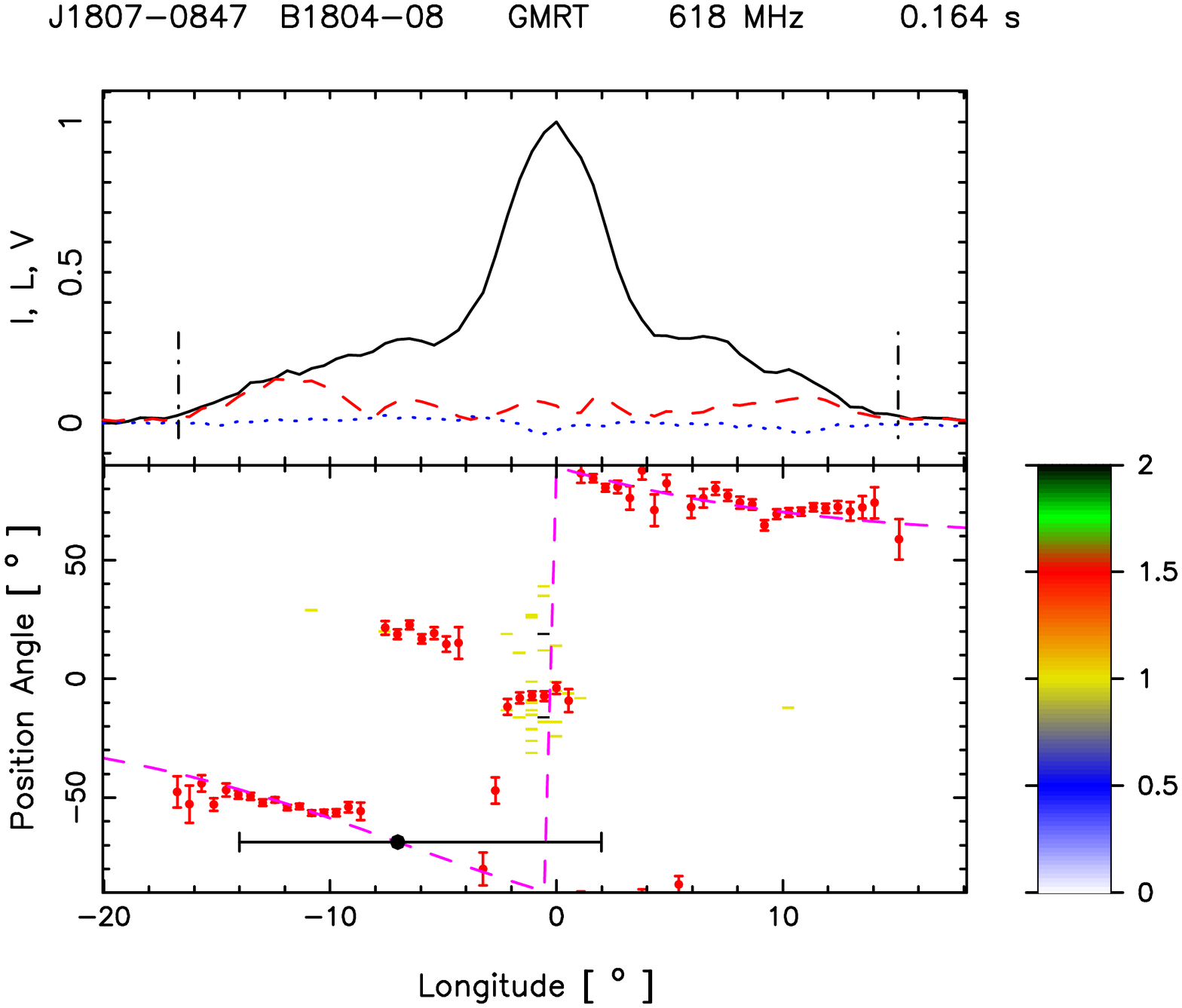}}}\\
{\mbox{\includegraphics[width=9cm,height=6cm,angle=0.]{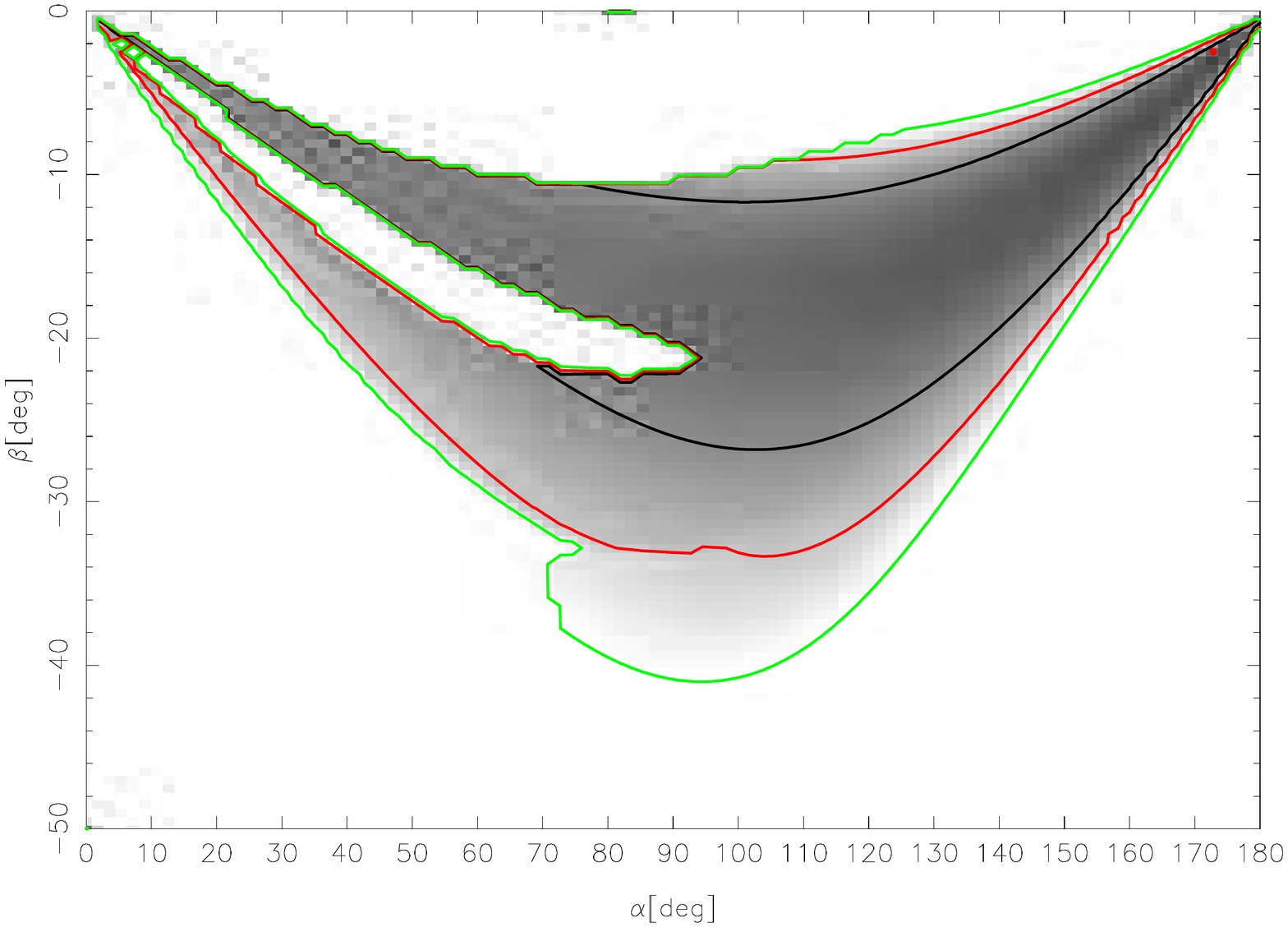}}}&
{\mbox{\includegraphics[width=9cm,height=6cm,angle=0.]{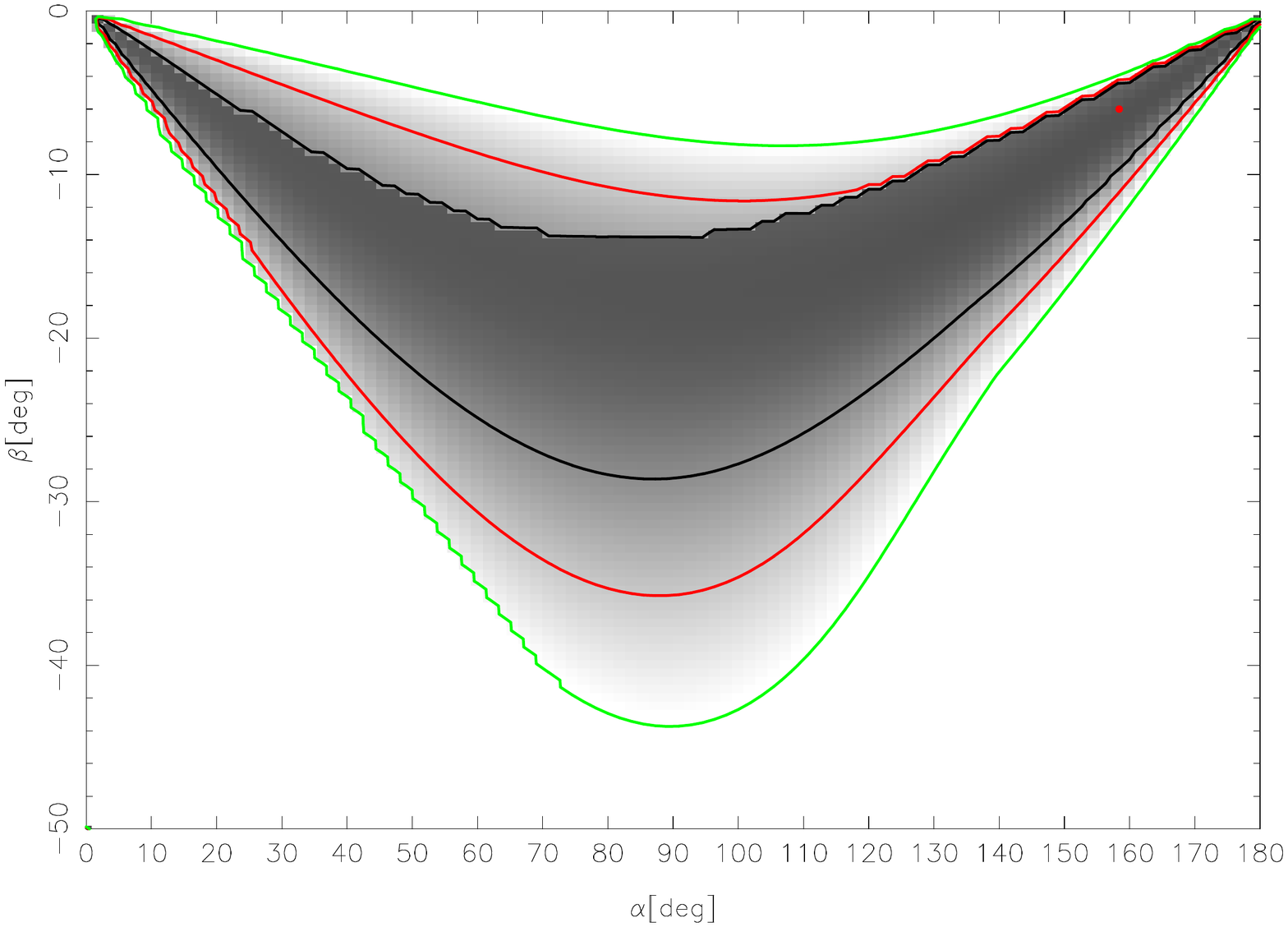}}}\\
&
\\
\end{tabular}
\caption{Top panel (upper window) shows the average profile with total
intensity (Stokes I; solid black lines), total linear polarization (dashed red
line) and circular polarization (Stokes V; dotted blue line). Top panel (lower
window) also shows the single pulse PPA distribution (colour scale) along with
the average PPA (red error bars).
The RVM fits to the average PPA (dashed pink
line) is also shown in this plot. Bottom panel show
the $\chi^2$ contours for the parameters $\alpha$ and $\beta$ obtained from RVM
fits.}
\label{a56}
\end{center}
\end{figure*}


\begin{figure*}
\begin{center}
\begin{tabular}{cc}
{\mbox{\includegraphics[width=9cm,height=6cm,angle=0.]{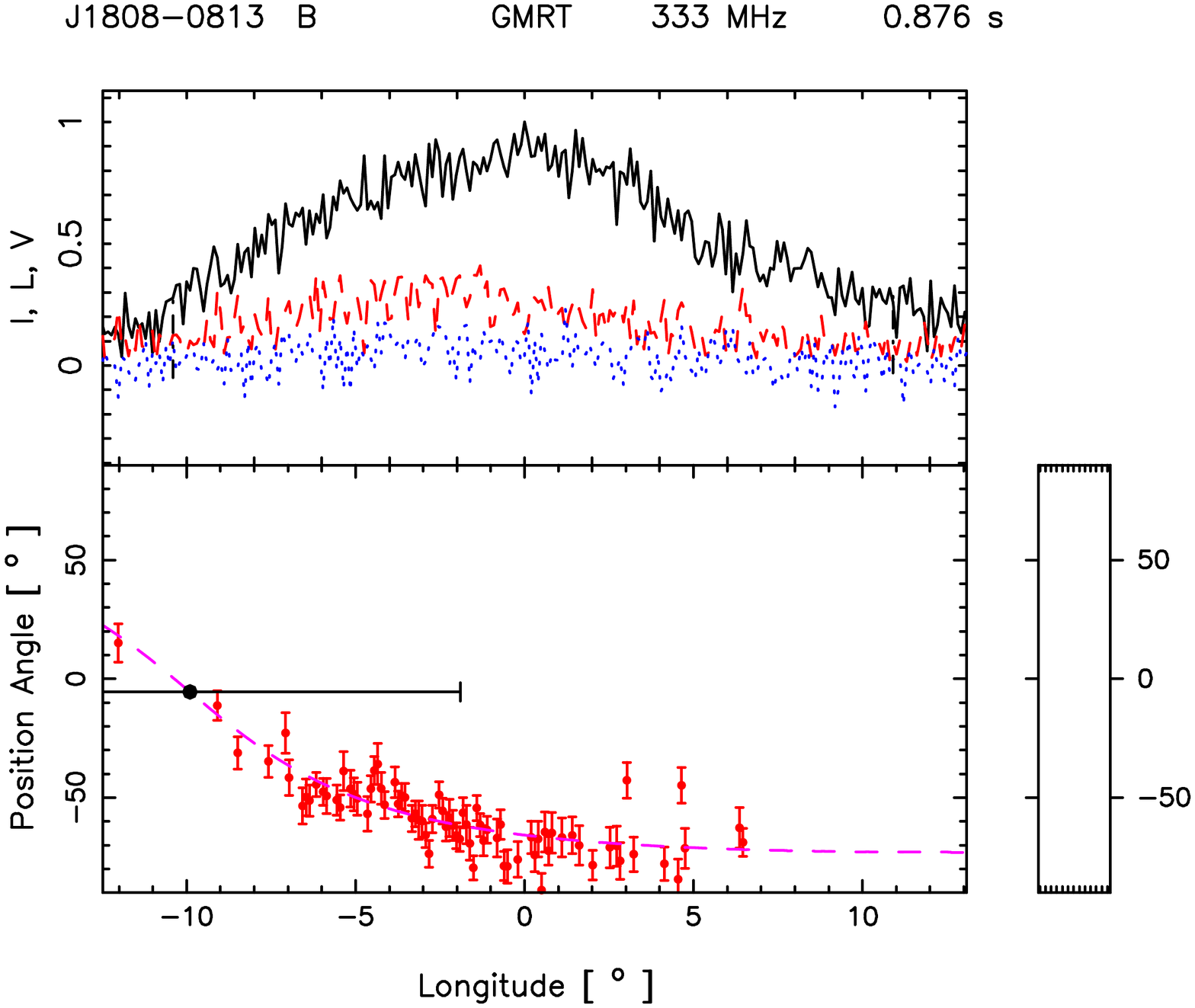}}}&
{\mbox{\includegraphics[width=9cm,height=6cm,angle=0.]{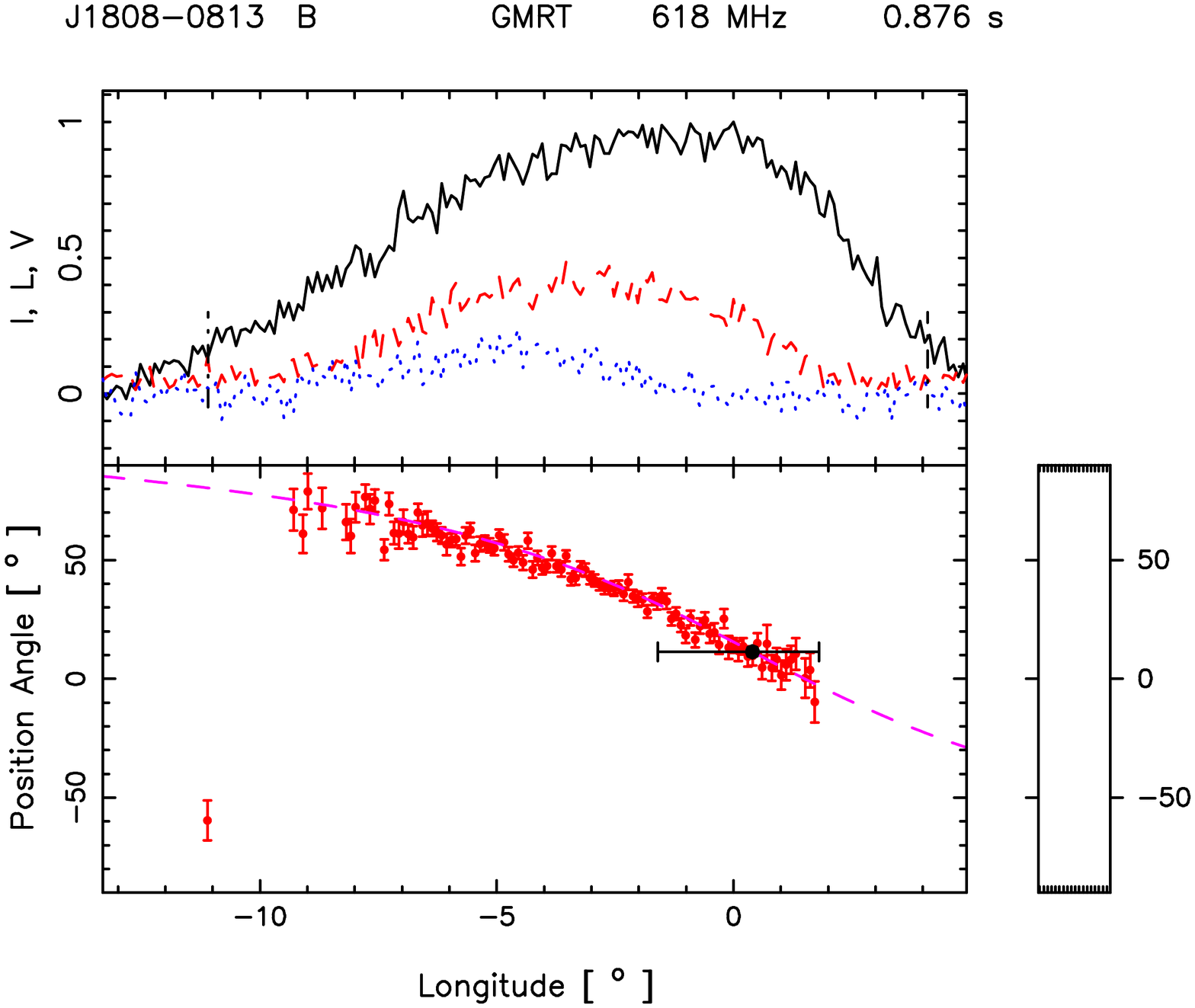}}}\\
{\mbox{\includegraphics[width=9cm,height=6cm,angle=0.]{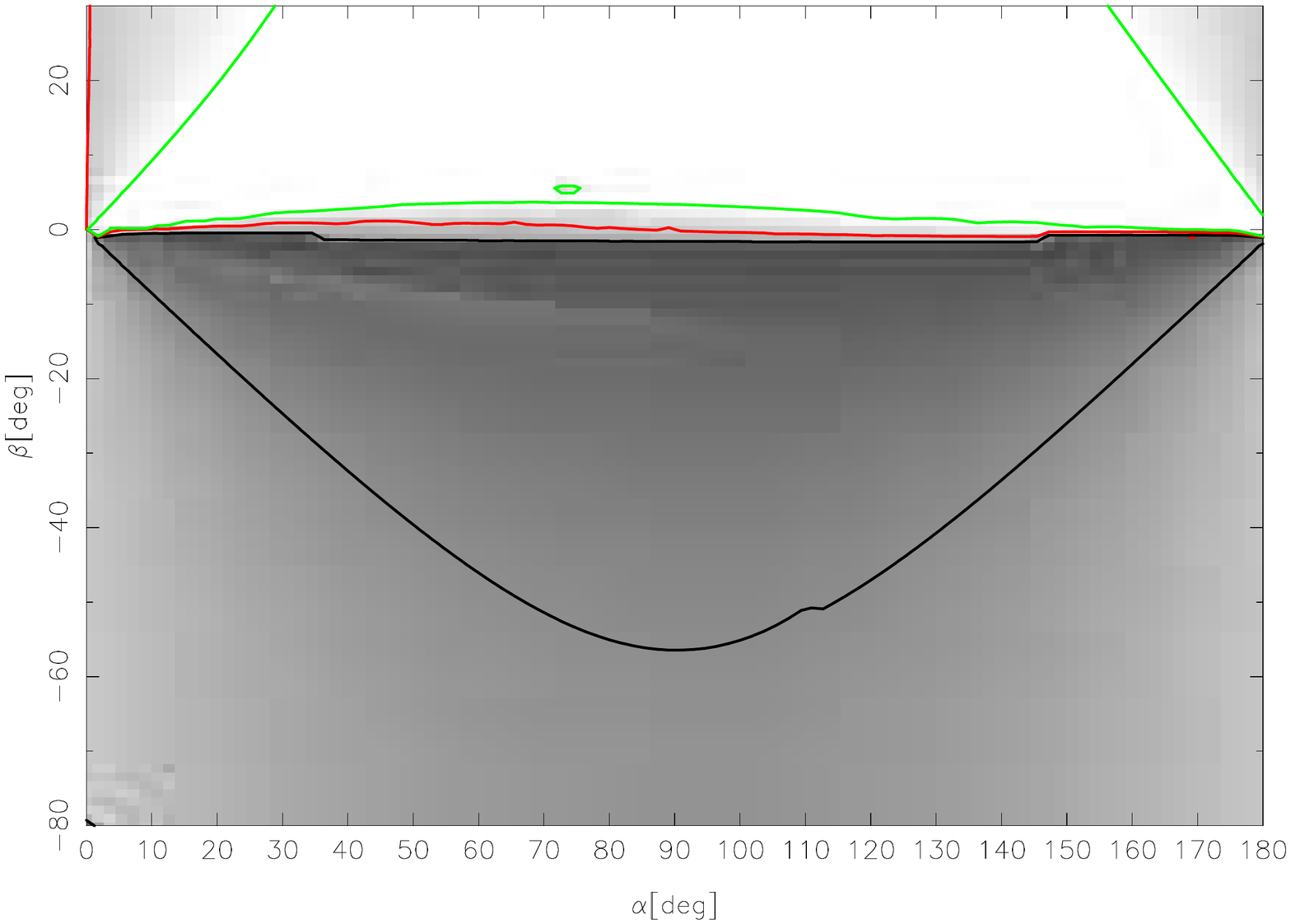}}}&
{\mbox{\includegraphics[width=9cm,height=6cm,angle=0.]{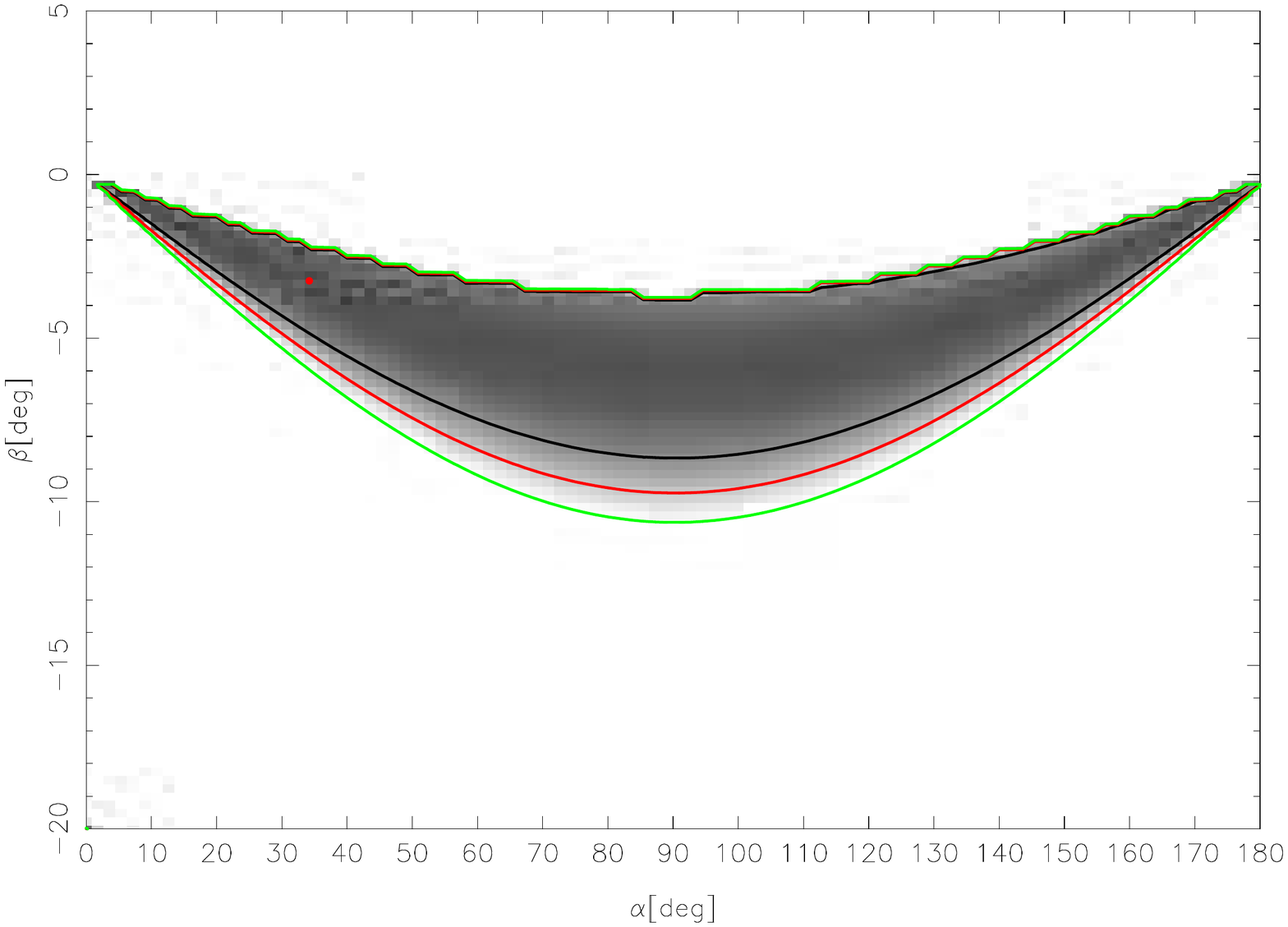}}}\\
&
\\
\end{tabular}
\caption{Top panel (upper window) shows the average profile with total
intensity (Stokes I; solid black lines), total linear polarization (dashed red
line) and circular polarization (Stokes V; dotted blue line). Top panel (lower
window) also shows the single pulse PPA distribution (colour scale) along with
the average PPA (red error bars).
The RVM fits to the average PPA (dashed pink
line) is also shown in this plot. Bottom panel show
the $\chi^2$ contours for the parameters $\alpha$ and $\beta$ obtained from RVM
fits.}
\label{a57}
\end{center}
\end{figure*}


\begin{figure*}
\begin{center}
\begin{tabular}{cc}
{\mbox{\includegraphics[width=9cm,height=6cm,angle=0.]{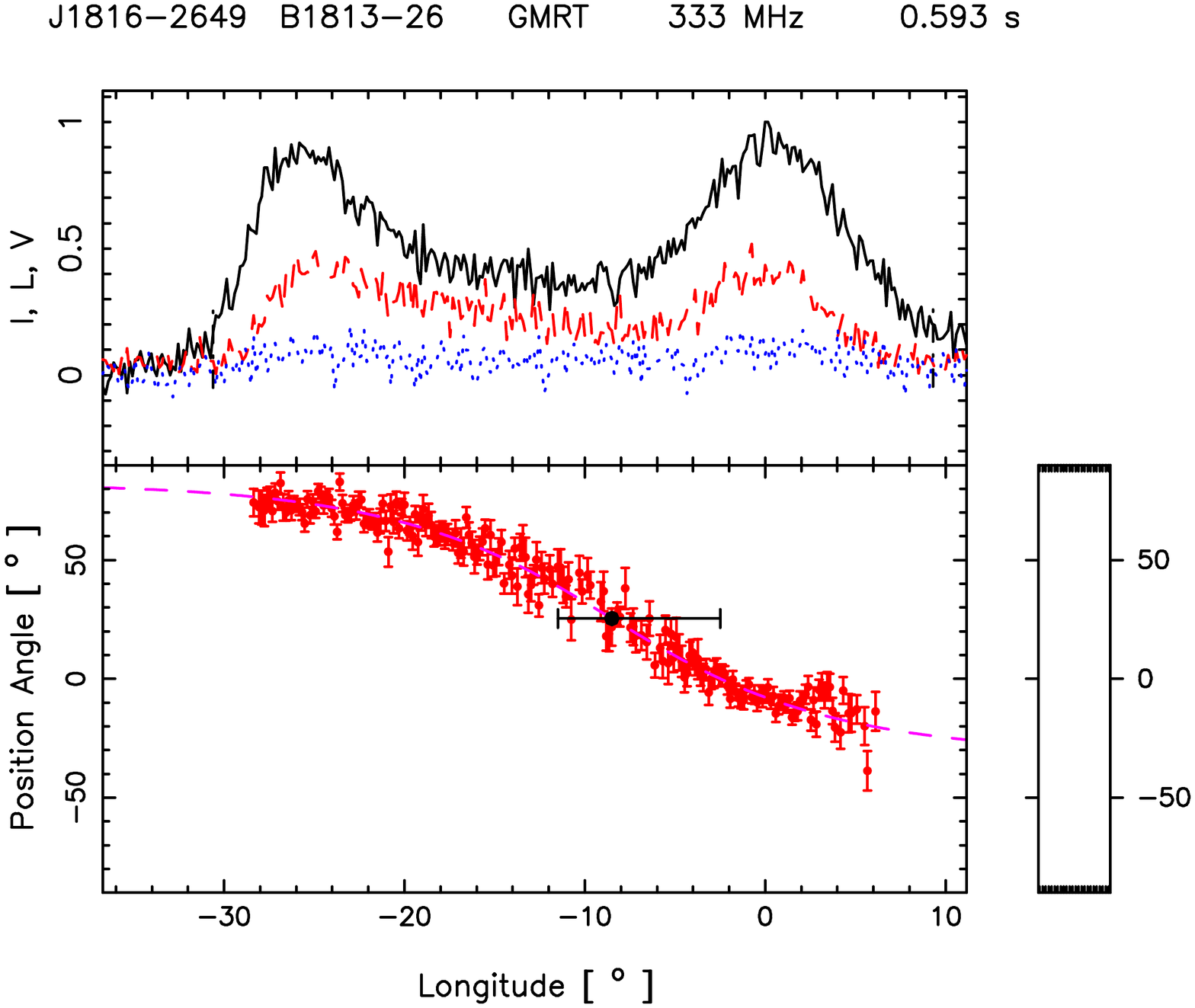}}}&
{\mbox{\includegraphics[width=9cm,height=6cm,angle=0.]{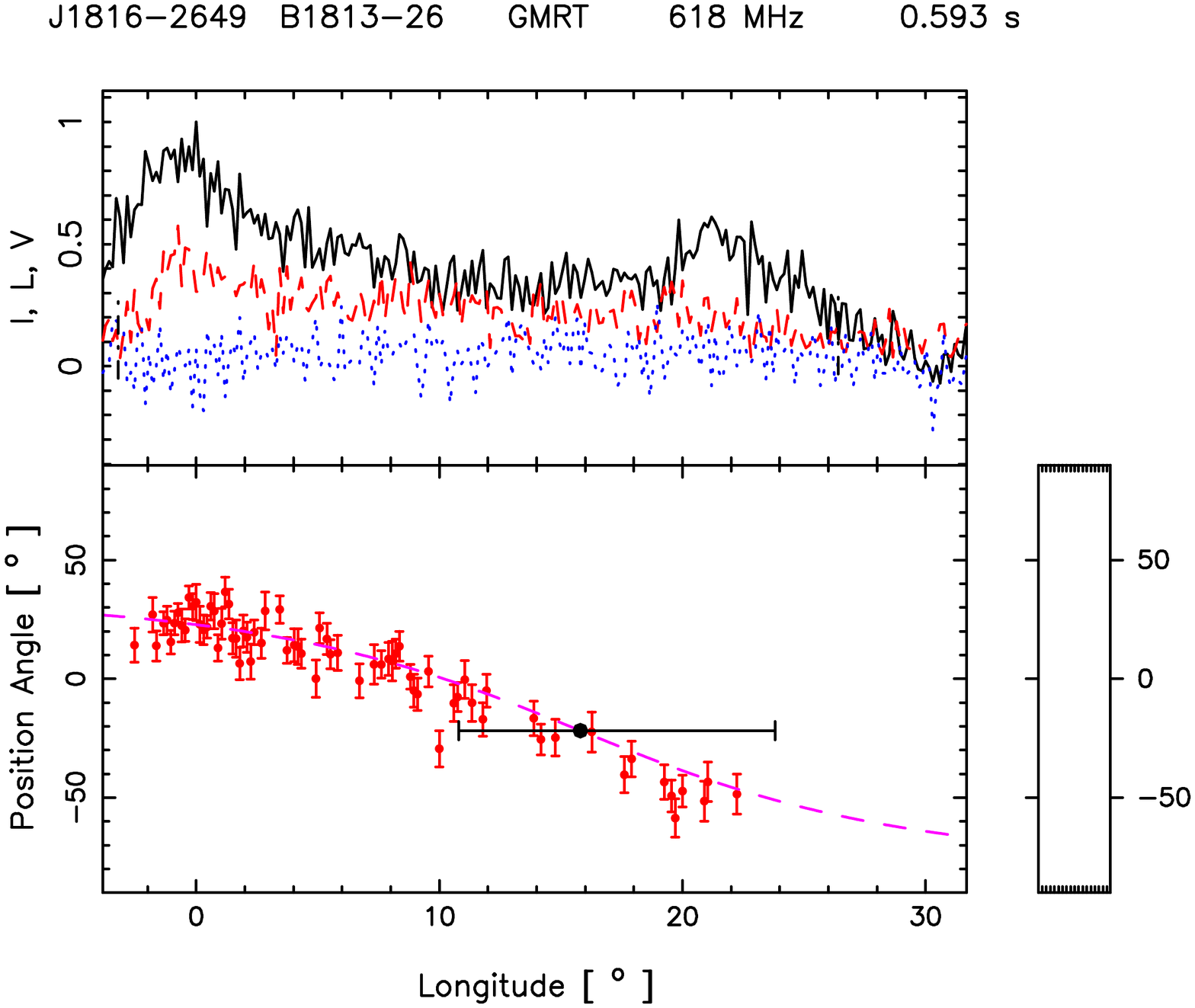}}}\\
{\mbox{\includegraphics[width=9cm,height=6cm,angle=0.]{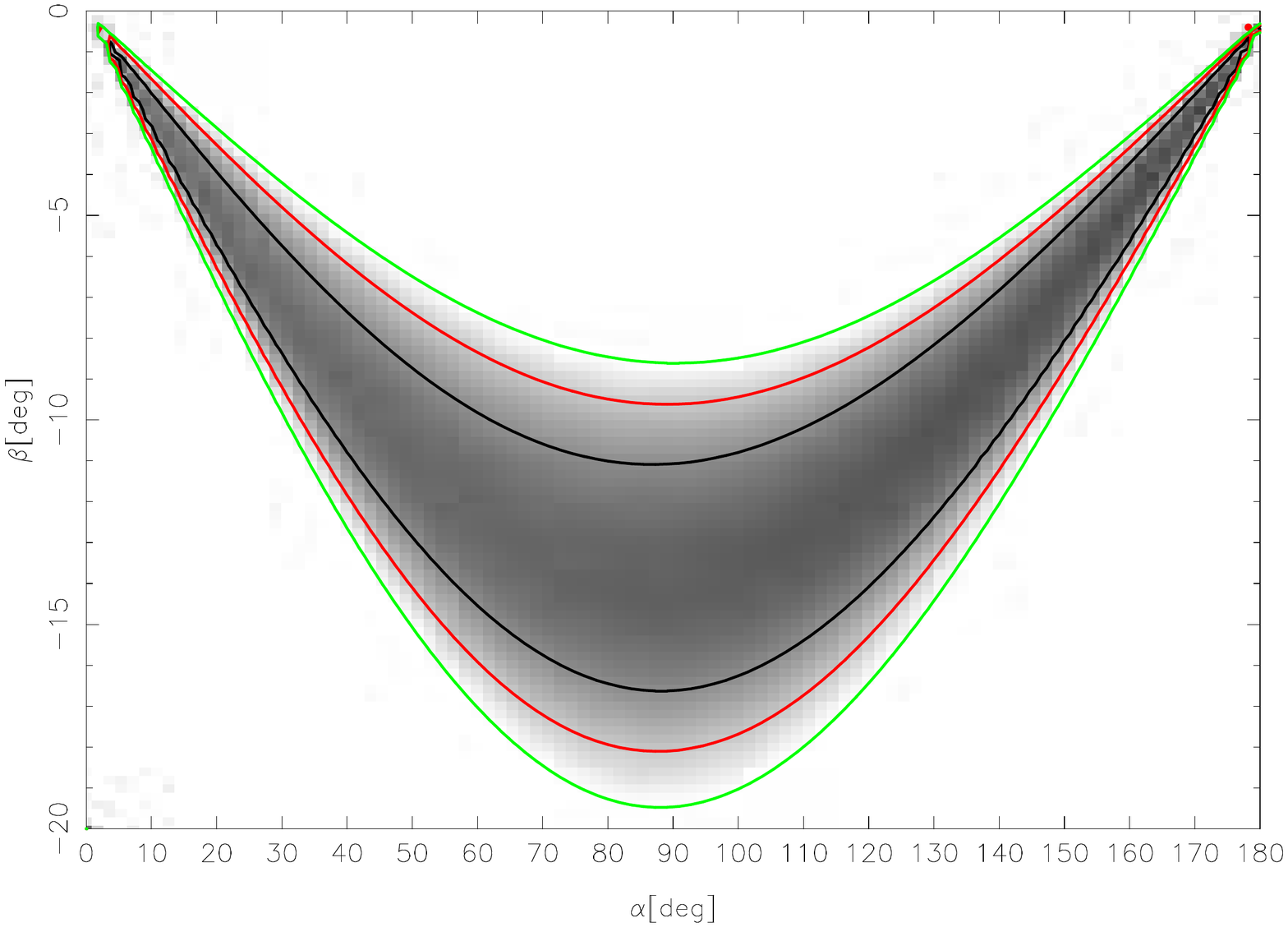}}}&
{\mbox{\includegraphics[width=9cm,height=6cm,angle=0.]{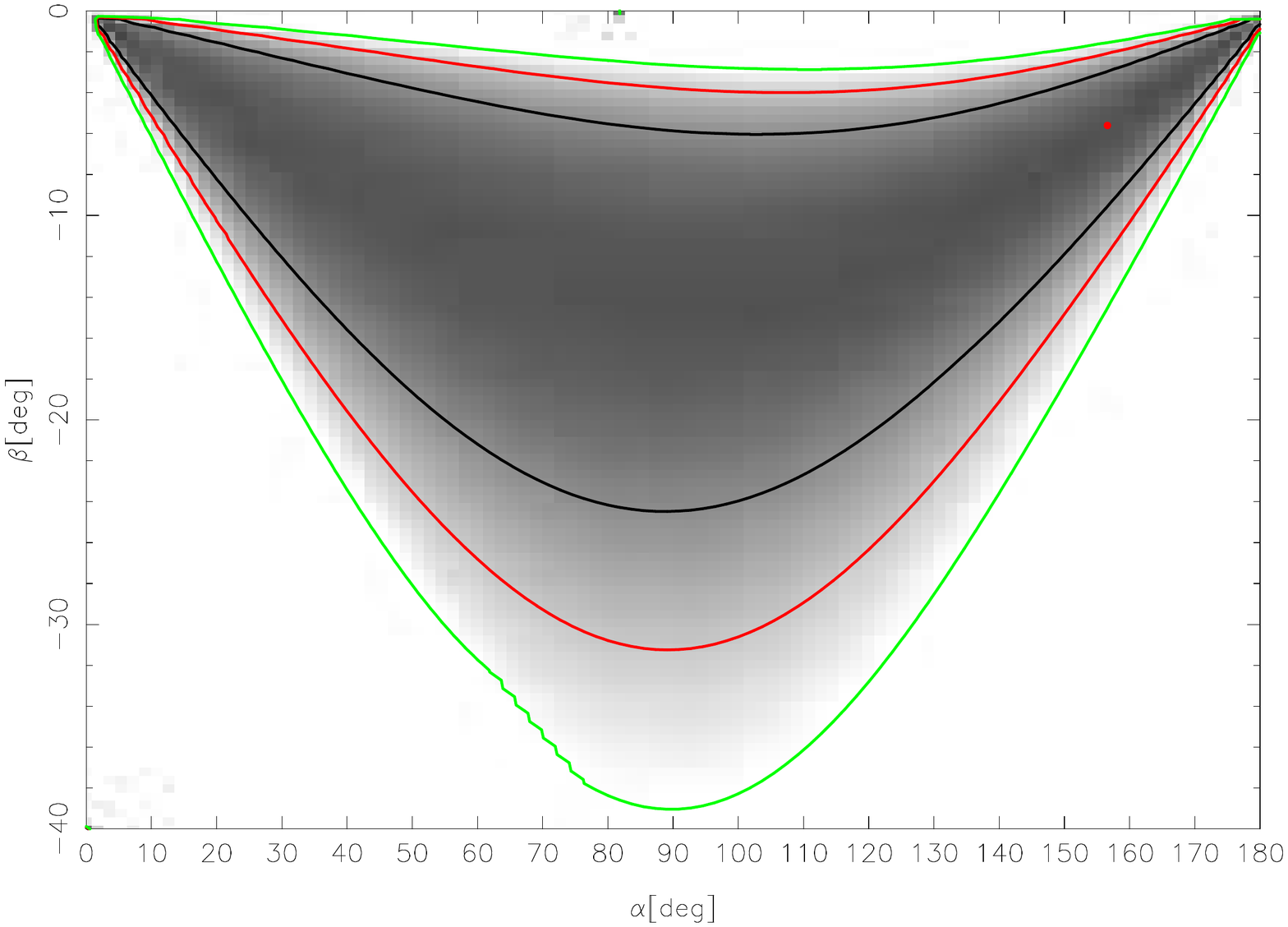}}}\\
&
\\
\end{tabular}
\caption{Top panel (upper window) shows the average profile with total
intensity (Stokes I; solid black lines), total linear polarization (dashed red
line) and circular polarization (Stokes V; dotted blue line). Top panel (lower
window) also shows the single pulse PPA distribution (colour scale) along with
the average PPA (red error bars).
The RVM fits to the average PPA (dashed pink
line) is also shown in this plot. Bottom panel show
the $\chi^2$ contours for the parameters $\alpha$ and $\beta$ obtained from RVM
fits.}
\label{a58}
\end{center}
\end{figure*}


\begin{figure*}
\begin{center}
\begin{tabular}{cc}
{\mbox{\includegraphics[width=9cm,height=6cm,angle=0.]{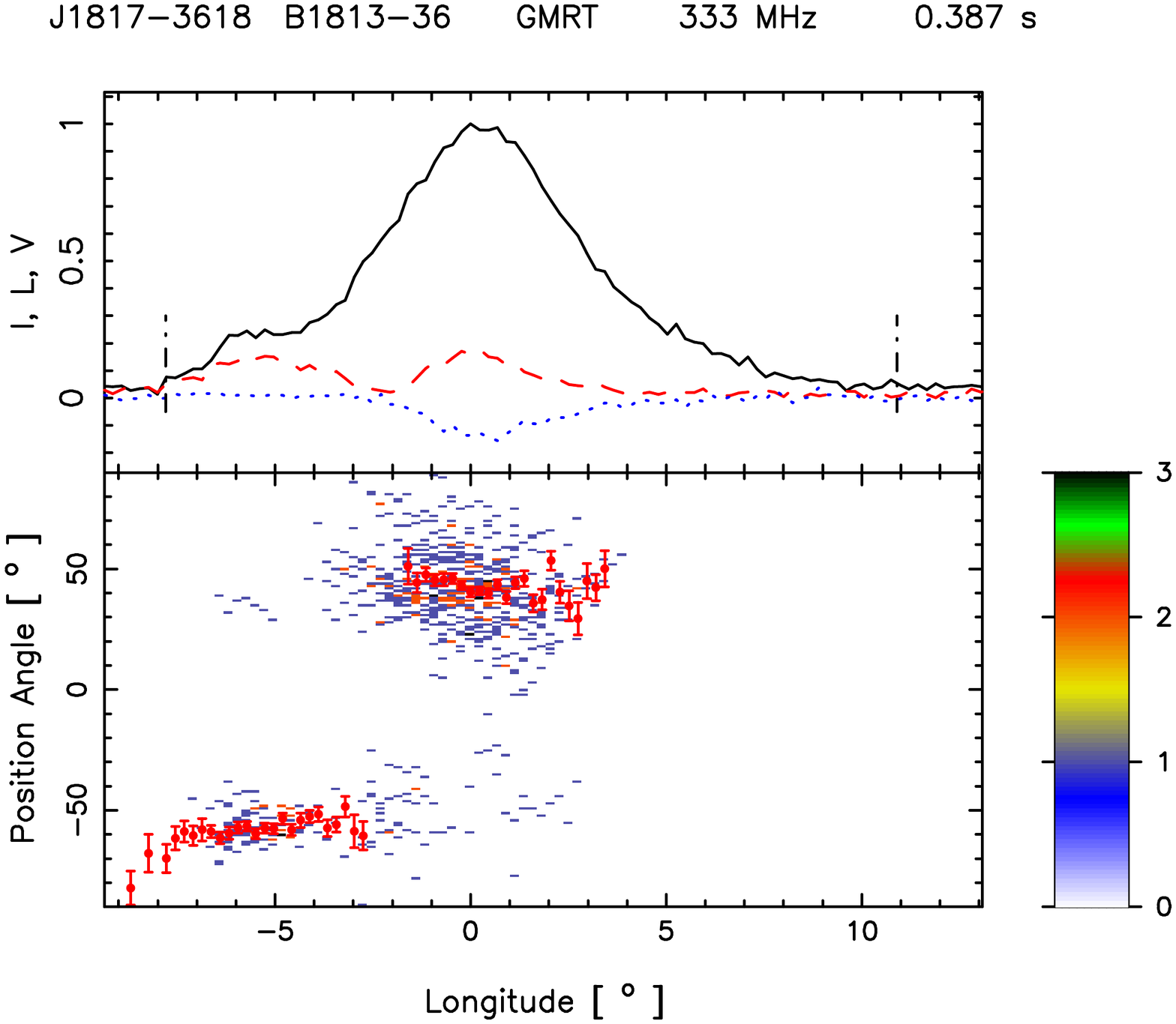}}}&
\\
&
\\
{\mbox{\includegraphics[width=9cm,height=6cm,angle=0.]{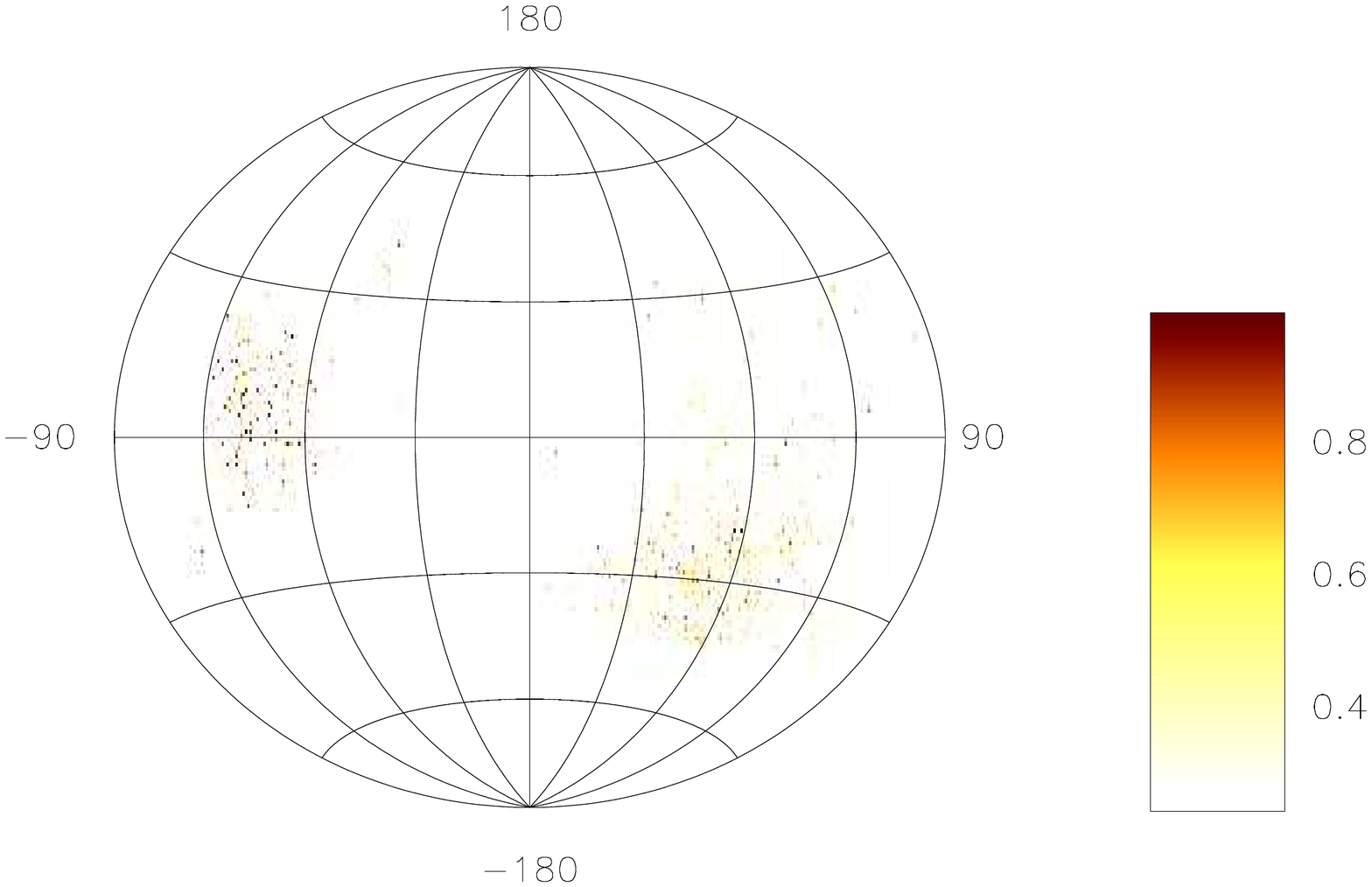}}}&
\\
\end{tabular}
\caption{Top panel (upper window) shows the average profile with total
intensity (Stokes I; solid black lines), total linear polarization (dashed red
line) and circular polarization (Stokes V; dotted blue line). Top panel (lower
window) also shows the single pulse PPA distribution (colour scale) along with
the average PPA (red error bars).
The RVM fits to the average PPA (dashed pink
line) is also shown in this plot. Bottom panel show
the $\chi^2$ contours for the parameters $\alpha$ and $\beta$ obtained from RVM
fits.}
\label{a59}
\end{center}
\end{figure*}


\begin{figure*}
\begin{center}
\begin{tabular}{cc}
{\mbox{\includegraphics[width=9cm,height=6cm,angle=0.]{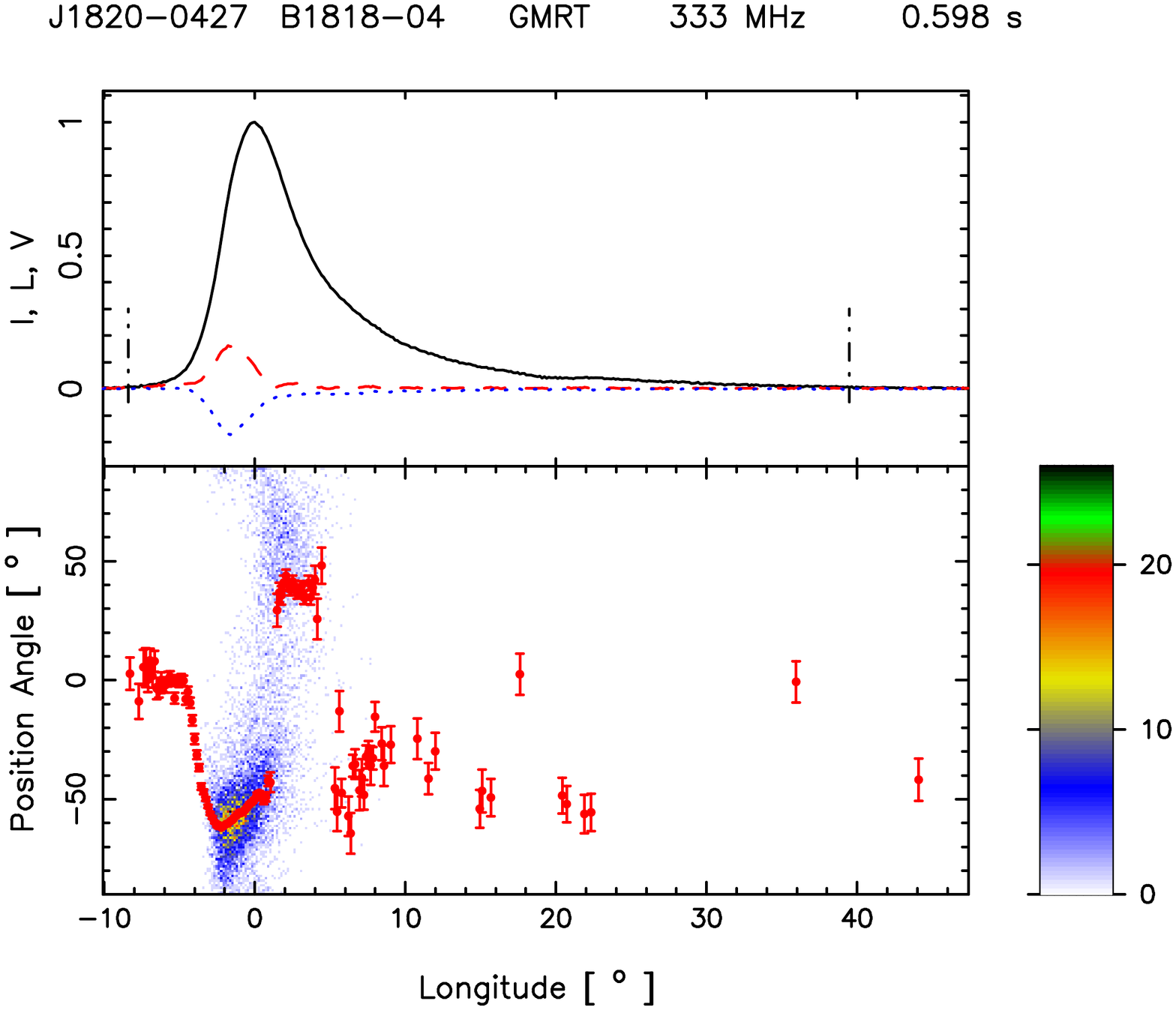}}}&
{\mbox{\includegraphics[width=9cm,height=6cm,angle=0.]{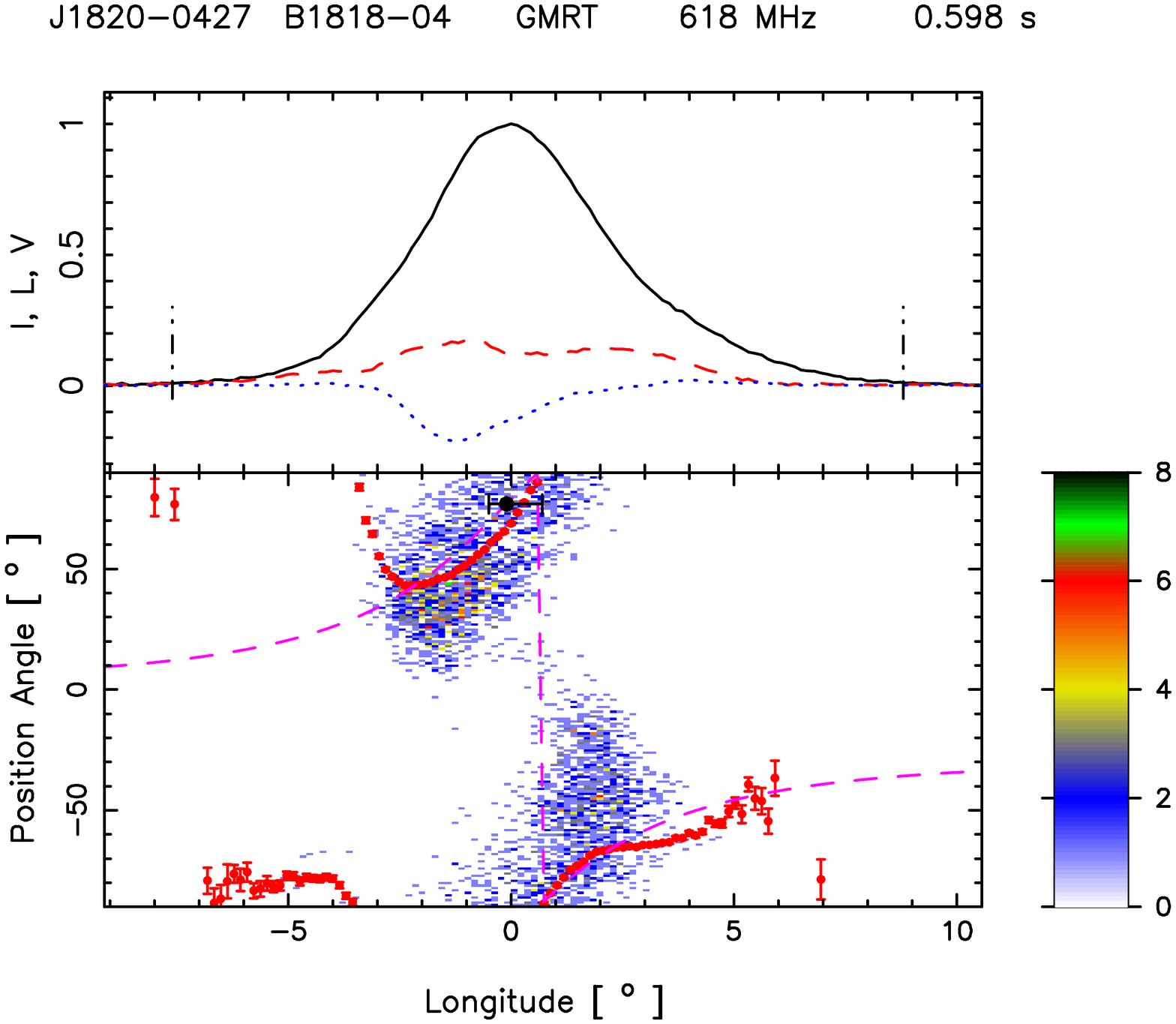}}}\\
&
{\mbox{\includegraphics[width=9cm,height=6cm,angle=0.]{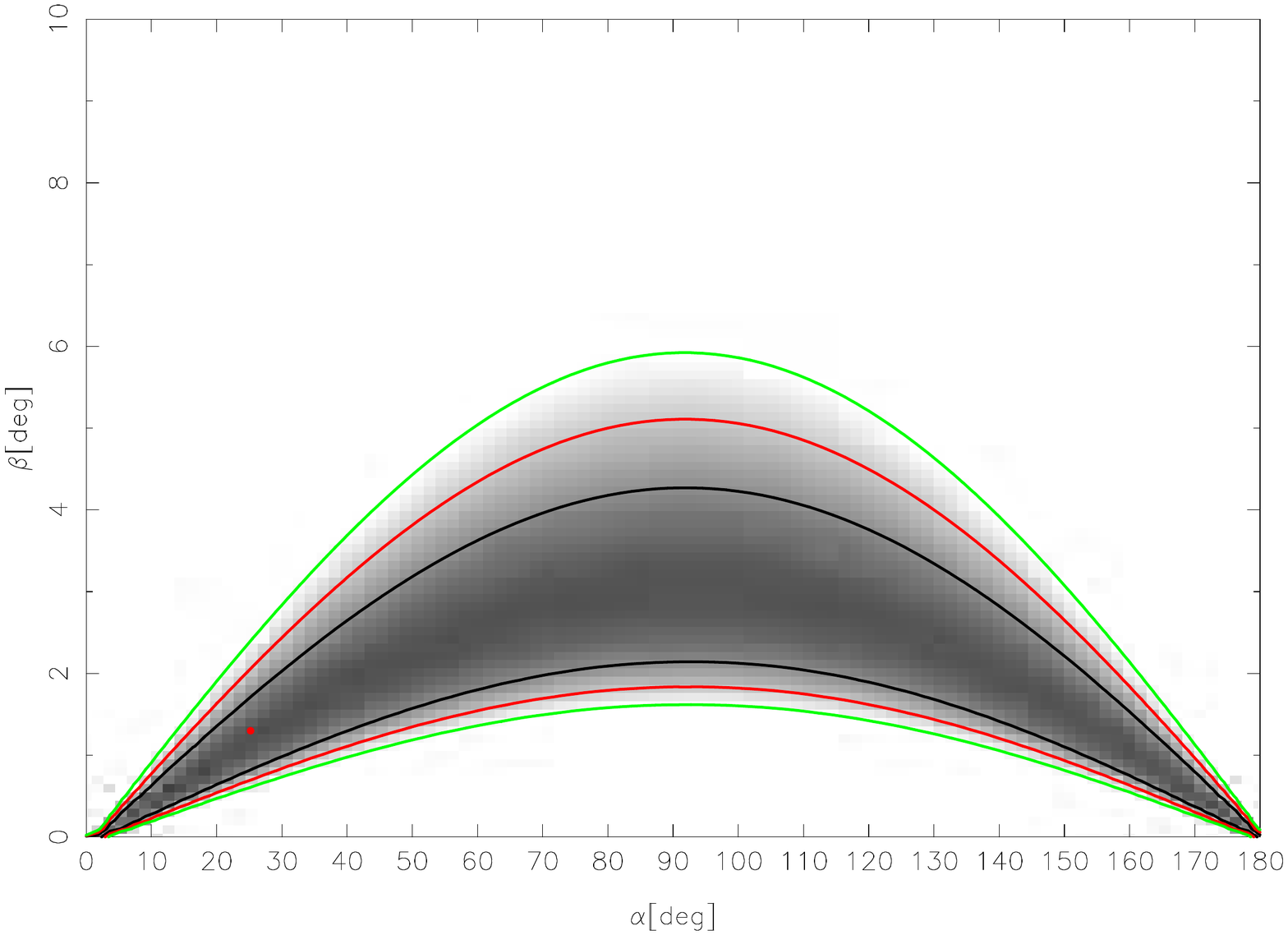}}}\\
{\mbox{\includegraphics[width=9cm,height=6cm,angle=0.]{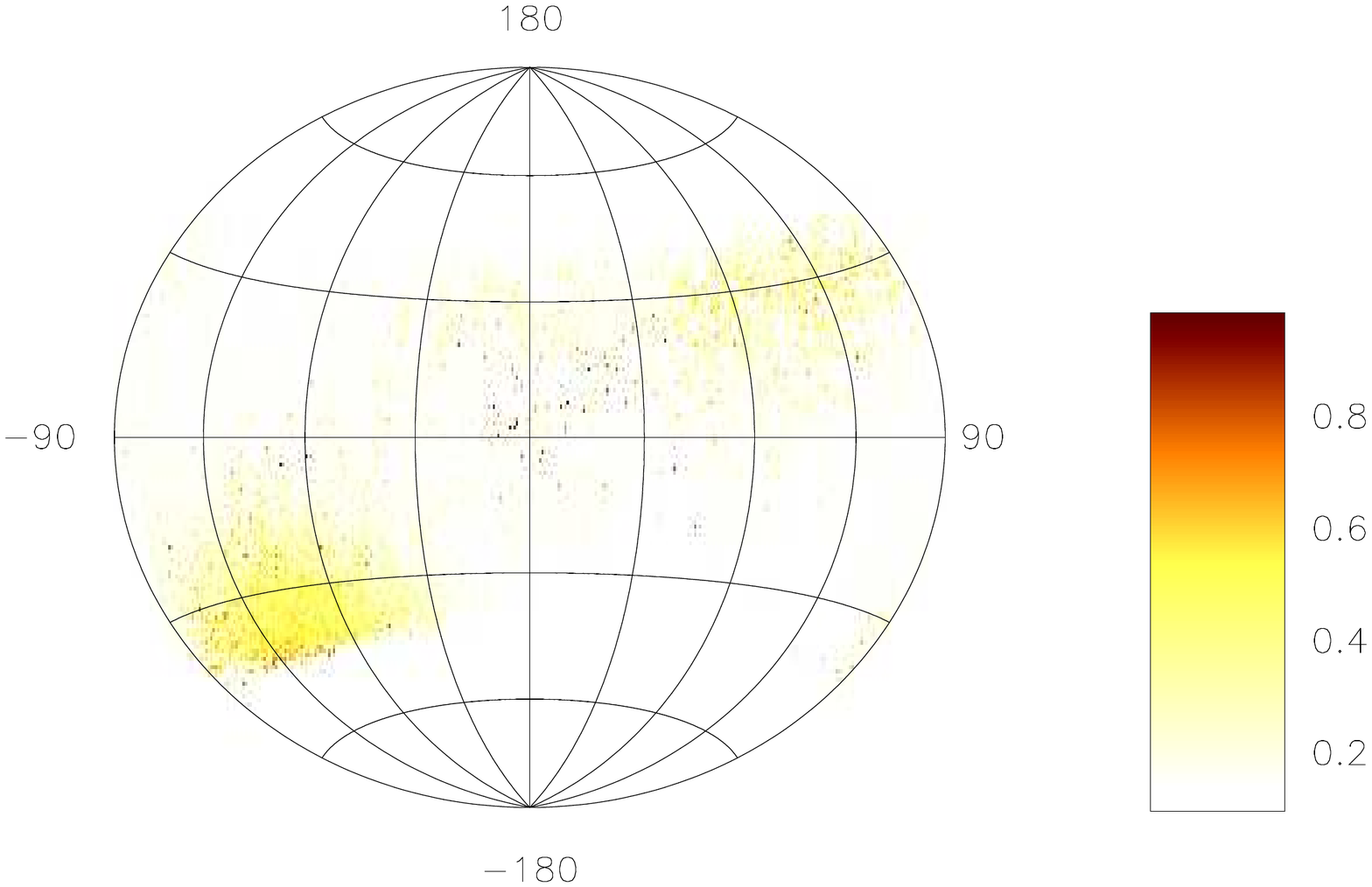}}}&
{\mbox{\includegraphics[width=9cm,height=6cm,angle=0.]{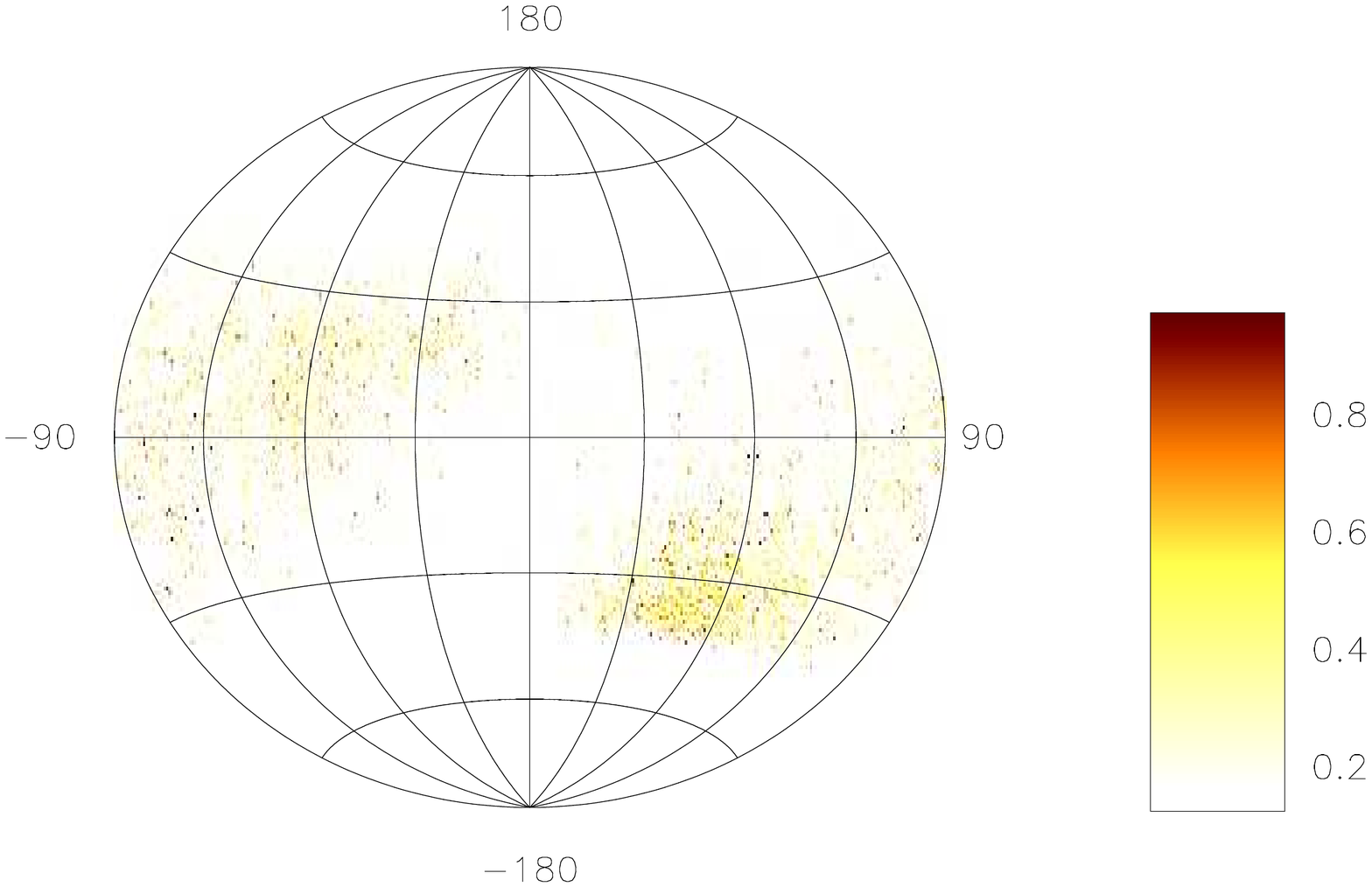}}}\\
\end{tabular}
\caption{Top panel (upper window) shows the average profile with total
intensity (Stokes I; solid black lines), total linear polarization (dashed red
line) and circular polarization (Stokes V; dotted blue line). Top panel (lower
window) also shows the single pulse PPA distribution (colour scale) along with
the average PPA (red error bars).
The RVM fits to the average PPA (dashed pink
line) is also shown in this plot. Middle panel only for 618 MHz show
the $\chi^2$ contours for the parameters $\alpha$ and $\beta$ obtained from RVM
fits.
Bottom panel shows the Hammer-Aitoff projection of the polarized time
samples with the colour scheme representing the fractional polarization level.}
\label{a60}
\end{center}
\end{figure*}


\begin{figure*}
\begin{center}
\begin{tabular}{cc}
{\mbox{\includegraphics[width=9cm,height=6cm,angle=0.]{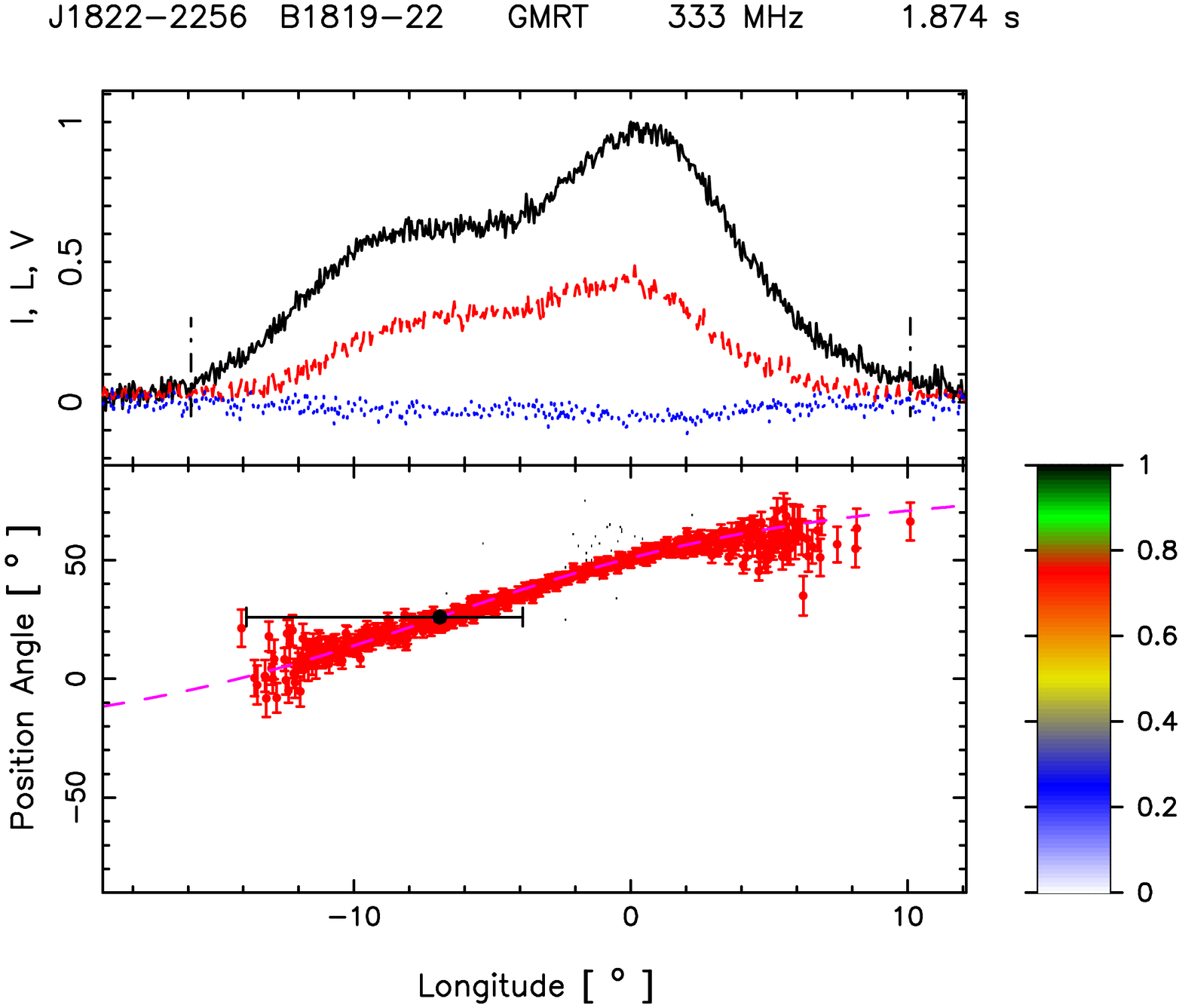}}}&
{\mbox{\includegraphics[width=9cm,height=6cm,angle=0.]{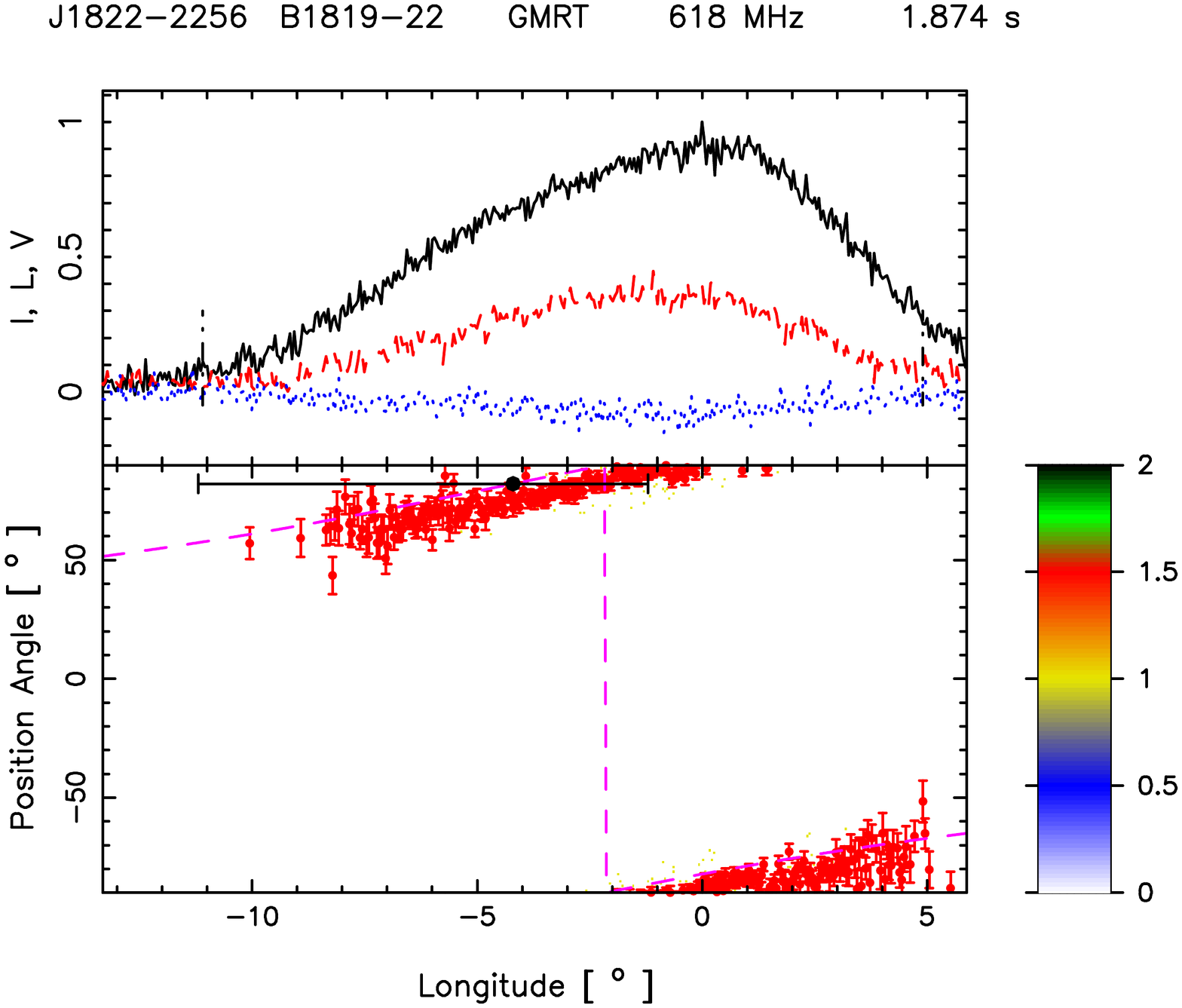}}}\\
{\mbox{\includegraphics[width=9cm,height=6cm,angle=0.]{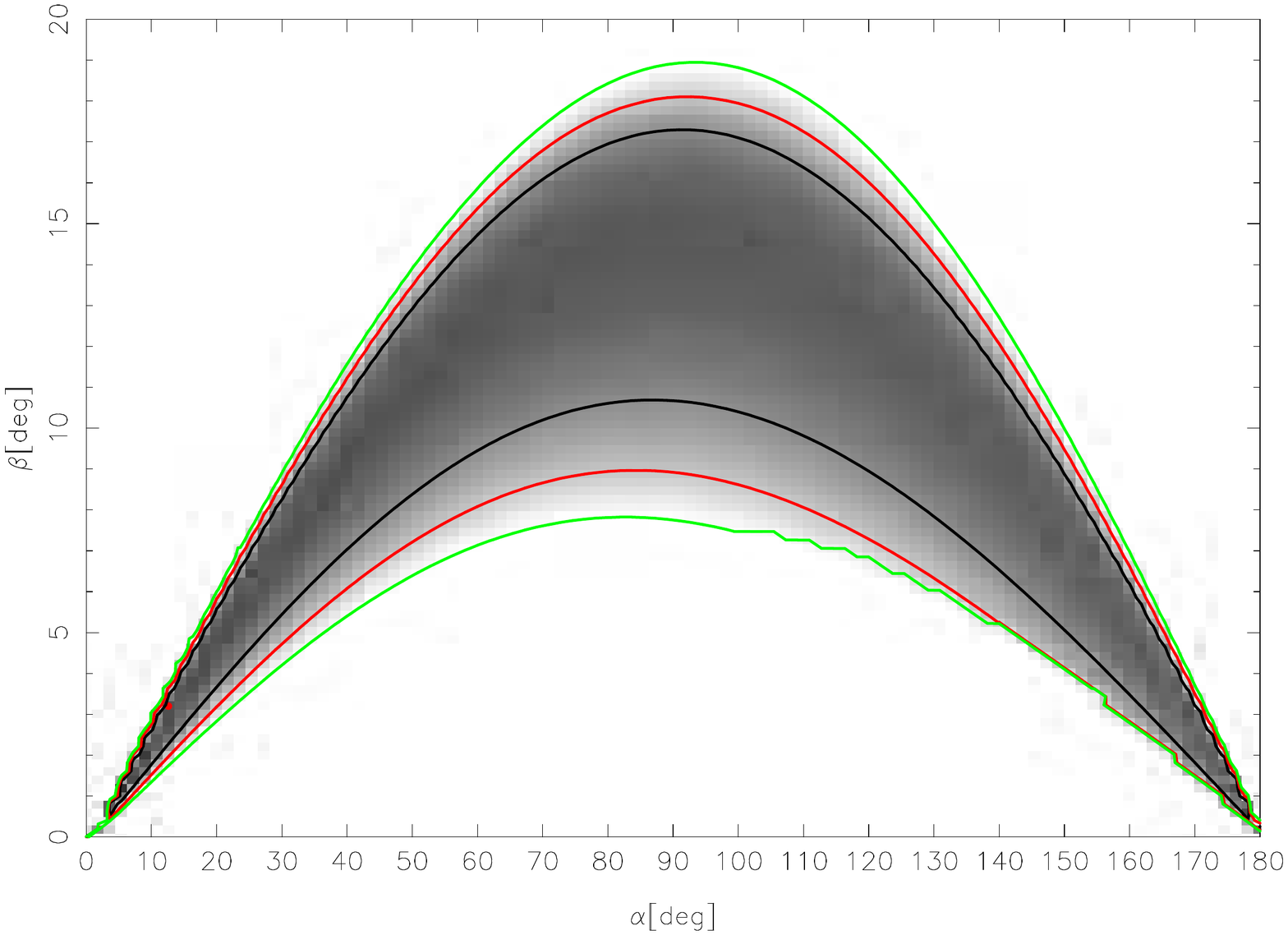}}}&
{\mbox{\includegraphics[width=9cm,height=6cm,angle=0.]{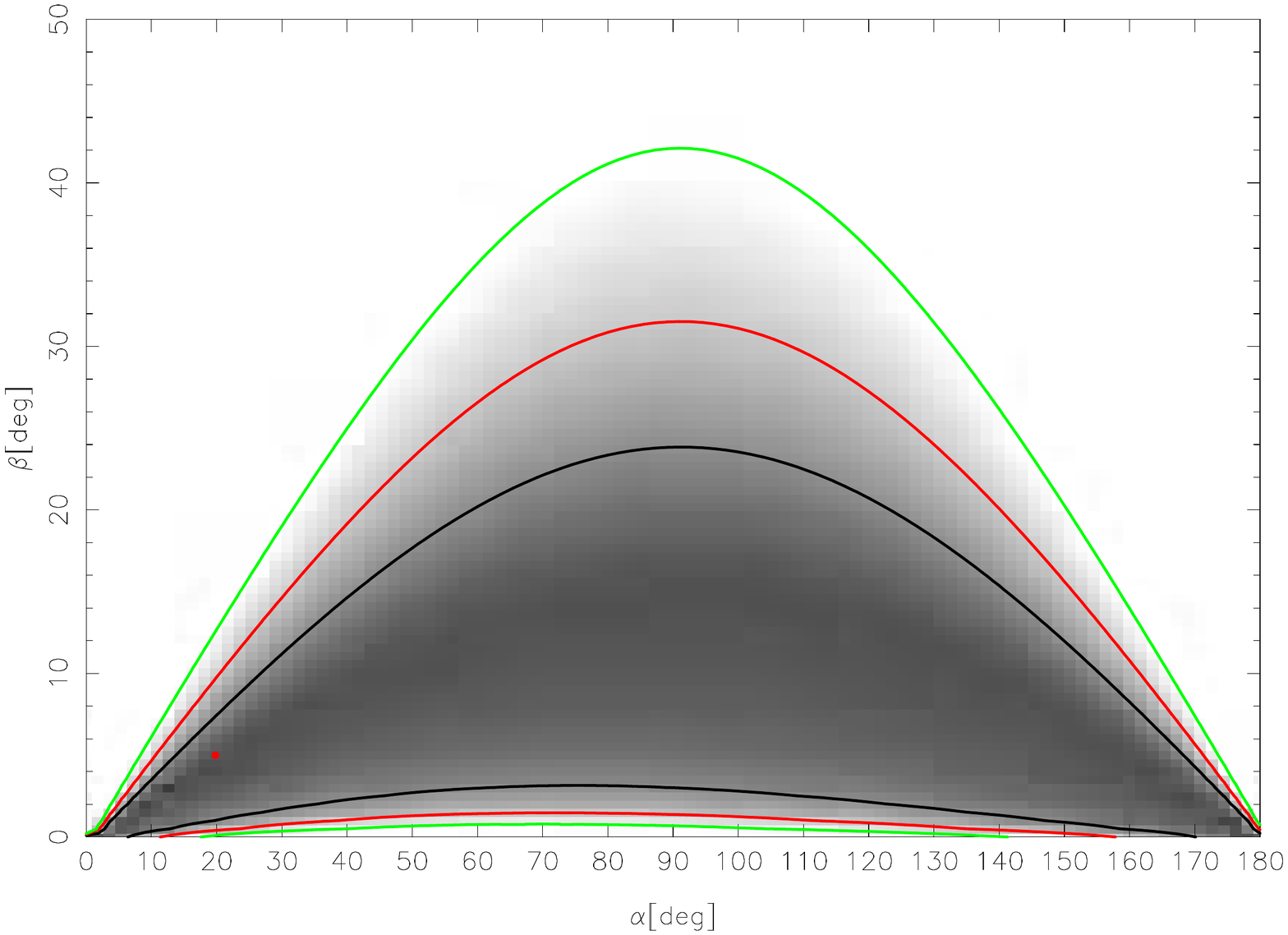}}}\\
&
{\mbox{\includegraphics[width=9cm,height=6cm,angle=0.]{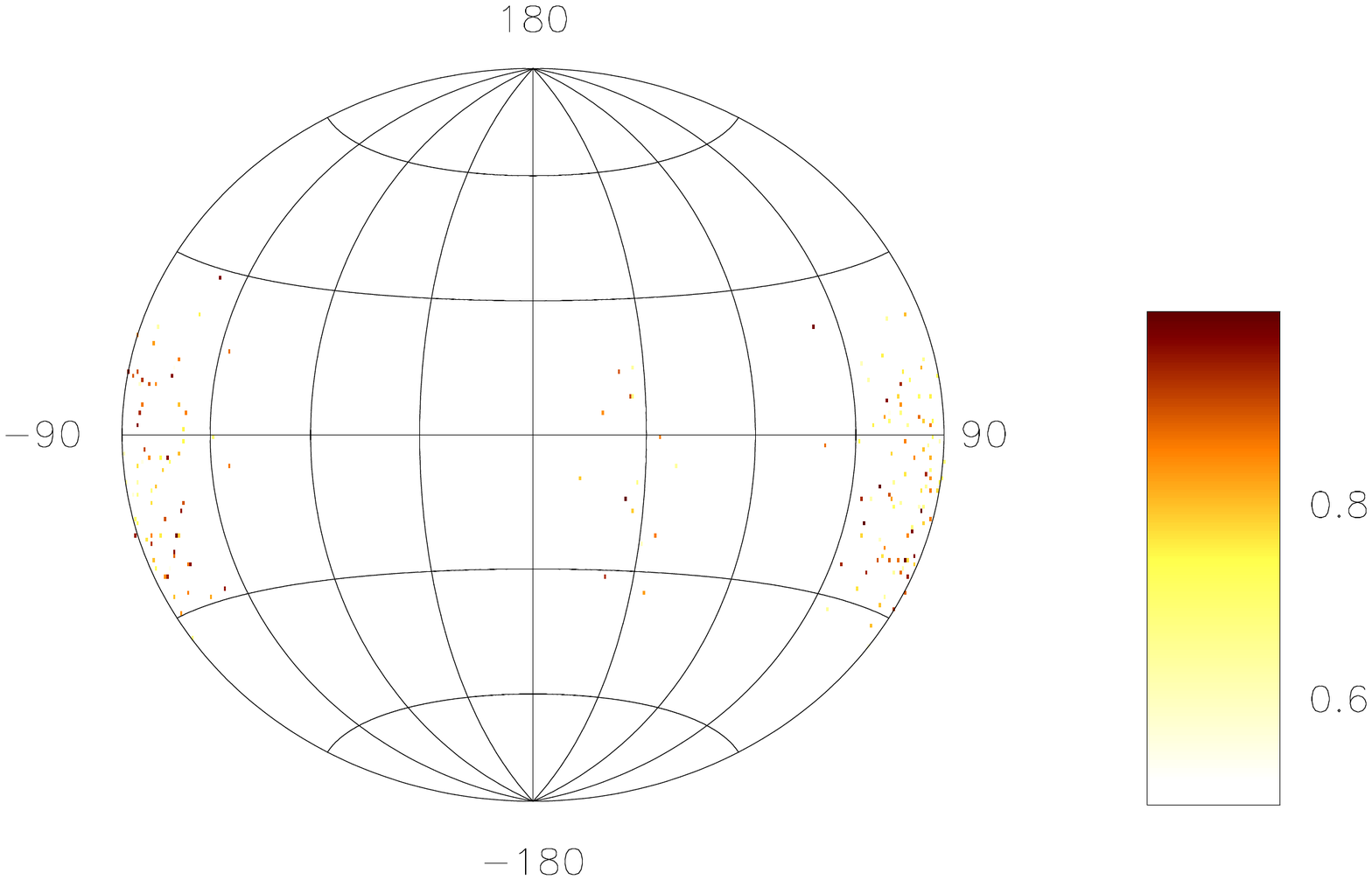}}}\\
\end{tabular}
\caption{Top panel (upper window) shows the average profile with total
intensity (Stokes I; solid black lines), total linear polarization (dashed red
line) and circular polarization (Stokes V; dotted blue line). Top panel (lower
window) also shows the single pulse PPA distribution (colour scale) along with
the average PPA (red error bars).
The RVM fits to the average PPA (dashed pink
line) is also shown in this plot. Middle panel show
the $\chi^2$ contours for the parameters $\alpha$ and $\beta$ obtained from RVM
fits.
Bottom panel only for 618 MHz shows the Hammer-Aitoff projection of the polarized time
samples with the colour scheme representing the fractional polarization level.}
\label{a61}
\end{center}
\end{figure*}


\begin{figure*}
\begin{center}
\begin{tabular}{cc}
{\mbox{\includegraphics[width=9cm,height=6cm,angle=0.]{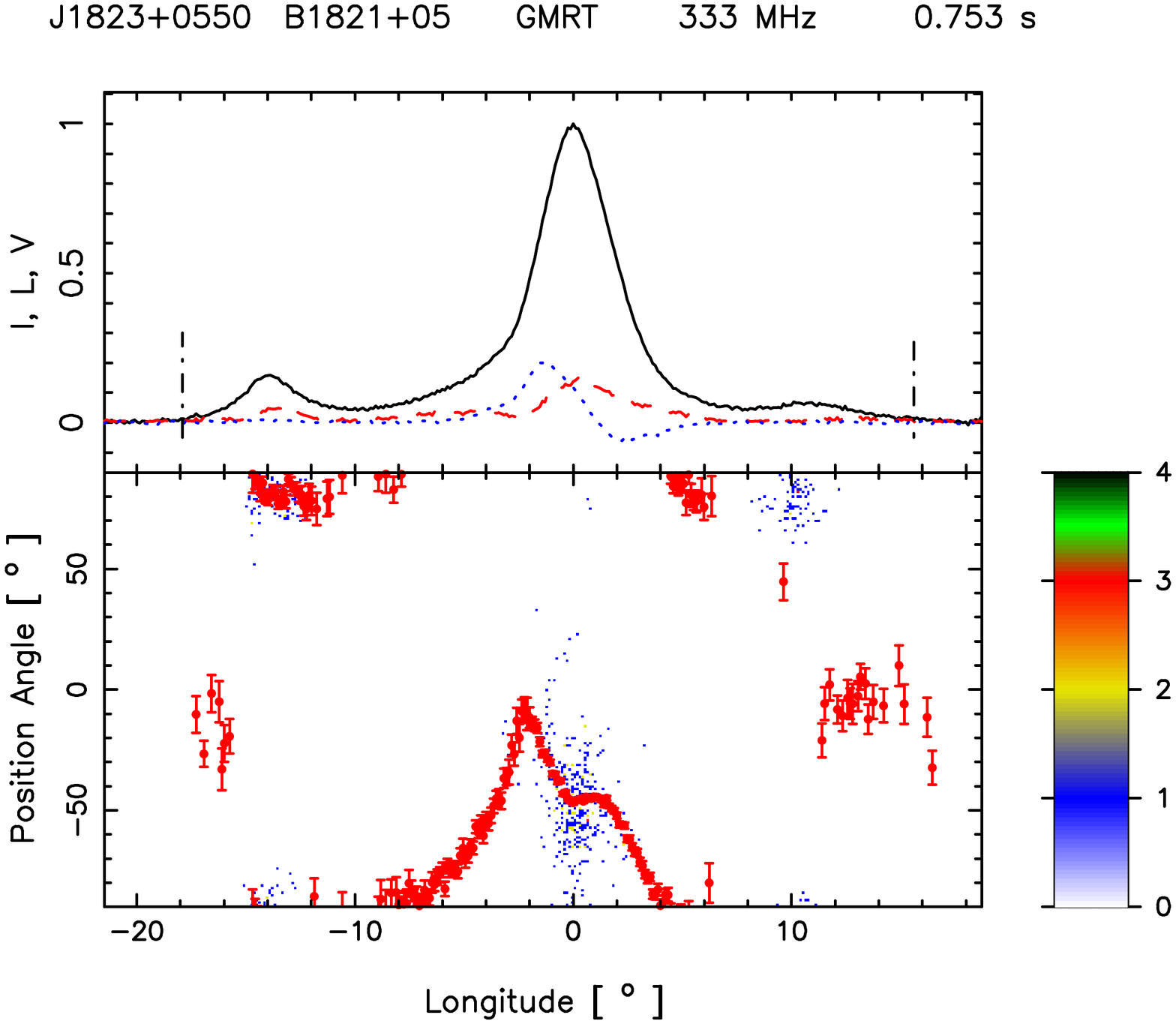}}}&
\\
&
\\
{\mbox{\includegraphics[width=9cm,height=6cm,angle=0.]{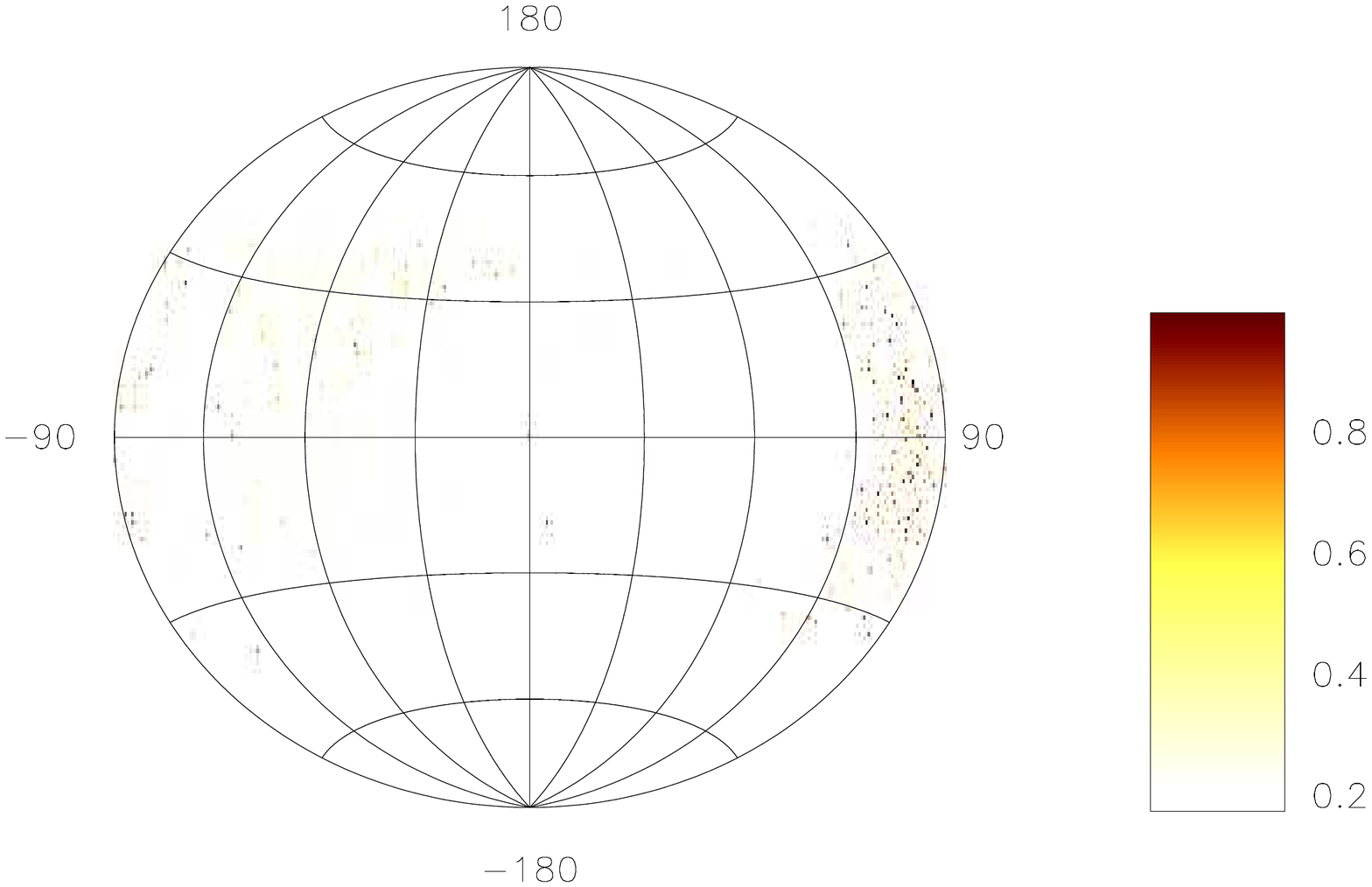}}}&
\\
\end{tabular}
\caption{Top panel only for 333 MHz (upper window) shows the average profile with total
intensity (Stokes I; solid black lines), total linear polarization (dashed red
line) and circular polarization (Stokes V; dotted blue line). Top panel (lower
window) also shows the single pulse PPA distribution (colour scale) along with
the average PPA (red error bars).
Bottom panel only for 333 MHz shows the Hammer-Aitoff projection of the polarized time
samples with the colour scheme representing the fractional polarization level.}
\label{a62}
\end{center}
\end{figure*}

\begin{figure*}
\begin{center}
\begin{tabular}{cc}
{\mbox{\includegraphics[width=9cm,height=6cm,angle=0.]{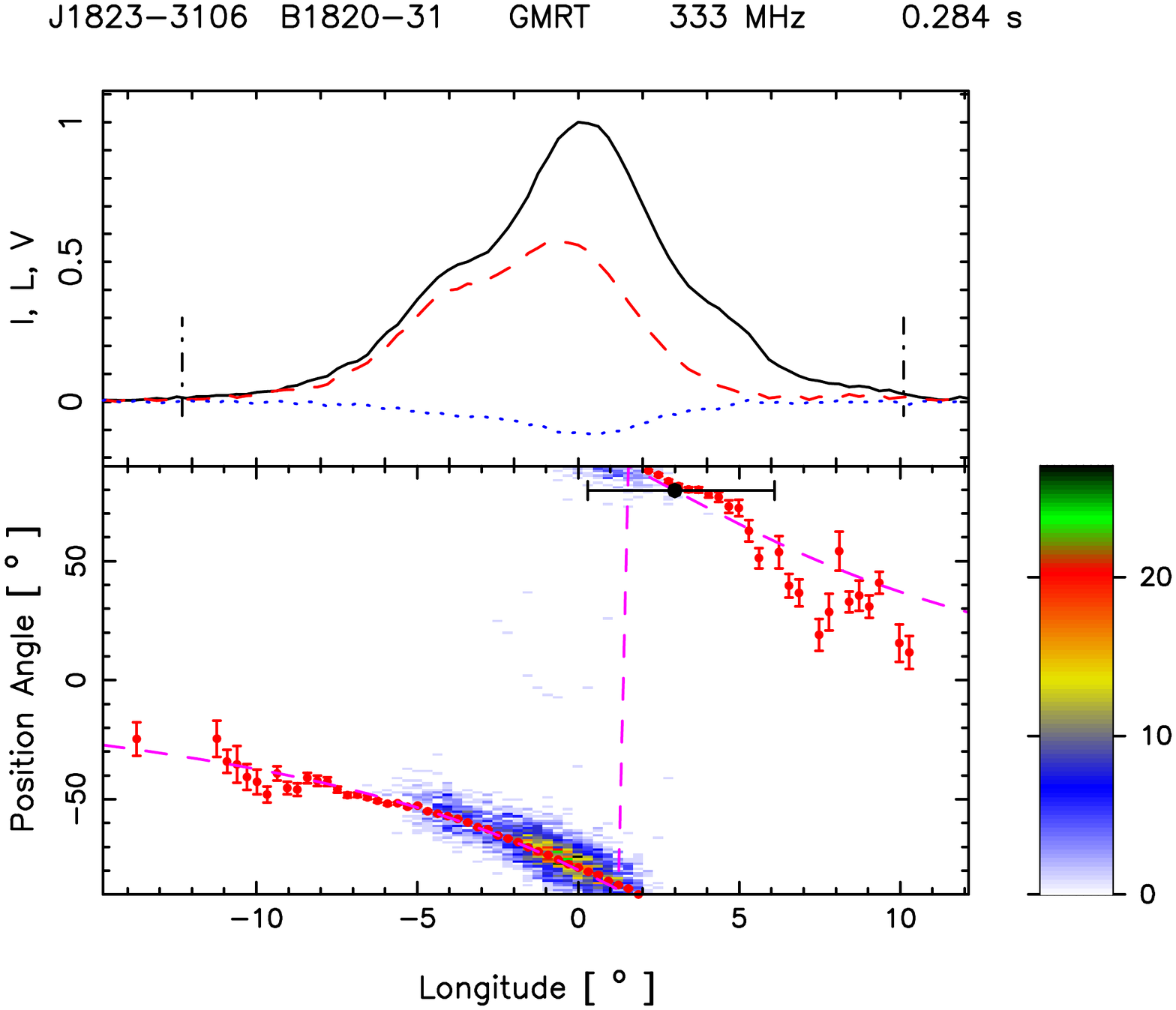}}}&
{\mbox{\includegraphics[width=9cm,height=6cm,angle=0.]{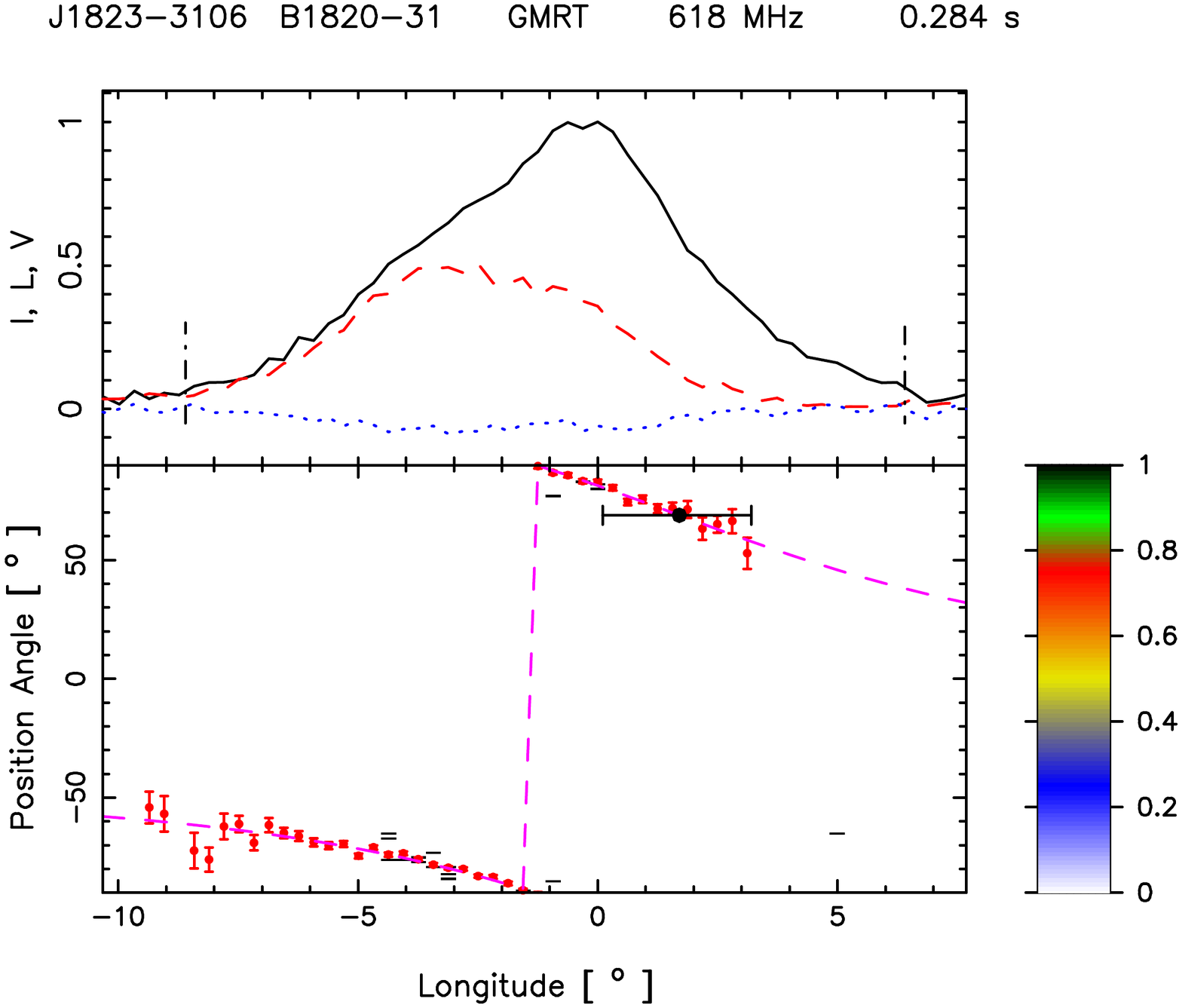}}}\\
{\mbox{\includegraphics[width=9cm,height=6cm,angle=0.]{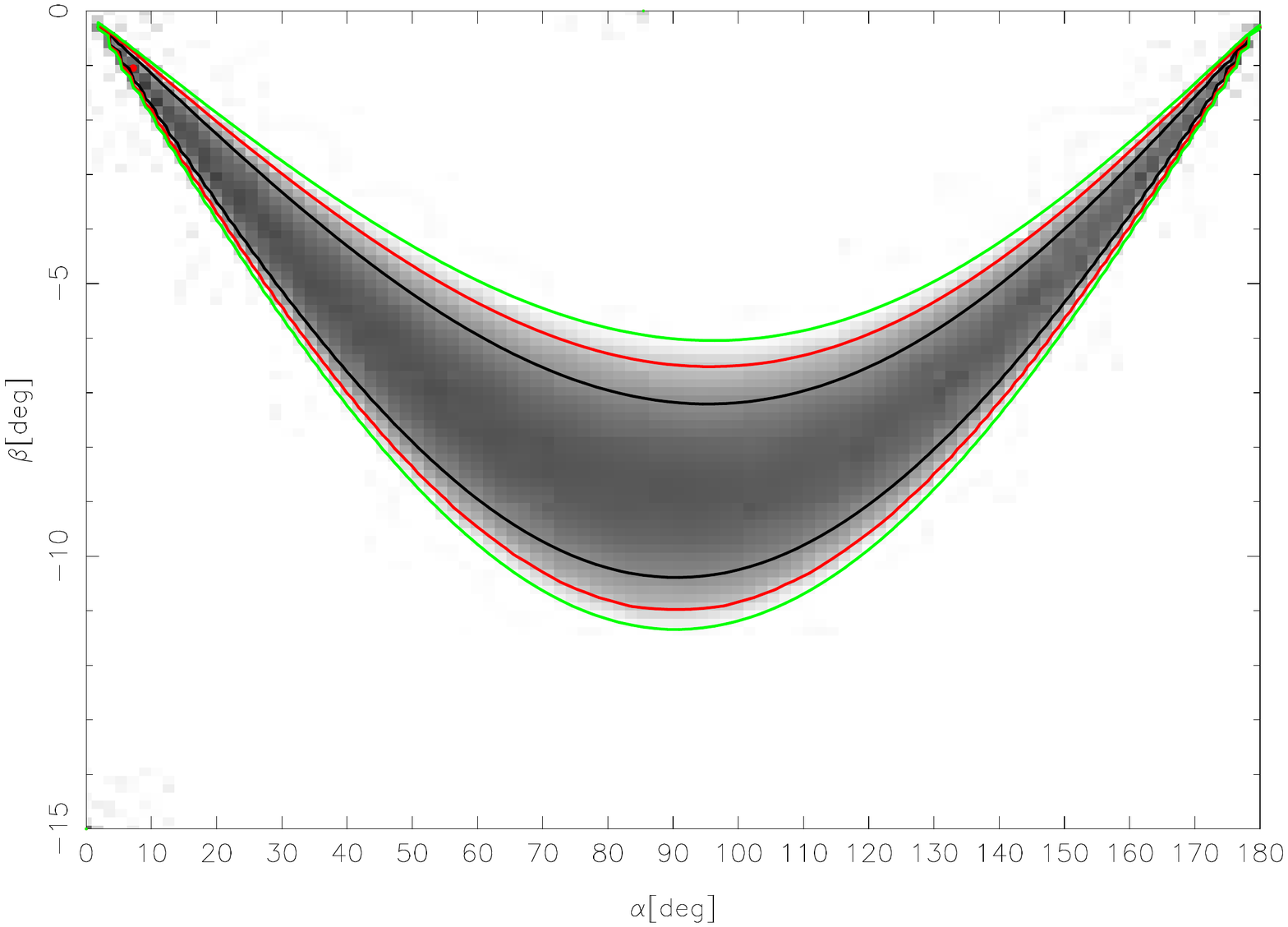}}}&
{\mbox{\includegraphics[width=9cm,height=6cm,angle=0.]{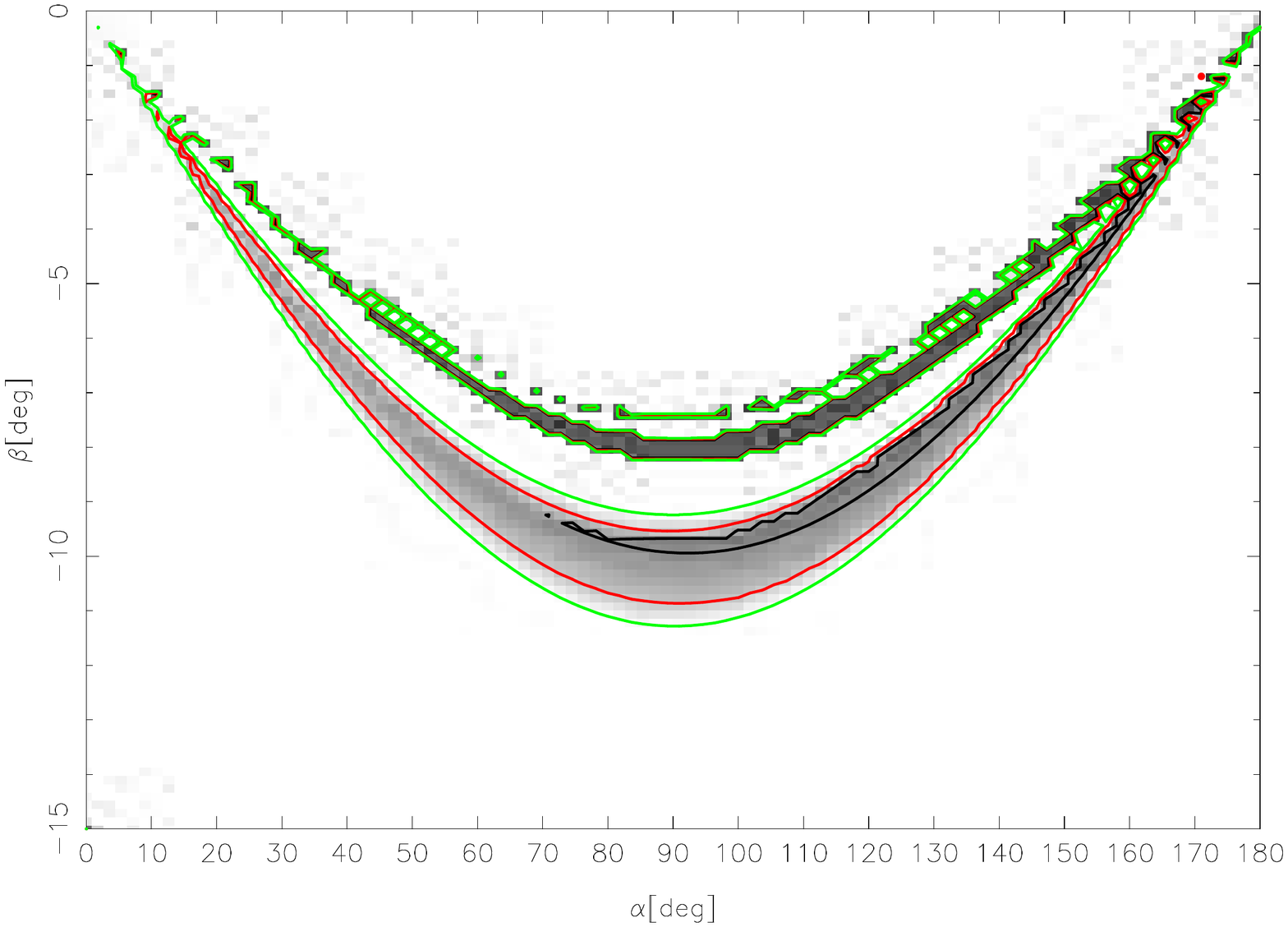}}}\\
{\mbox{\includegraphics[width=9cm,height=6cm,angle=0.]{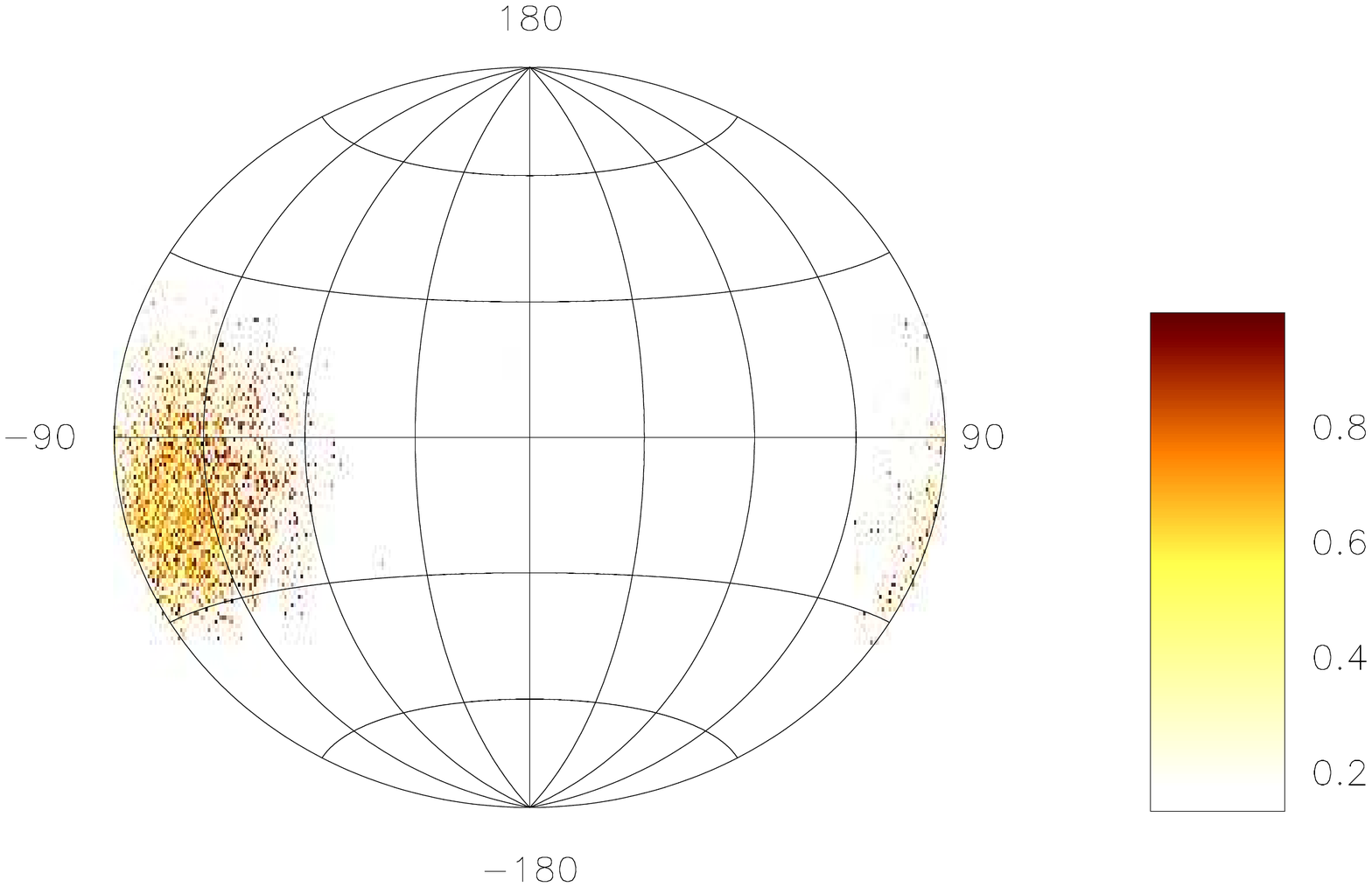}}}&
\\
\end{tabular}
\caption{Top panel (upper window) shows the average profile with total
intensity (Stokes I; solid black lines), total linear polarization (dashed red
line) and circular polarization (Stokes V; dotted blue line). Top panel (lower
window) also shows the single pulse PPA distribution (colour scale) along with
the average PPA (red error bars).
The RVM fits to the average PPA (dashed pink
line) is also shown in this plot. Middle panel show
the $\chi^2$ contours for the parameters $\alpha$ and $\beta$ obtained from RVM
fits.
Bottom panel only for 333 MHz shows the Hammer-Aitoff projection of the polarized time
samples with the colour scheme representing the fractional polarization level.}
\label{a63}
\end{center}
\end{figure*}


\begin{figure*}
\begin{center}
\begin{tabular}{cc}
{\mbox{\includegraphics[width=9cm,height=6cm,angle=0.]{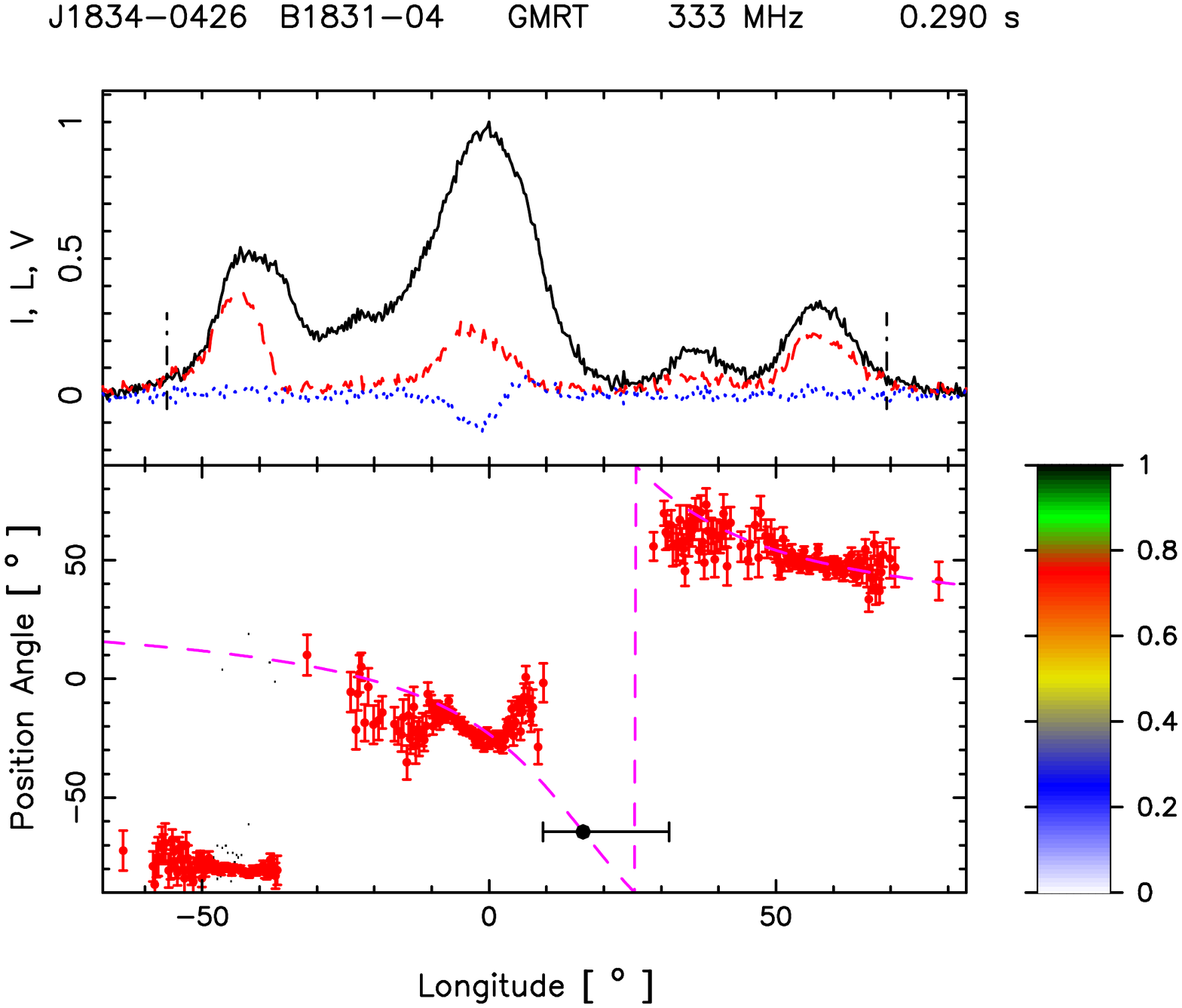}}}&
{\mbox{\includegraphics[width=9cm,height=6cm,angle=0.]{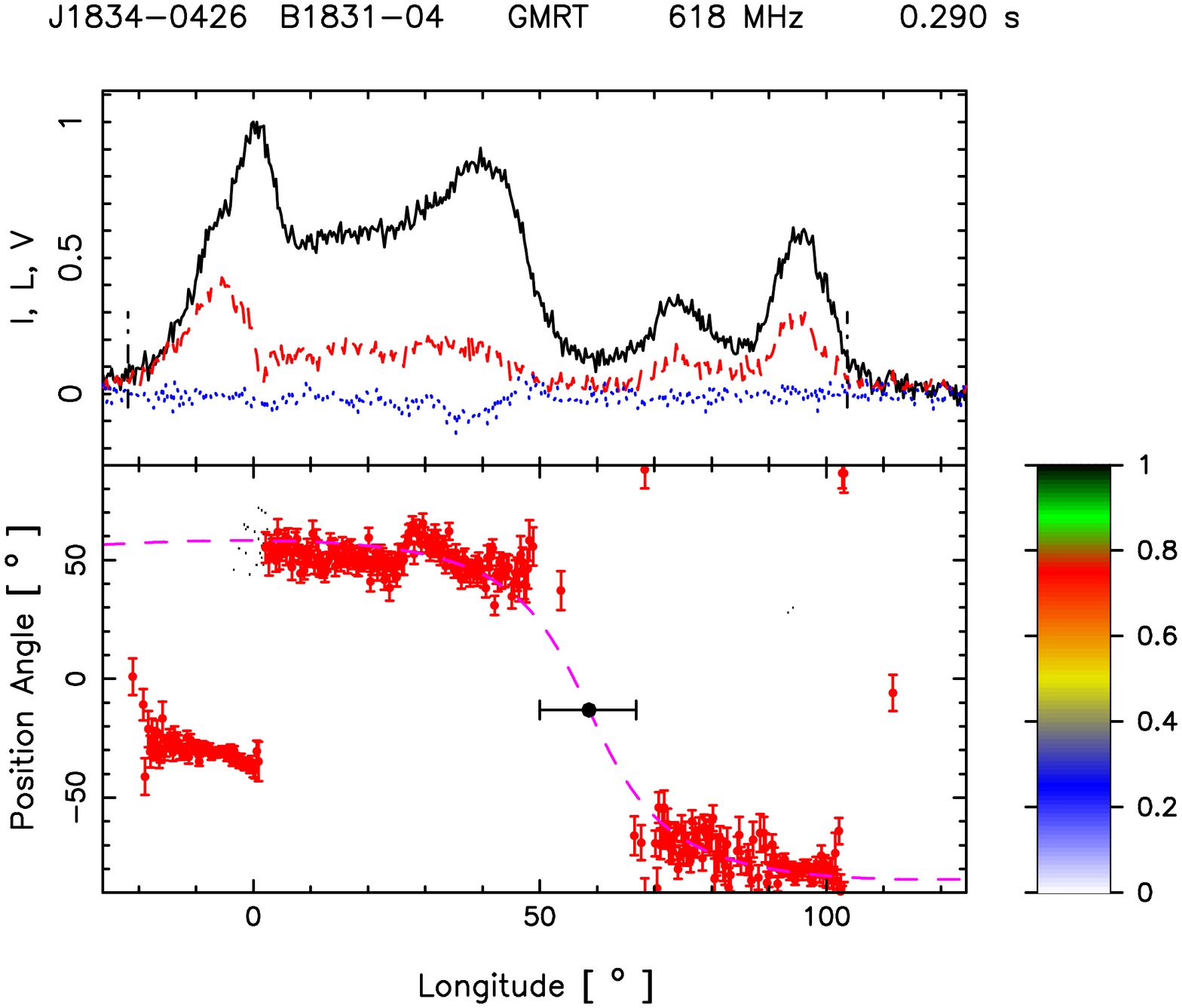}}}\\
{\mbox{\includegraphics[width=9cm,height=6cm,angle=0.]{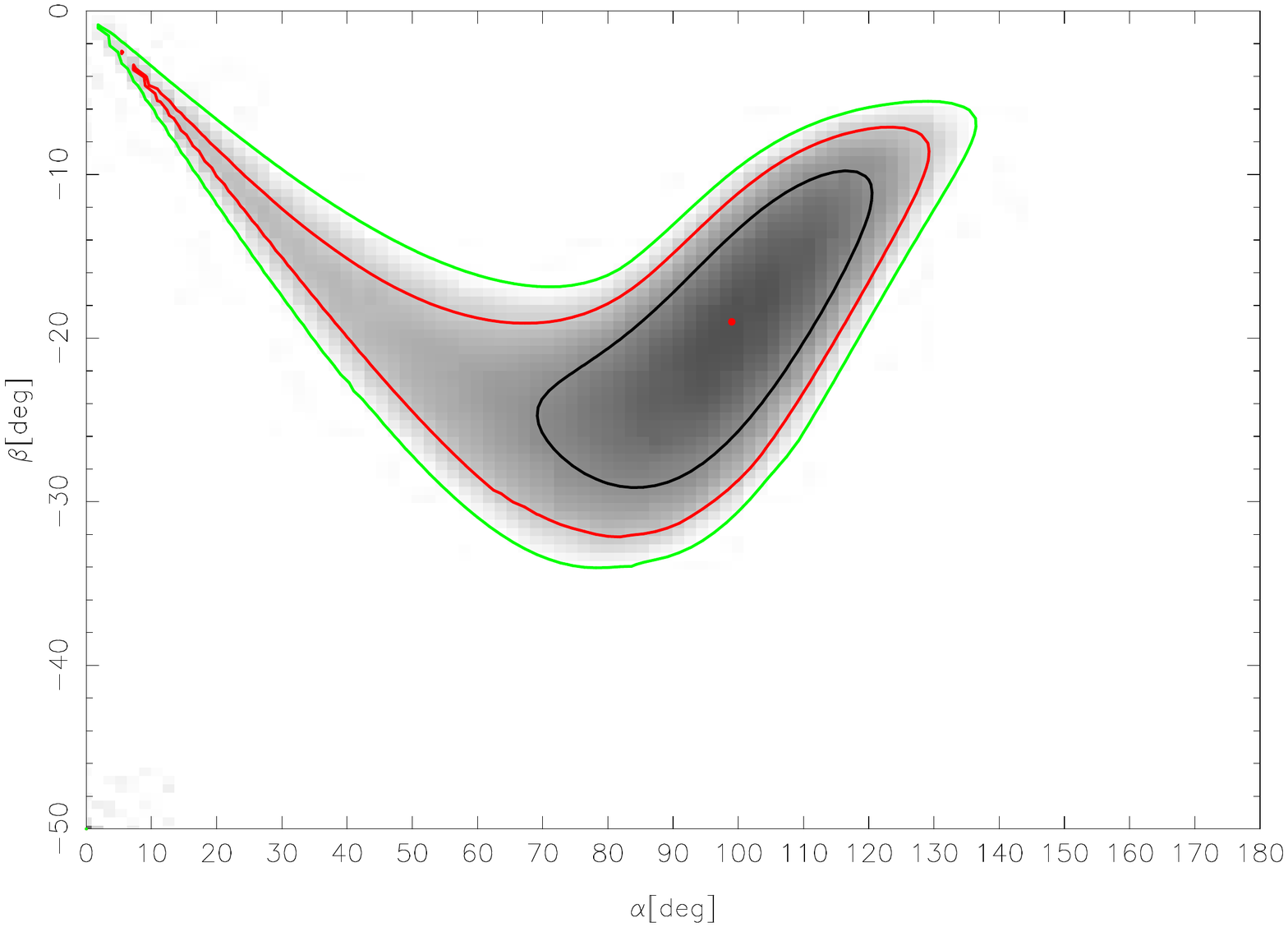}}}&
{\mbox{\includegraphics[width=9cm,height=6cm,angle=0.]{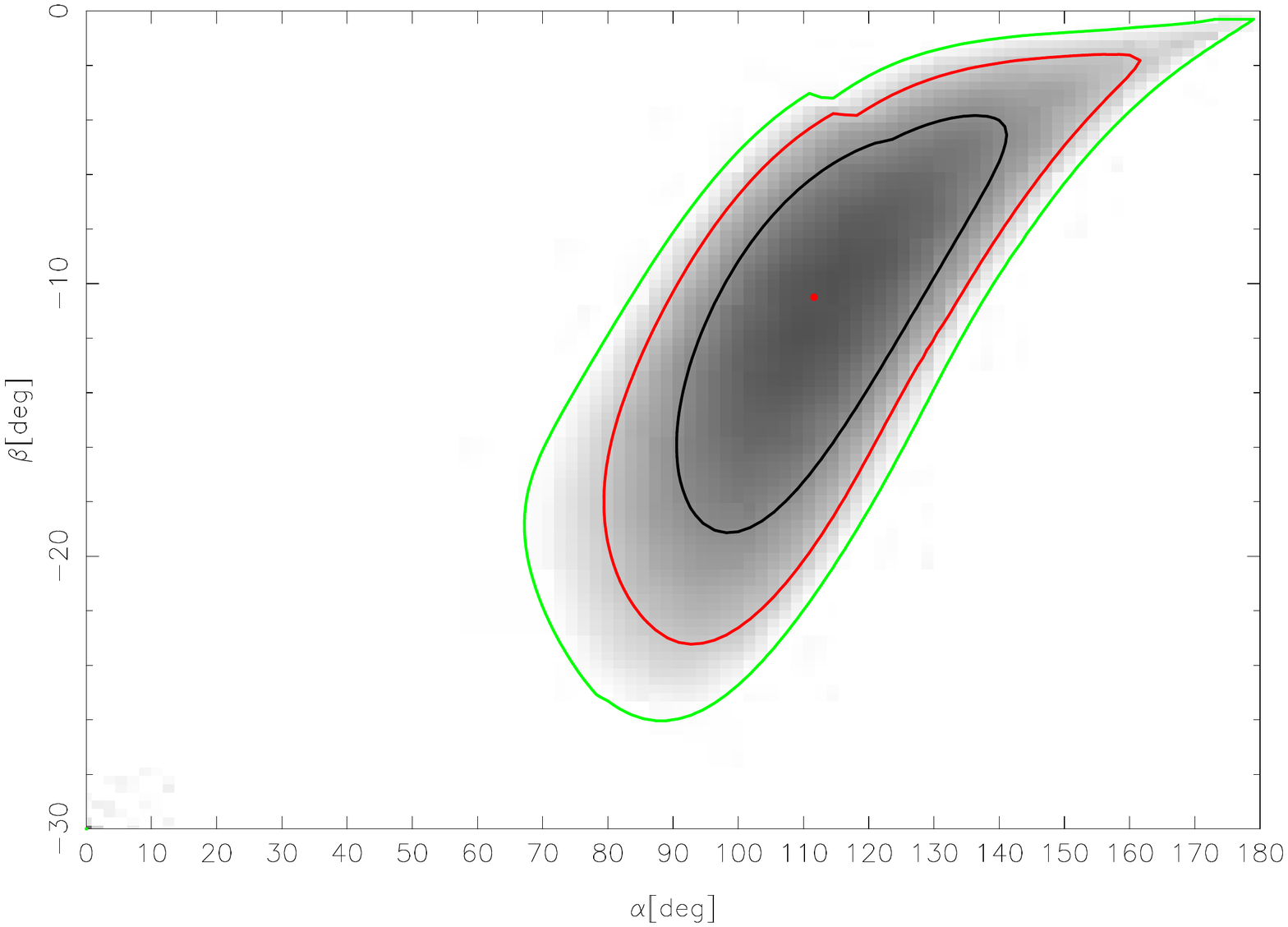}}}\\
&
\\
\end{tabular}
\caption{Top panel (upper window) shows the average profile with total
intensity (Stokes I; solid black lines), total linear polarization (dashed red
line) and circular polarization (Stokes V; dotted blue line). Top panel (lower
window) also shows the single pulse PPA distribution (colour scale) along with
the average PPA (red error bars).
The RVM fits to the average PPA (dashed pink
line) is also shown in this plot. Bottom panel show
the $\chi^2$ contours for the parameters $\alpha$ and $\beta$ obtained from RVM
fits.}
\label{a64}
\end{center}
\end{figure*}


\begin{figure*}
\begin{center}
\begin{tabular}{cc}
{\mbox{\includegraphics[width=9cm,height=6cm,angle=0.]{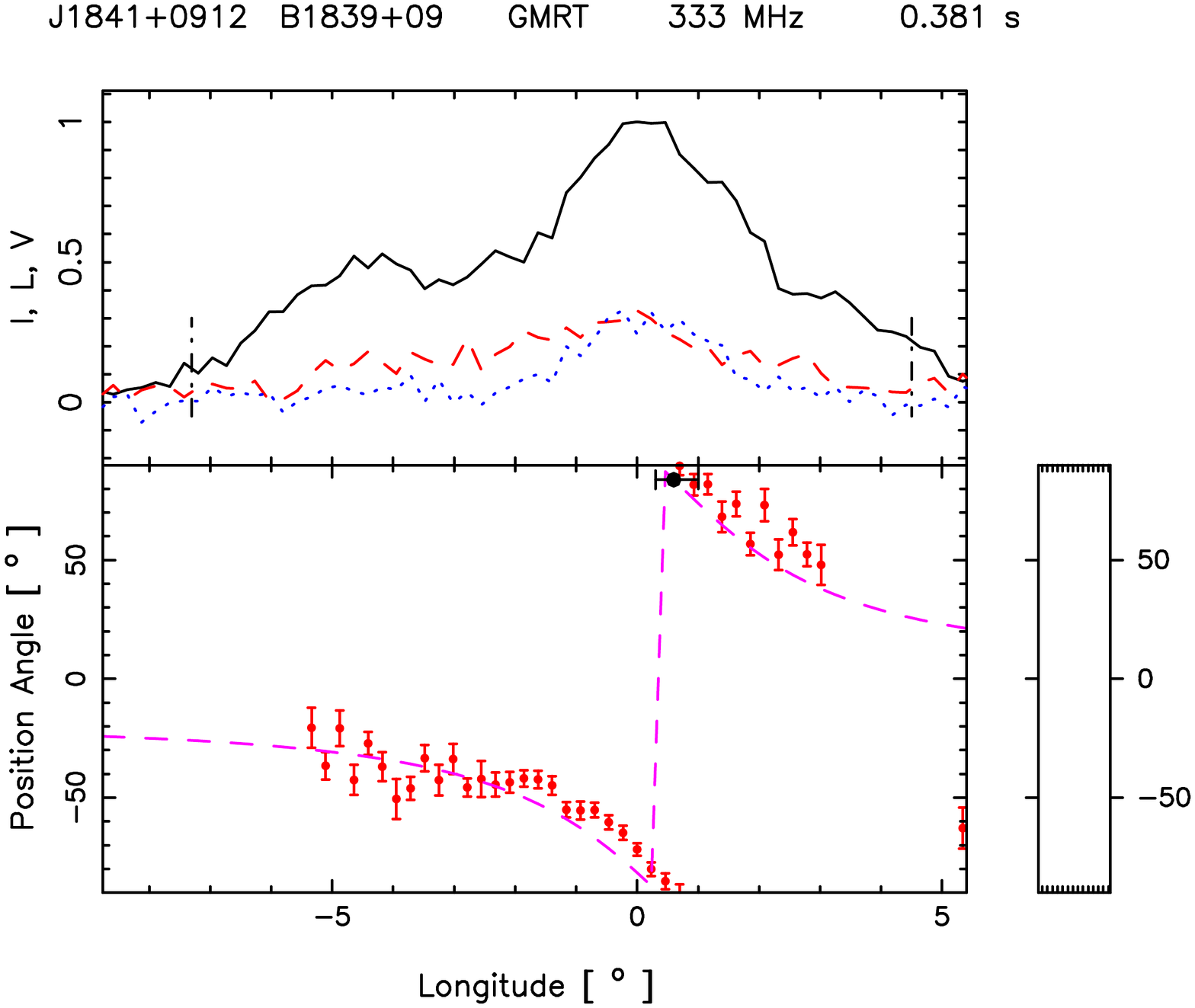}}}&
{\mbox{\includegraphics[width=9cm,height=6cm,angle=0.]{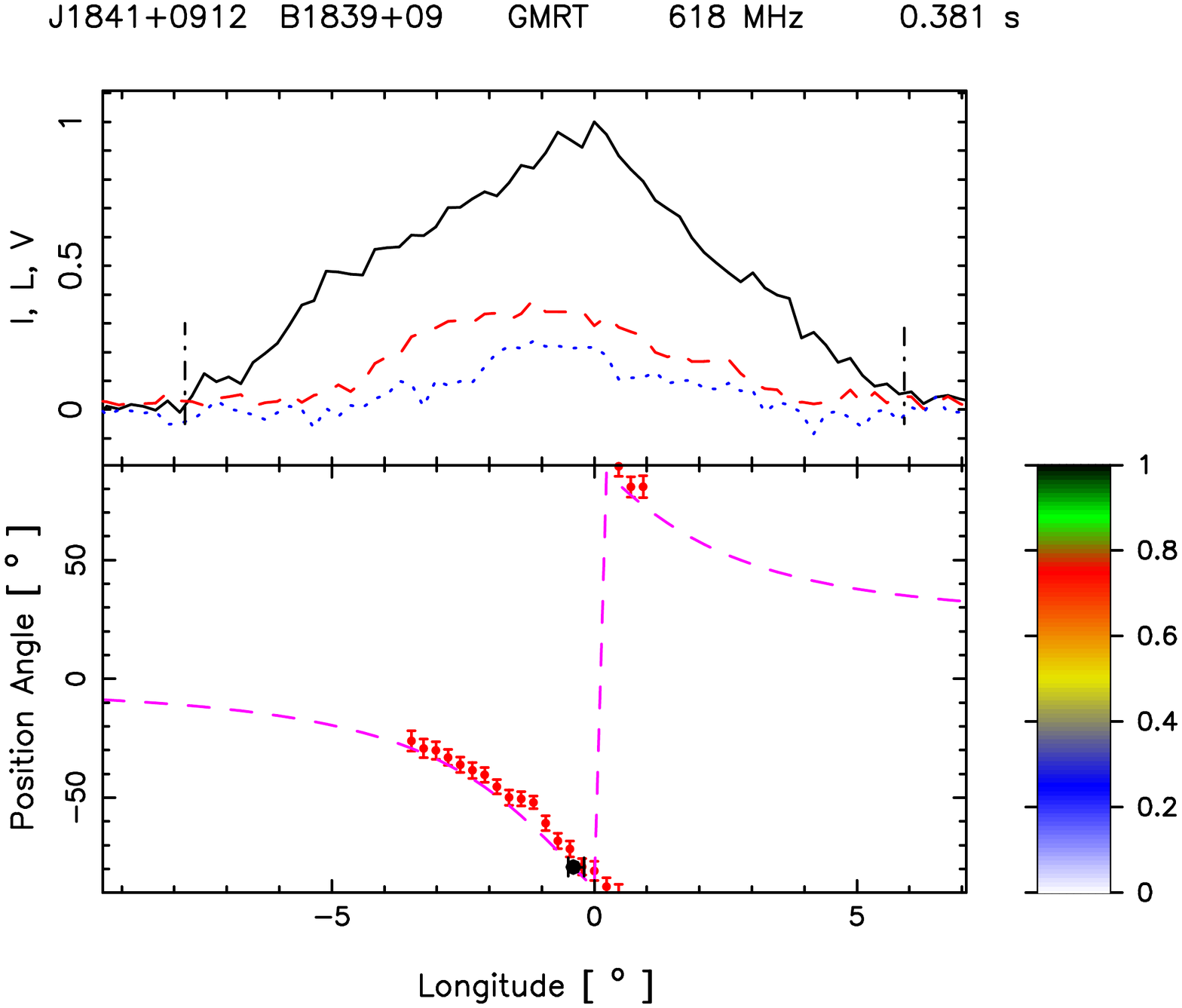}}}\\
{\mbox{\includegraphics[width=9cm,height=6cm,angle=0.]{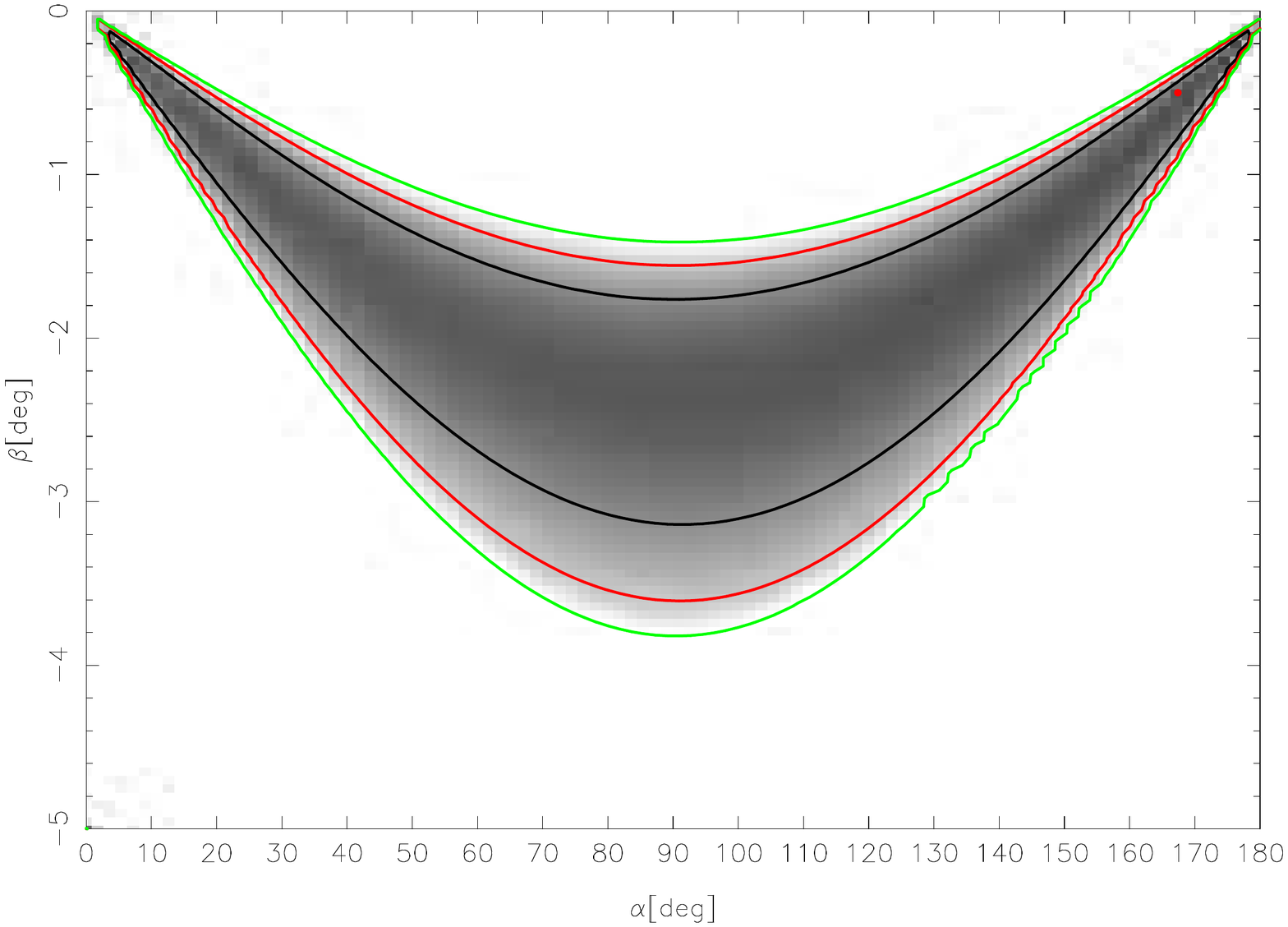}}}&
{\mbox{\includegraphics[width=9cm,height=6cm,angle=0.]{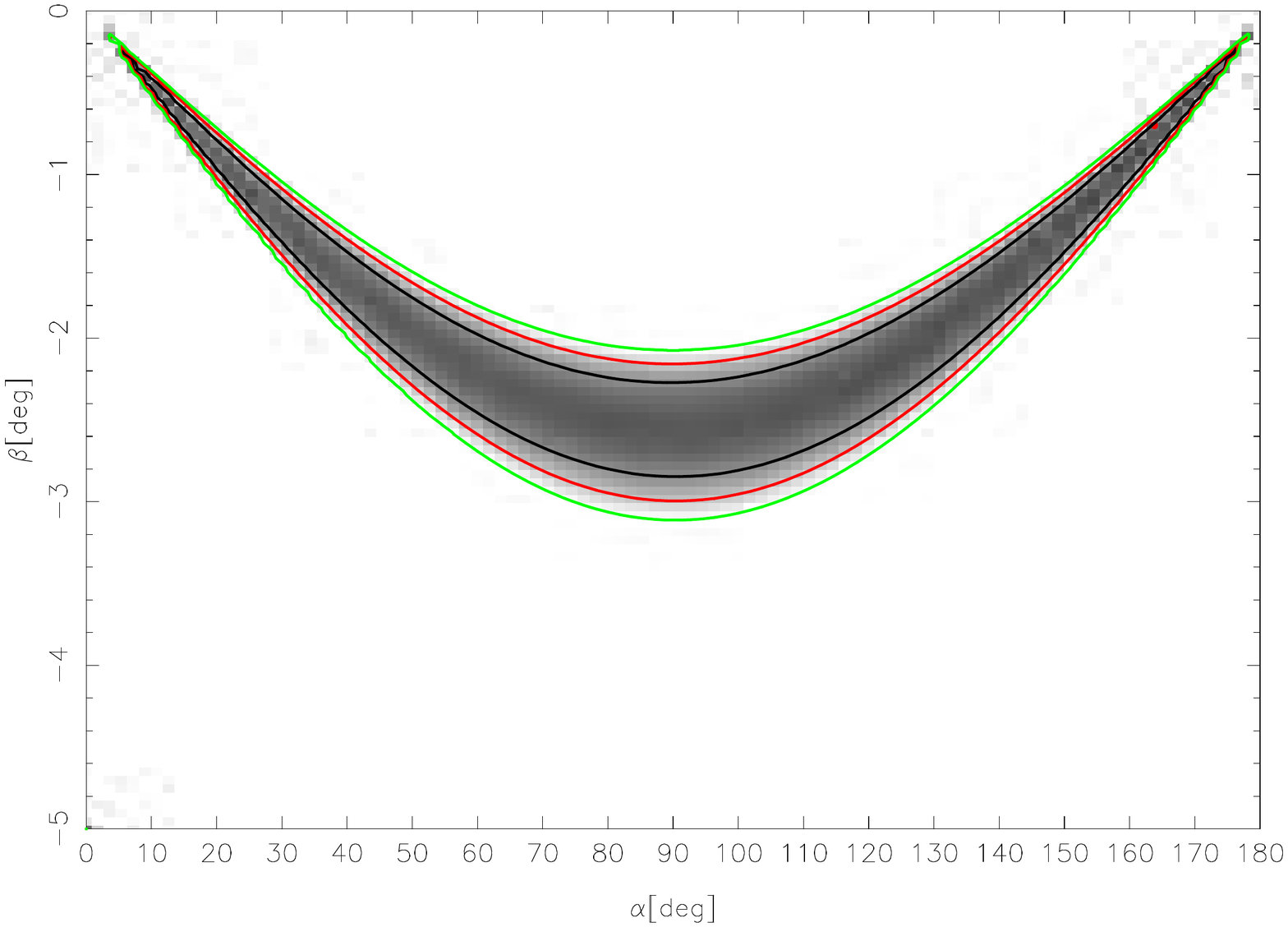}}}\\
&
\\
\end{tabular}
\caption{Top panel (upper window) shows the average profile with total
intensity (Stokes I; solid black lines), total linear polarization (dashed red
line) and circular polarization (Stokes V; dotted blue line). Top panel (lower
window) also shows the single pulse PPA distribution (colour scale) along with
the average PPA (red error bars).
The RVM fits to the average PPA (dashed pink
line) is also shown in this plot. Bottom panel show
the $\chi^2$ contours for the parameters $\alpha$ and $\beta$ obtained from RVM
fits.}
\label{a65}
\end{center}
\end{figure*}

\clearpage


\begin{figure*}
\begin{center}
\begin{tabular}{cc}
&
{\mbox{\includegraphics[width=9cm,height=6cm,angle=0.]{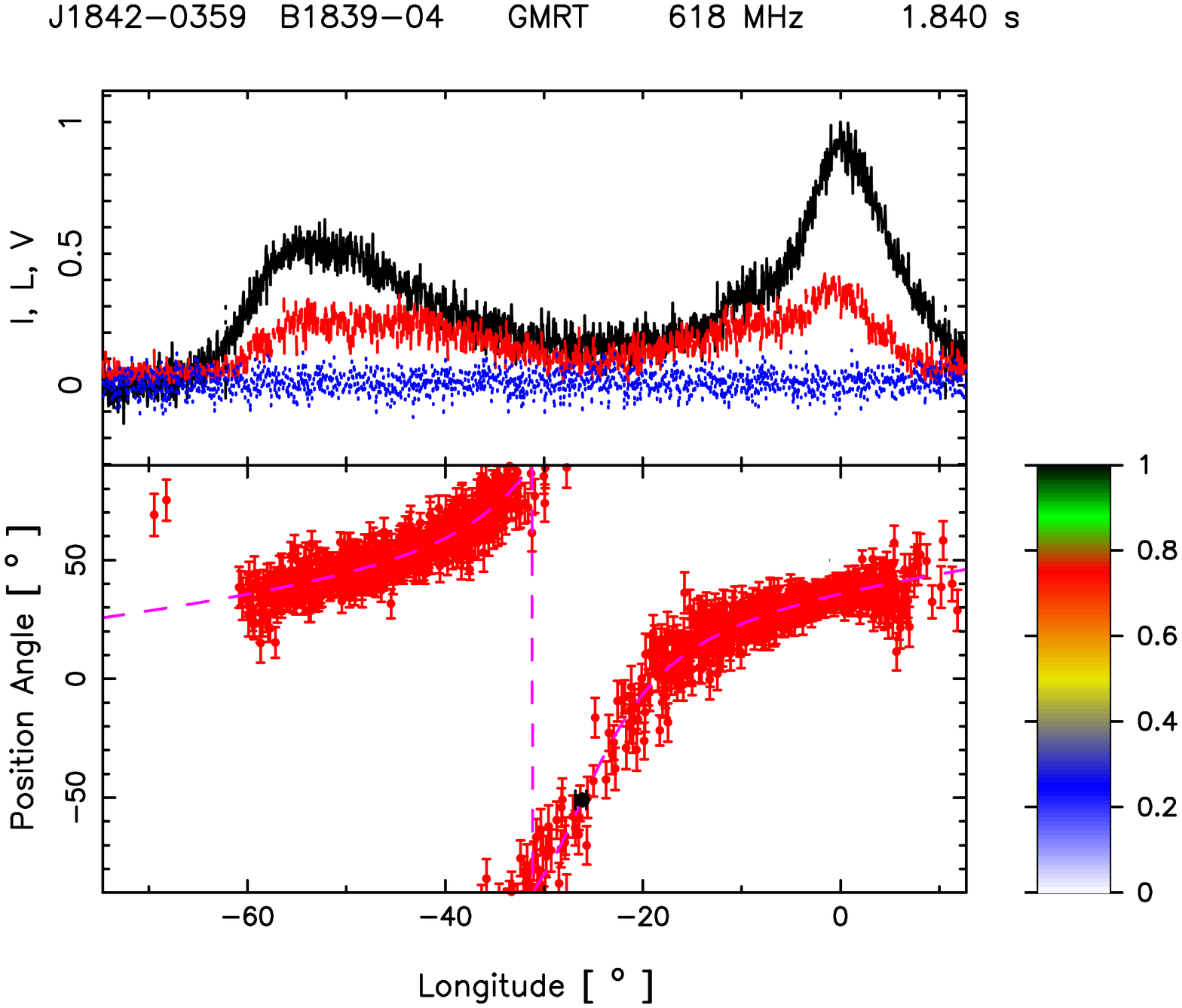}}}\\
&
{\mbox{\includegraphics[width=9cm,height=6cm,angle=0.]{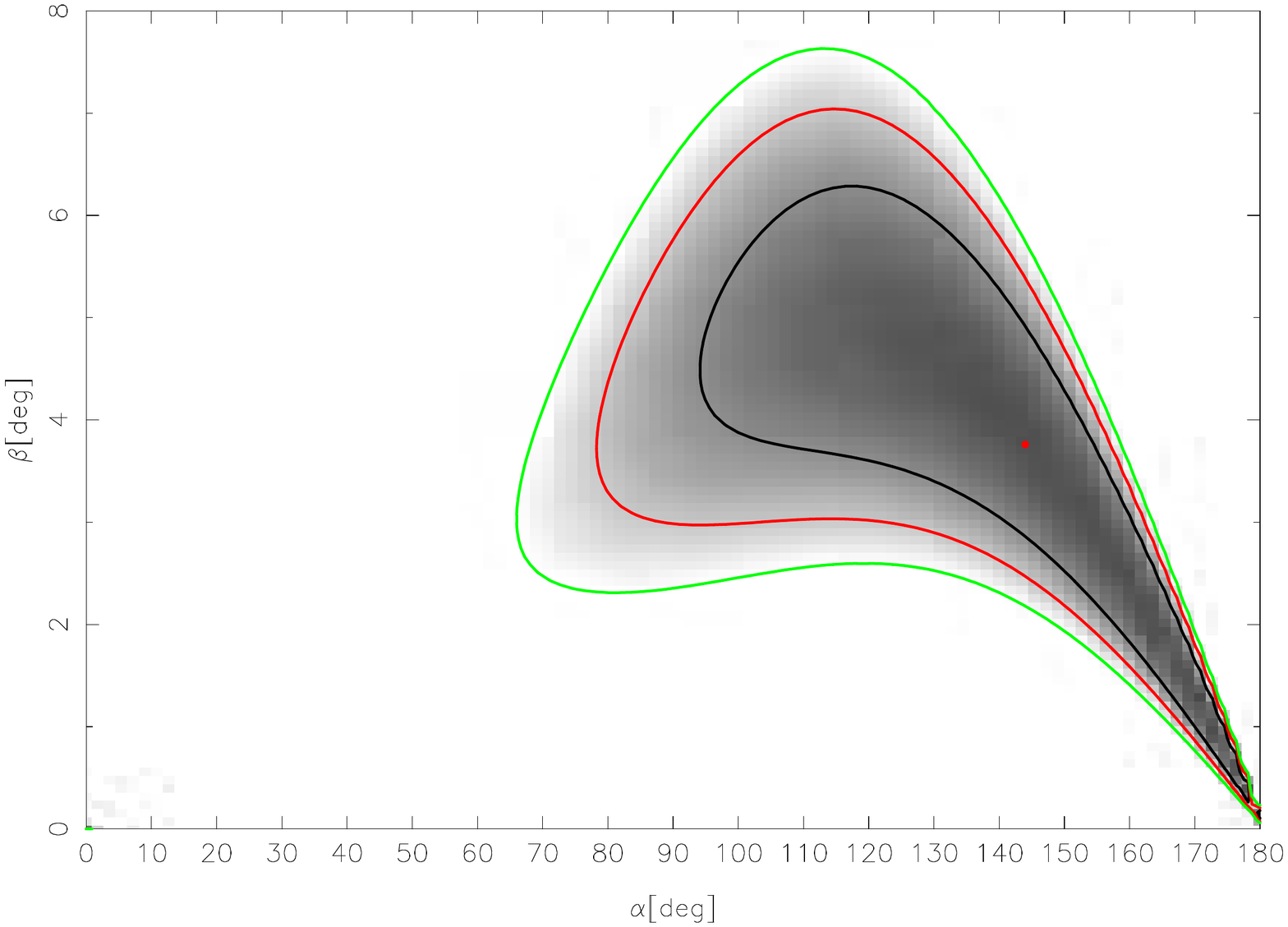}}}\\
&
\\
\end{tabular}
\caption{Top panel only for 618 MHz (upper window) shows the average profile with total
intensity (Stokes I; solid black lines), total linear polarization (dashed red
line) and circular polarization (Stokes V; dotted blue line). Top panel (lower
window) also shows the single pulse PPA distribution (colour scale) along with
the average PPA (red error bars).
The RVM fits to the average PPA (dashed pink
line) is also shown in this plot. Bottom panel only for 618 MHz show
the $\chi^2$ contours for the parameters $\alpha$ and $\beta$ obtained from RVM
fits.}
\label{a66}
\end{center}
\end{figure*}


\begin{figure*}
\begin{center}
\begin{tabular}{cc}
{\mbox{\includegraphics[width=9cm,height=6cm,angle=0.]{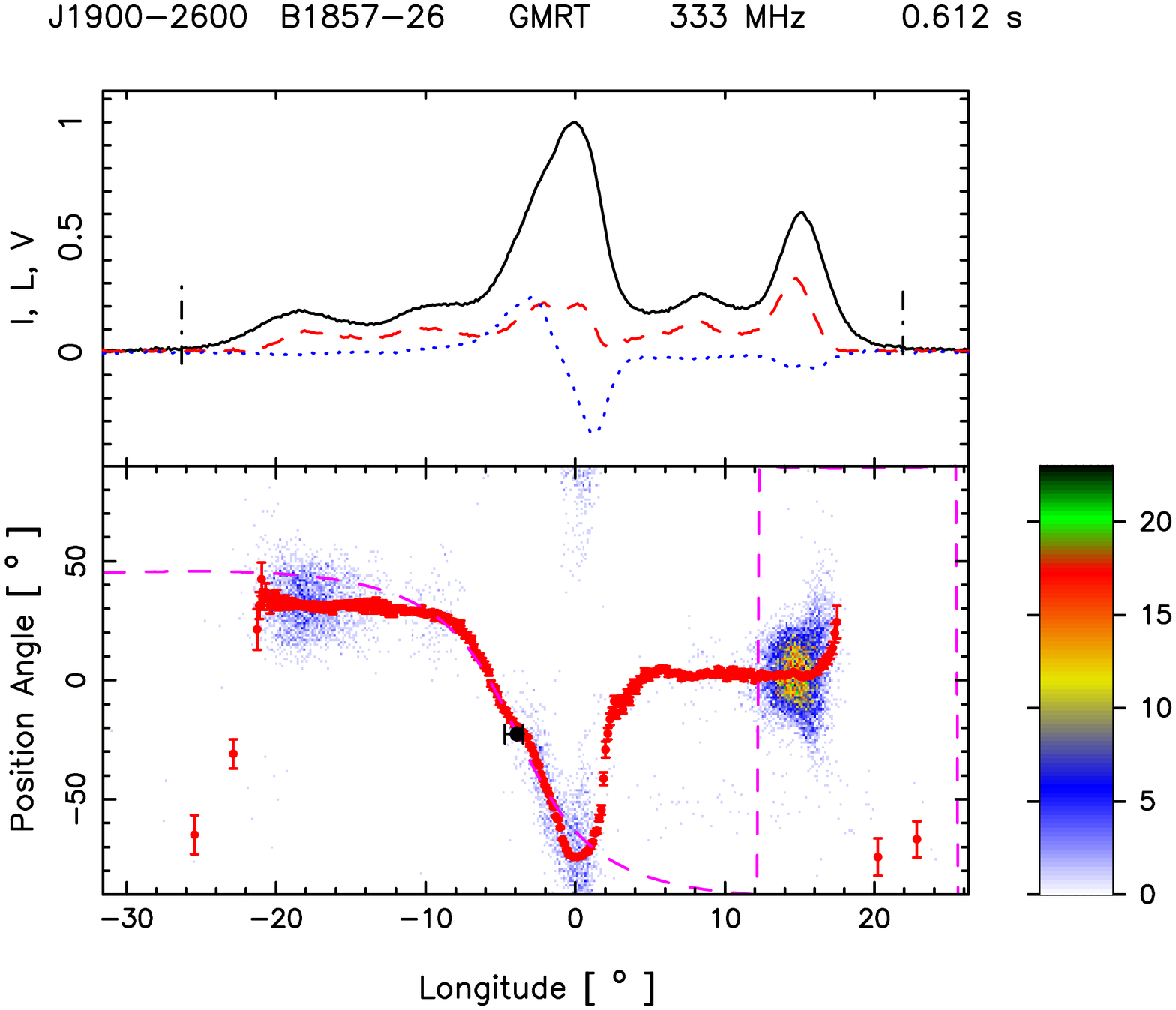}}}&
{\mbox{\includegraphics[width=9cm,height=6cm,angle=0.]{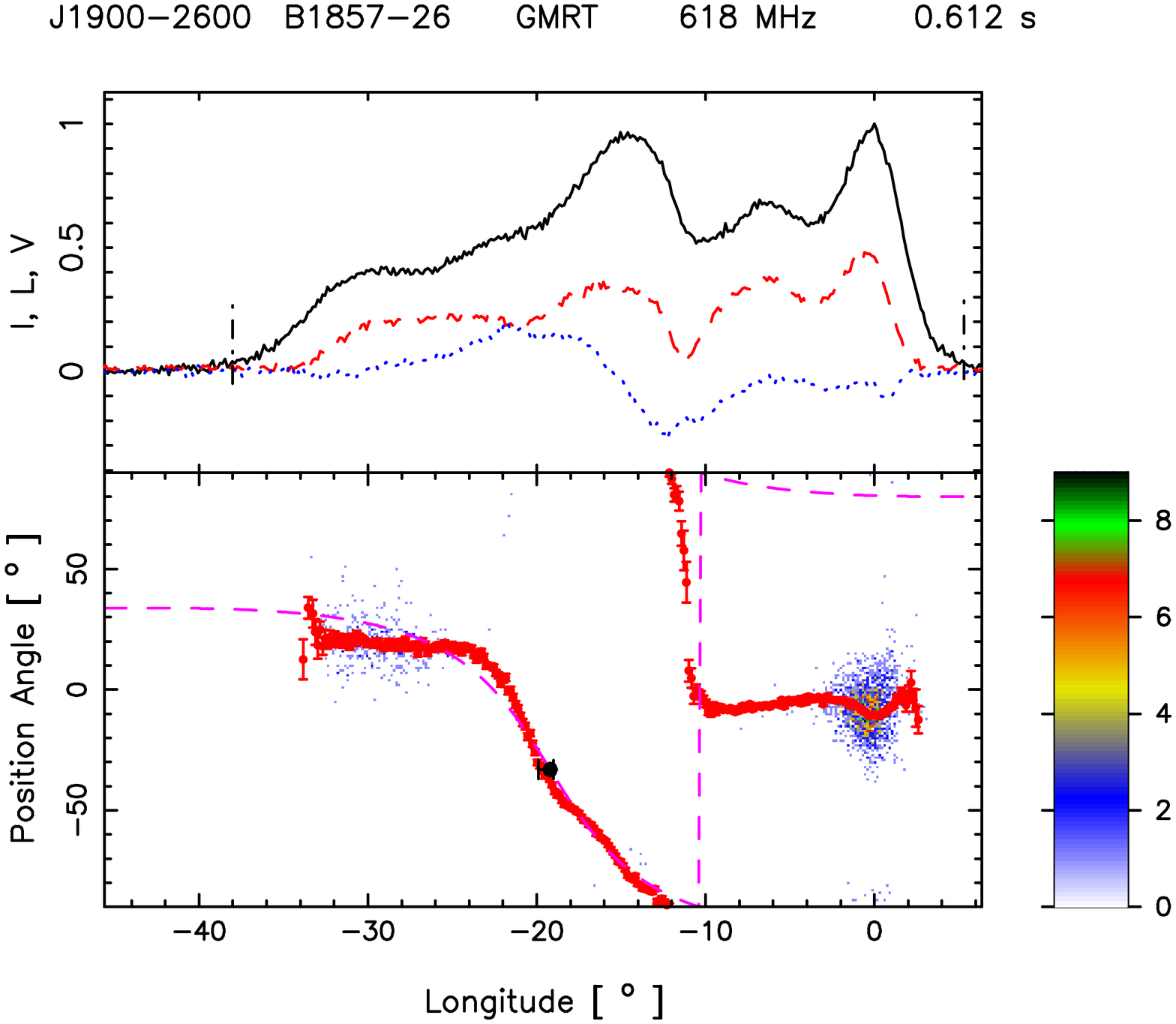}}}\\
{\mbox{\includegraphics[width=9cm,height=6cm,angle=0.]{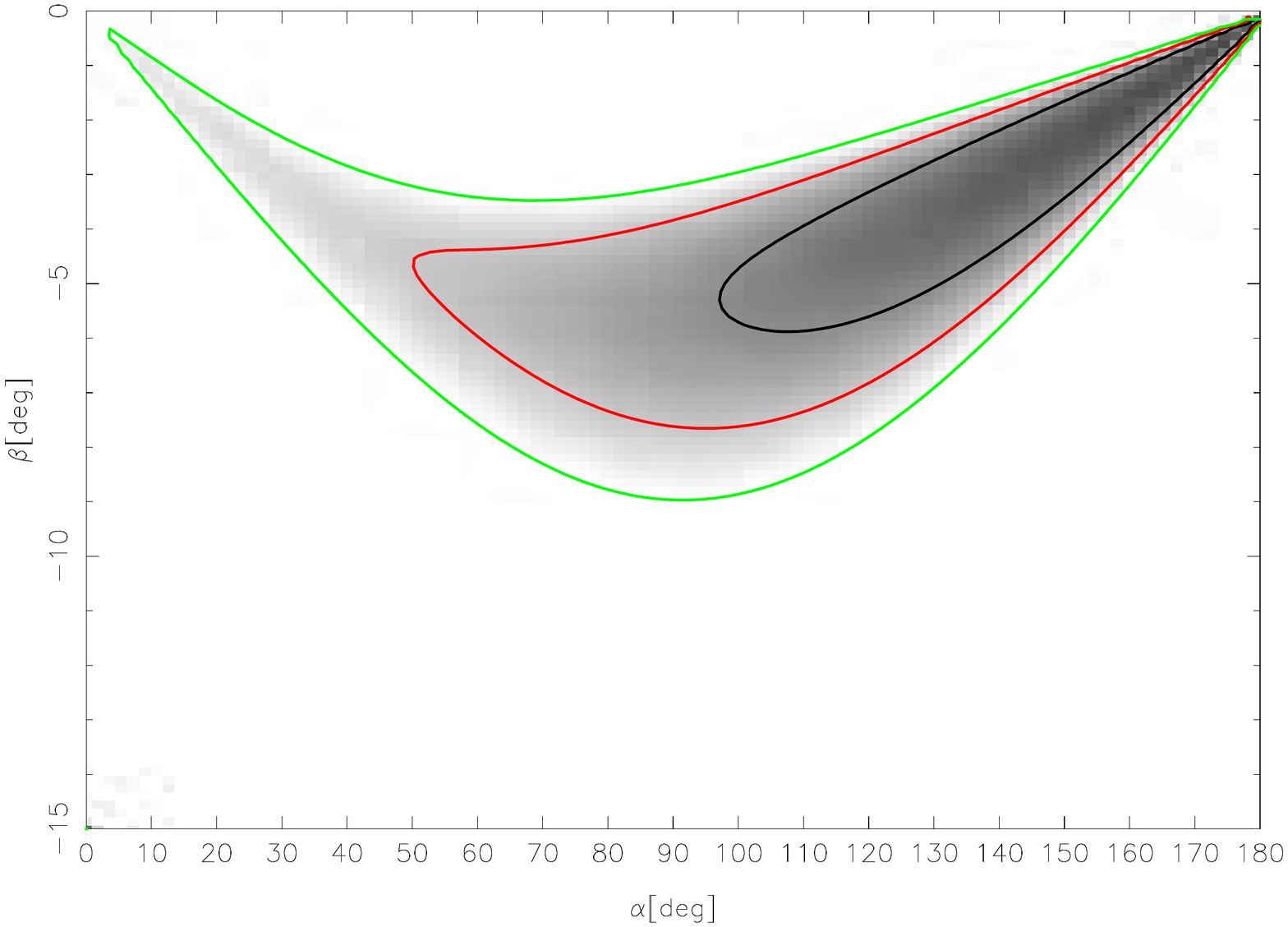}}}&
{\mbox{\includegraphics[width=9cm,height=6cm,angle=0.]{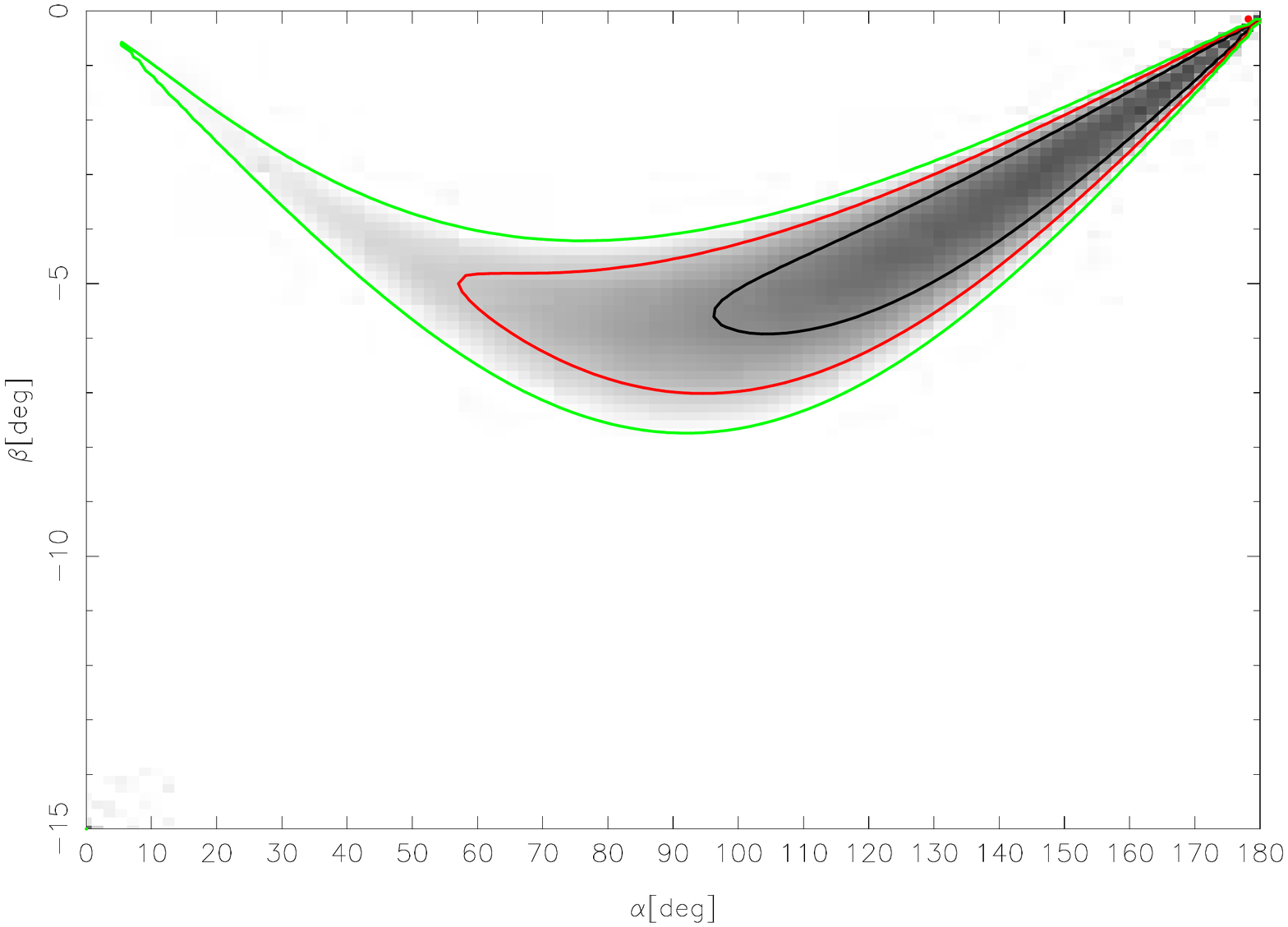}}}\\
{\mbox{\includegraphics[width=9cm,height=6cm,angle=0.]{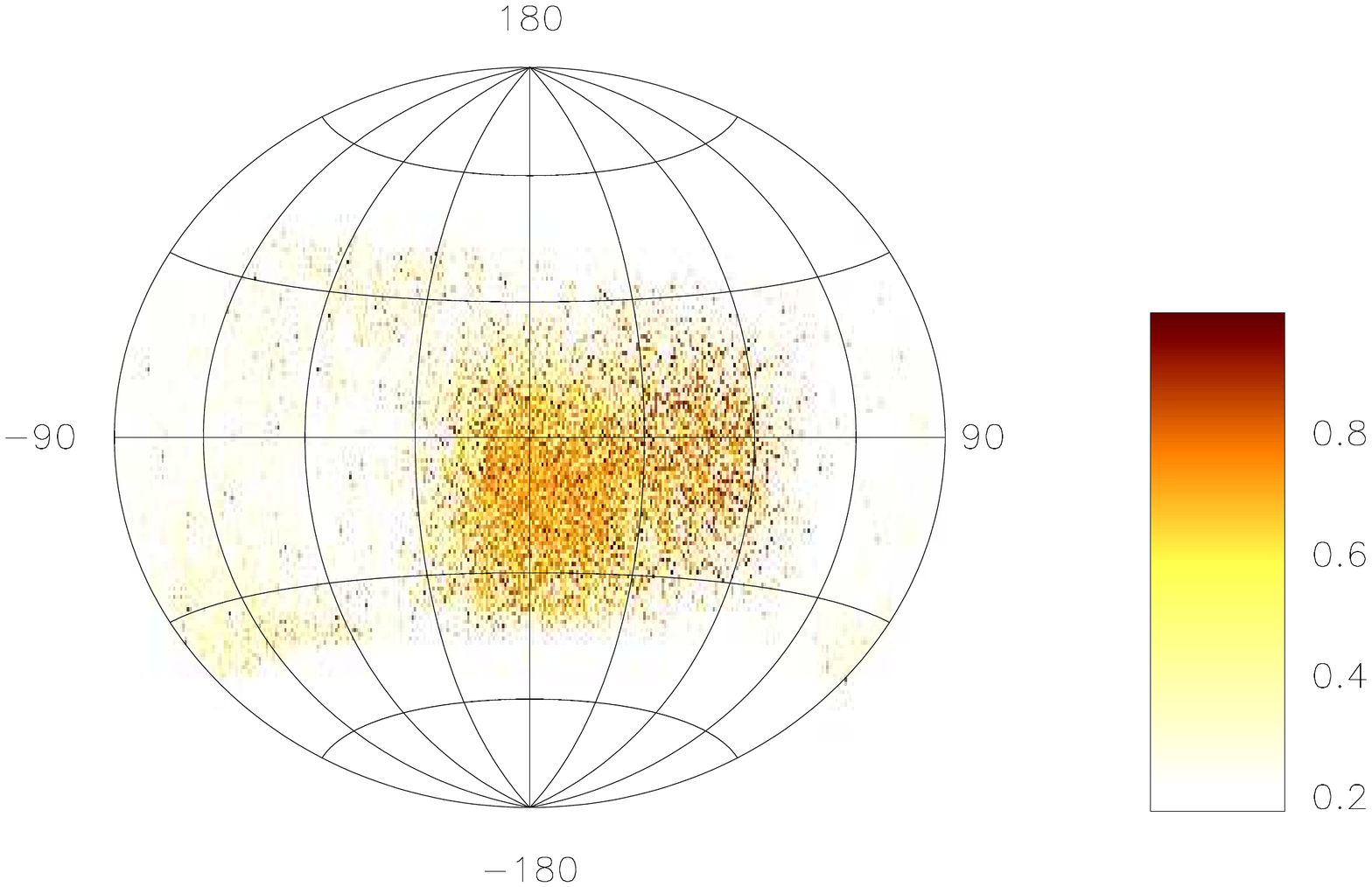}}}&
{\mbox{\includegraphics[width=9cm,height=6cm,angle=0.]{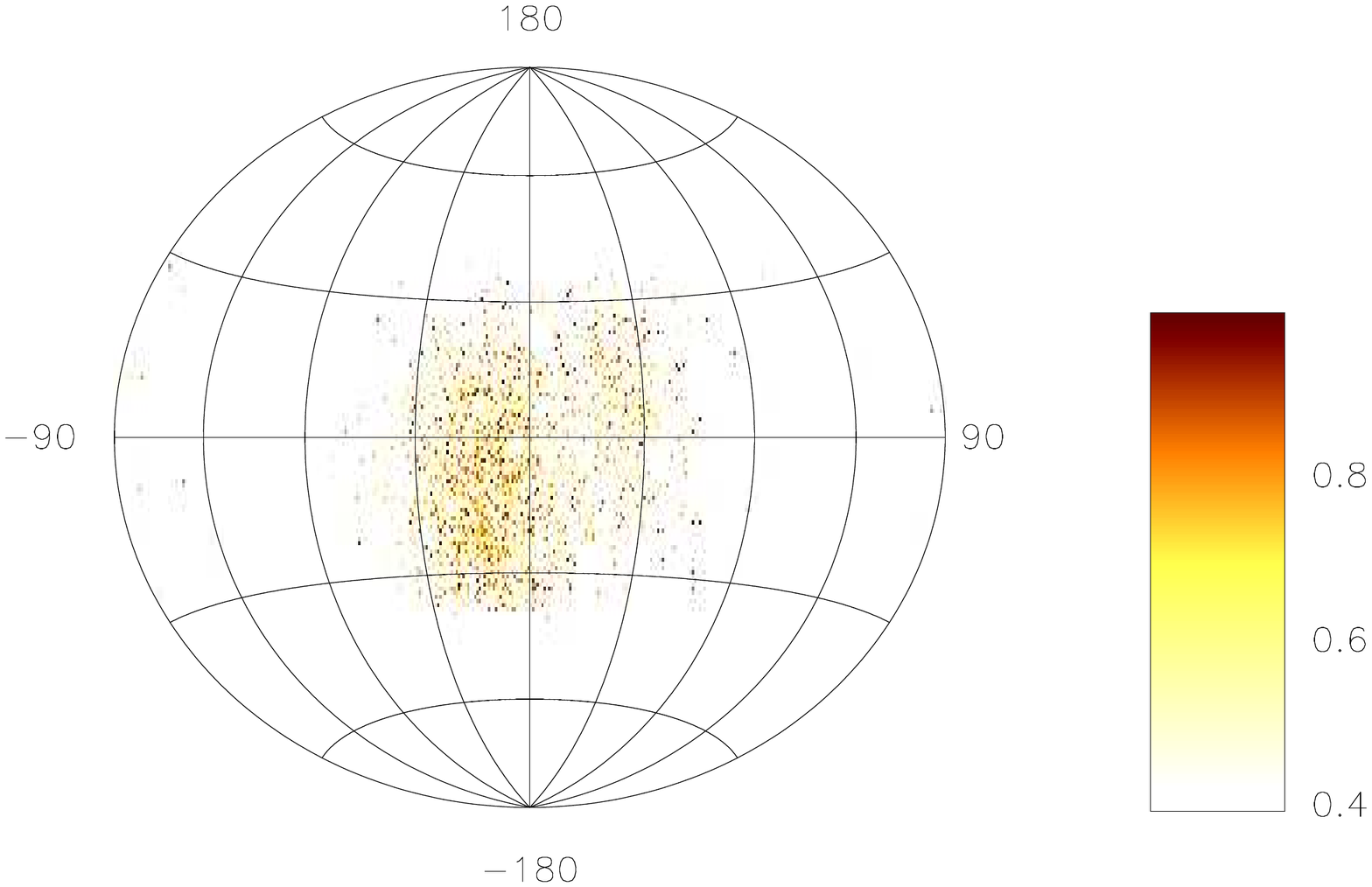}}}\\
\end{tabular}
\caption{Top panel (upper window) shows the average profile with total
intensity (Stokes I; solid black lines), total linear polarization (dashed red
line) and circular polarization (Stokes V; dotted blue line). Top panel (lower
window) also shows the single pulse PPA distribution (colour scale) along with
the average PPA (red error bars).
The RVM fits to the average PPA (dashed pink
line) is also shown in this plot. Middle panel show
the $\chi^2$ contours for the parameters $\alpha$ and $\beta$ obtained from RVM
fits.
Bottom panel shows the Hammer-Aitoff projection of the polarized time
samples with the colour scheme representing the fractional polarization level.}
\label{a67}
\end{center}
\end{figure*}


\begin{figure*}
\begin{center}
\begin{tabular}{cc}
{\mbox{\includegraphics[width=9cm,height=6cm,angle=0.]{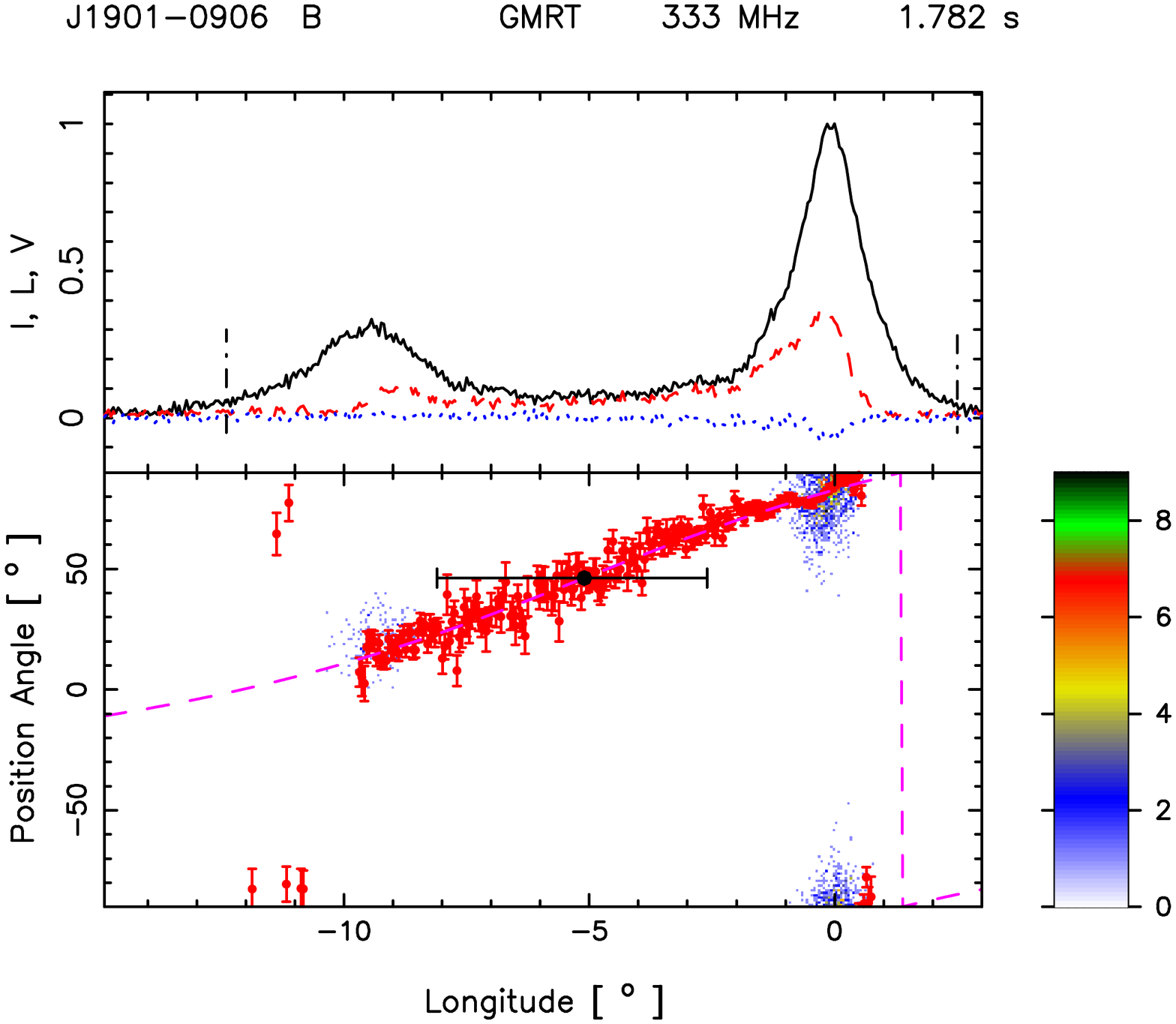}}}&
{\mbox{\includegraphics[width=9cm,height=6cm,angle=0.]{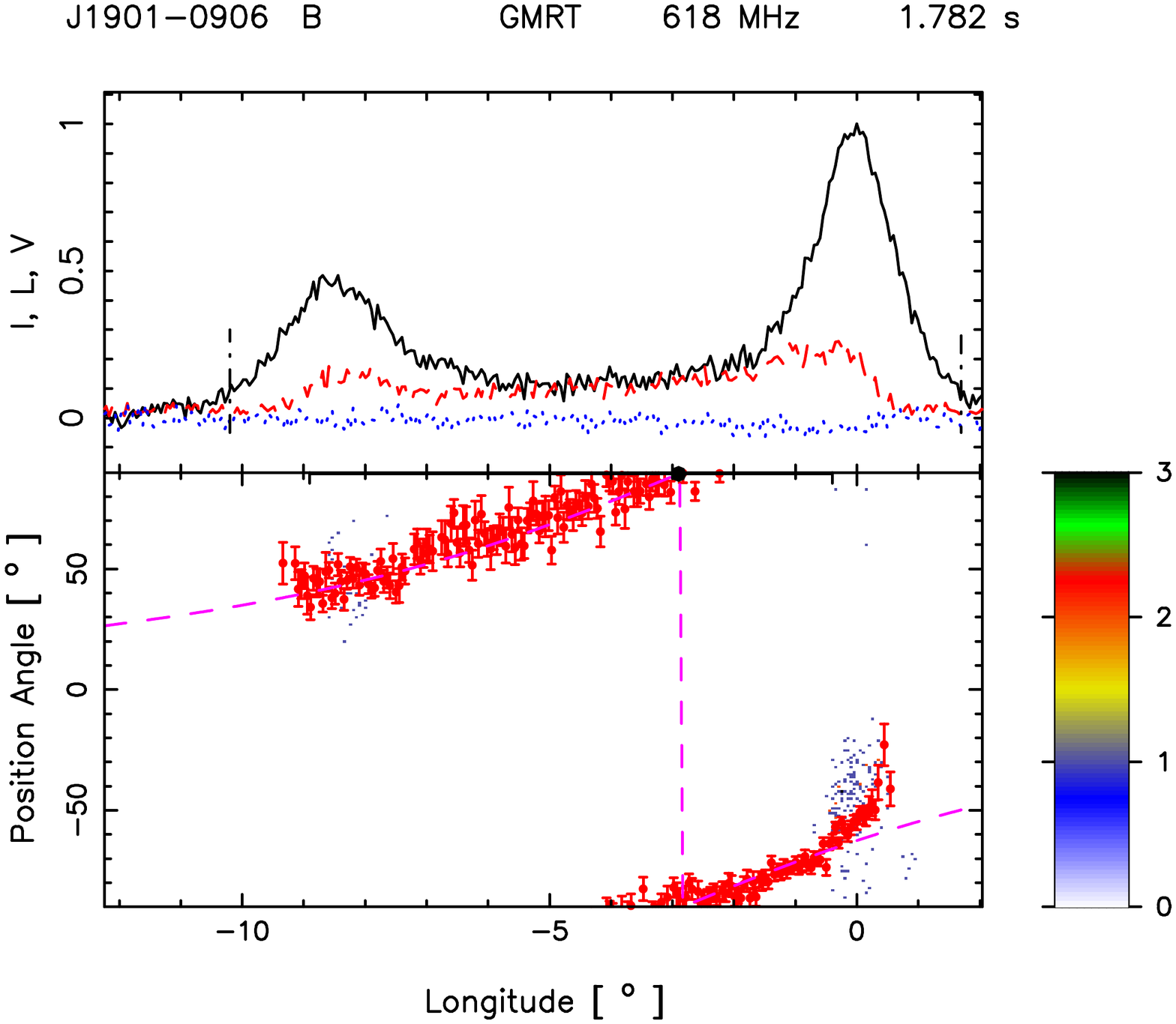}}}\\
{\mbox{\includegraphics[width=9cm,height=6cm,angle=0.]{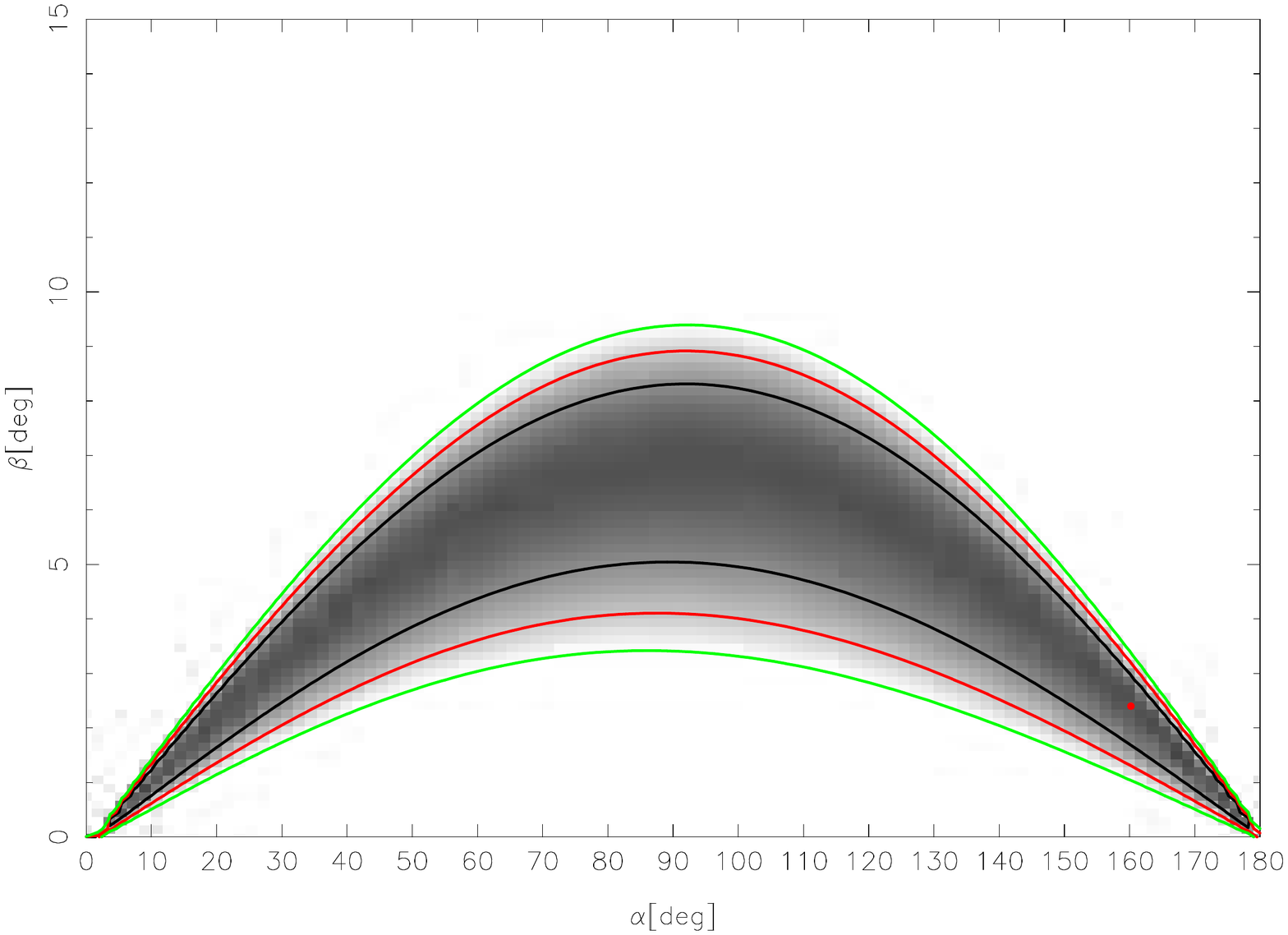}}}&
{\mbox{\includegraphics[width=9cm,height=6cm,angle=0.]{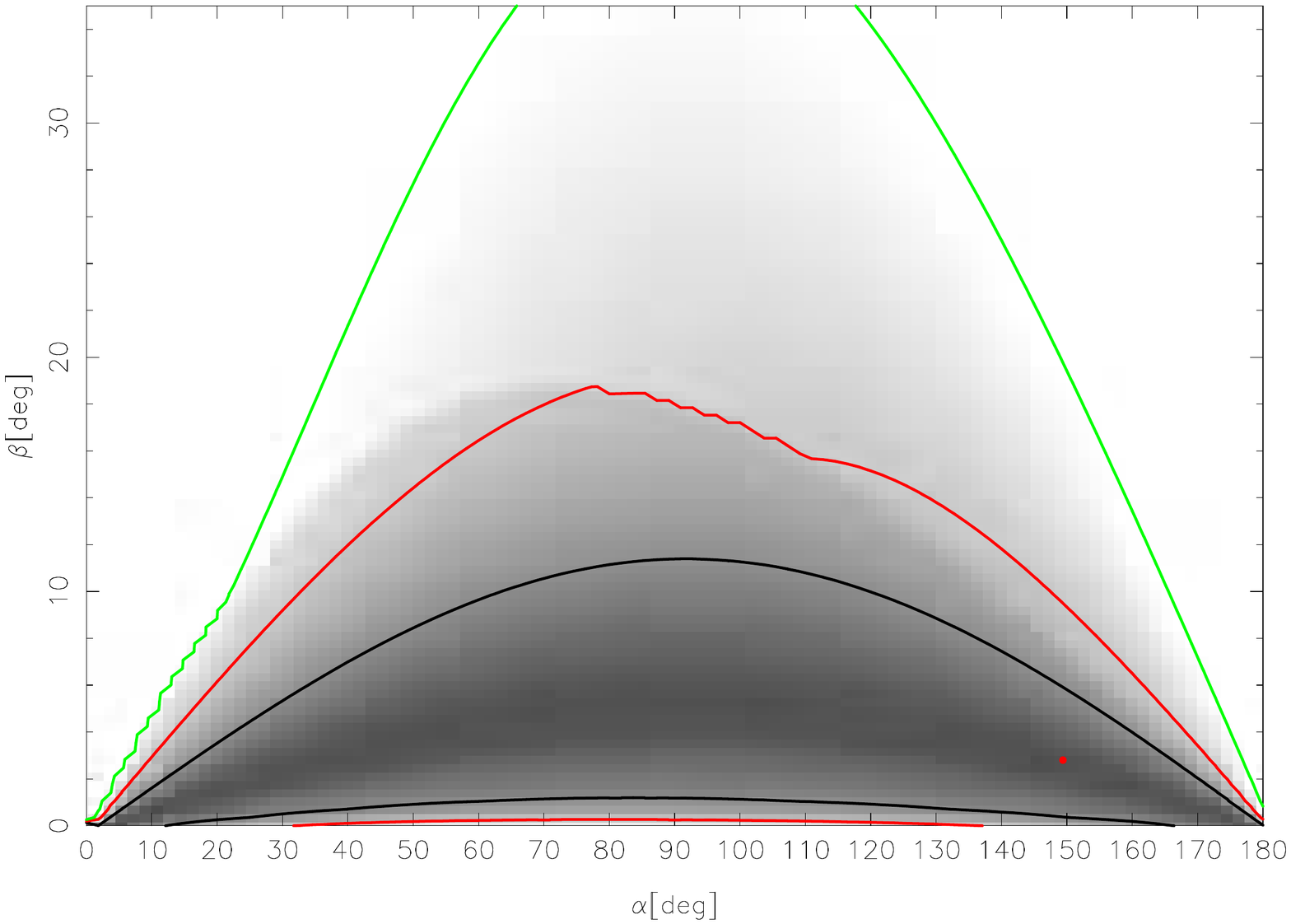}}}\\
{\mbox{\includegraphics[width=9cm,height=6cm,angle=0.]{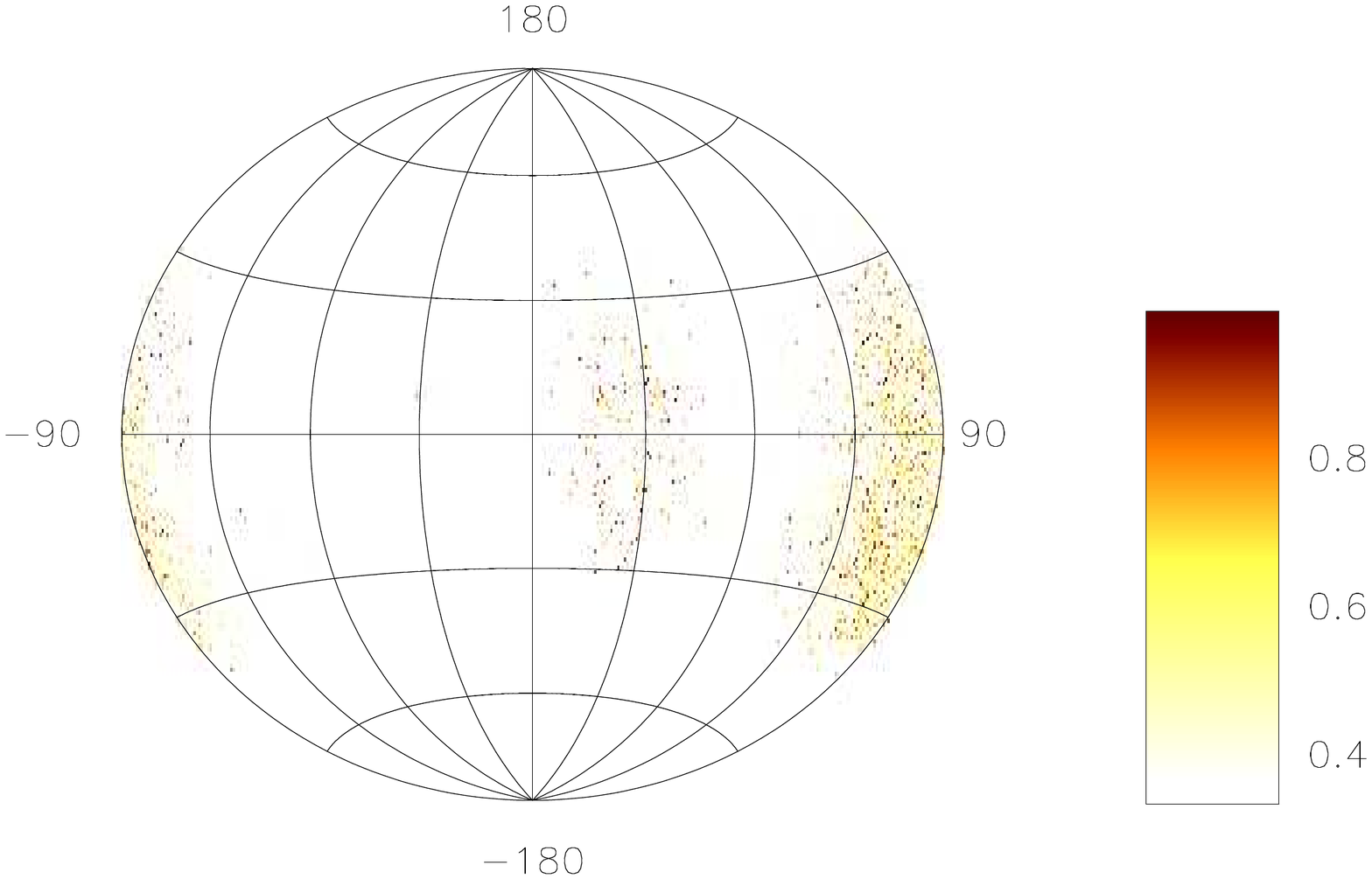}}}&
{\mbox{\includegraphics[width=9cm,height=6cm,angle=0.]{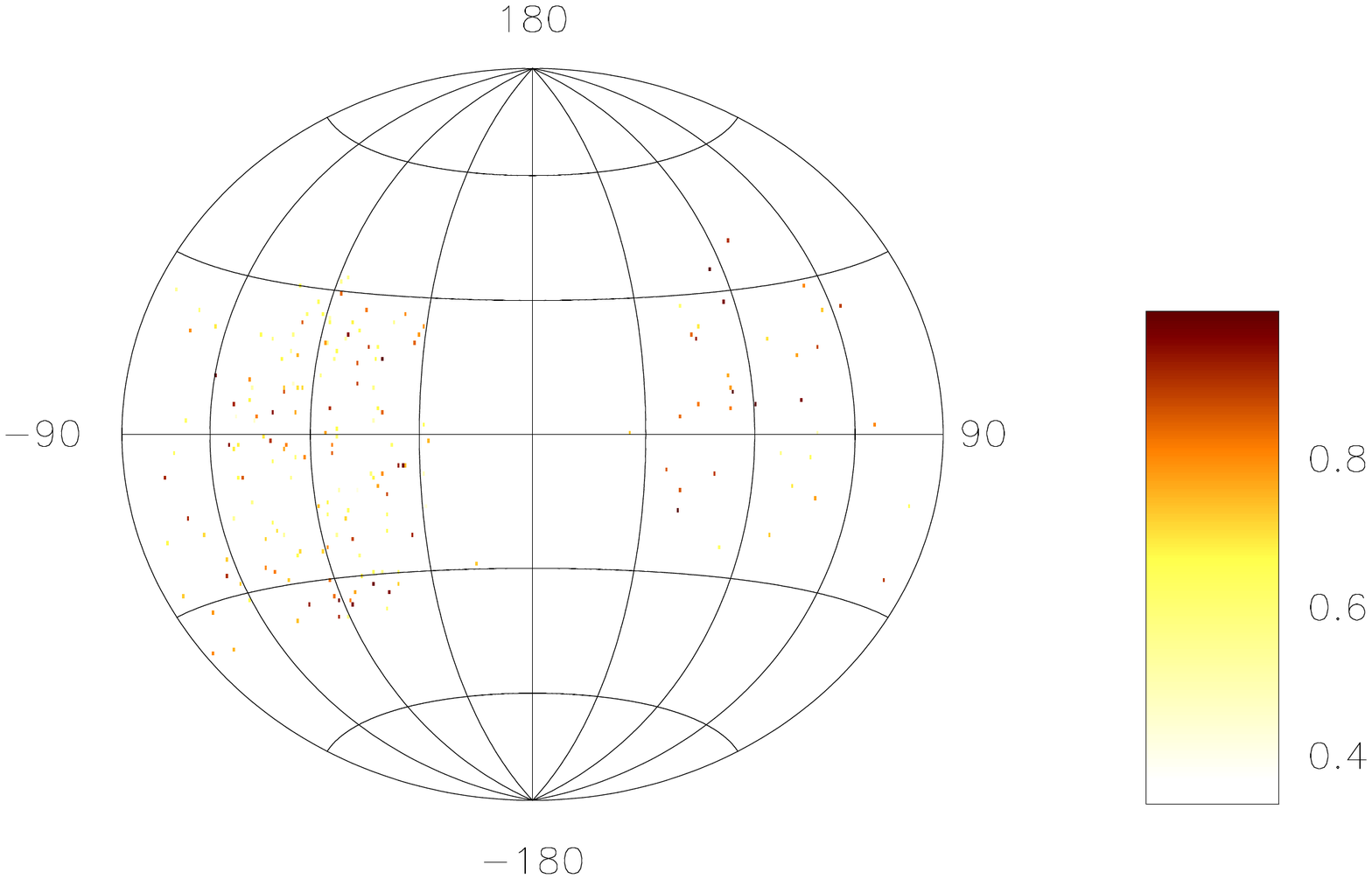}}}\\
\end{tabular}
\caption{Top panel (upper window) shows the average profile with total
intensity (Stokes I; solid black lines), total linear polarization (dashed red
line) and circular polarization (Stokes V; dotted blue line). Top panel (lower
window) also shows the single pulse PPA distribution (colour scale) along with
the average PPA (red error bars).
The RVM fits to the average PPA (dashed pink
line) is also shown in this plot. Middle panel show
the $\chi^2$ contours for the parameters $\alpha$ and $\beta$ obtained from RVM
fits.
Bottom panel shows the Hammer-Aitoff projection of the polarized time
samples with the colour scheme representing the fractional polarization level.}
\label{a68}
\end{center}
\end{figure*}


\begin{figure*}
\begin{center}
\begin{tabular}{cc}
{\mbox{\includegraphics[width=9cm,height=6cm,angle=0.]{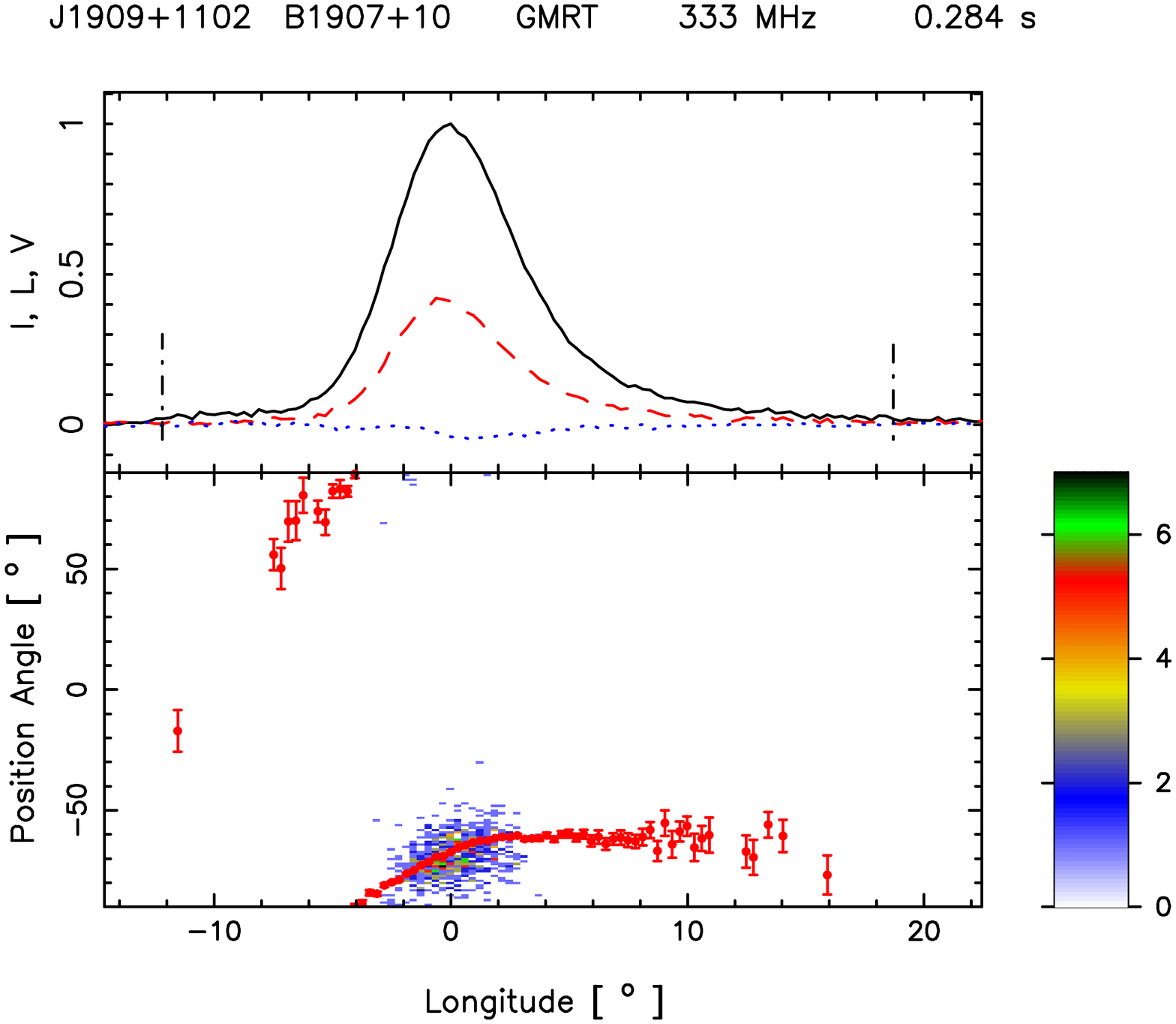}}}&
{\mbox{\includegraphics[width=9cm,height=6cm,angle=0.]{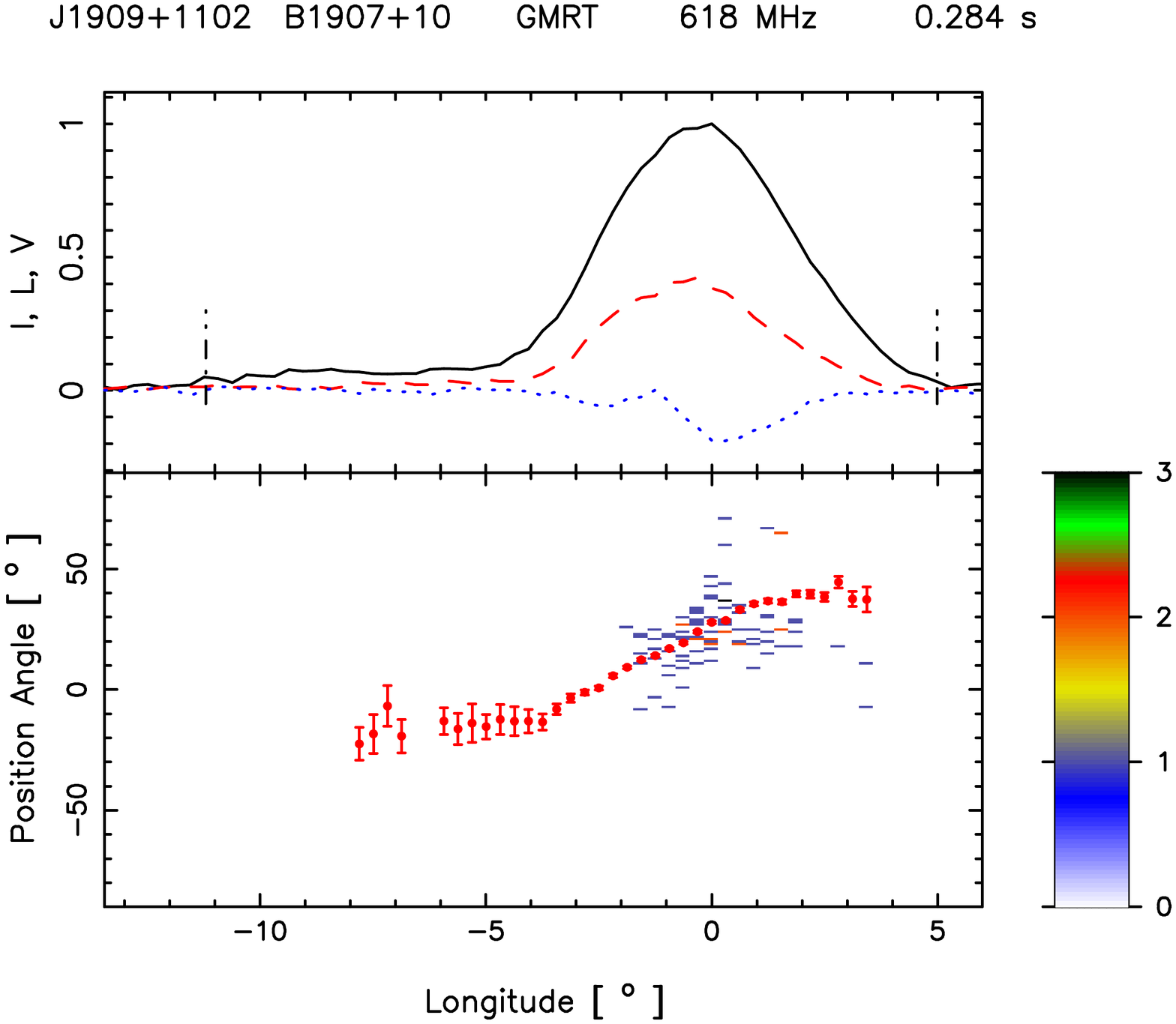}}}\\
&
\\
{\mbox{\includegraphics[width=9cm,height=6cm,angle=0.]{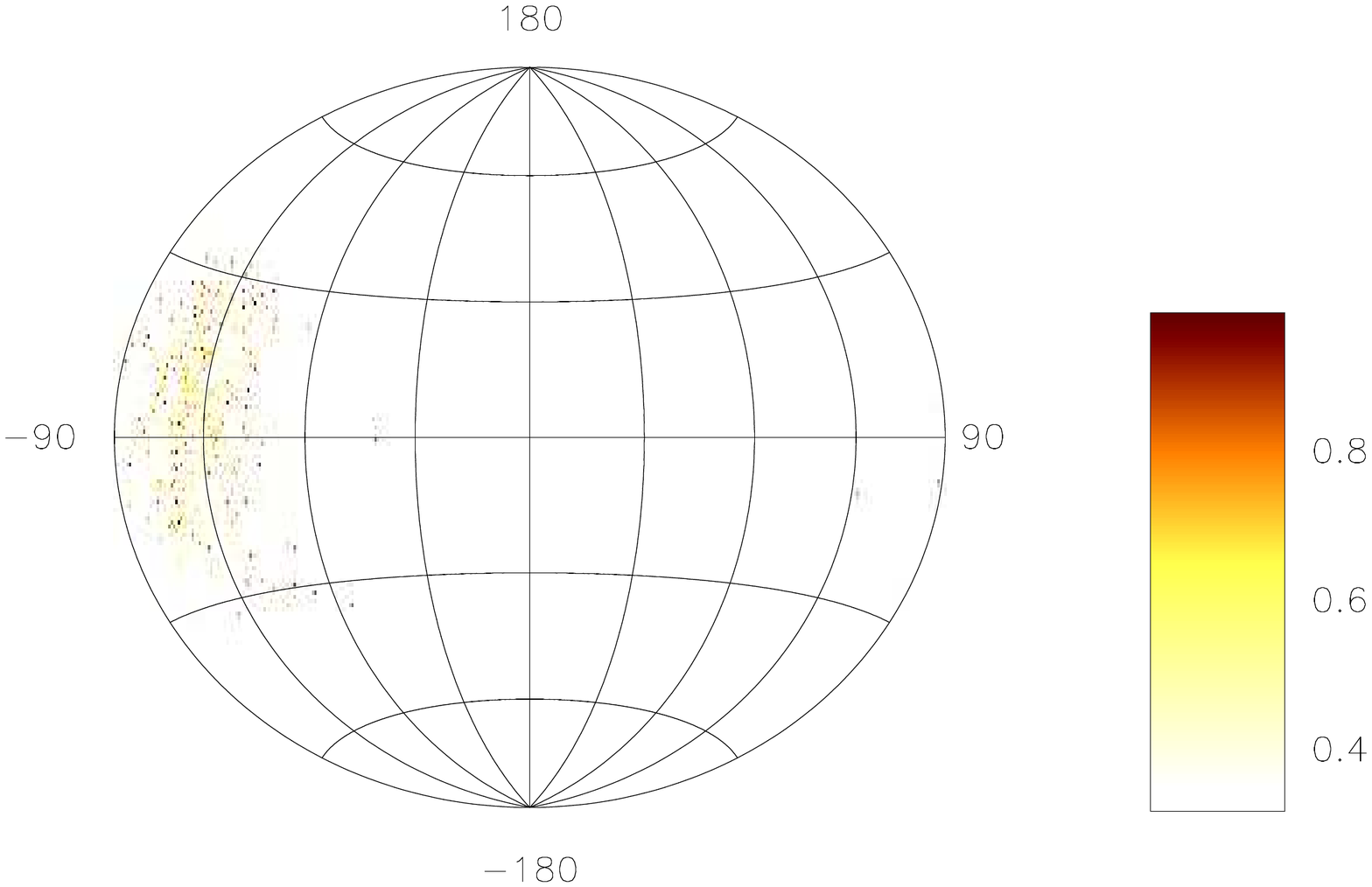}}}&
{\mbox{\includegraphics[width=9cm,height=6cm,angle=0.]{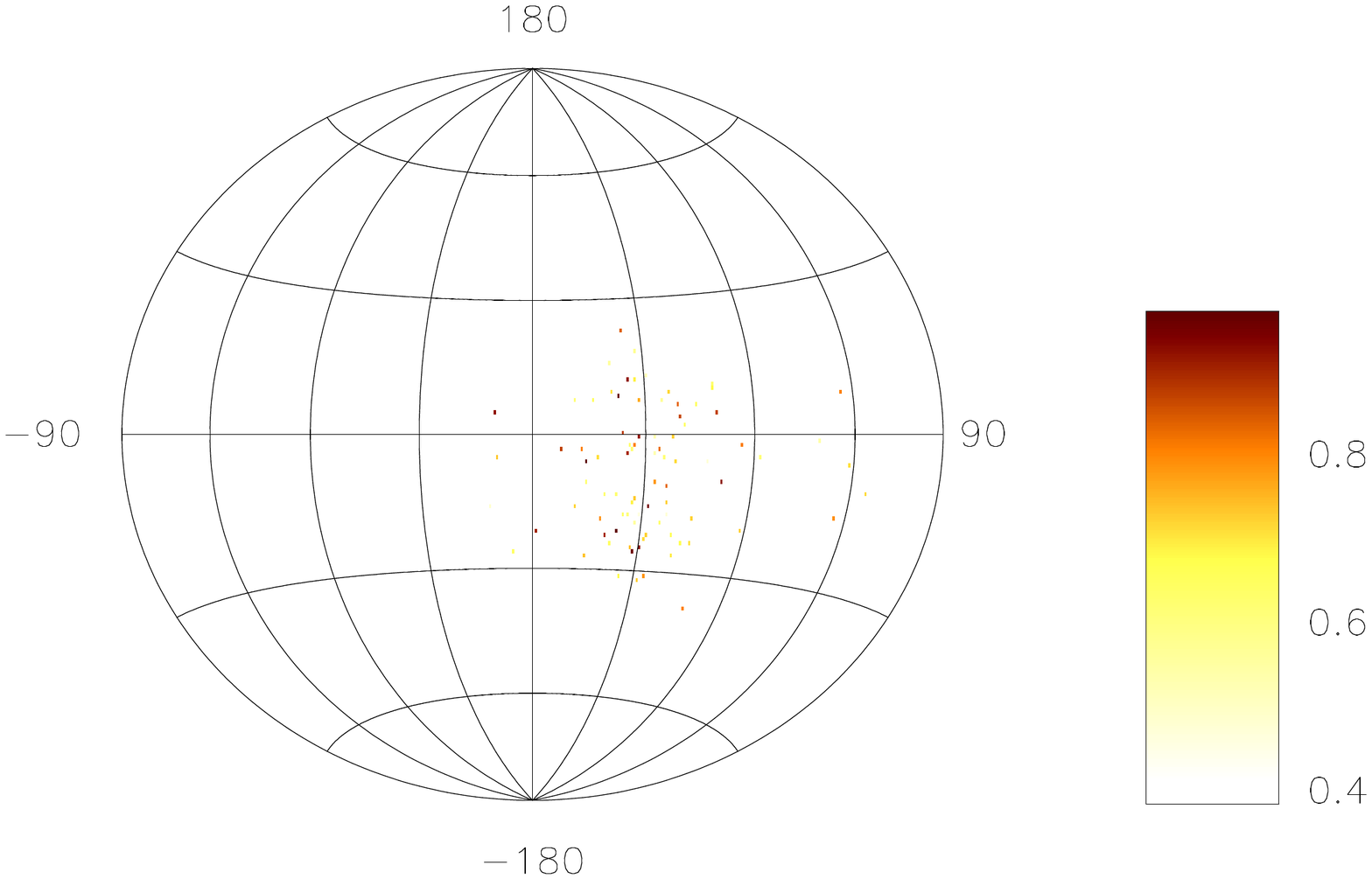}}}\\
\end{tabular}
\caption{Top panel (upper window) shows the average profile with total
intensity (Stokes I; solid black lines), total linear polarization (dashed red
line) and circular polarization (Stokes V; dotted blue line). Top panel (lower
window) also shows the single pulse PPA distribution (colour scale) along with
the average PPA (red error bars).
Bottom panel shows the Hammer-Aitoff projection of the polarized time
samples with the colour scheme representing the fractional polarization level.}
\label{a69}
\end{center}
\end{figure*}


\begin{figure*}
\begin{center}
\begin{tabular}{cc}
{\mbox{\includegraphics[width=9cm,height=6cm,angle=0.]{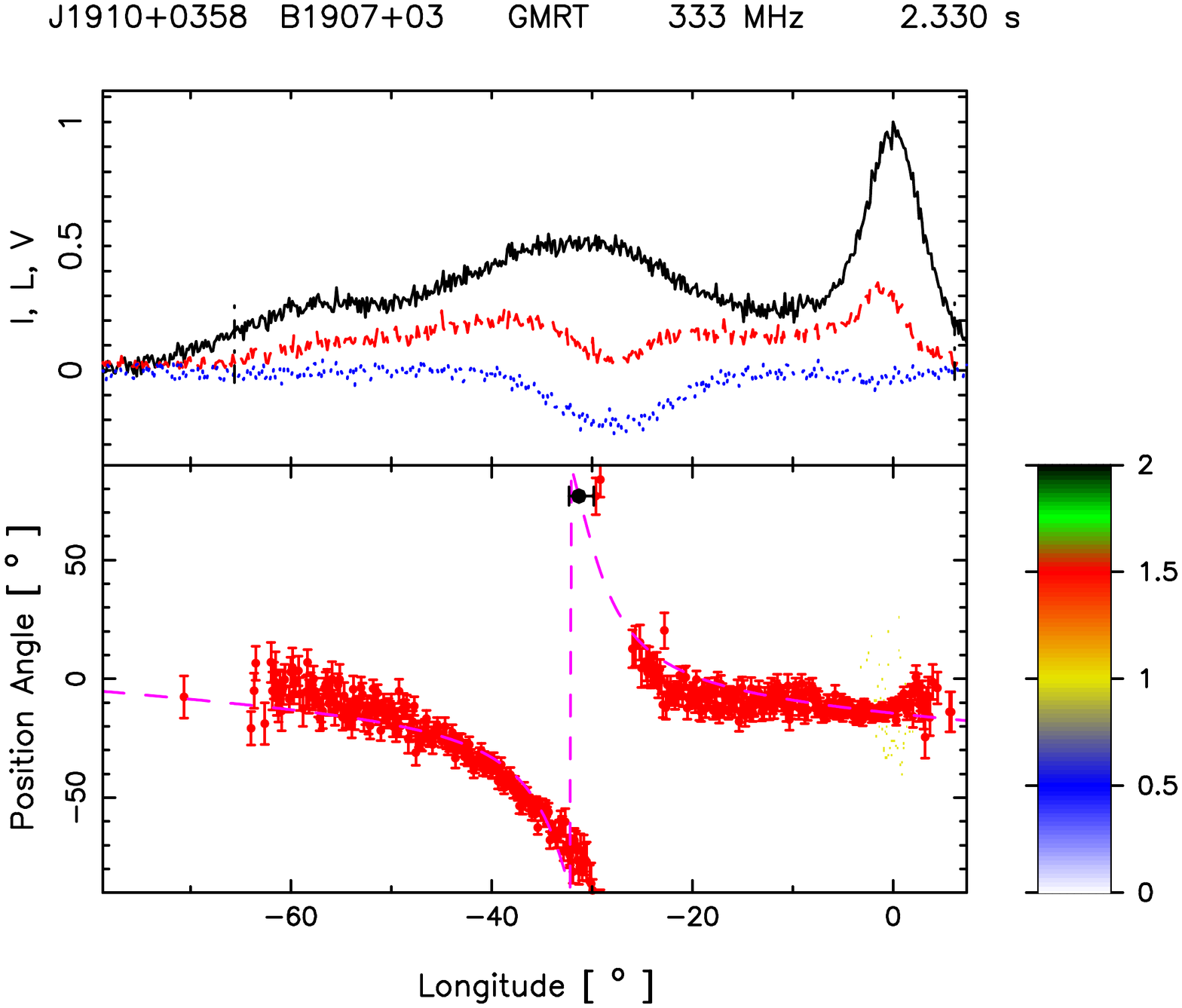}}}&
{\mbox{\includegraphics[width=9cm,height=6cm,angle=0.]{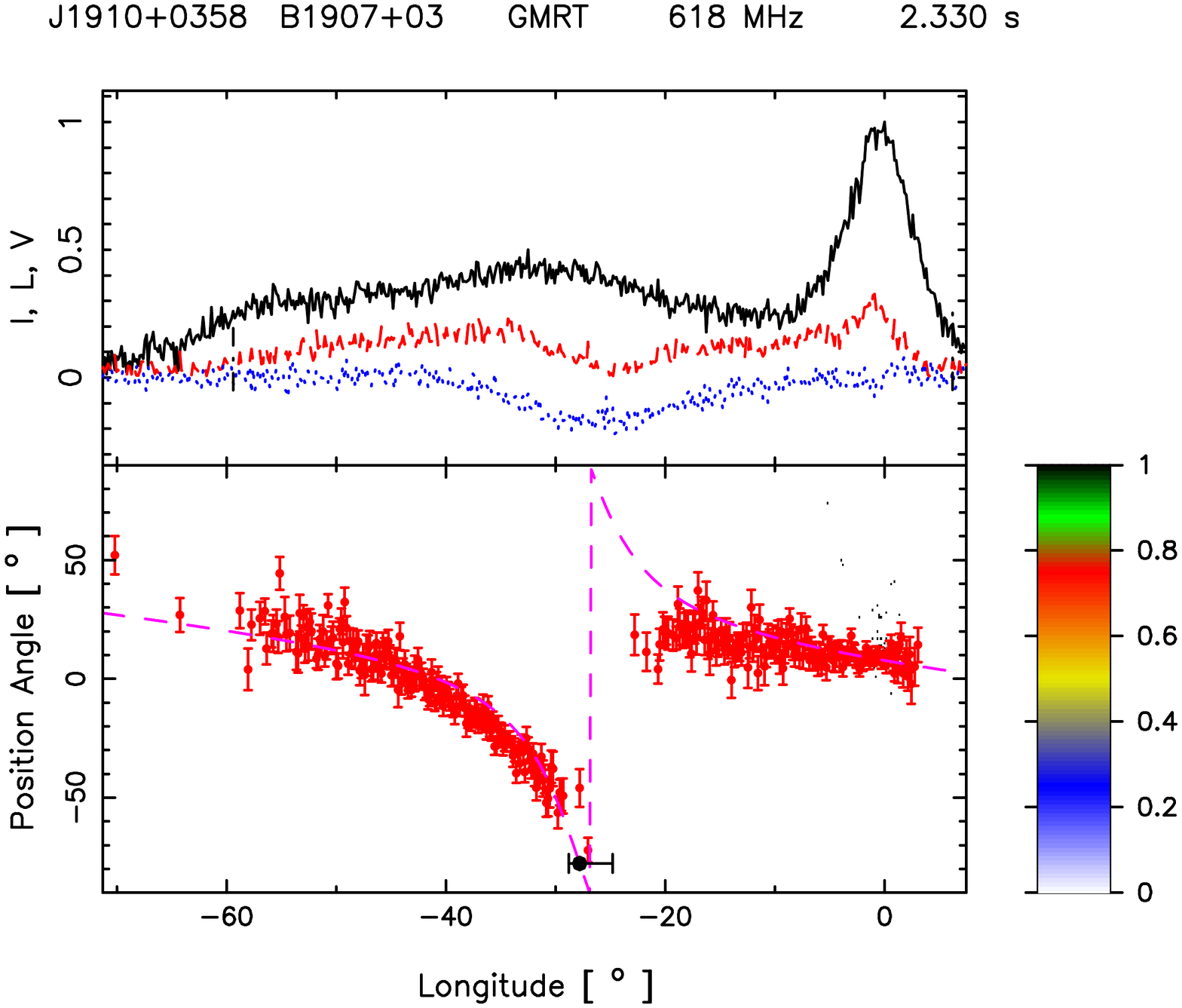}}}\\
{\mbox{\includegraphics[width=9cm,height=6cm,angle=0.]{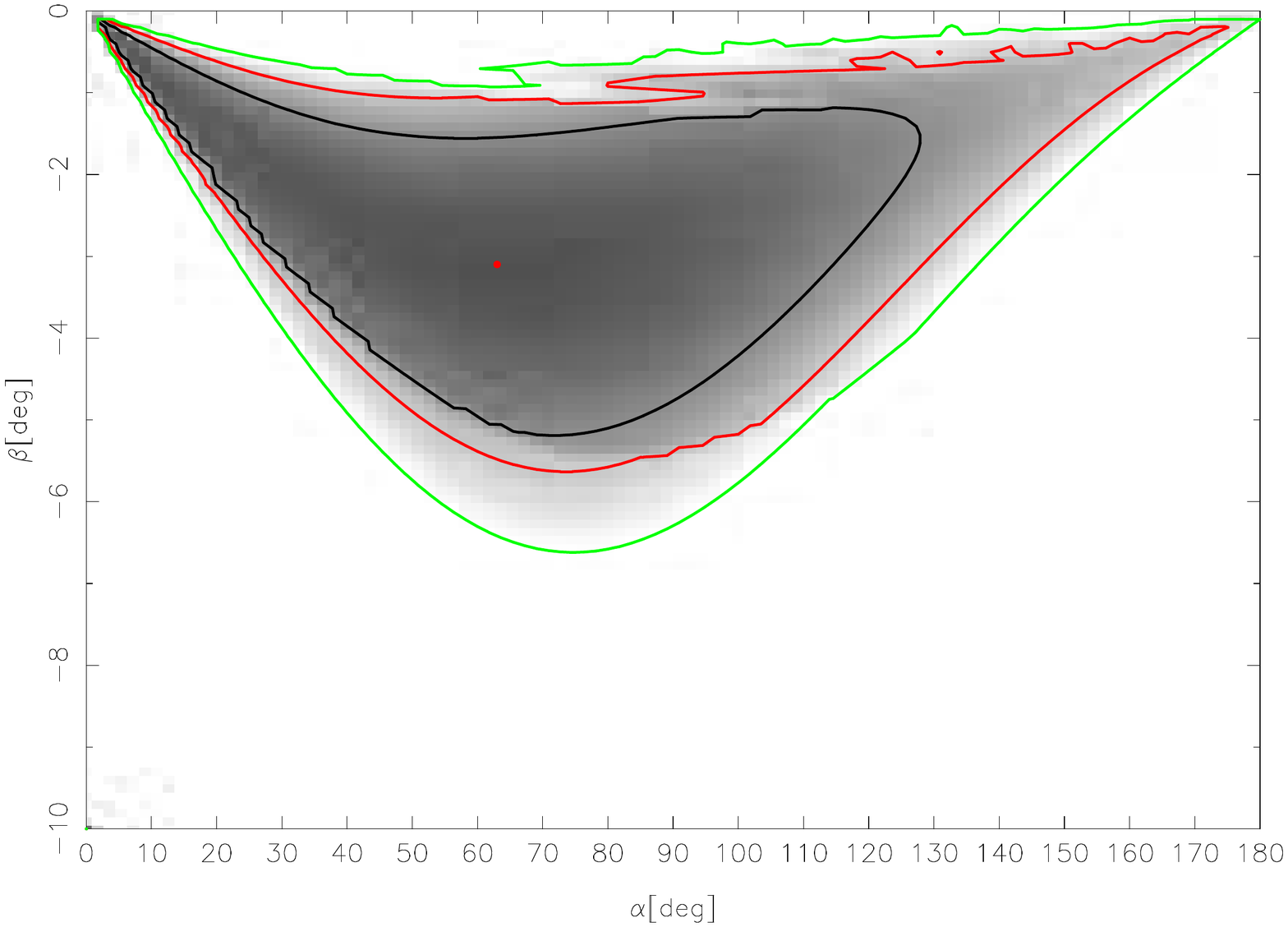}}}&
{\mbox{\includegraphics[width=9cm,height=6cm,angle=0.]{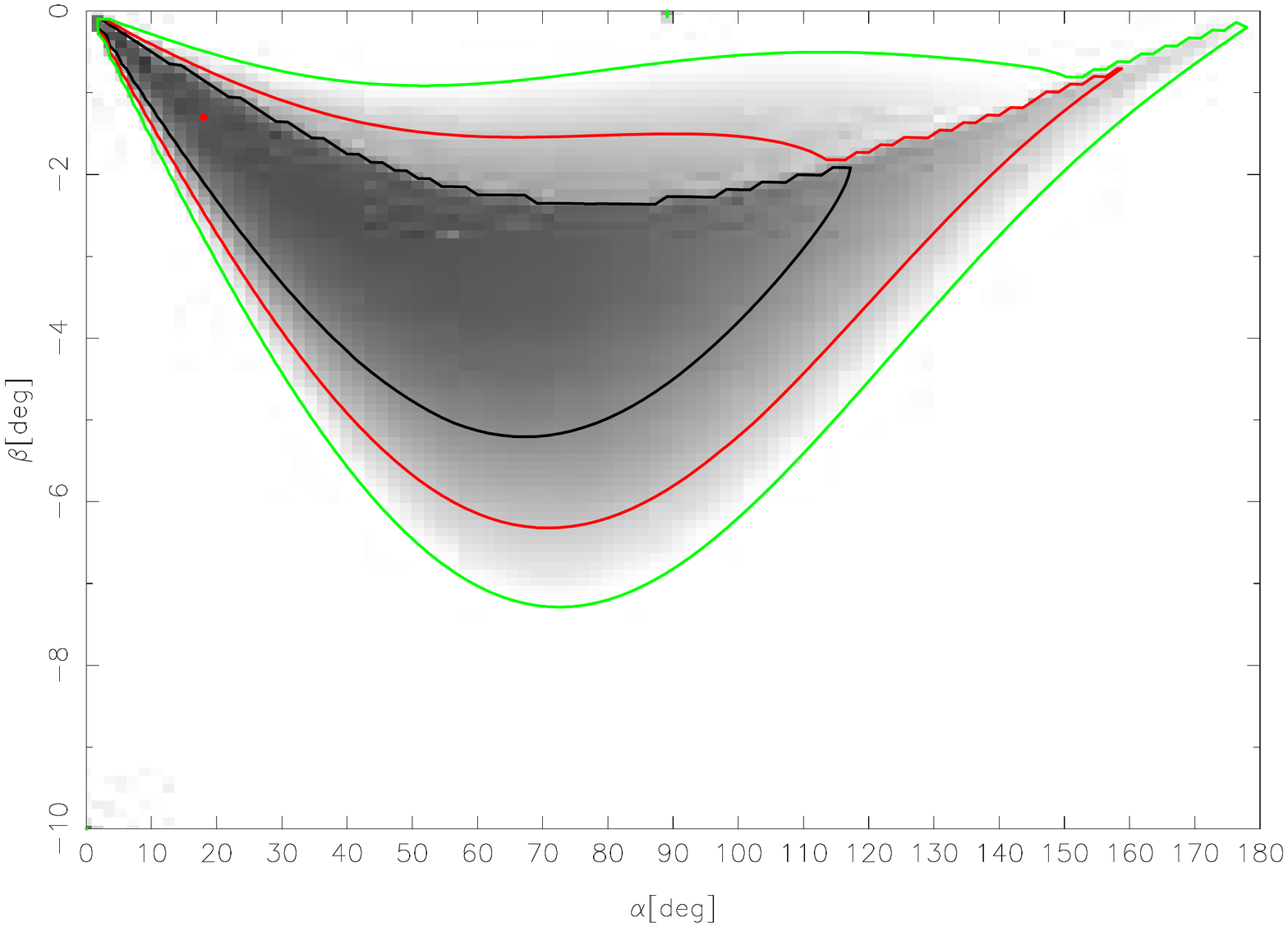}}}\\
&
\\
\end{tabular}
\caption{Top panel (upper window) shows the average profile with total
intensity (Stokes I; solid black lines), total linear polarization (dashed red
line) and circular polarization (Stokes V; dotted blue line). Top panel (lower
window) also shows the single pulse PPA distribution (colour scale) along with
the average PPA (red error bars).
The RVM fits to the average PPA (dashed pink
line) is also shown in this plot. Bottom panel show
the $\chi^2$ contours for the parameters $\alpha$ and $\beta$ obtained from RVM
fits.}
\label{a70}
\end{center}
\end{figure*}


\begin{figure*}
\begin{center}
\begin{tabular}{cc}
{\mbox{\includegraphics[width=9cm,height=6cm,angle=0.]{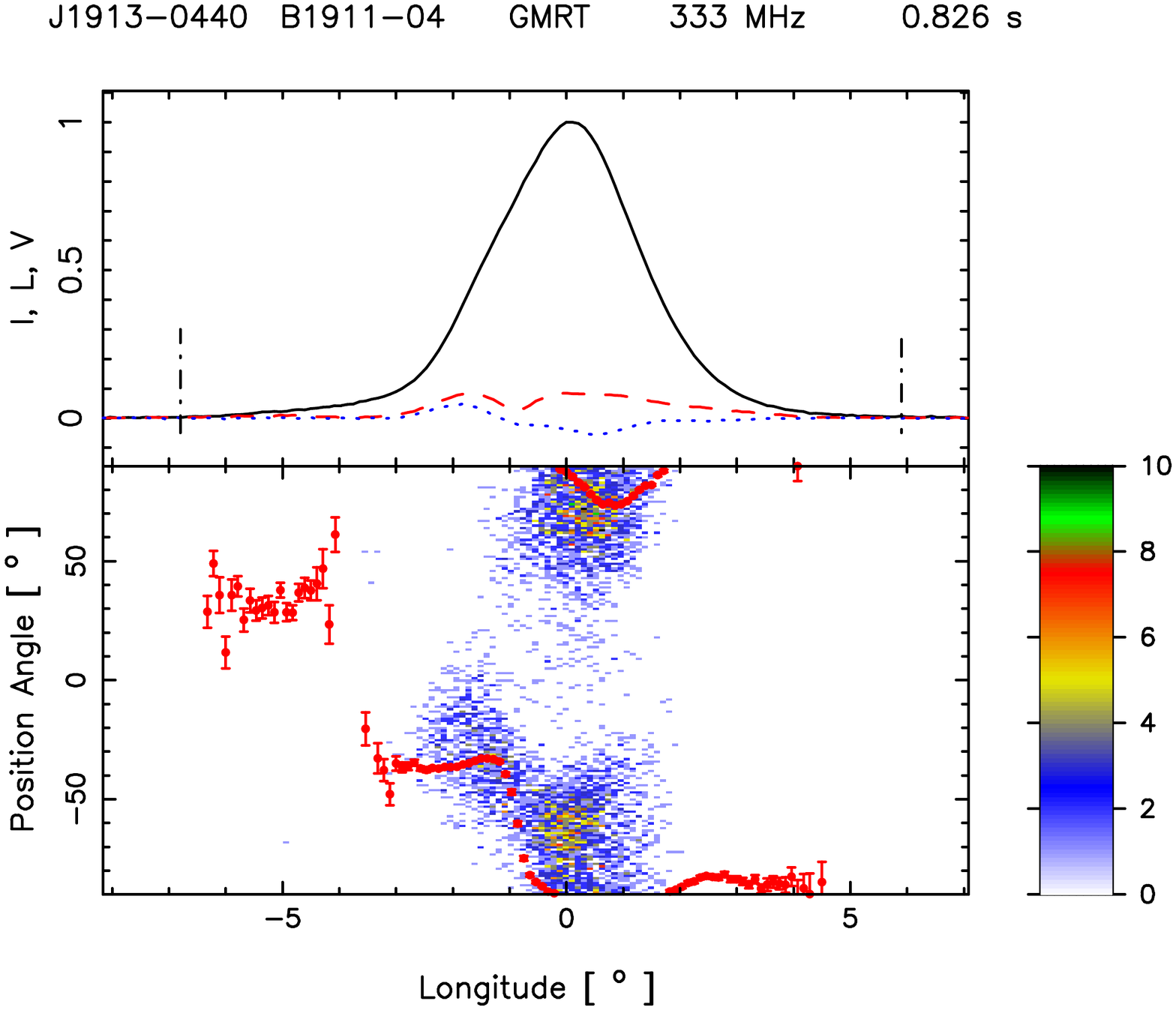}}}&
{\mbox{\includegraphics[width=9cm,height=6cm,angle=0.]{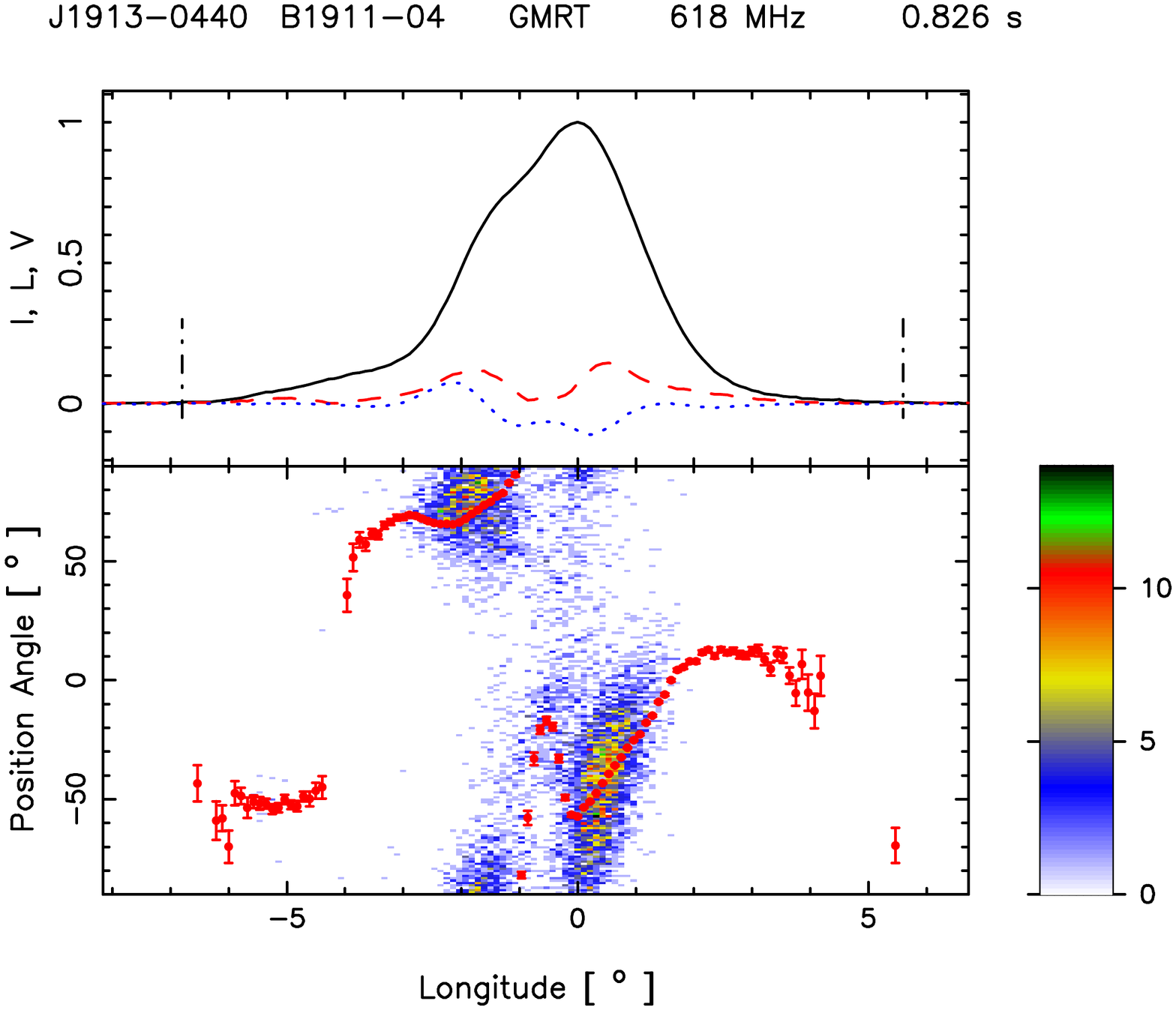}}}\\
&
\\
{\mbox{\includegraphics[width=9cm,height=6cm,angle=0.]{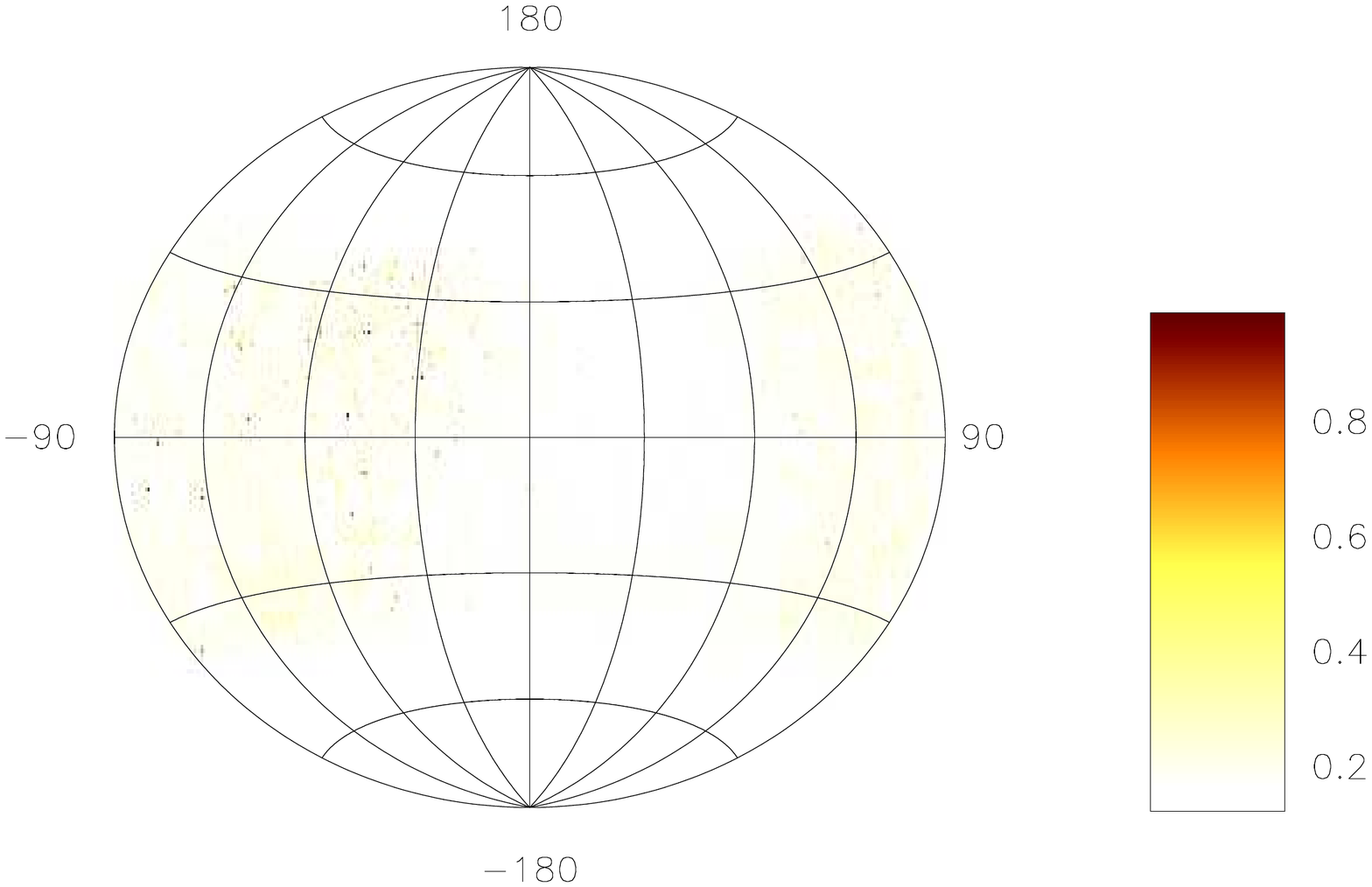}}}&
{\mbox{\includegraphics[width=9cm,height=6cm,angle=0.]{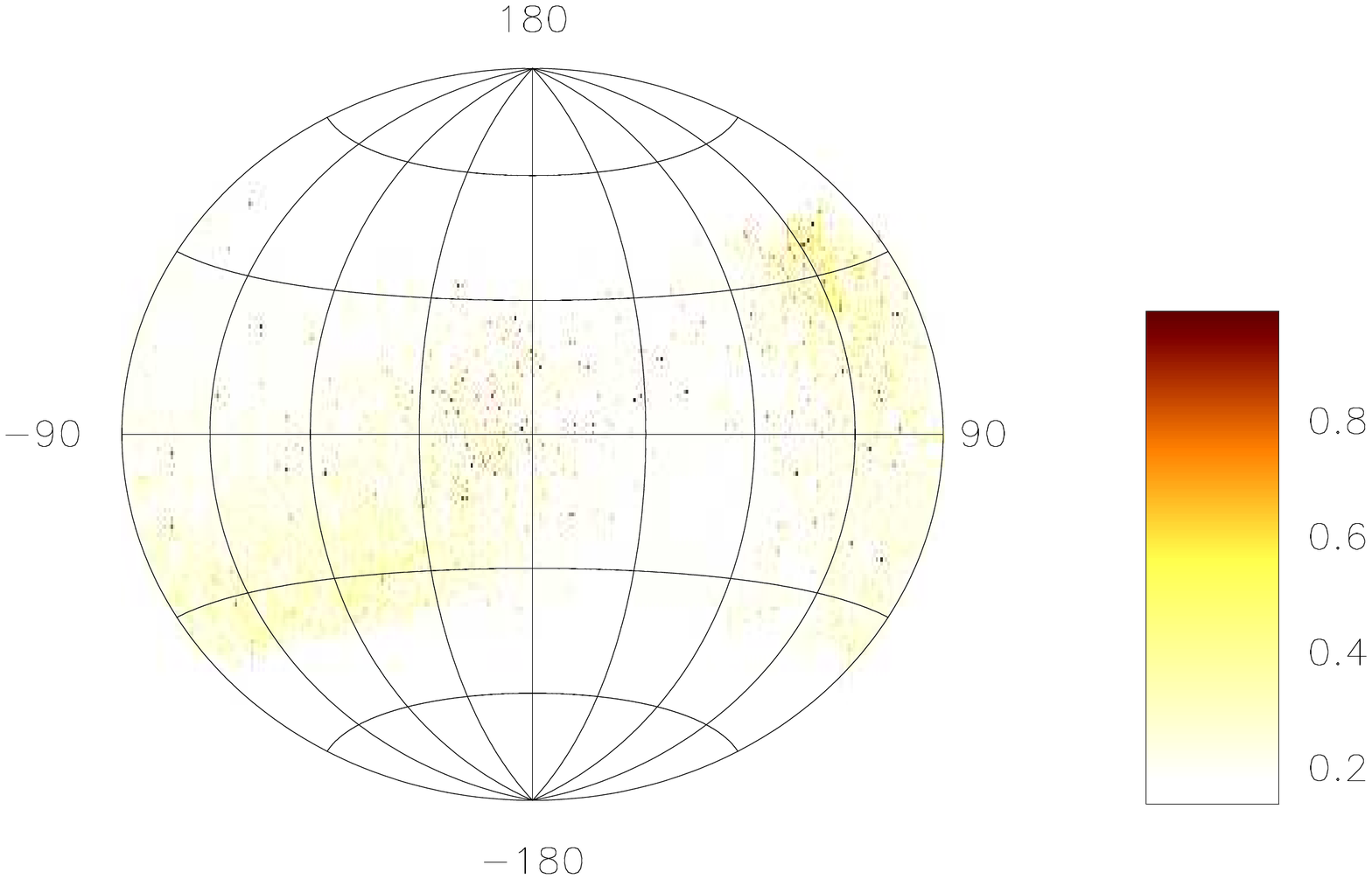}}}\\
\end{tabular}
\caption{Top panel (upper window) shows the average profile with total
intensity (Stokes I; solid black lines), total linear polarization (dashed red
line) and circular polarization (Stokes V; dotted blue line). Top panel (lower
window) also shows the single pulse PPA distribution (colour scale) along with
the average PPA (red error bars).
Bottom panel shows the Hammer-Aitoff projection of the polarized time
samples with the colour scheme representing the fractional polarization level.}
\label{a71}
\end{center}
\end{figure*}


\begin{figure*}
\begin{center}
\begin{tabular}{cc}
{\mbox{\includegraphics[width=9cm,height=6cm,angle=0.]{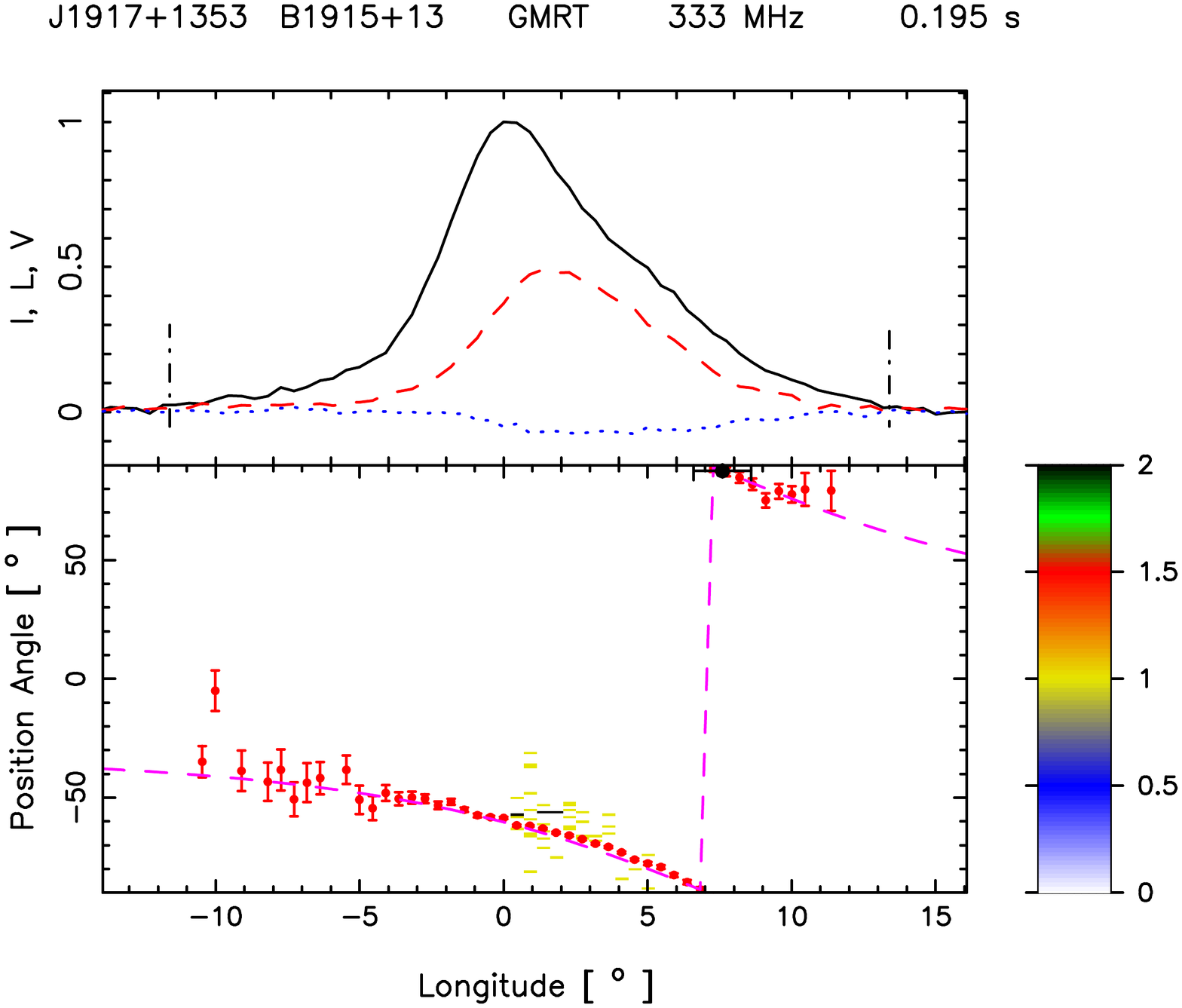}}}&
{\mbox{\includegraphics[width=9cm,height=6cm,angle=0.]{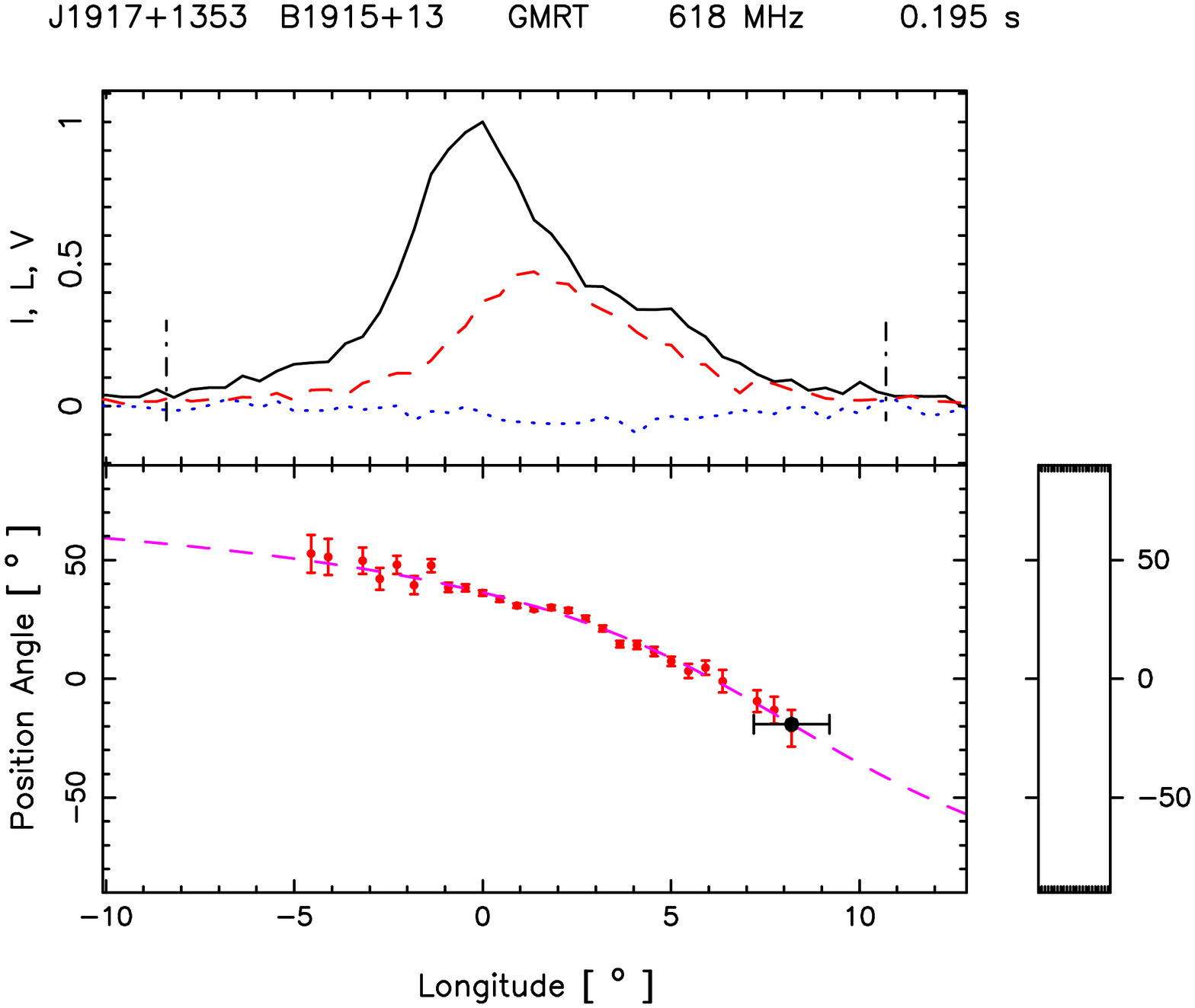}}}\\
{\mbox{\includegraphics[width=9cm,height=6cm,angle=0.]{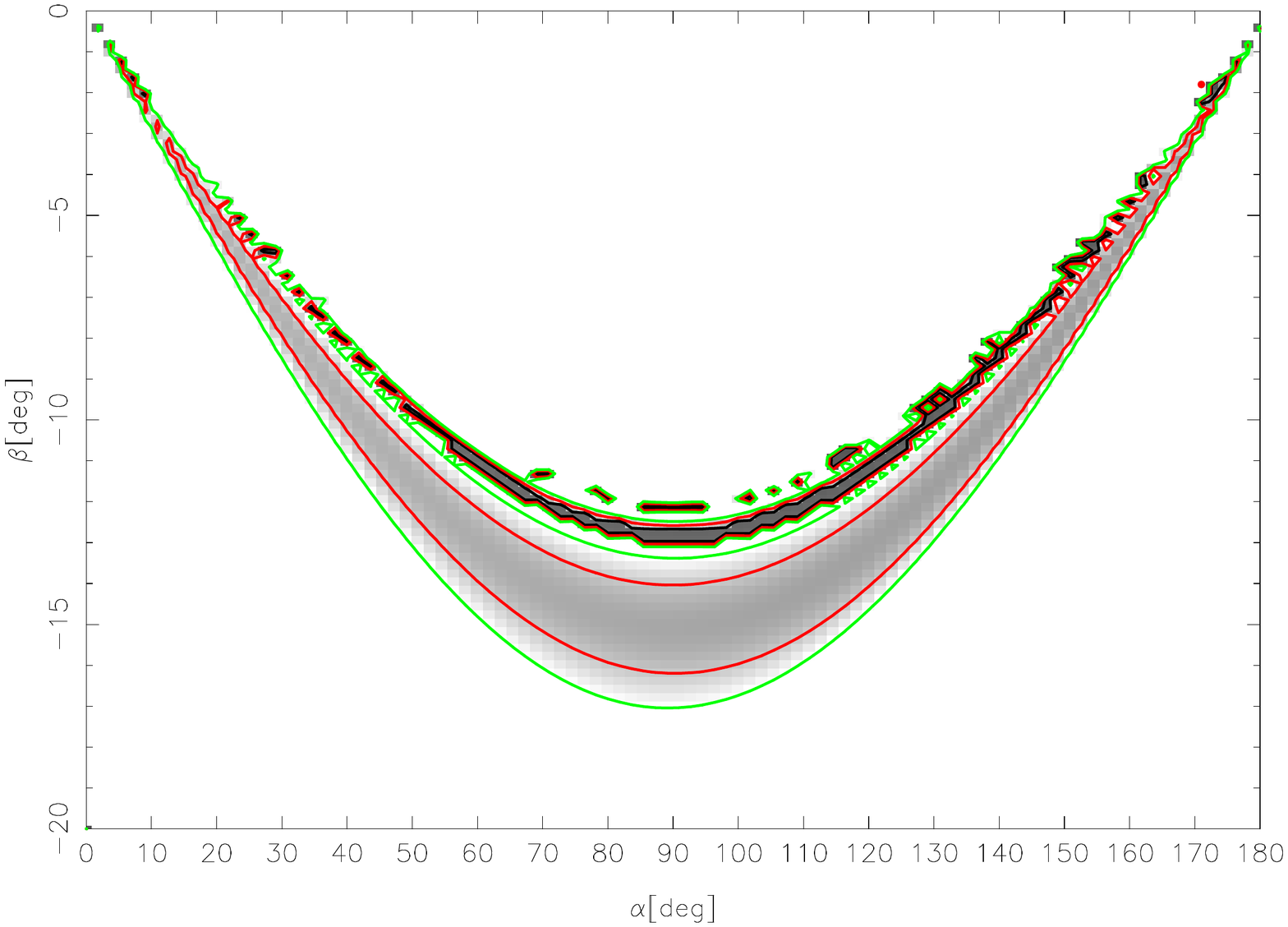}}}&
{\mbox{\includegraphics[width=9cm,height=6cm,angle=0.]{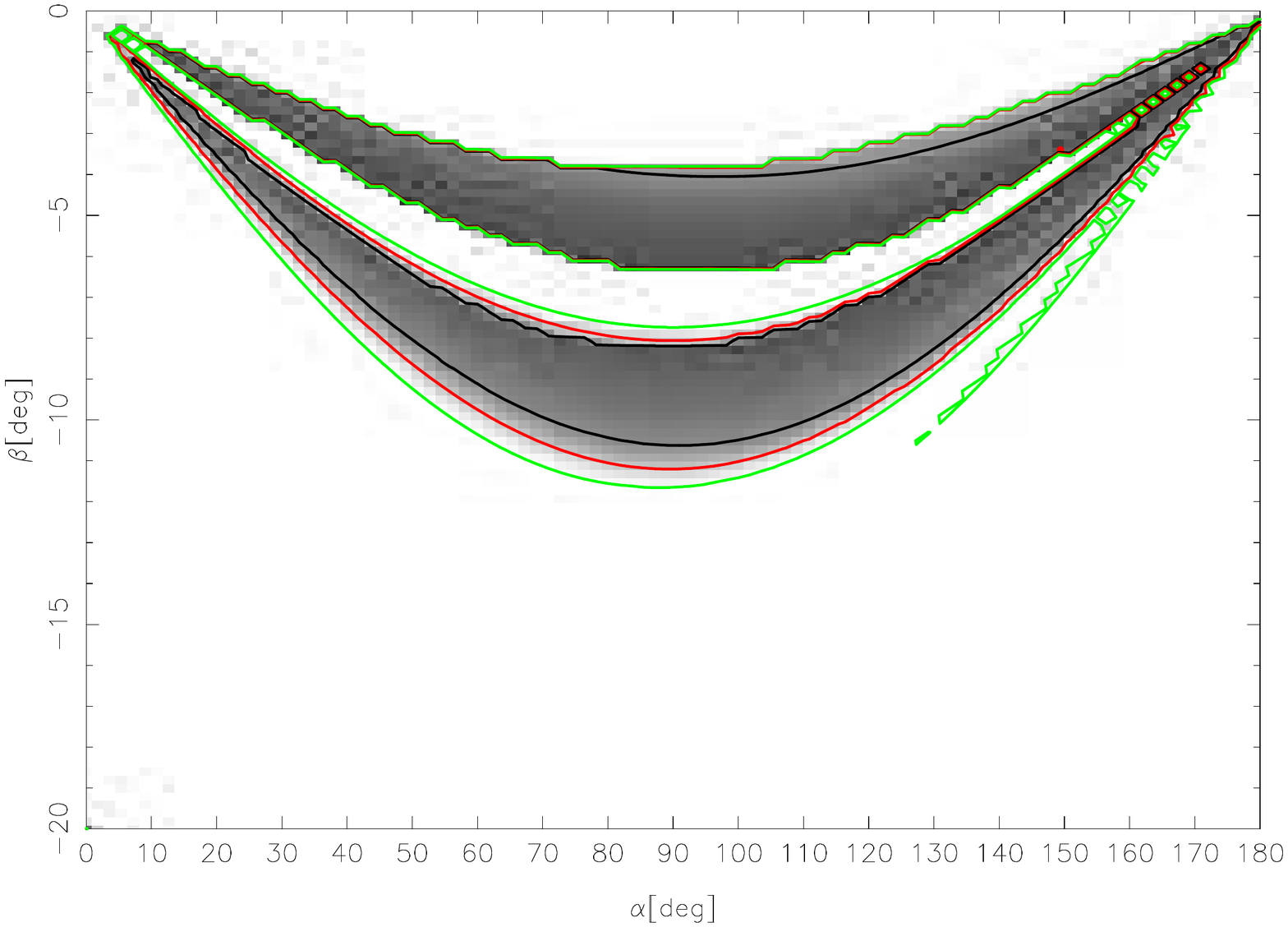}}}\\
&
\\
\end{tabular}
\caption{Top panel (upper window) shows the average profile with total
intensity (Stokes I; solid black lines), total linear polarization (dashed red
line) and circular polarization (Stokes V; dotted blue line). Top panel (lower
window) also shows the single pulse PPA distribution (colour scale) along with
the average PPA (red error bars).
The RVM fits to the average PPA (dashed pink
line) is also shown in this plot. Bottom panel show
the $\chi^2$ contours for the parameters $\alpha$ and $\beta$ obtained from RVM
fits.}
\label{a72}
\end{center}
\end{figure*}


\begin{figure*}
\begin{center}
\begin{tabular}{cc}
{\mbox{\includegraphics[width=9cm,height=6cm,angle=0.]{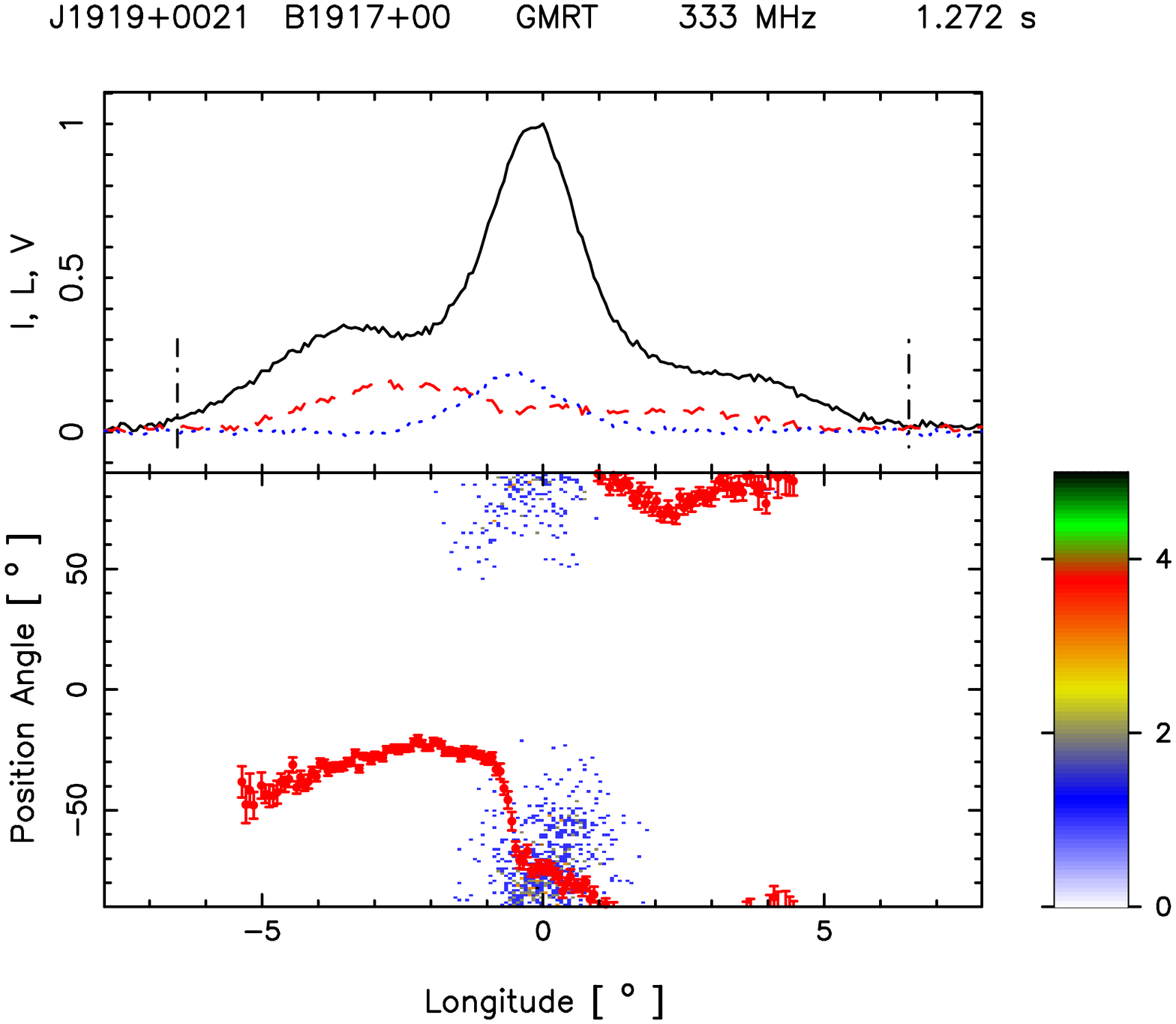}}}&
\\
&
\\
{\mbox{\includegraphics[width=9cm,height=6cm,angle=0.]{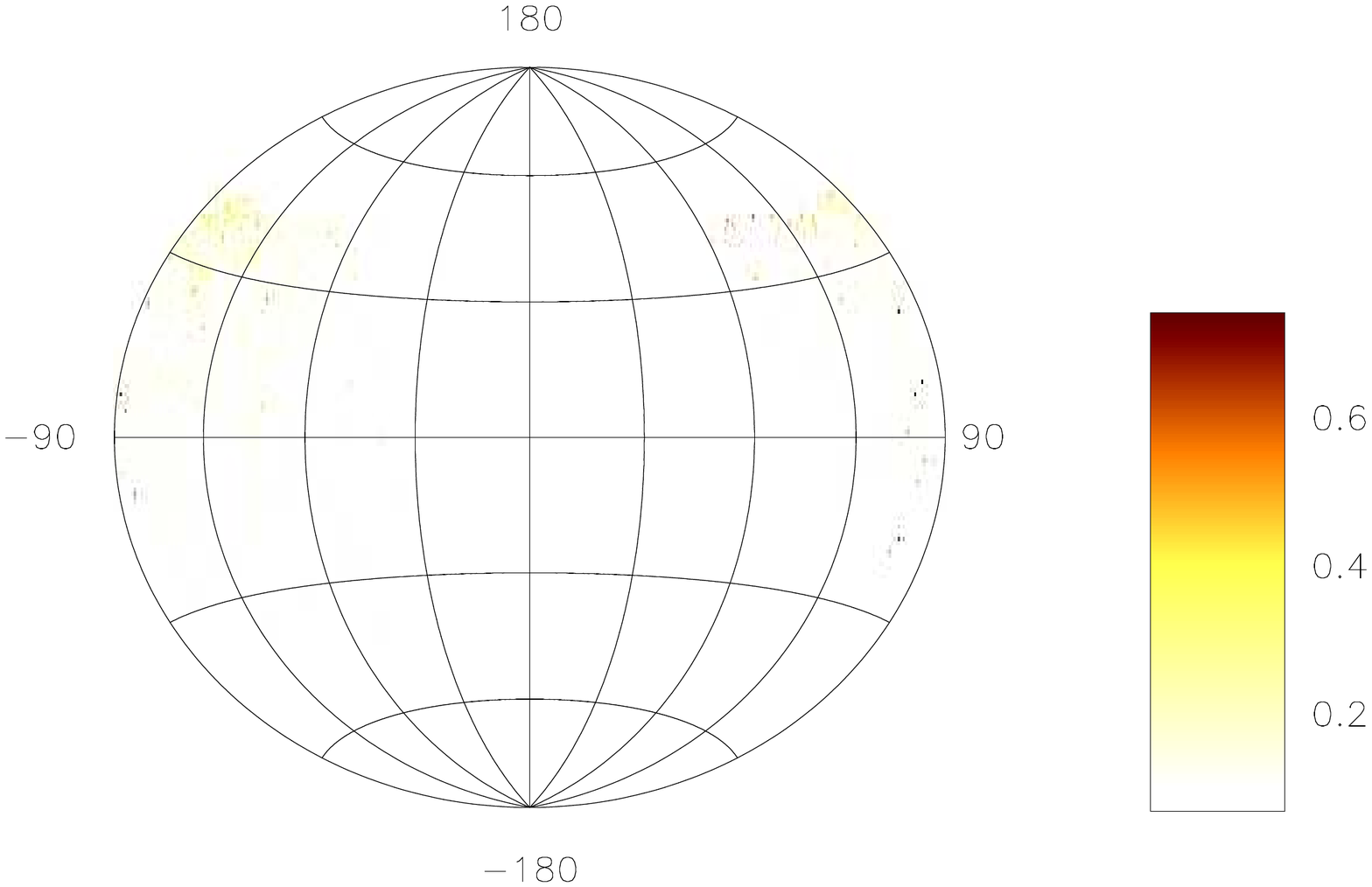}}}&
\\
\end{tabular}
\caption{Top panel only for 333 MHz (upper window) shows the average profile with total
intensity (Stokes I; solid black lines), total linear polarization (dashed red
line) and circular polarization (Stokes V; dotted blue line). Top panel (lower
window) also shows the single pulse PPA distribution (colour scale) along with
the average PPA (red error bars).
Bottom panel only for 333 MHz shows the Hammer-Aitoff projection of the polarized time
samples with the colour scheme representing the fractional polarization level.}
\label{a73}
\end{center}
\end{figure*}


\begin{figure*}
\begin{center}
\begin{tabular}{cc}
&
{\mbox{\includegraphics[width=9cm,height=6cm,angle=0.]{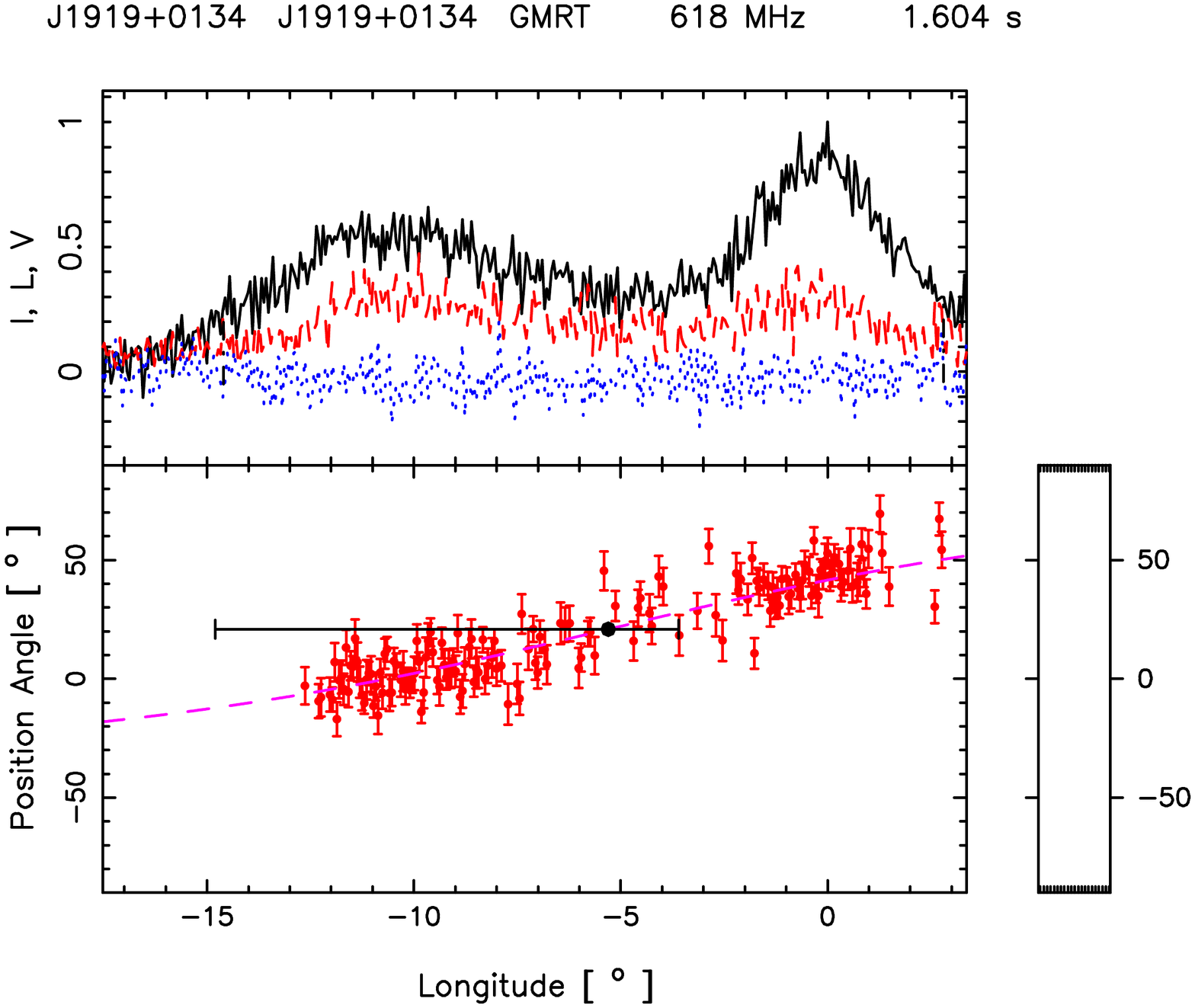}}}\\
&
{\mbox{\includegraphics[width=9cm,height=6cm,angle=0.]{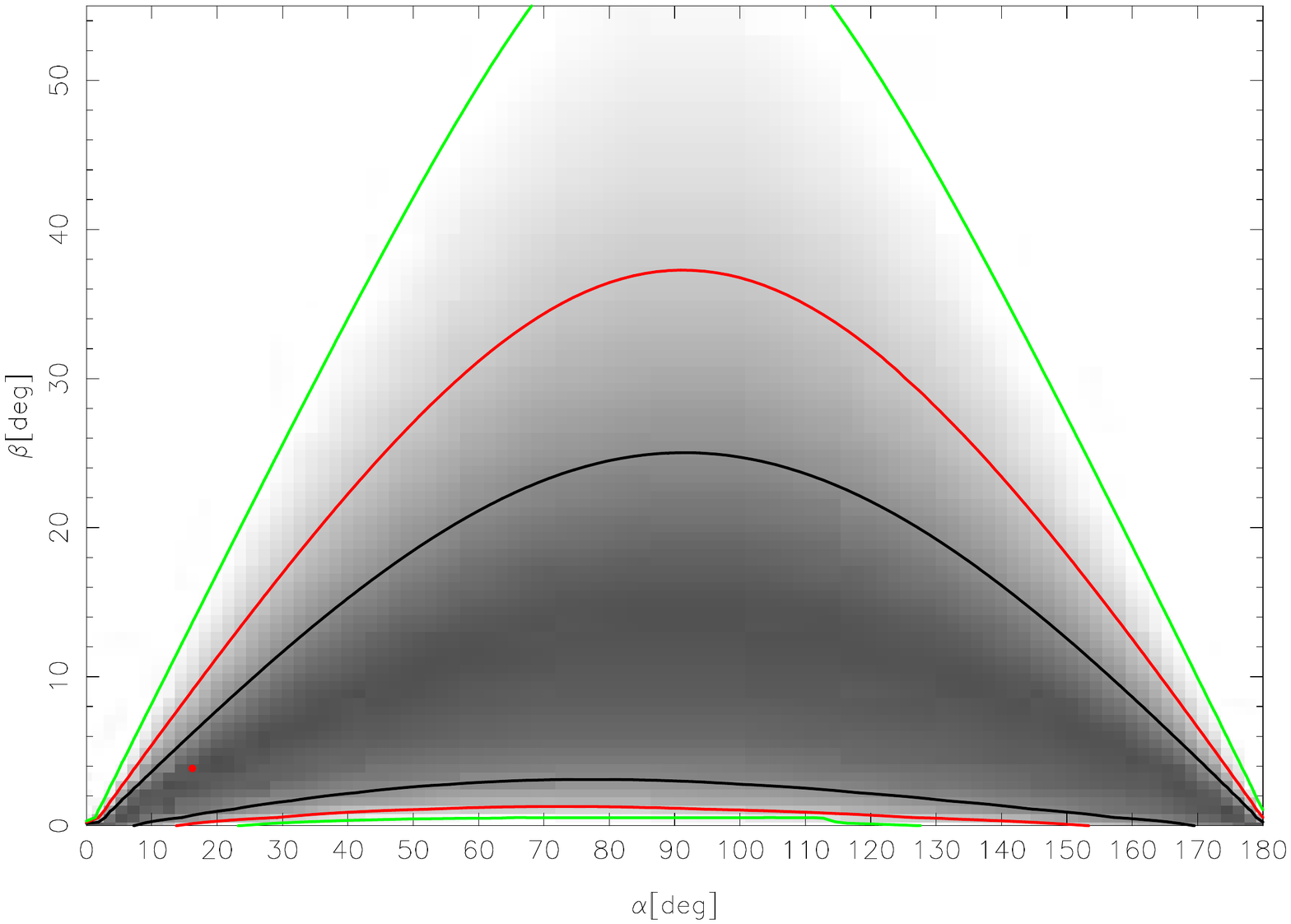}}}\\
&
\\
\end{tabular}
\caption{Top panel only for 618 MHz (upper window) shows the average profile with total
intensity (Stokes I; solid black lines), total linear polarization (dashed red
line) and circular polarization (Stokes V; dotted blue line). Top panel (lower
window) also shows the single pulse PPA distribution (colour scale) along with
the average PPA (red error bars).
The RVM fits to the average PPA (dashed pink
line) is also shown in this plot. Bottom panel only at 618 MHz  show
the $\chi^2$ contours for the parameters $\alpha$ and $\beta$ obtained from RVM
fits.}
\label{a74}
\end{center}
\end{figure*}


\begin{figure*}
\begin{center}
\begin{tabular}{cc}
&
{\mbox{\includegraphics[width=9cm,height=6cm,angle=0.]{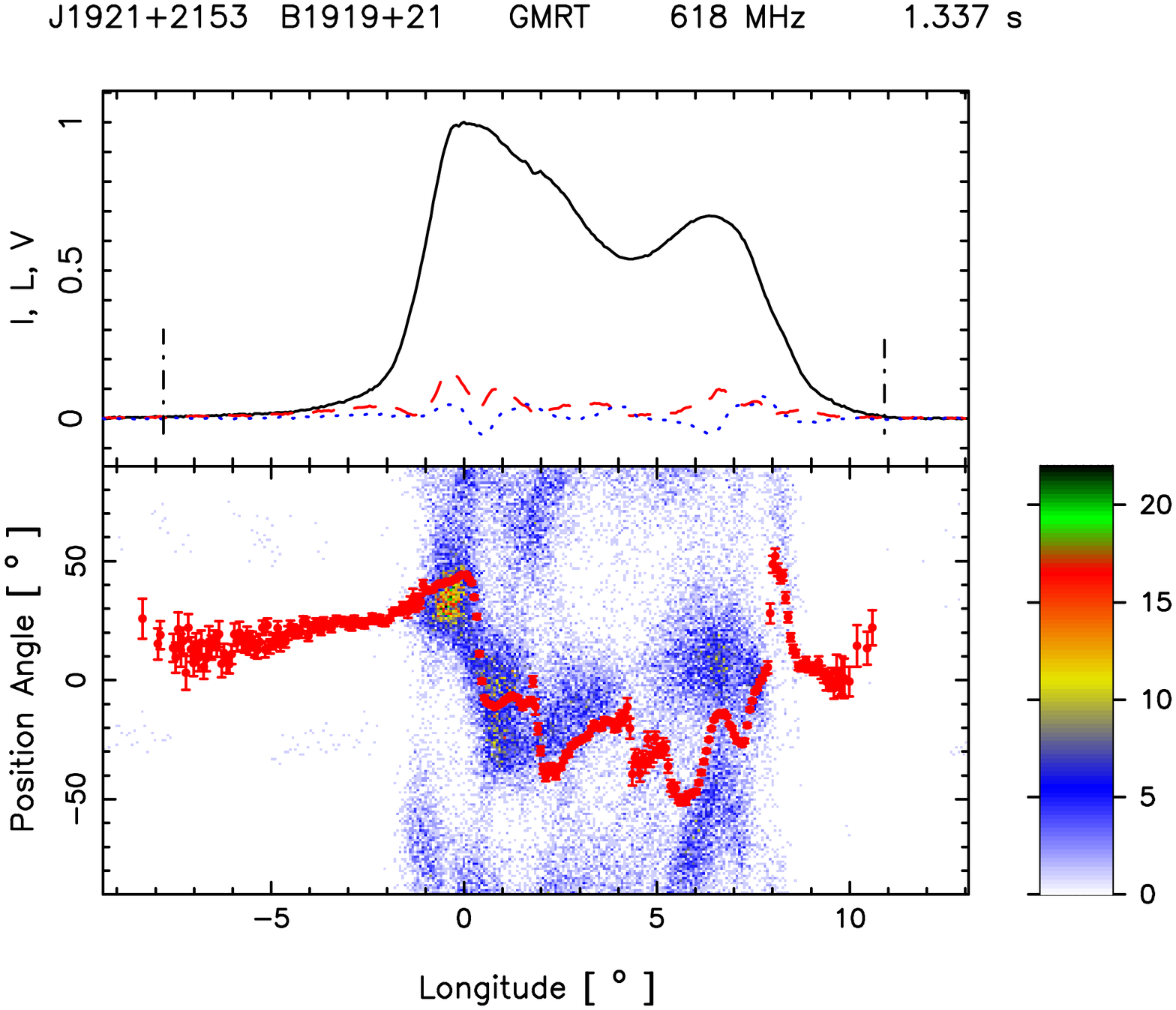}}}\\
&
\\
&
{\mbox{\includegraphics[width=9cm,height=6cm,angle=0.]{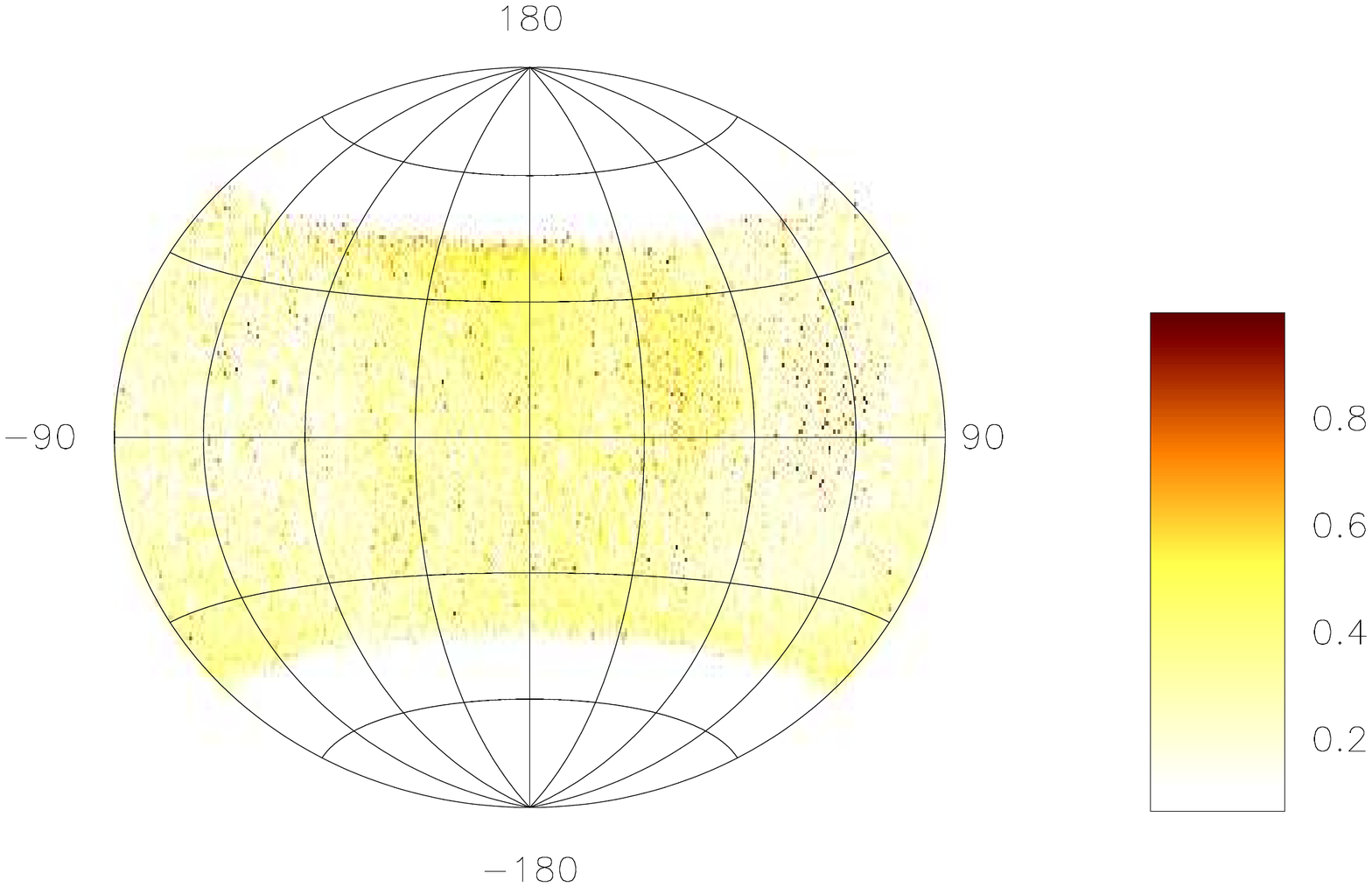}}}\\
\end{tabular}
\caption{Top panel only at 618 MHz (upper window) shows the average profile with total
intensity (Stokes I; solid black lines), total linear polarization (dashed red
line) and circular polarization (Stokes V; dotted blue line). Top panel (lower
window) also shows the single pulse PPA distribution (colour scale) along with
the average PPA (red error bars).
Bottom panel only at 618 MHz shows the Hammer-Aitoff projection of the polarized time
samples with the colour scheme representing the fractional polarization level.}
\label{a75}
\end{center}
\end{figure*}


\begin{figure*}
\begin{center}
\begin{tabular}{cc}
{\mbox{\includegraphics[width=9cm,height=6cm,angle=0.]{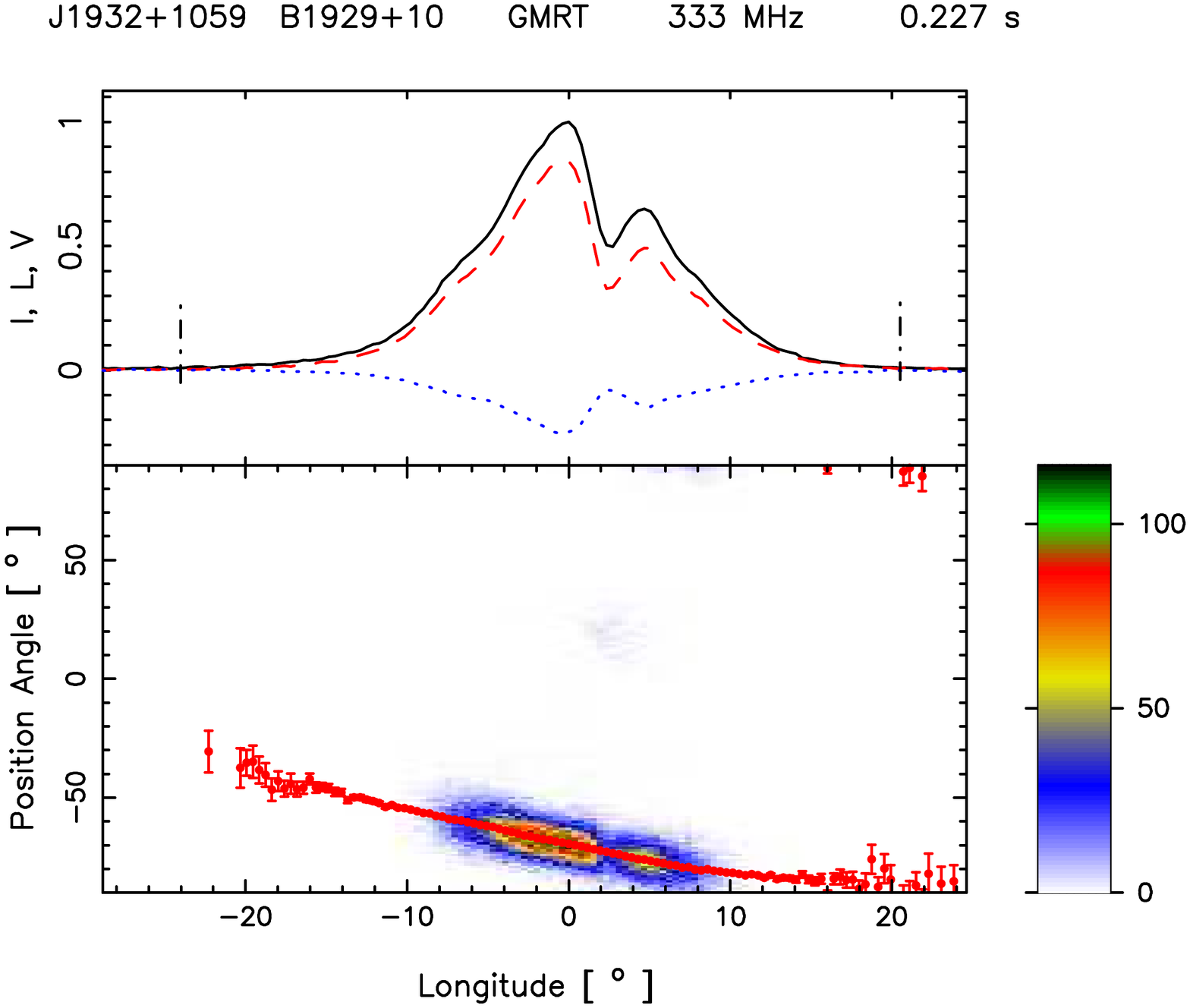}}}&
{\mbox{\includegraphics[width=9cm,height=6cm,angle=0.]{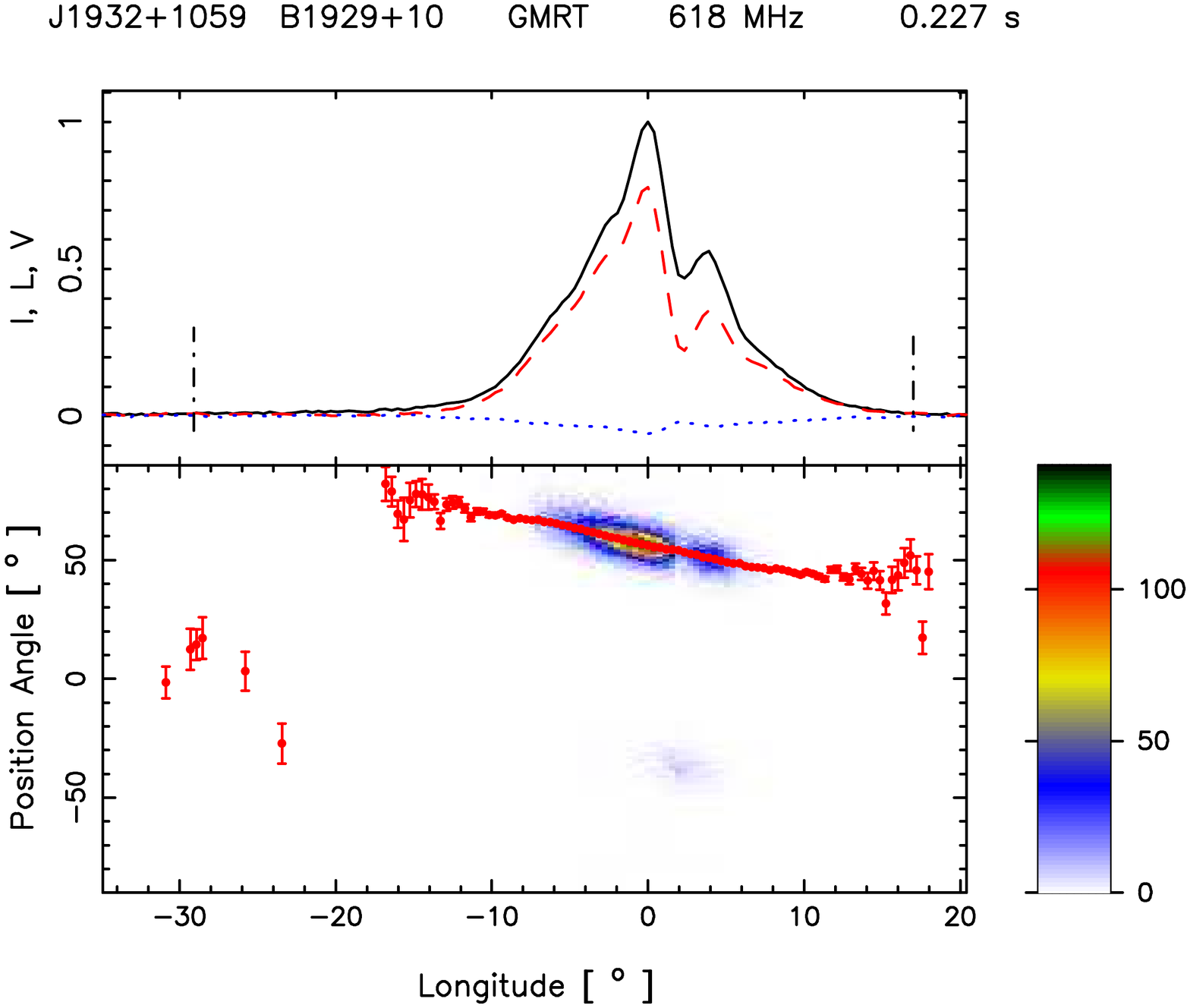}}}\\
&
\\
{\mbox{\includegraphics[width=9cm,height=6cm,angle=0.]{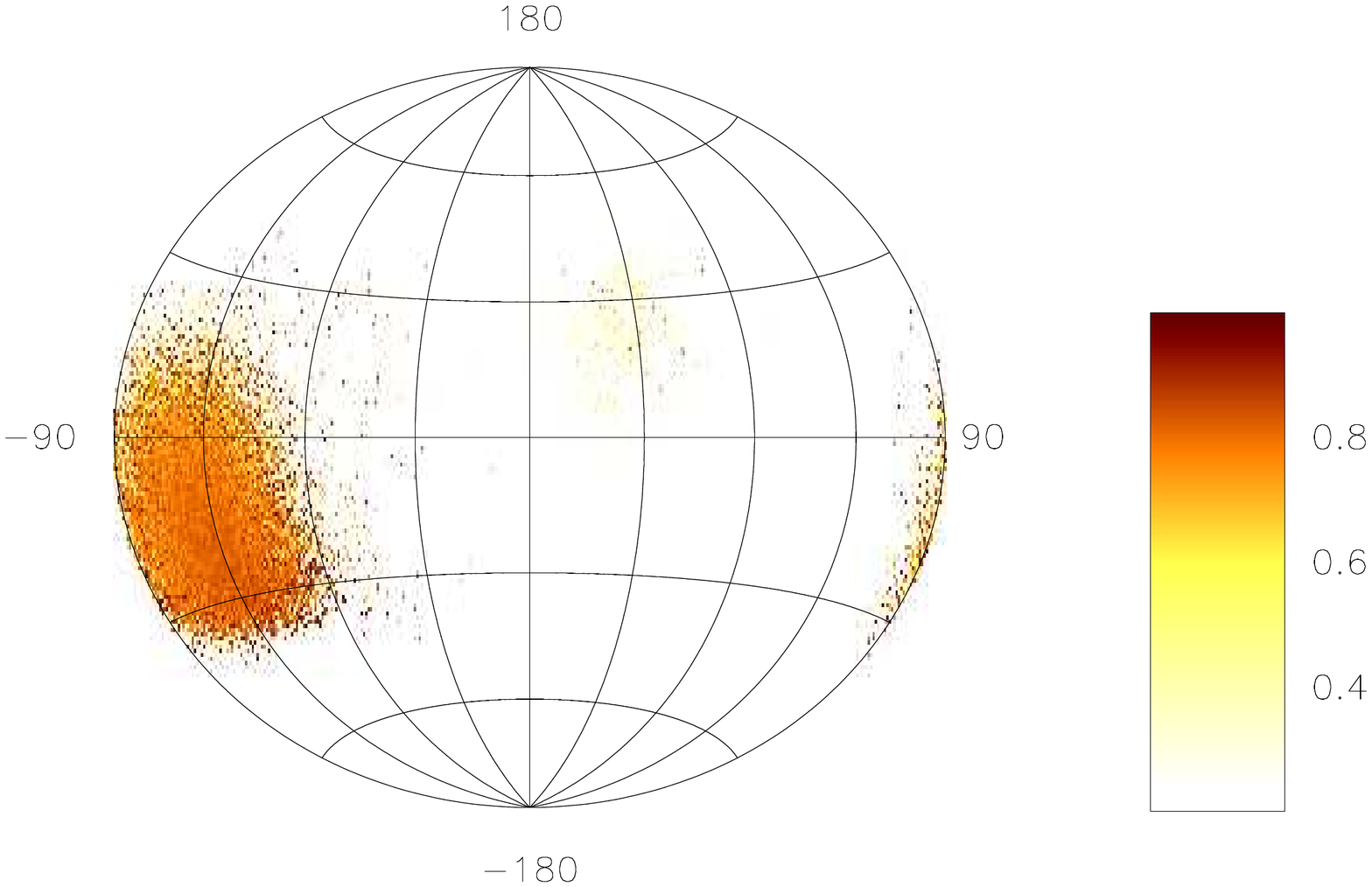}}}&
{\mbox{\includegraphics[width=9cm,height=6cm,angle=0.]{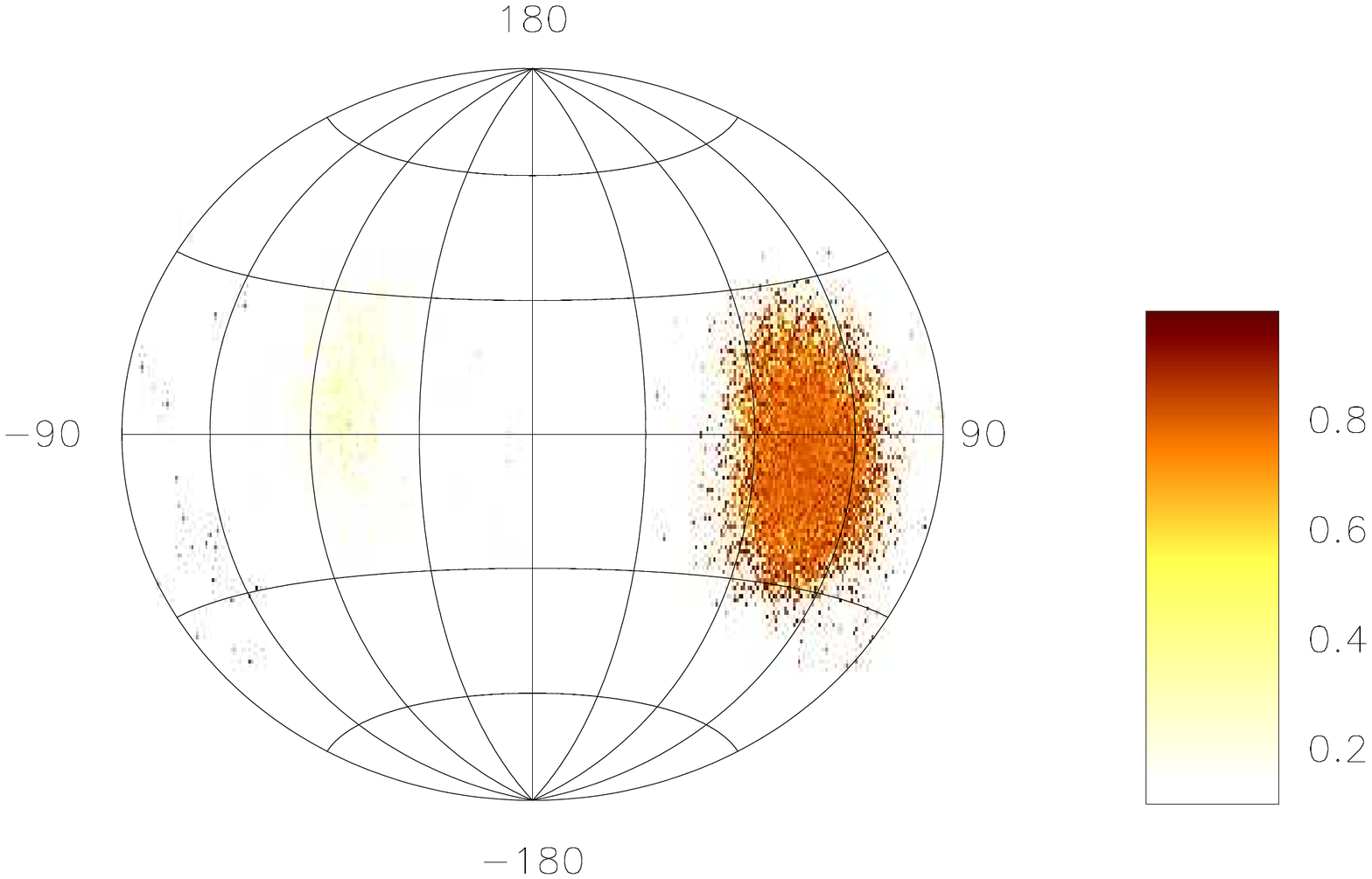}}}\\
\end{tabular}
\caption{Top panel (upper window) shows the average profile with total
intensity (Stokes I; solid black lines), total linear polarization (dashed red
line) and circular polarization (Stokes V; dotted blue line). Top panel (lower
window) also shows the single pulse PPA distribution (colour scale) along with
the average PPA (red error bars).
Bottom panel shows the Hammer-Aitoff projection of the polarized time
samples with the colour scheme representing the fractional polarization level.}
\label{a76}
\end{center}
\end{figure*}


\begin{figure*}
\begin{center}
\begin{tabular}{cc}
{\mbox{\includegraphics[width=9cm,height=6cm,angle=0.]{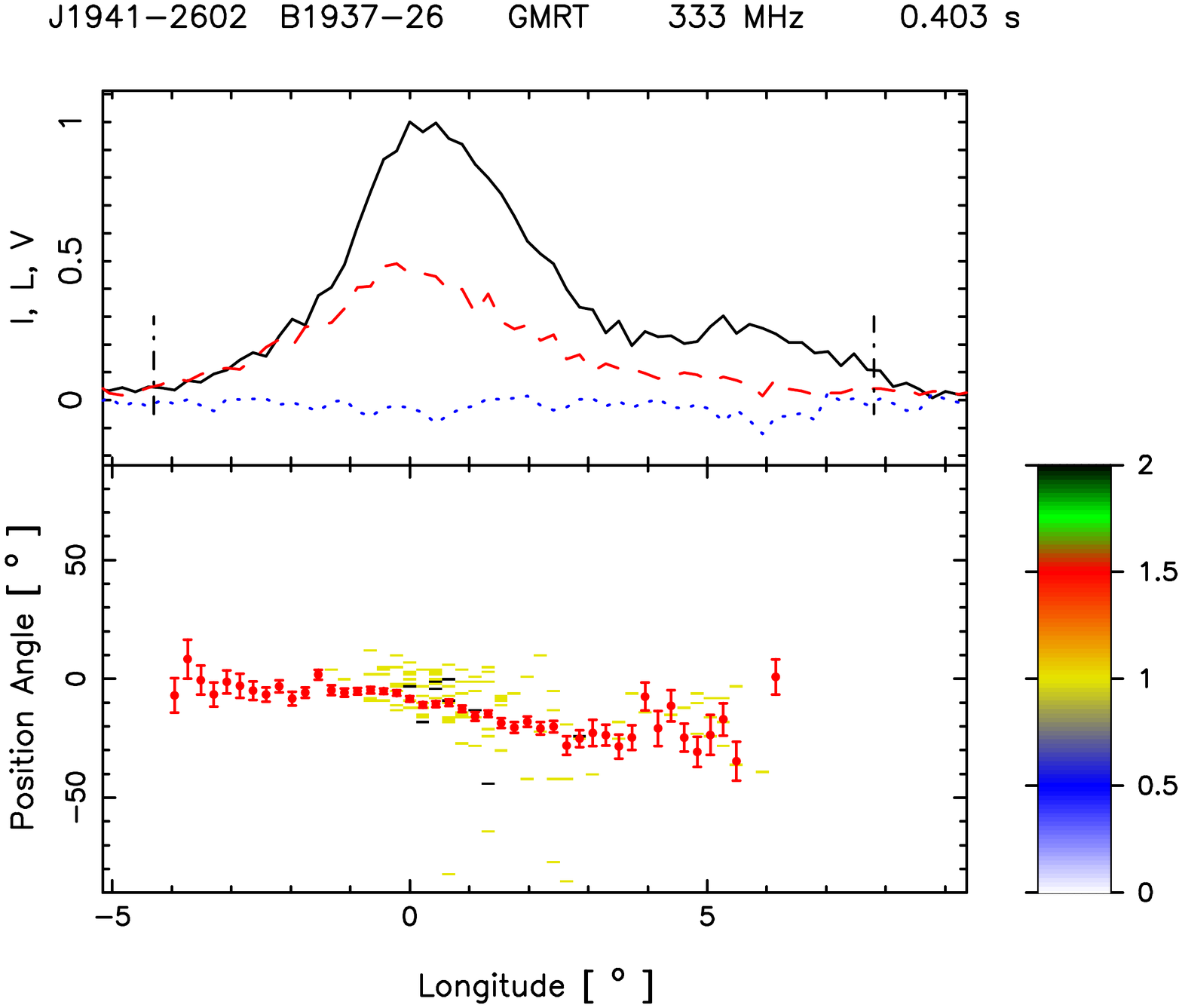}}}&
{\mbox{\includegraphics[width=9cm,height=6cm,angle=0.]{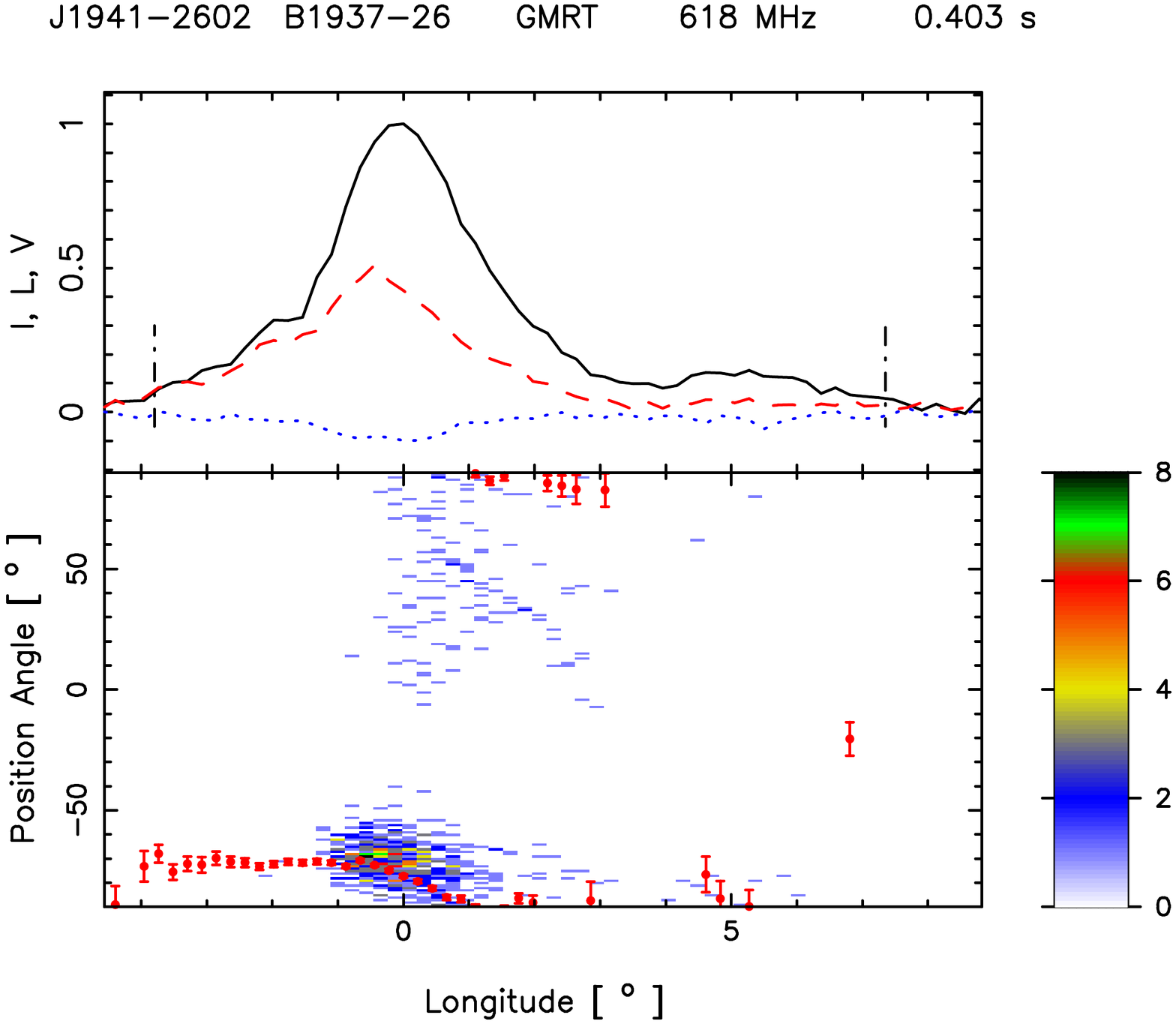}}}\\
&
\\
{\mbox{\includegraphics[width=9cm,height=6cm,angle=0.]{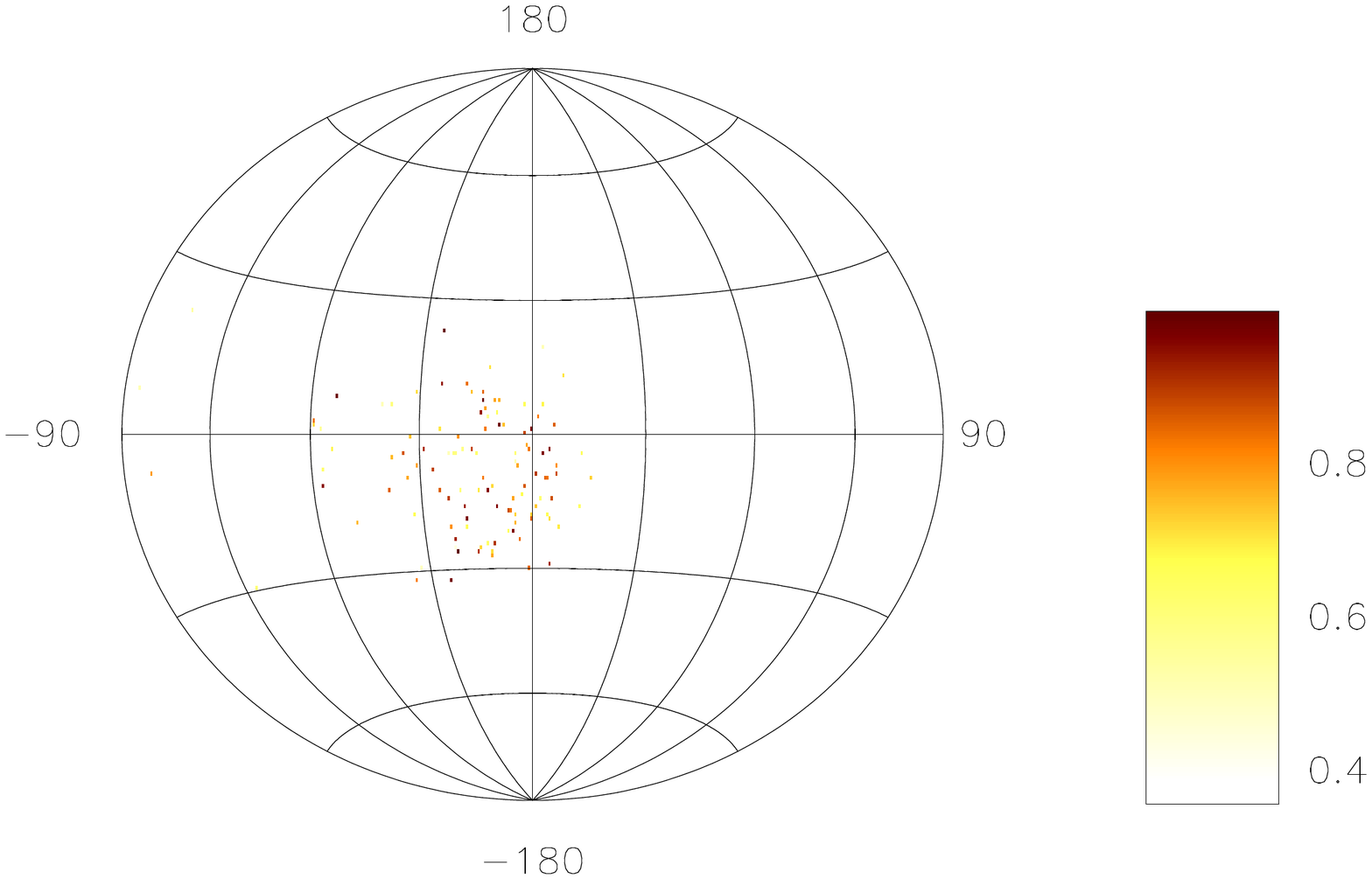}}}&
{\mbox{\includegraphics[width=9cm,height=6cm,angle=0.]{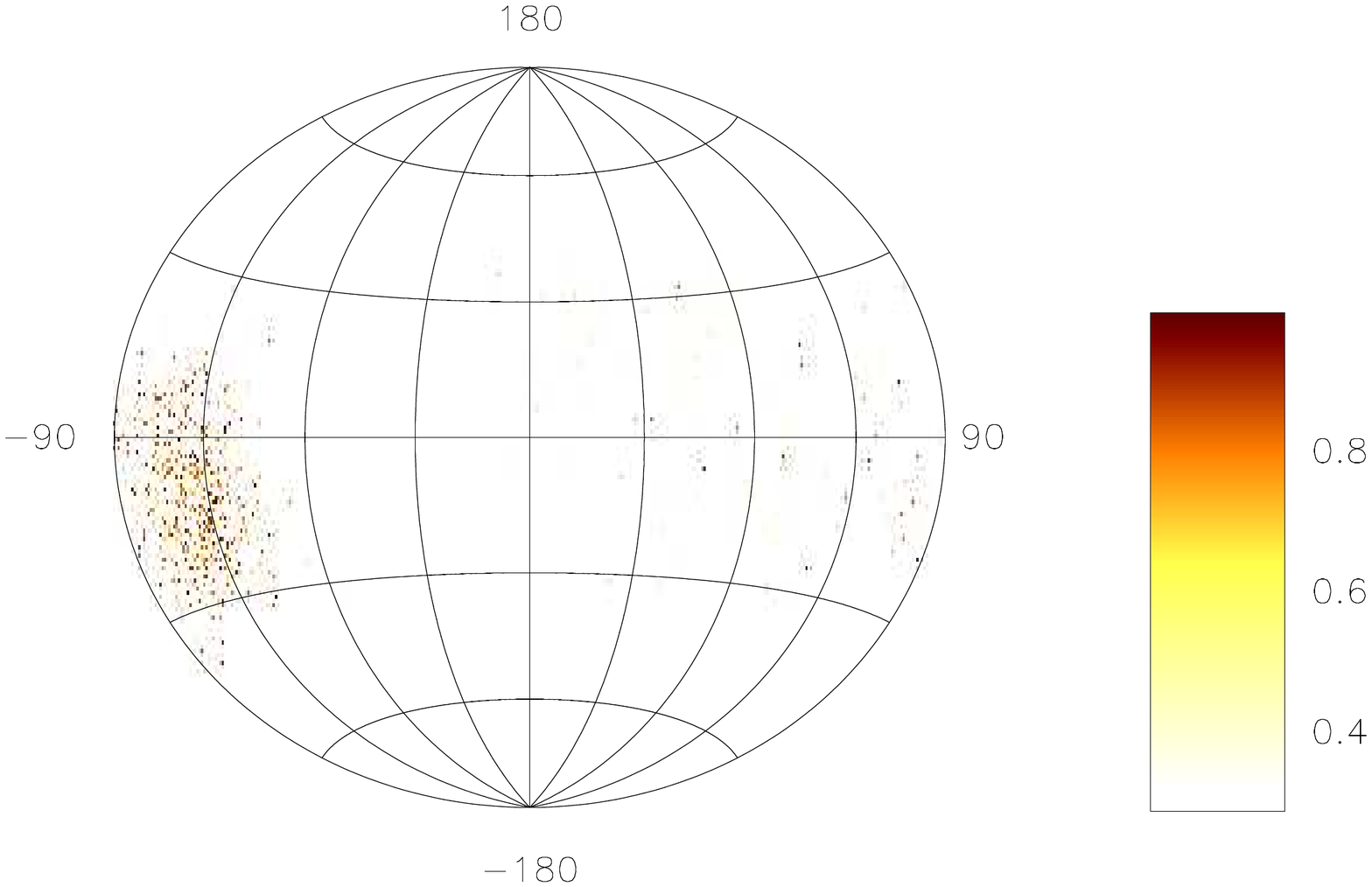}}}\\
\end{tabular}
\caption{Top panel (upper window) shows the average profile with total
intensity (Stokes I; solid black lines), total linear polarization (dashed red
line) and circular polarization (Stokes V; dotted blue line). Top panel (lower
window) also shows the single pulse PPA distribution (colour scale) along with
the average PPA (red error bars).
Bottom panel shows the Hammer-Aitoff projection of the polarized time
samples with the colour scheme representing the fractional polarization level.}
\label{a77}
\end{center}
\end{figure*}


\begin{figure*}
\begin{center}
\begin{tabular}{cc}
{\mbox{\includegraphics[width=9cm,height=6cm,angle=0.]{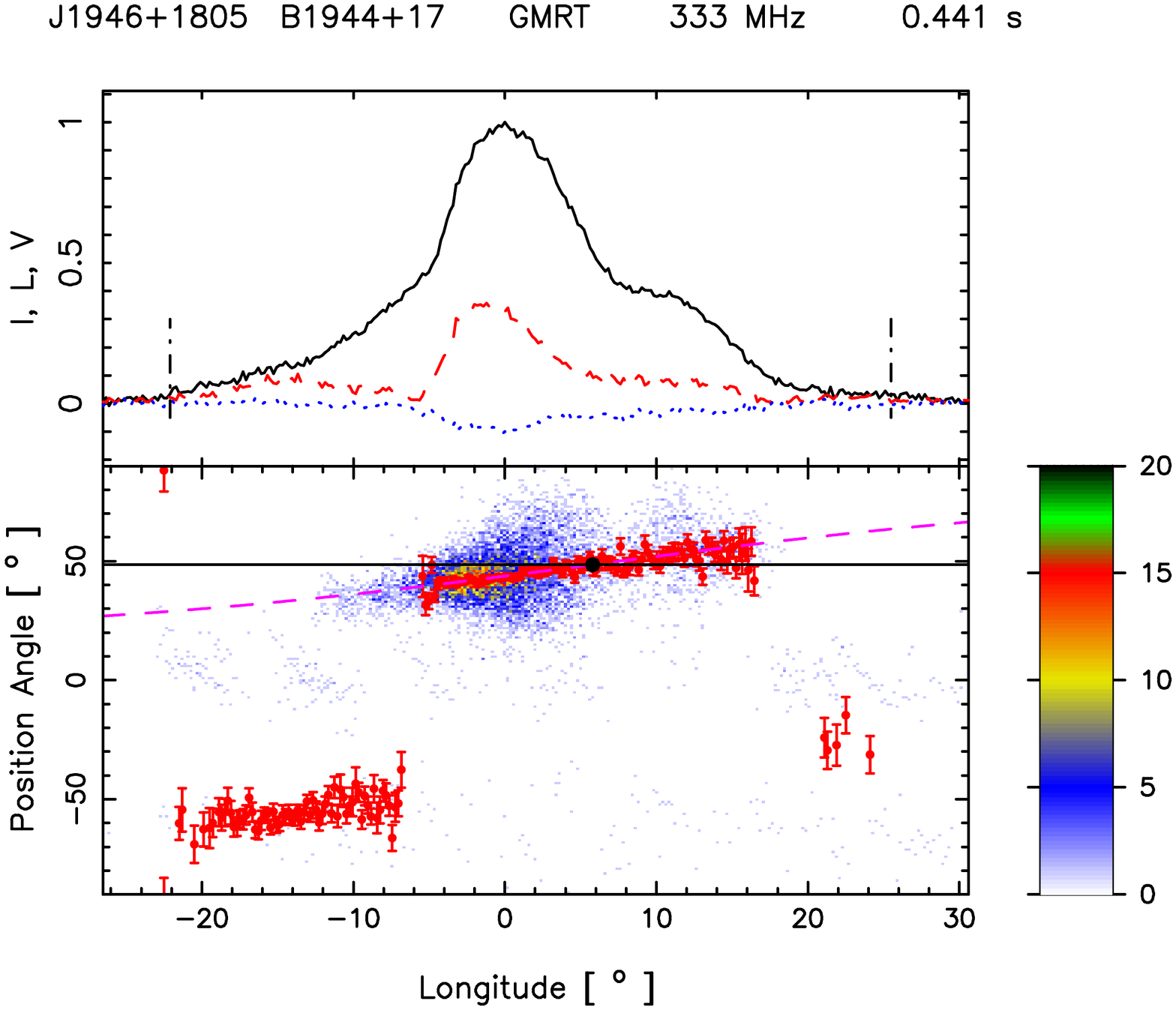}}}&
{\mbox{\includegraphics[width=9cm,height=6cm,angle=0.]{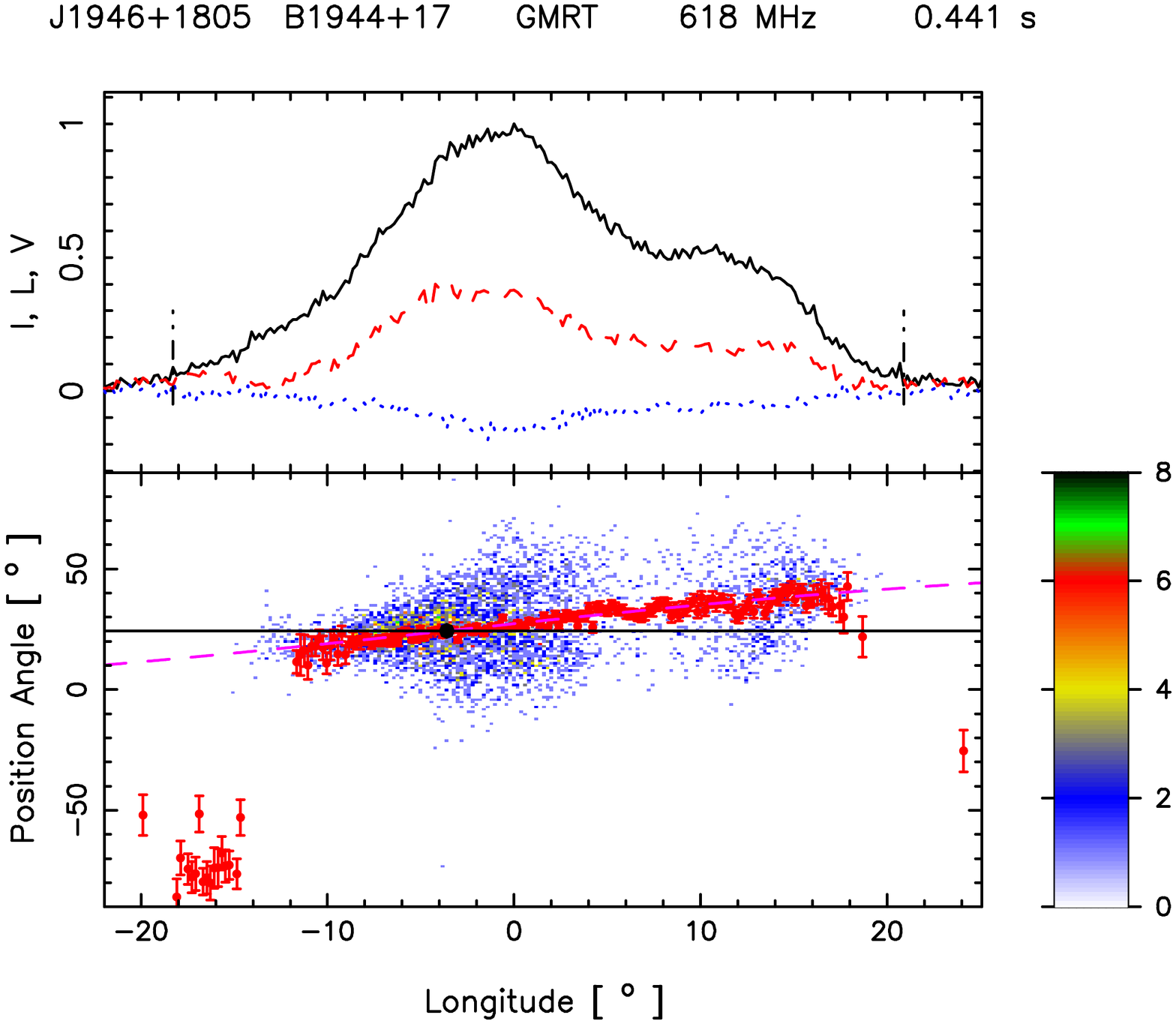}}}\\
{\mbox{\includegraphics[width=9cm,height=6cm,angle=0.]{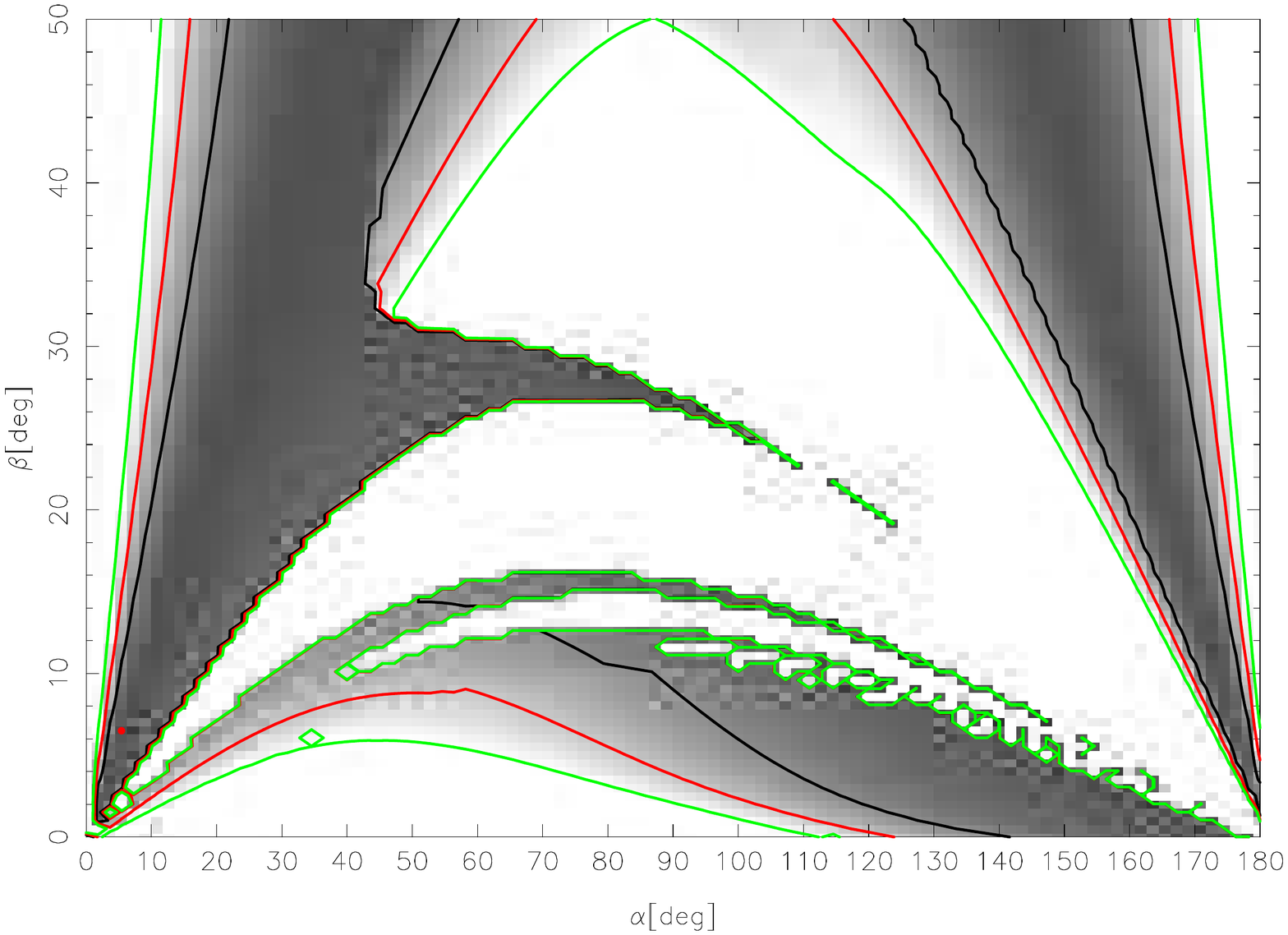}}}&
{\mbox{\includegraphics[width=9cm,height=6cm,angle=0.]{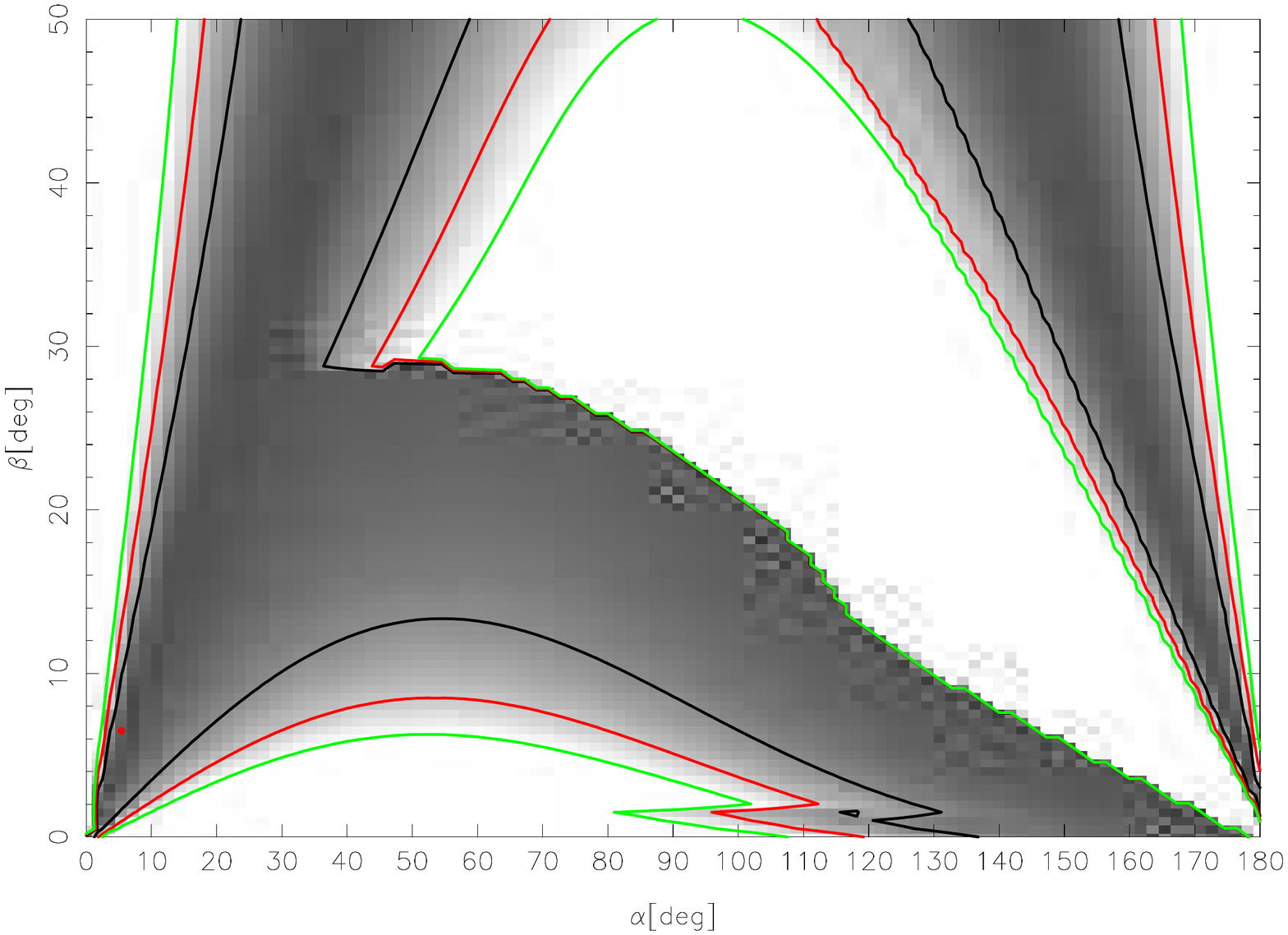}}}\\
{\mbox{\includegraphics[width=9cm,height=6cm,angle=0.]{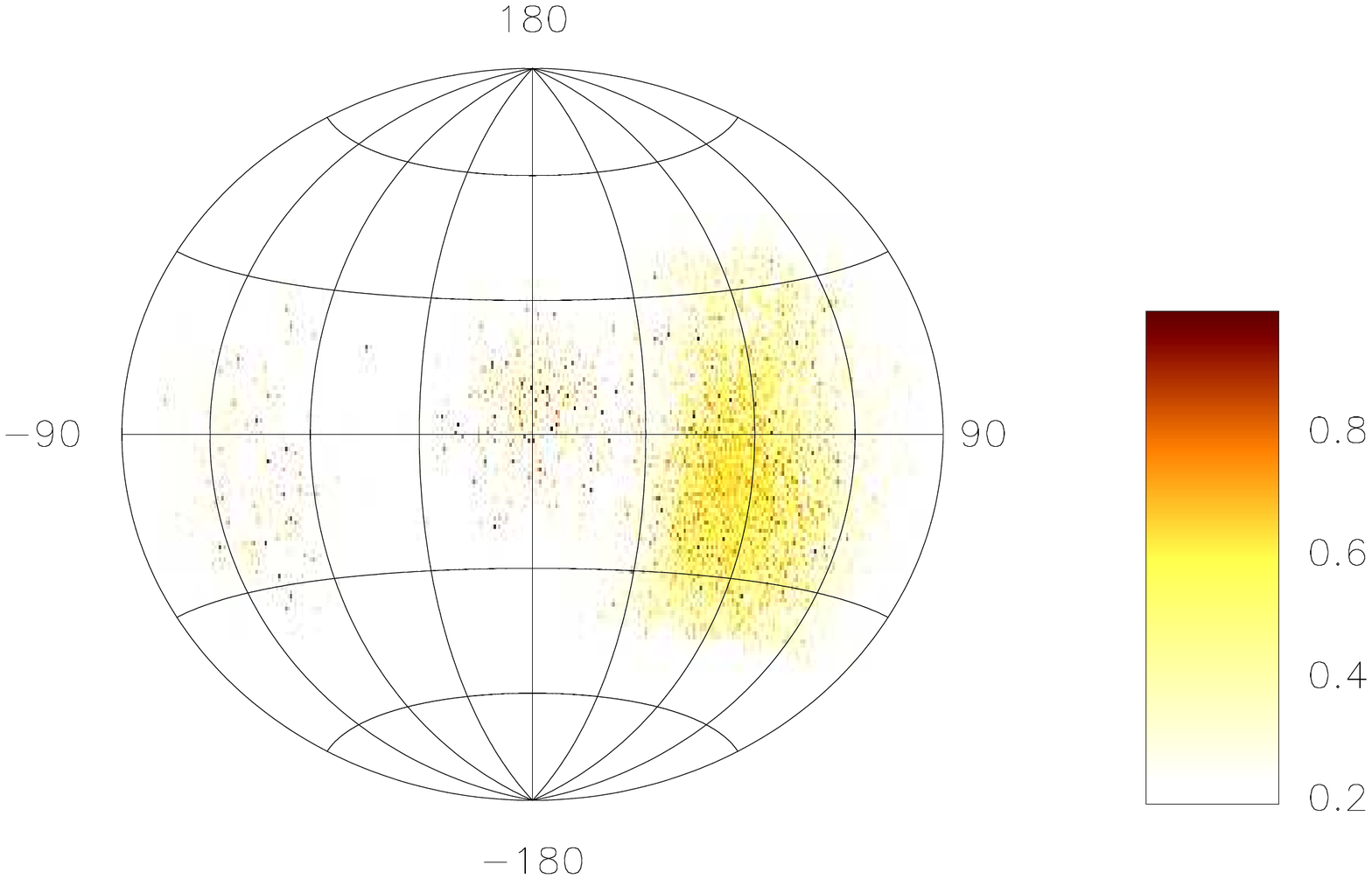}}}&
{\mbox{\includegraphics[width=9cm,height=6cm,angle=0.]{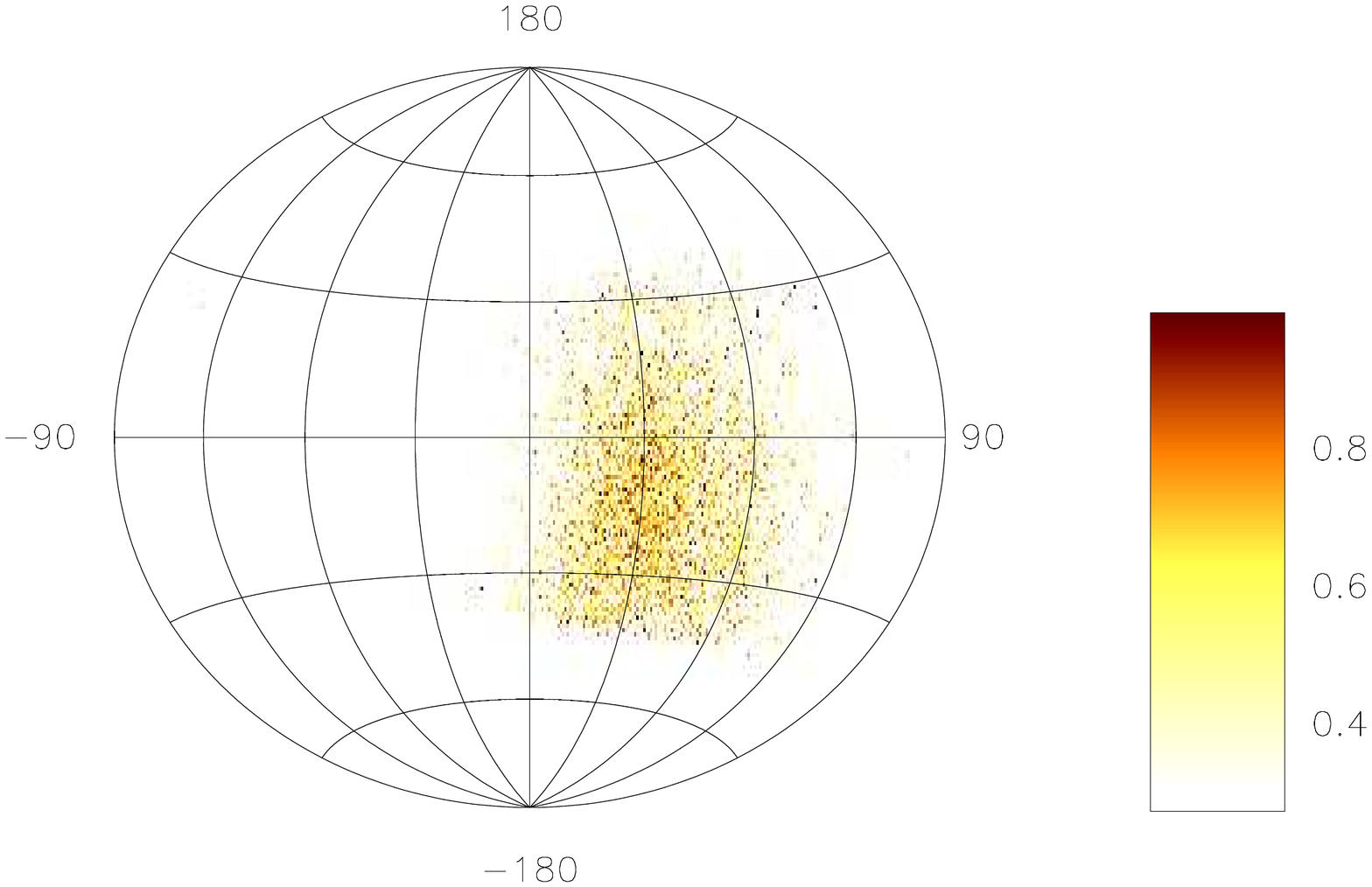}}}\\
\end{tabular}
\caption{Top panel (upper window) shows the average profile with total
intensity (Stokes I; solid black lines), total linear polarization (dashed red
line) and circular polarization (Stokes V; dotted blue line). Top panel (lower
window) also shows the single pulse PPA distribution (colour scale) along with
the average PPA (red error bars).
The RVM fits to the average PPA (dashed pink
line) is also shown in this plot. Middle panel show
the $\chi^2$ contours for the parameters $\alpha$ and $\beta$ obtained from RVM
fits.
Bottom panel shows the Hammer-Aitoff projection of the polarized time
samples with the colour scheme representing the fractional polarization level.}
\label{a78}
\end{center}
\end{figure*}


\begin{figure*}
\begin{center}
\begin{tabular}{cc}
{\mbox{\includegraphics[width=9cm,height=6cm,angle=0.]{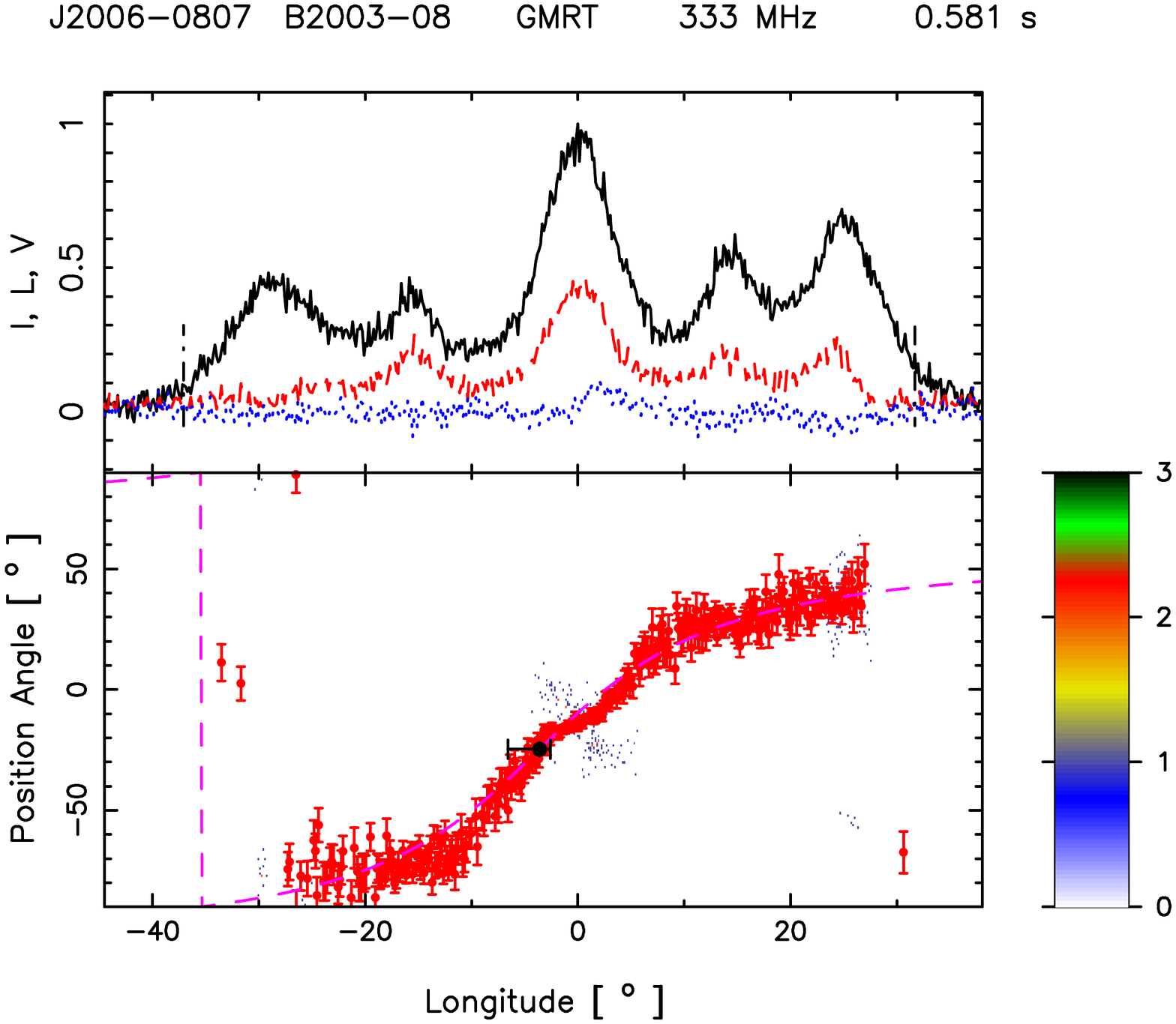}}}&
{\mbox{\includegraphics[width=9cm,height=6cm,angle=0.]{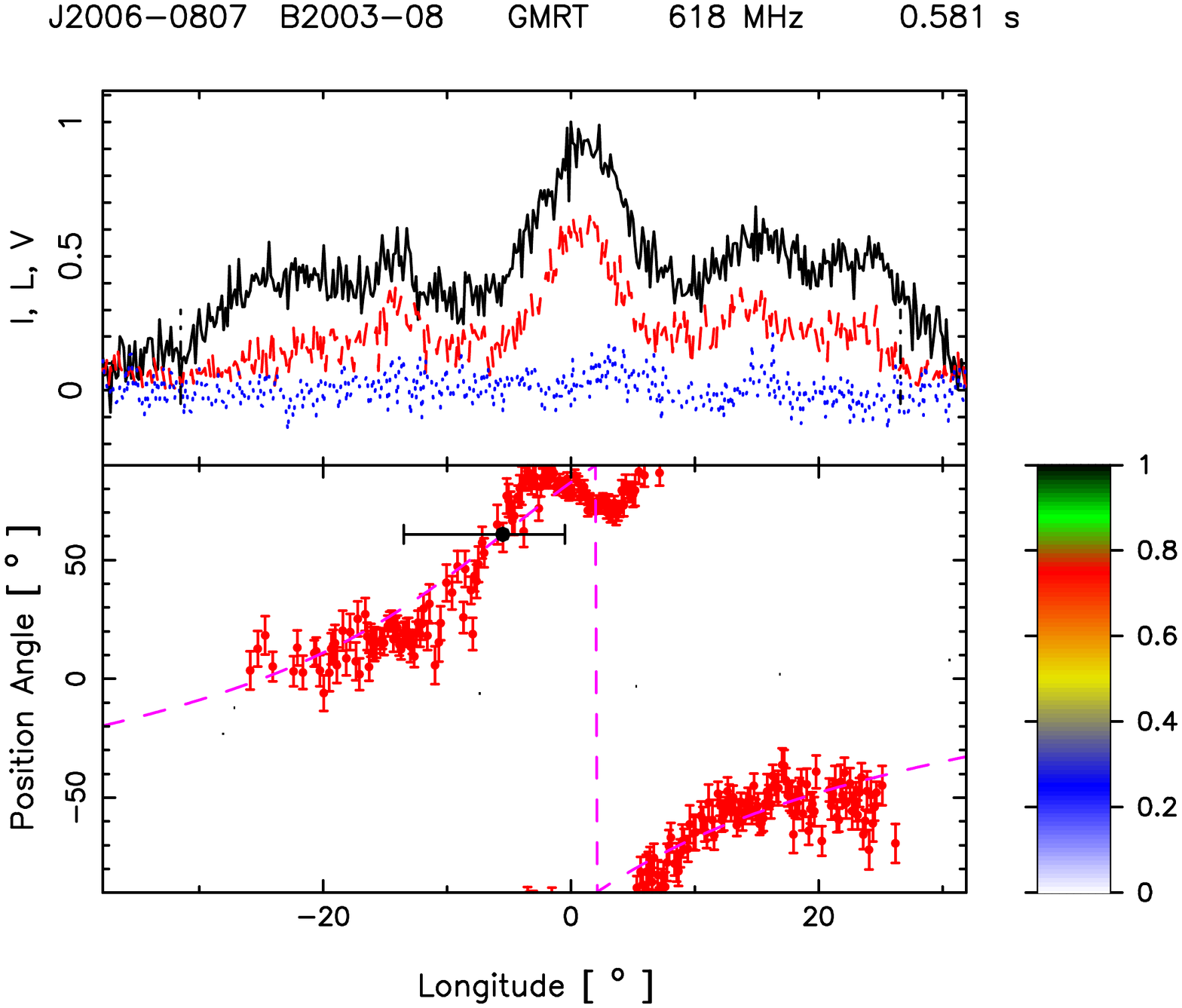}}}\\
{\mbox{\includegraphics[width=9cm,height=6cm,angle=0.]{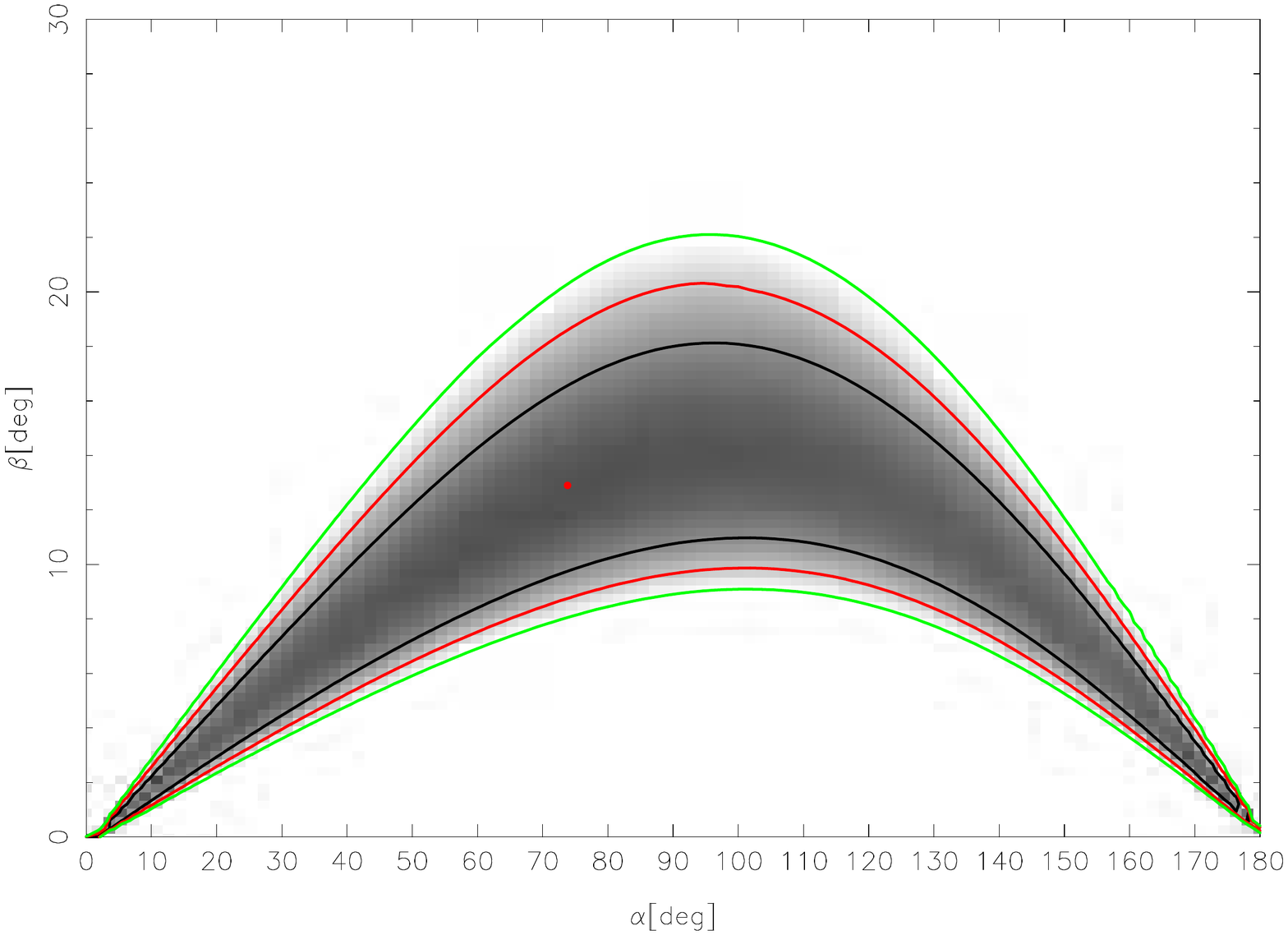}}}&
{\mbox{\includegraphics[width=9cm,height=6cm,angle=0.]{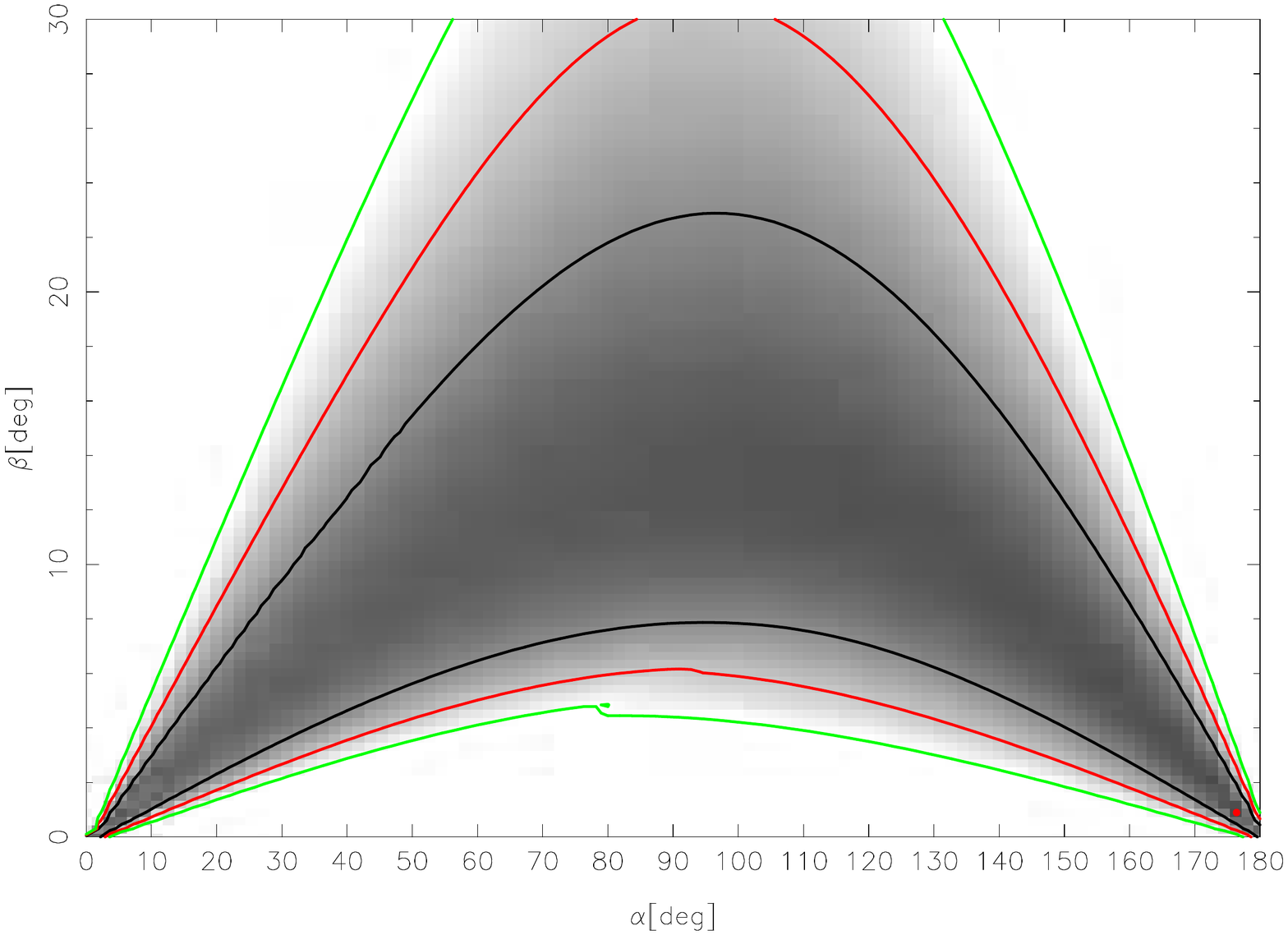}}}\\
{\mbox{\includegraphics[width=9cm,height=6cm,angle=0.]{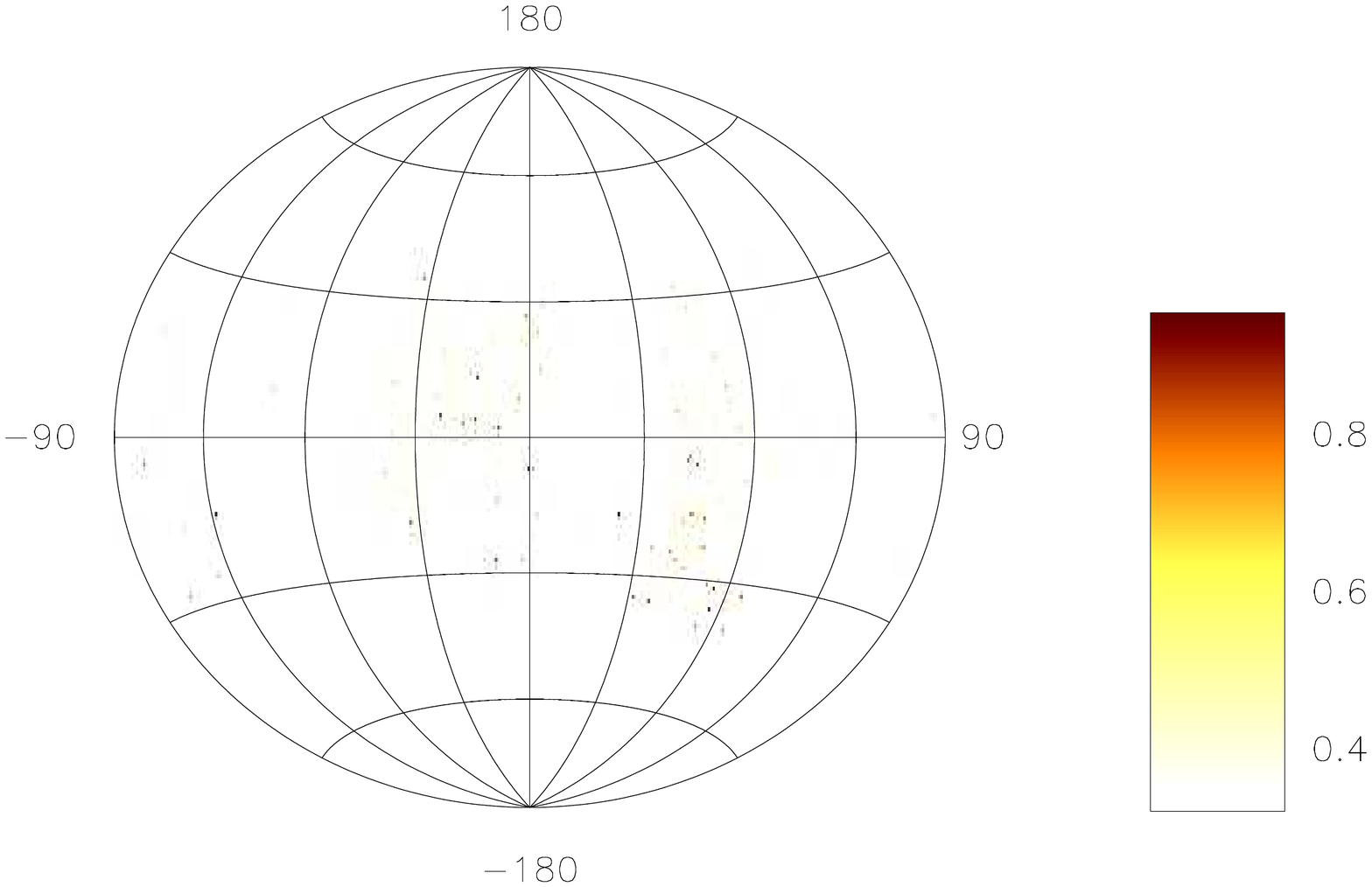}}}&
{\mbox{\includegraphics[width=9cm,height=6cm,angle=0.]{J1946+1805_602MHz_30Apr14.dat.epn2a.71.pdf}}}\\
\end{tabular}
\caption{Top panel (upper window) shows the average profile with total
intensity (Stokes I; solid black lines), total linear polarization (dashed red
line) and circular polarization (Stokes V; dotted blue line). Top panel (lower
window) also shows the single pulse PPA distribution (colour scale) along with
the average PPA (red error bars).
The RVM fits to the average PPA (dashed pink
line) is also shown in this plot. Middle panel show
the $\chi^2$ contours for the parameters $\alpha$ and $\beta$ obtained from RVM
fits.
Bottom panel only for 333 MHz shows the Hammer-Aitoff projection of the polarized time
samples with the colour scheme representing the fractional polarization level.}
\label{a79}
\end{center}
\end{figure*}


\begin{figure*}
\begin{center}
\begin{tabular}{cc}
&
{\mbox{\includegraphics[width=9cm,height=6cm,angle=0.]{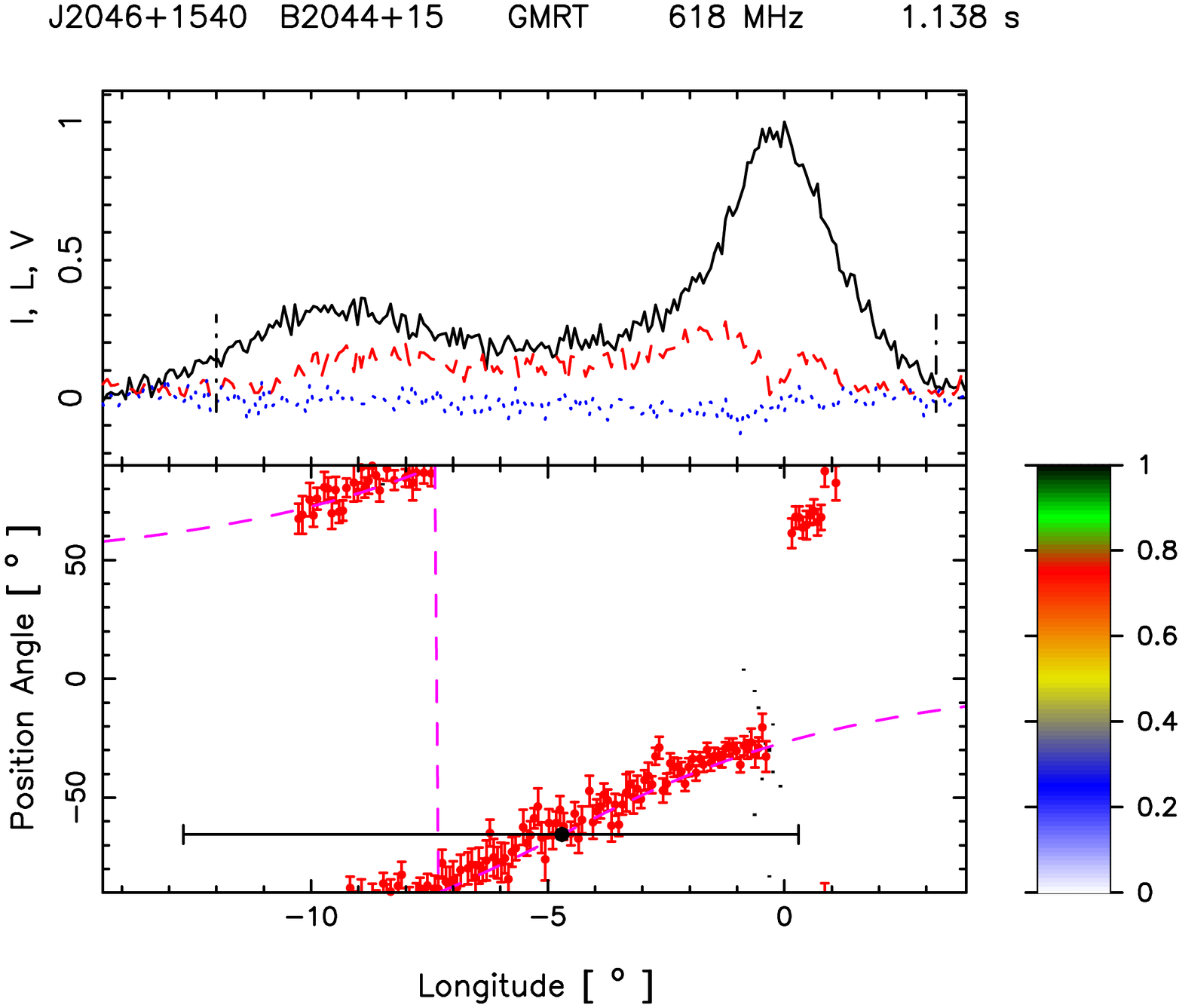}}}\\
&
{\mbox{\includegraphics[width=9cm,height=6cm,angle=0.]{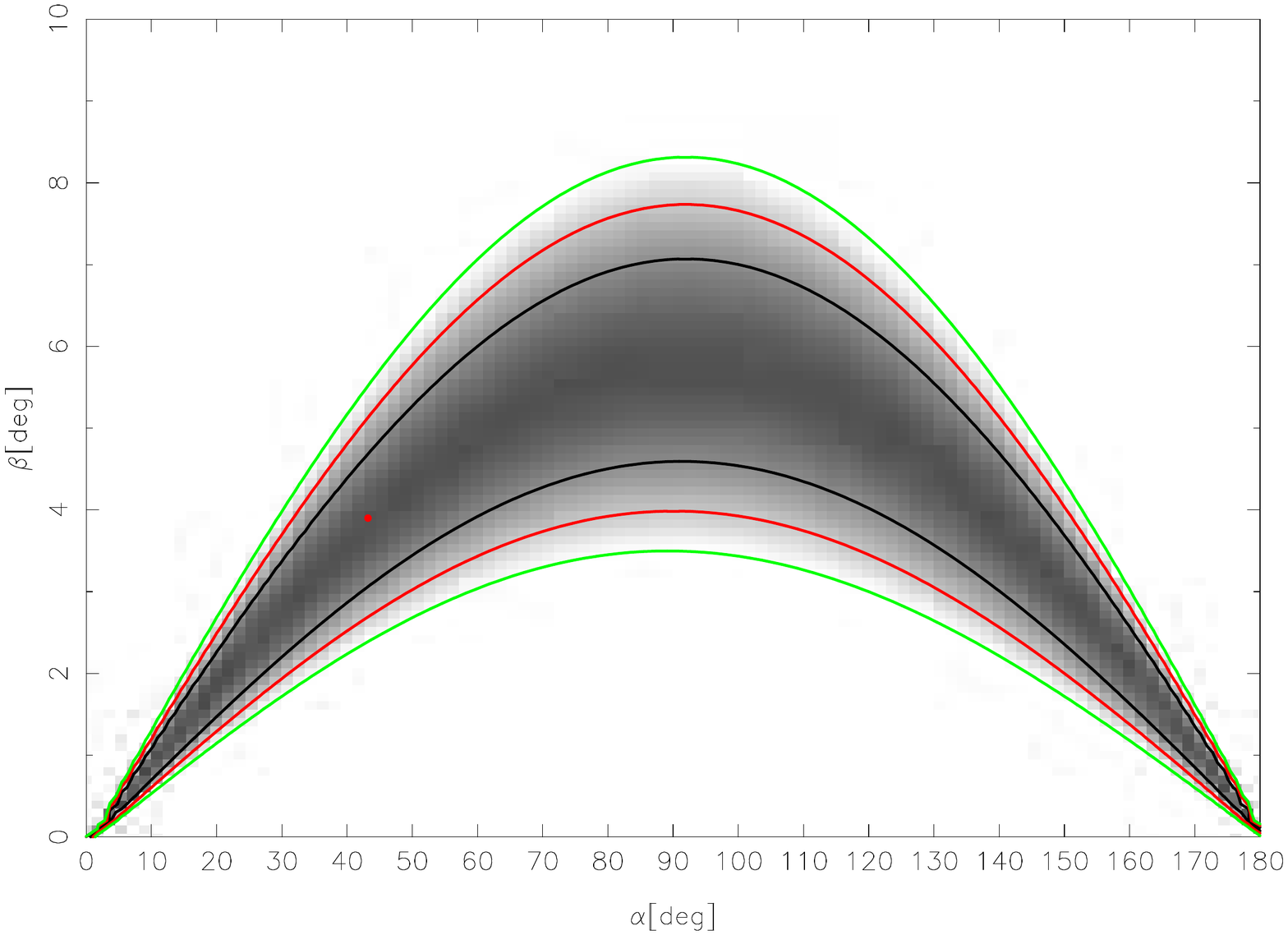}}}\\
&
\\
\end{tabular}
\caption{Top panel only at 618 MHz (upper window) shows the average profile with total
intensity (Stokes I; solid black lines), total linear polarization (dashed red
line) and circular polarization (Stokes V; dotted blue line). Top panel (lower
window) also shows the single pulse PPA distribution (colour scale) along with
the average PPA (red error bars).
The RVM fits to the average PPA (dashed pink
line) is also shown in this plot. Bottom panel only at 618 MHz show
the $\chi^2$ contours for the parameters $\alpha$ and $\beta$ obtained from RVM
fits.}
\label{a80}
\end{center}
\end{figure*}


\begin{figure*}
\begin{center}
\begin{tabular}{cc}
{\mbox{\includegraphics[width=9cm,height=6cm,angle=0.]{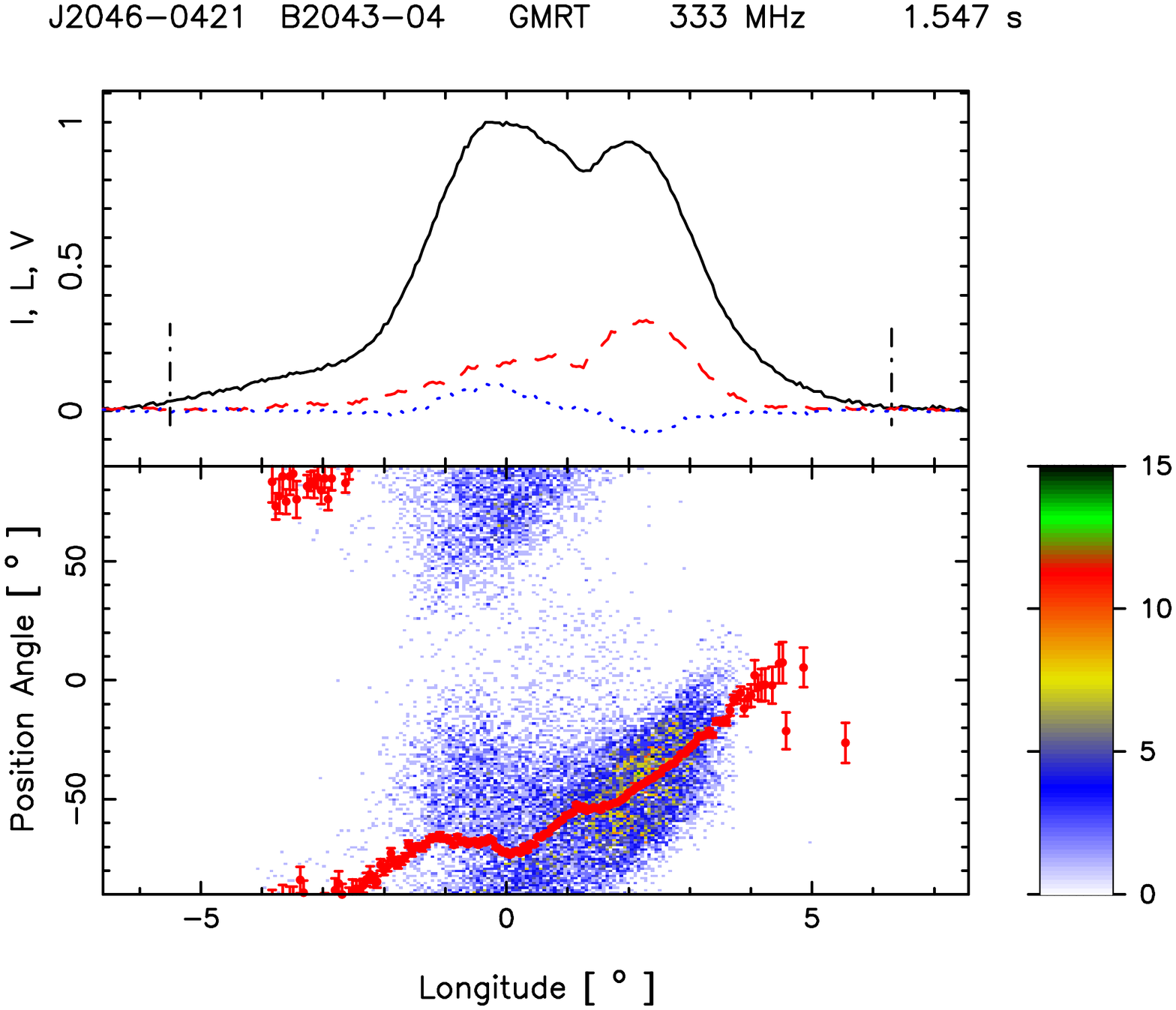}}}&
{\mbox{\includegraphics[width=9cm,height=6cm,angle=0.]{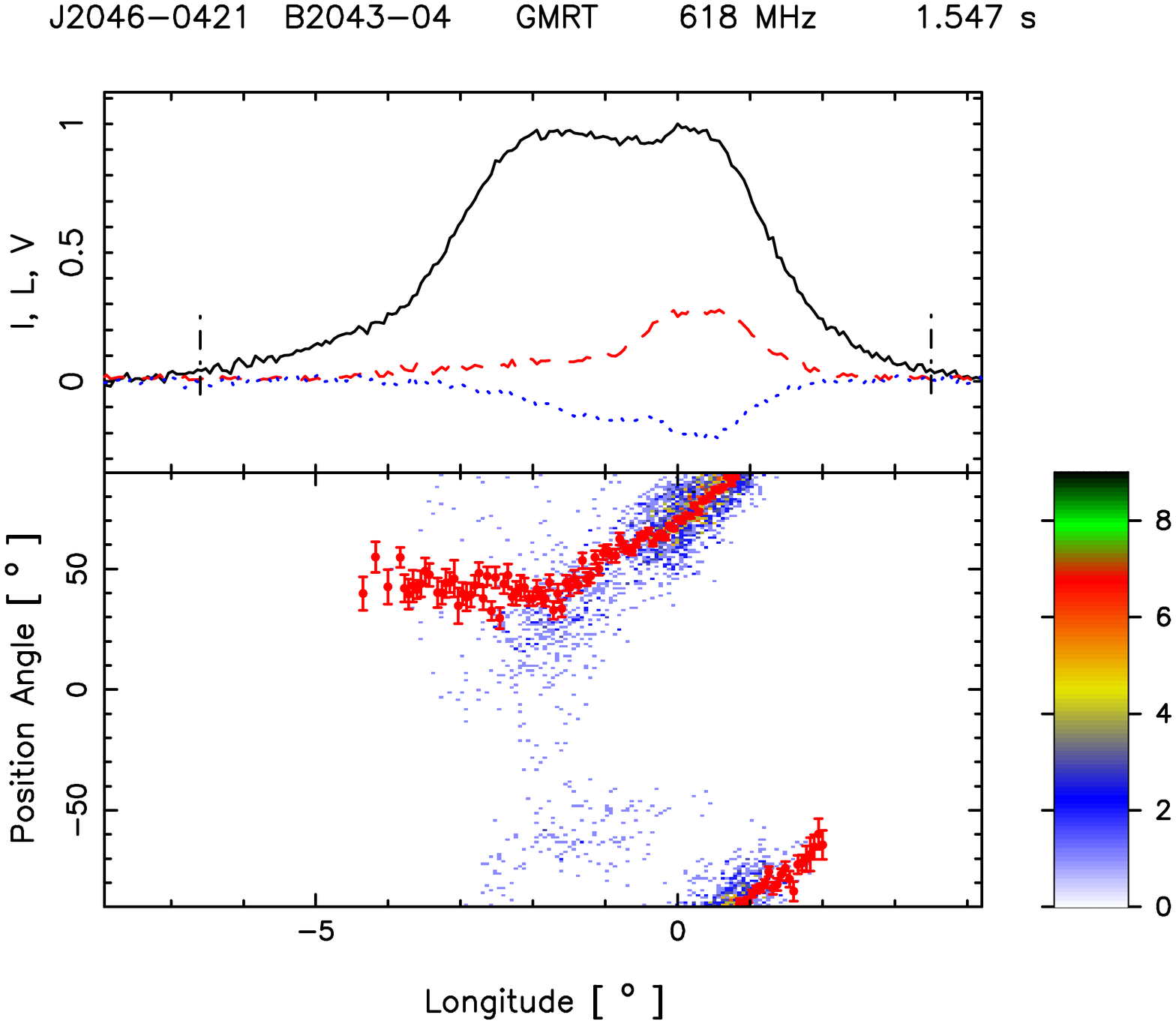}}}\\
{\mbox{\includegraphics[width=9cm,height=6cm,angle=0.]{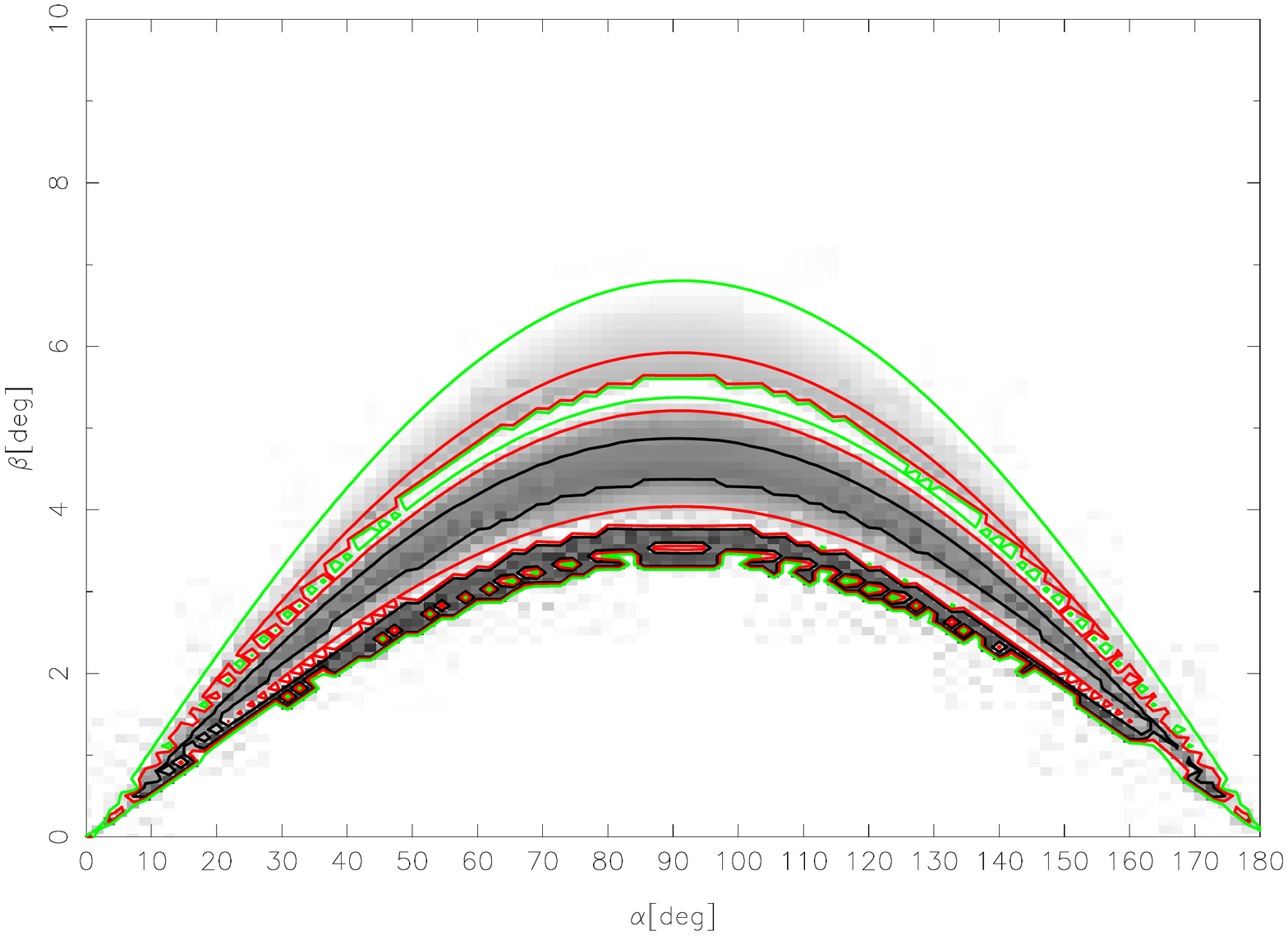}}}&
{\mbox{\includegraphics[width=9cm,height=6cm,angle=0.]{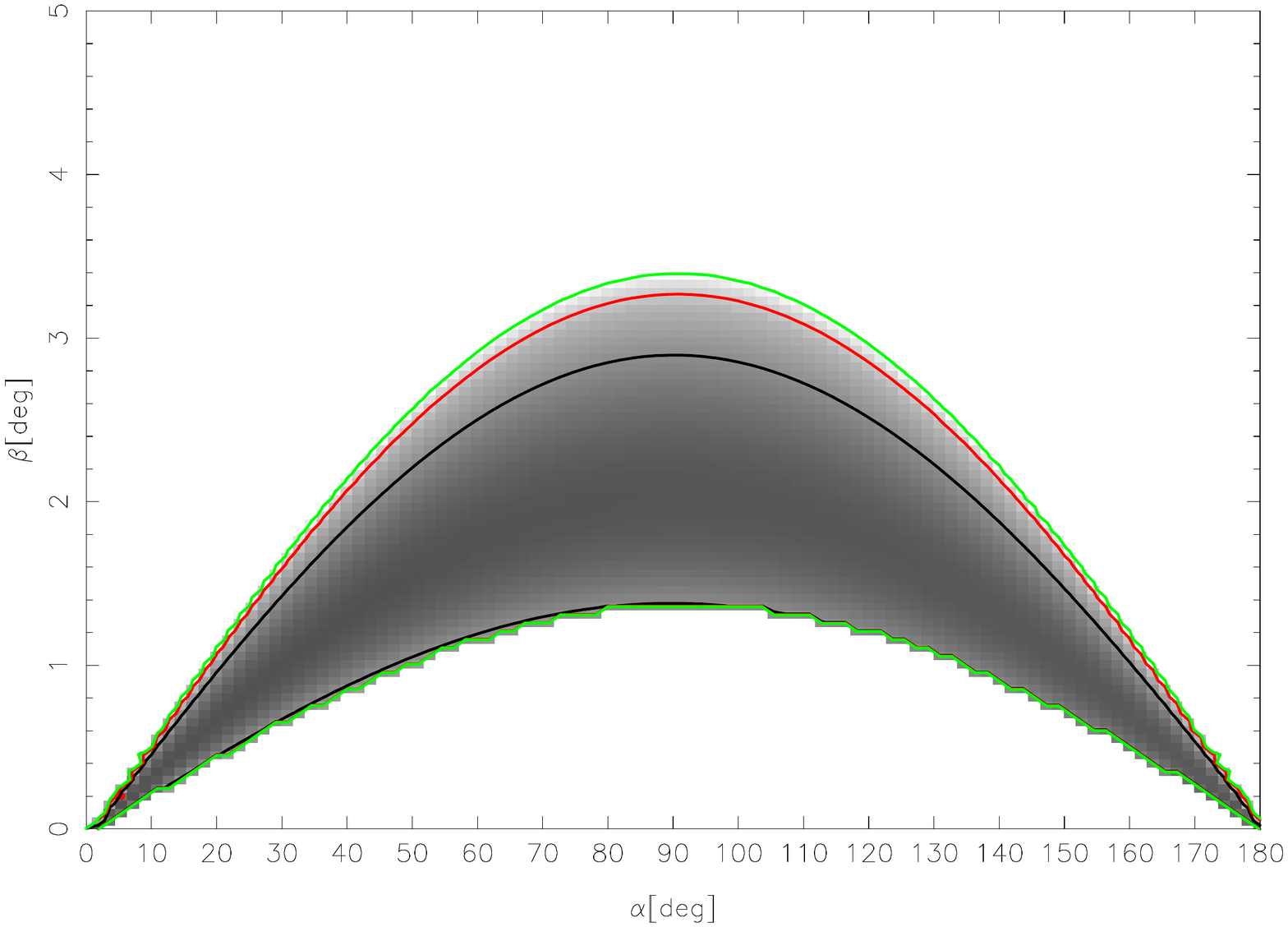}}}\\
{\mbox{\includegraphics[width=9cm,height=6cm,angle=0.]{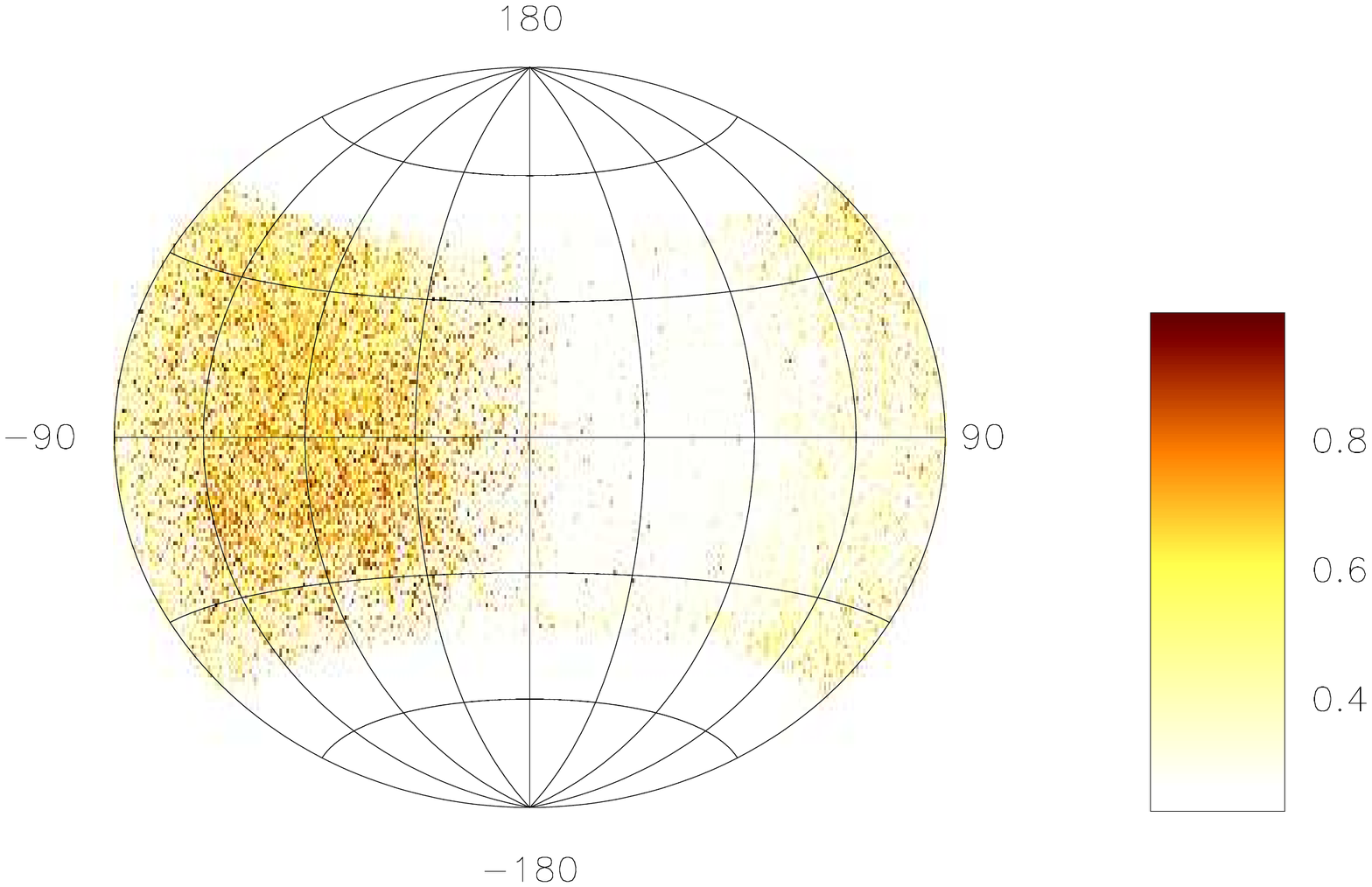}}}&
{\mbox{\includegraphics[width=9cm,height=6cm,angle=0.]{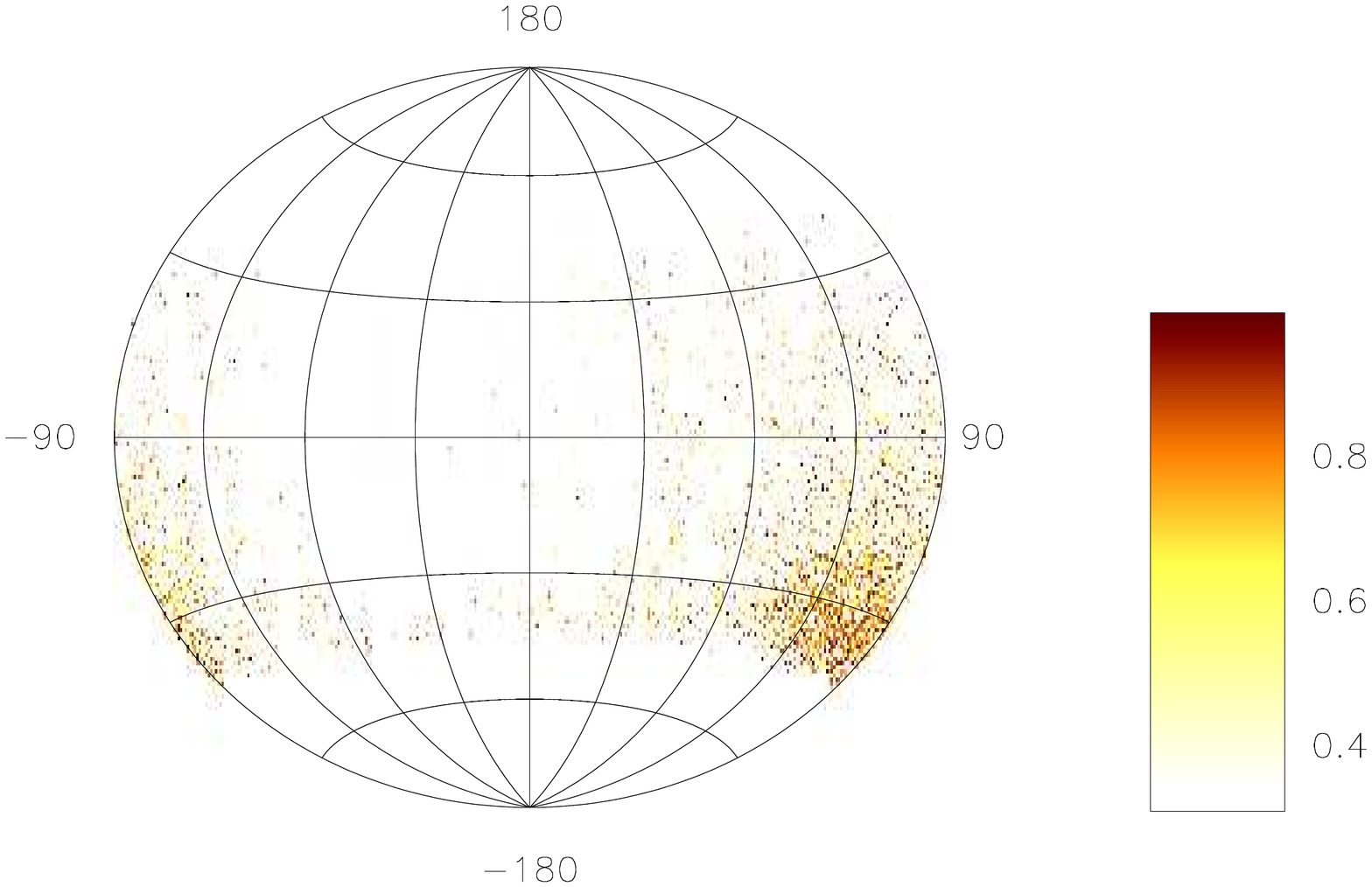}}}\\
\end{tabular}
\caption{Top panel (upper window) shows the average profile with total
intensity (Stokes I; solid black lines), total linear polarization (dashed red
line) and circular polarization (Stokes V; dotted blue line). Top panel (lower
window) also shows the single pulse PPA distribution (colour scale) along with
the average PPA (red error bars).
The RVM fits to the average PPA (dashed pink
line) is also shown in this plot. Middle panel show
the $\chi^2$ contours for the parameters $\alpha$ and $\beta$ obtained from RVM
fits.
Bottom panel shows the Hammer-Aitoff projection of the polarized time
samples with the colour scheme representing the fractional polarization level.}
\label{a81}
\end{center}
\end{figure*}


\begin{figure*}
\begin{center}
\begin{tabular}{cc}
{\mbox{\includegraphics[width=9cm,height=6cm,angle=0.]{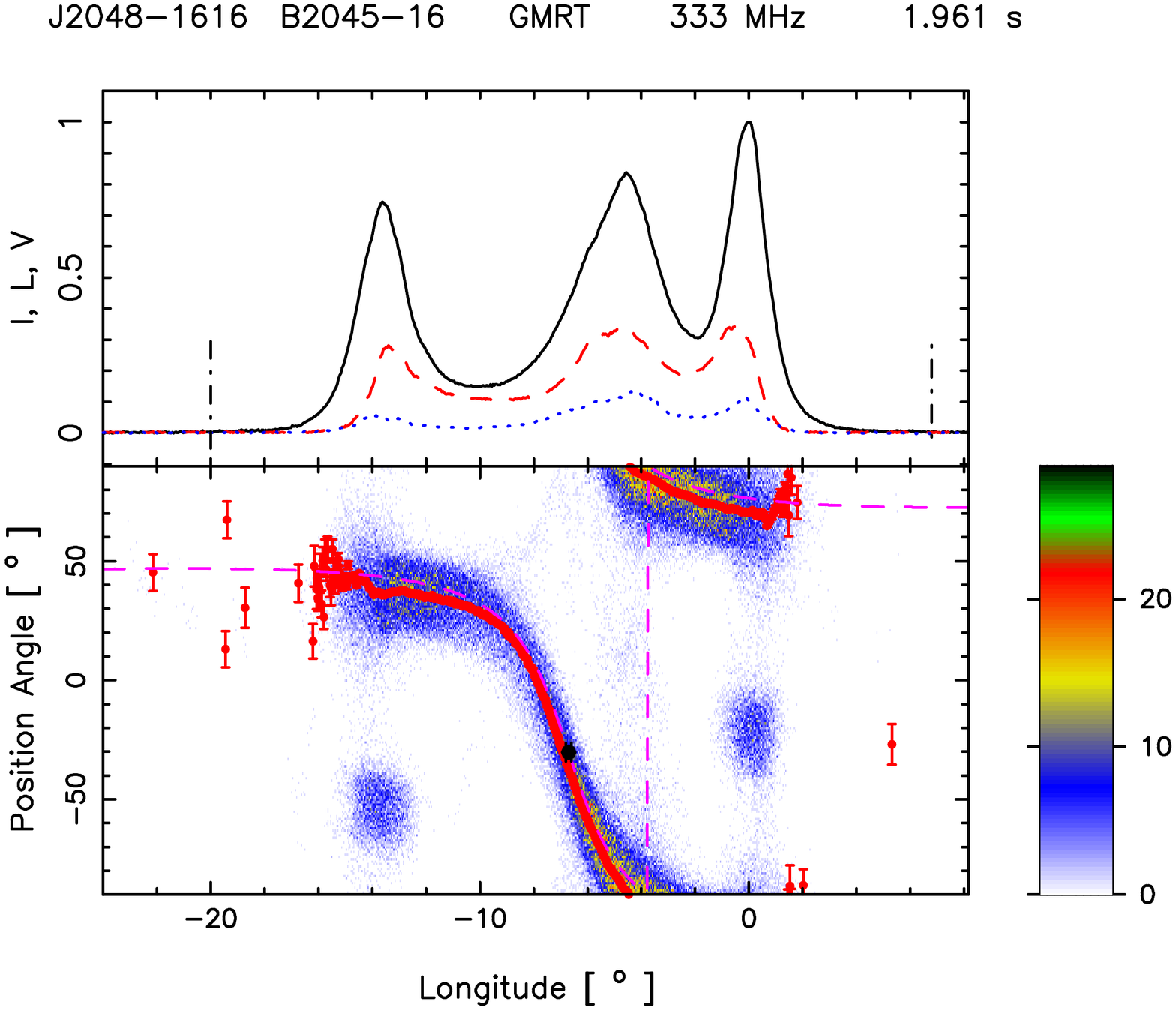}}}&
{\mbox{\includegraphics[width=9cm,height=6cm,angle=0.]{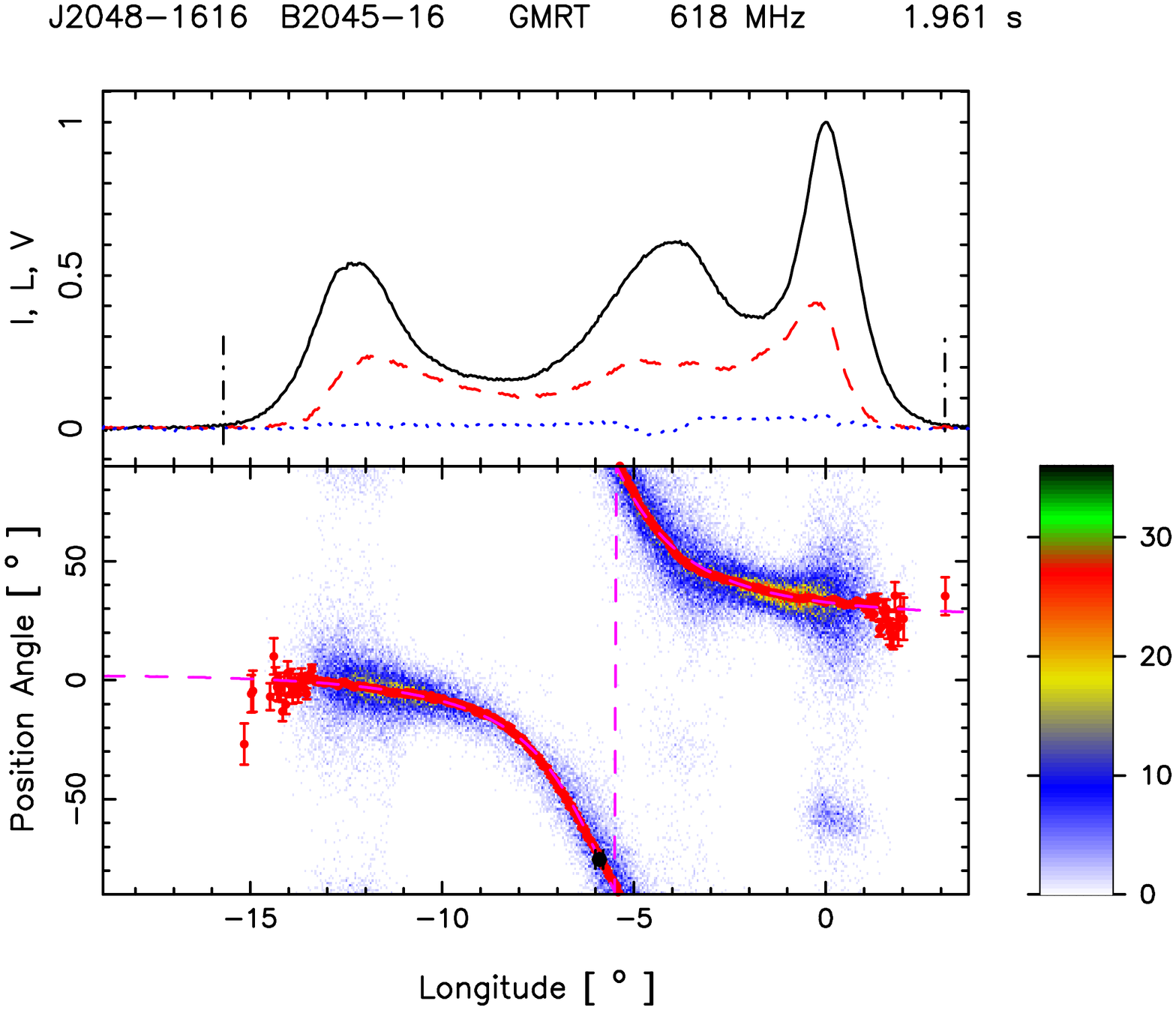}}}\\
{\mbox{\includegraphics[width=9cm,height=6cm,angle=0.]{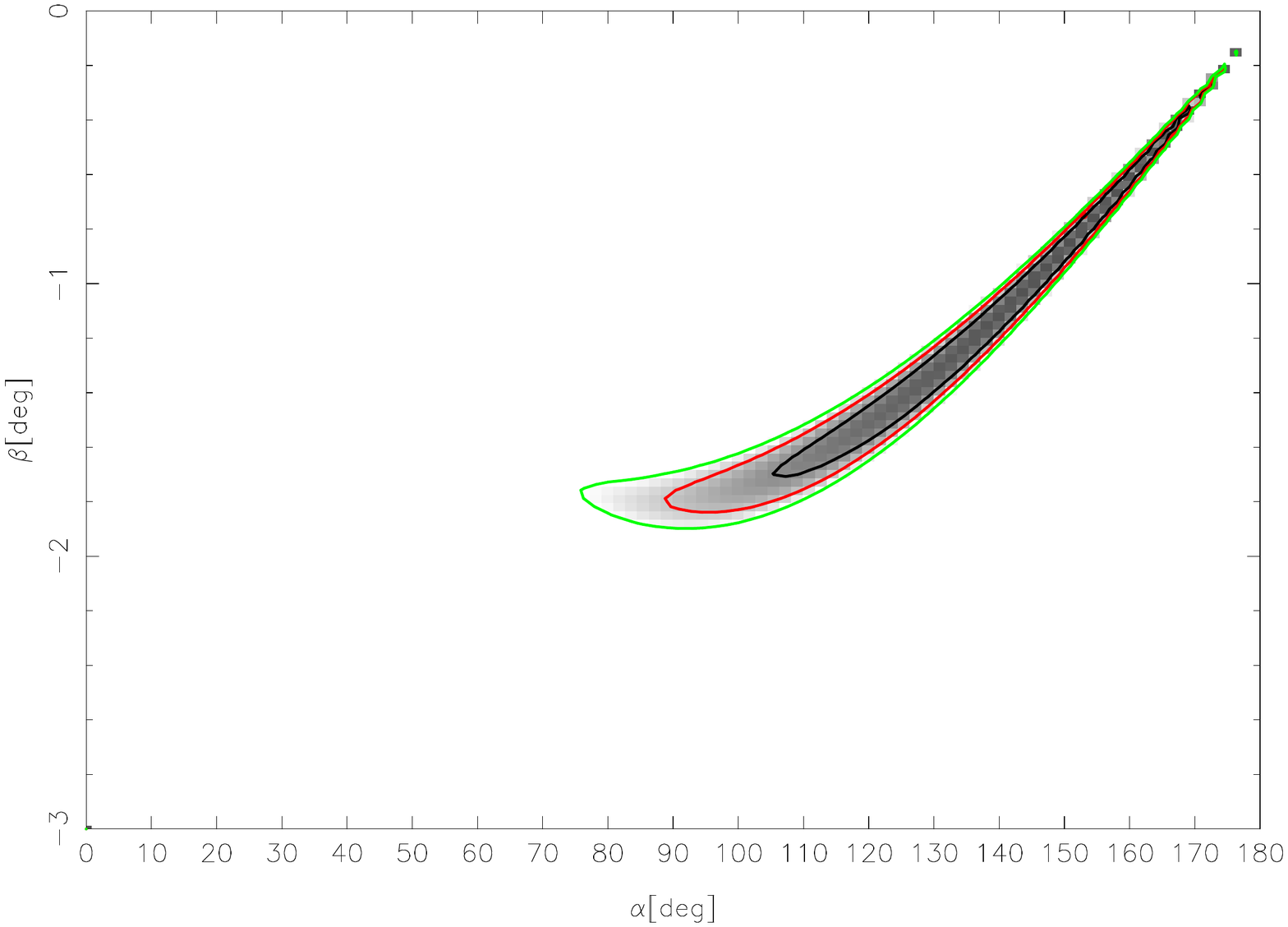}}}&
{\mbox{\includegraphics[width=9cm,height=6cm,angle=0.]{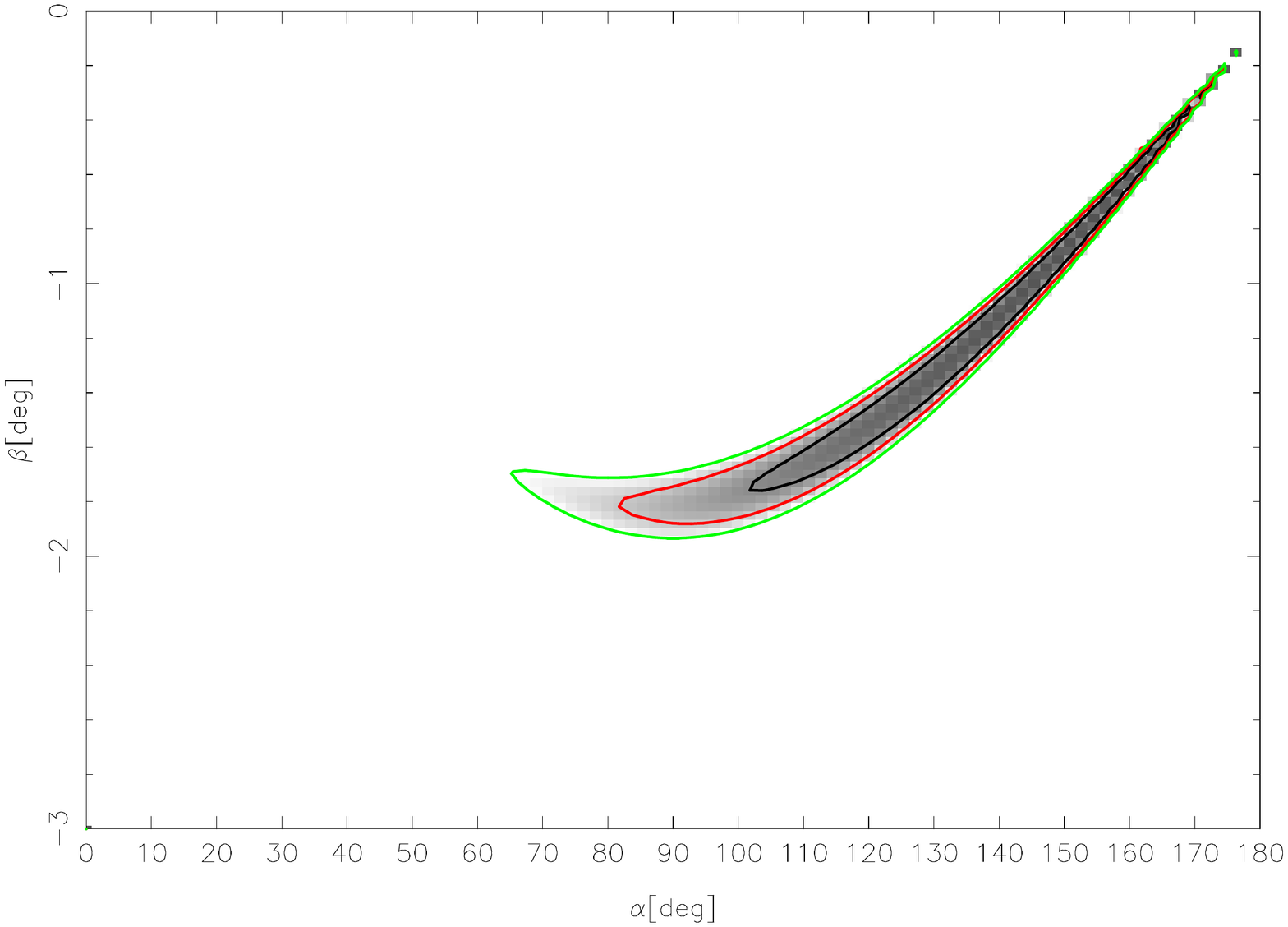}}}\\
{\mbox{\includegraphics[width=9cm,height=6cm,angle=0.]{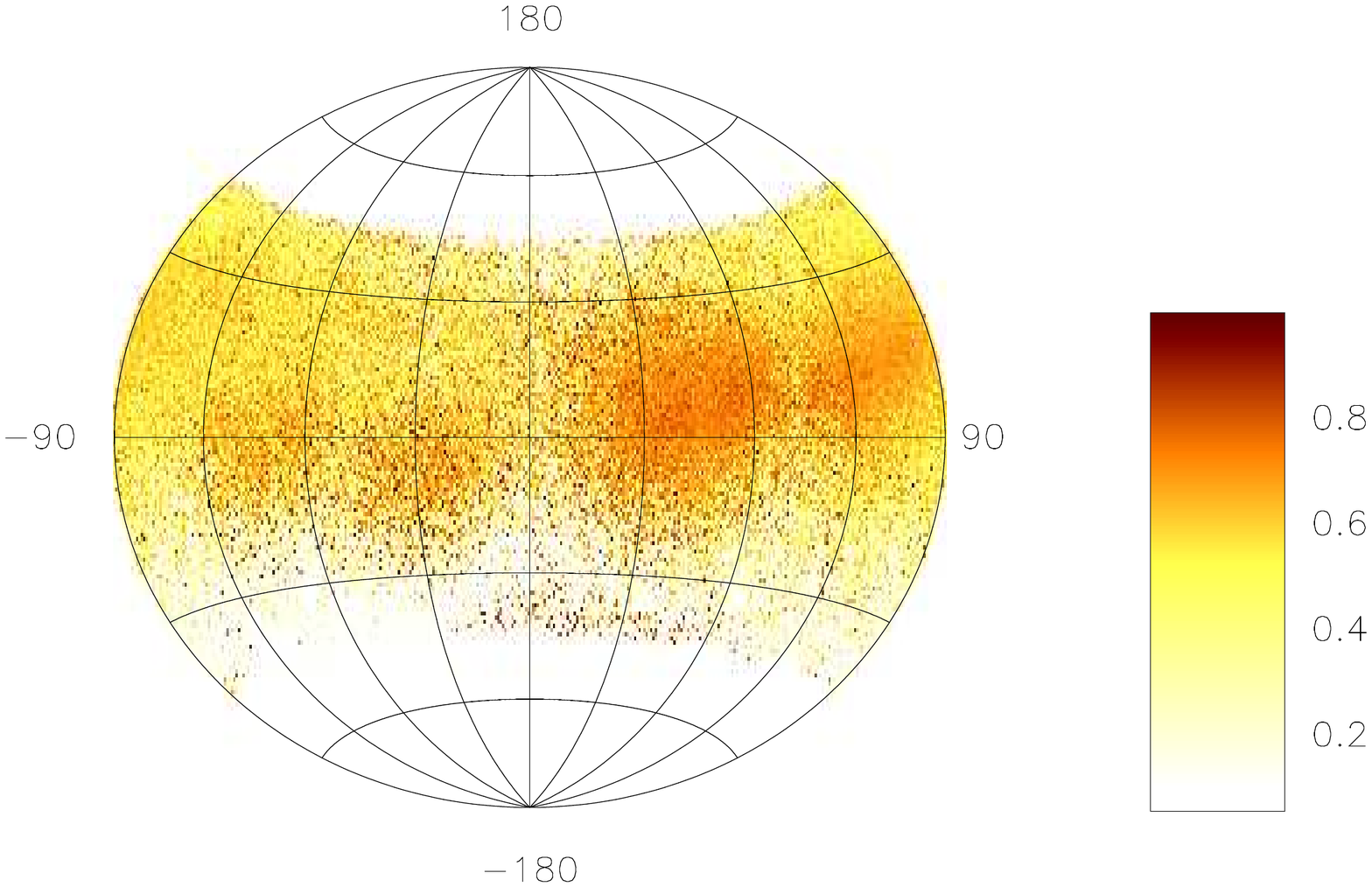}}}&
{\mbox{\includegraphics[width=9cm,height=6cm,angle=0.]{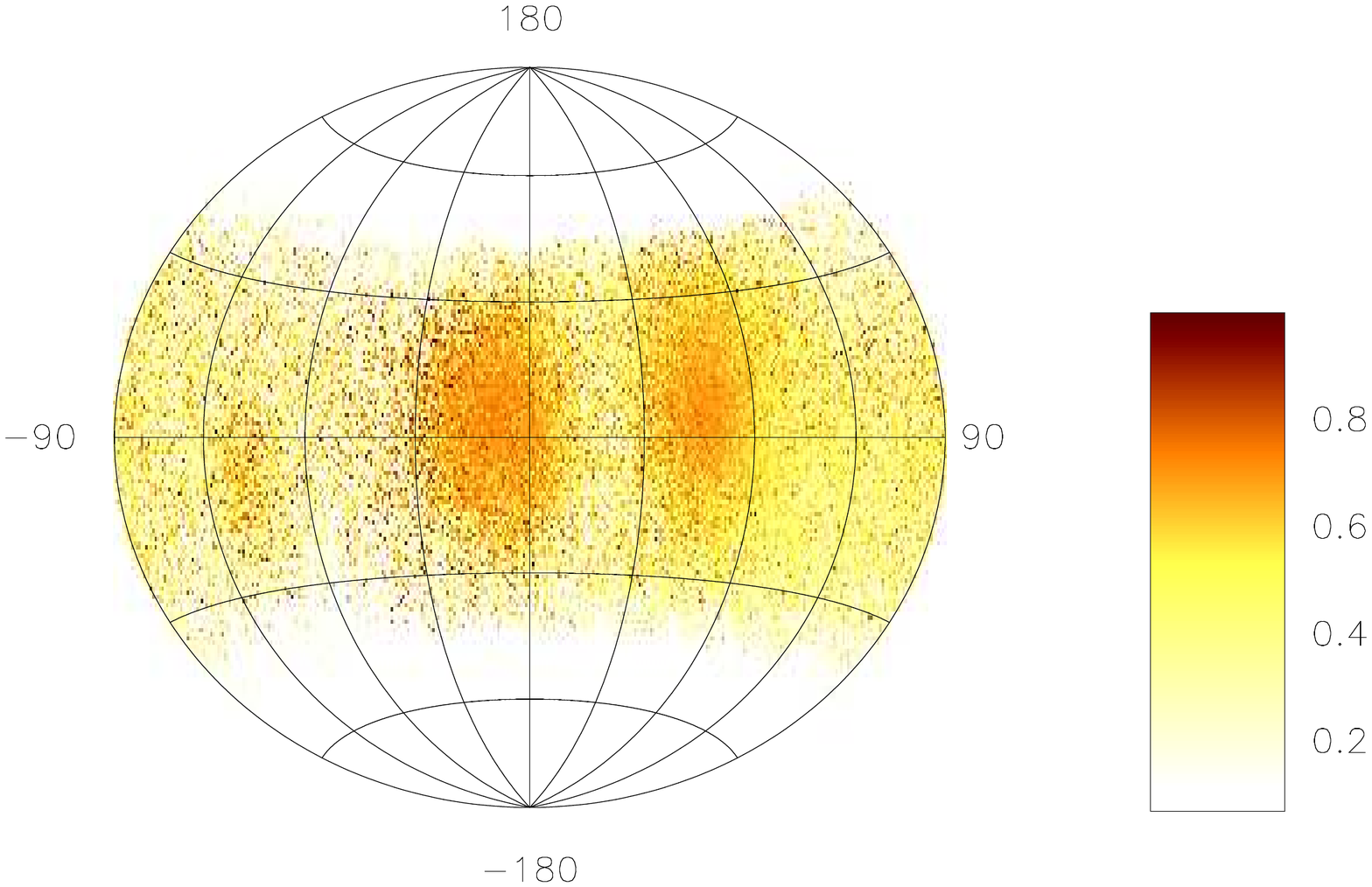}}}\\
\end{tabular}
\caption{Top panel (upper window) shows the average profile with total
intensity (Stokes I; solid black lines), total linear polarization (dashed red
line) and circular polarization (Stokes V; dotted blue line). Top panel (lower
window) also shows the single pulse PPA distribution (colour scale) along with
the average PPA (red error bars).
The RVM fits to the average PPA (dashed pink
line) is also shown in this plot. Middle panel show
the $\chi^2$ contours for the parameters $\alpha$ and $\beta$ obtained from RVM
fits.
Bottom panel shows the Hammer-Aitoff projection of the polarized time
samples with the colour scheme representing the fractional polarization level.}
\label{a82}
\end{center}
\end{figure*}


\begin{figure*}
\begin{center}
\begin{tabular}{cc}
{\mbox{\includegraphics[width=9cm,height=6cm,angle=0.]{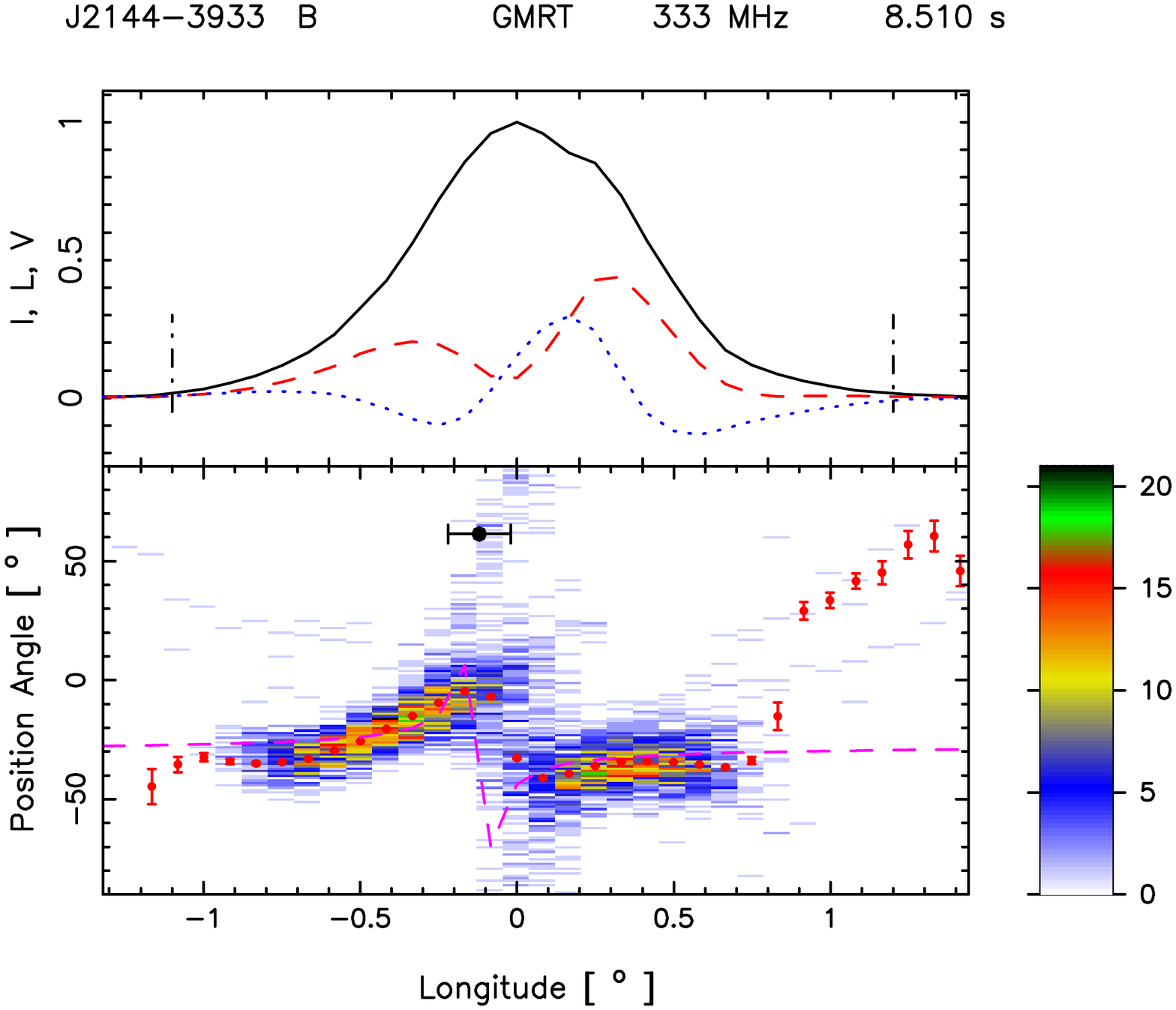}}}&
\\
{\mbox{\includegraphics[width=9cm,height=6cm,angle=0.]{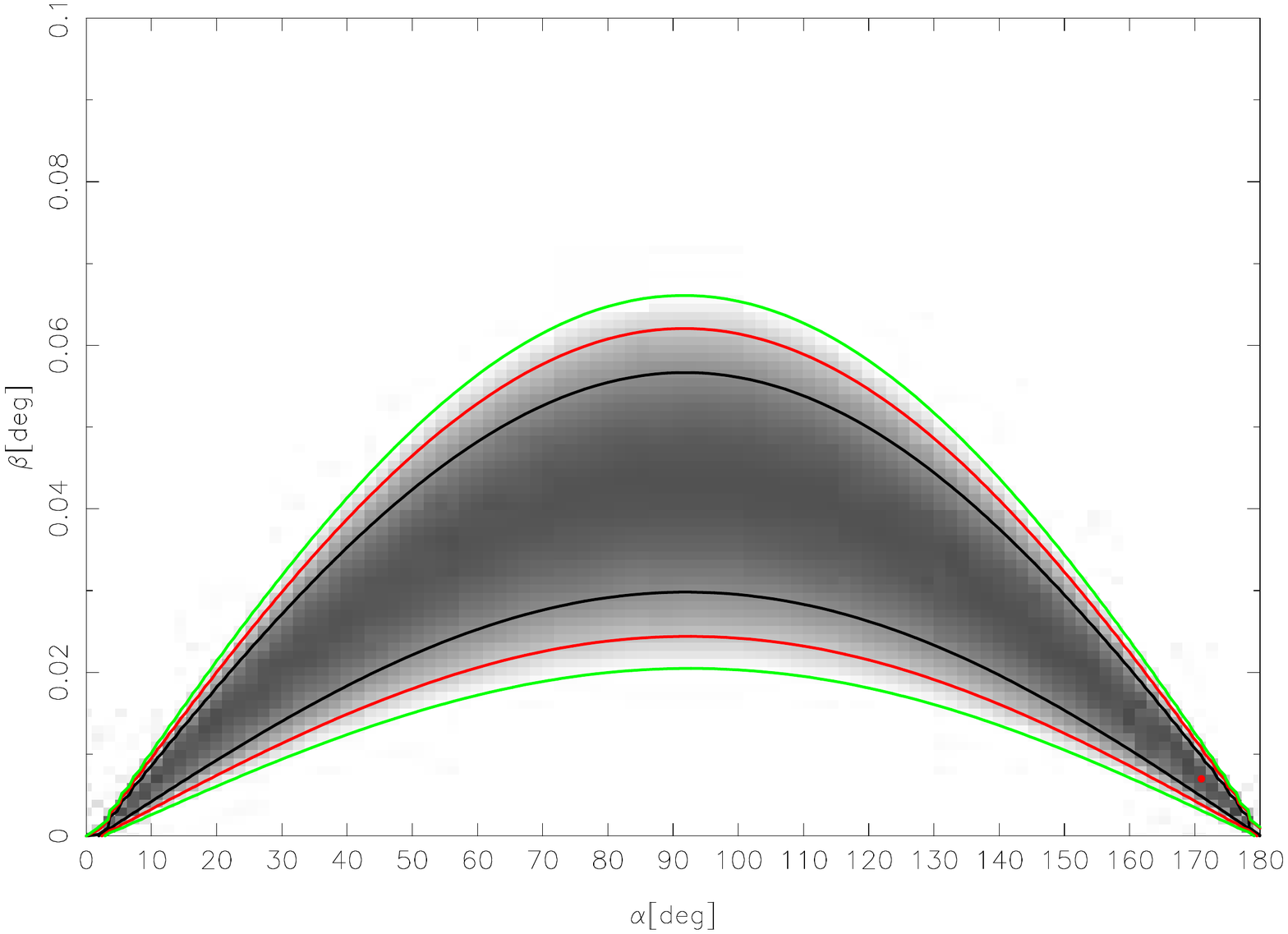}}}&
\\
{\mbox{\includegraphics[width=9cm,height=6cm,angle=0.]{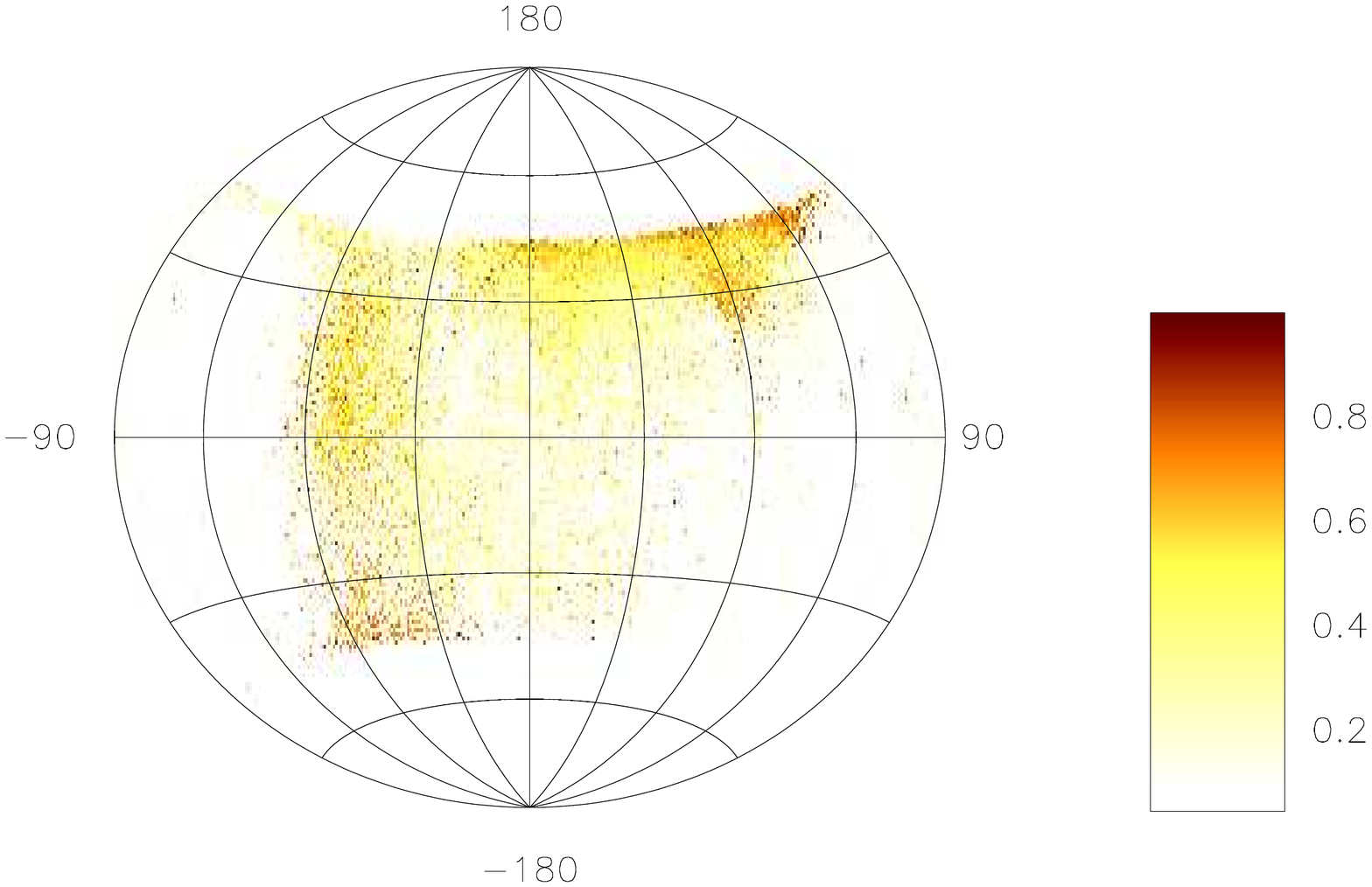}}}&
\\
\end{tabular}
\caption{Top panel only at 618 MHz (upper window) shows the average profile with total
intensity (Stokes I; solid black lines), total linear polarization (dashed red
line) and circular polarization (Stokes V; dotted blue line). Top panel (lower
window) also shows the single pulse PPA distribution (colour scale) along with
the average PPA (red error bars).
The RVM fits to the average PPA (dashed pink
line) is also shown in this plot. Middle panel only at 618 MHz show
the $\chi^2$ contours for the parameters $\alpha$ and $\beta$ obtained from RVM
fits.
Bottom panel only at 618 MHz shows the Hammer-Aitoff projection of the polarized time
samples with the colour scheme representing the fractional polarization level.}
\label{a83}
\end{center}
\end{figure*}


\begin{figure*}
\begin{center}
\begin{tabular}{cc}
{\mbox{\includegraphics[width=9cm,height=6cm,angle=0.]{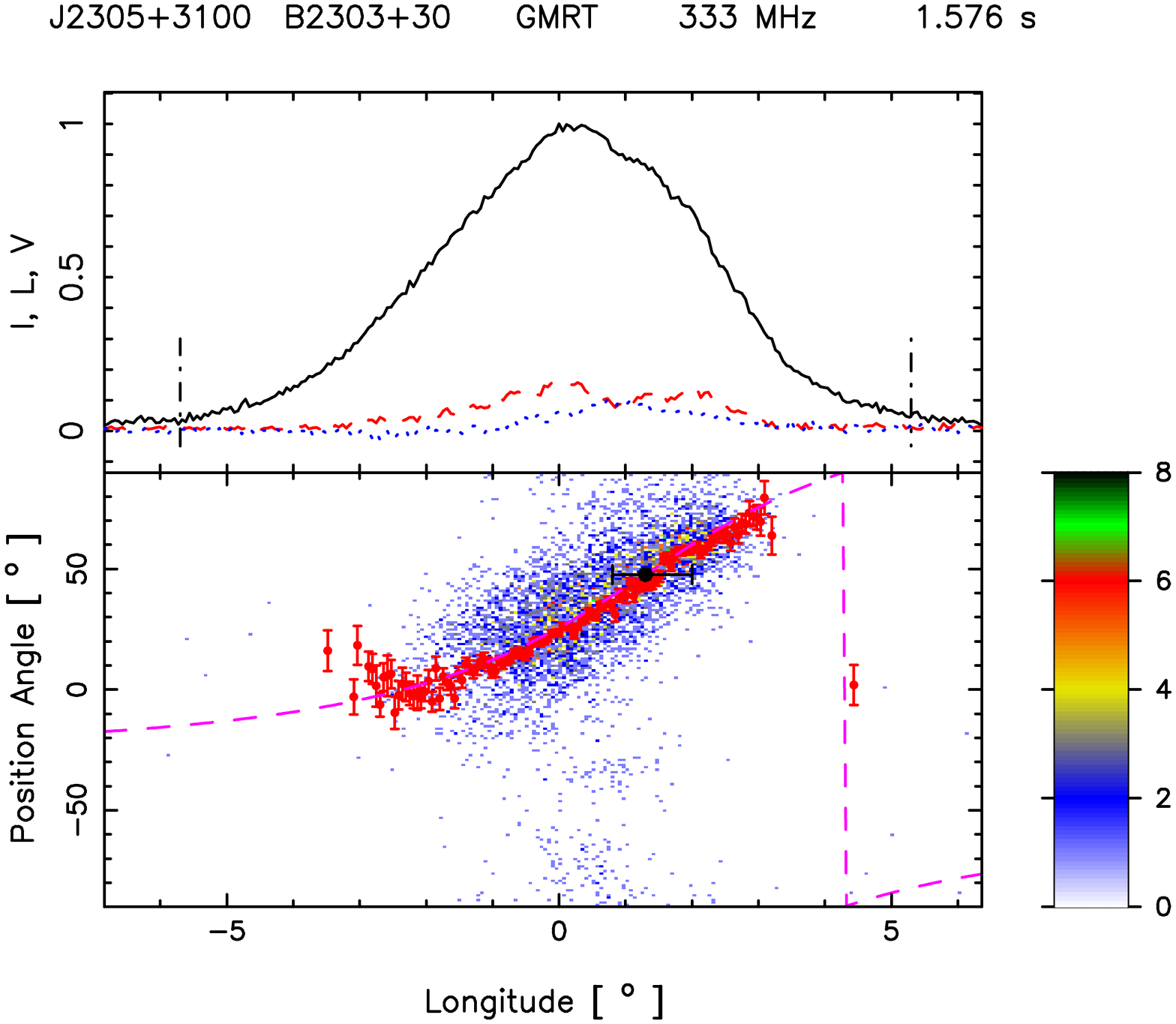}}}&
\\
{\mbox{\includegraphics[width=9cm,height=6cm,angle=0.]{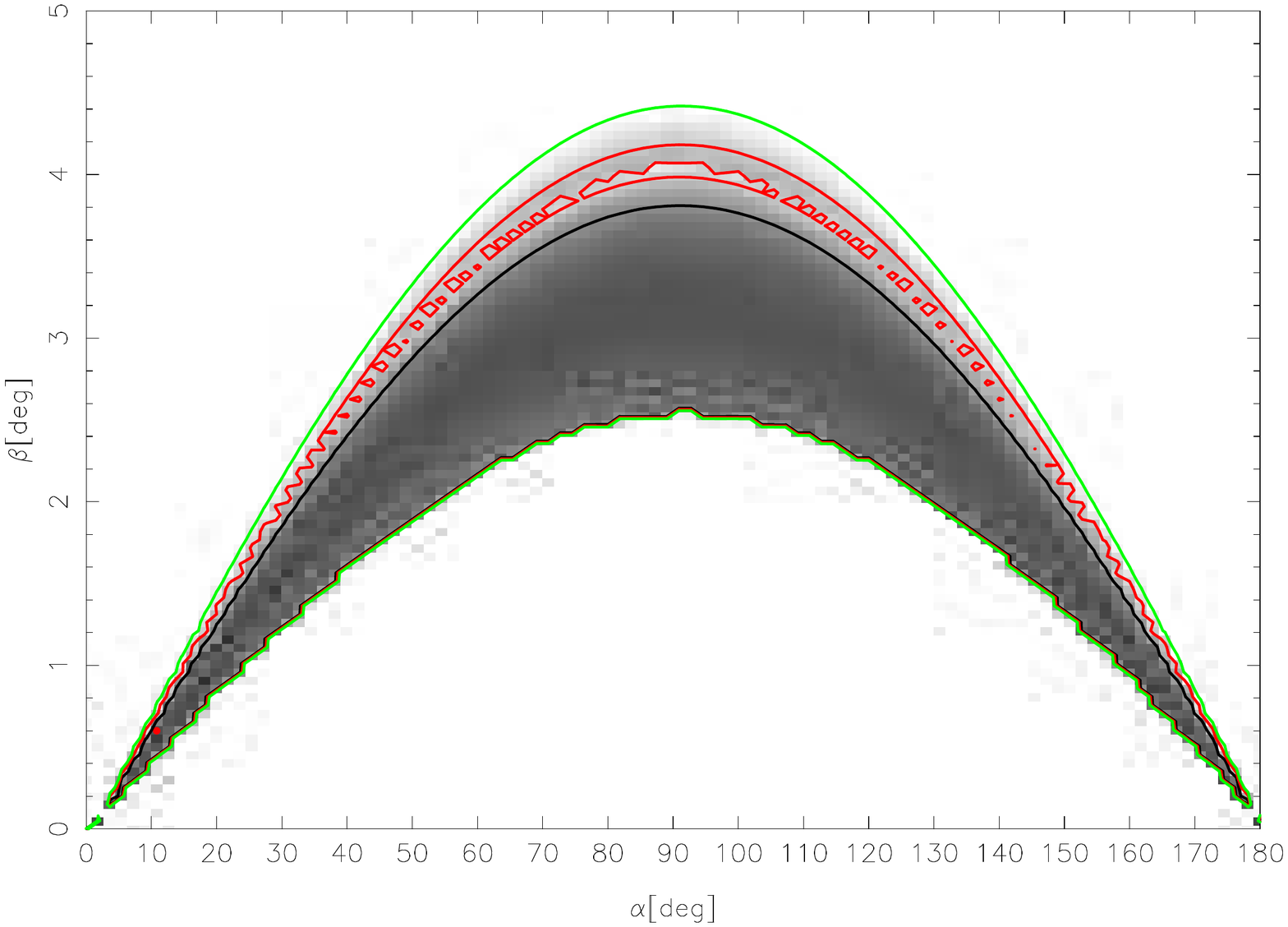}}}&
\\
{\mbox{\includegraphics[width=9cm,height=6cm,angle=0.]{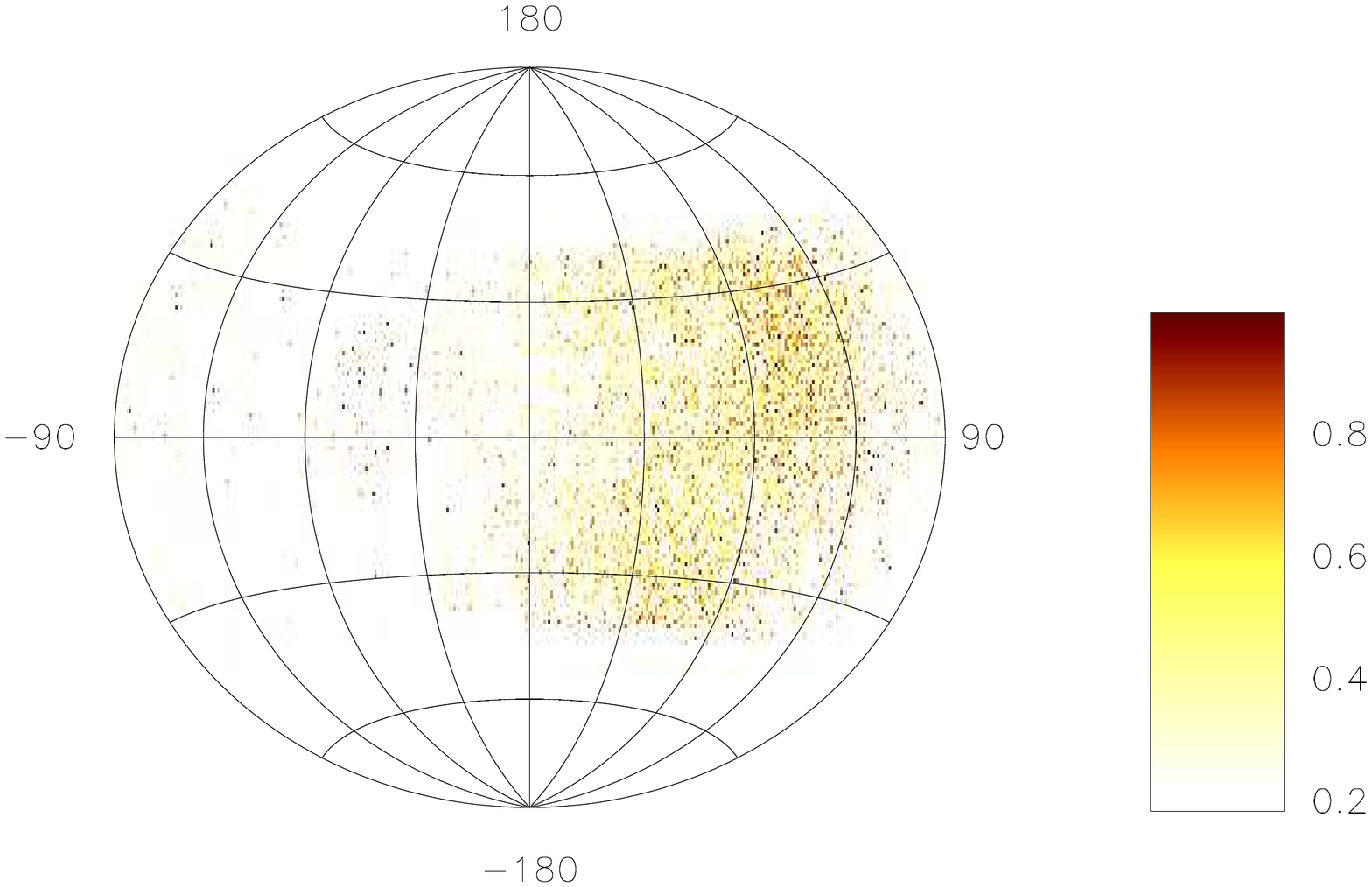}}}&
\\
\end{tabular}
\caption{Top panel only at 333 MHz (upper window) shows the average profile with total
intensity (Stokes I; solid black lines), total linear polarization (dashed red
line) and circular polarization (Stokes V; dotted blue line). Top panel (lower
window) also shows the single pulse PPA distribution (colour scale) along with
the average PPA (red error bars).
The RVM fits to the average PPA (dashed pink
line) is also shown in this plot. Middle panel only at 333 MHz show
the $\chi^2$ contours for the parameters $\alpha$ and $\beta$ obtained from RVM
fits.
Bottom panel only at 333 MHz shows the Hammer-Aitoff projection of the polarized time
samples with the colour scheme representing the fractional polarization level.}
\label{a84}
\end{center}
\end{figure*}


\begin{figure*}
\begin{center}
\begin{tabular}{cc}
{\mbox{\includegraphics[width=9cm,height=6cm,angle=0.]{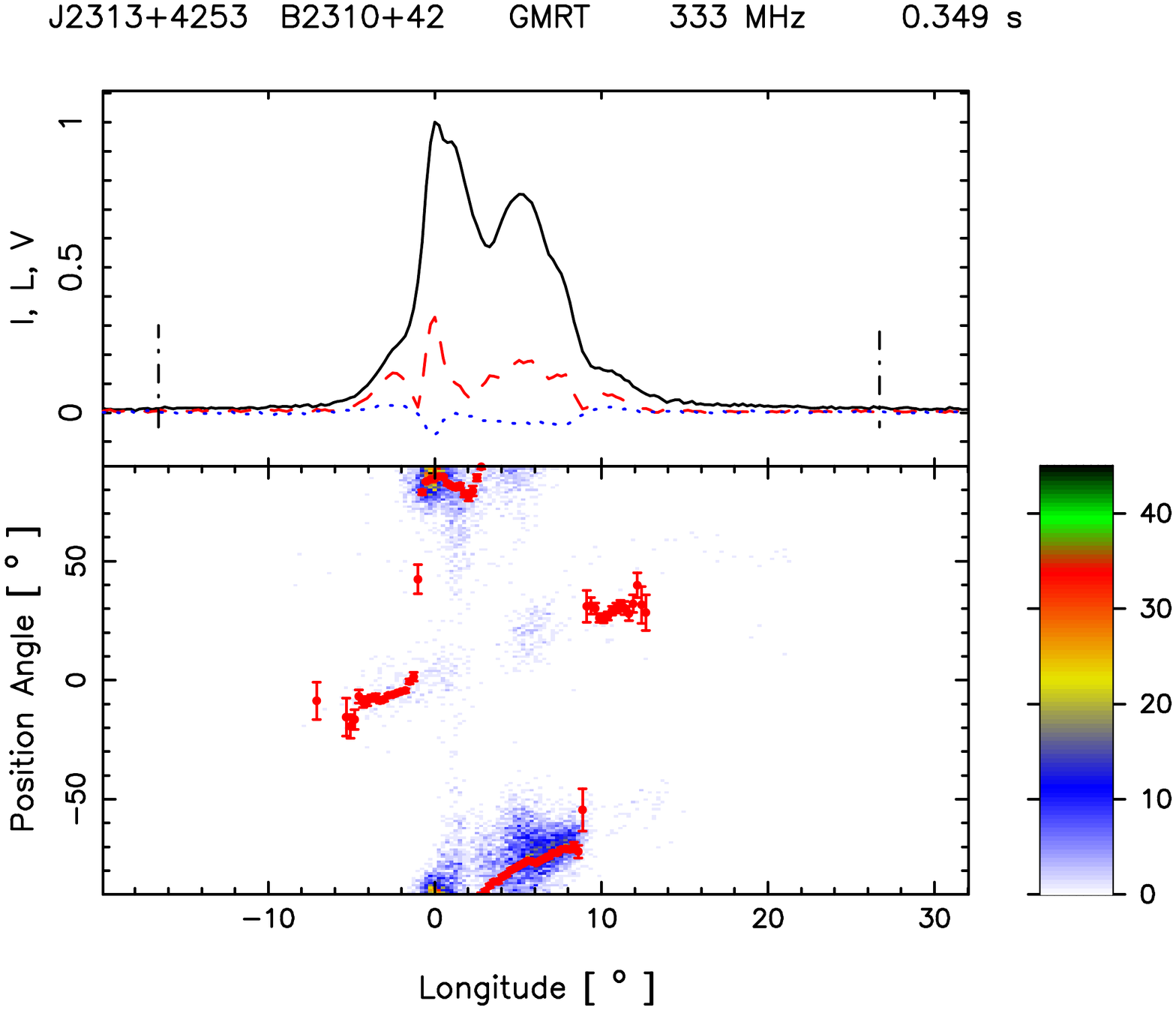}}}&
\\
&
\\
{\mbox{\includegraphics[width=9cm,height=6cm,angle=0.]{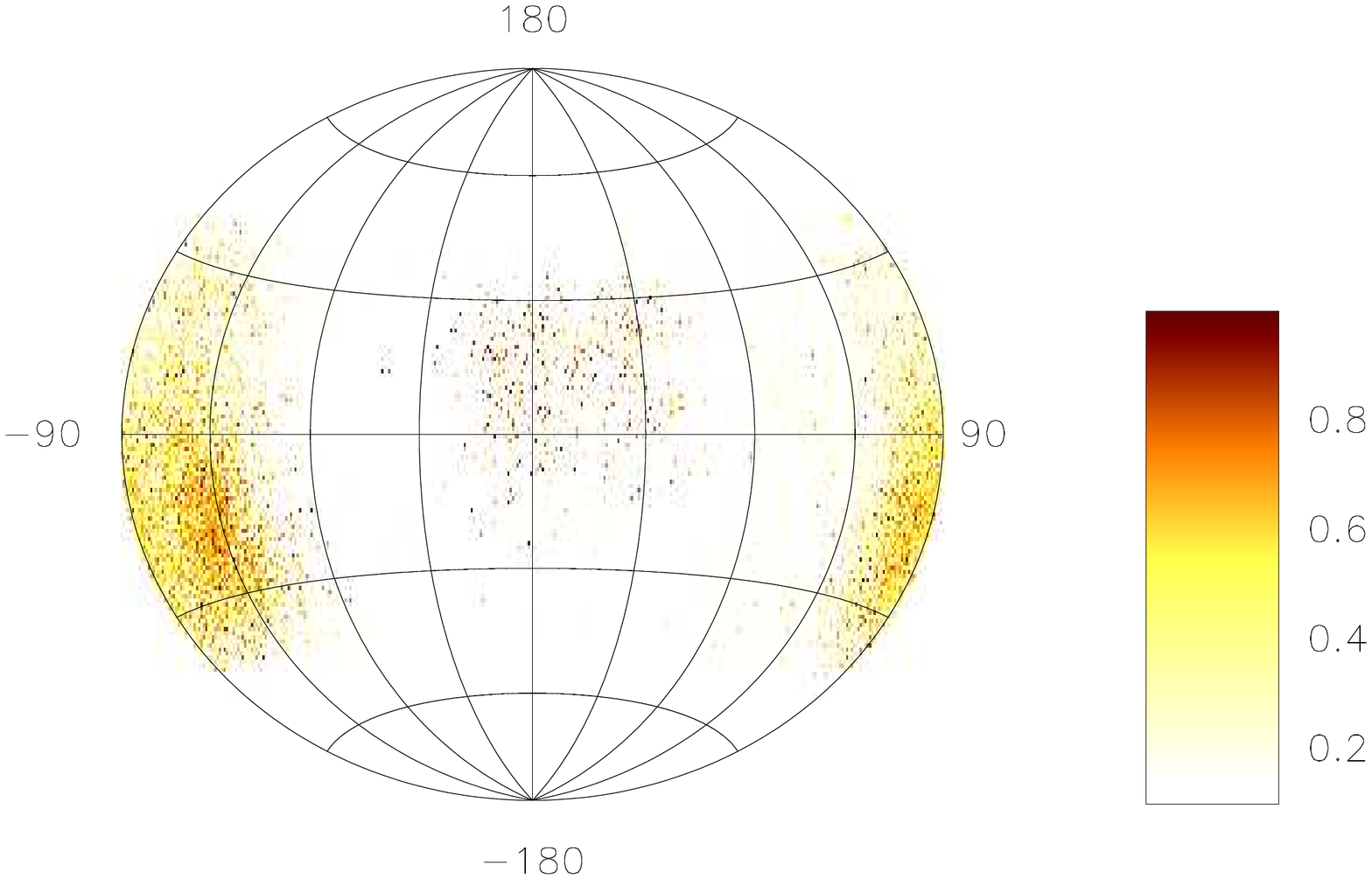}}}&
\\
\end{tabular}
\caption{Top panel only at 333 MHz (upper window) shows the average profile with total
intensity (Stokes I; solid black lines), total linear polarization (dashed red
line) and circular polarization (Stokes V; dotted blue line). Top panel (lower
window) also shows the single pulse PPA distribution (colour scale) along with
the average PPA (red error bars).
Bottom panel only at 333 MHz shows the Hammer-Aitoff projection of the polarized time
samples with the colour scheme representing the fractional polarization level.}
\label{a85}
\end{center}
\end{figure*}


\begin{figure*}
\begin{center}
\begin{tabular}{cc}
{\mbox{\includegraphics[width=9cm,height=6cm,angle=0.]{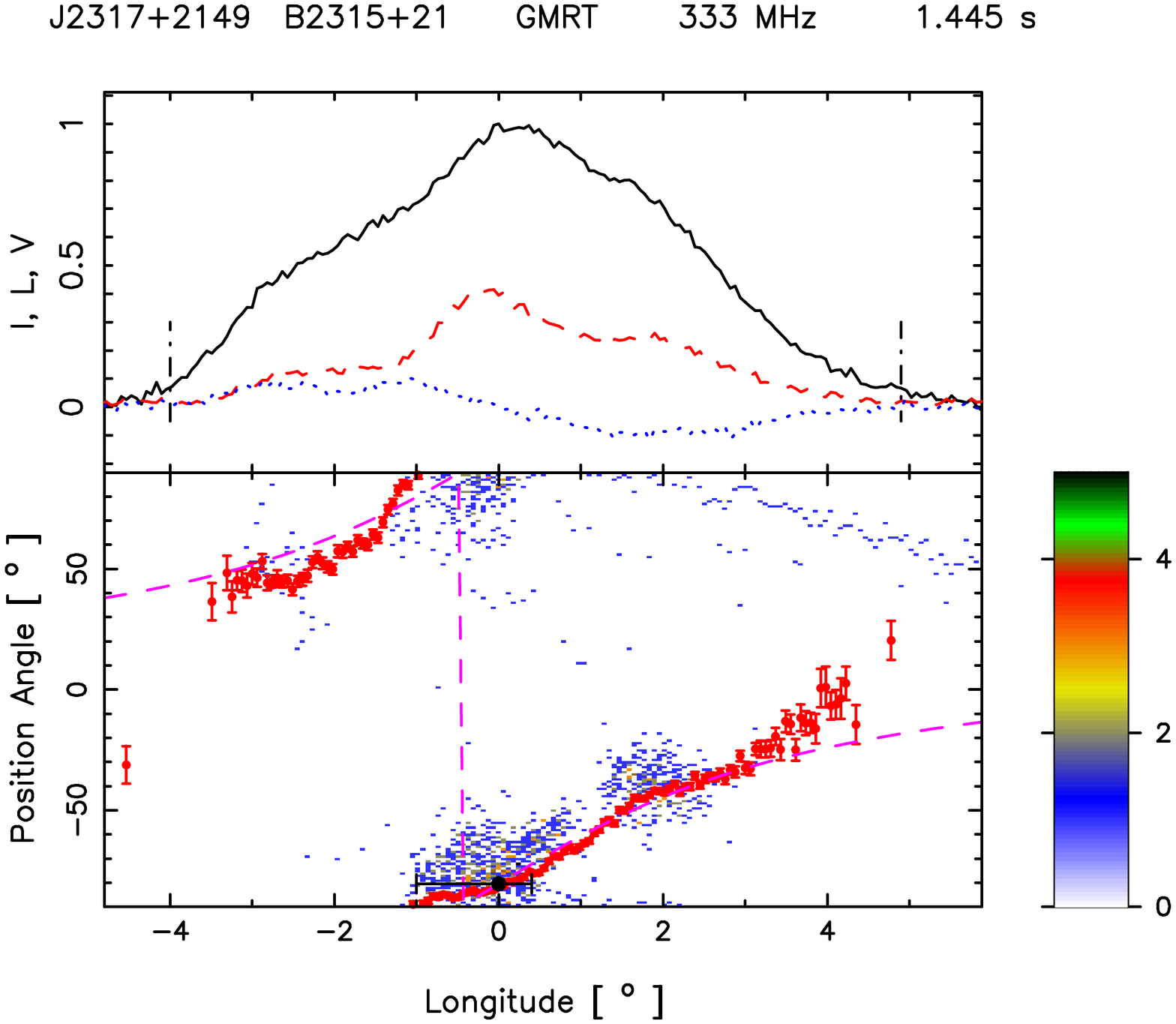}}}&
{\mbox{\includegraphics[width=9cm,height=6cm,angle=0.]{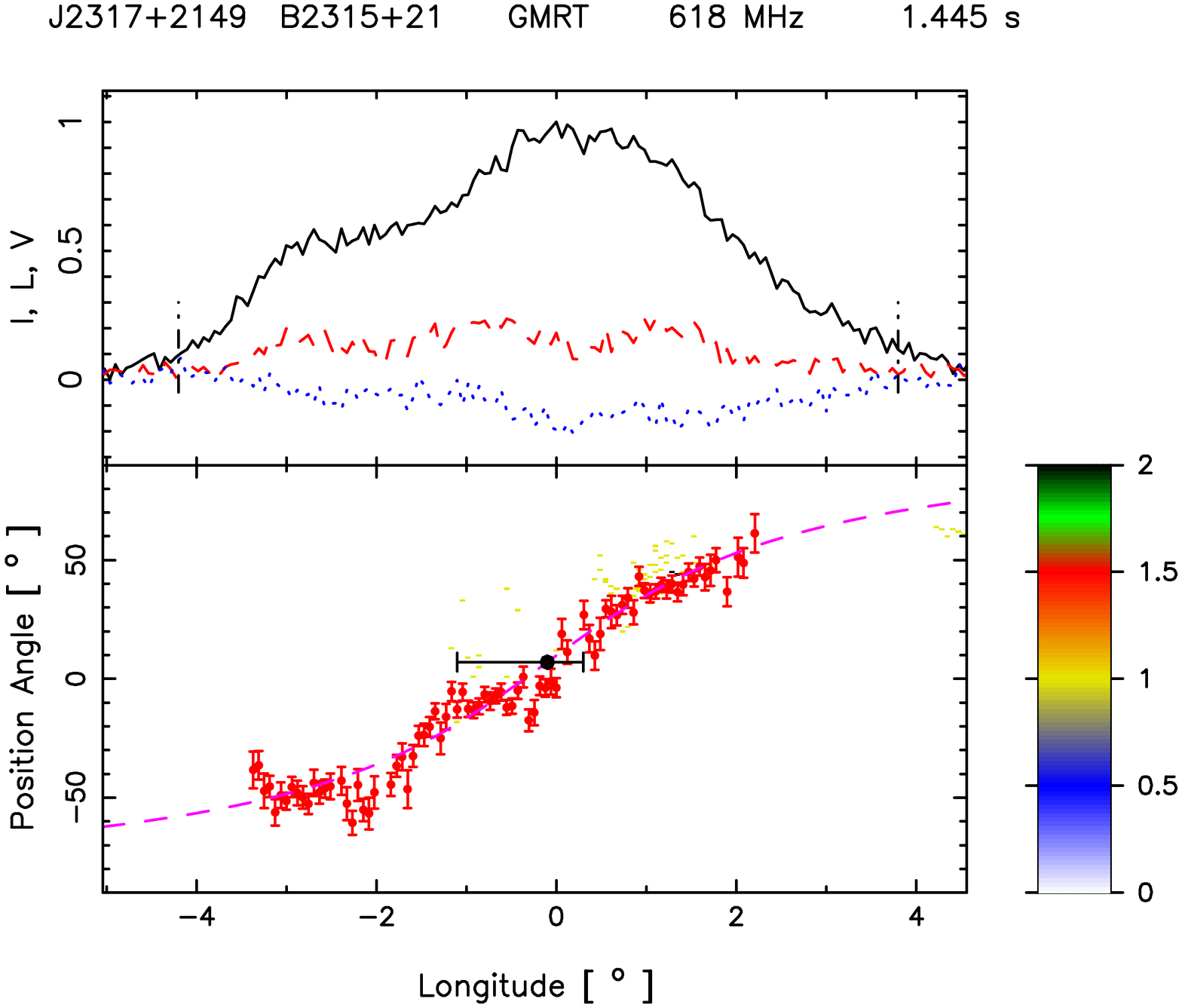}}}\\
{\mbox{\includegraphics[width=9cm,height=6cm,angle=0.]{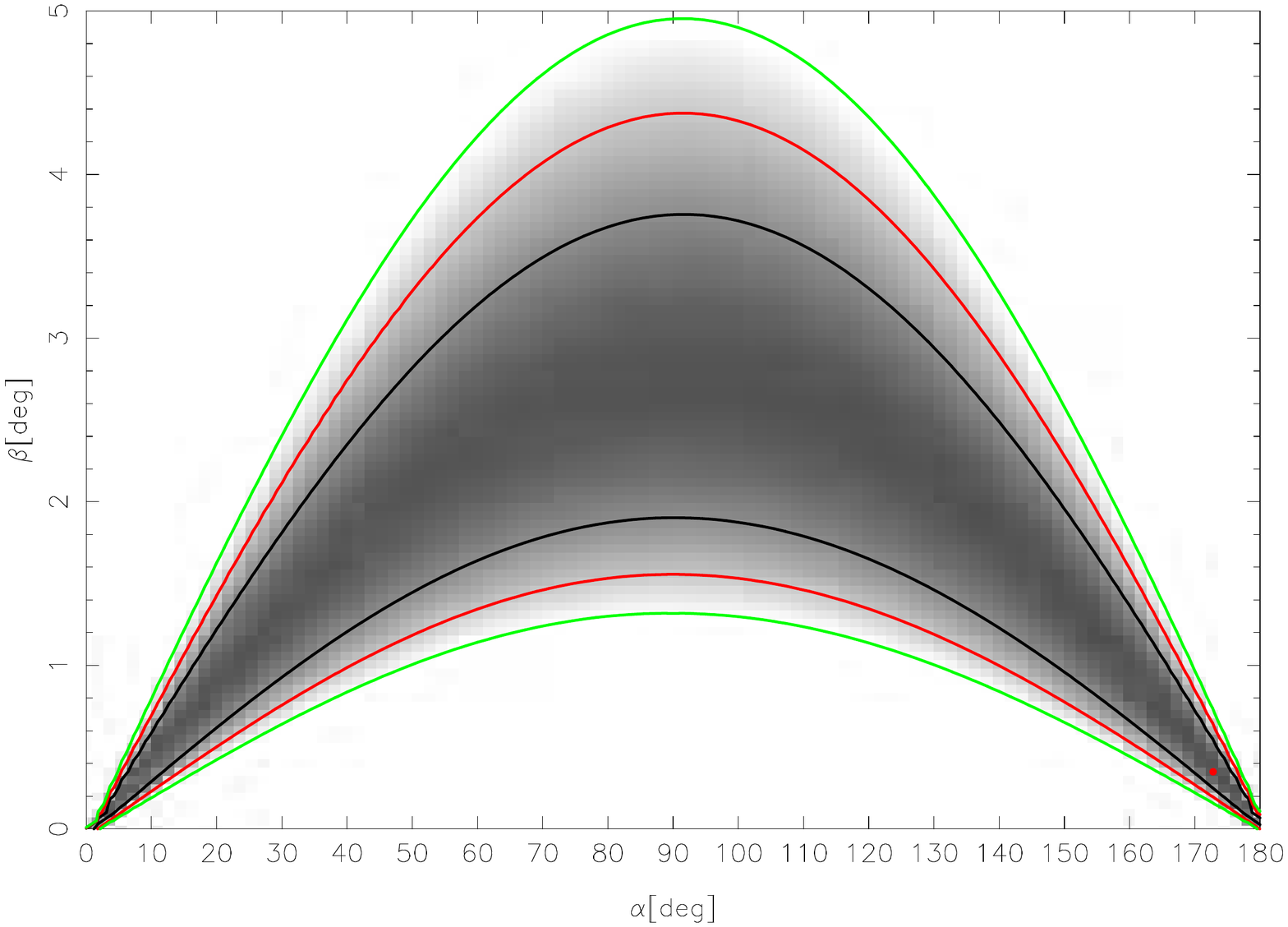}}}&
{\mbox{\includegraphics[width=9cm,height=6cm,angle=0.]{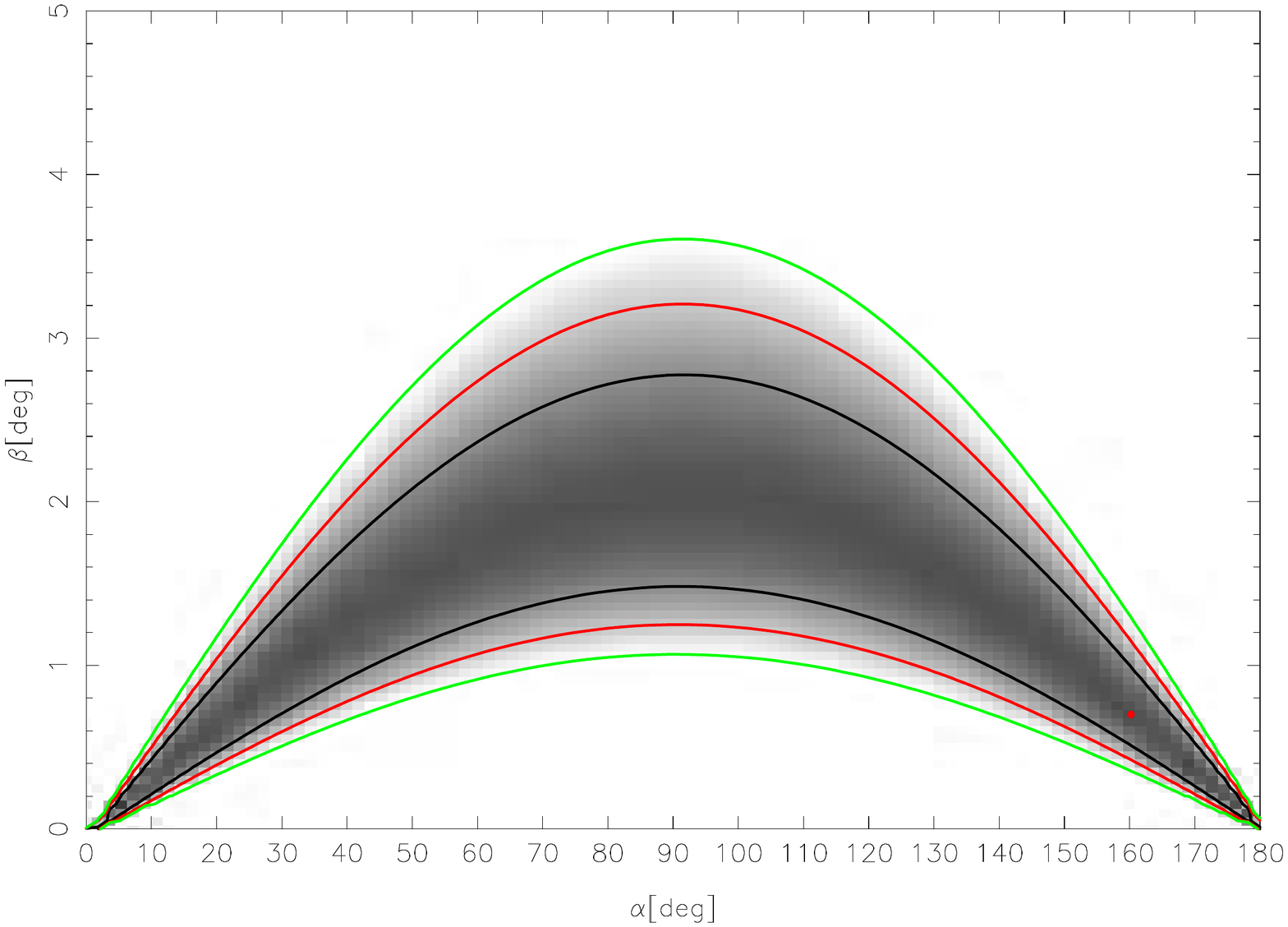}}}\\
{\mbox{\includegraphics[width=9cm,height=6cm,angle=0.]{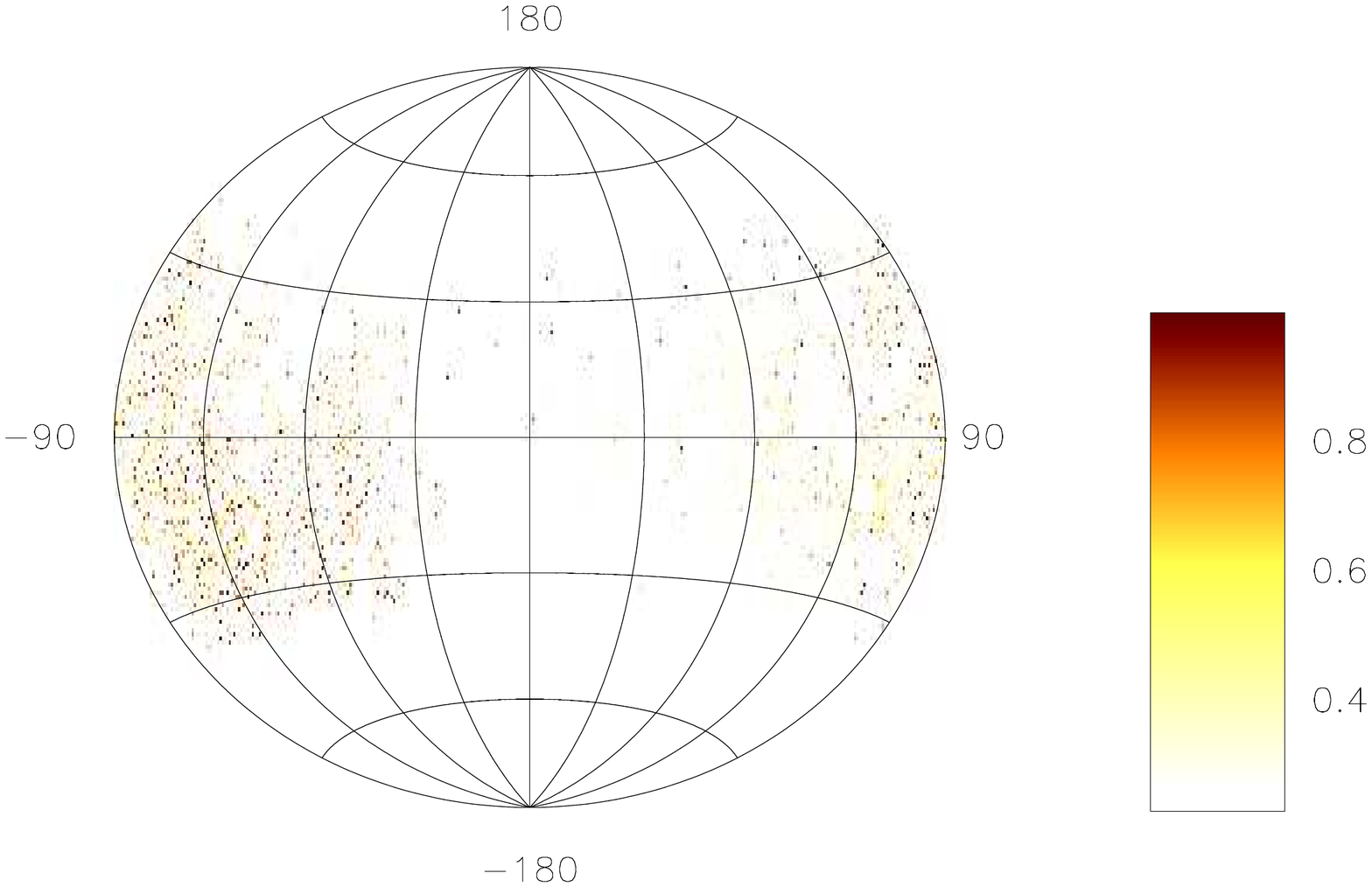}}}&
{\mbox{\includegraphics[width=9cm,height=6cm,angle=0.]{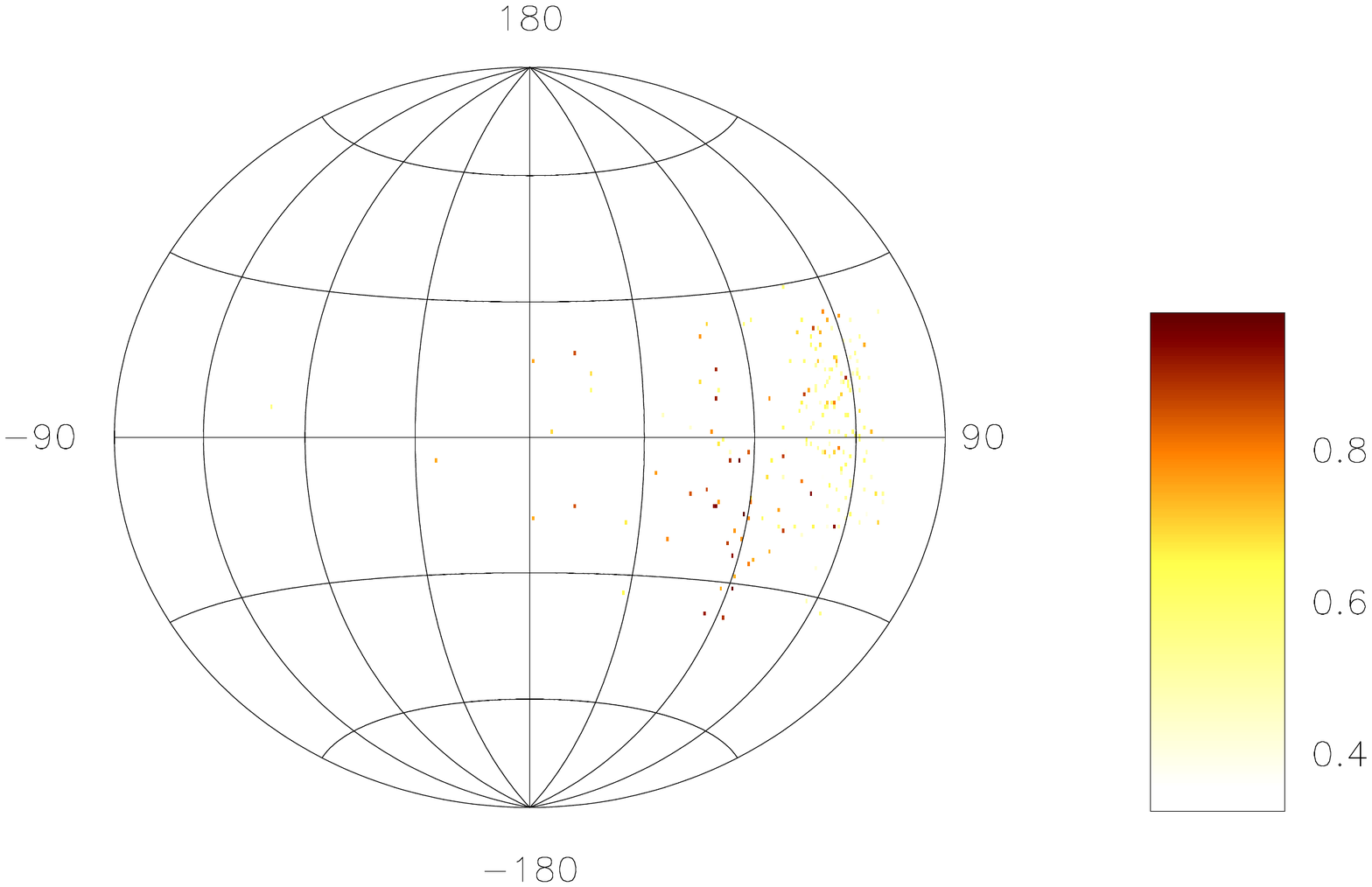}}}\\
\end{tabular}
\caption{Top panel (upper window) shows the average profile with total
intensity (Stokes I; solid black lines), total linear polarization (dashed red
line) and circular polarization (Stokes V; dotted blue line). Top panel (lower
window) also shows the single pulse PPA distribution (colour scale) along with
the average PPA (red error bars).
The RVM fits to the average PPA (dashed pink
line) is also shown in this plot. Middle panel show
the $\chi^2$ contours for the parameters $\alpha$ and $\beta$ obtained from RVM
fits.
Bottom panel shows the Hammer-Aitoff projection of the polarized time
samples with the colour scheme representing the fractional polarization level.}
\label{a86}
\end{center}
\end{figure*}


\begin{figure*}
\begin{center}
\begin{tabular}{cc}
{\mbox{\includegraphics[width=9cm,height=6cm,angle=0.]{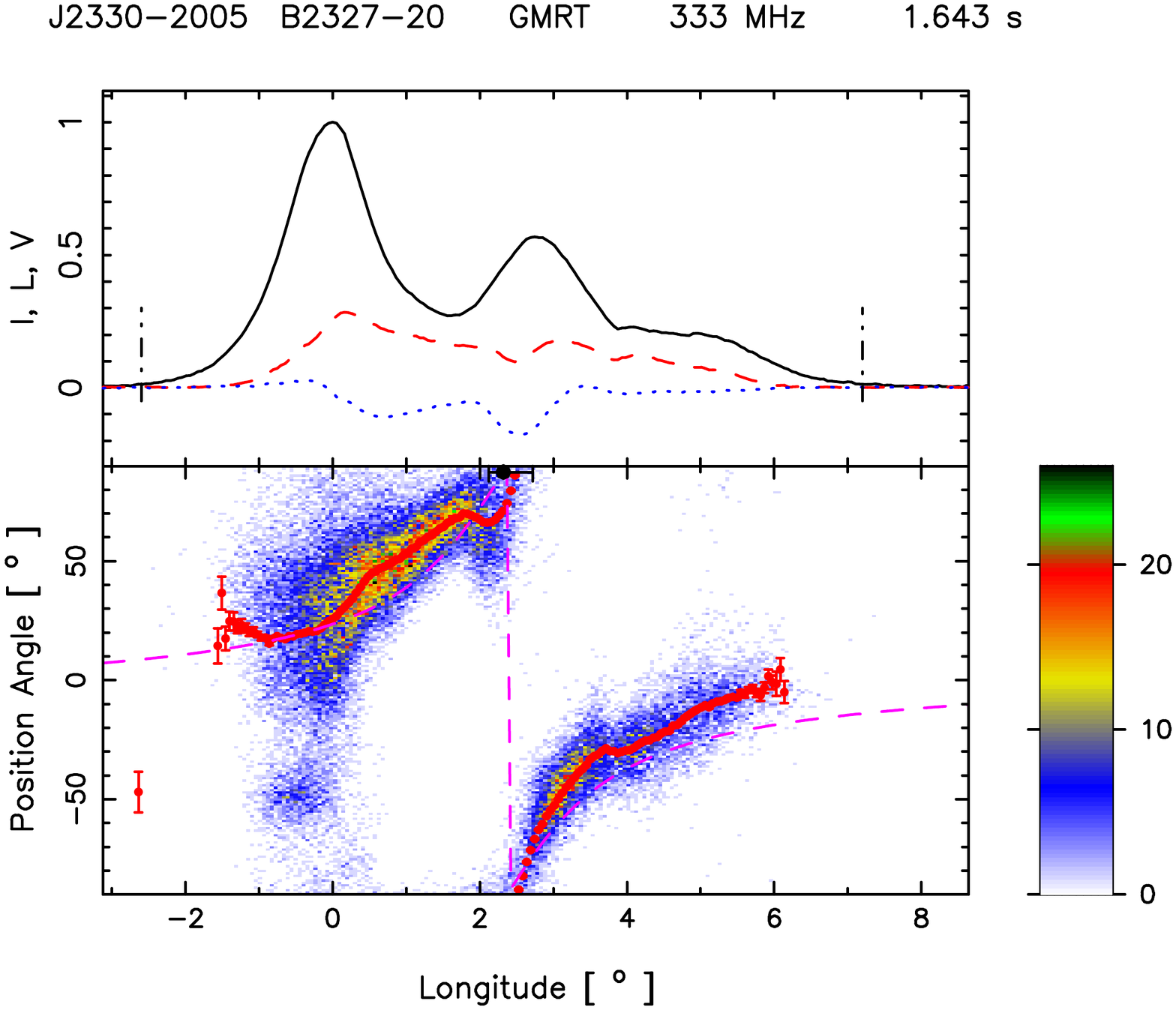}}}&
{\mbox{\includegraphics[width=9cm,height=6cm,angle=0.]{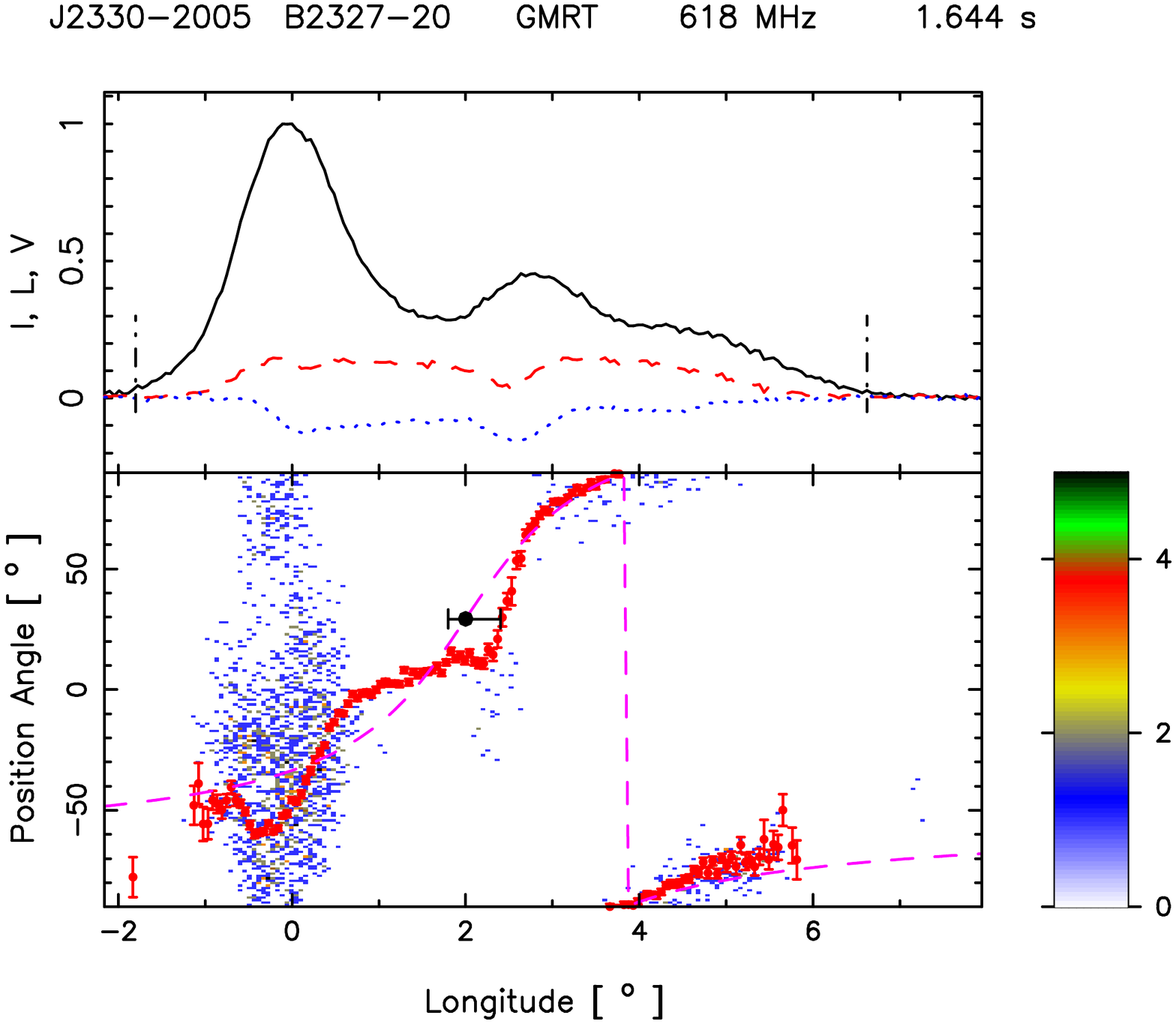}}}\\
{\mbox{\includegraphics[width=9cm,height=6cm,angle=0.]{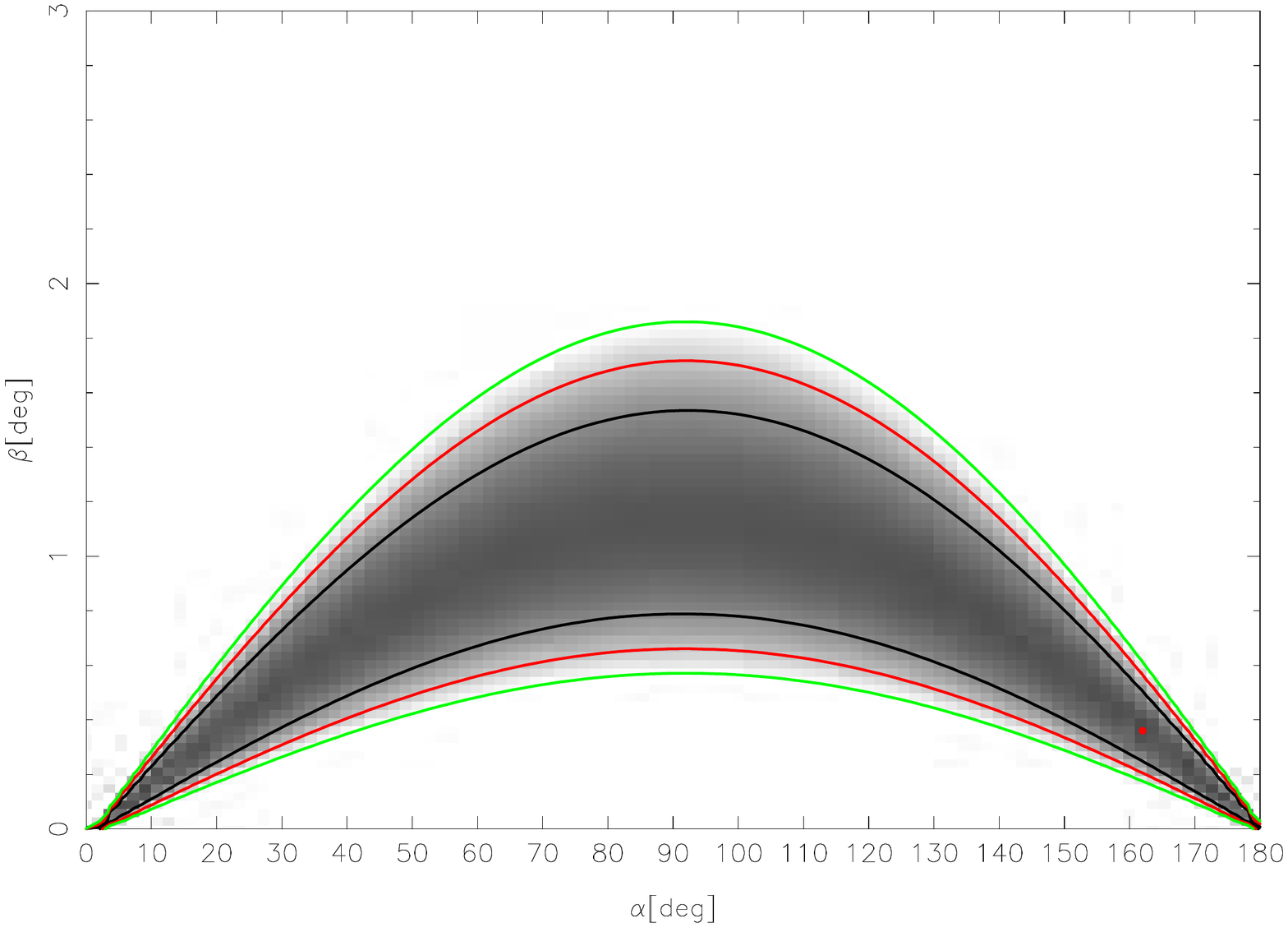}}}&
{\mbox{\includegraphics[width=9cm,height=6cm,angle=0.]{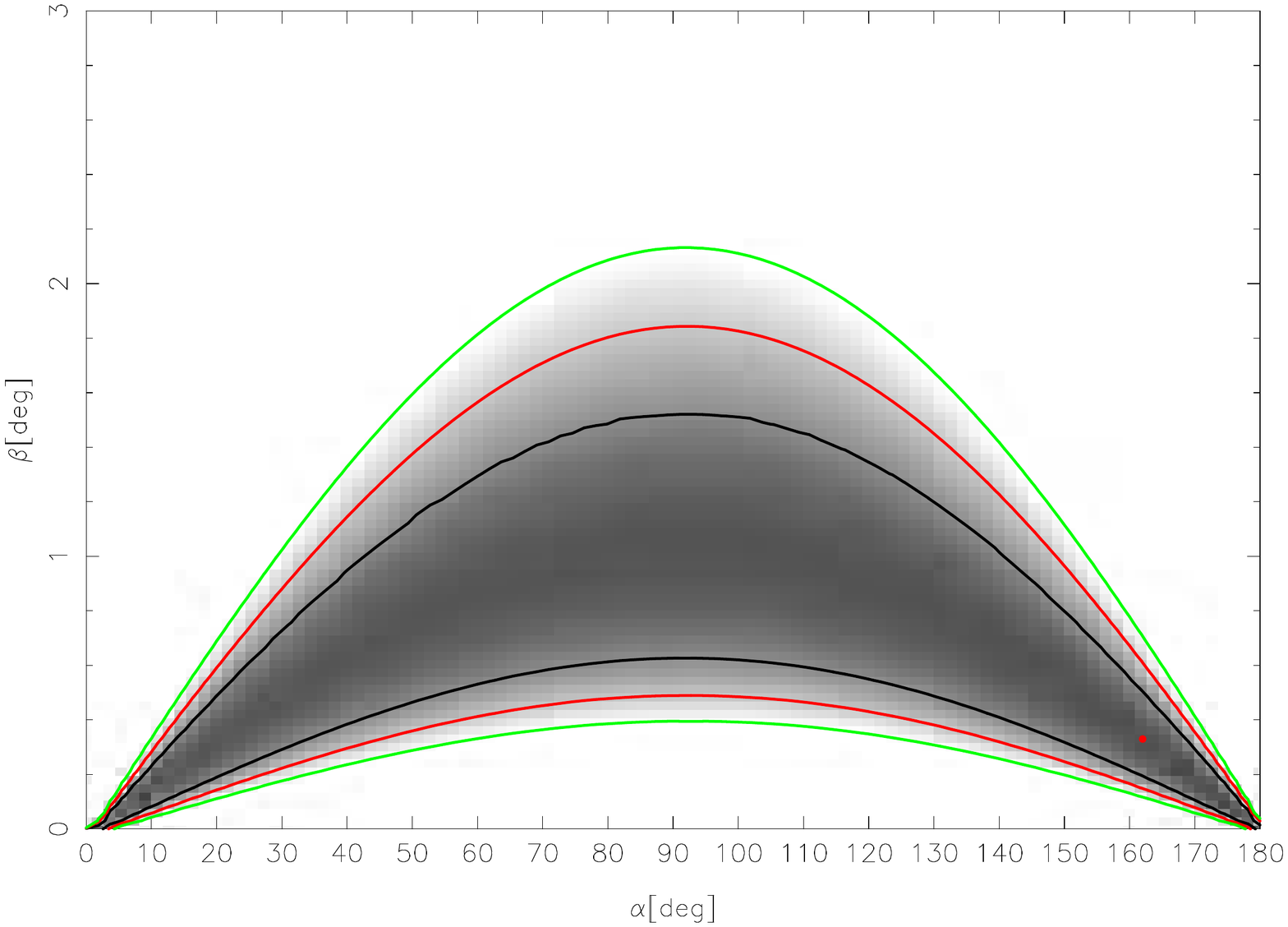}}}\\
{\mbox{\includegraphics[width=9cm,height=6cm,angle=0.]{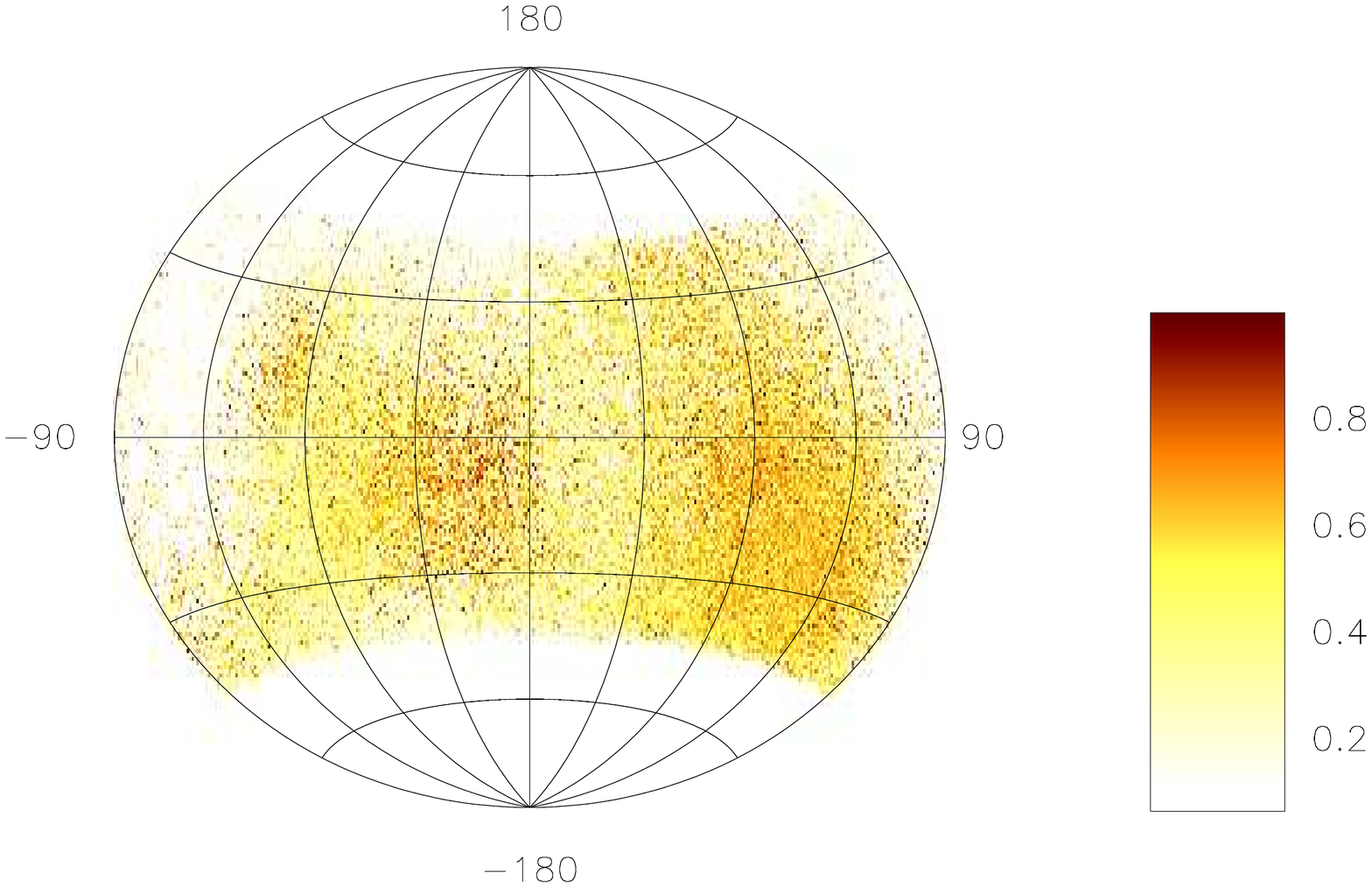}}}&
{\mbox{\includegraphics[width=9cm,height=6cm,angle=0.]{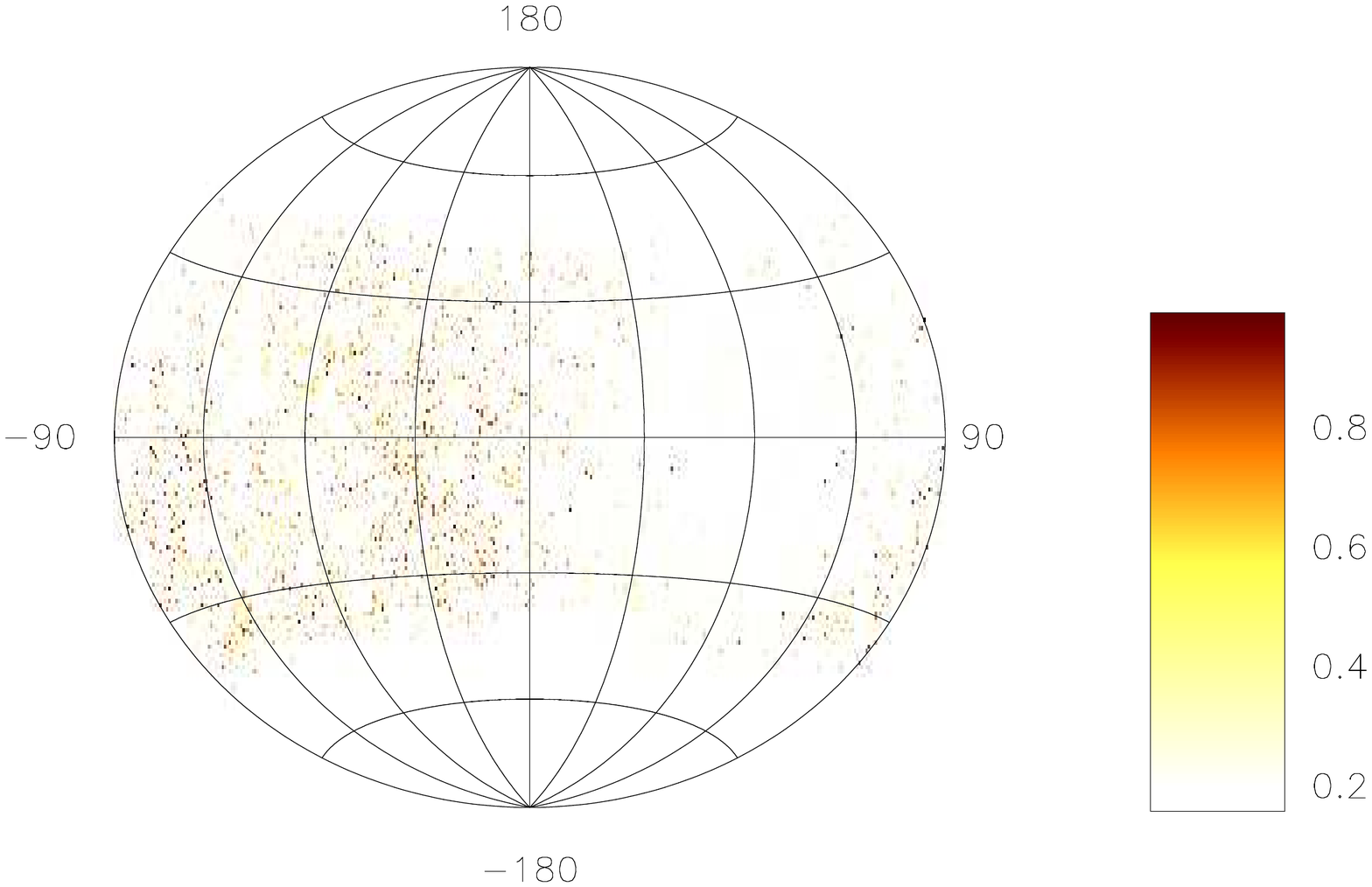}}}\\
\end{tabular}
\caption{Top panel (upper window) shows the average profile with total
intensity (Stokes I; solid black lines), total linear polarization (dashed red
line) and circular polarization (Stokes V; dotted blue line). Top panel (lower
window) also shows the single pulse PPA distribution (colour scale) along with
the average PPA (red error bars).
The RVM fits to the average PPA (dashed pink
line) is also shown in this plot. Middle panel show
the $\chi^2$ contours for the parameters $\alpha$ and $\beta$ obtained from RVM
fits.
Bottom panel shows the Hammer-Aitoff projection of the polarized time
samples with the colour scheme representing the fractional polarization level.}
\label{a87}
\end{center}
\end{figure*}


\begin{figure*}
\begin{center}
\begin{tabular}{cc}
{\mbox{\includegraphics[width=9cm,height=6cm,angle=0.]{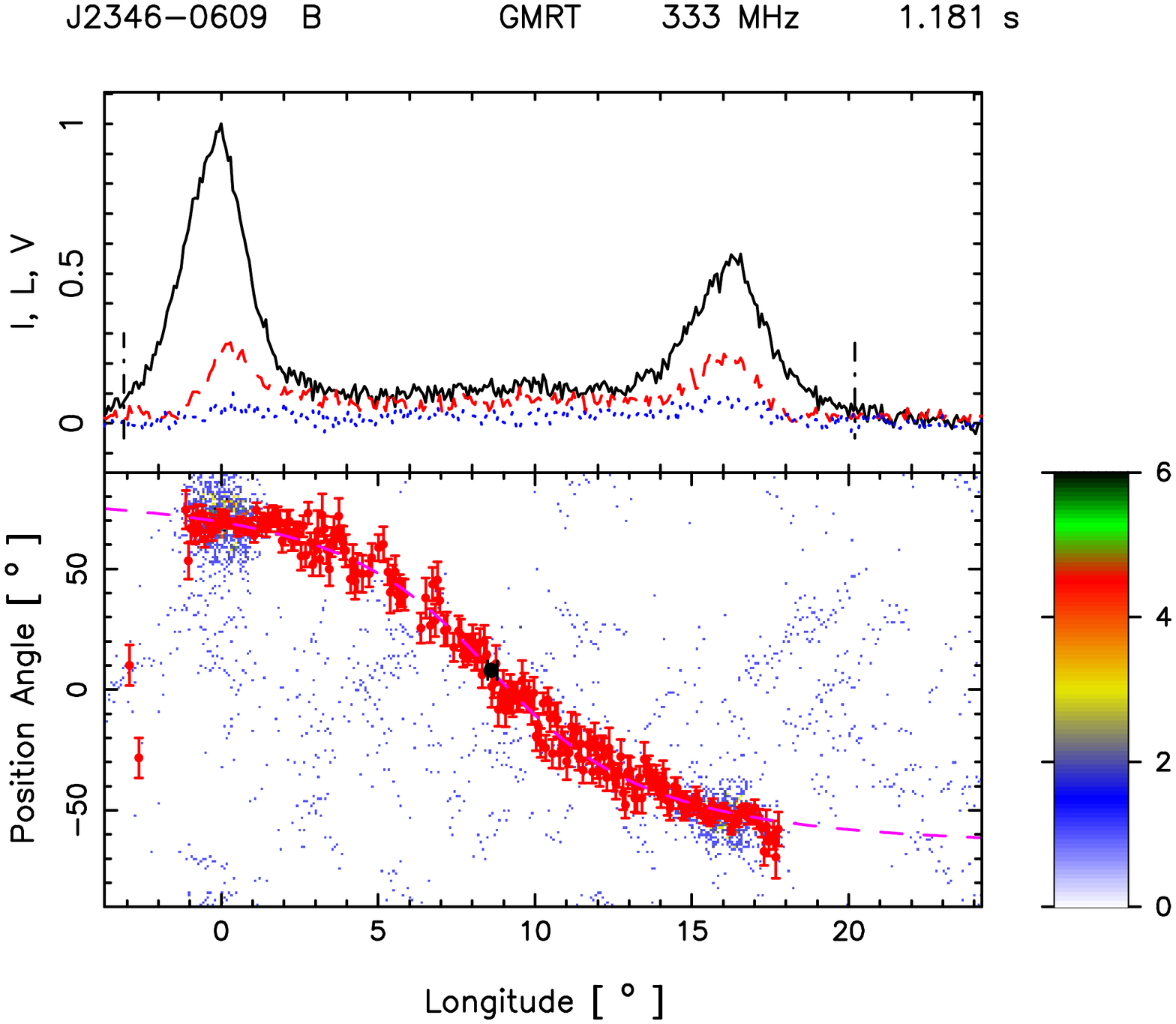}}}&
{\mbox{\includegraphics[width=9cm,height=6cm,angle=0.]{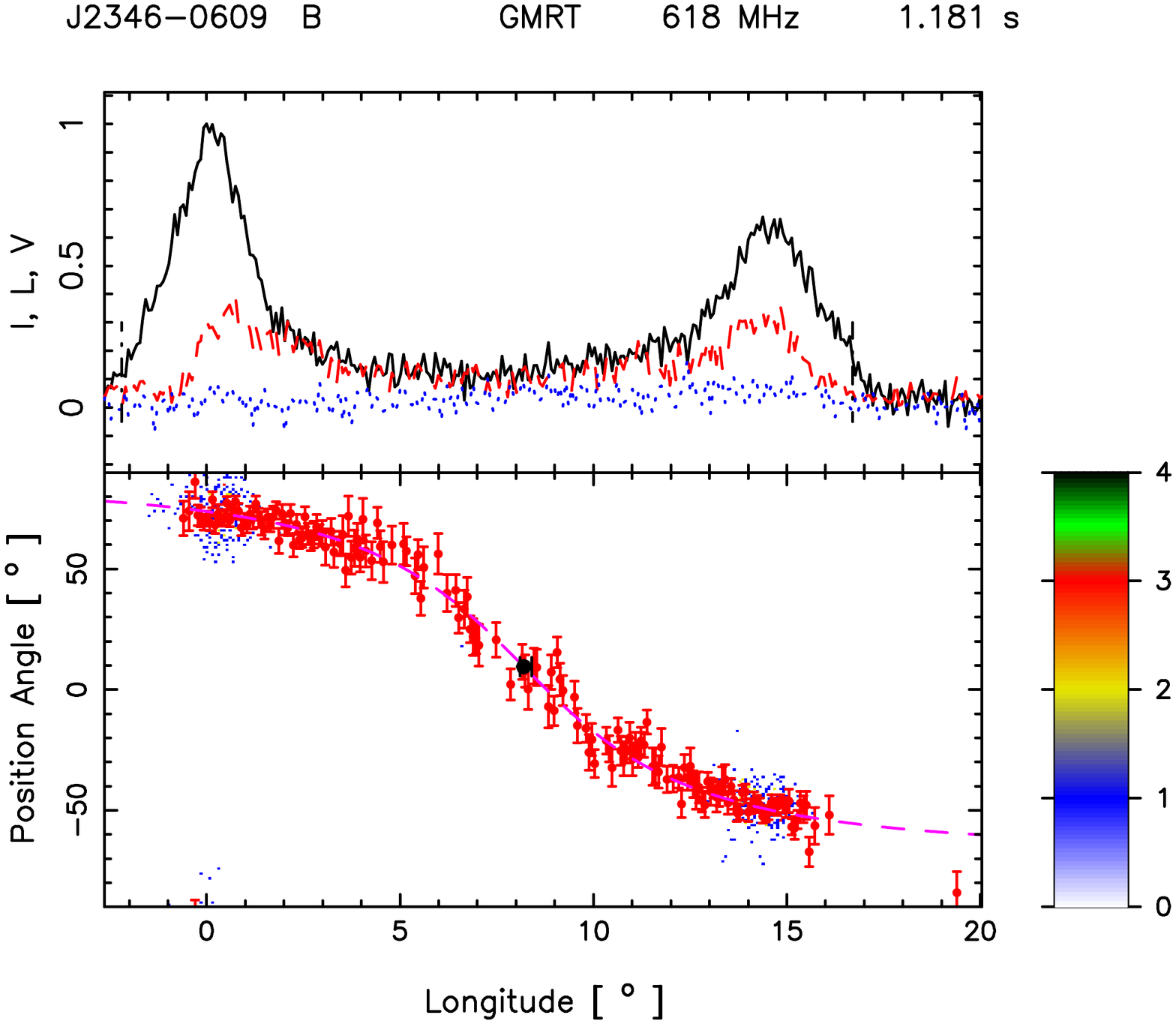}}}\\
{\mbox{\includegraphics[width=9cm,height=6cm,angle=0.]{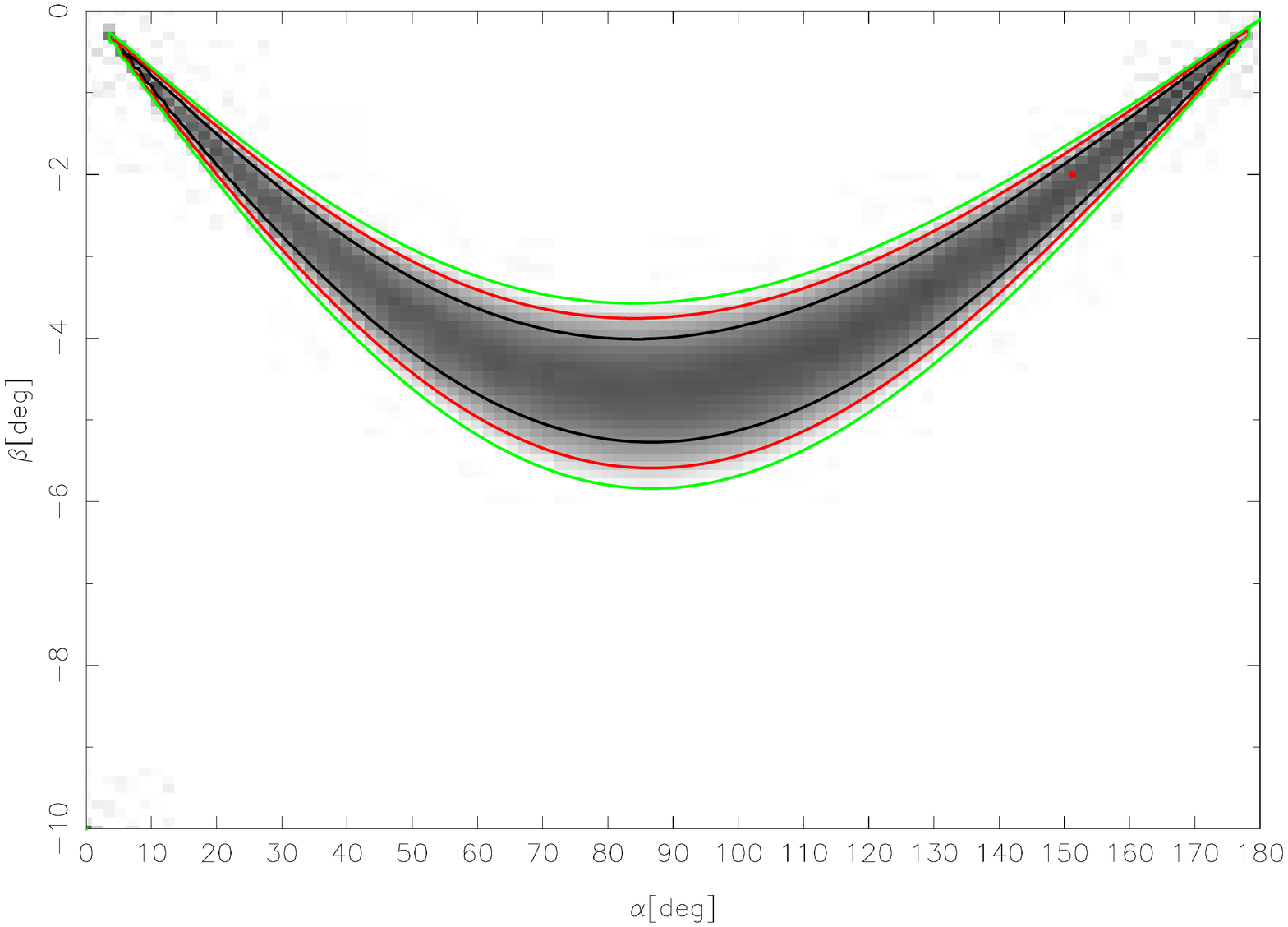}}}&
{\mbox{\includegraphics[width=9cm,height=6cm,angle=0.]{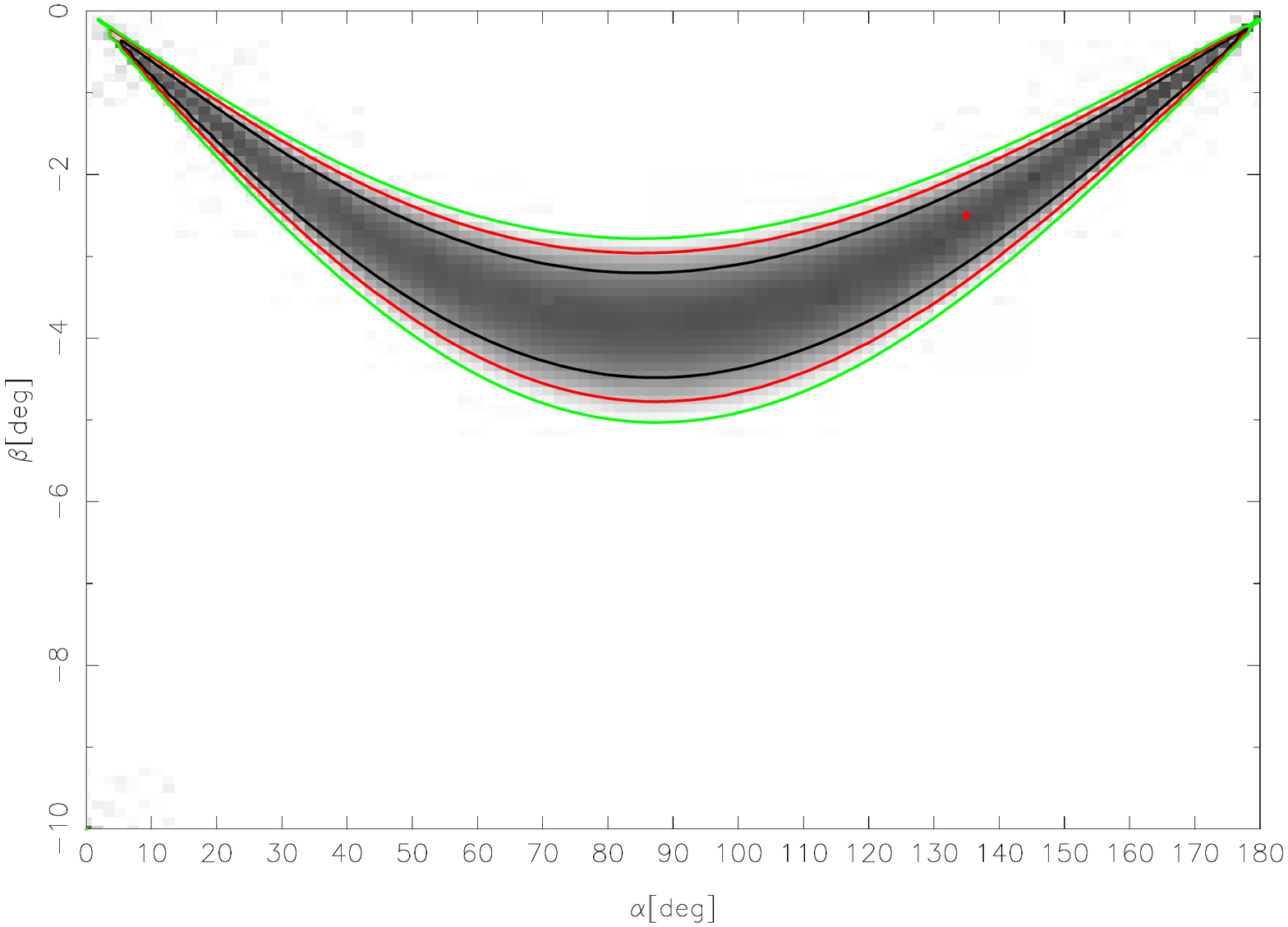}}}\\
{\mbox{\includegraphics[width=9cm,height=6cm,angle=0.]{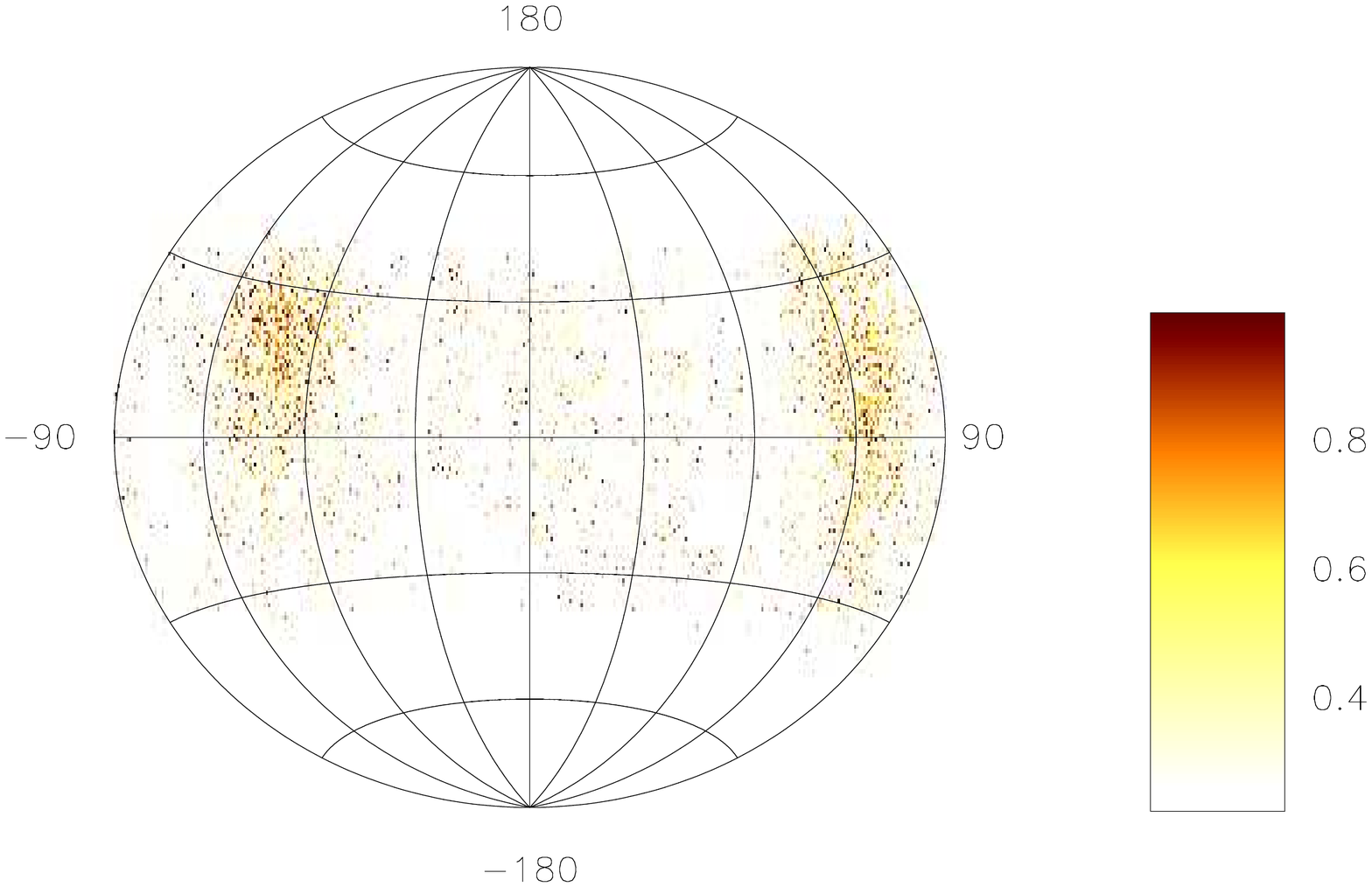}}}&
{\mbox{\includegraphics[width=9cm,height=6cm,angle=0.]{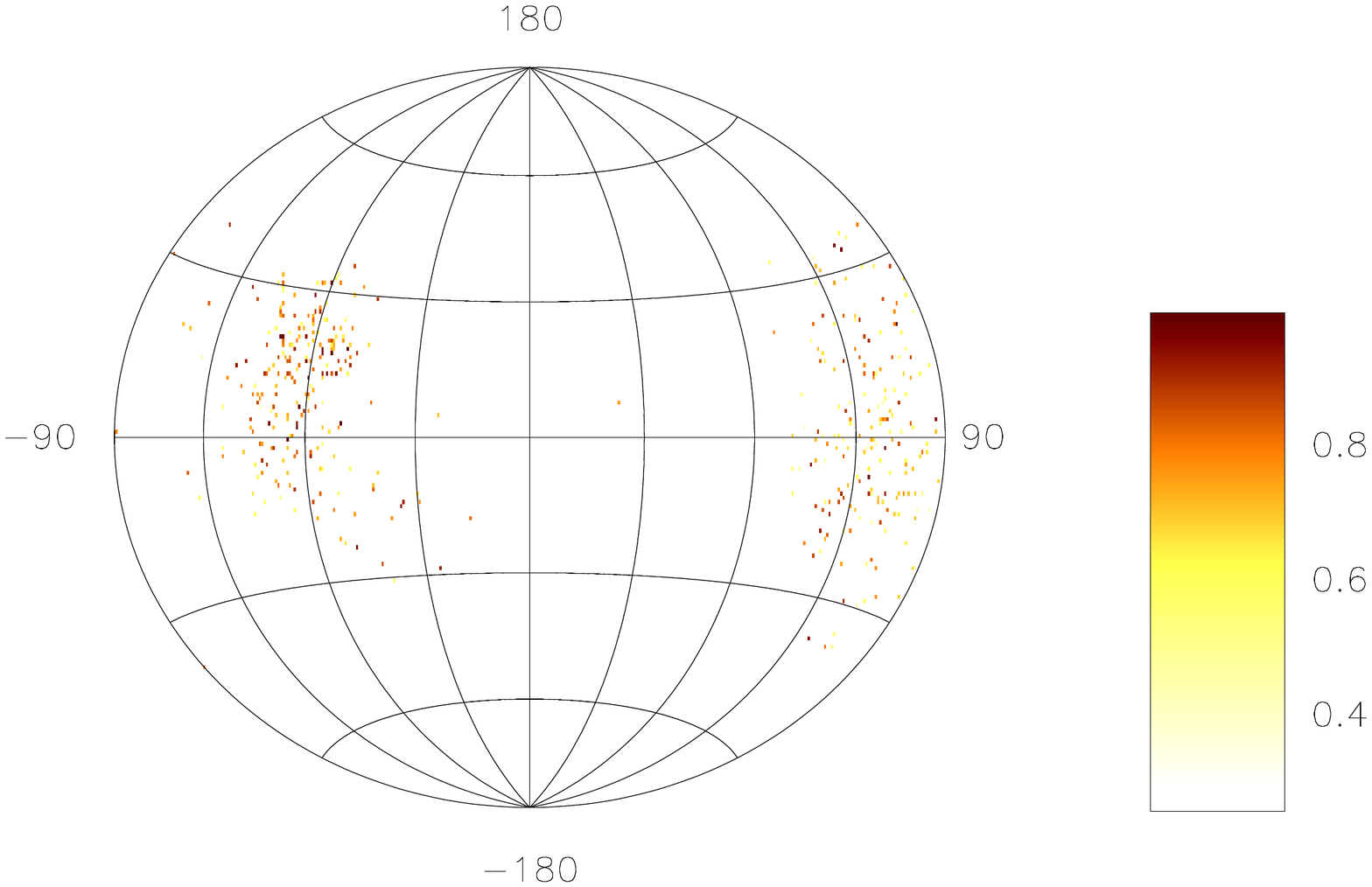}}}\\
\end{tabular}
\caption{Top panel (upper window) shows the average profile with total
intensity (Stokes I; solid black lines), total linear polarization (dashed red
line) and circular polarization (Stokes V; dotted blue line). Top panel (lower
window) also shows the single pulse PPA distribution (colour scale) along with
the average PPA (red error bars).
The RVM fits to the average PPA (dashed pink
line) is also shown in this plot. Middle panel show
the $\chi^2$ contours for the parameters $\alpha$ and $\beta$ obtained from RVM
fits.
Bottom panel shows the Hammer-Aitoff projection of the polarized time
samples with the colour scheme representing the fractional polarization level.}
\label{a88}
\end{center}
\end{figure*}



\end{document}